# InnerSource Circumplex Model: Mapping Cross-organizational Developer Collaboration Patterns with Insights from Japanese Corporate Experience

by

HATTORI, Yuki

A Master's Thesis

Submitted to Graduate School of Business

Department of Business Administration

Aoyama Gakuin University

In Fulfillment of the Requirements

for the Degree of Master of Arts in Business Administration

January 2025

# Abstract

In today's business environment, where the insourcing of software development and cross-organizational code sharing are becoming increasingly critical, the concept of "InnerSource"—applying open-source development methodologies within a corporate setting—has garnered significant attention. The purpose of this study is to elucidate both the process of introducing InnerSource and the manner in which it evolves within companies. First, a comparative analysis of Japanese and global enterprises highlights differences in the state of software sharing, perceptions of its importance, and barriers to implementation. In particular, delayed insourcing and organizational silos—commonly observed in Japanese firms—are extracted and empirically verified as factors that hinder early InnerSource adoption.

Next, this study demonstrates that InnerSource adoption involves multi-layered, topological evolution beyond conventional staged models of program evolution. The research proposes three theoretical frameworks: InnerSource Topologies, which conceptualizes collaborative structures and categorizes internal collaboration levels; the Multi-layered Incentive Model, which combines monetary and non-monetary rewards at individual and project levels; and the InnerSource Circumplex Model, which helps organizations define InnerSource forms based on their specific needs. By mapping InnerSource evolution as a circumplex rather than simple staged progression, leaders can better adjust their focus during implementation. These frameworks help refine the previously ambiguous concept of InnerSource from the perspectives of sharing scope and community growth.

These findings reaffirm that successful InnerSource adoption requires the parallel pursuit of top-down program structuring and bottom-up voluntary collaboration. They also contribute to fostering a sustainable innovation culture and enhancing software-sharing practices within enterprises. Furthermore, the newly proposed frameworks, particularly the Circumplex Model, offer versatile guidelines for organizations of varying cultural backgrounds and scales, enabling them to flexibly redefine and introduce InnerSource. This research is thus expected to advance corporate software sharing and spur innovation in diverse industrial contexts.

# Table of Content

# Preface

## 1 Introduction

In today's global business environment, code sharing and reuse in software development are increasingly regarded as pivotal strategic elements that contribute to more efficient development processes, higher product quality, and the creation of innovations. In particular, the use of open source and an internally open source-like approach known as "InnerSource" has been widely adopted, especially among multinational corporations. InnerSource refers to the application of open source development methods within an organization, thereby establishing software development processes that are highly transparent, reusable, and collaborative, ultimately supporting organizational learning and fostering innovation. According to Gartner's analysis, the implementation of InnerSource helps organizations avoid the pitfalls of siloed structures and achieve a more robust and tightly integrated software development lifecycle [1]. Indeed, in Gartner's "Hype Cycle for Software Engineering" (2023), InnerSource was highlighted as a major trend [2], reflecting its growing prominence in the industry.

Internationally, knowledge sharing and code reuse in software development are recognized as key drivers of productivity and innovation. Nevertheless, due to differences in organizational characteristics and cultural contexts, the degree of progress and the methods of implementation vary significantly by region and corporate group.

This study focuses on the implementation process of InnerSource—an approach that integrates open source methodologies within organizations—and empirically examines how different organizational characteristics and cultural backgrounds influence software sharing behavior. Specifically, by using Japanese corporations as case studies and comparing them with global enterprises, this research aims to elucidate the mechanisms by which institutional factors such as organizational culture, decision-making processes, and employment practices affect code sharing behavior.

Through empirical research employing predominantly qualitative methods, this study systematically analyzes the effects of organizational characteristics and cultural contexts on code sharing behavior, identifying both the factors that promote and hinder this behavior. Based on these findings, the study ultimately aims to construct a theoretical framework enabling companies with diverse organizational traits and cultural backgrounds to cultivate an effective code-sharing culture that leverages their unique attributes.

From an academic perspective, the significance of this research lies in extending and refining existing theoretical frameworks in the interdisciplinary field that spans organizational behavior, technology management, and software engineering by empirically analyzing the unique organizational and cultural contexts specific to each company. Moreover, in terms of practical implications, the proposed framework is expected to offer concrete insights for various organizations to effectively introduce and practice InnerSource, thereby taking advantage of their specific organizational characteristics.

## 1.1 Background – The Current Business Environment Surrounding Software

In recent years, as digital transformation (DX) progresses across various industries—including manufacturing—software has come to be positioned as a core source of competitive advantage. This trend reflects the widespread recognition that responding rapidly to changing customer needs, creating new business models, and optimizing business processes are all achievable through software. However, it has been pointed out that Japanese companies still lag behind in internalizing (i.e., insourcing) their software development. According to a survey by the Information-technology Promotion Agency (IPA) of Japan, 73.6% of Japan's IT professionals work for IT companies such as system integrators, whereas in the United States only 35.1% of IT professionals are employed in IT firms [3]. These figures suggest that non-IT companies in Japan have limited in-house engineering teams.

In response to this situation, large corporations have been spearheading a move toward insourcing; indeed, a survey by the same IPA showed that 40.4% of companies with over 1,001 employees reported that they are "promoting insourcing" [4]. Some of these companies have attempted to adopt speedy and flexible development approaches, such as agile methodologies and continuous delivery, which are already widely used overseas. However, in more traditional industries like manufacturing, the introduction of these new development methods faces various constraints, and the conventional waterfall development method remains deeply entrenched.

Moreover, the cross-organizational sharing of software technologies and components nurtured within a company is expected to enhance product competitiveness by avoiding redundant development and shortening development times. However, a serious challenge called "siloing" often arises, where organizational barriers hinder effective collaboration. When inter-departmental collaboration is insufficient and information asymmetry exists, different departments may redundantly redevelop the same functionalities (commonly known as “reinventing the wheel”), resulting in increased costs and lower development efficiency. These drawbacks are particularly pronounced in large enterprises, posing risks that undermine transparency and co-creation within the organization.

To address such organizational issues, various approaches have been proposed in the software industry concerning cultural reforms and developer work practices. One prominent example is the concept of “developer experience (DevEx),” which has garnered substantial attention. Developer experience refers to a holistic concept capturing the qualitative aspects of the environment, tools, processes, and culture that software engineers encounter within an organization. The SPACE framework provides an integrated method for measuring developer productivity, satisfaction, and effectiveness, and it highlights that excellent developer experience is crucial for boosting productivity, fostering innovation, and attracting and retaining top engineering talent [5].

Given this backdrop, reassessing organizational culture and structures to build a product development system that is agile, transparent, and well-equipped with in-house capabilities is believed to contribute to optimizing the developer experience as well as improving talent acquisition and employee engagement. Here, practicing InnerSource—whereby the principles and methods of open source are adopted internally—may prove extremely advantageous for companies looking to establish a shared base of code and knowledge and to encourage collaboration across teams and departments.

The adoption of InnerSource is steadily advancing across a wide range of global industries, including traditional sectors such as manufacturing and finance. This trend signifies the high degree of compatibility between the flexible collaboration model offered by InnerSource and the needs of companies striving to enhance efficiency and establish innovative development methods, thereby fostering a global software development ecosystem. Japanese companies also have the potential to benefit from introducing InnerSource, and their case studies may provide valuable insights for initiatives in other regions.

In this context, InnerSource goes beyond being merely a technical framework to function as a catalyst for organizational transformation. Particularly for the structural challenges faced by Japanese companies—such as slow progress in insourcing and siloed organizational structures—InnerSource offers a practical solution. By making development processes more transparent, promoting knowledge sharing, and fostering cross-organizational collaboration, it is expected to contribute to building a sustainable foundation for innovation.

## 1.2 The Conceptual Framework of InnerSource

In this study, the core focus is on InnerSource, which involves adopting the principles and methods of the open source development model (collaborative, transparent, and distributed) within corporate software development processes. Traditionally, open source software has been developed by voluntary communities spanning geographic and organizational boundaries, and its outputs have been accumulated worldwide as intellectual assets available for reuse. In contrast, InnerSource applies these principles to the internal environment of a company, aiming to break down “silos” between business units and organizational units, facilitate the sharing of code and knowledge, and strengthen development efficiency, quality, and insourcing capabilities.

The concept was first referenced by Tim O’Reilly in 2000 under the term “inner sourcing,” introducing the idea of applying open source development methods within corporations [6]. Over time, it has evolved through the efforts of organizations such as The InnerSource Commons Foundation and pioneering corporate practitioners. While a strict academic definition of InnerSource has yet to be established, The InnerSource Commons Foundation presents four core guiding principles [7]:

- **Openness**: Internal projects should be made “discoverable” across the entire internal community, supported by sufficient documentation, and easily accessible without relying on specific departments. This environment allows developers to voluntarily contribute and complement each other’s knowledge.
- **Transparency**: Project teams should clearly disclose project visions, unresolved feature requirements, progress, and decision-making processes so that contributors from other departments can actively and meaningfully participate.
- **Prioritized Mentorship**: The hosting team should provide appropriate training and mentoring to help guest contributors fully understand the project’s codebase and design principles, thereby enabling them to smoothly make modifications. This fosters individual developers’ growth and skill enhancement.

- **Voluntary Code Contribution**: Participation in InnerSource should not be mandatory. Developers and teams contribute autonomously, generating voluntary collaboration and enabling sustainable community building over the long term.

As these principles suggest, InnerSource goes beyond simply sharing code. It is positioned as a comprehensive initiative that invigorates internal knowledge flows, reforms organizational culture, and enhances the developer experience, thereby strengthening an enterprise's innovation base from within.

While there are various reasons for introducing InnerSource, the primary focus is on simultaneously strengthening in-house development capabilities and improving development efficiency in response to competitive environments and accelerating technological innovation that companies face. Key benefits include: (1) reducing redundant code development and improving software asset reusability to lower development costs; (2) effectively leveraging intellectual capital through cross-departmental collaboration; and (3) creating a transparent and flexible work environment that attracts and retains highly skilled talent.

These outcomes translate directly into a company's sustainable competitive advantage and hold promise from a return-on-investment perspective. For instance, shortening development cycles through a shared software foundation reduces time-to-market for new products and services, thereby expanding revenue opportunities. Additionally, by creating an internal developer community, the organization can accumulate expertise and technological excellence in-house, benefiting from less reliance on outsourcing and greater advantages in talent management. These aspects represent not merely cost-cutting but rather a strategic capital investment.

## 1.3 Research Objectives

The primary objective of this research is to construct a practical framework for the InnerSource implementation process. This framework will propose a step-by-step approach for effectively sharing and reusing software assets, taking into account the unique cultural characteristics of each organization.

Another key aim of this study is to clarify the conceptual definition of InnerSource. Given that existing InnerSource practices span a wide spectrum—from fully open initiatives to more narrowly defined collaborative areas—a universal definition or scope remains elusive. By presenting a phased introduction model aligned with organizational maturity and cultural context, this study seeks to reduce the conceptual ambiguity surrounding InnerSource.

Ultimately, this study aims to generate universal insights into the adoption of InnerSource. Although the analysis will begin with case studies of Japanese companies, its findings will be fed back into the global InnerSource community, offering practical implications that promote digital transformation in organizations worldwide.

## 1.4 Research Significance

By developing a practical framework for the InnerSource adoption process, this study yields significant academic and practical implications. One of its key contributions is a systematic framework suited to emerging markets and new industrial domains. Through comparing Japanese corporations in their early adoption stage with global enterprises at more mature stages, the study clarifies how InnerSource adoption evolves as organizations grow. It also proposes concrete implementation methods tailored to each stage, allowing companies to adopt a staged approach that aligns with their unique maturity level and organizational characteristics.

Another contribution is the elucidation of the organizational development stages and state transitions involved in InnerSource implementation. While existing research has posited ideal end states and incremental maturity levels, it offers limited understanding of the specific conditions at each stage and the processes that enable transitions between them. Through in-depth case analysis, this study identifies these conditions and processes, building a theoretical foundation that more closely reflects the actual realities of InnerSource adoption.

A further contribution is the enhancement and expansion of existing InnerSource maturity models. Although models such as the Inner Source Capability Maturity Model [8] and the InnerSource Patterns Maturity Model [9] offer a foundational staged approach, empirical findings highlight the need for more nuanced observation and theorization. This includes exploring organizational behavior at the boundary between InnerSource and conventional inter-departmental collaboration, along with informal, pre-strategic “champion” activities. By positioning these “intermediate states” and “transition processes” within a robust theoretical framework, this study deepens the practical understanding of InnerSource. It also shows that organizations may develop multiple distinct InnerSource configurations in parallel, thereby suggesting more flexible strategic pathways.

# Main

## 2 Software Sharing Practices in Japanese and Global Companies

This chapter aims to compare and analyze the current state of software sharing in Japanese companies and in global enterprises, thereby elucidating the key differences. Through this comparative analysis, it becomes possible to highlight the challenges faced by Japanese companies and the advanced approaches practiced by global enterprises.

Focusing on global research trends concerning “InnerSource,” a central concept of this study, it is evident that various strands of research have progressed in a global context. InnerSource, which applies open source practices within corporate software development, has been examined from multiple angles: research on transfer pricing systems [10], the development of capability maturity models for assessing InnerSource maturity [8], investigations into the realities of practical implementation, and analyses of the value that InnerSource can generate [11].

By contrast, the research landscape on InnerSource in Japan is still extremely limited, with only a few concrete examples of practice. Nonetheless, even if the term “InnerSource” is not explicitly used, there is a notable accumulation of studies in Japan on the sharing and reuse of source code. A notable example is the study of software component reuse in Japanese manufacturing industries. Research has shown that implementing software product lines [12] and developing reusable components leads to increased software productivity through reduced manufacturing costs and streamlined processes [13, 14]. These studies have primarily aimed at cost reduction and process optimization, thereby contributing to improved productivity in software development.

However, efforts within Japanese companies have predominantly focused on cost reduction and manufacturing process shortening. They have often lacked more comprehensive perspectives, such as co-creation, building competitive advantage, enhancing the developer experience, and fostering human resources. This observation implies an essential difference between the InnerSource practices promoted by global enterprises and the software reuse efforts long implemented by Japanese companies.

In addition, in Japanese companies, software sharing and reuse initiatives often remain confined within a single department or specific project, lacking the promotion of organization-wide knowledge sharing and collaboration. This situation likely reflects the fact that the organizational structure, corporate culture, and internal rules or accounting practices in Japanese companies have evolved differently from global standards.

In contrast, InnerSource goes beyond mere code sharing and reuse; it is positioned as a means for organizational transformation, knowledge sharing, innovation facilitation, and even developer experience improvement [15]. The companies that adopted InnerSource successfully strive to leverage their internal intellectual assets to the maximum extent, removing interdepartmental barriers through InnerSource practices and thereby enhancing overall competitiveness.

Accordingly, this chapter first organizes the background and characteristics of worldwide software sharing trends and then examines the current status of software sharing in Japanese companies. Doing so lays the groundwork for subsequent chapters, which present a more detailed comparative analysis of Japanese and global enterprises. Particular attention is paid to differences in organizational structures, cultural characteristics, in-house rules, and accounting practices, as well as to the historical background and developmental processes that have shaped them. The objective is to systematically identify the factors that lead to disparities in InnerSource practices.

## 2.1 Global Trends in Software Sharing

This section offers an overview of the international status of InnerSource adoption. The following subsections discuss concrete case examples in global enterprises and the insights gained from them, followed by trends and theoretical or analytic findings from scholarly research, and then survey results from the nonprofit organization The InnerSource Commons Foundation, including regional variations and general adoption levels. Finally, this section highlights prominent global trends as of 2024.

## 2.2 Preceding Industry Practices

InnerSource is attracting attention across various industries and corporate scales as a method for transforming software development into a more open and collaborative process. Notably, technology leaders such as SAP [16], Microsoft [17], and IBM [18] have embraced the InnerSource concept early on, promoting internal source code reuse and cross-department/cross-location collaboration among developers. These companies use clearly defined guidelines and internal portals to visualize ongoing projects and manage code repositories with open access, thereby establishing an open source–like environment within the organization.

Through InnerSource, these companies not only enhance the reusability of existing code assets but also elevate organizational knowledge management, enabling individual developers to swiftly disseminate insights—acquired both inside and outside the firm—across the organization. Consequently, internal communities conduct code reviews and peer support, which improve software quality and reduce development risk. Even departments with insufficient resources can shorten development lead times and reduce costs by reusing artifacts from other departments, ultimately reinforcing the firm's overall competitiveness.

Moreover, the summits and workshops organized by The InnerSource Commons Foundation showcase a diverse array of corporate case studies [19, 20], primarily from Europe and the United States. They have reported that industries such as finance, manufacturing, telecommunications, and healthcare—i.e., sectors beyond the core technology industry—also utilize InnerSource. For instance, Robert Bosch enables shared use of a common codebase between product development and service divisions, thus implementing a rapid feedback cycle from the field and promoting efficient internal collaboration and improved product quality [21]. In the financial sector, companies are balancing risk mitigation with increased development efficiency by building an internal code-sharing and review framework that adheres to strict security requirements and compliance regulations [22].

Nonetheless, issues remain in establishing InnerSource. Because it involves organizational and cultural changes, resistance from management or certain groups of developers can emerge in the initial stages. Additionally, it is necessary to establish multifaceted measures: mechanisms for visualizing and evaluating internal contributions, policies regarding licensing and intellectual property, and adaptations of quality assurance processes [23]. To address these challenges, The InnerSource Commons Foundation and related communities provide resources such as the InnerSource Learning Path [24] and the InnerSource Patterns [25], which summarize case studies in a pattern format to support corporate InnerSource implementations.

Overall, pioneering cases of InnerSource in industry serve as catalysts for shifting software development away from a department-based, siloed model toward a more open and collaborative organizational culture. Going forward, companies around the world may draw from these precedents to introduce and expand InnerSource in ways suited to their own contexts, potentially offering new directions for corporate innovation.

## 2.3 Research Trends from an Academic Perspective

From an academic standpoint, InnerSource has recently garnered interest in a wide range of interdisciplinary fields, including software engineering, knowledge management, and information systems. Historically, studies on open source software have accumulated a considerable body of knowledge regarding distributed development models, community governance, quality assurance processes, and innovation mechanisms. By contrast, InnerSource raises new research questions about how integrating open source principles internally can affect resource sharing and innovation processes within organizations.

The InnerSource Capability Maturity Model (IS-CMM) [8] evaluates the maturity of InnerSource adoption and suggests the conditions under which organizations can effectively deploy and expand InnerSource at each stage. This model clarifies the process by which InnerSource becomes progressively institutionalized: organizations typically begin with limited code sharing, then move toward forming internal communities, establishing governance models, standardizing quality management processes, and obtaining top management support.

Research has highlighted various benefits of InnerSource—from higher productivity and cost reduction to enhanced knowledge sharing [26]—while also presenting quantitative methods to assess its value creation [27]. These findings empirically demonstrate that introducing InnerSource goes beyond the mere formal adoption of a method and can significantly contribute to business outcomes.

Another research has analyzed how InnerSource adoption can address specific challenges in platform-based product development [28], focusing on problems such as delays, increased defect rates, and redundant software components that arise from structural separation between product units and platform teams. The findings suggest that promoting knowledge sharing and collaboration through InnerSource can improve the corporate software development process.

Furthermore, studies are emerging on how InnerSource contributes to productivity gains, quality improvements, and resolution of architectural problems in large-scale development organizations. Badampudi et al. examined how InnerSource and DevOps practices could be used to address such challenges within Ericsson. Their case study shows that, although developing reusable assets entails initial costs, in the long run, it yields benefits in terms of software quality, productivity, and customer experience [29].

As complex software systems expand, the rise in interdependencies and information silos can lead to technical debt and increased maintenance costs. InnerSource is regarded as a potential solution, as it encourages transparent and rapid issue identification and resolution by making corporate repositories visible and integrating open code reviews conducted by development communities.

However, introducing InnerSource can entail not only micro-level frictions—e.g., conflicts with existing development processes, challenges in resource allocation, and inadequate developer education—but also macro-level considerations essential for strategically rolling out and popularizing InnerSource across the organization. Tasks related to InnerSource projects have been analyzed, providing an in-depth discussion of project management issues and success factors [30]. In recent years, the establishment of governance structures—sometimes called "InnerSource Program Offices" (ISPO)—has gained attention as a way to systematically advance InnerSource throughout the organization. These program offices define foundational principles and guidelines for InnerSource, develop training and educational programs, coordinate team-level interests, select common tools, design incentive structures, and otherwise provide comprehensive support [31].

Organizational initiatives such as ISPOs create frameworks for top-down executive sponsorship, community building, metrics (KPIs) development, and the proper management of licenses and intellectual property, all of which are critical for fostering an InnerSource culture organization-wide. Consequently, success and failure examples—once confined to individual product teams or specific technical communities—can be disseminated across the organization, enabling a gradual reduction in barriers through standardized practices. The benefits of InnerSource therefore extend beyond a small subset of teams, ultimately improving corporate-wide software asset management, operational efficiency, and innovation capability.

Another research focusing on accounting and managerial processes emphasizes the necessity of building new communities within the enterprise. Investigation of computational tools and techniques for measuring InnerSource developments shows that existing tools are inadequate for effectively managing the InnerSource process [11].

Future research directions include (1) establishing quantitative evaluation methods to measure the organizational outcomes of InnerSource; (2) optimizing adoption strategies and governance models for uncertain business environments; (3) clarifying institutional barriers and adaptation requirements in large multinational and multicultural organizations; and (4) assessing the long-term impact on human resource development and career formation. In other words, effectively embedding and expanding InnerSource requires improvements at the micro (team/product) level as well as institutional and cultural adjustments, managerial structures, and organizational learning mechanisms at the macro (organizational/governance) level.

Synthesizing these perspectives, InnerSource represents a multifaceted subject of analysis extending from technical challenges to organization-wide managerial issues. As research increasingly adopts macro-level perspectives, there is growing anticipation that more comprehensive insights will accumulate regarding how InnerSource enhances organizational productivity and competitiveness.

## 2.4 Current Status and Challenges of Software Sharing in Japanese Companies

As discussed in the previous section, while global companies are actively adopting InnerSource practices, the current status of software sharing in Japanese firms shows distinct characteristics and remains relatively limited. While explicit efforts under the banner of "InnerSource" are rare, certain industries—most notably manufacturing—have built up substantial experience in software sharing and reuse. In fields such as automotive, consumer electronics, and industrial machinery, embedded software constitutes a key component of the end product, and companies have amassed extensive code assets. Historically, Japanese firms, which have grown through hardware-oriented manufacturing, have managed these assets through rigorous quality assurance and standardized procedures, reusing them primarily within specific departments or among partner companies in the supply chain. In this sense, one could argue that Japanese companies already possess a unique internal software-sharing foundation. However, this form of sharing often lacks cross-departmental transparency and open contribution models, diverging from the organizational ethos of "applying open source community principles internally" that InnerSource aims to cultivate. In particular, stringent quality and security requirements tend to restrict the free viewing and editing of code, leaving only limited room for an InnerSource-style collaborative culture to flourish.

Several factors underlie this situation. First, many Japanese companies tend to rely on vertical and hierarchical approaches to information sharing and decision-making [32], more so than their Western counterparts. The Ringi system (a bottom-up proposal and approval process) and seniority-based personnel systems often reinforce hierarchical relationships and departmental boundaries. These cultural dimensions can hinder cross-functional knowledge sharing and spontaneous contributions, both crucial to InnerSource adoption. Moreover, firms inclined to be risk-averse are often cautious about new development methods and organizational change, creating psychological and institutional barriers to implementing an open, collaborative model like InnerSource.

Second, Japanese software development has long been characterized by a vertically integrated subcontracting structure [33]. Typically, the main contractor is responsible for requirement definition and other upstream processes, whereas subcontractors and sub-subcontractors handle implementation and testing. In such an externally oriented development culture, code and knowledge are segmented along corporate boundaries, making it difficult to form an open internal ecosystem. While Western companies often nurture diverse expert communities in-house to foster an OSS-like culture, Japanese companies rely heavily on supplier networks, with structural limitations on building an integrated "community" within the organization itself.

Third, the degree of acceptance for innovative software development methodologies is an issue. Although agile development and DevOps have gained global traction, some Japanese companies continue to favor waterfall-based models and traditional processes [34]. These environments may struggle to accommodate InnerSource, which presupposes open contributions and continuous improvement cycles.

Accounting also poses structural challenges for software sharing in Japanese enterprises. In manufacturing companies especially, software is treated as an intangible asset incorporated into final products. Its valuation and cost management often follow hardware-centric accounting principles, which may not adequately capture the value of cross-departmental sharing or continuous improvement activities. Research indicates that Japanese accounting standards fail to reflect the properties of reuse in software accounting, potentially leading to duplicated recognition of costs. Moreover, current standards do not differentiate between varying degrees of software customization, which may misrepresent actual development conditions. Another major shortcoming is the lack of case distinctions for reusable commercial software [35].

Modern software development typically involves ongoing improvements and functional additions post-delivery, creating discrepancies with conventional fixed asset accounting frameworks. This is particularly problematic in the context of subscription-based or cloud services, where relevant accounting standards are still evolving [36]. Such accounting restrictions can hinder efforts to foster open software sharing and collaboration within organizations.

Nevertheless, driven by the acceleration of digital transformation (DX) and intensifying global competition, Japanese companies increasingly face the urgent need to enhance speed, quality, and flexibility in software development. In the automotive industry, for example, connected cars, autonomous driving technology, and smart factories require a higher level of software-centric development capabilities. Similarly, in non-manufacturing sectors such as finance, distribution, retail, and healthcare, the rapid advance of IT has heightened the demand for shared software assets and more efficient contribution processes within organizations.

The SECI model (Socialization, Externalization, Combination, and Internalization) [37] resonates with the community-building philosophy underlying InnerSource by emphasizing the iterative exchange of tacit and explicit knowledge. In practice, the adoption of distributed version control tools (e.g., GitHub Enterprise, GitLab), the establishment of documentation processes, and the introduction of collaboration platforms are on the rise—changes that can facilitate InnerSource-like practices. GitHub is becoming mainstream in Japan as well, with GitHub reporting that as of November 2024, Japan surpassed 3.5 million GitHub users, an increase of 23% year-over-year [38]. These developments can complement conventional top-down governance models by enabling engineers to autonomously and cross-functionally contribute to the codebase—effectively forming an "in-house OSS community."

Additionally, domestic policy trends, open innovation support measures, and increased participation by Japanese firms in international standard-setting organizations may provide a more conducive environment for InnerSource. For instance, growing involvement in the Linux Foundation’s OpenChain project and stronger ties with local OSS communities help Japanese companies integrate themselves into the global open source ecosystem. Several leading examples exist, such as Toyota [39] and Hitachi [40], and multiple firms have attained ISO/IEC 5230:2020 (OpenChain 2.1) certification. By learning from global best practices and actively engaging with international communities, Japanese firms may find it fully feasible to incorporate InnerSource approaches into their strategies.

In summary, the degree of maturity of software sharing practices within Japanese enterprises remains in an early stage compared to the global InnerSource movement, influenced by unique cultural and industrial structures that have slowed adoption. Still, given the push for DX, the demands of global markets, and the broader influence of open source communities, there is ample potential for InnerSource methods to take root.

Ultimately, the introduction of InnerSource into Japanese firms is likely not just a technological reform, but an overarching transformation of organizational culture, governance structures, and approaches to human resource development. Its success or failure may be a critical factor determining whether Japanese companies can effectively participate in the global software development ecosystem and sustain a competitive edge.

## 2.5 Summary of the International Comparison of InnerSource Adoption

This chapter has examined worldwide InnerSource trends alongside the current state of software sharing in Japanese companies, drawing from industry case studies, academic research, and factors unique to the Japanese context. From a global perspective, particularly among major technology companies, InnerSource is increasingly recognized as a method for reshaping internal software development culture while simultaneously driving innovation and efficiency gains. The InnerSource Commons Foundation serves as a pivotal international community, offering a platform for exchanging best practices and providing implementation guidance.

From an academic standpoint, studies on InnerSource have approached it from multiple angles, including organizational learning, knowledge management, and community formation. Research findings have included the establishment of maturity models, quantitative effect measurements, and explorations of organizational change mechanisms.

Turning to Japanese companies, the manufacturing sector has developed its own software-sharing culture, but this remains distinct from global InnerSource approaches. Hierarchical organizational structures, vertically integrated development processes, and risk-averse corporate mindsets represent significant barriers. Nevertheless, the progression of digital transformation and shifts in global competitive environments offer windows of opportunity for InnerSource implementation in Japan.

These insights provide the foundation for the quantitative analyses presented in the following chapters and offer critical perspectives for the international comparative study of InnerSource.

# 3 Global and Japanese Trends in InnerSource

This chapter presents a comparative analysis of international InnerSource trends and the current state of Japanese enterprises, providing crucial insights into regional characteristics and cultural factors affecting the adoption of global software development methodologies. From Chapter 3 through 6, this study contrasts the findings from the global State of InnerSource Survey with independently conducted research in Japan, highlighting commonalities and differences in InnerSource implementation. The analysis utilized anonymous survey data from the 2024 report, obtained with permission from lead author Clare Dillon, exclusively for research purposes with appropriate data handling protocols and confidentiality measures in place.

The primary objective of this chapter is to establish a foundation for subsequent analyses by elucidating global InnerSource trends, survey methodologies, and data characteristics. Particular attention is paid to methodological differences and sample characteristics between global and Japanese surveys to enable more precise comparative analysis.

The InnerSource Commons Foundation, a non-profit organization, conducts the State of InnerSource Survey to systematically track international InnerSource trends. Since its inception in 2016, the survey has been conducted in 2020, 2021, 2023, and 2024, with various research institutions and corporate sponsors leading each iteration. The survey encompasses implementation status, success factors, challenges, tools, processes, and organizational support structures, providing comprehensive insights into international best practices and trends.

According to direct inquiries with the foundation, there was an extremely low participation rate of Japanese companies in this international survey. Only one respondent from Japan participated in the 2021 survey, with zero participation in 2016, 2023, and 2024. Conversely, the 2024 survey showed 55% of respondents from Europe and 40% from the Americas, highlighting the widespread adoption of InnerSource in Western regions[41].

This geographical disparity suggests limited awareness and interest in InnerSource outside Western regions. In Japan particularly, unique organizational culture and practices may present barriers to InnerSource adoption, potentially widening the gap with global software development trends. This situation underscores the importance of investigating InnerSource applicability in Japanese companies and identifying specific impediments.

## 3.1 Global InnerSource Trends and Main Challenges as of 2024

As of 2024, global InnerSource initiatives exhibit notable diversity and increasing maturity. According to the report, InnerSource continues to be recognized primarily as a concept centered on "software/source code reuse," but it has evolved into an approach that applies open source culture and techniques within organizations and has been extending its scope and depth [41].

From a conceptual standpoint, InnerSource is highlighted not merely as a matter of tool adoption or code sharing but also as an encompassing approach that fosters cultural transformation in development organizations, nurtures a learning-oriented environment, and strengthens networks among engineers. The 2024 survey indicates that although the percentage of measurable progress in knowledge sharing had been at 68% in 2023, it decreased significantly to 33% in 2024. This finding implies that, despite the continued expansion of InnerSource practices, the measurement and evaluation of resulting outcomes are still in the process of being established. Furthermore, the main project types to which InnerSource is applied include libraries, internal tools, DevOps initiatives, and platform projects, underscoring an ongoing trend toward building cross-organizational, widely reusable components and toolchains. Notably, documentation projects have also gradually increased, suggesting that companies in the early stages of InnerSource adoption often use documentation as a foothold to facilitate cultural transformation [41].

Although these developments indicate the potential for further evolution of InnerSource, several persistent challenges have been identified. For instance, establishing reliable indicators for evaluating knowledge sharing, gaining buy-in from middle management, balancing quality with speed, and navigating cultural differences among globally distributed engineering teams remain areas with room for improvement. In addition, given the rapidly advancing nature of contemporary software development, it is critical to organize development tools, architectural guidelines, standardized documentation, and community-building measures that support InnerSource activities.

Overall, global InnerSource trends in 2024 can be positioned as a "transitional phase," characterized by the continued expansion and maturation of practices but accompanied by the need to develop measurable outcome indicators and to overcome cultural and organizational obstacles. Further progress is expected through the integration of academic research and industry expertise, leading to the establishment of InnerSource environments that promote sustainable organizational learning and innovation.

## 3.2 Current State of InnerSource Adoption in Japanese Enterprises as of 2024

In order to gain a comprehensive understanding of the current state of InnerSource in Japanese enterprises, a survey was conducted that considered Japan's cultural context and organizational traits. This survey aimed not only to capture the technical aspects of InnerSource but also to examine its organizational and cultural dimensions.

The survey was designed based on feedback from leading practitioners in various industries, resulting in a comprehensive questionnaire. In addition, a conference targeting early adopters and highly interested stakeholders was organized, offering an opportunity to gather direct feedback from the field.

### 3.2.1 Design and Implementation of the Domestic InnerSource Trends Survey

While referencing the globally oriented State of InnerSource Survey, which targets enterprises worldwide, a more comprehensive and detailed survey was devised to reflect the circumstances in Japan. A key consideration in constructing this survey was the recognition that while many companies in the United States and Europe have already implemented InnerSource, Japan remains in a relatively early stage of adoption.

On this basis, the Japanese survey was designed to enable comparison with global surveys up to 2024. While retaining concrete practical perspectives, the survey paid special attention to barriers to InnerSource implementation, thereby enabling collection of data relevant to organizations that have a strong interest in adopting this approach. Specifically, the survey comprised 26 questions addressing the following topics:

- Profile of survey respondents
- Characteristics of the respondents' organizations
- Obstacles to InnerSource adoption
- Actual conditions of cross-organizational collaboration
- Status of InnerSource practices

### 3.2.2 Organization of "InnerSource Gathering Tokyo 2024" and Response Collection

On August 8, 2024, "InnerSource Gathering Tokyo 2024" was held as the first InnerSource-focused conference in Japan. This conference was planned and executed with the dual objectives of providing a venue for knowledge sharing and discussion among early adopters and highly interested participants and serving as a platform for survey data collection.

The InnerSource Commons Foundation hosted this conference, and a member of its board played a central role in its organization. Sponsorship was provided by The Linux Foundation, and operational support was received from KDDI Corporation and Nifty Corporation, enhancing the conference's planning and operational base. The 74 participants ranged from small companies to large corporations, including enterprises that have already implemented InnerSource in Japan, highly innovative user companies strongly interested in the concept, manufacturers, and major IT "mega-venture" companies [42]. The diverse participant base allowed for cross-organizational comparisons at different stages of InnerSource adoption and enabled knowledge transfer across different fields and industries, thereby broadening and enriching discussions.

In order to avoid making the conference a one-way flow of information, a design was adopted that promoted mutual knowledge exchange among participants. Specifically, in addition to sessions showcasing real-world examples from various companies, group discussions were integrated to provide opportunities for direct dialogue on success factors, challenges, and strategic approaches for InnerSource implementation. These initiatives fostered deeper understanding of InnerSource, facilitated the sharing of practical insights, and offered opportunities for participants to explore future directions.

### 3.2.3 Survey Response Collection and Data Aggregation Method

In this study, participants in the above-mentioned conference were asked to respond to a survey. Among the 74 participants, 52 submitted responses, resulting in a response rate of approximately 70%. Considering that the respondents were individuals already implementing InnerSource in Japan or those with a strong interest in this field, the sample size is deemed sufficient to ensure the reliability of the findings. An analysis of respondent attributes revealed several noteworthy tendencies among those in Japan who exhibit a strong interest in InnerSource. Chapter 4 and beyond will examine these findings in the context of international comparisons.

## 3.3 Methodological Considerations for Comparative Analysis with the Global Survey

For the purposes of this study, the 2024 State of InnerSource Survey served as the primary data source, comprising 59 valid responses. To qualify as valid, a response had to be complete for at least 50% of the questionnaire items and contain sufficiently concrete and detailed comments in the free-text sections. These criteria were established to maintain qualitative rigor and ensure the reliability of the analysis.

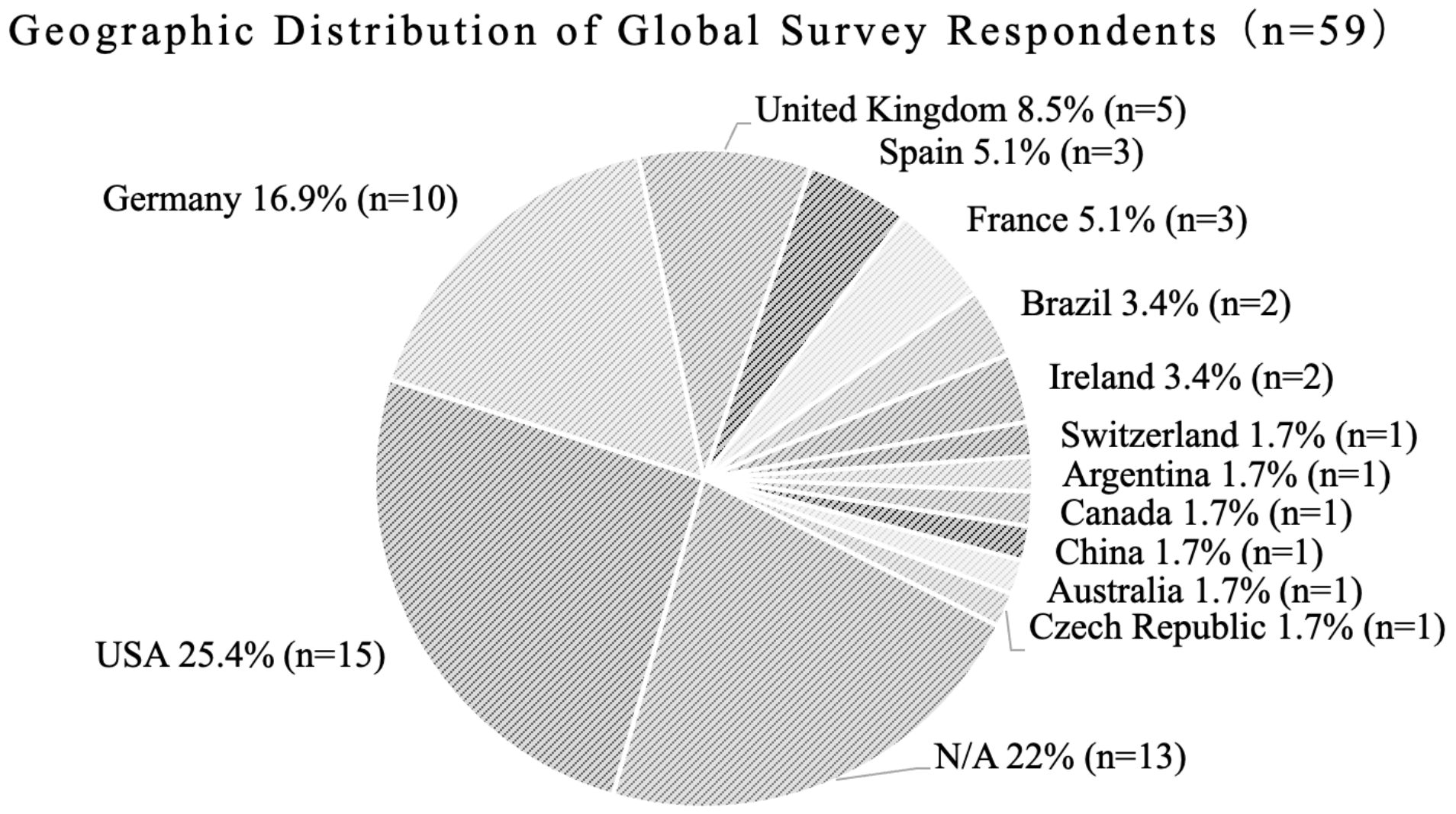

*Figure 3.1 Geographic Distribution of Respondents in the 2024 State of InnerSource Survey*

The distribution of global survey respondents includes 25% from North America, 30% from Europe, 5% from South America, and 4% from Asia/Oceania, demonstrating broad geographic diversity. This finding suggests that InnerSource is becoming a widespread phenomenon that transcends regional boundaries. By contrast, the Japanese survey data predominantly reflect responses from organizations that are at the forefront of InnerSource implementation, and the quality of the responses thus exhibits distinctive characteristics.

From the standpoint of survey methodology, the Japanese sample was drawn from participants in a domestic conference through a direct approach, whereas the global sample was collected online through an international invitation process. Although both samples captured respondents strongly interested in InnerSource, the global sample appears to include a more mature practitioner community.

Despite partial structural differences in the survey instruments, both surveys address fundamental topics such as the basic concept of InnerSource, core practices, and organizational challenges. This shared analytical framework facilitates the alignment of Japanese cultural and organizational traits with their position in a global context.

## 3.4 Overview of the Japan–Global Comparison

This chapter examined international InnerSource survey data and a Japan-specific study, focusing on each survey's methodology, data characteristics, and points requiring attention when making comparisons. In particular, the chapter clarified the attributes of data collected by The InnerSource Commons Foundation's worldwide survey and by the first InnerSource-focused conference in Japan, laying the methodological groundwork for comparative analysis.

Although the global survey and the Japanese survey differ in their sampling methods and respondent pools, they share a common analytical foundation concerning the core elements of InnerSource, including its fundamental concept, core practices, and organizational challenges. Notably, both surveys were conducted in 2024, an overlap that enhances the reliability of regional comparisons.

Chapter 4 will compare respondent attributes in order to analyze the characteristics of InnerSource in Japan and globally, illuminating its level of advancement in "emerging regions" versus "more established communities." Chapters 5 and 6 will then integrate the two datasets (N=111) to attempt a more comprehensive synthesis, paying careful attention to differences arising from variations in the maturity stage of InnerSource adoption. Through these analyses, the objective is to elucidate the features and challenges of introducing InnerSource in Japanese enterprises within a global context.

# 4 A Comparative Analysis of InnerSource Adoption in Japan and Globally

In analyzing the current state of InnerSource practice, comparative research between Japan and the global market provides critical insights for understanding how software development methods are introduced in emerging markets. This chapter presents a systematic examination of the characteristics and differences between Japan-based samples (n = 52) and globally sourced samples (n = 59) by focusing on practical experience, stage of adoption, years of experience, distribution of roles, job functions, industry sectors, and organizational and developer scales.

From an organizational maturity standpoint, clear distinctions are observed between features characteristic of early-stage Japanese companies and the more mature InnerSource adoption observed in global companies. These differences shed light on how regional and cultural factors influence the introduction of InnerSource and offer important perspectives for analysis.

The findings of this study have the potential to offer practical guidance to emerging markets and developing regions that are considering the adoption of InnerSource. In particular, it may serve as a practical reference for how to adapt and evolve InnerSource practices in regions such as the Asia-Pacific and Latin America, where organizational structures and development cultures differ significantly from those in Europe and North America.

Moreover, this comparative analytical approach may prove useful as a reference model for organizations aiming to introduce InnerSource in a variety of regional contexts. By deepening the understanding of how factors such as organizational scale, industry characteristics, and role distribution among developers influence successful InnerSource adoption, it becomes possible to craft strategies suited to region-specific challenges.

The analytical framework adopted here can also serve as a knowledge-transfer model within the global software development community. In particular, it can provide empirical insights into the challenges faced during the process of organizational transformation in emerging markets, as well as concrete approaches to overcoming those challenges.

## 4.1 International Comparisons of InnerSource Maturity

Notable differences in maturity can be observed between Japan and the global market regarding InnerSource practice. This section provides a detailed analysis of these differences from three perspectives: individual-level practical experience, organizational adoption stage, and years of professional experience.

In global markets, especially in Europe and North America, InnerSource practice has largely reached a more mature stage, featuring organizational-level rollouts and strategic utilization. In these regions, systematic knowledge transfer led by experienced practitioners and well-established governance frameworks are the norm.

By contrast, InnerSource adoption in Japan remains in an early stage, with many organizations taking an exploratory approach. Rather than viewing this gap solely as a disadvantage, it may also be seen as an opportunity to create a model of InnerSource adoption tailored to the unique organizational cultures and development practices found in Japan.

Drawing on quantitative survey data, this chapter provides a comparative analysis of current practices in both markets. Through this analysis, the study clarifies the barriers and opportunities for InnerSource adoption in Japan while offering insights for further development.

### 4.1.1 Differences in Individual-Level Maturity Between Japan and the Global Context

According to the Japanese sample (n = 52), 53.8% responded “N/A of the above/Not sure” regarding direct experience with InnerSource projects. However, 46.2% reported having some type of InnerSource experience. Specifically, 23.1% had contributed to an InnerSource project outside their own team; 26.9% had worked on a project that accepted guest contributions; 19.2% had experience rolling out or scaling InnerSource within their organization; and 11.5% had advised or coached InnerSource practitioners.

In contrast, 89.8% of respondents in the global sample reported engaging in some aspect of InnerSource practice, with 66.1% of that group having been involved in organizational-level rollouts, scaling initiatives, or providing coaching/consulting support. This figure is notably more than three times the comparable figures of 19.2% and 11.5% in Japan, suggesting the existence of robust InnerSource leadership and strategic engagement roles in global environments.

These findings illustrate both that InnerSource is still at an early or exploratory stage in Japan and that many global organizations have already achieved a considerable level of maturity. While preliminary steps such as accepting external contributions from other teams and sharing knowledge across the organization are becoming more widespread in Japan, there remains a limited pool of practitioners who play major coaching or strategic leadership roles.

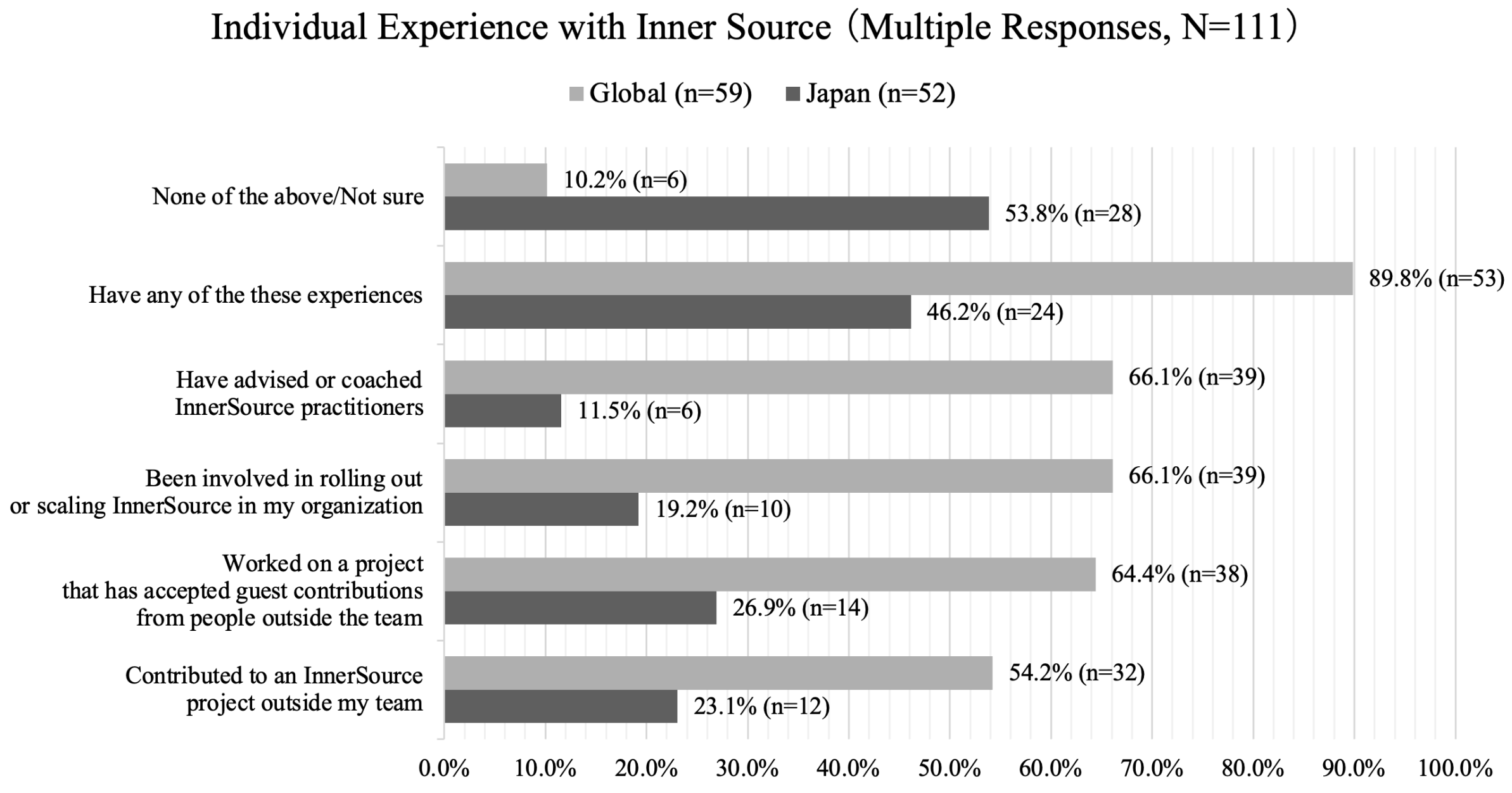

*Figure 4.1 Comparison of InnerSource Experience*

Despite the high level of interest in InnerSource among survey respondents in both samples, the global sample includes a greater proportion of organizations and individuals who are already in advanced stages of practice. Nevertheless, it should be noted that even within the global sample, some organizations are still at an early stage of adoption, just as in Japan there are signs of emerging leadership potential within organizations.

Overall, these findings confirm that InnerSource adoption in Japan remains in its infancy and that there is a marked maturity gap compared with global contexts, particularly those in Europe and North America. Most notable is that, while usage and best practices are well established at a practical level in many global settings, even highly interested Japanese organizations often have limited experience in fully operationalizing InnerSource. This gap highlights key obstacles to be addressed for the expansion and institutionalization of InnerSource in Japan.

### 4.1.2 Comparing Adoption Stages: Japan and Global

In Japan, 44.2% of organizations fall into the “Ideas stage,” followed by the “Pilot stage” (19.2%). These two early stages collectively account for the majority, indicating that many organizations remain at the initial stages of InnerSource adoption, and that practical implementation generally retains a nascent or exploratory character.

A subset of advanced organizations did report the “Early adoption” stage (13.5%), “Growth stage” (5.8%), and “Mature stage” (3.8%). For these organizations, systematic deployment and integration of InnerSource in corporate structures are underway. However, very few organizations in Japan have reached maturity, underscoring the reality that InnerSource remains far from fully established across Japanese industries.

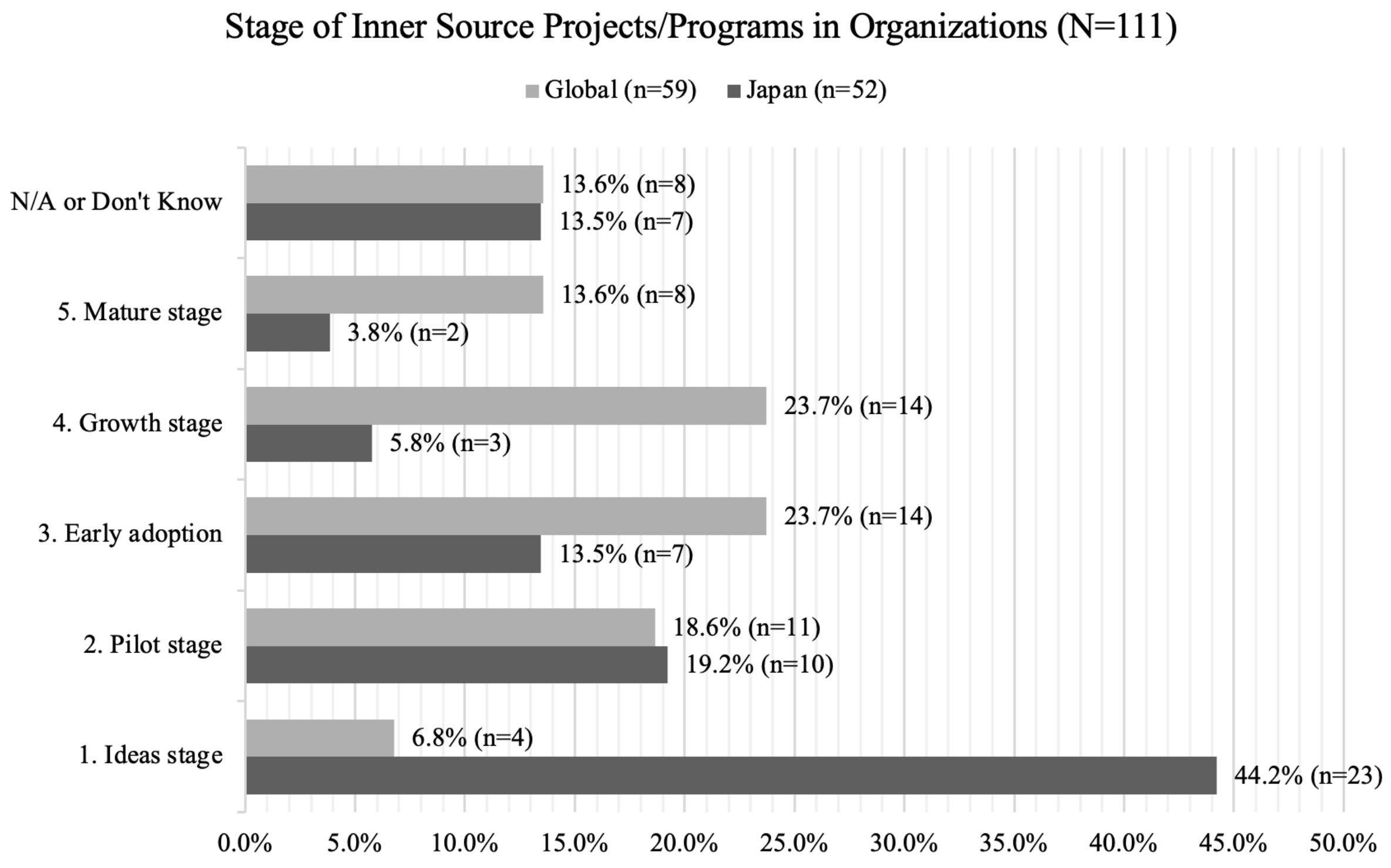

*Figure 4.2 Stage of InnerSource Projects/Programs Within Organizations*

Comparative analysis with the global sample reveals significant differences in maturity levels. Within the common population of organizations interested in InnerSource, global companies show a distinctly higher proportion of organizations in the early adoption to growth stages compared to Japanese companies. However, even among global companies, organizations at the pilot stage outnumber those at the mature stage. These results suggest that InnerSource adoption is currently evolving worldwide, with organizations at various developmental stages coexisting in the maturation process.

### 4.1.3 Distribution of Years of Experience

The results also indicate notable differences in respondents' overall years of work experience. In the global sample, 14.3% of participants reported 11–15 years, 40.5% reported 16–20 years, and 28.6% reported 26 years or more of professional experience. This suggests that mid-level to senior professionals play a central role in leading InnerSource adoption in global organizations.

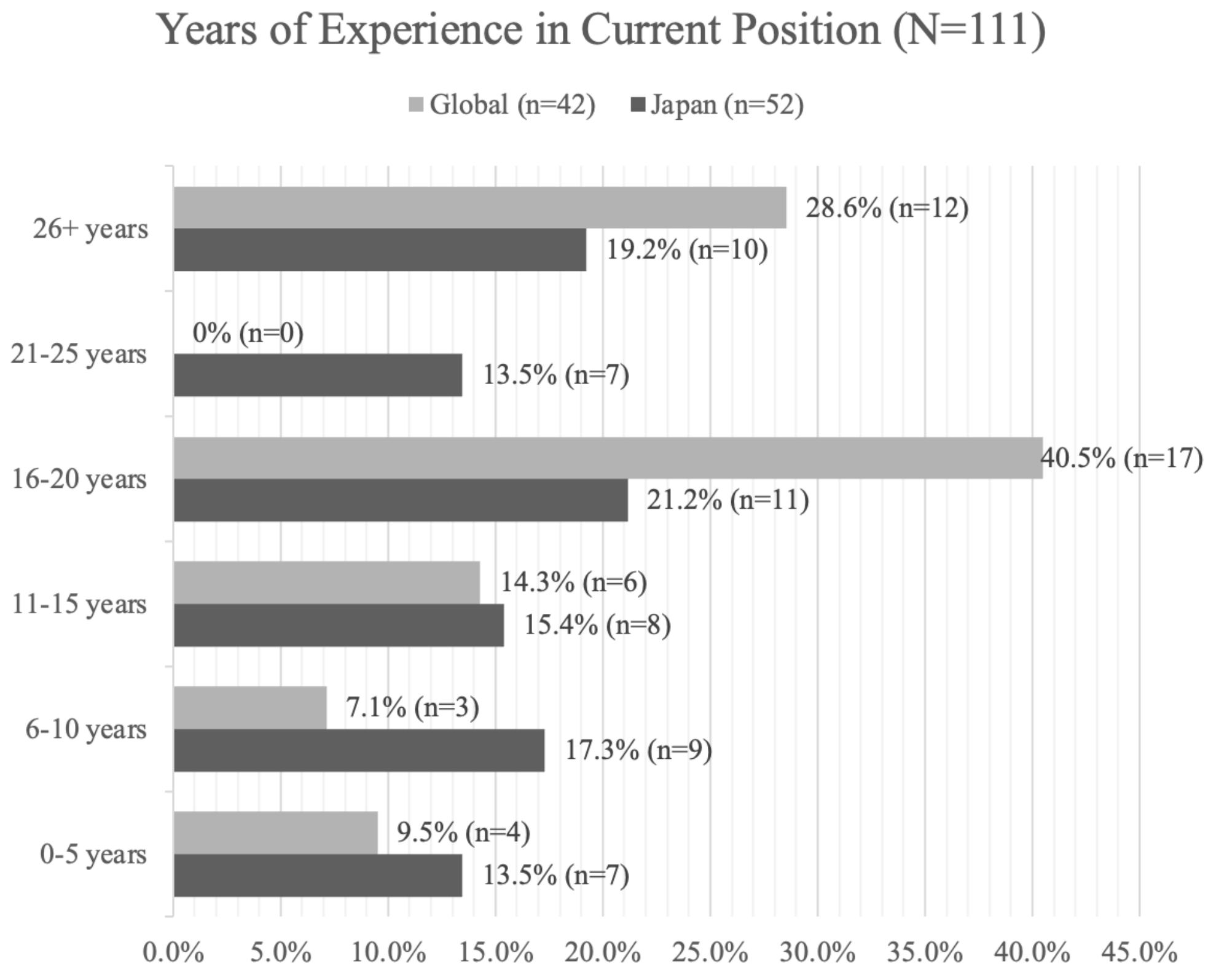

*Figure 4.3 Total years of work experience*

Conversely, the Japanese sample exhibits a more balanced distribution of work experience, ranging from fewer than five years to over 26 years. This broad range of experience implies that interest in InnerSource is not restricted to a specific career stage or hierarchy. Younger employees and relatively early-career developers appear to have comparable interest levels to more senior staff, suggesting the potential for innovation to emerge from various parts of the organization without relying exclusively on a small cadre of senior leaders.

At the same time, despite the presence of senior professionals, Japanese InnerSource initiatives remain in the early stages on an organizational level. Unlike in global organizations, where the involvement of senior experts often accelerates adoption by establishing governance and supportive communities, Japanese organizations have yet to fully leverage experienced personnel for InnerSource scale-out.

Taken together, the global sample benefits from a cohort of veteran practitioners who drive governance and community formation, while Japan's strength lies in its diverse range of participants—from junior to senior levels—interested in adopting InnerSource. However, many Japanese organizations are still navigating how to integrate the experiences of seasoned professionals with the flexibility and innovative ideas of younger staff.

## 4.2 Comprehensive Analysis of InnerSource Practitioners' Roles and Industry Characteristics

A clear understanding of the roles of individuals who drive InnerSource within their organizations, as well as the industries they represent, is crucial for assessing the current state of InnerSource adoption and predicting its future developments. This section examines the roles, occupational domains, and industries of InnerSource practitioners in Japan and the global market.

In Japan, developers represent a significant portion of InnerSource practitioners, but managers and agile-related roles also participate actively. On the global side, specialized positions explicitly dedicated to InnerSource, or open source are prevalent, reflecting a higher level of organizational maturity in many cases.

In terms of industry distribution, technology-oriented enterprises dominate in both regions, although Japan features a notable presence of manufacturing industries, while the global sample includes higher proportions of financial services and healthcare.

This section presents a detailed analysis of these distinctions, offering insights into the regional specificity and overall adoption patterns of InnerSource.

### 4.2.1 The Distribution of Roles and Regional Traits in InnerSource Practice

Within the Japanese sample (n=52), the largest group of respondents identified themselves as “Developers (28.8%)”, followed by “Management (25%)” and “Agile-related” roles (15.4%). This suggests that InnerSource garners considerable interest among those directly involved in the software development process while also attracting a notable portion of individuals in decision-making positions. The agile-related cohort’s relatively high proportion (15.4%) underscores the potential synergy between InnerSource and agile practices, as both emphasize transparency, continuous feedback, and collaborative problem-solving.

In contrast, the global sample (n=59) showed a higher proportion of respondents in roles explicitly labeled “InnerSource-related (23.7%)” and “Developer (18.6%)”, as well as “Architect (13.6%)” and “Open Source-related (10.2%)”. Management (6.8%) and program/project manager roles (5.1%) were less common, and there were no respondents reporting “Agile-related” positions. These differences suggest that many global organizations have already formalized specialized InnerSource or open source roles, reflecting well-established governance and a relatively mature environment for open collaboration.

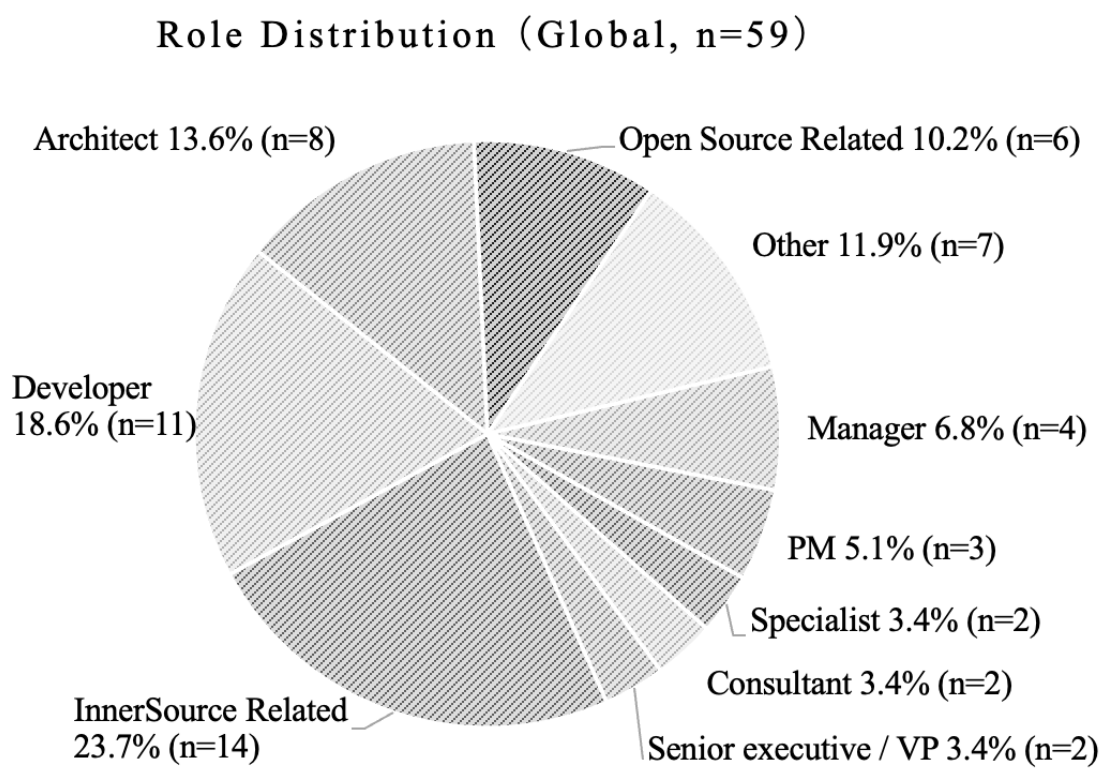

*Figure 4.4 Distribution of roles within organizations (Global)*

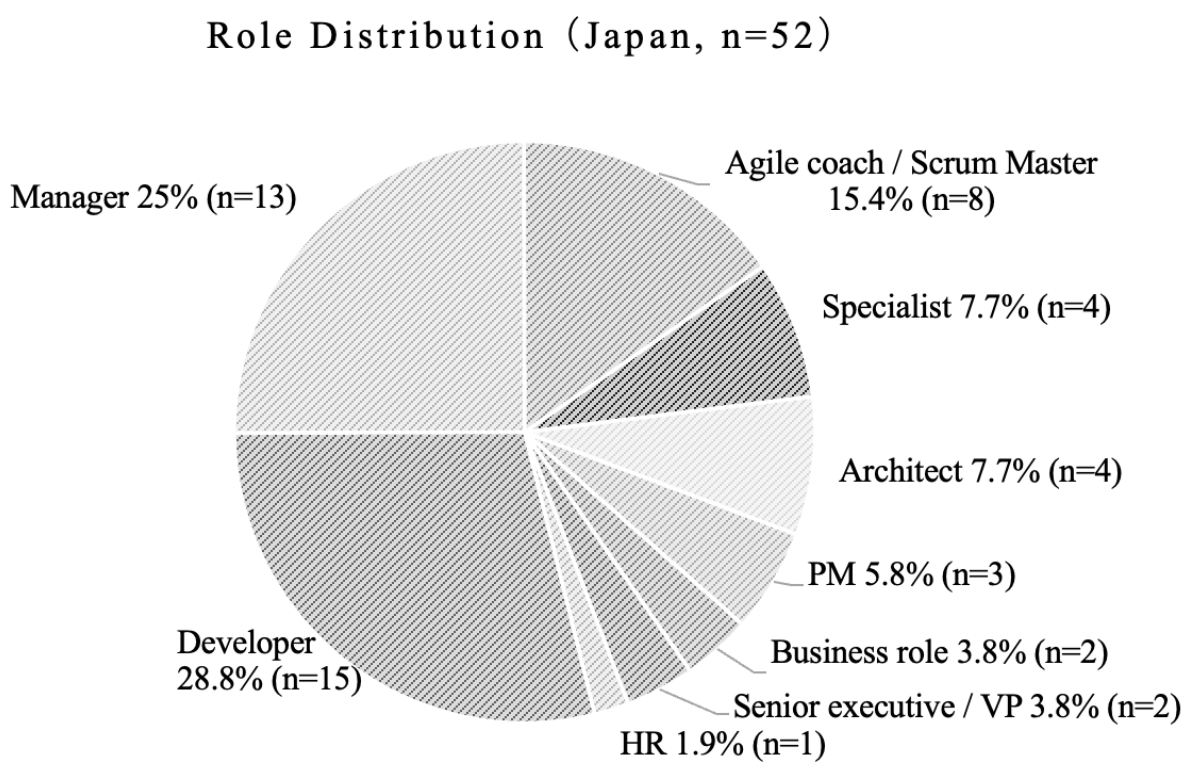

*Figure 4.5 Distribution of roles within organizations (Japan)*

In the Japanese sample, it is notable that interest in InnerSource comes from a wide range of roles, including agile-related positions. This appears to reflect Japanese companies' transition from traditional development processes to agile methodologies and open knowledge-sharing practices. While "management" accounts for 25% of respondents in Japanese companies, a distinctive feature is the absence of explicitly "InnerSource-related" job classifications. In Japan, due to geographical and cultural backgrounds as well as long-term employment practices, role distribution, organizational structures, and job definitions have evolved uniquely. For example, career paths such as job rotation and generalist tracks are common, and there may be less emphasis on specializing in specific fields compared to overseas. Additionally, awareness of InnerSource is still growing. As a result, interest in InnerSource appears to be at a stage prior to being distinguished as clear InnerSource-related roles, suggesting it spans relatively broad job domains including developers, agile specialists, and management layers.

In contrast, the global sample already includes job classifications such as InnerSource-related and Open Source-related, suggesting that mature knowledge-sharing and governance models are being driven by groups of specialists. This indicates that in global companies and communities where OSS culture has historically taken root, InnerSource is viewed not merely as a development style but as a strategic differentiator and foundation for cross-organizational collaboration, leading to the placement of dedicated specialists. Moreover, in environments where InnerSource has become relatively well-established, individuals with specific expertise and knowledge take on roles, forming foundations such as ISPOs and OSPOs that facilitate cross-organizational innovation and technology transfer.

From this analysis, while Japan has few explicit InnerSource specialist positions, there is broad involvement from both field-level practitioners and management, and combined with its affinity with agile culture, InnerSource is expected to take root as a new development methodology. Meanwhile, in the global context, specialized roles and clear governance models are already established, and InnerSource is becoming highly institutionalized as a mechanism supporting cross-organizational open collaboration. These differences in role distribution suggest that not only the implementation stage and historical background but also corporate culture and job definition differences significantly influence the implementation and diffusion process of InnerSource.

### 4.2.2 InnerSource Interests by Job Function in Japanese Organizations

While global-level data on specific occupational definitions remain somewhat sparse, Japan's survey included additional items clarifying respondents' "position of work" or "business domain" to gain deeper insights into how InnerSource garners interest in day-to-day operations. The following results help us understand how InnerSource is recognized as useful in diverse functional areas across Japanese organizations—and potentially guide global comparisons in future studies.

Software development roles comprised the largest portion, indicating that InnerSource is strongly associated with improving internal development processes, standardization, and productivity enhancement through code reuse. Additionally, 18% of respondents came from research and development departments, suggesting that InnerSource can foster an ecosystem for innovation and accelerate internal implementation of new technologies. R&D professionals can share new ideas and experimental projects through InnerSource-style repositories and communities, potentially integrating technical insights and achieving rapid practical implementation through cross-departmental feedback and contributions.

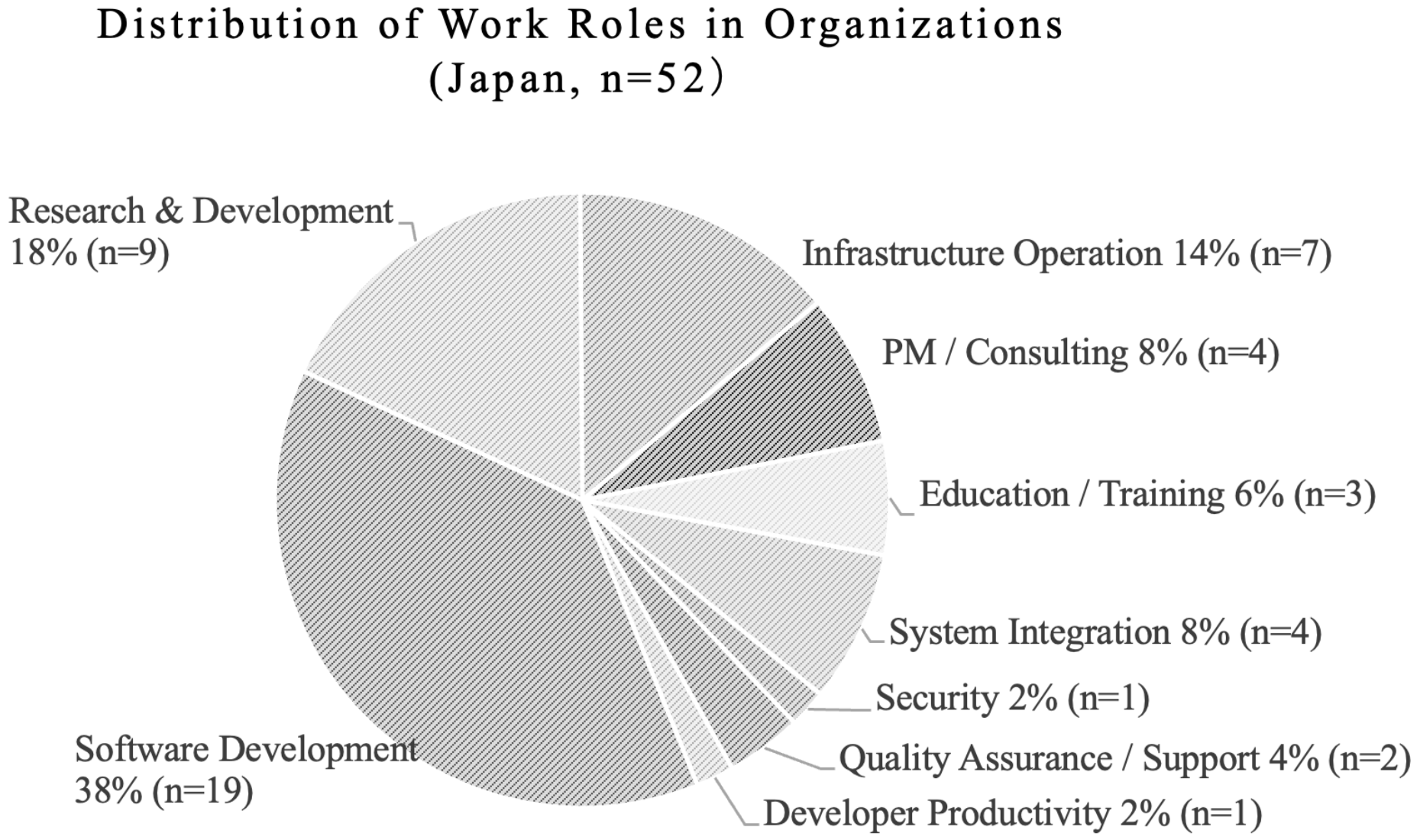

*Figure 4.6 Distribution of work domains/positions within organizations (Japan)*

Although this analysis is limited to the Japanese sample and does not directly compare with global data, the detailed functional domain analysis reveals that InnerSource generates interest and engagement across a wide range of job categories beyond development teams. In Japan, job roles span diverse domains, suggesting that InnerSource can function as a complex ecosystem that connects and facilitates mutual influence among various organizational functions. These findings could serve as an important reference point for future global sample analyses that comprehensively examine factors such as regional cultural influences, differences in job definitions, and international labor market trends.

The establishment of specialized roles such as ISPO and OSPO professionals in mature stage companies, as mentioned in the previous subsection, provides valuable insights into the organizational positioning of InnerSource. Comparative analysis with these precedent cases would lead to a more systematic understanding of the establishment and development processes within organizations, as well as how regional characteristics influence job distribution.

### 4.2.3 Industry-Specific Trends and the Spread of InnerSource Adoption

In Japan, technology enterprises represent 65% of the organizations surveyed, reinforcing the perception that InnerSource remains most prominent in the IT sector, where software development and innovation are core elements of the business. It is therefore unsurprising that these organizations find high strategic value in harnessing InnerSource for knowledge sharing and code reuse within their internal development landscape.

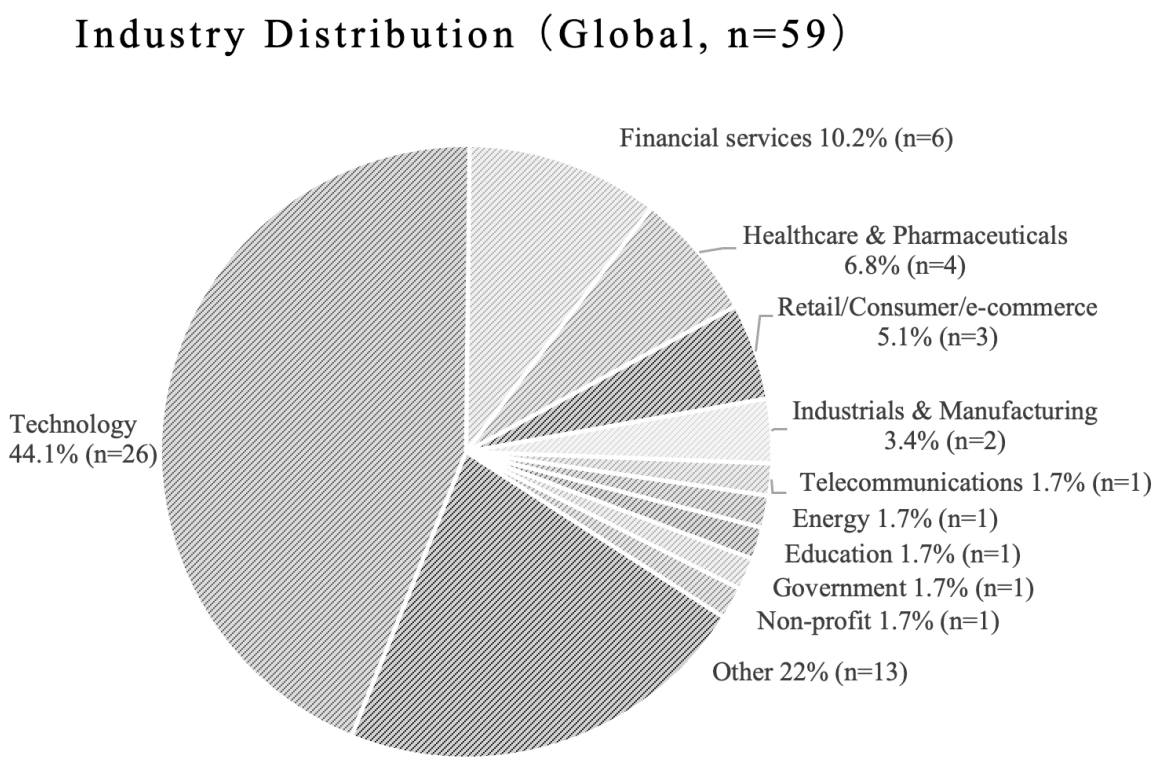

Figure 4.7 Industry Distribution (Global)

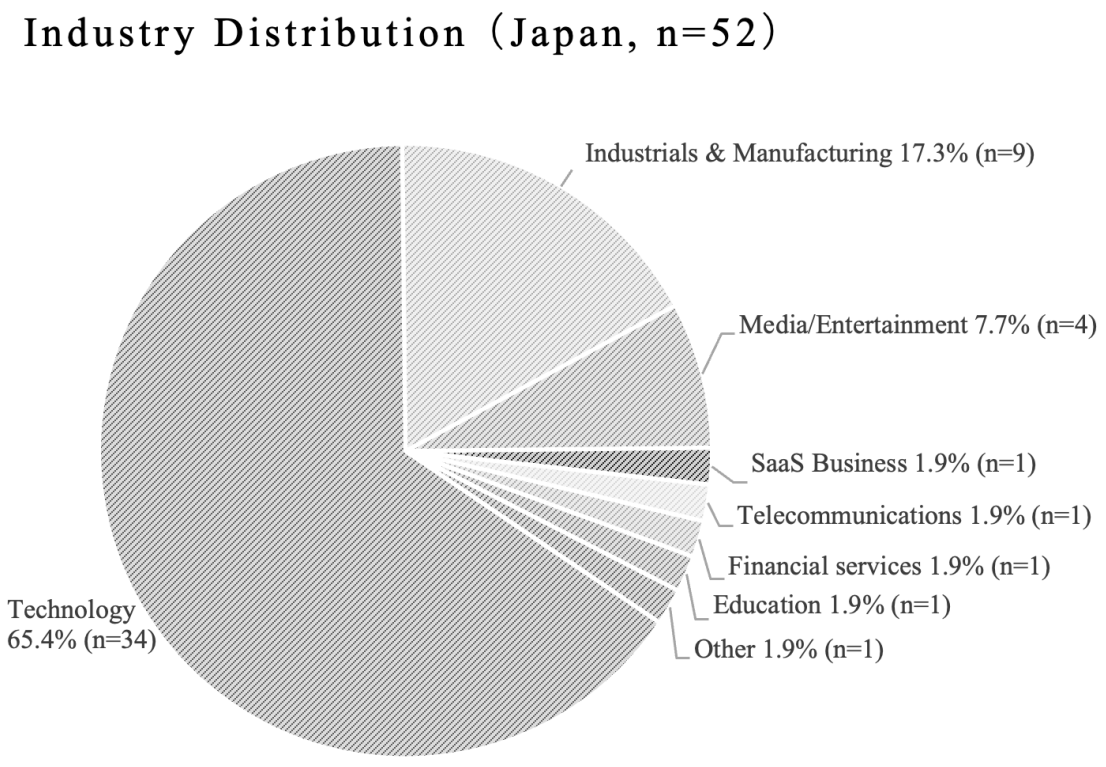

Figure 4.8 Industry Distribution

However, the 17% representation of industrial and manufacturing companies in the Japanese sample illustrates the broader digital transformation trend wherein traditional hardware-centric business models are integrating software more intensively into their product lifecycles. Here, InnerSource can be viewed as a means to optimize cross-organizational collaboration and accelerate the reuse of development resources.

Globally, the data also reveals a high portion of technology-driven firms (44%), but financial services (10%) and healthcare (7%) are similarly noteworthy. These sectors face stringent regulations and diverse customer requirements, prompting a need to internalize software capabilities. By adopting InnerSource, they enhance organizational knowledge sharing, standardize their development infrastructure, and ultimately create added value in their business models.

In summary, although technology companies continue to lead the way, InnerSource is steadily permeating other industries in both Japan and elsewhere. In Japan, the trend is particularly evident in manufacturing [43, 44]. In global contexts, regulated industries [45] are beginning to show tangible progress in adopting InnerSource as a strategic method for enabling software-driven innovation. These cross-sector expansions affirm InnerSource's flexibility as a broadly applicable approach for fostering collaboration and accelerating value creation in software-based processes.

## 4.3 Organizational Scale and InnerSource Practices

This section explores how organizational size influences InnerSource implementation, analyzing both the total number of employees and the total number of developers. Notable differences between Japanese and global companies were observed, stemming from their distinct organizational structures and scales. These differences are potentially closely related to the objectives and expected benefits of InnerSource adoption. The following analysis examines characteristic trends from the perspective of organizational size and considers their implications.

### 4.3.1 Differences in Adoption Context by Total Number of Employees

In the Japanese sample, "100–499 employees" comprised about 28.8% of respondents, with a high overall representation of small- to medium-sized organizations. Additionally, Japan has a marked tendency toward forming subsidiary companies in which each entity—such as an IT subsidiary or domain-specific affiliate—operates with relative autonomy, often with fewer employees in each organization. Consequently, "removing silos between teams or departments" (a central motivation in many large corporations) is not always the primary driver for adopting InnerSource in Japan.

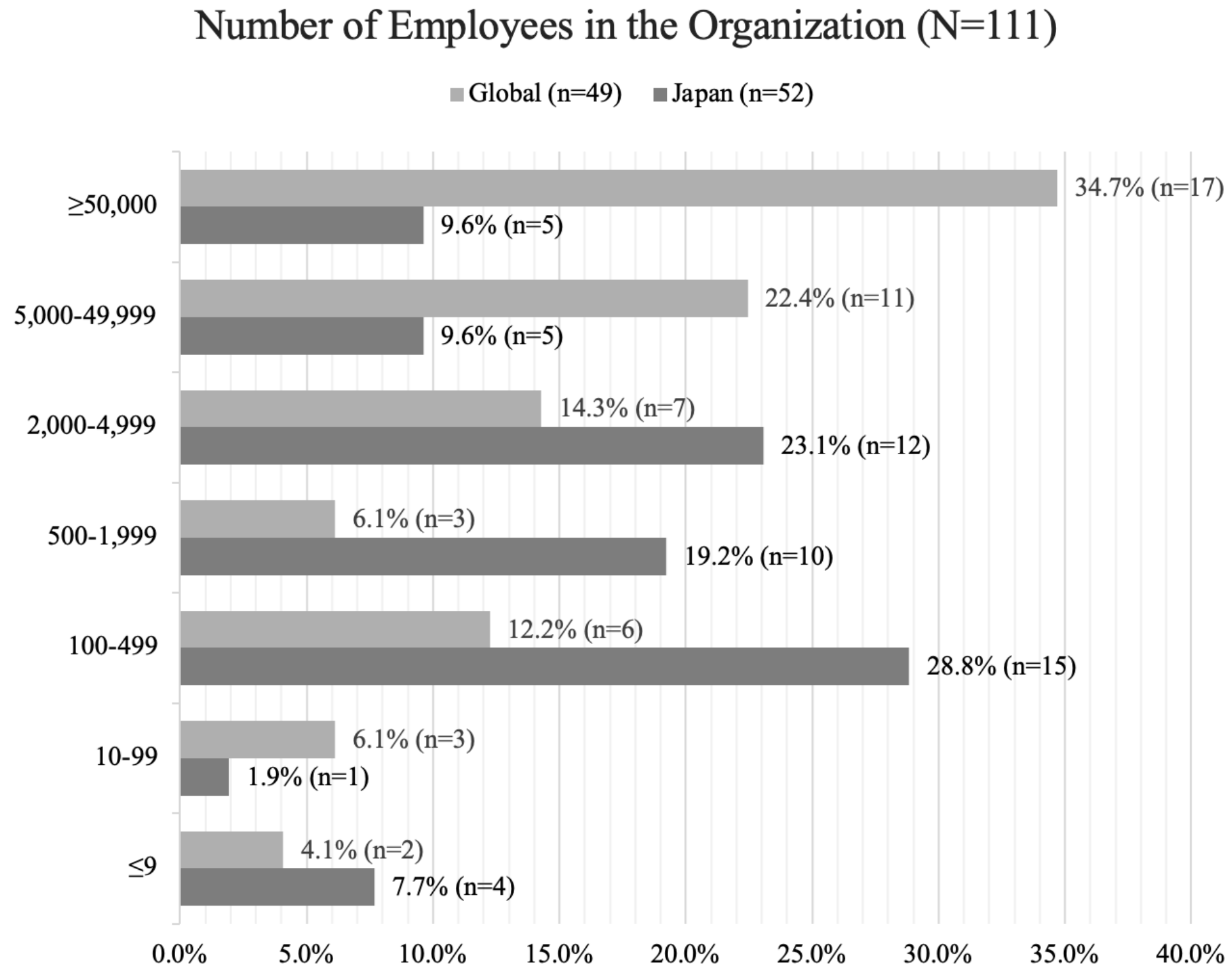

*Figure 4.9 Number of employees in the organization*

In contrast, a significant number of organizations in the global sample (n=59) exceed 5,000 employees, with many surpassing 50,000. In such extremely large organizations, departmental silos are likely to become more severe. It can be inferred that these organizations have a higher need for InnerSource adoption primarily aimed at breaking down silos.

Given that the global sample shows a notable bias toward extremely large enterprises, their background and motivation for InnerSource adoption potentially tend to focus on "breaking down silos in geographically and organizationally expanded development environments." In contrast, Japanese companies may not necessarily implement InnerSource primarily for silo elimination but might place more emphasis on functional aspects such as resource limitations and knowledge sharing. This point will be discussed in detail in Chapter 5 when analyzing adoption backgrounds and barriers.

Furthermore, in Japan, various group companies, such as system subsidiaries and domain-specific subsidiaries, tend to form relatively independent organizational units. Therefore, it may be difficult to accurately evaluate the objectives and effects of InnerSource practices without considering not only the nominal employee size but also the clear division of project domains and decision-making independence of each company. While Japanese companies may constitute part of large conglomerates when viewed on a consolidated basis, individual organizational units tend to be smaller compared to global giants. Consequently, it would be worth examining how different challenges and motivations might emerge when utilizing InnerSource, as these may differ from those of global enterprises.

### 4.3.2 Differences in Adoption Context by Number of Developers

A comparison of developer headcount reveals yet another clear divergence between Japan and the global sample. Among Japanese respondents, the largest group (40.4%) reported "100–499 developers," followed by "500–1,999 developers (27.7%)" and "10–99 developers (17%)." Conversely, the global sample included organizations with 5,000–9,999 developers (28.9%), 2,000–4,999 developers (13.3%), and even over 50,000 developers (8.9%).

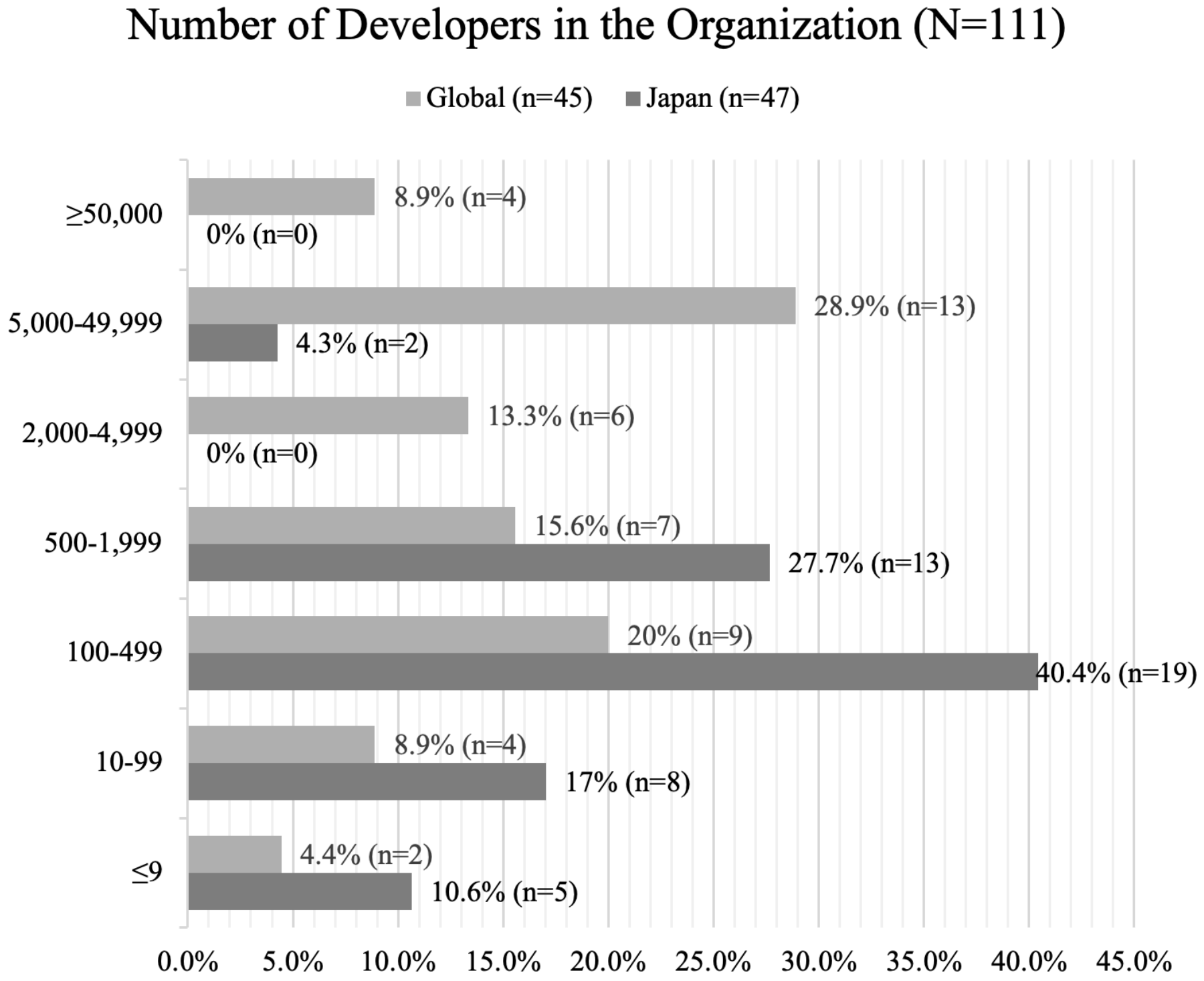

*Figure 4.10 Number of developers in the organization*

This difference suggests consistency with the aforementioned organizational scale and cultural background. In Japanese samples, where relatively many small to medium-sized enterprises exist including group subsidiaries with segmented organizational structures, developer numbers typically range from several hundred to around 2,000. As a result, InnerSource may function effectively as a means to strengthen information sharing and knowledge circulation within relatively compact development communities. Even if silos emerge, their scale and complexity may not become as severe as in large organizations, suggesting that the introduction of InnerSource could potentially yield results more quickly and directly.

In contrast, the global sample includes numerous massive engineering organizations exceeding 5,000 developers, where InnerSource is expected to serve as a "structural problem-solving measure" for silo elimination and efficient knowledge reuse against the backdrop of complex development systems spanning multiple regions and departments. In other words, the primary challenge becomes how to make enormous development organizations function as "one community," and InnerSource may likely become established as a strategic solution to this challenge.

Comprehensively, the comparison of developer numbers highlights the contextual differences in InnerSource adoption between Japan and global organizations. When analyzed alongside organizational roles, it appears that small to medium-sized development organizations in Japan tend to utilize InnerSource as a means to establish and strengthen new cultural values, backed by agile culture and emphasis on transparency. Meanwhile, in large and complex global development organizations, some companies establish specialized open source and InnerSource positions, and InnerSource tends to be positioned as a strategic "problem-solving tool" for silo elimination and large-scale knowledge integration, building upon an already incorporated open source culture. These differences may be attributed not only to variations in cultural maturity and historical acceptance of open source culture across countries and regions but also to the disparity in adoption stages between the global sample's "early adoption interest group" and Japan's "late adoption group." This serves as an excellent example of how the significance and effects of InnerSource adoption can vary significantly depending on organizational scale and historical growth stages.

## 4.4 Summary of Differences by Attribute

Through comparative analysis between Japan and the global market, it appears that InnerSource adoption exhibits differences according to regional and organizational attributes. In Western enterprises,

InnerSource may function as a means to further internalize and strengthen existing community thinking and open development practices, building upon their well-established open source culture. In contrast, many Japanese organizations, where open source culture is less mature and corporate groups tend to be subdivided, often contain multiple medium-sized organizations within a single large enterprise. Consequently, InnerSource tends to be adopted as a "packaged methodology," potentially serving as an initiative to create a new culture emphasizing agile practices and development transparency.

It is noteworthy that interest in InnerSource remains high even among medium-sized and smaller enterprises, as departmental isolation and insufficient information sharing can emerge as universal issues regardless of organizational size. In global cases, very large organizations utilize InnerSource to overcome increased communication costs due to multi-site operations and siloed structures. Meanwhile, Japanese small to medium-sized enterprises, even with relatively small silos, appear to have strong intentions to dissolve these barriers while nurturing agile culture and reforming processes through collaboration between development teams and management. These motivational differences can be observed in survey results categorized by region and scale.

Regional and organizational cultural differences are also strongly reflected in the attributes and roles of those interested in InnerSource. In global survey subjects, specialized positions such as "InnerSource-related" and "Open Source-related" roles clearly exist, with experienced engineers and coaches driving large-scale cross-organizational implementation. In Japan, while management and agile-related roles show interest, the position of "InnerSource specialist" is not yet common, and while developers from junior to senior levels are broadly involved, the number of practitioners remains limited. This situation suggests that the implications of "InnerSource" may subtly differ depending on regional and corporate InnerSource maturity levels.

However, it would be premature to definitively categorize InnerSource into two distinct types: " internalization of open source practices" and "packaged methodology for collaboration." Rather, it should be understood as a situation where implementation purposes, corporate maturity, and historical background interact in complex ways. A closer examination of actual implementation backgrounds, challenges, and types of barriers may provide clearer context for how InnerSource functions.

The following chapters analyze how these regional and organizational scale differences influence implementation barriers and examine the specific success models they generate, developing a more comprehensive discussion about the functions and positioning of InnerSource.

# 5 A Multifaceted Analysis of InnerSource Recognition and Adoption Motivations

InnerSource has the potential to bring significant transformation to organizational software development processes and cultures. Since the early 2020s, efforts to introduce InnerSource have been accelerating, particularly among global companies. However, perceptions and motivations for its adoption vary depending on regional and organizational characteristics, as well as on the level of acceptance of InnerSource culture and processes. This chapter presents a comprehensive analysis of the awareness and drivers of InnerSource adoption in Japanese and global enterprises.

The discussion first explores common understandings and diverse interpretations of InnerSource definitions, followed by a comparative examination of awareness patterns in Japanese versus global organizations. Subsequently, motivations for participation in InnerSource are analyzed in detail, and the changes in adoption motivations across different implementation stages are tracked. Through these analyses, this chapter aims to elucidate organizational and cultural challenges in InnerSource adoption and to provide insights for more effective promotion.

## 5.1 Basic Concepts of InnerSource and the Diversity in Definition

Analysis of free-text responses regarding the definition of InnerSource revealed several intriguing tendencies concerning perceptions and anticipated benefits. While most respondents indicated an understanding of the basic framework— "applying open source practices internally"—there was a noticeable skew in how InnerSource was interpreted and the outcomes expected from it.

*Table 5.1 Frequently Observed Words in Definitions of InnerSource*

| Normalized Word | Frequency |
|---|---|
| open source | 41 |
| organization | 29 |
| collaboration | 25 |
| practices | 21 |
| development | 19 |
| code | 13 |
| company | 12 |
| contribution | 12 |
| InnerSource | 12 |
| software | 12 |
| culture | 10 |
| internal | 10 |

Specifically, frequent references to keywords such as "collaboration," "practices," "contribution," "culture," and "internal" were identified, whereas mentions of "innovation" or "creativity" were relatively few. This tendency suggests that InnerSource is primarily understood in a practical context focused on leveraging existing resources and strengthening in-house collaboration, whereas the creative value-creation aspect typically associated with open source communities is not necessarily regarded as central to InnerSource.

Such a perception bias implies that the element of spontaneous problem-solving and the generation of novel value—hallmarks of open source communities—may not be sufficiently emphasized when InnerSource is introduced within enterprises. Because interpretations of the InnerSource concept can vary widely by organization or individual, systematically identifying its essential elements is an urgent task. In this study, co-occurrence network analysis based on text mining was employed to visualize and structure various definitions and objectives for InnerSource. The analysis identified five primary conceptual clusters: "Applying Open-Source Methodologies," "Applying Methods Internally," "Knowledge Sharing," "Preventing Wheel Reinvention," and "Cultural Best Practices."

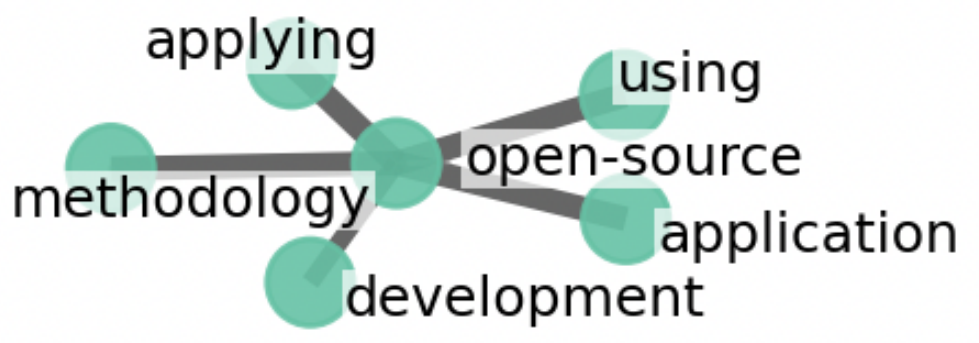

*Figure 5.1 Co-occurrence Analysis Result for "open source"*

**Applying Open-Source Methodologies**

e.g. InnerSource is enterprise in-house open source methodology on product development and collaboration.

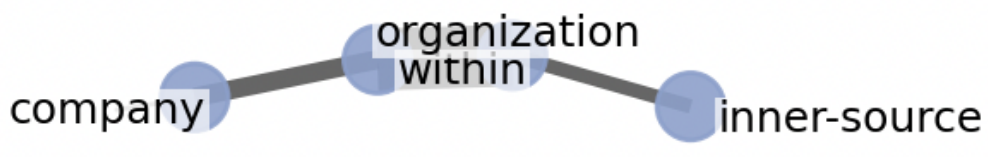

*Figure 5.2 Co-occurrence Analysis Result for "within organization"*

**Applying Methods Internally**

e.g. It's like open source, just everything stays within the organization.

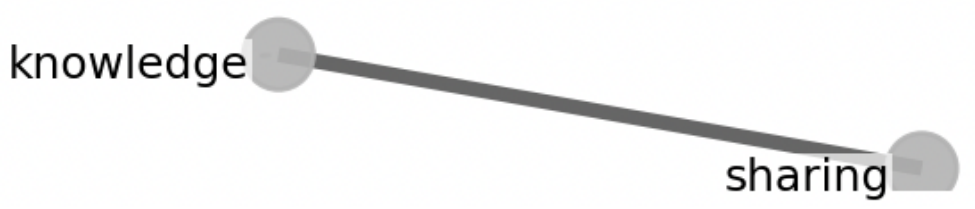

*Figure 5.3 Co-occurrence Analysis Result for "knowledge-sharing"*

**Knowledge Sharing**

e.g. Promoting culture of collaboration, fostering voluntary contribution within the organization, and enabling knowledge sharing.

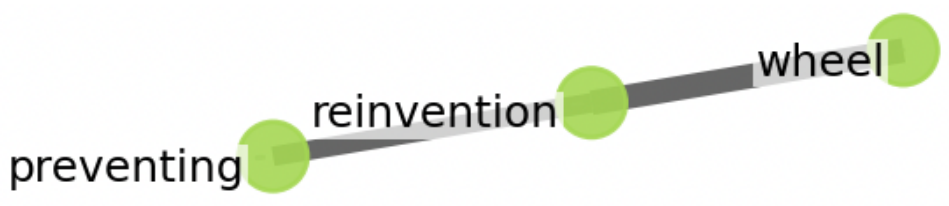

*Figure 5.4 Co-occurrence Analysis Result for "preventing wheel reinvention"*

**Preventing Wheel Reinvention**

e.g. Collaboration across the organization and preventing the reinvention of the wheel.

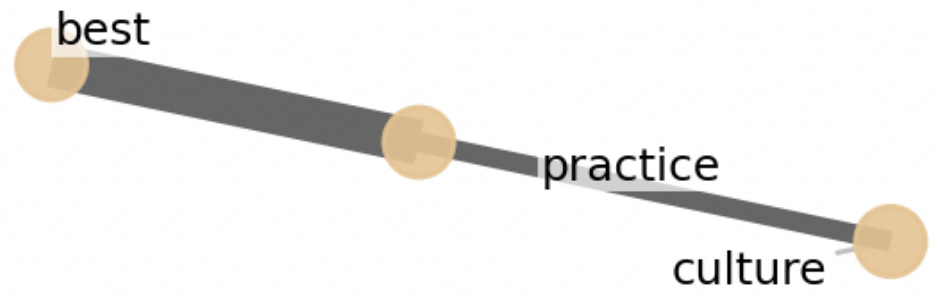

*Figure 5.5 Co-occurrence Analysis Result for "culture best practice"*

**Cultural Best Practices**

e.g. Application of open source best practices, culture and methodology for Enterprise-internal projects and collaboration.

Thus, InnerSource is expected to function as a comprehensive mechanism for organizational transformation—extending beyond the mere technical adaptation of open source practices—encompassing cultural change and innovative approaches to knowledge management. However, it should be noted that the findings may be limited by sample size constraints, which could affect the exhaustiveness of the extracted co-occurrence relationships.

## 5.2 Awareness Structure of InnerSource and Regional Differences

An analysis of perceptions of InnerSource revealed notable differences between Japan and other global regions. These differences reflect regional characteristics in the understanding of and expected value from InnerSource.

Respondents to the global survey tended to view InnerSource as a practical and concrete means of generating value. Interest in "software/source code reuse" was remarkably high at 98.3%, and there was pronounced emphasis on practical aspects such as contributing to projects on which an organization depends, even without formal involvement. This suggests that InnerSource has taken root as a specific mechanism to improve software development efficiency within organizations, evolving naturally from existing open source practices.

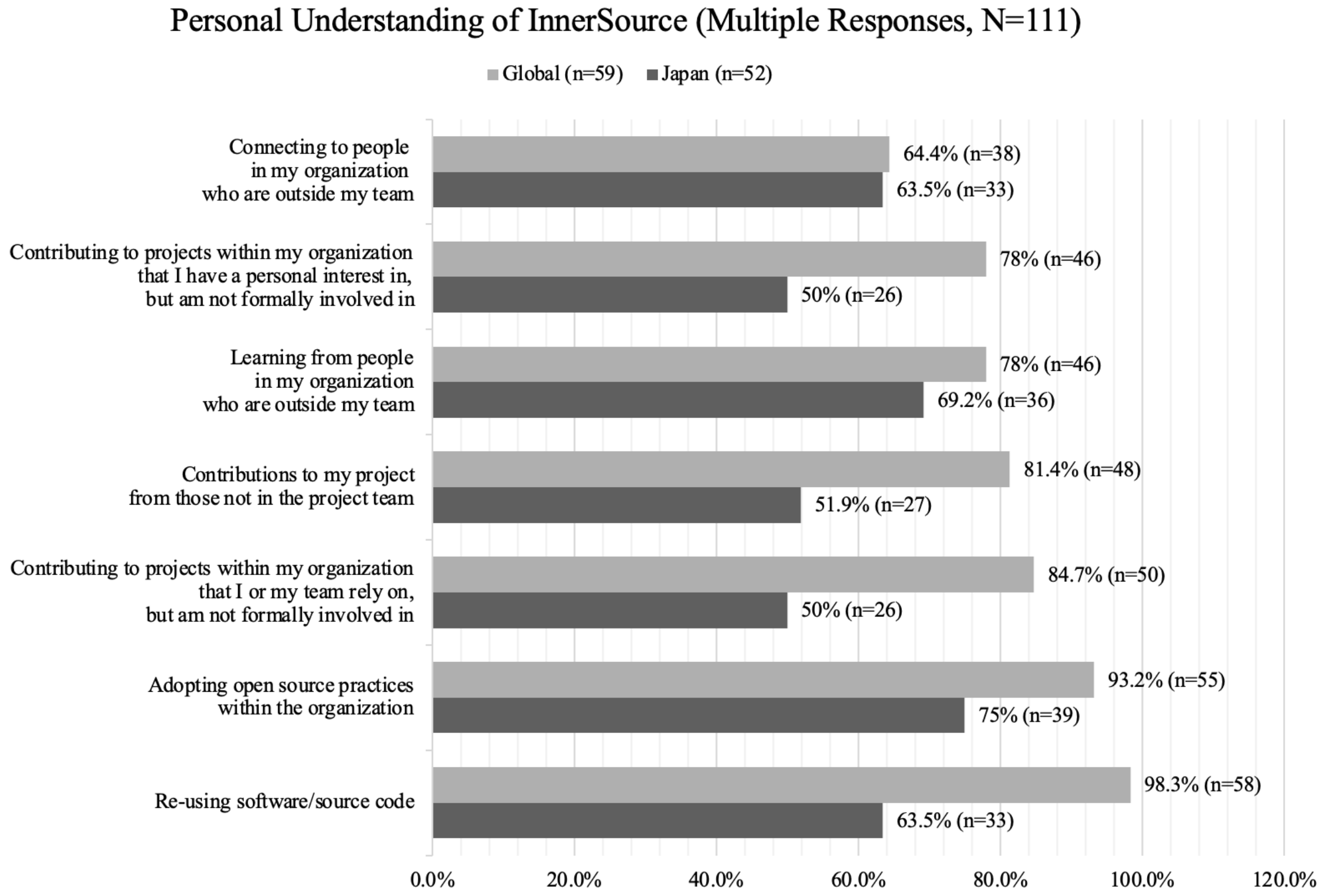

*Figure 5.6 Personal Recognition of InnerSource*

In contrast, Japanese respondents particularly emphasized elements related to “learning” and “connections,” reflecting an emphasis on intangible and relational aspects. Notably, interest in “connecting with people within the organization” was almost at the same level as that found in the global survey. However, the perceived value of “contributing to projects” was relatively low. This outcome implies that, even if there is a desire to remove organizational barriers, neither clear images nor established practices for actual contribution and collaboration are sufficiently developed.

As described at the end of Chapter 4, this phenomenon supports the existence of two phases: InnerSource as an “internalization of open source practices,” and InnerSource as a “packaged methodology.” The recognition patterns in Japan strongly reflect the latter, aligning with the timeline after 2022, when translations and the dissemination of InnerSource-related information accelerated.

Particularly noteworthy is the difference in recognition of “applying open source practices internally,” an aspect closely related to the core definition of InnerSource. Although this factor fundamentally underpins InnerSource, perceptions of its importance varied greatly among Japanese respondents. This finding suggests that in Japan, there is a strong tendency to conceive of InnerSource more abstractly, as a tool for organizational transformation and innovation.

Such differences underscore the necessity of a staged approach to InnerSource adoption. In the Japanese context, dismantling organizational silos and fostering an open culture of collaboration may constitute an initial priority before focusing on specific quantitative outcomes. This can be interpreted as a foundational phase of cultural development preceding the pursuit of direct productivity outcomes.

## 5.3 Comprehensive Analysis of Motivation for Participating in InnerSource

A quantitative analysis on motivations for participating in InnerSource revealed both universal tendencies across regions and region-specific characteristics. The most prominent common feature was a strong emphasis on "knowledge sharing" in both Japanese and global groups. “Remove silos & bottlenecks” also received high ratings in both regions, indicating that InnerSource’s fundamental value proposition of fostering cross-organizational information exchange is widely recognized across geographical and cultural boundaries.

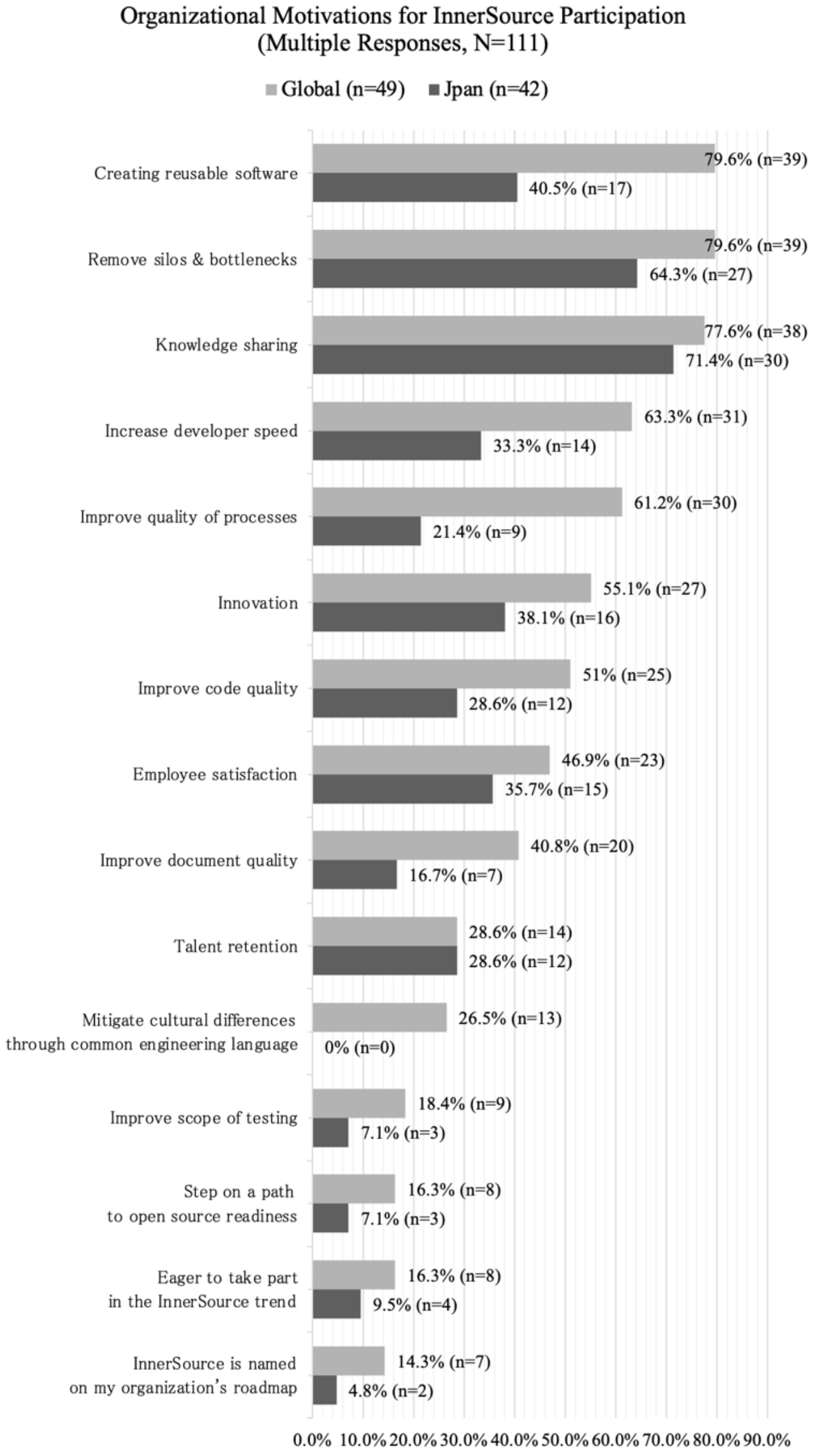

*Figure 5.7 Motivations for Organizations to Participate in InnerSource*

A notable difference was observed in the evaluations related to “creating reusable software.” Respondents from mature-stage global companies were distinctly inclined to select motivations tied to promoting reuse of software, accelerating development speed, and improving code quality—all tangible and quantifiable technical outcomes. This suggests that InnerSource is perceived as a concrete method for optimizing technical assets and enhancing productivity. In contrast, respondents from early-stage Japanese companies placed a comparatively stronger emphasis on the cultural and organizational benefits of knowledge sharing and the elimination of silos, rather than on measurable technical advantages.

These findings align with the earlier analysis of “Personal Recognition of InnerSource” (Figure 5.6). In other words, those at an early stage see InnerSource less as a direct means of improving productivity metrics and more as a catalyst for deepening communication and collaboration within the organization and instigating cultural change. This discrepancy likely stems from differences in organizational maturity, development culture, and strategic priorities, highlighting the need for diverse approaches to InnerSource adoption.

Another point of interest is that only a limited number of respondents in both groups cited "stepping on a path to open source readiness" as a primary motivation for introducing InnerSource. This indicates that InnerSource is not necessarily positioned as an intermediate step toward participation in open source communities. Rather, it can be interpreted as an independent, strategic initiative for organizational efficiency and cultural improvement.

Overall, although knowledge sharing and the enhancement of cross-organizational collaboration form the common foundation of InnerSource motivations, early-stage groups prioritize cultural and relational improvements, while mature-stage groups focus on technical and operational efficiency metrics. This difference reflects variations in organizational maturity, development culture, and managerial priorities. The findings suggest that a flexible approach, tailored to the maturity level and cultural environment of each organization, is needed for successful InnerSource adoption.

## 5.4 Changes in InnerSource Adoption Motivations by Implementation Stage

The adoption of InnerSource within organizations appears to exhibit a pattern where motivations evolve through distinct stages—Ideas Stage, Pilot Stage, Early Adoption Stage, and Growth/Mature Stage—corresponding to organizational maturity. This analysis attempts to capture the comprehensive dynamics of InnerSource adoption motivations in 2024 by examining both Japanese and global response data in a complementary manner. While the Japanese dataset contains limited responses from the Growth and Mature stages, and the global dataset shows fewer responses from the Ideas stage, the integration of both datasets enables a continuous overview across all adoption stages. The findings suggest that early stages tend to focus on expectations of fundamental benefits such as knowledge sharing and silo removal, while later stages demonstrate increased awareness of more specific and practical advantages, such as improvements in code quality and documentation quality.

This transitional process may reflect how the role of InnerSource gradually expands through progressive adoption stages, potentially permeating both corporate culture and the overall development process. The following sections provide detailed explanations of both global and Japanese data, concluding with an integrated analysis. The analysis focuses on top-ranked items, excluding those that did not exceed 40% in any of the overall data.

### 5.4.1 Characteristics of Adoption Stages and Motivations in Global Enterprises

An overview of global responses (n=45) indicates that motivations for introducing InnerSource significantly increase from the Pilot Stage through Early Adoption and on to the Growth/Mature Stage. Although only three responses were gathered for the Ideas Stage (precluding statistical generalization), the transition from Pilot to Growth/Mature Stage reveals a consistent emphasis on “remove silos & bottlenecks,” “creating reusable software,” and “knowledge sharing,” alongside a rising focus on technical aspects such as “improve quality of processes,” “improve code quality,” and “improve document quality.” These trends suggest that while immediate benefits, such as operational efficiency and leveraging existing resources, are prioritized in the early stages, organizations with accumulated InnerSource experience increasingly seek mechanisms to improve development quality over the long term.

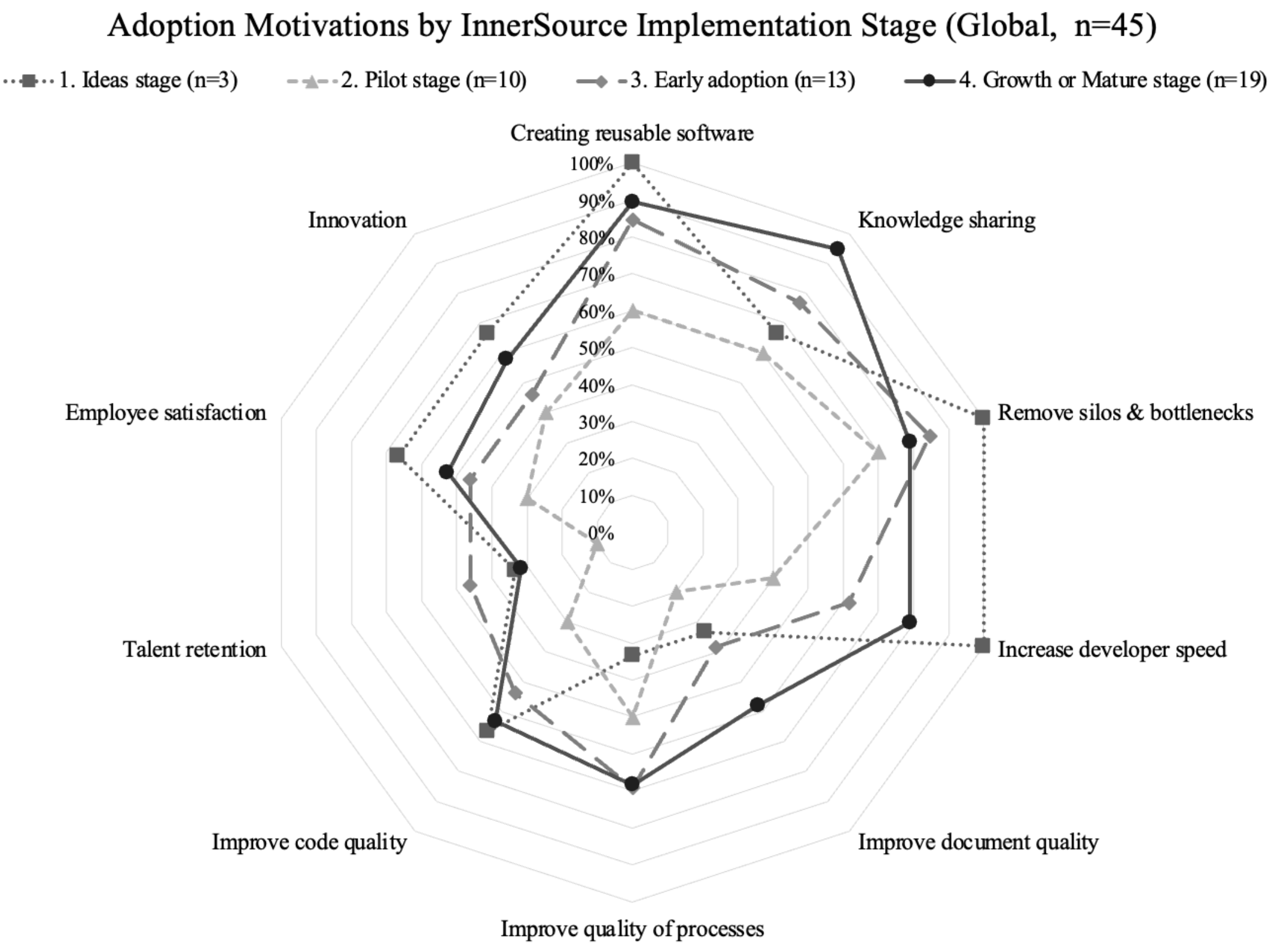

*Figure 5.8 Adoption Motivations by InnerSource Stage (Global)*

*Table 5.2 Adoption Motivations by InnerSource Stage (Global)*

| Ideas Stage (n=3) | Pilot Stage (n=10) | Early Adoption Stage (n=13) | Growth/Mature Stage (n=19) |
|---|---|---|---|
| Remove silos & bottlenecks (100%) | Remove silos & bottlenecks (70%) | Remove silos & bottlenecks (84.6%) | Knowledge sharing (94.7%) |
| Creating reusable software (100%) | Creating reusable software (60%) | Creating reusable software (84.6%) | Creating reusable software (89.5%) |
| Increase developer speed (100%) | Knowledge sharing (60%) | Knowledge sharing (76.9%) | Remove silos & bottlenecks (78.9%) |
| Innovation (66.7%) | Improve quality - processes (50%) | Improve quality - processes (69.2%) | Increase developer speed (78.9%) |
| Eager to take part in the InnerSource trend (66.7%) | Innovation (40%) | Increase developer speed (61.5%) | Improve quality - processes (68.4%) |
| Improve code quality (66.7%) | Increase developer speed (40%) | Improve code quality (53.8%) | Improve code quality (63.2%) |
| Employee satisfaction (66.7%) | | Innovation (46.2%) | Innovation (57.9%) |
| Knowledge sharing (66.7%) | | Talent retention (46.2%) | Improve document quality (57.9%) |
| | | Employee satisfaction (46.2%) | Employee satisfaction (52.6%) |

Table 5.2 shows selected motivations that garnered more than 40% of responses at each stage in the global data. The absolute percentages for each item tend to rise from the Pilot to Growth/Mature Stage, and the scope of interests also appears to widen.

When comparing the global sample across stages, radar charts reveal that the area of evaluated items expands as organizations progress from Pilot to Early Adoption and eventually to the Growth/Mature Stage. Specifically, while the Ideas Stage may be characterized by rather vague expectations for InnerSource's potential benefits, during the Pilot Stage, practical validation efforts tend to concentrate on particular business challenges (such as removing silos or software reuse). In the subsequent Early Adoption and Growth/Mature Stages, successful implementation projects may spread to other departments, potentially facilitating organization-wide process improvements and the establishment of code review cultures. This progression suggests that awareness of quality enhancement and process optimization becomes more pronounced.

It is noteworthy that some respondents in the Ideas Stage expressed broad expectations for InnerSource. In cases where top-down directives seek to enhance organizational agility and foster innovation, the Pilot Stage may demonstrate a tendency toward narrowing the focus to more demonstrable objectives, thereby temporarily constraining motivations. As projects expand, however, awareness appears to broaden again, potentially leading to a phase where multiple benefits are valued. To better understand these early-stage fluctuations, comparing differences between implementation stages using Japanese sample data may prove beneficial.

### 5.4.2 Actual Adoption and Characteristic Motivations in Japanese Enterprises

Japanese responses (n=42) showed an opposite distribution, with a large proportion in the Ideas Stage and very few in the Growth/Mature Stage. Consequently, although a general level of interest in InnerSource is evident at the earliest stage, only a small number of organizations have reached Growth or Mature stages. Moreover, in the Pilot or Early Adoption Stage, Japanese respondents reported relatively lower motivation rates compared to the global sample, with high marks on specific items like “knowledge sharing” and “remove silos & bottlenecks,” but not necessarily a strong drive toward technical or operational motivations such as “improve code quality,” “improve document quality,” or “increase developer speed.”

This outcome likely stems from the tendency among many Japanese companies to prioritize organizational culture and existing quality management processes, conducting only minimal trial deployments of InnerSource. Under organizational structures common in Japan—characterized by clearly defined roles, team formations, and sometimes long-term job rotations—the benefits of InnerSource may be more difficult to measure, prompting companies to start with small-scale implementations aimed at improving knowledge sharing or interdepartmental collaboration. Consequently, pilot or early adoption stages may see a certain level of acceptance for knowledge sharing and silo removal, yet momentum toward recognizing broader technical benefits appears to stall.

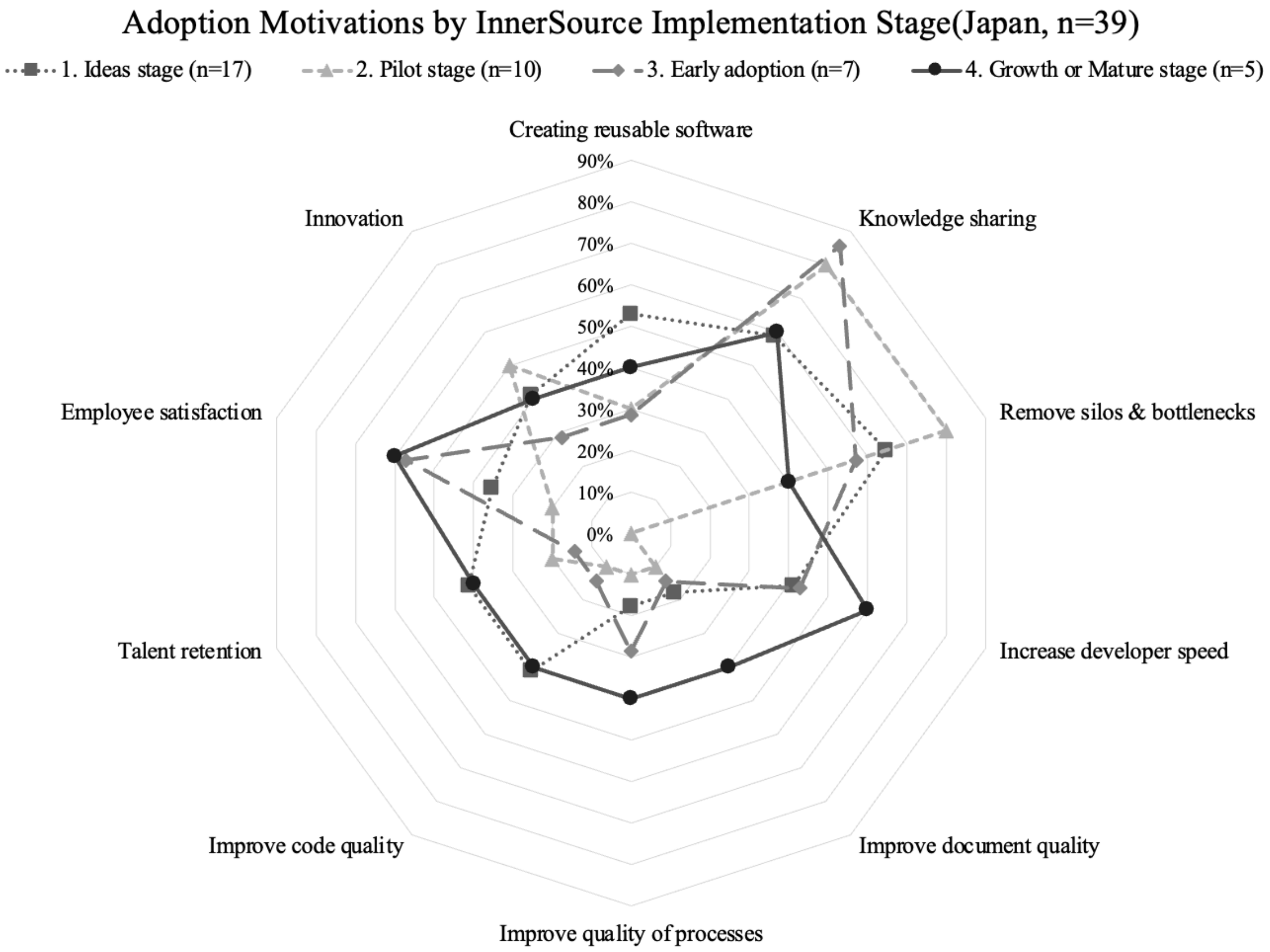

*Figure 5.9 Adoption Motivations by InnerSource Stage (Japan)*

Table 5.3 provides an overview of the stages in Japanese data. Compared alongside global trends, it suggests that although InnerSource is not yet widely understood in Japan, there is a tendency to narrow down key motivations during the early stages of introduction. Moreover, in Growth or Mature organizations, there is an emphasis on enhancing employee satisfaction, yet the potential technical advantages of reuse do not necessarily come to the forefront.

*Table 5.3 Adoption Motivations by InnerSource Stage (Japan)*

| Ideas Stage (n=17) | Pilot Stage (n=10) | Early Adoption Stage (n=7) | Growth/Mature Stage (n=5) |
|---|---|---|---|
| Remove silos & bottlenecks (64.7%) | Remove silos & bottlenecks (80%) | Knowledge sharing (85.7%) | Employee satisfaction (60%) |
| Knowledge sharing (58.8%) | Knowledge sharing (80%) | Remove silos & bottlenecks (57.1%) | Knowledge sharing (60%) |
| Creating reusable software (52.9%) | Innovation (50%) | Employee satisfaction (57.1%) | Increase developer speed (60%) |
| Innovation (41.2%) | | Mitigate against cultural differences through shared engineering language (42.9%) | Innovation (40%) |
| Improve code quality (41.2%) | | Increase developer speed (42.9%) | Improve code quality (40%) |
| Talent retention (41.2%) | | | Remove silos & bottlenecks (40%) |
| Increase developer speed (41.2%) | | | Improve scope of testing (40%) |
| | | | Improve document quality (40%) |
| | | | Improve process quality (40%) |

An analysis of responses from organizations that have reached the Growth/Mature Stage indicates a marked recognition of social or cultural benefits, such as improving employee satisfaction and facilitating corporate cultural transformation. This suggests that in organizations where InnerSource adoption is significantly advanced, it is evaluated more in terms of organizational change and talent management than from purely technical perspectives. However, caution is warranted in generalizing these findings due to the small number of Growth or Mature enterprises in the Japanese sample. Moreover, considering that the technical aspects of InnerSource are not always thoroughly recognized even at the Pilot or Early Adoption Stage, it may be necessary to establish mechanisms to fully capitalize on these technical advantages once adoption progresses further. Additional case studies would offer more detailed insights into these factors.

### 5.4.3 Overall Trend Analysis from Combined Japanese and Global Data

By integrating both Japanese and global data, it is possible to capture adoption motivations continuously from the Ideas Stage to the Growth/Mature Stage. The abundance of Japanese data in the Ideas Stage, combined with the global data in the Growth/Mature Stage, illustrates a consistent pattern: InnerSource adoption initially centers on basic organizational benefits such as knowledge sharing and silo removal. Over time, recognition of advanced technical factors—including process quality, code quality, and documentation quality—rises as InnerSource expands to additional departments and projects, and its effectiveness is verified in diverse use cases.

The observation that the radar chart area increases with each progressing stage suggests an evolution from focusing on localized needs early on to broader contributions to quality control and development velocity. In Japan, organizational-cultural indicators like employee satisfaction and innovation incentives are relatively more prominent, while in the global data, engineering indicators such as developer speed and code quality become pronounced in later stages. These variations likely stem from differences in corporate culture and organizational management philosophies. Global enterprises with substantial large-scale development experience may find it easier to perform quantitative evaluations of technical factors.

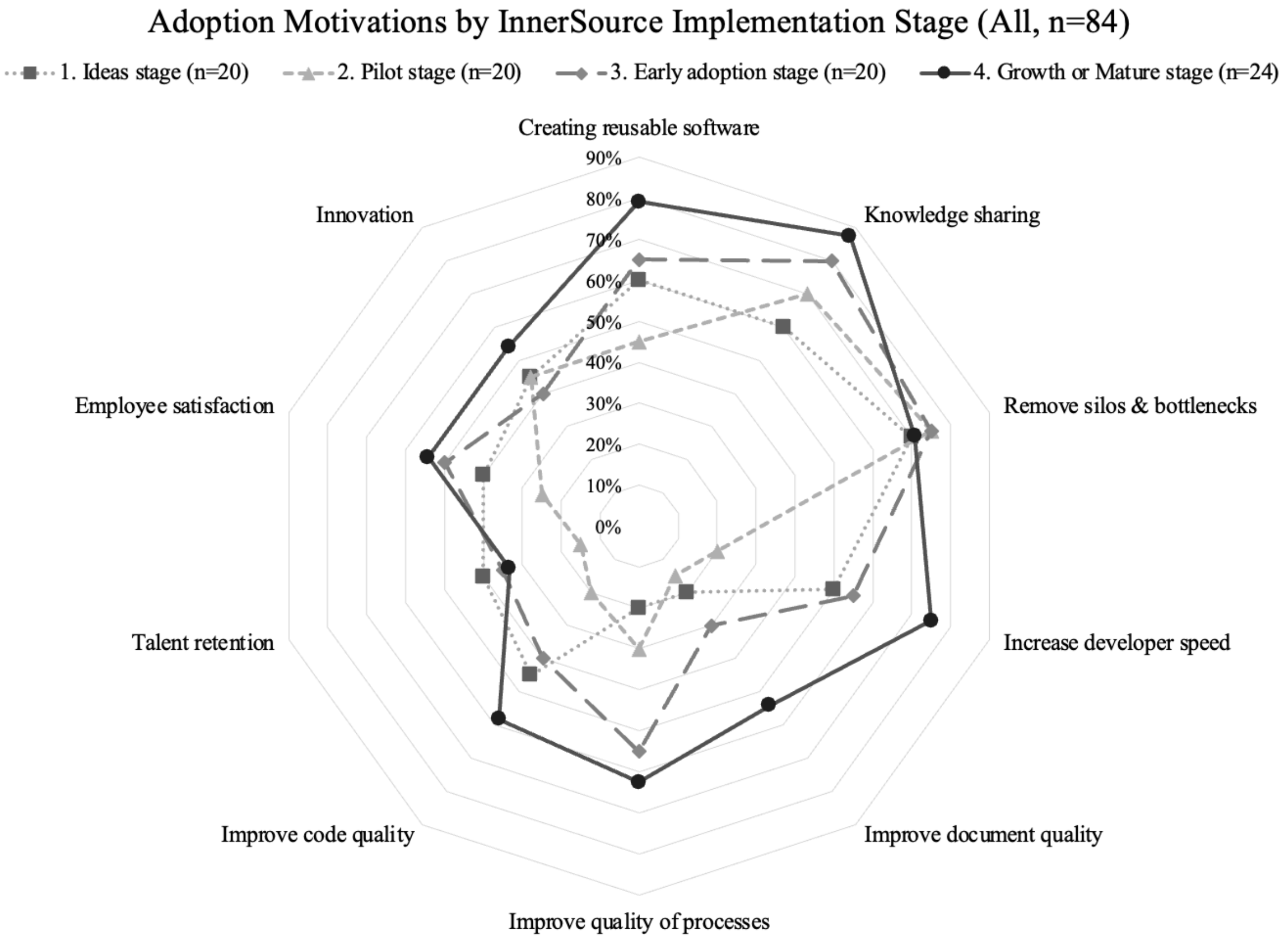

*Figure 5.10 Adoption Motivations by InnerSource Stage (Overall)*

Table 5.4 provides stage-wise data after combining both sets. Organizations that have ultimately reached the Growth/Mature Stage tend to perceive a wider range of benefits, and as InnerSource gains traction, its tangible impacts appear to extend to development speed and quality improvements.

*Table 5.4 Adoption Motivations by InnerSource Stage (Combined)*

| Ideas Stage (n=20) | Pilot Stage (n=20) | Early Adoption Stage (n=20) | Growth/Mature Stage (n=24) |
|---|---|---|---|
| Remove silos & bottlenecks (70%) | Remove silos & bottlenecks (75%) | Knowledge sharing (80%) | Knowledge sharing (88%) |
| Creating reusable software (60%) | Knowledge sharing (70%) | Remove silos & bottlenecks (75%) | Creating reusable software (79%) |
| Knowledge sharing (60%) | Creating reusable software (45%) | Creating reusable software (65%) | Increase developer speed (75%) |
| Increase developer speed (50%) | Innovation (45%) | Increase developer speed (55%) | Remove silos & bottlenecks (71%) |
| Improve code quality (45%) | | Improve quality - processes (55%) | Improve quality - processes (63%) |
| Innovation (45%) | | Employee satisfaction (50%) | Improve code quality (58%) |
| Talent retention (40%) | | Improve code quality (40%) | Improve document quality (54%) |
| Employee satisfaction (40%) | | Innovation (40%) | Employee satisfaction (54%) |
| | | | Innovation (54%) |

Overall, the combined analysis suggests that InnerSource provides increasingly diverse benefits spanning technical and organizational domains as adoption progresses through different stages. Nevertheless, the specific motivations and areas of emphasis vary markedly across countries and organizational cultures. Therefore, while focusing on a narrow set of objectives during the transition from the Ideas Stage to the Pilot Stage and systematically accumulating successful projects may be crucial, adopting a broader view of development quality and organizational reform when moving from Early Adoption to Growth/Mature Stage is key to maximizing the benefits of InnerSource. In Japan, employee satisfaction and human resource strategies tend to be relatively more stressed, whereas globally there is often a sharper emphasis on technical efficiency and quality management. Thus, it is advisable to formulate introduction plans that consider both organizational goals and unique cultural attributes.

## 5.5 Examination of the Relationship Between Adoption Motivations and Attributes

Cross-tabulations were created for both the global (n=49) and Japanese (n=42) samples to clarify how motivations for InnerSource adoption relate to organizational attributes such as company size and development team size. Each motivation item (e.g., “remove silos & bottlenecks,” “innovation”) was placed in rows, and organizational attributes (e.g., total number of employees, size of the development team) were placed in columns. Chi-square tests were performed to evaluate the independence of distributions across these categories, with p-values determining whether any motivation was strongly linked to specific organizational characteristics.

In the global sample, “remove silos & bottlenecks” was found to be significantly related to organizational size. Specifically, organizations with more than 5,000 employees displayed a statistically higher likelihood ($\chi^2=8.39$, $p<.01$) of emphasizing this motivation, suggesting a greater perceived need to address information gaps between departments and promote organization-wide knowledge circulation.

In addition, the global analysis indicated a modest correlation ($p<.10$ level) between “innovation” and both prior InnerSource experience (Pilot Stage or beyond) and development team size (500 or more). This suggests that enterprises with large development teams and cumulative experience in InnerSource may be more inclined to use InnerSource as a catalyst for fostering innovation.

In contrast, the Japanese sample showed a weaker relationship between organizational size and “remove silos & bottlenecks” ($\chi^2=3.02$, $p\approx.08$). Although large Japanese firms also recognize the importance of interdepartmental collaboration, they may not adopt a quantitatively driven approach to utilize InnerSource for solving organizational issues, or adoption stages might differ significantly among firms.

A notable finding in Japan was that “innovation” was significantly correlated with development team size (500 or more) ($\chi^2=7.17$, $p<.05$). Larger development teams are thus more likely to view InnerSource as a means of generating new technologies or new lines of business. This result underscores the possibility that some Japanese enterprises regard InnerSource not only as a tool for organizational reform and personnel development but also as a driving engine for direct business innovation.

**Evaluating independence between InnerSource adoption motives and attributes using chi-square ($\chi^2$) test (N=111)**

| | HR | | Organizational | | | Productivity | | | Quality | | | Strategy | | | |
|---|---|---|---|---|---|---|---|---|---|---|---|---|---|---|---|
| | Talent retention | Employee satisfaction | Remove silos & bottlenecks | Knowledge sharing | Mitigate cultural differences through a shared engineering language | Creating reusable software | Increase developer speed | Improve scope of testing | Improve code quality | Improve quality of processes | Improve document quality | Step on a path to open source readiness | Innovation | Eager to take part in the InnerSource trend | InnerSource is named on my organization's roadmap |
| **Global (n=49)** | | | | | | | | | | | | | | | |
| Organization Size *1 | 0.54 | 1.05 | 8.39 | 0.23 | 0.52 | 0.32 | 0.01 | 0.04 | 1.18 | 1.04 | 0.64 | 4.76 | 0.03 | 0.56 | 0.65 |
| (sig P,* ≤ .05, † ≤ .10) | 0.46 | 0.30 | **0.00*** | 0.63 | 0.47 | 0.57 | 0.91 | 0.83 | 0.28 | 0.31 | 0.42 | **0.03*** | 0.86 | 0.46 | 0.42 |
| Development Organization Size *2 | 0.04 | 0.00 | 0.13 | 0.13 | 0.01 | 1.41 | 1.22 | 0.13 | 0.04 | 0.04 | 1.16 | 2.80 | 2.78 | 0.20 | 0.36 |
| (sig P,* ≤ .05, † ≤ .10) | 0.83 | 1.00 | 0.72 | 0.72 | 0.91 | 0.23 | 0.27 | 0.72 | 0.85 | 0.84 | 0.28 | 0.09† | 0.10† | 0.65 | 0.55 |
| Developer Status *3 | 0.22 | 1.72 | 0.02 | 0.81 | 4.77 | 0.02 | 3.11 | 0.11 | 0.09 | 1.27 | 0.99 | 0.22 | 0.60 | 0.28 | 0.09 |
| (sig P,* ≤ .05, † ≤ .10) | 0.64 | 0.19 | 0.88 | 0.37 | **0.03*** | 0.88 | 0.08† | 0.74 | 0.76 | 0.26 | 0.32 | 0.64 | 0.44 | 0.60 | 0.76 |
| **Japan (n=42)** | | | | | | | | | | | | | | | |
| Organization Size *1 | 1.71 | 0.91 | 3.02 | 0.23 | 0.30 | 0.07 | 0.64 | 0.27 | 0.23 | 0.72 | 0.25 | 0.27 | 1.48 | 1.21 | 1.02 |
| (sig P,* ≤ .05, † ≤ .10) | 0.19 | 0.34 | 0.08† | 0.63 | 0.58 | 0.78 | 0.43 | 0.60 | 0.63 | 0.39 | 0.61 | 0.60 | 0.22 | 0.27 | 0.31 |
| Development Organization Size *2 | 1.30 | 0.13 | 1.10 | 0.06 | 1.08 | 2.06 | 1.61 | 0.02 | 1.30 | 0.29 | 0.13 | 0.02 | 7.17 | 0.33 | 0.16 |
| (sig P,* ≤ .05, † ≤ .10) | 0.25 | 0.72 | 0.29 | 0.81 | 0.30 | 0.15 | 0.20 | 0.90 | 0.25 | 0.59 | 0.71 | 0.90 | **0.01*** | 0.56 | 0.69 |
| Developer Status *3 | 0.79 | 8.28 | 0.00 | 0.01 | 0.01 | 1.08 | 0.06 | 1.15 | 2.77 | 0.09 | 0.62 | 1.15 | 0.74 | 0.00 | 0.75 |
| (sig P,* ≤ .05, † ≤ .10) | 0.37 | **0.00*** | 0.96 | 0.91 | 0.92 | 0.30 | 0.80 | 0.28 | 0.10† | 0.76 | 0.43 | 0.28 | 0.39 | 0.95 | 0.39 |

*1 Organization has 5000 or more employees
*2 Development team has 500 or more members
*3 Role is "Developer" or not

In summary, while large global organizations emphasize removing organizational silos as a key motivator, large Japanese organizations and development teams appear to prioritize the innovation potential of InnerSource. These differences highlight how Japanese firms, which often place strong emphasis on employee satisfaction and organizational culture, also see significant potential for generating new development fields when large technical teams are involved.

These quantitative findings serve as critical supplementary information for understanding how companies decide to adopt InnerSource. In particular, large organizations or those with sizable development teams often have distinct needs—such as removing silos or promoting innovation—that strongly drive early stages of InnerSource adoption.

### 5.5.1 Summary of InnerSource Adoption Drivers

Motivations for adopting InnerSource gradually shift as enterprises progress from the ideas stage, through pilot and early adoption, and eventually into the growth and mature stages. Initial concerns focus on addressing urgent organizational issues—such as removing silos, bottlenecks, and promoting knowledge sharing. Once pilot projects have yielded tangible results, attention turns to development quality and speed. Finally, in the growth and mature stages, the range of perceived benefits expands to include innovation and employee satisfaction, highlighting the broader organizational and cultural impact of InnerSource.

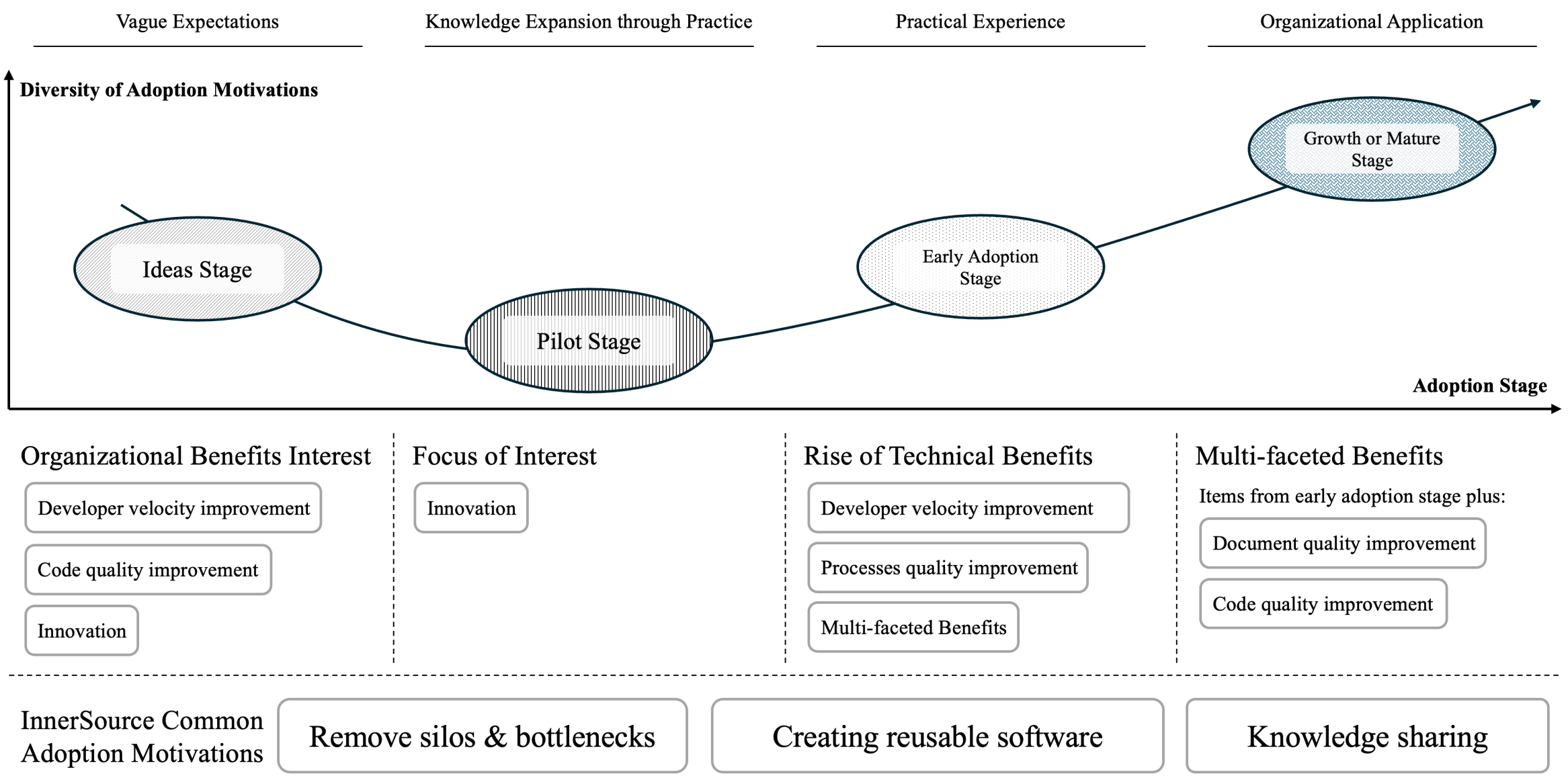

*Figure 5.11 Changes in Motivations for InnerSource Adoption and Emergence of Diverse Drivers with Accumulated Experience*

These stage-based transformations also differ depending on company attributes. Larger organizations often prioritize removing silos, while enterprises with large development teams are more likely to regard InnerSource as a means of innovation. In Japan, managers tended to place importance on improving employee satisfaction, although this concern was not always mirrored among rank-and-file developers.

Cross-tabulations and chi-square tests further showed that large enterprises and organizations with extensive development teams prominently consider preparing for open source engagement in the global dataset, whereas no direct link with company size was observed in Japan. This discrepancy mirrors the differential priorities of foreign corporations (actively promoting open source and external collaboration) versus domestic corporations (not necessarily prioritizing open source participation). Nevertheless, organizations in both regions reported pronounced changes in motivations and expectations depending on their stage of adoption.

Ultimately, InnerSource adoption is driven by a complex interplay of organizational size, development team size, the perspectives or responsibilities of the individuals leading the initiative, and various other factors. Clear objectives and expected outcomes should be defined, followed by the creation of stage-appropriate strategies. Matching these differences in organizational characteristics and maturity levels with consensus-building within the organization is key to maximizing the multifaceted benefits of InnerSource.

# 6 Organizational Barriers to InnerSource Adoption

InnerSource adoption often encounters a variety of organizational barriers, and quantitatively analyzing these barriers provides valuable insights for formulating effective implementation strategies. This chapter presents a comparative analysis of adoption barriers observed in Japanese and global companies, highlighting characteristic patterns and organizational implications.

To ensure the validity of measurement methods, a data standardization process was conducted. Specifically, five-point Likert scale responses collected from Japanese companies were converted into binary data for comparability with binary-format data directly obtained from global companies. In this study, responses categorized as “quite large” or “decisively large” were operationally defined as indicating major adoption barriers.

Statistical tests suggest that these datasets are sufficiently comparable. The mean values for Japanese and global enterprises (0.351 vs. 0.397) and their respective medians (0.372 vs. 0.398) exhibited similar distributions, although a difference in standard deviation (0.116 for Japan vs. 0.180 for global) indicates a more constrained range in the Japanese data. This difference presumably reflects the information-loss effect during the conversion from a Likert scale to binary data.

Despite these differences in measurement characteristics, there was no pronounced divergence in the numeric ranges of the two datasets. Consequently, the ensuing statistical analyses adopt both sets of data to compare and evaluate organizational barriers in each stage of the InnerSource adoption process. This analytical framework proves effective in revealing the commonalities and contrasts in adoption barriers across organizations that operate under distinct cultural contexts.

## 6.1 Overview of Major Barriers in InnerSource Adoption

The survey results suggest that cultural and organizational factors predominate as barriers to InnerSource adoption, regardless of an organization’s size or geographic region. Notably, in the initial stage of adoption, challenges related to changing organizational culture and mindset tend to overshadow purely technical issues. This observation implies that transitioning from traditional hierarchical structures to more open and collaborative organizational forms involve inherent difficulties.

One reason these cultural and organizational factors exert a particularly strong influence is that adopting InnerSource is not merely a matter of introducing new tools or methods. Rather, InnerSource often requires reconfiguring existing organizational structures and business processes. For instance, maintaining a traditional top-down command-and-control hierarchy while simultaneously encouraging the independent contributions and collaboration of individual engineers is not straightforward. Open code-sharing among developers also necessitates revising evaluation criteria and incentive systems, thus necessitating close coordination not only with the team spearheading InnerSource adoption but also with departments such as human resources and legal.

### 6.1.1 Major Barriers in Global Enterprises

In global companies, “organizational culture and silo mentality,” “lack of familiarity with InnerSource principles,” and “lack of middle management buy-in” emerge as the primary barriers to InnerSource adoption. These findings suggest that the rigid boundaries among organizational divisions and hierarchical layers can undermine efforts to internalize an open development culture.

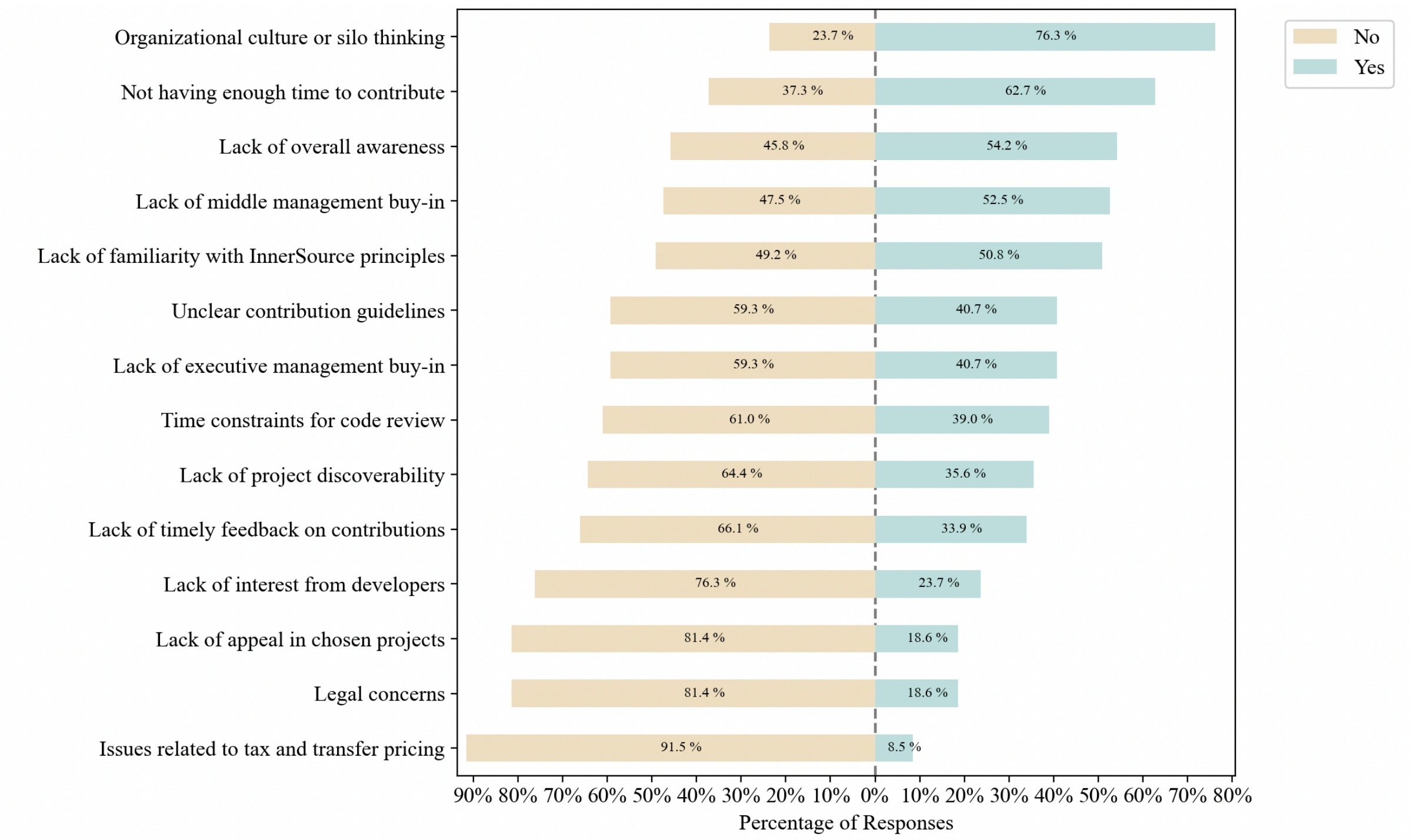

*Figure 6.1. Major Barriers to InnerSource Success (Global)*

One distinct feature among global enterprises is that “lack of interest from developers” is not widely recognized as a key barrier. A possible explanation lies in the higher mobility of engineers and a cultural emphasis on proactively engaging with new technologies and methodologies to advance individual career development. Unlike Japan, where there is a historical tendency toward lifetime employment and rigid career paths, developers in global enterprises often view InnerSource as a means to enhance their skill sets and track record. This environment tends to preserve motivation for voluntary contributions.

Although legal and tax-related risks (e.g., intellectual property rights, transfer pricing) are generally viewed as relatively minor barriers, industries subject to stricter international regulations may nonetheless face serious constraints. Indeed, legal complexities surrounding intellectual property and transfer pricing can introduce significant uncertainty into InnerSource activities in certain regions or industrial sectors.

### 6.1.2 Major Barriers in Japanese Enterprises

Likert-scale analyses indicate that the most substantial barriers to InnerSource adoption in Japanese enterprises are “lack of overall awareness,” “organizational culture and silo mentality,” “lack of interest from developers,” and “lack of middle management buy-in,” each cited as quite large or decisively large by over half of respondents.

Notably, “lack of interest from developers” ranked third and included 26% of respondents who viewed it as “decisively large.” This contrasts sharply with global results. It suggests that engineers in Japan often remain disinterested in collaborative approaches or the adoption of new methods in the context of compartmentalized organizational structures. Contributing factors may include the prevalence of long-term, project-based development and the prominence of system integration service models. In these environments, companies typically undertake projects for client or parent organizations rather than developing in-house products, leaving little room for contribution to InnerSource-based activities. Even when individual developers are highly motivated, a lack of organizational frameworks and processes to support open collaboration can lead to decreased interest and create a vicious cycle.

The prominence of “lack of middle management buy-in” appears rooted in the reliance of Japanese enterprises on consensus building at the mid-management level for decision-making and project execution. However, “lack of executive management buy-in” was also seen as a significant obstacle, underscoring a more fundamental organizational shortfall in overall awareness.

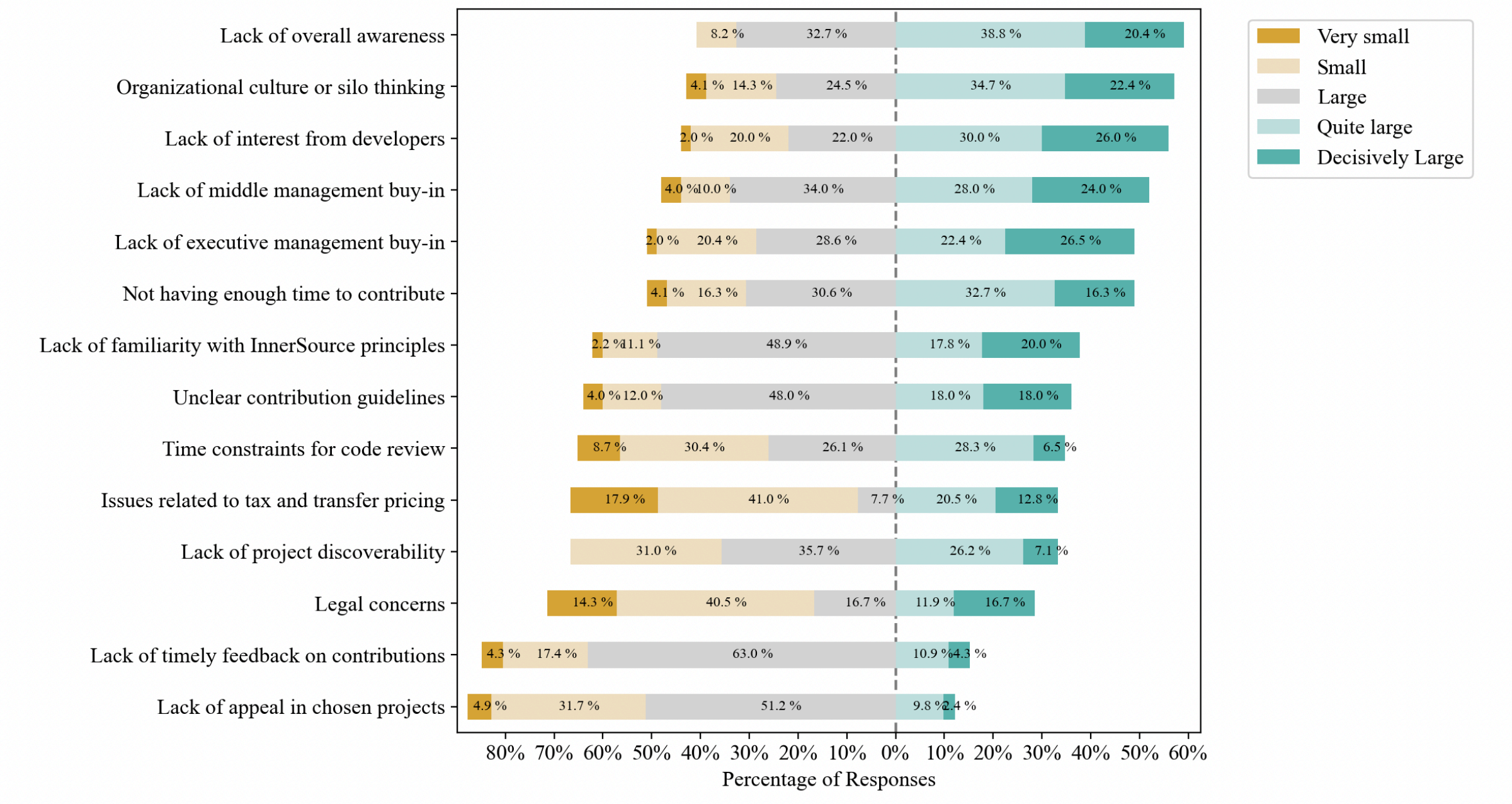

*Figure 6.2. Major Barriers to InnerSource Success (Japan)*

"Not having enough time to contribute" also emerged as a relatively large barrier, whereas "time constraints for reviewing code" and "not getting timely feedback on contributions" garnered lower percentages. This trend supports the inference that InnerSource implementation in Japanese enterprises remains at an early stage. In these initial stages, developers may struggle even to begin contributing, such that downstream issues related to code review and feedback have yet to surface.

Although tax and legal concerns are regarded as relatively minor on average, they cannot be overlooked. If InnerSource activities expand globally and collaboration increases among domestic and international branches, risks involving intellectual property and transfer pricing may intensify. Notably, respondents' assessments of these concerns showed a degree of polarization: some regarded them as critical obstacles, while others viewed them as trivial.

## 6.2 Shifting Barriers Across Different Stages of Adoption

Organizational barriers to InnerSource adoption are not static; they evolve as the adoption process progresses. This study compared Japanese (n=43) and global (n=51) datasets to examine how barriers change in different stages of InnerSource adoption.

The Japanese sample was skewed toward the ideas stage, while the global sample contained more cases in growth and mature stages, rendering a straightforward comparison difficult. Nevertheless, analyzing the shared and distinct features in each stage can elucidate how barriers shift over time. Although the Japanese and global datasets have different compositions, their complementary perspectives facilitate the identification of issues that must be sequentially addressed and those that remain persistent barriers over the entire adoption process.

Overall findings indicate that early-stage adoption is often marked by conceptual hurdles, such as organizational culture and limited awareness. As organizations progress through the adoption process, more operational and logistical challenges emerge. Moreover, although both datasets suggest that cultural factors remain influential, each stage exhibits different points of emphasis in Japan versus global enterprises. The following subsection presents radar charts showing the transition of top items. The analysis focuses on top-ranked items, excluding those that did not exceed 40% in any of the overall data.

### 6.2.1 Stage-Specific Barriers in Global Enterprises

Although the number of respondents in the global sample at the ideas stage is small, the absence of visible benefits of InnerSource often leads to significant barriers, such as incomplete understanding of the concept and lack of internal rules. At this stage, no concrete project has begun, and uncertainty persists regarding which operational guidelines will be established and what tangible outcomes can be expected. Consequently, a variety of potential difficulties are foreseen.

During the pilot stage, "lack of executive management buy-in" and "lack of familiarity with InnerSource principles" become more apparent. While challenges at the ideas stage may have been abstract, the initiation of actual pilot projects transforms them into concrete organizational issues, exposing insufficient endorsement from upper management and inadequate communication or evangelism efforts.

At the early adoption stage, a variety of issues can arise in parallel. "Unclear contribution guidelines," "lack of interest from developers," and "lack of middle management buy-in" may all surface concurrently. Expansion without full coordination with mid-level managers risks diluting motivation to adopt InnerSource on a broader scale.

As organizations transition to growth/mature stages, operational burdens, such as "time constraints for reviewing code" and "not getting timely feedback on contributions," become more prevalent. When an organization manages multiple InnerSource projects at once, collaboration—one of the major benefits of InnerSource—increases, but so do constraints on resources and the complexity of task coordination.

Meanwhile, barriers like “lack of executive management buy-in,” “lack of middle management buy-in,” and “organizational culture and silo mentality” decrease in prevalence. As the value of InnerSource becomes visible, acceptance among executive and middle management tends to grow, mitigating those barriers. InnerSource appears to follow a “learn by doing” dynamic; organizational resistance wanes as tangible benefits materialize, although operational challenges intensify as the number of projects increases.

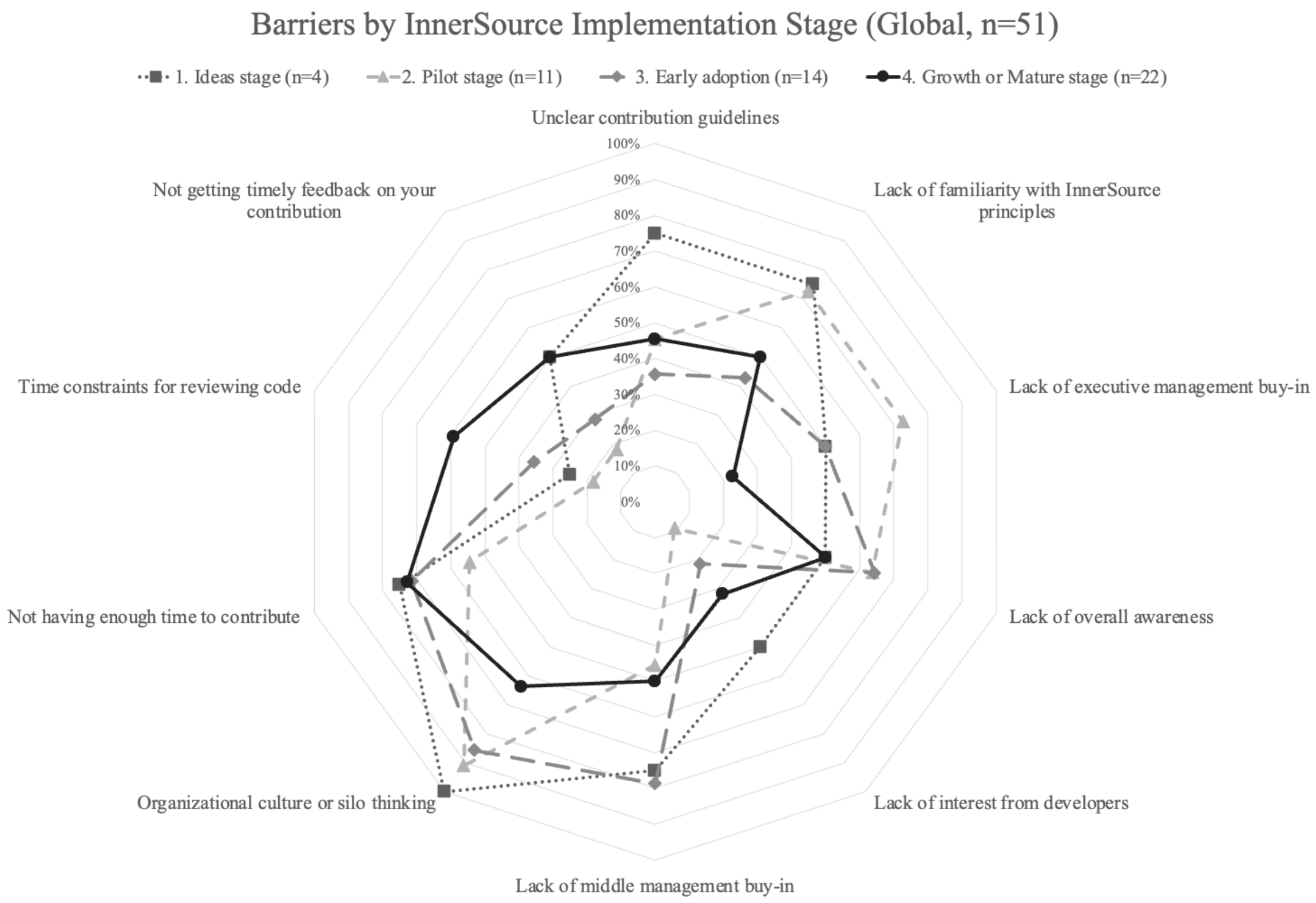

*Figure 6.3. Stage-Specific Barriers in Global InnerSource Adoption*

*Table 6.1. Stage-Specific Barriers in Global InnerSource Adoption*

| Ideas Stage (n=4) | Pilot Stage (n=11) | Early Adoption Stage (n=14) | Growth/Mature Stage (n=22) |
|---|---|---|---|
| Org Culture or silo thinking (100%) | Org Culture or silo thinking (91%) | Org Culture or silo thinking (86%) | Not having enough time to contribute (73%) |
| Unclear contribution guidelines (75%) | Lack of familiarity with InnerSource principles (73%) | Lack of middle management buy-in (79%) | Org Culture or silo thinking (64%) |
| Lack of familiarity with InnerSource principles (75%) | Lack of executive management buy-in (73%) | Not having enough time to contribute (71%) | Time constraints for reviewing code (59%) |
| Lack of middle management buy-in (75%) | Lack of overall awareness (64%) | Lack of overall awareness (64%) | Lack of familiarity with InnerSource principles (50%) |
| Not having enough time to contribute (75%) | Not having enough time to contribute (55%) | Lack of executive management buy-in (50%) | Lack of overall awareness (50%) |
| Lack of executive management buy-in (50%) | Unclear contribution guidelines (45%) | Lack of familiarity with InnerSource principles (43%) | Lack of middle management buy-in (50%) |
| Lack of overall awareness (50%) | Lack of middle management buy-in (45%) | | Not getting timely feedback on your contribution (50%) |
| Lack of interest from developers (50%) | Lack of project discoverability / findability (45%) | | Unclear contribution guidelines (45%) |
| Not getting timely feedback on your contribution (50%) | | | |
| Lack of project discoverability / findability (50%) | | | |
| Chosen projects lack appeal (50%) | | | |

In summary, the global data suggest the following process: at the idea and pilot stages, buy-in from executives, middle managers, and developers remains incomplete. As adoption becomes concrete, operational challenges and guideline shortcomings come to the fore. By the growth/mature stages, feedback loops and time constraints pose significant barriers. Hence, each stage surfaces distinct hurdles that may either be resolved step by step or replaced by new issues as the organization evolves. Once executives and managers begin seeing real outcomes, cultural resistance tends to diminish quickly, while operational issues become more pressing. However, resolving these operational challenges ultimately depends on adequate commitment from both developers and managers, ensuring that management decisions allocate the resources necessary for sustainability.

### 6.2.2 Stage-Specific Barriers in Japanese Enterprises

When Japanese responses are analyzed by adoption stage, the “Ideas stage” is characterized by high-level barriers such as “overall lack of awareness,” “lack of interest from developers,” “organizational culture and silos," and "not having enough time to contribute." Because InnerSource is an externally introduced concept, organizations unaccustomed to it may display unconscious resistance or apathy. This suggests that the main bottleneck at this stage is a poor understanding of what InnerSource actually entails and why it is beneficial, rather than specifics around code review or guidelines. Additionally, Japanese employment practices and hierarchical workplace norms may weaken incentives for developers to voluntarily collaborate across projects.

In the "Pilot stage," the total volume of perceived obstacles decreases relative to the "Ideas stage," potentially because the scope remains limited and serious problems have yet to surface. However, "lack of middle management buy-in" and "not having enough time to contribute" persist as major issues. In hierarchical Japanese organizations, initiatives will not scale successfully unless middle management offers support. Securing top-down endorsement at this early stage is therefore a critical inflection point.

In the "Early adoption stage," challenges that were not fully visible during the pilot stage, such as "unclear contribution guidelines" and "lack of interest from developers," resurface. Even developers who were initially enthusiastic may encounter confusion around ambiguous rules as InnerSource expands across broader organizational units. Because Japanese corporate culture tends to prioritize formal agreements and procedural consistency, insufficient guidelines or accountability mechanisms may undermine sustainability. Overcoming organizational resistance requires not only top-down directives but also deeper collaboration with on-site development teams.

Although the Japanese sample in the "Growth" and "Mature" stages is limited, the most pressing issues at this stage concern operational optimization, such as code reviews and feedback cycles. A unique characteristic of the Japanese data is that "unclear contribution guidelines" and "lack of middle management buy-in" often persist even at these advanced stages. Despite accumulated evidence of success, formalized guidelines and updates to evaluation systems have not always kept pace, leaving organizational structures in which partial success is difficult to scale enterprise wide. Institutionalizing InnerSource in performance reviews or other official frameworks could become a crucial tipping point in the future.

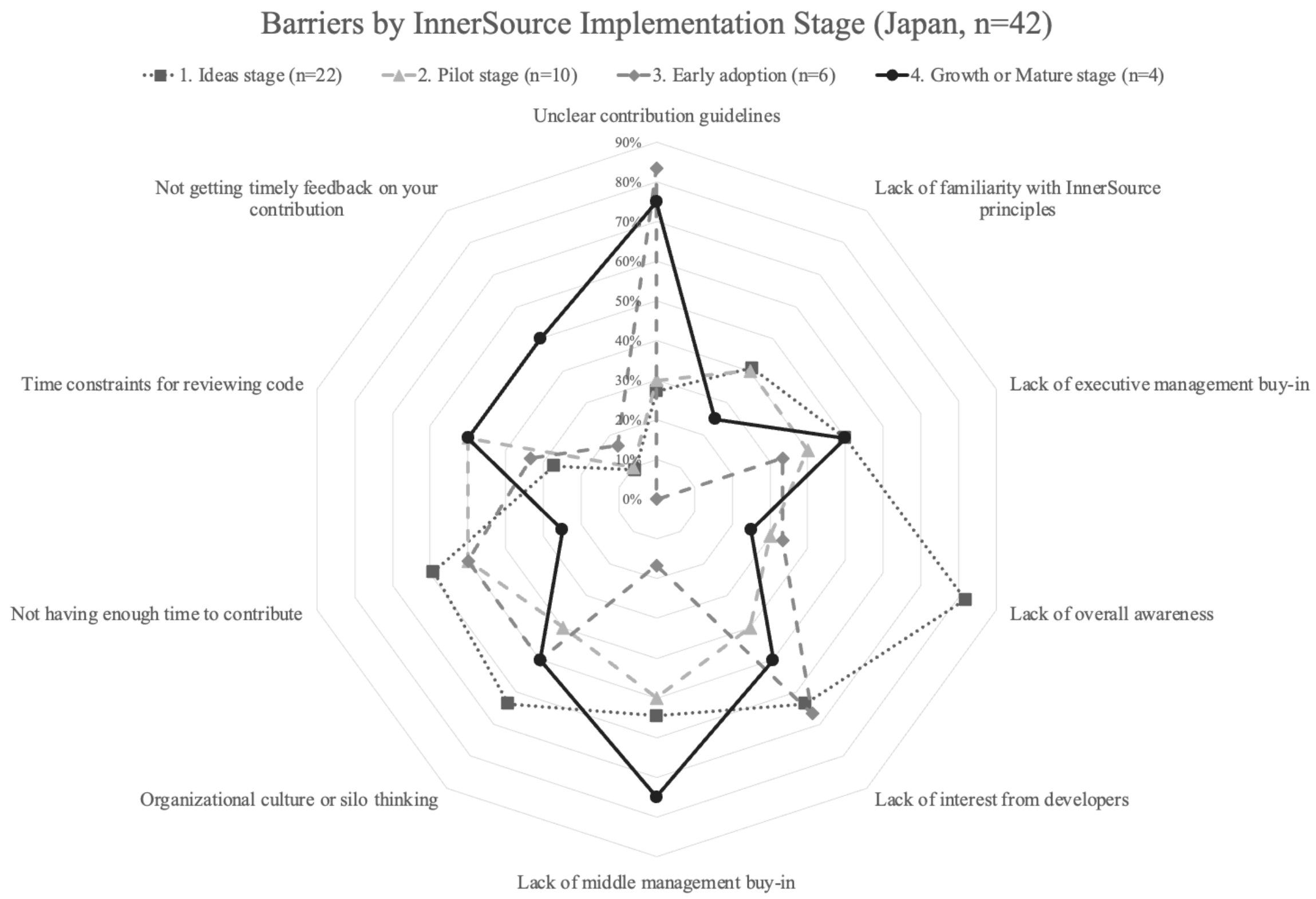

*Figure 6.4. InnerSource Barriers by Adoption Stage (Japan)*

*Table 6.2. Major Barriers to InnerSource by Adoption Stage (Japan)*

| Ideas Stage (n=22) | Pilot Stage (n=10) | Early Adoption Stage (n=6) | Growth/Mature Stage (n=4) |
|---|---|---|---|
| Lack of overall awareness (82%) | Lack of middle management buy-in (50%) | Unclear contribution guidelines (83%) | Unclear contribution guidelines (75%) |
| Lack of interest from developers (64%) | Not having enough time to contribute (50%) | Lack of interest from developers (67%) | Lack of middle management buy-in (75%) |
| Org Culture or silo thinking (64%) | Time constraints for reviewing code (50%) | Org Culture or silo thinking (50%) | Lack of executive management buy-in (50%) |
| Not having enough time to contribute (59%) | Legal concerns (50%) | Not having enough time to contribute (50%) | Lack of interest from developers (50%) |
| Lack of middle management buy-in (55%) | Lack of familiarity with InnerSource principles (40%) | | Org Culture or silo thinking (50%) |
| Lack of executive management buy-in (50%) | Lack of executive management buy-in (40%) | | Time constraints for reviewing code (50%) |
| Lack of familiarity with InnerSource principles (41%) | Lack of interest from developers (40%) | | Not getting timely feedback on your contribution (50%) |
| | Org Culture or silo thinking (40%) | | Lack of project discoverability / findability (50%) |
| | Issues related to transfer pricing (40%) | | |

### 6.2.3 Commonalities and Differences Across Adoption Stages

An integrated analysis of all data (n=93) from both Japan and global samples confirms that each stage of InnerSource adoption exhibits its own distinctive barriers. In the "Ideas stage," overall lack of awareness and developer apathy constitute key challenges. In moving to the "Pilot" and "Early adoption" stages, guideline-related shortcomings and persuading middle management become paramount. Upon reaching the "Growth" and "Mature" stages, operational concerns—such as time constraints for reviewing code, feedback cycles, and resource allocation—predominate.

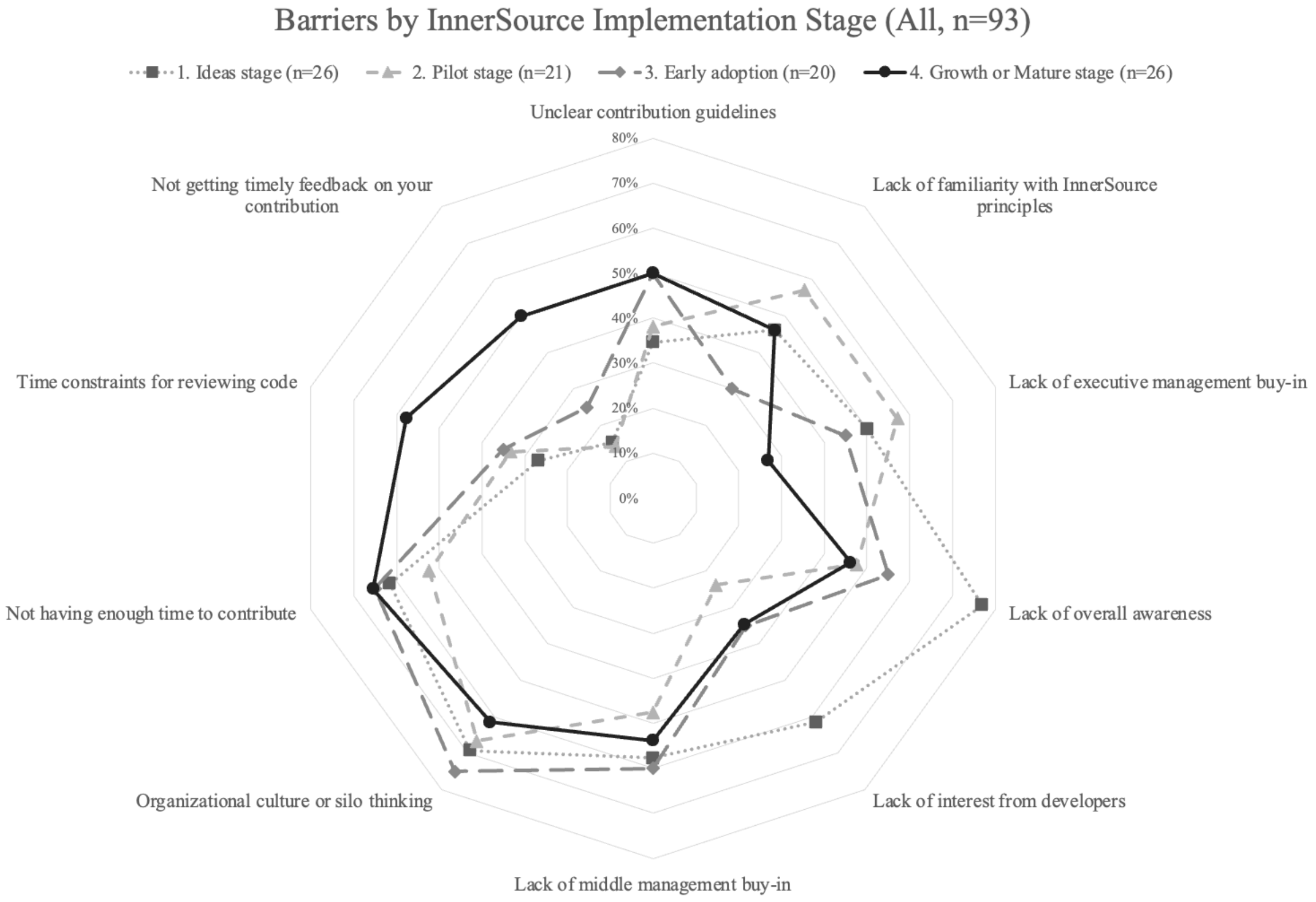

*Figure 6.5. InnerSource Barriers by Adoption Stage (Overall)*

*Table 6.3. Major Barriers to InnerSource by Adoption Stage (Overall)*

| Ideas Stage (n=20) | Pilot Stage (n=20) | Early Adoption Stage (n=20) | Growth/Mature Stage (n=24) |
|---|---|---|---|
| Lack of overall awareness (77%) | Org Culture or silo thinking (67%) | Org Culture or silo thinking (75%) | Not having enough time to contribute (65%) |
| Org Culture or silo thinking (69%) | Lack of familiarity with InnerSource principles (57%) | Not having enough time to contribute (65%) | Org Culture or silo thinking (62%) |
| Lack of interest from developers (62%) | Lack of executive management buy-in (57%) | Lack of middle management buy-in (60%) | Time constraints for reviewing code (58%) |
| Not having enough time to contribute (62%) | Not having enough time to contribute (52%) | Lack of overall awareness (55%) | Lack of middle management buy-in (54%) |
| Lack of middle management buy-in (58%) | Lack of overall awareness (48%) | Unclear contribution guidelines (50%) | Unclear contribution guidelines (50%) |
| Lack of executive management buy-in (50%) | Lack of middle management buy-in (48%) | Lack of executive management buy-in (45%) | Not getting timely feedback on your contribution (50%) |
| Lack of familiarity with InnerSource principles (46%) | | | Lack of familiarity with InnerSource principles (46%) |
| | | | Lack of overall awareness (46%) |

As a whole, the nature of barriers evolves from conceptual and cultural resistance in the early stages to more operational and execution-related problems in later stages. To achieve the benefits of InnerSource on an organizational scale, it is critical to address both sets of challenges via comprehensive measures and agile learning at each step. In Japan, “unclear contribution guidelines” and “lack of middle management buy-in” tend to persist into later stages. This tendency suggests that simply relying on natural organizational learning may not suffice; deeply rooted cultural and institutional factors can remain even after other barriers have been resolved. In contrast, global enterprises sometimes pursue more decisive early-stage reforms, whereas Japanese enterprises often adopt a more gradual, step-by-step approach, which can prolong or delay the resolution of certain barriers.

It is important to note that direct comparisons between Japanese and global samples must be interpreted carefully, as the two groups do not comprise enterprises with identical levels of organizational maturity. However, the overall pattern of barrier progression—moving from initial cultural resistance to subsequent operational problems—emerges consistently, reinforcing the importance of proactive management at each stage.

Solutions to these challenges will be presented through frameworks detailed in Chapter 7 and beyond. It is essential to develop guidelines that promote correct understanding of InnerSource among stakeholders, including management, which serves as the starting point for practical implementation. Simply understanding InnerSource and organizational transformation in abstract terms may result in proceeding without effectively removing barriers. Therefore, it is necessary to accurately redefine InnerSource and enable each stakeholder to identify scope and set goals.

Ultimately, overcoming these barriers necessitates top-down action to establish institutional frameworks, combined with bottom-up innovation on the development floor. From cultural resistance to operational burdens, there are multiple layers of hurdles in InnerSource adoption. In the Japanese context, ensuring that guidelines and middle-management support are not neglected in the later stages of growth and maturity appears essential. Early planning and policy formation, including the integration of InnerSource into organizational evaluation systems, represent critical factors for success.

## 6.3 Statistical Analysis of Adoption Stages and Related Factors

To examine the relationship between the stage of InnerSource adoption and perceived barriers, chi-square tests were conducted on the independence between organizational and respondent attributes and various barrier factors. Cross-tabulation was performed to visualize whether respondents' organizations had progressed beyond the pilot stage, alongside organizational size, development team size, occupational roles, and other attributes.

Analysis revealed that "legal concerns" regarding "tax or transfer pricing" exhibited a statistically significant relationship with organizational size in both Japanese and global enterprises. Specifically, in Japanese enterprises, organizational size ($\chi^2$=5.85, $p<.05$) and development team size ($\chi^2$=6.47, $p<.01$) correlated with concerns about tax and transfer pricing. A similar association ($\chi^2$=4.18, $p<.05$) was observed in global enterprises.

*Table 6.4. Independence Tests for Attributes Related to InnerSource Success Barriers*

**Evaluating independence between InnerSource adoption barriers and attributes using chi-square (χ²) test (N=111)**

| | Interest & Understanding | | | | | Resource Allocation | | | Environment | | | Guidelines | | |
|---|---|---|---|---|---|---|---|---|---|---|---|---|---|---|
| | Lack of familiarity with InnerSource principles | Lack of overall awareness | Lack of executive management buy-in | Lack of middle management buy-in | Lack of interest from developers | Not having enough time to contribute | Time constraints for reviewing code | Not getting timely feedback on your contribution | Organizational culture or silo thinking | Lack of project discoverability / findability | Chosen projects lack appeal | Unclear contribution guidelines | Legal concerns | Issues related to transfer pricing / tax |
| **Global (n=59)** | | | | | | | | | | | | | | |
| Organization's InnerSource Experience *1 | 0.51 | 0.96 | 0.34 | 2.23 | 0.01 | 2.85 | 1.24 | 0.53 | 0.01 | 0.74 | 0.40 | 0.34 | 0.40 | 1.30 |
| (sig P,* ≤ .05, † ≤ .10) | 0.48 | 0.33 | 0.56 | 0.14 | 0.91 | 0.09† | 0.27 | 0.47 | 0.91 | 0.39 | 0.53 | 0.56 | 0.53 | 0.25 |
| Organization Size *2 | 10.89 | 5.74 | 1.36 | 7.04 | 0.59 | 2.87 | 0.85 | 0.19 | 6.70 | 1.31 | 0.73 | 0.34 | 1.92 | 4.18 |
| (sig P,* ≤ .05, † ≤ .10) | **0.00*** | **0.02*** | 0.24 | **0.01*** | 0.44 | 0.09† | 0.36 | 0.66 | **0.01*** | 0.25 | 0.39 | 0.56 | 0.17 | **0.04*** |
| Development Team Size *3 | 1.11 | 1.61 | 0.05 | 1.14 | 0.00 | 1.21 | 0.05 | 4.73 | 4.11 | 0.45 | 0.63 | 0.00 | 2.50 | 2.20 |
| (sig P,* ≤ .05, † ≤ .10) | 0.29 | 0.20 | 0.83 | 0.29 | 1.00 | 0.27 | 0.83 | **0.03*** | **0.04*** | 0.50 | 0.43 | 1.00 | 0.11 | 0.14 |
| Developer Role *4 | 0.07 | 0.00 | 1.08 | 0.27 | 0.23 | 2.11 | 3.45 | 2.57 | 0.23 | 0.41 | 0.66 | 1.08 | 3.10 | 1.25 |
| (sig P,* ≤ .05, † ≤ .10) | 0.79 | 0.98 | 0.30 | 0.60 | 0.63 | 0.15 | 0.06† | 0.11 | 0.63 | 0.52 | 0.42 | 0.30 | 0.08† | 0.26 |
| **Japan (n=52)** | | | | | | | | | | | | | | |
| Organization's InnerSource Experience *1 | 1.84 | 11.91 | 1.09 | 1.21 | 1.77 | 1.04 | 1.11 | 0.63 | 3.06 | 0.79 | 0.04 | 3.34 | 1.16 | 0.46 |
| (sig P,* ≤ .05, † ≤ .10) | 0.18 | **0.00*** | 0.30 | 0.27 | 0.18 | 0.31 | 0.29 | 0.43 | 0.08† | 0.37 | 0.85 | 0.07† | 0.28 | 0.50 |
| Organization Size *2 | 0.09 | 0.99 | 0.19 | 0.06 | 0.59 | 0.19 | 0.46 | 2.01 | 0.01 | 0.64 | 0.00 | 0.34 | 4.09 | 5.85 |
| (sig P,* ≤ .05, † ≤ .10) | 0.76 | 0.32 | 0.66 | 0.81 | 0.44 | 0.66 | 0.50 | 0.16 | 0.92 | 0.43 | 0.98 | 0.56 | **0.04*** | **0.02*** |
| Development Team Size *3 | 1.34 | 0.42 | 1.61 | 0.10 | 0.19 | 0.04 | 0.00 | 1.11 | 0.00 | 0.25 | 0.23 | 0.02 | 5.35 | 6.47 |
| (sig P,* ≤ .05, † ≤ .10) | 0.25 | 0.52 | 0.20 | 0.75 | 0.67 | 0.85 | 0.97 | 0.29 | 1.00 | 0.61 | 0.63 | 0.89 | **0.02*** | **0.01*** |
| Developer Role *4 | 0.00 | 0.01 | 1.05 | 0.55 | 2.61 | 0.01 | 0.58 | 0.06 | 0.00 | 1.39 | 2.60 | 0.15 | 0.53 | 0.92 |
| (sig P,* ≤ .05, † ≤ .10) | 0.95 | 0.94 | 0.31 | 0.46 | 0.11 | 0.93 | 0.45 | 0.80 | 1.00 | 0.24 | 0.11 | 0.70 | 0.47 | 0.34 |

*1 Whether InnerSource implementation stage is "Pilot stage" or later
*2 Organization has 5000 or more employees
*3 Development team has 500 or more members
*4 Role is "Developer" or not

Note: Barriers are binarized (present/not present) for Japan data, where levels 4 and 5 are treated as "yes"

Large organizations, notably publicly traded corporations or those with over 50,000 employees, viewed tax and transfer pricing issues as major obstacles. This pattern reflects the growing need for institutional strategies to address legal risks related to intellectual property and corporate tax codes as organizations scale. Large multinational corporations with complex subsidiary structures attach particular importance to managing legal and tax liabilities.

*Table 6.5. Organizational Size and Perceived Obstacles in Tax or Transfer Pricing (Japan and Global, n=88)*

| **Organizational Size** | No | Yes |
|---|---|---|
| **≤9** | 5 | 0 |
| **10-99** | 3 | 1 |
| **100-499** | 14 | 0 |
| **500-1,999** | 8 | 3 |
| **2,000-4,999** | 14 | 3 |
| **5,000-49,999** | 12 | 3 |
| **≥50,000** | 14 | 8 |
| Total | 70 | 18 |

Interestingly, no significant relationship was found between these legal/tax concerns and an organization's level of InnerSource adoption experience. Even organizations with more extensive InnerSource implementation may continue to harbor anxieties about tax and legal issues, likely because such risks often remain dormant until flagged by regulators. Although collaboration with in-house finance and legal teams can mitigate some concerns, perfect certainty is difficult to achieve. Excessively strict compliance measures may also compromise flexibility, underscoring the need to strike a balance.

Supplementary analyses revealed notable differences by job type in the perception of tax and transfer pricing issues' importance. In the Japanese sample, approximately 46% of respondents who identified these problems as major barriers worked in research and development roles, while only about 8% of those rating them as moderate or lower barriers (3 or below on a 5-point scale) were in R&D. Furthermore, Japanese respondents who recognized these issues as significant were concentrated in technology and manufacturing sectors, suggesting the possibility that such challenges become more prominent in industries where technology serves as a source of competitive advantage.

Existing research addresses many of these complexities in detail, positioning transfer pricing issues as key challenges to be overcome when advancing InnerSource. The present study recognizes that matters of transfer pricing and in-house accounting rules are highly context-specific; hence, Chapter 8 discusses them more comprehensively. Overemphasis on a single issue risk obscuring the broader picture of InnerSource practices and governance models, which must be examined holistically.

Regarding differences by job role, global data indicate only a weak tendency ($\chi^2=3.45$, $p<.10$) for "time constraints for reviewing code" to vary with occupation. Developers perceive review-related time constraints as more pressing than management or business personnel do. This likely arises from developers' daily, hands-on involvement in code quality assurance.

These findings highlight the multilayered nature of barriers to InnerSource. In addition to general barriers—such as lack of executive or middle management buy-in—there are specialized challenges, especially in large enterprises, related to tax and legal compliance. Understanding these interlocking factors and tailoring responses to each organization's characteristics is crucial for successful InnerSource promotion.

## 6.4 Summary of Barrier Progression and Organizational Impact

During InnerSource adoption, the barriers organizations face evolve over time. As illustrated in Figure 6.6, the earliest stages tend to expose primarily technical challenges, but as adoption proceeds, organizational concerns such as obtaining middle management buy-in, clarifying contribution guidelines, securing resources, and demonstrating tangible operational benefits move to the forefront.

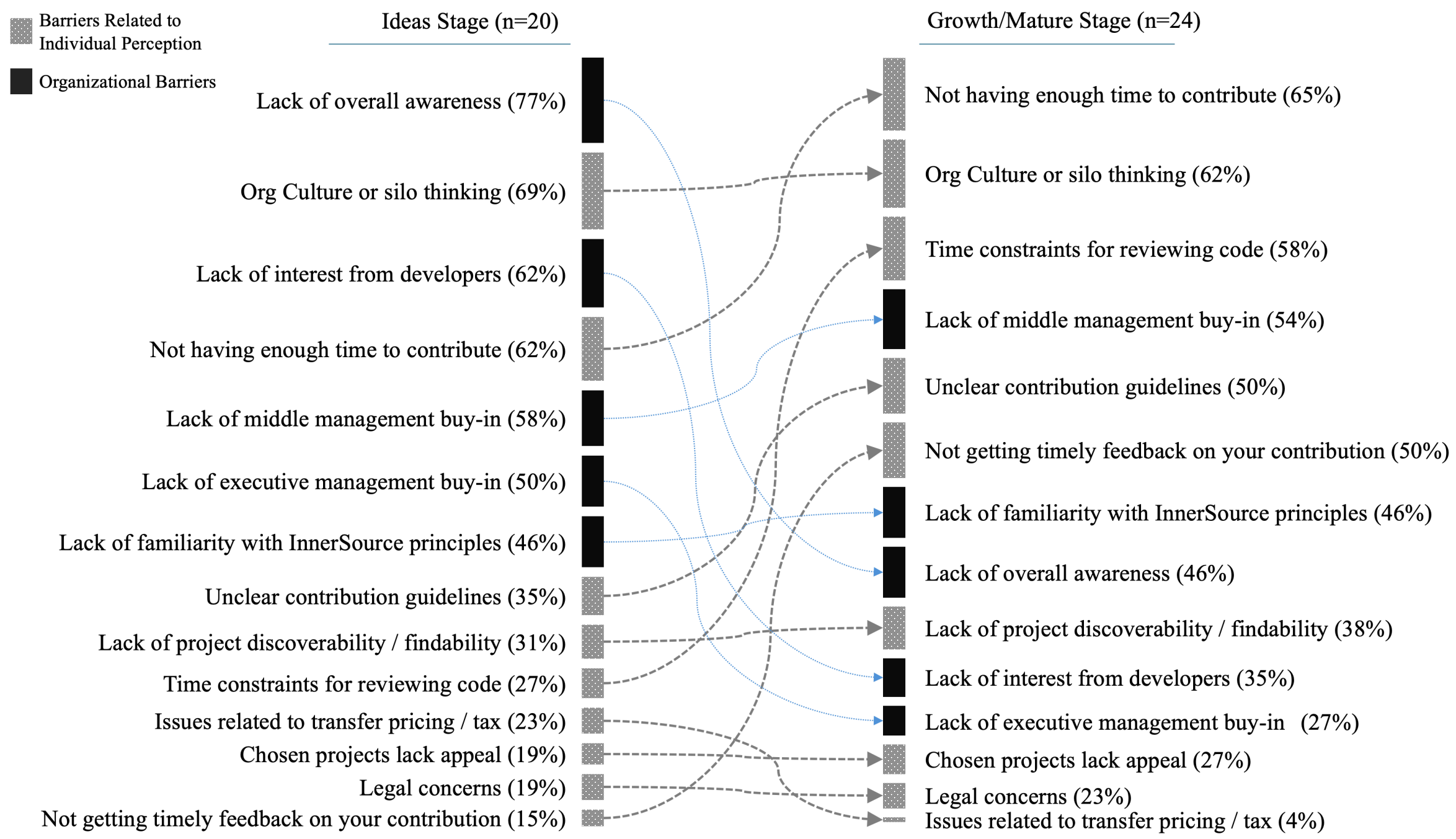

*Figure 6.6. Changes in Barrier Types by InnerSource Adoption Stage*

Existing InnerSource maturity models can be valuable as indicators of evolutionary progress, but they do not specifically aim to identify distinct barriers in each stage or to elaborate how those barriers can be surmounted in practice. Previous InnerSource studies have predominantly focused on particular projects or technical frameworks (e.g., governance models or specific strategies) rather than on the overall implementation process. Consequently, comprehensive analyses addressing mid-development initiatives, transitions within the organization, or the integration of multiple scattered InnerSource activities remain scarce. In particular, there is limited guidance for enterprises with little existing InnerSource infrastructure, for organizations still working to formalize programs or secure executive approval, or for individuals seeking to demonstrate InnerSource's value at an early stage without broad organizational backing.

Furthermore, InnerSource champions are not always formal appointees such as OSPO staff, developer productivity engineers, or GitHub administrators. Rather, individuals from various roles often start grassroots efforts, motivated by belief in InnerSource's potential value. This study aims to complement existing frameworks by constructing a new analytical approach that includes the perspectives of such diverse champions. The following chapters—beginning with Chapter 7—explore in detail the specific barriers these champions encounter and the methods they employ to overcome them.

# 7 Comprehensive Reexamination of Existing Maturity Models

According to the analyses conducted in this study, cultural and organizational barriers tend to surface more prominently than technological hurdles in Japanese companies during the idea and early adoption stages of InnerSource. In the growth and maturation stages, a lack of well-defined contribution rules appears to hinder expansion.

In practice, although many Japanese companies have begun implementing InnerSource with the aim of accelerating in-house development, only a limited number of cases have achieved organizationally mature adoption. Several interrelated factors are presumed to underlie this situation, including delays in reaching consensus with middle management and senior leadership, the absence of organization-wide evaluation criteria, and structural issues that prevent companies from moving away from established outsourcing practices.

Meanwhile, leading global examples often maintain dedicated teams, such as ISPO, and community management structures to balance voluntary contribution cultures with proper governance. These organizations integrate mechanisms for evaluation models, license management, and reward incentives, thereby establishing a solid foundation at the level of corporate strategy to support emergent collaboration. However, Japanese companies frequently lack an effective system to connect top-down decision-making with bottom-up voluntary contributions, which poses a significant risk of confining success to isolated projects rather than propagating InnerSource across the entire organization.

Furthermore, this study highlights that Japanese companies tend to emphasize “organizational culture transformation” as a principal motivation for InnerSource adoption, whereas fewer cases clearly articulate technical advantages such as software reuse and improved development efficiency. This tendency may contribute to a scarcity of quantitative engineering metrics, complicating internal persuasion efforts and investment decision-making processes.

In addition, substantial knowledge gaps persist regarding specific incentive designs, internal accounting rules, and visualization methods for large-scale organizations. Empirical research remains limited on how significantly InnerSource champions influence adoption outcomes. Examining these factors in an integrated manner is considered a crucial step toward establishing success patterns in InnerSource implementation.

Based on these findings, the current chapter organizes the issues and considerations revealed thus far and lays the groundwork for subsequent chapters (Chapters 8, 9, and 10), in which the concepts of "InnerSource Topologies," "Multilayered Incentive Model," and the "InnerSource Circumplex Model" are explored in detail. This approach will deepen the discussion on the scope, evaluation methods, and promotion frameworks for InnerSource, ultimately clarifying how sustainable InnerSource strategies can be formulated to meet company-specific needs.

## 7.1 Strengthening and Expanded Application of Existing Maturity Models

The present research focuses on visualizing priority issues at each stage from the ideas stage through the growth and maturation stages, and on proposing concrete processes for addressing these issues. By outlining the sequential tasks and discussion points encountered in InnerSource implementation projects, this study offers a practical framework for systematically planning organizational transformation.

Notable InnerSource Maturity Models include Inner Source Capability Maturity Model (IS-CMM) [8] and the assessment framework provided by the InnerSource Commons Foundation [9]. Although such models function as indicators for measuring progressive maturity, they do not sufficiently clarify priorities at the early stages of adoption—namely, "where to start and what scope to initially undertake"—nor do they explicitly address organizational-level transformation.

In existing models, priorities in the adoption stage remain ambiguous, leaving open the question of how to differentiate among organizational, product-related, and process-oriented factors that drive or hinder motivation. A lack of scope definition can also lead to confusion between project-specific characteristics and organization-wide strategic considerations, and existing models tend to lack concrete governance schemes.

Moreover, discussions on governance often remain too general, and there is no clearly articulated framework for addressing accounting management, licensing perspectives, and the alignment of evaluation systems. Because each company's internal practices, accounting systems, and security guidelines do not generally require external disclosure, best-practice examples in these areas are limited. Organizations with little experience in open source adoption face similar obstacles even before implementing InnerSource, and a systematic approach for removing these initial barriers has not been established. These challenges cannot be addressed merely by "creating rules." They require iterative dialogue and negotiation that consider operational realities and corporate cultural contexts, necessitating an understanding of how consensus is formed.

In addition, previous research has revealed diverse motivations for InnerSource adoption, with notable differences between Japanese and global companies, and even within individual organizations, where InnerSource may be viewed either as a "comprehensive package" or as an "internal phase of OSS development." This suggests the need for a maturity model framework that can clearly define the "mature" state desired by each organization. Simply setting an abstract goal of "implementing open source culture" may lead to diverse interpretations and loss of concrete direction.

In this context, understanding how InnerSource is practiced within organizations becomes crucial. The positioning and influence of champions within organizations are particularly difficult to assess through surveys alone. The governance and incentive structures needed to overcome adoption challenges may vary significantly depending on whether champion activities are based on voluntary enthusiasm or formally defined organizational responsibilities.

To gain deeper practical insights, cross-sectional interviews with key stakeholders - including early-stage developers, managers, executives, pilot project participants, and deployment coordinators - prove valuable. This approach enables examination of critical organizational transformation aspects, such as decision-making processes, successful negotiation strategies, and pivotal moments in InnerSource institutionalization. Equally important is understanding what stakeholders fundamentally aim to achieve through "InnerSource."

Field observations indicate that while many projects initially progress through dedicated volunteer engineers, they often lose momentum due to institutional inadequacies and organizational coordination challenges.

These circumstances clearly demonstrate the need for a model that enables promoters to define "where they are actually heading" at each adoption stage. Since initiatives during ideation and pilot stages significantly impact subsequent expansion and maturation, it is essential to design a clear process for creating successful cases through initial small-scale projects and engaging both executive and middle management.

The significance of analyzing InnerSource adoption processes, particularly in their initial phases, holds both practical and academic value. The "mature states" described in conventional maturity models allow for various interpretations; complementing these with specific measures for initial implementation and stabilization enables a more comprehensive understanding of InnerSource and potentially enhanced utilization of existing tool sets.

## 7.2 Pathways for Emergence of InnerSource Champions and Organizational Dynamics

According to the survey results, InnerSource adoption in Japanese companies tends to be driven from the bottom up. At first glance, this might suggest that grassroots innovation is thriving. However, in reality, hierarchical decision-making processes that require broad consensus often slow approval, compelling small units to proceed on their own. Larger companies in particular face more complex coordination with administrative departments and upper management, raising the barriers to unanimous agreement. At the working level, however, engineers find it relatively easy to experiment with new methods to address daily challenges, which manifests as bottom-up InnerSource initiatives.

In contrast, global samples demonstrated consistent top-down initiatives across all InnerSource implementation stages, indicating executive buy-in from the early phases. Analysis of adoption barriers in the chapter 6 analysis revealed a shift in primary challenges from "securing top and middle management consensus" to "time constraints" as implementation progressed, suggesting relatively smooth organizational advancement of InnerSource in global samples.

The chapter 6 revealed that Japanese companies, continued to face middle management resistance as a significant barrier even in later implementation stages. Addressing this challenge requires early organizational stakeholder engagement. However, a purely top-down approach risks lacking ground-level buy-in, potentially leading to implementation stagnation when cultural barriers or resource constraints emerge. Conversely, a purely bottom-up approach may face authority limitations when institutional reforms or company-wide rule establishment becomes necessary. To overcome these challenges, a "mixed" implementation process combining top-down and bottom-up approaches appears effective, potentially enhancing organizational flexibility and accelerating InnerSource benefits realization.

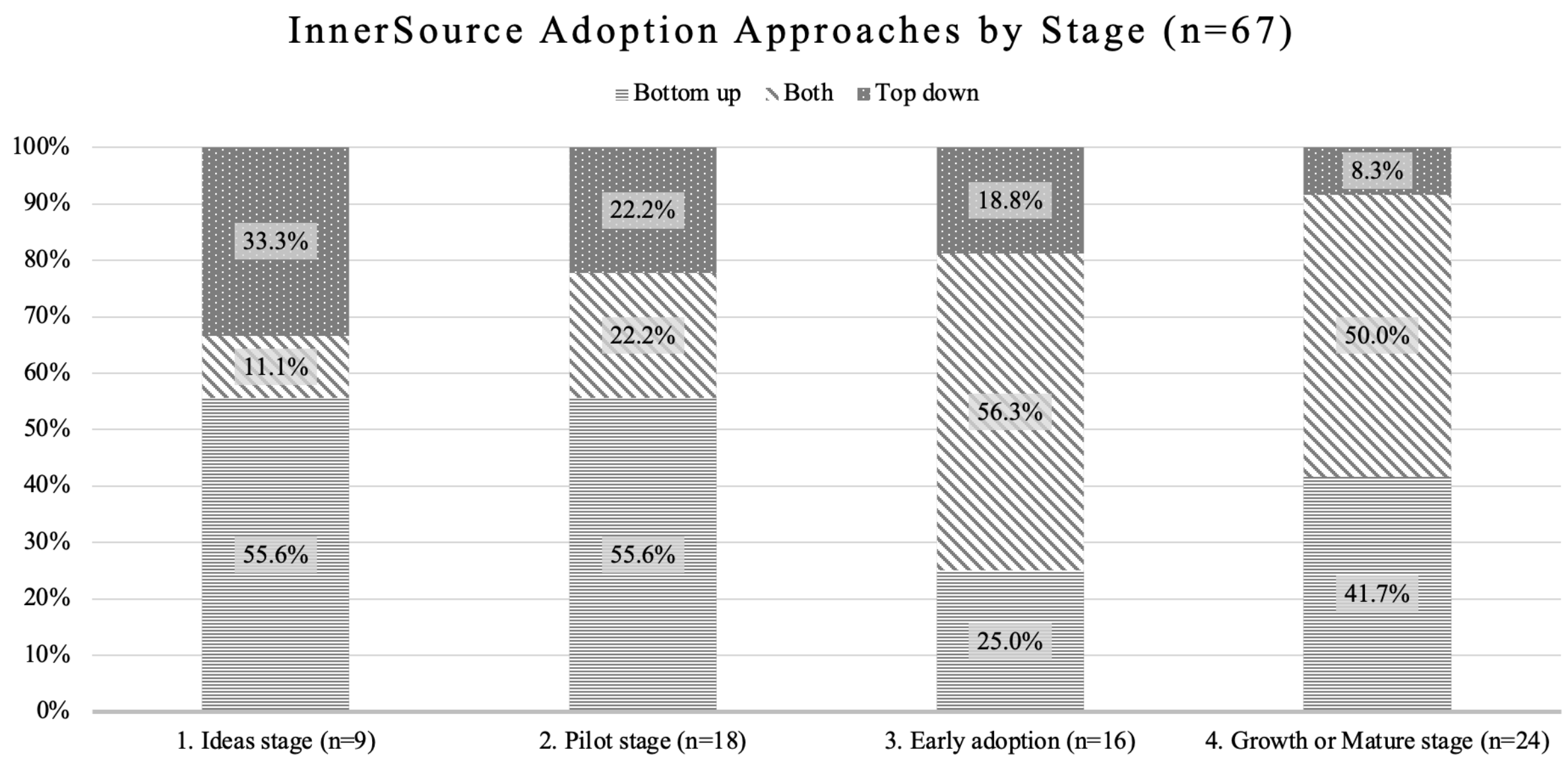

*Figure 7.1 Adoption Stage and Introduction Method of InnerSource (All Respondents)*

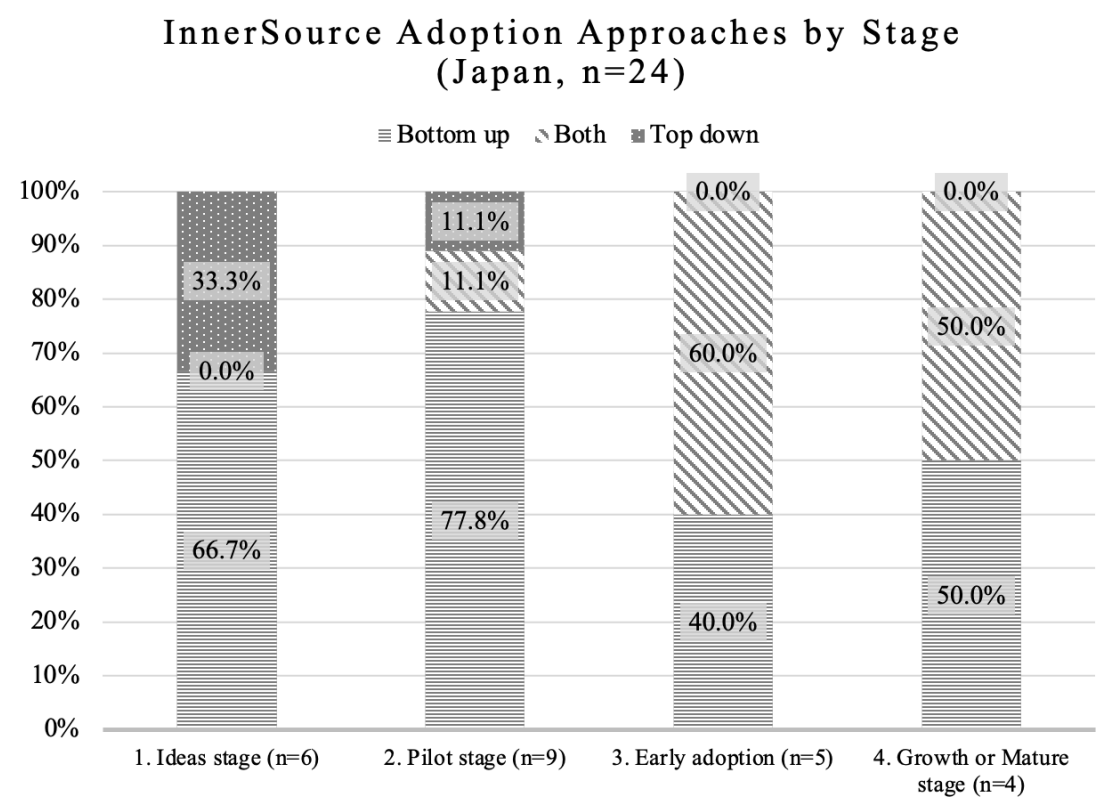

*Figure 7.2 Adoption Stage and Introduction Method of InnerSource (Japan)*

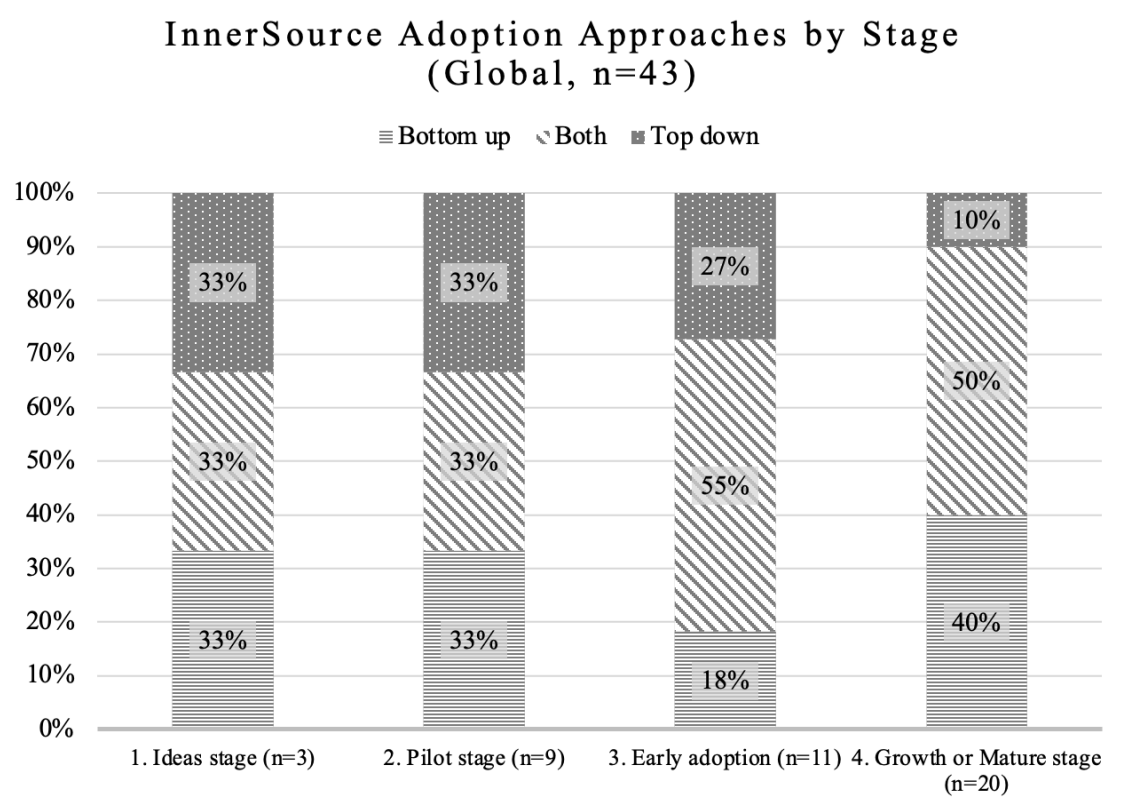

*Figure 7.3 Adoption Stage and Introduction Method of InnerSource (Global)*

If champions are limited to volunteer-based activities, it becomes difficult to address institutional reforms and achieve organization-wide consensus. For instance, adjustments to licensing and intellectual property management inherited from open source software must align with existing company rules. Without sufficient authority and resources, InnerSource initiatives can stall in the early stages. In consensus-based Japanese organizations, unofficial champions are less likely to succeed in mobilizing legal, accounting, and executive departments, which risks undermining early enthusiasm.

To mitigate these risks, it is crucial to treat InnerSource not as a mere “extension of OSS usage in-house,” but rather as a “comprehensive organizational reform package” attuned to a company’s unique human resources, quality management, and accounting frameworks. Champions must lead this package design, coordinating horizontally among diverse stakeholders in technology, corporate planning, human resources, and compliance. They must guide revisions to necessary systems and rules at each implementation stage and secure consensus on these changes.

## 7.3 Interview-Based Investigations for Practical Insights

As part of this research, detailed interviews were conducted with multiple companies representing a range of adoption outcomes, including large-scale successful deployments, projects that failed during the pilot stage, and companies that have completed initial implementation but are struggling with accounting procedures or evaluation models. Interviewees included personnel in charge of InnerSource, managers, and in some cases senior leadership, ensuring the collection of a broad spectrum of perspectives.

During the interviews, attention was directed to previously identified barriers and working hypotheses, addressing issues such as how obstacles were removed or where failures occurred; how champions carried out their roles in practice; when and how evaluation models were established, and how incentives were designed; and how companies overcame challenges related to accounting or compliance. The responses provided valuable data for systematizing concrete strategies for implementation and barrier removal that have not always been adequately covered in existing maturity models or global case studies. Appendix materials at the end of this manuscript include detailed information on the interviews, preserving anonymity for the companies while offering sufficient detail for objective and in-depth analysis. By covering industries ranging from manufacturing and internet services to the information and communications sector, the interviews achieve a high degree of comprehensiveness and balance in representing the actual state of InnerSource adoption.

Interviews were conducted with the following six companies:

*Table 7.1. Overview of Interview Participants*

| Company Name | Industry | Company Size | Affiliation | Title | Implementation Stage |
|---|---|---|---|---|---|
| **Company A** | Not disclosed | Over 50,000 employees | Technology-related Division | Not disclosed | Early Adoption |
| **Company B** | Manufacturing | Over 50,000 employees | Common Development Division (Holding Company) | Senior Technical Expert | Early Adoption |
| **Company C** | Internet | Several thousand | Development Productivity Division | Leader | Mature |
| **Company D** | Internet | Several thousand | Development HQ (Core Group Co.) | Lead Engineer | Mature |
| **Company E** | Internet | Several thousand | Accounting Department | Person in charge | Mature |
| **Company F** | Information & Communication | About 500 employees | Engineering Department | Engineering Manager / Engineer | Early Adoption |

These companies collectively span the manufacturing, internet, and information/telecommunications industries, providing coverage sufficiently broad to ensure balanced data from an industry-distribution standpoint. Detailed information was gathered on each firm's InnerSource initiatives, including adoption background, implementation status, current challenges, and realized benefits.

Subsequent chapters will organize specific examples and insights gleaned from these interviews in an effort to refine InnerSource adoption models for Japanese companies. This includes identifying factors driving success or failure, clarifying champion roles, designing accounting and evaluation models, and proposing solutions to various adoption-stage challenges that have received limited attention in previous research.

## 7.4 Primary Issues and Practical Implications by Introduction Stage

This section provides a broad overview of main points to consider in each stage of the InnerSource introduction process, followed by more detailed analysis in the ensuing chapters.

In the ideas stage, it is important to foster an internal appreciation for the value of collaboration while refraining from attempting to overhaul accounting or intellectual property management rules all at once. Instead, the recommended approach is to implement incremental measures within a limited scope. By initially restricting the sphere of influence, multiple small-scale activities can run in parallel with minimal risk. These initial steps then lay the foundation for subsequent large-scale deployment.

In the pilot stage, collecting empirical data on repository usage and making it visible are critical for building the evidence needed to persuade managers and cross-functional departments. If a pilot project succeeds and can demonstrate tangible benefits, the groundwork is laid for incorporating InnerSource into accounting processes and performance assessments. According to the Company F case, once pilot success stories became widely recognized, the organizational mindset shifted from “it might be okay to try” to actively embracing it. Further, engineering managers reported that formalized collaboration guidelines and cross-departmental linkages simplified consensus-building significantly.

Nevertheless, it is important to note that challenges may resurface when moving from the pilot stage to the early adoption stage. Attempting to expand adoption in the absence of clear evaluation criteria can dissipate momentum gained during the pilot. Therefore, leveraging pilot data to define concrete evaluation models and authority structures is critical for smoothing the transition to the early adoption stage.

Once the early adoption stage is on track, more robust governance, security, and compliance frameworks are needed. Large enterprises often have intricately intertwined evaluation mechanisms, accounting rules, and internal regulations. Failing to involve managerial departments at an early stage risks stagnation before the community transitions into a major expansion phase. In Company F, for instance, participants initially expressed uncertainty over whether contributing to other teams would be recognized. However, once the evaluation metrics and cost-allocation rules were formulated, engineers reported feeling more confident about engaging in collaboration.

A major challenge during early adoption is how to build cooperative relationships among engineers and middle managers. In Company A, for example, institutional support lagged behind the enthusiasm of key individuals, leading engineers to “work in secret” to advance InnerSource. It is therefore imperative to strengthen coordination with department heads and administrative functions, ensuring that engineers can confidently contribute within a formal framework of institutional support and recognition.

Upon overcoming the early adoption stage and entering the growth/mature stage, organizational operating structures and authority management systems tend to become more firmly established. The number of projects and the extent of collaboration expand, making it easier to foster broader participation. In the Company F example, even though the company had already allowed employees to view repositories, no one took action initially. However, once the organization widely disseminated guidelines (like CONTRIBUTING.md) explaining proper collaboration procedures for InnerSource, approximately 19 new projects sprang from the first experimental initiatives. As these successes multiply, cross-departmental cooperation tends to strengthen at an accelerating pace.

Taken together, the ability to continuously motivate core participants and develop fair evaluation models in a gradual, step-by-step manner emerges as a crucial pathway to eventual maturity. Evidence from both Company A and Company F suggests that opaque assessment criteria make it difficult to secure agreement among engineers and middle managers, potentially leading to stagnation in the early expansion phase. Hence, it is vital to synchronize institutional design efforts with broader inter-departmental awareness-building initiatives throughout each adoption stage.

## 7.5 Summary of This Chapter

This chapter examined multiple approaches to InnerSource adoption, focusing on the scale and culture of the enterprise. In particular, it highlighted the characteristics and limitations of both top-down and bottom-up approaches while analyzing effective strategies under each scenario.

To explore these issues in greater detail, extensive interview research was conducted across a wide range of industries, including manufacturing, internet services, and telecommunications. This research covered companies experiencing success with InnerSource, organizations currently grappling with challenges, and others at various adoption stages. The interviews yielded granular information on the specific initiatives undertaken, the obstacles encountered, and the processes by which problems were resolved.

Based on these interviews, the chapter classified the prominent features of InnerSource at each adoption stage and systematically identified the measures and considerations required at each stage. From the ideation stage to the pilot, early adoption, and eventual growth/mature stages, the analysis generated actionable insights regarding the organizational tasks to address and practical solutions for overcoming barriers.

In the ideation stage, ambiguity regarding whether InnerSource is truly permissible often impedes progress. Early on, enthusiastic champions can accelerate internal awareness, create small-scale success stories, and introduce lightweight guidelines, thus fostering an environment of “try it and see.” The pilot stage is best served by test-running cross-departmental repositories to gather both quantitative (e.g., number of commits, review status) and qualitative evaluations, thereby constructing an evidence-based framework to build consensus with management. The early adoption stage requires stronger alignment with administrative units—accounting, legal, and human resources—along with official recognition of a champion or other point of accountability, thus laying the institutional groundwork for broader adoption. Developing draft versions of accounting and evaluation rules and codifying governance guidelines allows the organization to shift its mindset from “it is acceptable to do InnerSource” to “it is beneficial to do InnerSource.” Even after growth and maturity are achieved, increases in code-review and operational workloads often spark new questions about how to properly evaluate individual contributions. Establishing a dedicated community management or ISPO function and introducing comprehensive evaluation models—combining quantitative and qualitative metrics—helps manage this increased load. Standardized guidelines and tools can also reduce management complexity.

By progressing incrementally in this manner, companies entering the mature stage can nurture a culture that enables autonomous collaboration across multiple departments, thereby fostering innovation and accelerating technology development. One of the central aims of this paper is thus to clarify the obstacles that arise in the early implementation stages and to systematize methods of overcoming them. Continuous recognition of ongoing challenges, coupled with cross-organizational consensus-building, helps define a clear path from the initial stages of adoption through maturity.

Chapter 8 introduces the "InnerSource Topologies," focusing on setting boundaries for collaboration and providing contribution guidelines. Chapter 9 discusses the design of incentives, classifying them into six types for InnerSource promotion. Chapter 10 proposes the "InnerSource Circumplex Model," which reframes InnerSource growth as a multidimensional process, thereby constructing a framework for more flexible promotion strategies. Finally, Chapter 11 synthesizes these findings to offer a comprehensive perspective on overcoming the barriers to staged adoption and establishing a sustained collaboration infrastructure. This overall framework is intended to propose diversified InnerSource adoption strategies aligned with organizational scale and culture.

# 8 Representation as a Network: InnerSource Topologies

This chapter focuses on the inherently polysemous definitions of InnerSource and the diverse scope of its application. Although previous research and practical knowledge often characterize InnerSource in a singular manner—namely, as an "open source development methodology implemented within an organization"—in reality, each company has its own unique interpretations and operational requirements. Multiple concepts, ranging from the broad to the narrow sense of InnerSource, may coexist. The existence of these multilayered definitions implies that companies can flexibly adopt InnerSource practices depending on factors such as organizational size, business structure, and risk management systems.

Furthermore, this chapter's analysis addresses not only the notion that InnerSource maturity progresses in a linear, stage-based manner but also the topological evolution that occurs as unexpected network structures emerge within communities. Even though an InnerSource program may expand step by step during organizational transformation, new forms of collaboration arise through the complex interplay among individual developer communities and inter-team linkages.

The aim of this chapter is to clarify the network structure facilitated by InnerSource. By examining how InnerSource is implemented within companies, how nodes (individuals or organizational units) are interconnected, and how contracts and accounting treatments are established as edges (i.e., relationships), this chapter seeks to provide guidelines for building sustained and efficient collaboration.

Introducing InnerSource often entails numerous "reasons not to proceed," such as difficulties in driving large-scale organizational transformation all at once or mismatches with existing accounting and top-down instruction systems. Consequently, in practice, it is often necessary to carefully decide the boundaries of InnerSource adoption and the degree of transparency to be permitted, based on each company's circumstances. If a company views the options solely in terms of "complete openness or none at all," it may ultimately abandon InnerSource altogether.

In addition, the scope of InnerSource can vary significantly depending on the size and structure of the workforce across different companies or divisions. For example, an InnerSource initiative involving 5,000 employees within a company of 50,000 may be deemed as valid an application of InnerSource as an initiative in which a 2,000-person company promotes InnerSource in its entirety. The definition is not simply binary—namely, whether "everything is open or not at all." Therefore, setting flexible levels of transparency at the adoption stage is critical for long-term, sustainable development. Depending on scale and business structure, some companies may limit themselves to a single subsidiary or conduct pilot programs in the research and development department, implementing partial InnerSource.

Generally, InnerSource refers to “open source within an organization,” characterized by transparent collaboration. Nonetheless, it need not extend across the entire company. While a form that closely resembles open source may be viewed as the narrow sense of InnerSource, it is equally valid to pursue what might be called the broader sense of InnerSource, in which transparency is safeguarded only within specific domains. Under this broader definition, companies can flexibly choose a form consistent with their institutional requirements and risk tolerance.

In light of these considerations, the degree of transparency and visibility regarding contributions serves as the fundamental criterion for evaluating inner-source-style collaboration. In actual practice, there are examples such as the “individual contract model,” employed by Company B, which restricts access to certain members only, or the “consortium model,” used by company A, which grants access solely to registered members. Even if collaboration is confined to a small group and relies on contract-based or budget-secured approaches, it can still be regarded as a broad sense of InnerSource if it contains inner-source-related elements.

For instance, initiatives under an individual contract model may at first glance appear merely to be formal joint projects. However, installing a dedicated InnerSource portal or releasing SDKs and interface information can generate benefits associated with InnerSource. Thus, even when circumstances related to accounting or security prevent the full disclosure of source code, establishing a situation in which partial transparency is ensured, and new contributors are welcomed still qualifies as a form of InnerSource. Within the same company, the scope of code visibility and contribution takes various forms.

Empirical observations also indicate that the primary reason for promoting InnerSource has, in some cases, shifted from the “internal application of open source” to the “introduction of a new collaboration style that fosters transparency.” In such companies, the goal is to build an efficient and flexible environment for cross-unit collaboration by sharing only what is necessary, rather than disclosing all code internally. Accordingly, there is a tendency to adopt models in which only some aspects of InnerSource are implemented.

Strictly defining InnerSource or judging whether certain practices meet the criteria of InnerSource are beyond the scope of this paper. However, it is a fact that such discussions and forms of intermediate collaboration do arise in the course of InnerSource's maturity. Considering the possibility that organizations and employees may pass through these stages as they develop InnerSource capabilities, the present study integrates a range of collaboration forms, including the partial adoption of InnerSource techniques. Ideally, complete visibility and universal editing privileges are desirable, but in actual enterprise settings characterized by accounting, organizational, and security constraints, diverse collaboration forms emerge. Even if only a portion of the source code is released for reasons related to accounting or security, if the code is open to contributor participation and collaboration, this paper treats it as a form of InnerSource.

This paper proposes the concept of "InnerSource Topologies" to systematize the multiple layers of access permissions and rights that arise during the introduction of InnerSource. Methods that grant individual teams or developers bidirectional repository access can create an intersection between bottom-up development demands and managerial strategic objectives, enabling multiple topologies to coexist simultaneously.

## 8.1 Defining Organizational Collaboration Using an InnerSource Topologies

It is difficult to reduce the expansion of InnerSource to a simple tree structure. Rather, diverse forms of interconnectedness arise as a result of inter-team collaboration, return on investment (ROI) considerations, and the interplay of management strategies. This structure, which is not wholly bottom-up, has a semilattice-like linkage reminiscent of Christopher Alexander's argument [46], implying that an organization can become interconnected along multiple axes.

As bottom-up activities stimulate demand, the addition of more structured, plan-based approaches creates a network distinctively combining autonomy and control. From this perspective, it becomes possible to visualize one-way or two-way collaboration patterns by treating individuals or teams as nodes and viewing repository access rights—along with the directionality of collaboration—as edges.

Various attempts have been made to represent InnerSource initiatives and open source dependencies as networks. This research not only aims to capture inter-team collaboration direction in InnerSource but also focuses on developing a more precise understanding of these connections—the "edges." While previous research has primarily focused on simple connectivity such as dependencies and access permissions, there is significant value in systematically organizing the definitions and types of connections to understand actual collaboration patterns.

Table 8.1 categorizes the topologies discovered through interview research (see Appendix for interview details). It outlines how patterns such as point-to-point, star, partial mesh, and full mesh configurations manifest in InnerSource practice. Notably, all these topologies were recognized as either "InnerSource" or "InnerSource-like" practices. While fundamentalist interpretations might presume InnerSource requires complete openness, the research revealed that InnerSource-style communication can occur even with partial openness or limited collaboration between two teams.

In practical implementation cases, these patterns rarely exist in isolation; instead, hybrid forms where multiple patterns coexist within the same company are more common. While many companies ultimately develop complex team networks resulting in hybrid formations, individual stakeholders may differ in their recognition of which topology areas (clusters) constitute InnerSource.

*Table 8.1 Overview of InnerSource Topologies*

| Name | Description |
|---|---|
| **Point-to-Point** | Involves only individual contracts and does not amount to full openness, but partially enhancing communication and discoverability can yield inner-source benefits. Company B has templates for such individual contracts. |
| **Star** | Multiple individuals contribute to a specific repository, and it functions in both small and large-scale settings. This configuration often evolves into partial mesh or full mesh in later stages. |
| **Partial Mesh** | Multiple individuals or teams maintain mutual access without fully opening everything. It enables relatively extensive collaboration under existing constraints. |
| **Full Mesh** | Every individual and team hold mutual access rights, allowing unrestricted collaboration. This configuration is most similar to a genuine open source environment. |
| **Hybrid** | A common InnerSource unit or team provides a platform for collaboration, while other domains retain distinct modes of contribution. This pattern is frequently observed in practice. |

### 8.1.1 Point-to-Point Type

A point-to-point topology refers to situations in which collaboration occurs between teams or individuals operating under specific contracts. Absent elements that convey an inner-source-like openness, this remains limited to narrow joint development. Nonetheless, setting up a dedicated portal and sharing README files, documentation, SDKs, or interface information within a defined boundary can serve as an entry point to InnerSource.

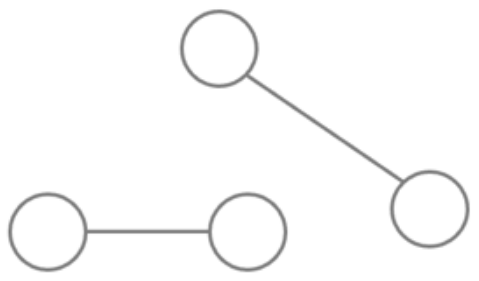

*Figure 8.1 Point-to-Point Type*

Its key benefits include ease of initial implementation without conflicting with established accounting practices or management procedures, as well as a safe framework for controlling confidential information within a restricted scope. On the other hand, because visibility is limited to the direct parties involved, the broader organizational impact of these efforts tends to be low, and wider InnerSource diffusion is frequently difficult. An example is the individualized contract template used by Company B, which restricts mutual sharing to only the targeted project and maintains small-scale participating communities. Although this format offers a high level of psychological safety and an easy starting point, expansions to broader collaboration require additional governance and portal support.

### 8.1.2 Star Type

A star topology entails multiple teams or individuals—consumers—converging around a central repository or platform, each providing contributions in turn. On a small scale, a foundational team might supply libraries or tools to several teams that offer feedback and requests for modifications. In mid-sized cases, a separate foundational team (the host) maintains the development platform while product teams (the guests) deploy services on top of it and submit improvement proposals. On a large scale, numerous engineers may work on a vast monolithic system, which often transitions toward a partial mesh or full mesh configuration.

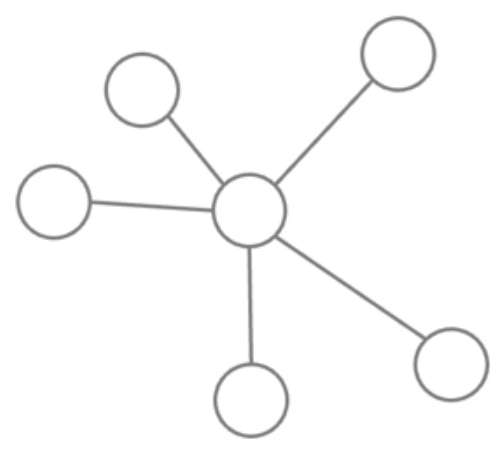

*Figure 8.2 Star Type*

The principal advantage of the star topology is that once a central repository is set up, guest teams can easily begin contributing. However, inter-guest collaboration is initially limited, making the host the primary linkage point. In Company E, for instance, many teams once centered on a single monolithic platform repository prior to migrating to a microservices architecture. As teams became more adept at making self-directed improvements, the arrangement gradually evolved from a localized star to a broader mesh.

### 8.1.3 Mesh Type

A mesh topology involves teams or individuals holding reciprocal access rights across multiple repositories, which are interconnected in all directions. Such mesh arrangements may arise spontaneously as bottom-up initiatives gain momentum or may be deliberately fostered through organizational planning to connect specific teams. The main benefits include reduced duplication of work and innovative outcomes derived from diverse perspectives. However, not all teams have identical access privileges in a mesh setting, necessitating complex permission design and potentially increasing coordination costs.

One example is Company F that, while formally granting company-wide repository access, initially had only a small group actually collaborating. Over time, small success stories accumulated and led to a gradual expansion toward a mesh arrangement. A key feature is that mesh configurations transform dynamically over time: partial or "virtual" meshes may form and, under certain circumstances, approximate a full mesh.

### 8.1.3.1 Partial Mesh / Virtual Mesh Type

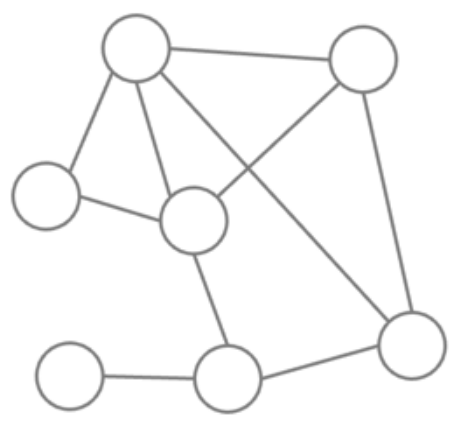

*Figure 8.2 Partial Mesh Type*

A partial mesh exists where only certain parts of an organization or specific teams mutually grant each other repository access, while overall visibility remains short of fully open. The advantage here is that code containing sensitive or proprietary information remains protected, even as multiple teams collaborate extensively. The disadvantage lies in the difficulty of determining precise access boundaries, the increase in departmental partitions, and the added administrative burdens for permission management and policy coordination.

Company F sector exemplifies a gradual expansion in which collaboration incrementally shifted from small pockets toward a broader mesh. Although partial mesh arrangements align well with actual corporate demands, achieving the deeper benefits of a full mesh requires additional consensus-building. Conversely, an originally small enterprise that operated under a full mesh may, upon growing larger, impose certain hidden areas and thus revert to partial mesh.

In some corporate environments, a "virtual company-wide team" may be granted comprehensive access privileges within source code management settings, effectively forming a "virtual mesh." This allows near-full-mesh collaboration within a designated InnerSource team or project space without significantly overhauling either the organization's structure or its primary source code management system. The advantage lies in enabling high-level collaboration in a defined domain, preserving the possibility of scaling successful practices later. However, beyond that domain, constraints from legacy systems remain, limiting the broader network effect. Setting up a dedicated "InnerSource team" with repository access is one practical example of such a virtual mesh approach. Because these configurations rely on platform-specific permission controls, they function like partial mesh arrangements for the purposes of this discussion.

#### 8.1.3.2 Full Mesh Type

A full mesh configuration grants every individual or team direct access to one another's repositories, permitting fully bidirectional collaboration and representing the arrangement most closely resembling open source. It entails disclosing corporate information in its entirety, which poses significant challenges regarding accounting rules and security. Once established, a full mesh can drive rapid cross-departmental collaboration. Although the full mesh topology may be viewed as the ideal form of InnerSource from a purist perspective, it is not necessarily the only goal.

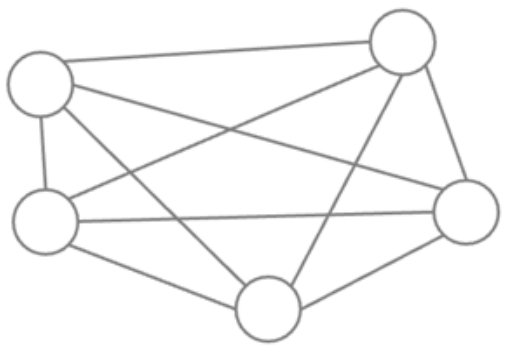

*Figure 8.2 Full Mesh Type*

In practice, a "special full-mesh organization" can be created in a source code management system to enable partial full-mesh functionality, effectively establishing a hybrid arrangement at the operational level.

### 8.1.4 Hybrid Type

A hybrid model emerges when two or more of the topologies coexist within the same company and are chosen flexibly to fit particular circumstances or objectives. Rather than enforcing a universal full mesh across the entire organization, different collaborations—including point-to-point and partial mesh—are used selectively according to security requirements and business domain considerations.

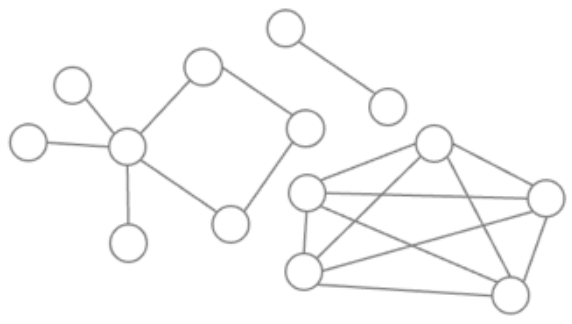

*Figure 8.5 Hybrid Type*

Hybrid arrangements can optimize each domain to its own needs, thereby reducing friction while gradually advancing InnerSource. However, the coexistence of multiple permission and contractual models can complicate management, rendering the overall vision less transparent to the organization. Company A, for example, maintained an environment that approached a full mesh in some areas while continuing point-to-point contract collaborations elsewhere. Because many companies end up adopting a hybrid approach in pursuit of wide-ranging collaboration, the key issue becomes how to coordinate between an overarching strategy and localized optimization.

## 8.2 Community Formation and Incremental Expansion in InnerSource

When promoting InnerSource, it is essential to clarify the preferred mode of collaboration and the extent to which relevant stakeholders or "champions" can exert influence. In Company A, for instance, a key person with broad authority oversaw cross-divisional interests, enabling an approach in which a consortium-style partial mesh was established. By contrast, in Company F, InnerSource was initiated by an engineering manager and began with a small-scale star-type collaboration. This gradually expanded laterally to partial-mesh and hybrid structures, as coordination with other departments and managers evolved. Sometimes large parts of the workforce already had nominal repository access, leading to a latent partial-mesh environment with minimal practical contribution until formal InnerSource adoption and the accumulation of small successes helped it flourish.

Krebs et al. have argued that communities are built upon connections, and that new opportunities emerge as network structures evolve [47]. Concretely, they define a process in which networks progress from "scattered clusters," to "hub-and-spoke (star)," to "multi-hub," ultimately forming a "core-periphery" structure. InnerSource communities expand similarly over time. InnerSource is not merely about "creating a place for collaboration"; rather, it fundamentally seeks to revitalize collaboration among teams. Even the most sophisticated InnerSource program will fail to deliver results in the absence of community or team involvement in building the requisite complex networks. Hence, deliberate and incremental efforts to connect people and weave the network are vital.

Company B offers a valid example, promoting point-to-point collaboration that gradually disrupts entrenched organizational silos. Even if the company as a whole has not reached full maturity, smaller collaborations can stimulate the internal network and pave the way for future expansion of InnerSource. This process is particularly relevant in Japanese enterprises, where sweeping organizational change is often difficult to achieve quickly, and there is a common preference for iterative steps that secure management and security approvals.

Moreover, treating InnerSource purely as a voluntary endeavor for individual contributors tends to limit its potential. Formalizing it institutionally and linking it to business activities prove key success factors. Decisions on how much code should be made open and when the scope of access should be broadened demand a balance between top-down mandates and bottom-up initiatives. The implementation of stage-by-stage governance and evaluation systems emerges as a critical requirement. Although this paper does not strictly define InnerSource, building advanced collaboration structures—ranging from full mesh to virtual mesh—generally benefits from repeated agile experimentation at each stage. Some companies run multiple topologies in parallel: for instance, point-to-point contracts for a small set of projects, star relationships for certain platforms, and partial mesh for cross-functional teams. The final topology and level of openness are determined by company culture, accounting mechanisms, and personnel evaluation systems. The primary objective for InnerSource advocates is to strategically identify these elements and guide incremental scaling.

## 8.3 Distinguishing Access Rights from Utilization Rights

It is difficult to classify a particular mesh structure under one label when corporate InnerSource environments may exhibit varying types of edges linking nodes. Even in tightly integrated forms of InnerSource, stringent licensing or usage restrictions might confine collaborations to point-to-point–like operations.

The present investigation of Company A and Company B revealed that practical concerns extend beyond simple "read/write" privileges. Three usage modes—access only, R&D use, and commercial use—form significant operational distinctions. These modes relate not only to the demarcation in Japanese accounting systems between software categorized as research and development expenses versus capitalized commercial software but also to the scenario in which code is accessed merely for reference through a marketplace without further use.

Organizing these distinctions yields the structure shown in Table 8.2, illustrating how usage objectives can substantially alter the real constraints, even under the same read/write setting in a tool. Sometimes write privileges are granted on the technical platform, yet accounting or contractual restrictions make them practically unusable—a symbol of InnerSource's inherent complexity. Detailed discussion of related contracts appears in the following section.

*Table 8.2 Classification of Access and Utilization Rights*

| | Read | Read/Write |
|---|---|---|
| **Access Rights Only** | Enterprise repositories are viewable but cannot be modified or commercially exploited, sharing information under minimal risk. Although knowledge expands through experimental mutual reference, it rarely leads to active contribution. Utilization remains unavailable; for actual use, the R&D or commercial route must be chosen. | Basic viewing and minor editing are allowed, but further usage is disallowed. Commercial or R&D usage requires separate permission. |
| **R&D Use** | Viewing and utilizing source code for research and development. This may include examining ideas or assessing technical feasibility. If the software moves toward commercialization, separate contractual or budget approvals may be required. | Viewing and modifying code for research and development, allowing prototype implementations that facilitate technical validation or lead to future commercial use. Feedback is often requested in return for providing code access. However, transitioning to commercial use typically involves reevaluating cost allocations and additional corporate processes. |
| **Commercial Use** | Granted permission to view code for commercial purposes, making it possible to embed source code in business processes. Modifications may be disallowed, so bug fixes or feature additions can require other privileges. Individual contracts are often necessary, and this stage may involve support agreements with the host team. | Full privileges for commercial purposes, covering both viewing and modifying source code. This stage typically entails rigorous risk management and cost allocation. Individual contracts are common, and support agreements with the host team may be required. |

Because the hierarchical nature of usage permissions extends beyond the superficial permission settings in source code management tools, careful consideration of multilayered controls is essential when evaluating mesh structures.

When preparing an InnerSource environment, initial focus tends to be on the system permissions column in Table 8.2, as tools like GitHub primarily offer basic Read or Write visibility options. However, selecting these permissions alone does not resolve the fundamental issues, as they fail to provide clear criteria for accounting, tax compliance, or visibility scope considerations. This often leads to defaulting to restricted access.

Conversely, providing meaningful InnerSource guidance requires attention to the usage permissions row in Table 8.2. This is particularly relevant in contexts like Japan, where clear distinctions exist between research and development costs versus commercial use assets in accounting practices. While Table 8.2 specifically highlighted the cost versus commercial use dichotomy, the usage permission rows could be expanded to accommodate various conditions such as export control restrictions, transfer pricing regulations, and prevention of profit provision.

Consequently, even in environments where teams appear to have full mesh connectivity through system-level Read/Write permissions, actual usage may be significantly constrained by other factors, resulting in partial star-type or point-to-point configurations beneath the surface. It is crucial to recognize that organizational requirements often extend beyond the simple Read/Write permissions offered by source code management tools.

From an evolutionary standpoint, some organizations begin with modest point-to-point contracts for experimental collaboration. After achieving certain outcomes, multiple teams might adopt partial mutual access, moving toward star or partial mesh modes. Conversely, an organization might initially implement large-scale full mesh only to revert partially to point-to-point in restricted areas of secrecy. Such developments become increasingly diversified with organizational change, the birth of new team-based collaborations, or the emergence of overlapping virtual mesh layers.

Ultimately, detailed analysis of contractual relationships and technical permission settings can illuminate obstacles to introducing InnerSource as well as its potential benefits. While such analysis can help identify essential attributes of InnerSource, the central question is how organically participants can cultivate collaboration that transcends the mechanical allocation of tool permissions.

Licensing or individual contractual arrangements can govern these usage rights. The subsequent section discusses contracts in more detail.

## 8.4 InnerSource Contracts: Integrated Analysis of Licensing, Accounting, and Taxation

As outlined in the previous section, in actual InnerSource implementation, complex linkages develop among nodes (teams and individuals) with differing usage modes. The contractual relationships that emerge in these linkages can be grouped broadly into three categories:

- Implicit InnerSource License
- Explicit InnerSource License
- Individual Contract

An InnerSource license is a legal framework designed to address liabilities and accounting challenges when sharing software source code among different entities within the same corporate group [48]. By providing a reusable legal standard for internal source code sharing, it offers fresh collaboration opportunities and clarifies the rights and responsibilities among group companies.

Although InnerSource licensing is not a novel concept, practical examples from Japanese enterprises—and their alignment with accounting and licensing requirements—remain insufficiently documented. This study surveys relevant cases to illustrate how contractual forms support InnerSource environments.

### 8.4.1 Implicit InnerSource License: Tacit Agreements and Established Practices

Many enterprises promote InnerSource collaboration without establishing explicit licensing documents, relying instead on practical consensus or custom. In digital-native companies or online service providers such as Internet-industry Company C or Company D, source code sharing and modifications are informally permitted among engineers before meticulous accounting or licensing rules have been finalized. In such firms, accounting "proceeds smoothly," and "organizational consensus" is assumed, enabling cross-project collaboration to arise organically. This approach is particularly observable in startups and internet-oriented companies, where development costs are frequently expensed and commercial applicability remains uncertain, reducing the immediate need for formal license agreements.

Nevertheless, since these implicit licenses do not clearly indicate permissible uses or contributions, engineers sometimes remain unsure about their scope of freedom. On the other hand, explicitly documenting which repositories can be utilized might encourage developers to pursue collaboration. Thus, while implicit licensing alone is not always ideal, drafting a minimal explicit license that clarifies usage and modification parameters can boost willingness to contribute when an organization seeks to institutionalize InnerSource.

Some enterprises, such as Company F in the information and communications sector, have formalized the connection between implicit licenses and the accounting of developers' time by forging explicit agreements on how to credit contribution hours. Company F manages working time in 30-minute increments by assigning order numbers to specific products or features. Although engineers often record their work at the end of each day or the following day, the system still fosters a clear and positive environment in which cross-departmental contributions are visible and more likely to be rewarded. This approach relies on a SaaS-based time-tracking system capable of registering InnerSource activities.

Meanwhile, large-scale global corporations in the internet industry (e.g., Company E) must carefully observe transfer pricing rules and capitalize software assets. Nonetheless, such firms do not necessarily adopt rigid license documents in advance. Instead, they may simply distinguish development activities that do not directly generate revenue—expensing them accordingly—while capitalizing software linked to core business logic. Managers then rely more on qualitative assessments, facilitating a bottom-up culture with few formal licensing rules. This indicates that a specialized license framework is not always critical to success in InnerSource endeavors.

Ultimately, some organizations find that collaboration flourishes under an unspoken agreement supported by engineering culture, while others benefit from an explicit license specifying usage parameters to encourage engagement. Regardless of the strategy, the reality is that bottom-up development in InnerSource is often sustained by tacit consensus within the company. Over time, as InnerSource matures, implicit and explicit licensing may coexist and become intertwined with refinements in accounting and evaluation systems.

### 8.4.2 Explicit InnerSource License: The Consortium Model Example

Large corporations that encompass multiple legal entities often face challenges in offering free access to valuable source code without violating rules against bestowing unearned benefits or breaching transfer pricing regulations. Overseas facilities may also trigger additional export controls or tax obligations. In such situations, certain restrictions may be placed on how far InnerSource can be extended, making the creation of an InnerSource license a viable solution. In practice, organizations frequently craft simple statements resembling GPL-based language but with limited discussion of accounting, taxation, and inter-company collaboration schemes.

Merely adopting a publicly available “InnerSource license” is not necessarily feasible for Japanese enterprises if it conflicts with local accounting systems, tax policies, or corporate culture. Moreover, typical license texts rarely cover the underlying rationale—such as which specific accounting requirements or conditions are addressed. Thus, it is necessary to consider not only the text of the license itself but also the overall scheme and intended collaboration format.

One example is Company A's “consortium model,” which loosely links multiple corporate entities within a group. A particular corporate division manages the central platform, and participating enterprises pay a subscription fee to gain access to a defined subset of source code and tools. Company A had once examined a "source code acquisition model," in which the parent company would centrally purchase all source code assets and distribute user costs, but this proved too difficult to implement and was abandoned. The adopted approach features flexibility in dividing responsibilities and benefits.

#### 8.4.2.1 Three-Tier Consortium Model: Viewing, Contributing, and Commercial Use

In this consortium model, usage is defined across three tiers— “viewing,” “contribution,” and “commercial usage”—to accommodate free browsing and development activities in the R&D phase, while imposing licensing fees when commercialization ensues.

*Table 8.3 Consortium Model’s Three Tiers*

| | **Viewing Tier** | **Contribution Tier** | **Commercial Use Tier** |
|---|---|---|---|
| **Viewing** | Permitted (for R&D) | Permitted (for R&D) | Permitted |
| **License** | InnerSource License (Non-Commercial) | InnerSource License (Non-Commercial) | Separate agreement with the development department |
| **Usage Type** | Trial use | Free use (limited to R&D purposes) | Commercial use |
| **Freedom to Use and Modify** | N/A (view only) | Modification and sharing allowed (limited to R&D) | Modification permitted (subject to contract) |
| **Support** | N/A | N/A | Available if specified in the agreement |
| **Maintenance** | N/A | N/A | Maintenance offered (subject to agreement) |
| **Financial Flows** | N/A | Payment of membership fees to the consortium | License fees incurred (distinct costs) |
| **Value Contributed by Users** | Feedback, testing (non-monetary) | Feedback, testing, project contributions (non-monetary) | N/A (Only Monetary) |

In the viewing tier, source code access is restricted to research and development (e.g., for prototyping technology prior to patent filing or to investigate emerging ideas). At this stage, member firms freely view, test, and give feedback on code. Because it is limited to R&D, it is accounted for internally as “research and development expenses,” and non-financial exchanges such as “tester feedback” form the principal reciprocity mechanism. Company A found it infeasible to open all internal source code at once, so restricting the scope to pre-commercial domains minimized risk and facilitated seamless technical validation. Prolonging the R&D phase also allowed extensive internal use of the asset, effectively allowing cost treatment as research and development.

The contribution tier continues to limit usage to non-commercial endeavors but grants members the right to contribute code to the project. Light membership fees strengthen these relationships, creating a consortium-like environment for reciprocal code usage. Financially, this stage is still classified as R&D, but members jointly enhance the asset with bug fixes and additional features, leading to "mutual asset enrichment." Such structures encourage more active collaboration than mere viewing and increase the quality and utility of the asset across the consortium. Because the relevant activities remain within an R&D scope, it offers financial flexibility for further developing key technologies.

Finally, the commercial-use tier covers the transition into marketable software. License fees are charged at this point, and usage of the software or components in revenue-generating products requires entering an individual or paid license. This ensures that once a project shifts from exploratory research to actual business use, financial returns for the asset's creators are guaranteed. It also clarifies the transfer pricing problem, rationalizing the imposition of usage fees as standard market transactions.

#### 8.4.2.2 Advantages of the Consortium Model

The consortium model excels in three respects: (1) legal and tax adaptability among multiple group entities; (2) cultivation of an internal collaborative culture; and (3) capacity to establish a revenue model at scale. Even among separate corporations within the same group, non-commercial collaboration during the R&D phase can be justified, while license fees in the commercial phase provide a path to fair compensation. The high degree of freedom during the viewing and contribution stages aligns well with hackathons or training programs, thereby fostering an inner-source culture. Scaling up is facilitated by the clarity with which assets move from R&D to commercial licensing, ensuring the initiative does not stall at mere free sharing. Consequently, the consortium model serves as a robust framework for large enterprises seeking to promote InnerSource strategically and maintain internal controls.

### 8.4.3 Individual Contracts: An Outsourcing-Type Collaboration Model

Company B, one of the interview targets, is a large corporate group with multiple subsidiaries under a holding-company headquarters. The headquarters houses a common development division that, in cooperation with the internal OSPO secretariat, leads joint development with subsidiaries and promotes open source-style methods throughout the group. Not all development activities take place in completely open repositories; the environment accommodates both broad-scale InnerSource collaboration and more strictly defined "outsourcing-type collaboration." The latter operates on formal outsourcing contracts with compensation based on work hours.

While previous InnerSource cases have reported instances where employees from certain departments are treated as contractors with internal monetary exchanges [49], this case has evolved into a more comprehensive framework enabling value transfer between group companies. The basic structure follows a two-stage approach: during the R&D phase, feedback is considered as compensation, while in the productization phase, license and support fees are collected.

#### 8.4.3.1 The Two-Phase Outsourcing Model Linking R&D to Commercialization

This outsourcing-type collaboration model includes the following features. First, teams within the group sign a contract that calculates compensation based on “time spent × labor cost.” For example, subsidiary engineers may access an enterprise repository owned by headquarters, providing development hours for which the subsidiary receives fees from headquarters. The resulting deliverables generally become the property of headquarters or a designated project owner, which distinguishes between R&D and commercialization phases in its compensation formula.

Because the viability of software functionalities remains uncertain during the R&D stage, providing monetary remuneration is tricky. Hence, activities such as requirements definition, bug reporting, and QA support—collectively referred to as “feedback”—are treated as a form of compensation. Feedback is deemed equivalent to testing labor. Thus, it is not purely free of charge but rather a reciprocal arrangement.

During commercialization, a project transitions to a formal product, and explicit monetary terms—license and support fees—are introduced under a product supply agreement. Through this two-phase approach, uncertainties in R&D can be absorbed, and subsequent commercial use offers a path to revenue generation. Standardized contract templates specify rights ownership, scope of reuse, interface definitions, and so forth, minimizing negotiation costs and avoiding future conflicts.

*Table 8.4 Phase-by-Phase Compensation in Individual Contracts*

| Phase | Compensation Type | Contract Format Example | Accounting/Tax Treatment |
|---|---|---|---|
| **R&D** | Feedback (non-monetary) | R&D License Agreement | Counted as effort for design/QA; can be explained as labor |
| **Commercialization** | Monetary (license fees, etc.) | Product Supply License, Outsourcing Agreement | Involves transfer pricing; further requirements for global transactions may apply |

#### 8.4.3.2 Transfer Pricing and High-Value Code Sharing: Practical Perspectives on Global Expansion

In global business operations, free sharing of source code across international units often raises transfer pricing issues. High-value or marketing-related core functionalities are difficult to explain solely through labor costs, so licensing or buyout/subscription fees become necessary. One example is features with high commercial or marketing impact. Complex cases that exceed the scope of ordinary contracts may require specialized agreements defining detailed approaches for transfer pricing and intellectual property allocations.

In some instances, domestic (e.g., Japan-based) enterprises concentrate core development locally, with overseas branches primarily providing labor input, thereby mitigating transfer pricing complexities. In other words, addressing the tax risks triggered by InnerSource–oriented code sharing may require tightening contractual structures as well as adjusting organizational design (e.g., localizing high-value development).

#### 8.4.3.3 Standardization and Template-Driven Implementation: Integrated Management of Contracts, Accounting, and Tax

As software transitions from research (an intangible “development cost”) to a commercial product, the enterprise shifts from an R&D license agreement to a product supply license. Adopting standardized contract templates provided by a central management team allows individual departments to ensure compliance with accounting and tax requirements by simply following established procedures, thereby reducing the burden of intrafirm negotiation. Table 8.5 offers examples of such template types.

*Table 8.5 Major Types of Contract Templates*

| Type | Target Phase | Compensation | Key Provisions |
|---|---|---|---|
| **R&D License** | R&D phase | Non-monetary (feedback) | Obligation for requirements definition, QA reporting, permissible scope of joint R&D |
| **Product Supply License** | Commercial phase | Monetary (license fee) | Fees for post-release support, ownership and reuse provisions |
| **Special Contract** | Complex/high-value projects | Case-by-case | Rights ownership, transfer pricing, confidentiality, allowances for further reuse |

When expanding overseas, sharing source code with foreign offices may trigger export controls if the technology is subject to military or sanction-related regulations. Generally, it is necessary to check whether the repository includes restricted technologies, possibly performing repeated compliance reviews during repository creation or user registration. Even when promoting InnerSource, thoroughly documenting repository functionality and usage objectives is crucial for clearing export controls.

#### 8.4.3.4 Advantages of the Outsourcing-Type Collaboration Model

This model is useful for several reasons. During the R&D phase, accepting feedback as compensation accommodates high uncertainty, while in the commercialization phase, specific licensing or support fees ensure a robust monetization route. Moreover, standardized contract formats, architecture diagrams, and consistent accounting/tax explanations reduce compliance costs related to transfer pricing or export regulations. Ultimately, the framework supports both bottom-up innovation and top-down oversight, enabling enterprises to incorporate potential collaboration opportunities across the group in a systematic manner.

Rather than simply implementing “in-house open source,” this outsourcing-type collaboration model constitutes a comprehensive system that accounts for legal, accounting, and tax considerations. In future applications, it may prove increasingly beneficial in industries where stricter guidelines on high-value software transfers are required, prompting comparative studies and accumulated cases. As a result, it offers a powerful means for inner-source–oriented companies to maintain internal governance while pursuing flexible collaboration.

Its treatment of code in the accounting context resembles Company A’s approach. The main difference is whether collaboration is driven by elaborate contract templates for each project or by membership fees in a consortium setting. In both cases, these practices extend beyond a simple adoption of open source approaches internally and instead constitute a systematic framework rooted in accounting, tax, and legal prerequisites. On the other hand, from the standpoint of internal openness, portals for information sharing, documentation disclosure, and techniques for stimulating collaboration "beyond individual contracts" remain vital. Although the precise scope of InnerSource is not the primary concern of this paper, additional efforts to determine how individual contracts can be integrated into wider collaboration may further enhance InnerSource adoption.

## 8.5 Summary: Flexible Governance via InnerSource Guidelines

When implementing InnerSource within an enterprise, merely establishing rules for accounting or tax compliance is insufficient. It is more important to share a cohesive vision of the desired collaboration model across the entire organization and to create additional rules only as necessary. Understanding the diversity of InnerSource topologies, as described in the first half of this chapter, and designing mechanisms that adjust levels of transparency and editing rights according to organizational circumstances are essential.

Preliminary surveys of Japanese firms have revealed a clear polarization between companies that are extremely sensitive to accounting, tax, and legal concerns, and those that show little concern for such matters. While not all companies require strict management frameworks like Company A or international manufacturer Company B, interviews with Internet companies E and F demonstrate that many organizations have already established implicit protocols for accounting practices and inter-departmental coordination. Therefore, basic guidelines for value and resource transfer should be considered essential elements when implementing InnerSource. If psychological safety and collaboration boundaries remain ambiguous, middle management and audit departments are likely to impede implementation. In this context, it is necessary to consider a comprehensive management framework that encompasses not only source code access rights but also the actual scope of usage permissions.

InnerSource fundamentally aims to promote new value creation via shared codebases and cross-departmental contributions, dismantling organizational silos in the process. However, in Japanese corporate cultures with strong emphasis such as "Ringi" (written approval loop processes) and quality assurance, having external engineers modify code can generate resistance from oversight teams—especially if accompanied by further legal concerns over accounting or export controls.

Paradoxically, addressing all these concerns at the outset can enlarge the stakeholder pool unnecessarily, hindering the adoption process. Harnessing InnerSource effectively often calls for introducing it on a small scale, within psychologically safe boundaries, building early successes, and only later refining accounting and security frameworks. It is usually more effective to begin with minimal guidelines and then gradually develop more sophisticated regulations once the field has demonstrated concrete results.

As noted in reference to Company A, there is also an approach that treats accounting standards, compliance, and intellectual property considerations as "part of an early-phase rulebook," potentially facilitating stepwise dissemination of InnerSource culture. Although it may appear counterintuitive, making institutional details explicit from the start—such as specifying how copyrights or internal accounting will be managed—can reassure line managers that "there are mechanisms in place to prevent chaos." For instance, clarifying a licensing system for internal use, modification, or redistribution can remove ambiguity and smooth the introduction phase. Given the case such as Japanese business norms of implicit agreement, close coordination with key stakeholders can help gather each department's preferences early on, enabling champions to share information enterprise-wide. Although this coordination requires careful communication, once consensus is reached, it tends to remain stable, supporting broad-scale expansion.

Hence, in drafting InnerSource guidelines, it is crucial to delineate which aspects should be standardized enterprise-wide, and which can be adapted by individual departments or projects. Overly uniform requirements risk stifling innovation, while the complete absence of guidelines can unsettle managers enough to reject adoption. Deciding which guidelines to formalize at each stage, while balancing risk tolerance, performance appraisal systems, and technology stacks, requires a thoughtful plan.

In summary, formulating InnerSource guidelines clarifies what network structures and collaborative breadth an enterprise ultimately aims to achieve. It also entails defining how much transparency and editing freedom to grant at each stage. This process goes beyond mere compliance with accounting or tax rules. By elevating psychological safety and encouraging voluntary contributions within the enterprise, a scenario emerges in which bottom-up creativity and top-down governance can harmoniously cultivate a mature InnerSource environment. This is the fundamental significance of devising guidelines for InnerSource adoption.

# 9 Multi-layered Incentive Design and Key Considerations for Corporate Adoption

This chapter classifies and analyzes six types of incentive mechanisms used to establish and sustain InnerSource activities within corporations. These typologies are formed through complex interactions with each organization's culture, business model, and employee motivation design. The selection and combination of these mechanisms significantly influence the vitality and continuity of InnerSource initiatives. The classification framework consists of two primary dimensions: the focus of incentives (individual versus project-based) and the nature of rewards (monetary compensation versus merit recognition/activity acknowledgment). As illustrated in the following table, these dimensions combine to form six distinct models. Through interviews, the actual adoption patterns of these models have been revealed.

The effectiveness of incentive programs in InnerSource projects has been empirically demonstrated through Huawei's case study [50]. This chapter extends this understanding by systematically analyzing incentive programs through case studies of Japanese corporations. Furthermore, it examines the incentive formation process prior to institutionalization, analyzing applicable incentive models at each developmental stage. Notably, it reveals that different incentive models are required during initial and mature phases.

While the effectiveness of incentive models is widely recognized, implementing systematic reward programs and other institutional measures immediately is challenging when organizational adoption of InnerSource is still in its early stages. Therefore, this chapter categorizes feasible incentive models for companies in the initial implementation phase and presents a graduated pathway for their introduction. The aim is to provide guidelines for appropriate incentive design that aligns with the developmental stages of InnerSource activities.

*Table 9.1. Six Types of Incentive Models*

| | Individual-Based | Project-Based |
|---|---|---|
| **Monetary (Financial Return)** | **Individual Performance-Linked Model**<br>Direct reflection in salary or bonus. In the case of Company F in the information and communications industry, individual contributions are discussed with managers and linked to regular performance evaluations. | **Inter-departmental Financial Transaction**<br>Based on the actual use of tools or libraries developed through InnerSource activities, funding is returned to—or exchanged among—the relevant departments. An example from a global manufacturing company (Company B) illustrates an explicit contractual arrangement. |
| **Acknowledgement (Recognition of Accomplishments/Activities)** | **Awards Program Model**<br>Individual engineers are honored for their achievements through personal awards or by having their names listed in a game's end credits. At Company D, the custom of "listing names" increases engineers' motivation. | **Formalization of Activities Model**<br>Not only are individual engineers acknowledged, but the relevant activities become officially recognized, and budgets are often allocated behind the scenes to support them. Company D also provides an example of how acknowledging names can significantly enhance motivation. |
| **Hybrid (Both Axes)** | **Awards Program Accompanied by Bonuses**<br>Corporate awards or internal accolades are publicized, and monetary rewards are provided. Company F in the information and communications industry offers an "Inventive Creativity Prize" and the "President's Prize." | **Foundation Model**<br>Similar to the Cloud Native Computing Foundation (CNCF), this model is adopted internally to honor individual or team technical contributions. It also provides expanded budgets or additional resources. Company C in the internet industry exemplifies this approach. |

Each of these six types has advantages and disadvantages. In actual corporate practice, it is common to combine multiple models in ways that best match the situation. To acknowledge collaboration and cultural contributions that cannot be evaluated solely through a single metric, incorporating a variety of reward factors into the evaluation design is indispensable.

Regardless of the chosen model, it is crucial to share evaluation criteria and operational processes across the organization so that engineers clearly understand what kind of return to expect from contributing to InnerSource. If the evaluation system remains ambiguous, there is a risk that highly motivated engineers will become overburdened, and that the system itself will lose substance. Therefore, careful attention must be paid to incentive design and operation during InnerSource promotion.

### 9.1.1 Individual Monetary Return / Individual Performance-Linked Model

The approach of providing financial returns to individuals or reflecting contributions in an existing personnel evaluation system tends to be relatively easy to implement in Japanese corporate human resource frameworks. Indeed, Company F in the information and communications sector demonstrates how this method plays a significant role in sustaining InnerSource.

When an individual engineer actively contributes to other departments or teams, that contribution is reflected in compensation or bonuses under the existing performance evaluation framework. This mechanism helps sustain engineers' motivation. However, determining how to award additional points for InnerSource-specific contributions depends heavily on managerial policy and discretion. At Company F, engineering managers coordinate semiannual goals by explicitly discussing "the extent to which engineers contribute to other departmental repositories" in advance. At the end of each review cycle, these contributions are quantified and reflected in compensation.

A crucial element for making this system function is the ability to objectively track and visualize InnerSource contributions, typically through visualization tools or workload management systems. At Company F, engineers enter work details into an internal web application, and engineering managers then reference this data to create evaluation materials. As a result, engineers gain a clear sense that "contributions are reflected numerically."

Nevertheless, relying solely on an approach that ties contributions to existing evaluation systems may overlook qualitative achievements such as solving highly complex issues or demonstrating the coordination skills needed to bridge departments. Therefore, Company F places emphasis on qualitative, subjective input as part of the final managerial assessment, including interviews with engineers and relevant stakeholders.

During the early phase of InnerSource adoption, managers frequently worry whether such contributions might interfere with primary job responsibilities. At Company F, this concern is mitigated by explicitly clarifying that team goals can coexist with InnerSource activities. A two-tier goal system is in place so that engineers can participate in InnerSource after meeting main project targets, and there is a formal procedure for consulting with superiors to prioritize activities.

An important advantage of the financial-return model is that it can pilot InnerSource initiatives without necessitating major reforms to the company-wide HR system. In contexts where a goal management system is already established, as at Company F, simply adding "InnerSource-related contributions" to the existing evaluation criteria keeps implementation costs and executive resistance relatively low.

However, an evaluation model heavily reliant on managerial discretion bears the risk of inconsistent assessments. One manager may rate InnerSource contributions highly, while another may not, potentially leading to perceptions of unfairness among engineers. Company F aims to mitigate such discrepancies by fostering stronger collaboration among engineering managers and striving to harmonize evaluation standards, but minor variations persist at the operational level.

To ensure fairness, it is important to balance quantitative assessments (e.g., the number of pull requests or issue resolutions) with qualitative aspects (e.g., contributions to communication or appreciation expressed by other departments). At Company F, a multidimensional approach ensures that not only workload, but also innovative ideas and leadership skills remain subject to evaluation.

In conclusion, combining monetary returns with existing evaluation systems constitutes an effective strategy for recognizing InnerSource activities without radically altering a company's underlying culture. However, operational considerations—such as dependence on managerial attitudes and how to incorporate qualitative factors—must be addressed carefully. Rather than relying exclusively on key performance indicators (KPIs), it is beneficial to create mechanisms that encourage engineers' learning, creativity, and autonomy.

### 9.1.2 Individual Acknowledgement / Awards Program Model

The individual acknowledgement or awards program model emphasizes non-monetary returns, prioritizing praise and status symbols for InnerSource contributors. At Company D, known for its core gaming business, listing names in the end credits represents a symbolic way of recognizing achievements, effectively enhancing developers' sense of self-affirmation and pride.

In internet-related industries, such as Company D's environment, the motivation often lies in "creating something interesting." In game development, being part of a successful product and building a personal reputation are culturally significant. Consequently, officially documenting an engineer's involvement in a project through InnerSource activities can profoundly boost motivation. In game development specifically, having one's name listed in the end credits is regarded as directly tied to future opportunities and career progression.

At Company D, developers who contributed to shared libraries or infrastructure tools are publicly acknowledged in games or internal documentation, thereby making visible the sentiment that "this product was completed thanks to certain individuals." This type of recognition serves a different purpose than financial compensation by elevating engineers' pride and mutual respect. Over the long term, it may raise the organization's overall technical expertise and willingness to learn.

One notable advantage of an acknowledgement-centric approach is that organizations can stimulate engineers' motivation without allocating substantial budgets. Nonetheless, as projects expand to include tens or hundreds of contributors, questions may arise regarding how to manage the granularity and scope of recognition. If names are added without clear criteria, there is a risk of diluting the value of inclusion over time.

On the other hand, excessive emphasis on recognition may incentivize developers to pursue visibly striking commits or short-term achievements that overshadow review processes, thereby undermining genuine quality improvement and learning benefits.

The priority placed on recognition can vary significantly among organizations. At Company D, recognition-based models such as listing names in the end credits and corporate awards function effectively due to a deeply ingrained mindset of valuing originality and creativity. Because engineers in such organizations often strongly seek recognition, these models can prove more effective than monetary or budget-based schemes. Thus, it is not an overstatement to suggest that alignment between organizational culture and the chosen evaluation model significantly influences InnerSource success.

In summary, the awards-based model has considerable potential to invigorate InnerSource activities without requiring major financial investments. Nevertheless, careful attention is required regarding operational challenges and balancing with other metrics. Overlooking elements beyond recognition could impede engineers' long-term career paths and growth, potentially leading to talent attrition or declining motivation. It is therefore essential to align the framework with the broader corporate culture.

### 9.1.3 Hybrid Individual Model / Awards Program Accompanied by Bonuses

The hybrid model that integrates both financial and recognition-based incentives provides a two-pronged approach to rewarding engineers. In this model, awards programs and honorary titles are supplemented by monetary benefits such as salary and bonus adjustments. At Company F, where an awards system is already well established, a nomination mechanism (recommendation system) for InnerSource achievements has been introduced, enabling smooth operation of this hybrid model.

Company F regularly holds award ceremonies (monthly or quarterly) to publicize "high-value contributions." When outstanding InnerSource activities are recognized, a small bonus or meal allowance is provided, combining both the esteem of an award and a financial benefit. While recognition remains at the core, outstanding engineers also benefit from improved salary evaluations or promotional opportunities. Multiple awards can potentially lead to recognition as a lead engineer, which in turn paves the way for higher compensation or advancement in managerial ranks.

One distinctive feature of this model is the nomination process. Project managers and team leaders at Company F hold nomination authority, lending credibility to recipients' achievements. This fosters a sense of fairness among engineers. However, if the nomination process becomes perfunctory, there is a risk of overshadowing genuine contributors.

Offering both monetary and recognition-based rewards accommodates the diverse values held by engineers. Some place greater emphasis on professional fulfillment and acknowledgment rather than compensation. This aspect is particularly appealing to junior employees and engineers with strong learning aspirations.

Nevertheless, using two parallel incentive mechanisms can complicate operations. If the schedule for awarding recognition or incorporating it into salaries and bonuses is ambiguous, confusion may arise, potentially undermining trust in the system. Company F addresses these concerns by synchronizing the award process with its quarterly performance-evaluation cycle.

The recommendation and awards systems also facilitate a culture in which engineers openly praise each other. At Company F, award recipients are announced through various channels, and messages of congratulations and appreciation can be exchanged among engineers. This fosters a positive environment surrounding InnerSource activities and enhances psychological safety within engineering teams.

Still, the risk of developers chasing recognition for personal prestige or managers exerting excessive control cannot be overlooked. Company F moderates these tendencies by integrating team goals and relative assessments into the incentive system, thereby curbing overzealous competition for recognition.

This hybrid evaluation model reflects corporate expectations of harnessing diverse motivational factors among the engineering workforce, while still collecting tangible corporate benefits. As InnerSource activities grow, they contribute not only to technological advancements but also promote a culture of voluntary collaboration and mutual support among engineers. Consequently, this model can serve as fertile ground for corporate innovation.

### 9.1.4 Monetary Return at the Project Level / Inter-departmental Financial Transactions

This model provides budgetary feedback based on the outcomes of InnerSource activities or on the extent of shared platform usage. Alternatively, it may involve licensing fees for usage. When a shared library or tool is developed internally and used by other departments—resulting in explicit cost reductions or other measurable advantages—a portion of the gained benefits is redistributed to the contributing team or department, or formal transactions take place among departments. Examples include Company A and Company B, which have introduced variations of consortium-based models, individual contractual agreements, or fee-based support contracts.

The key principle of this model is the clear quantification of the economic impact from InnerSource efforts, accompanied by visible intra-organizational transfers of value. Departments responsible for maintaining shared libraries or infrastructure gain a strong rationale to continue improvements when there is tangible financial feedback. Departments that once felt like “support teams in the background” might also find that clear financial returns elevate their standing and influence within the organization.

Such initiatives often align with cooperative policies in corporate structures. However, as suggested by Company A’s “Give and Given” philosophy, it is essential to position financial transactions within the broader InnerSource program context rather than treating them as isolated monetary exchanges.

Because project-level monetary feedback expands the scope of evaluation beyond individual contributions to departmental or project-based impacts, it can invigorate cross-functional collaboration. For instance, if a department desires a particular function, it might offer to share development costs, thereby strengthening the culture of collective ownership within InnerSource.

Designing how financial returns are allocated requires the integration of technical and financial indicators. Reliance solely on metrics such as commit counts can lead to superficial results, while focusing exclusively on final profit contributions may undervalue the foundational work needed at the initial stages. Additionally, the distribution method and timing of payments can significantly influence engineer motivation. A year-end lump-sum distribution may be out of sync with project progress cycles, so careful consideration is needed when implementing such mechanisms.

Implementing financial return models can also raise issues related to corporate law and accounting, especially in large enterprises. Concerns about transfer pricing and compliance necessitate meticulous determination of cost estimates and contractual forms.

Thus, making budgetary feedback a central strategy requires close collaboration with corporate finance and legal teams, as well as cross-organizational metric design. Although this can be costly in time and resources, successful implementation can create robust communities at both individual and departmental levels. In the long term, this approach can significantly enhance the sustainability of InnerSource while promoting more efficient resource utilization and accelerating new business development.

In summary, this model delivers a clear organizational message that "collaboration drives economic returns," but seamless operation demands sophisticated coordination. Moreover, these incentives target projects and departments, so the recognition of individual engineers may rely on each department's internal evaluation. Additional considerations for individual recognition may thus be necessary. Ultimately, organizational maturity and the strong commitment of senior management often determine successful adoption.

### 9.1.5 Acknowledgement at the Project Level / Formalization of Activities

This model goes beyond recognizing cross-departmental achievements with commendations. It involves formally institutionalizing such collaboration within the organization, providing continuing support and evaluation as an official process. A global manufacturing company, Company B, exemplifies such a mechanism: bottom-up initiatives or informal communities gain upper-level organizational recognition and become integrated into the enterprise-wide infrastructure.

Interviews indicate that Company B has a multi-layered collaboration structure. At the lowest level, engineers freely coordinate among themselves as needed, while upper layers include formal projects or technology-focused communities that require inter-departmental consensus. The concept of "formalizing activities" is pivotal, ensuring that contractual agreements and budget allocations are in place at the higher-level layers.

The “technology strategy community” at Company B focuses on corporate-wide challenges such as software development and open source adoption. Engineers from various departments voluntarily join these communities to solve problems. When an initiative produces outstanding outputs, recognition is granted throughout the group, and additional budgets or resources are allocated by upper management. This progression is known internally as “formalizing the activities.”

Participation in these cross-departmental communities incentivizes engineers by enabling them to contribute without negatively affecting their evaluations of primary job performance. For instance, when an internal community tackles open source compliance or standardization of the development environment, each engineer’s contributions are documented. Departments can then easily reference these contributions and evaluate them more accurately.

Clarifying the relationship between cross-department activities and formal performance assessments allows engineers to participate without fear of harming their main responsibilities. According to interviews, “the community reports which engineer contributed what,” thereby enhancing that individual’s evaluation and motivation.

Moreover, Company B holds an annual event to showcase bottom-up initiatives. Selected teams receive monetary rewards, additional funding, or recognition as formal projects. This progression from grass-roots efforts to official status in a higher-level technology strategy community or department contributes to the broader proliferation of InnerSource. One example is a small continuous integration/continuous deployment (CI/CD) library that evolved, through intermediate working groups, into a formal proposition: “Let this be the company-wide infrastructure,” eventually becoming managed by the platform engineering department. This bottom-up to top-down elevation lies at the heart of the formalization approach.

Company B’s human resource policies explicitly include the evaluation of cross-departmental activities, which broadens opportunities for younger engineers. By participating in external events, internal communities, or collaborating with an internal Open Source Program Office (OSPO), new chains of technical collaboration are initiated, ultimately extending InnerSource networks internationally.

Overall, this project-level acknowledgement and formalization approach actively embraces bottom-up initiatives while channeling high-value outcomes into upper-level organizational endorsement and support. The technology strategy community provides a core platform to unify such cross-functional work and tie it to performance evaluations. As a result, previously hidden or incremental improvements become more visible, encouraging sustained collaboration across multiple layers. Engineers are thus empowered to continue challenging conventional norms, and the organization gains fresh technical foundations and opportunities for innovation.

### 9.1.6 Acknowledgement at the Project Level / Foundation Model

Another model of project-level acknowledgement, often referred to as “budget and recognition reflection,” acknowledges InnerSource contributions not only with recognition but also by allocating budgets and awarding public commendations at the group or project level. Company C in the internet industry offers a prime example, often described as a “foundation model,” which parallels the mission of external organizations such as the Cloud Native Computing Foundation (CNCF). Multiple projects and engineers collaborating across different business lines receive both financial and honorary support.

At Company C, engineering teams that implement open source–style development can gain official recognition and dedicated funding for the following fiscal year if they demonstrate notable achievements. This approach was initially introduced to address a problem: although engineers at Company C had the freedom to develop and share new technical ideas, these activities were often overlooked in management by objectives (MBO) and key performance indicators (KPI), leading to low formal recognition.

Under the Foundation model, teams submit proposals to an internal system (known as “Program A”), which evaluates the number of users, user satisfaction, and technical maturity, among other indicators. According to interviews, these criteria are refined with each iteration of the program to achieve greater transparency and acceptance. The evaluation takes into account GitHub star counts, quality of documentation, collaboration across business units, and various other factors to ensure that outcomes are not judged solely by the number of commits.

*Table 9.2. Overview of the Foundation Model Incentive Program Stages*

| Phase | 1. Preparation & Application | 2. Review | 3. Evaluation & Ranking | 4. Periodic Renewal (Reassessment) |
|---|---|---|---|---|
| **Summary** | Prepare and submit the project for Program A, ensuring prerequisites are met. | Assess submitted material based on defined criteria (e.g., user base, satisfaction, GitHub stars). Conduct applicant interviews as needed. | Determine a ranking based on results, then allocate incentives and support (including technical marketing assistance). | Accept renewal applications, perform reassessments, and decide on rank upgrades or continuations. |
| **Primary Activities** | Confirm usage records and documentation, then submit the project proposal. (Examples include small- or medium-sized libraries, as well as emerging AI models.) | Evaluate the project according to user base, satisfaction, GitHub stars, etc. (Continuous refinements occur in subsequent rounds of review. | Assign a ranking on a five-tier scale, finalize the scope of support (technical marketing, funding). | Accept updated applications, perform assessments, and decide whether to renew or adjust rankings. |
| **Relevant Stakeholders** | Applicants | Board members (engineers from relevant units) | Board members, technical marketing teams | Applicants, board members |
| **Stakeholder Interests** | Increase project visibility, secure financial or logistical support, pursue career development | Ensure fair technical evaluation | Raise internal and external visibility, promote engineering culture | Secure continued or increased support |

The Foundation program has proven to be a strong motivational driver for engineers at Company C. Interview data suggest that individuals apply for a variety of reasons, including the desire for incentives or greater internal visibility. The program uses a five-tier grading system, allowing teams to reapply for higher levels once certain conditions are met, thus supporting ongoing project growth and providing avenues for career development.

From a recognition standpoint, those passing the review process are extensively featured in internal events and blogs, allowing projects and foundational technologies to become known across the organization's diverse domains (e.g., gaming, advertising, media). This elevated status for individual developers and their brands aligns with a broader corporate culture that encourages experimentation, including the open dissemination of proprietary OSS initiatives.

Nonetheless, the dual emphasis on funding and accolades raises questions about maintaining transparency and fairness in the review process. Company C endeavors to address this by forming a cross-departmental review committee for larger projects, aiming for more objective and balanced assessments.

Once granted funding, teams can experiment more freely, such as through pilot releases or proof-of-concept initiatives. Senior management also benefits, since “if there is corporate backing, platforms can be grown and maintained,” further accelerating InnerSource expansion.

Program A follows an annual review cycle, and deadlines for rank updates are clearly defined. Although direct financial payouts are limited, support often includes technical marketing, conference travel allowances, and server infrastructure usage. The partial coverage of costs for attending international conferences, for instance, represents an alternative reward mechanism reminiscent of CNCF’s sponsorship programs.

In summary, Company C’s Foundation model exemplifies how to apply a CNCF-inspired community-development strategy within a single corporation, mitigating silos across multiple business lines and motivating engineers to undertake proactive challenges. Maintaining fairness and consistency in the review process requires continuous updates to evaluation standards and a balanced committee approach. However, once these hurdles are overcome, cross-functional collaboration and technological innovation accelerate further, anchoring InnerSource more deeply within the corporate culture.

## 9.2 Multilayered Support Systems and Mechanisms for Sustaining Activity

Incentive models for InnerSource activities demonstrate enhanced effectiveness when implemented through a spread out web-like multilayered support structure. At Company B (global manufacturing), a "middle-layer community" structure facilitates cross-departmental collaboration, establishing an interconnected support network across various organizational levels. Similarly, Company F in the information and communications sector complements managerial alignment with a formal recognition system, while Company D implements diverse acknowledgment mechanisms, including personal appreciation and end-credit recognition.

This web-like multi-layered incentive structure enables engineers to participate in projects of interest beyond their primary departmental responsibilities. However, this framework does not constitute unrestricted autonomy. Instead, it incorporates formal roles and community frameworks, providing designated platforms for activity reporting and progress sharing.

For example, Company B eschews top-down mandates for cross-departmental collaboration. Instead, mid-level champions and voluntary communities initiate programs and projects independently, attracting engineers based on specific interests. This approach facilitates a system where "voluntary participation and demonstrated results naturally lead to recognition," eliminating the need for extensive organizational mandates. This structure also mitigates potential resistance to major organizational changes.

For the fundamental internal development of InnerSource, a foundation-type model, as implemented by Company C, emerges as an effective option. During the initial implementation stage, maintaining existing personnel evaluation models while adopting mechanisms based on individual assessment represents a practical approach. However, even in such cases, it is essential to establish systematic organizational support mechanisms rather than relying solely on individual discretion and enthusiasm. Notably, the recognition systems in these foundation-type models ultimately relate to products contributing to core business operations, which can potentially reflect in individual performance evaluations. Thus, incentive design requires comprehensive consideration of various interconnection points and inclusive handling of multiple factors.

## 9.3 Conclusion and Practical Guidelines

The preceding sections have examined six incentive models—monetary rewards, acknowledgements, budget allocation, and hybrid approaches—that underpin InnerSource. It has become evident that the optimal approach depends on each corporation’s culture and business orientation; there is no universal solution.

Individual-focused incentives (e.g., Individual Performance-Linked Models, Awards Program Models, and Awards Plus Bonus Models) effectively stimulate creativity and motivation on an engineer-by-engineer basis but may not fully support broader organizational collaboration. Conversely, group-focused approaches (e.g., Project-Level Monetary Returns, Formalization of Activities, Foundation Models) can be more conducive to large-scale or cross-departmental initiatives, though they generally require increased coordination with various stakeholders. Balancing these approaches is crucial.

The growing popularity of hybrid models that encompass financial, honorary, and budgetary rewards reflects the reality that engineers’ motivations extend beyond salaries. Emphasizing only recognition may offer insufficient tangible returns, while emphasizing only financial benefits may fail to fully unleash engineers’ passion and creativity. Designing flexible systems that incorporate multiple types of rewards can address these challenges.

Large enterprises may face added complications, such as inter-company accounting and transfer-pricing regulations, making it essential to engage legal and finance teams when implementing budget-oriented models. Strong leadership from senior management is often required to navigate such complexities successfully.

In InnerSource promotion, the level of “freedom” granted to engineers (in terms of time and communication) and the institutionalization of “formal approvals” (such as official support for exceptional outcomes) are equally vital considerations. Paying attention not only to direct rewards like monetary compensation or recognition but also to an environment that allows engineers to secure development time and build networks across departments is essential for sustaining motivation.

Among Japanese companies, where InnerSource is still in an early developmental stage, it may be advisable to start with lower-barrier options—for example, integrating InnerSource contributions into existing performance reviews or instituting awards programs—then gradually expand toward resource-intensive approaches such as budget-based or foundation-like models once initial benefits become evident. This incremental strategy helps cultivate an InnerSource culture while mitigating operational disruptions.

Where InnerSource reaches a growth stage, incentive models may evolve beyond mere quantitative assessments to form the foundation of community-building, learning, and innovation. Once a cycle emerges in which engineers comfortably contribute to repositories beyond their own departments—and receive appropriate recognition—development speed can accelerate, potentially generating new products and services for the organization.

In summary, corporations must combine, adapt, and continually update these six models according to their unique human resources systems and business environments, taking into account feedback from on-the-ground practices. Because each model offers distinct strengths and weaknesses, a phased approach that blends top-down leadership with bottom-up grassroots initiatives is recommended, developing an incentive framework that aligns with the organization’s growth stage.

Finally, it is crucial not to lose sight of InnerSource’s core value: promoting learning and innovation, rather than merely sharing code or reducing costs. Incentive design plays a central role in balancing engineers’ creativity with broader corporate interests; thus, enterprises should continuously refine their models, guided by long-term strategy and an engineer-centric culture.

# 10 The InnerSource Circumplex Model: A Two-Dimensional Approach to Complexity

This study confirms that diverse champions engage in InnerSource with varying motivations. As indicated by both the qualitative and quantitative investigations presented in the previous chapters, interpretations of InnerSource and the associated sense of purpose are not fixed but exhibit considerable variability. Furthermore, each champion envisions different outcomes when employing the term "InnerSource," implying that the objectives within an organization are likewise heterogeneous. Consequently, InnerSource evolves through a complex process that cannot be explained by a simple, stage-by-stage growth model.

Previous research has tended to conceptualize InnerSource adoption and maturity using single-stage maturity models. However, this chapter argues that a more complex, two-dimensional matrix perspective is necessary. Building upon this viewpoint, the concurrent use of maturity models alongside an "InnerSource circumplex model" facilitates a higher-resolution analysis of InnerSource progression. This approach is significant because it enables precise identification of multiple champions' activities and motivations, especially when they operate under distinct organizational cultures and motivations.

In order to deepen this discussion, this study applies the Olson's Circumplex Model [51] to the context of InnerSource. By drawing an analogy between an enterprise and a family, it is demonstrated that a circumplex representation of inter-team and interpersonal relationships can accurately capture organizational collaboration in terms of both flexibility and cohesion.

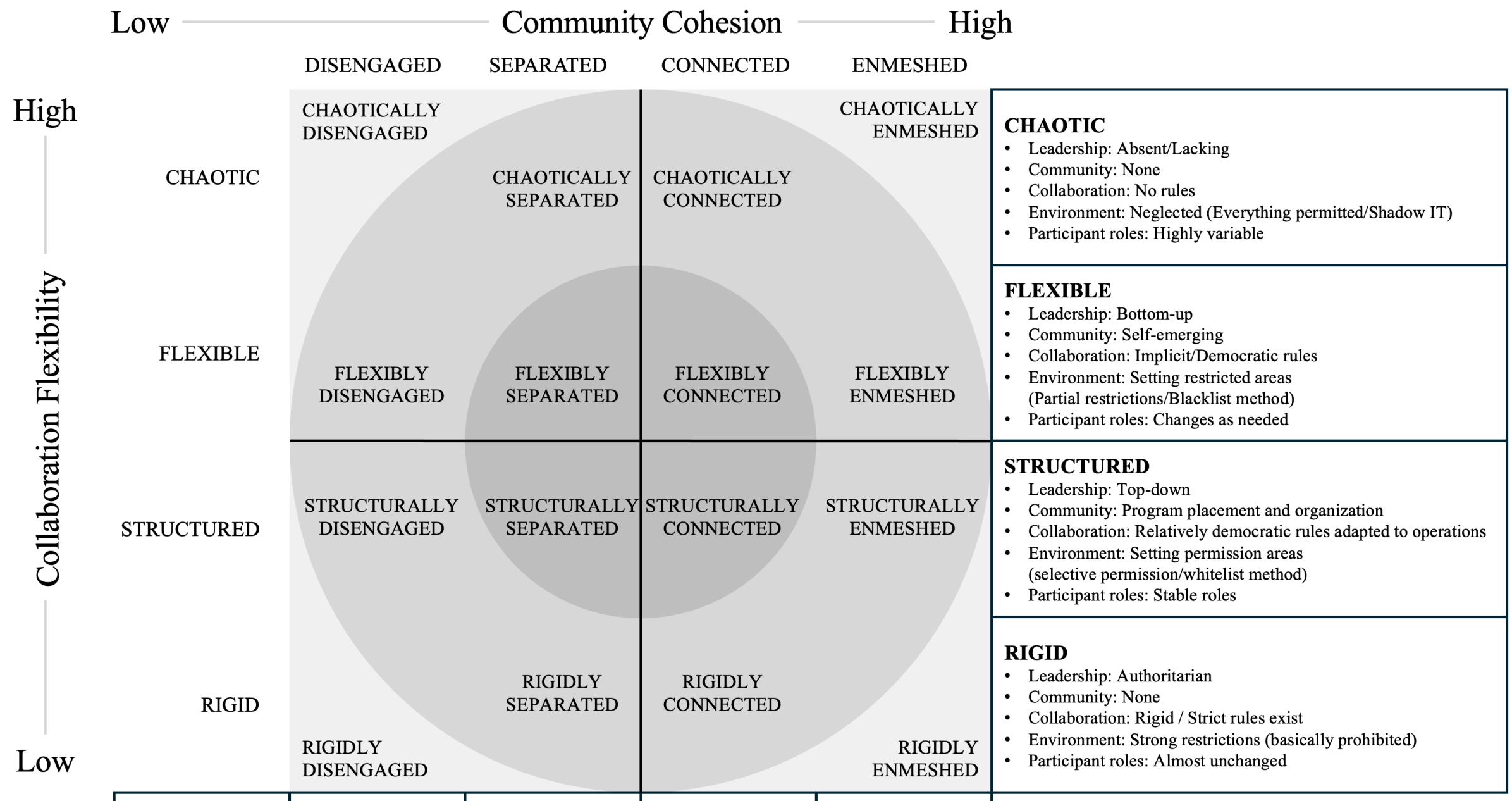

| | | | | |
|---|---|---|---|---|
| Cohesion | Disengaged | Separated | Connected | Enmeshed |
| My/Our Balance | My code<br>Strong code ownership | My/Our code<br>Weak code ownership | My/Our code<br>Weak code ownership | Our code<br>Collective code ownership |
| Dependency | Independent | Cooperative | Collaborative | Integrated |

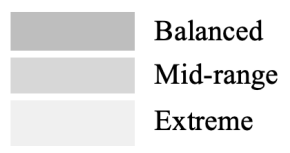

*Figure 10.1 InnerSource Circumplex Model*

Figure 10.1 depicts the InnerSource Circumplex Model. The intentions of InnerSource champions trace varied paths on this circumplex diagram, yet do not necessarily encompass the complete sharing of all code or resources within the organization. Certain champions apply only selected benefits of InnerSource to reflect practical constraints and business demands.

For example, in a state of RIGIDLY DISENGAGED (low cohesion, low flexibility), a lack of leadership and inter-team collaboration can result in repeated reinvention of the wheel, implying a high risk that even an advanced open source-style approach will fail due to immature talent and a shortage of supporters.

Under such circumstances, it is more realistic to establish a STRUCTURALLY DISENGAGED environment by introducing specific processes and rules to gradually encourage teamwork among different teams. Although the degree of organizational constraint remains considerable, this approach can still foster incremental cultural change and collaboration, ultimately moving closer to the FLEXIBLY CONNECTED state (high flexibility, high cohesion). This potential progression underscores the value of the proposed Circumplex Model.

In addition, a distinct feature of this model is that simply advancing along a linear path to higher levels of maturity is not the only possible route. In other words, an organization with low cohesion and relatively rigid practices may introduce a specialized environment in a stepwise manner while simultaneously establishing a highly flexible environment for specific projects or departments, thereby operating multiple InnerSource modes, which has different flexibility and cohesion level, in parallel.

Early-stage or newly appointed InnerSource champions sometimes imagine "opening everything at once," much like open source. In reality, however, achieving full openness in a short timeframe is challenging. A gradual transformation of corporate culture and processes is essential, and some members may remain resistant to complete openness. Such complexities represent significant hurdles to InnerSource adoption.

Consequently, selectively extracting the core essence of InnerSource and applying it in stages can constitute a practical strategy. The Circumplex Model proposed in this study visualizes such incremental and varied modes of engagement, enabling champions to accurately identify their respective positions. Subsequent sections examine specific transition pathways through illustrative case studies.

### 10.1.1 Transition from Low Cohesion, Low Flexibility

Large enterprises often adopt InnerSource primarily to drive existing process reform, using InnerSource to overcome obstacles related to accounting, legal compliance, or multidepartment decision-making structures. Although such initiatives may appear top-down, empirical evidence shows that open source-savvy engineers and middle managers frequently establish the groundwork and negotiate persistently with finance and legal departments, thereby advancing InnerSource transformation.

In large enterprises, it is necessary to obtain approvals from finance, legal, and strategic planning departments to modify existing rules and institutionalize new forms of in-house collaboration. The barriers associated with transfer pricing or other complex administrative procedures exceed the capabilities of engineers alone. Overcoming these issues represents a central challenge of the process-reform-oriented approach to InnerSource adoption.

In such cases, the design of institutional frameworks and regulations becomes the main task, requiring not only a focus on engineering issues but also a thorough understanding of consensus-based organizational cultures and the ability to balance the interests and resource allocations of various divisions. In the case of Company B, for instance, it was confirmed that a tenacious negotiator played a key role in working with the finance and legal divisions to achieve cross-departmental collaboration. Although the technological aspects of InnerSource serve as an entry point, the fundamental goal of such process-reform adoption is to radically transform organizational decision-making processes, necessitating extensive coordination skills. Based on interviews, both Company A and Company B are likely to operate multiple InnerSource environments in parallel when viewed through the lens of the Circumplex Model.

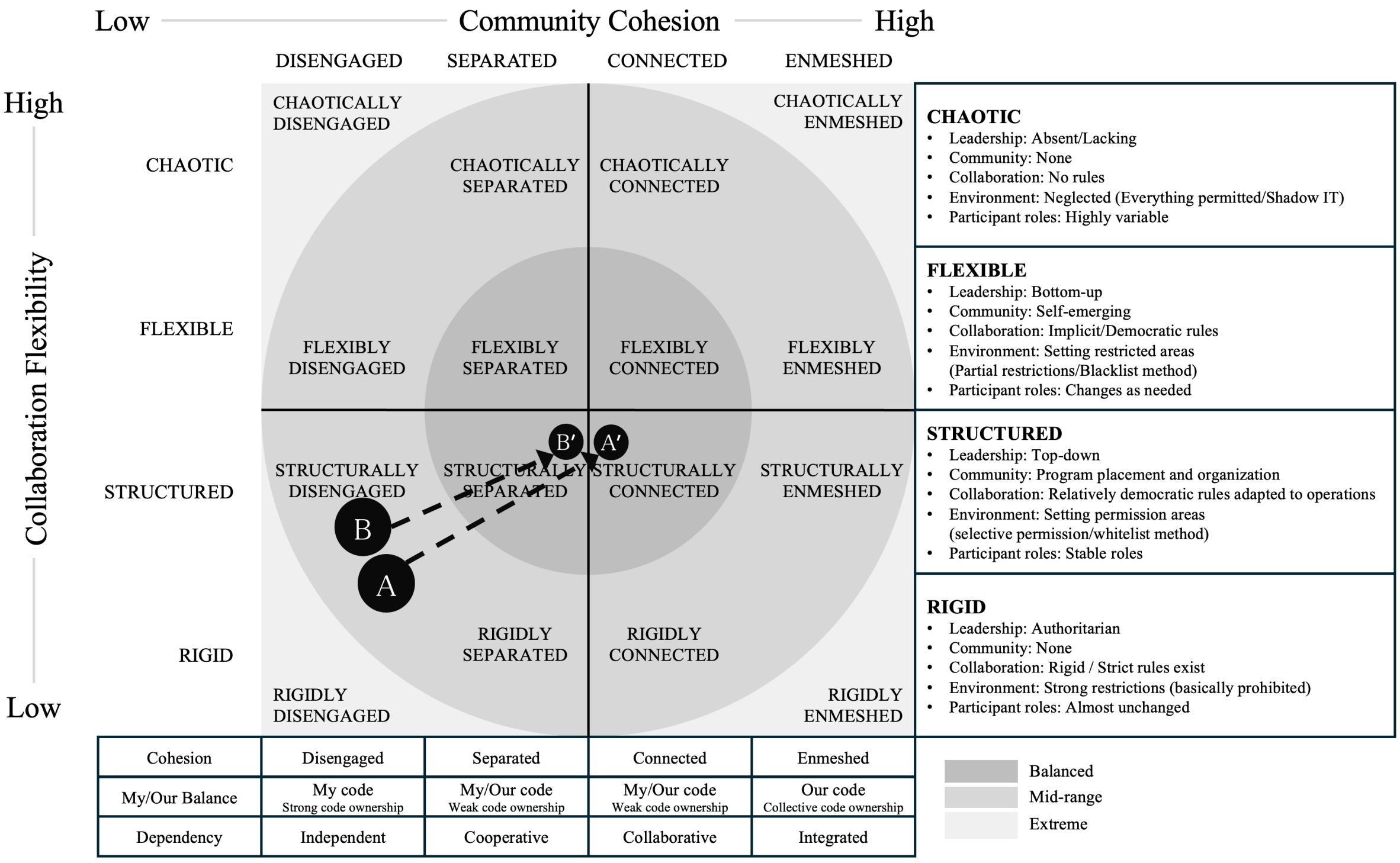

*Figure 10.2. Mapping of Company A and Company B in the InnerSource Circumplex Model*

As described in Chapter 9, Company A adopts a “consortium model” in which a membership fee system grants selective access privileges, thereby cultivating an environment conducive to collaboration. Meanwhile, Company B is progressively expanding its existing technical community, introducing tools such as standardized contract templates to facilitate seamless collaboration. However, certain consortium features have yet to take root, as evidenced by project-specific configuration of source-code management tools.

While Company A is using its consortium model to progress toward a new, flexible environment (A'), Company B appears to continue emphasizing individualized contractual arrangements, suggesting a more separated state (B'). Although both companies implement similar InnerSource practices from an accounting perspective, the contextual details may differ slightly depending on the chosen implementation method.

Deeper analysis highlights the broad influence that changes in regulations and rules can exert on corporate culture and departmental interests. When transitioning from the pilot phase to the early adoption stage, it is imperative to extend the successes of small-scale collaboration to company-wide system design, requiring commensurate awareness shifts and system adjustments proportional to the level of organizational barriers. A typical approach in large enterprises is illustrated by Company A, which demonstrated positive results through pilot projects before establishing formalized schemes and incorporating management and finance departments in institutional reforms. This methodology epitomizes the process-reform-oriented adoption of InnerSource.

### 10.1.2 Transition from Low Flexibility, High Cohesion

Enterprises characterized by low flexibility yet high cohesion are often found among smaller and medium-sized organizations. In such entities, teamwork is already functioning, but stringent rules introduced at an early stage may reduce the freedom of engineers.

For instance, in some system integration or staffing enterprises, projects and work hours are managed under strict guidelines, with tightly predefined roles that can hinder spontaneous emergence of InnerSource-driven collaboration. Such companies may also exhibit strong top-down authority and a pervasive adherence to fixed rules. Consequently, although they own the code, horizontal collaboration and mutual contributions often fail to take root, and communication among teams remains limited.

In these circumstances, it is generally advisable to promote stepwise collaboration aimed at achieving STRUCTURALLY CONNECTED on the Circumplex Model. As the organization expands, the risk of becoming RIGIDLY DISENGAGED increases. Therefore, the term “InnerSource” can be used to advance collaboration and strengthen internal coordination and transparency before cohesion is entirely lost.

Company F, which has approximately 500 employees, provides a clear example of this phenomenon. Despite its relatively small size, the company has faced an increase in factors that impede spontaneous collaboration as it grows, situating it in a RIGIDLY CONNECTED state. Nevertheless, its pursuit of heightened flexibility suggests an intention to transition toward FLEXIBLY CONNECTED (F').

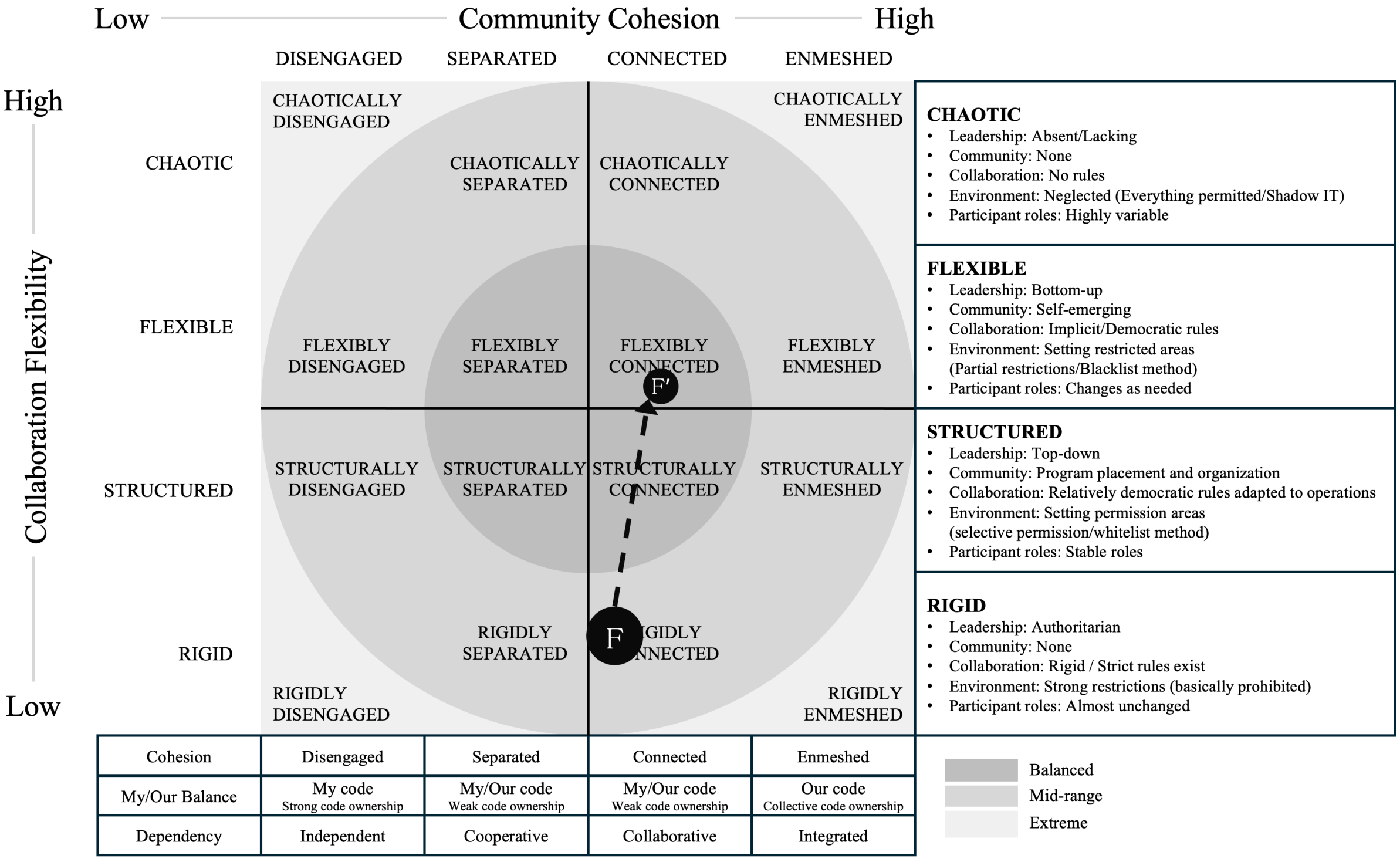

*Figure 10.3. Mapping of Information and Communications Company F in the InnerSource Circumplex Model*

Repository viewing and contribution privileges Company F were already relatively open, and the organization did not require complex inter-company accounting adjustments, allowing engineers to collaborate more easily without large-scale process reforms. The company's distinguishing feature lies in its primary motivation to cultivate an environment in which engineers can engage in creative activities and to enhance overall employee satisfaction, using InnerSource as one of the strategic tools.

In such organizations, cultural development is a central driver of implementation, and procedural barriers related to finance and legal compliance are less likely to be significant. Therefore, during the initial adoption stage, pilot projects demonstrating the value of inter-engineer collaboration serve as the primary vehicle for eventually scaling success across the entire organization. Company F validated the utility of its platforms and tools in small-scale contexts, advocating these successes to management and thereby expanding adoption incrementally.

Motivation and incentive design are critical for cultural transformation. Company F leverages existing commendation systems—such as awards for creativity or outstanding contributions—and employs internal interviews as well as monetary prizes to enhance both personal fulfillment and organizational synergy. Yet a dedicated evaluation framework for InnerSource has not yet been fully established, resulting in continued reliance on managerial discretion. Short-term incentives dominate, highlighting a future need for more robust institutional measures if the organization aims to scale InnerSource practices and transform the existing evaluation process.

Company F's adoption process began with discussions among managers, followed by active involvement from engineers who built specific use cases. These pilots then informed subsequent integration into the evaluation system, thereby fueling company-wide adoption. Going forward, it appears likely that bottom-up activities will occur more spontaneously and that InnerSource may expand into larger products.

Many Japanese companies still face ambiguity regarding whether InnerSource contributions affect compensation or promotions if top-down reforms to evaluation systems are not explicitly carried out. Even if an environment is conducive to engineers' enjoyment of collaborative tasks, consensus-driven decision-making and inflexible HR systems may hinder the recognition and resource allocation required for widespread acceptance. Company F counters this risk by generating visible pilot achievements, raising awareness across the organization, nurturing champions, and securing corporate support. This stepwise approach coordinates top-down advocacy with bottom-up engagement, mitigating the potential loss of momentum often observed in the early adoption stage.

In organizational culture change, the decisive factor is how extensively employee initiative is fostered and how that initiative is evaluated by management. The challenge is how to make bottom-up enthusiasm among engineers visible and how to link that enthusiasm to tangible support from executives and middle managers. Champions need leadership skills that can effectively mobilize people. Once a communication-driven culture is well established, InnerSource may evolve into a core engine of organizational innovation rather than simply a development methodology.

This aligns with findings on the motivations for InnerSource adoption. During the transition to the early adoption stage, employee satisfaction emerges anew as a key driver. That is, cultural transformation and improving workplace conditions become a central focus. Testimony from Company D also indicates that active feedback and communication among teams is highly desirable for advancing from growth to mature stages, and introducing InnerSource in such an environment can potentially increase both employee satisfaction and innovation. In this organizational-culture-driven approach, the critical issue is how to harness the enthusiasm of developers and encourage open communication, meaning champions require not only technical expertise but also skills in personnel management and the facilitation of dialogue.

### 10.1.3 Transition from High Flexibility, Low Cohesion

As InnerSource matures, it may be adopted for the primary goal of improving development efficiency in specific products or services. In Company C, for instance, where internal projects have grown large and complex, resource shortages and repeated redundant implementations of identical features have posed significant challenges. InnerSource has been used as a solution to these problems. In such scenarios, issues related to accounting or corporate culture are already partially resolved, so intensifying technical collaboration quickly yields tangible benefits.

Moreover, in Company E, an InnerSource-like development style spontaneously emerged as former teammates continued to fix bugs or add features to each other's projects even after moving to different departments. This pattern suggests that shared interest in productivity arose organically among engineers, fostering mutual assistance without formalized licenses or policies.

Company D exemplifies a strategy of reinforcing cooperation between platform teams and user-facing teams to improve overall service quality and accelerate development. The productivity-driven approach is more likely to take root where legal and accounting risks have been partly addressed and the necessity of platform integration is widely acknowledged, thus promoting rapid InnerSource expansion. However, even if a specific product is successful, its benefits may not automatically extend across the entire organization. Intentional sharing of success stories becomes critical, as highlighted by the dissemination efforts of Company C and Company D.

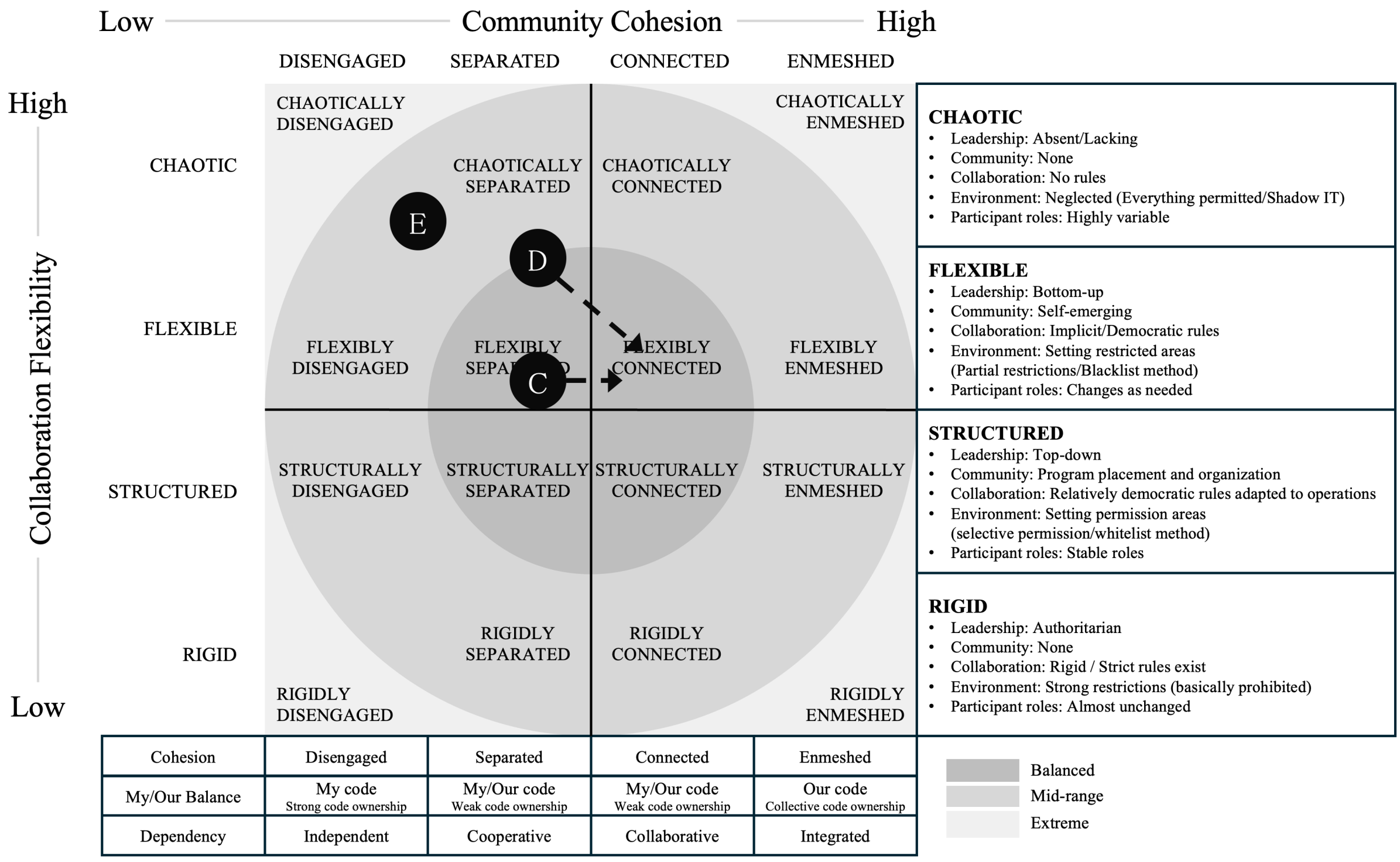

*Figure 10.4. Mapping of Internet Companies C, D, and E in the InnerSource Circumplex Model*

According to interviews with Company C, there are established evaluation programs and networks for integrated development projects that encourage cross-department collaboration. Company D does not use a formalized program yet achieves results through strong leadership and naturally emerging interactions. Company E has certain rules in place but limits its collaboration to relatively small pockets. Further acceleration will likely require additional leadership and systematized guidelines.

Even in settings with high flexibility, collaboration may be left to engineers' personal initiative, leaving room to increase cohesion. Positioning InnerSource as an official program, as in the case of Company C, can be beneficial. While conditions that encourage horizontal ties among engineers can produce visible outcomes early on, confining efforts to product-level achievements may limit broader company-wide impact. Accordingly, top-down support—through updated evaluation systems and broader consensus-building—becomes essential as the organization transitions to the growth and mature stages. During that process, clearly recognizing the value of InnerSource organization-wide and supporting champions with appropriate structures is of considerable importance.

### 10.1.4 Transition from High Flexibility, High Cohesion

A state of high flexibility and high cohesion is often observed in startup environments, which tend to have established source-code management tools, a single core product, and a small-scale community already engaged in seamless collaboration. Such organizations are akin to open source, featuring minimal rules yet maintaining a high level of transparency and rapid decision-making. It is not uncommon in such companies to rarely employ the term "InnerSource," since their small teams adopt practices resembling open source by default.

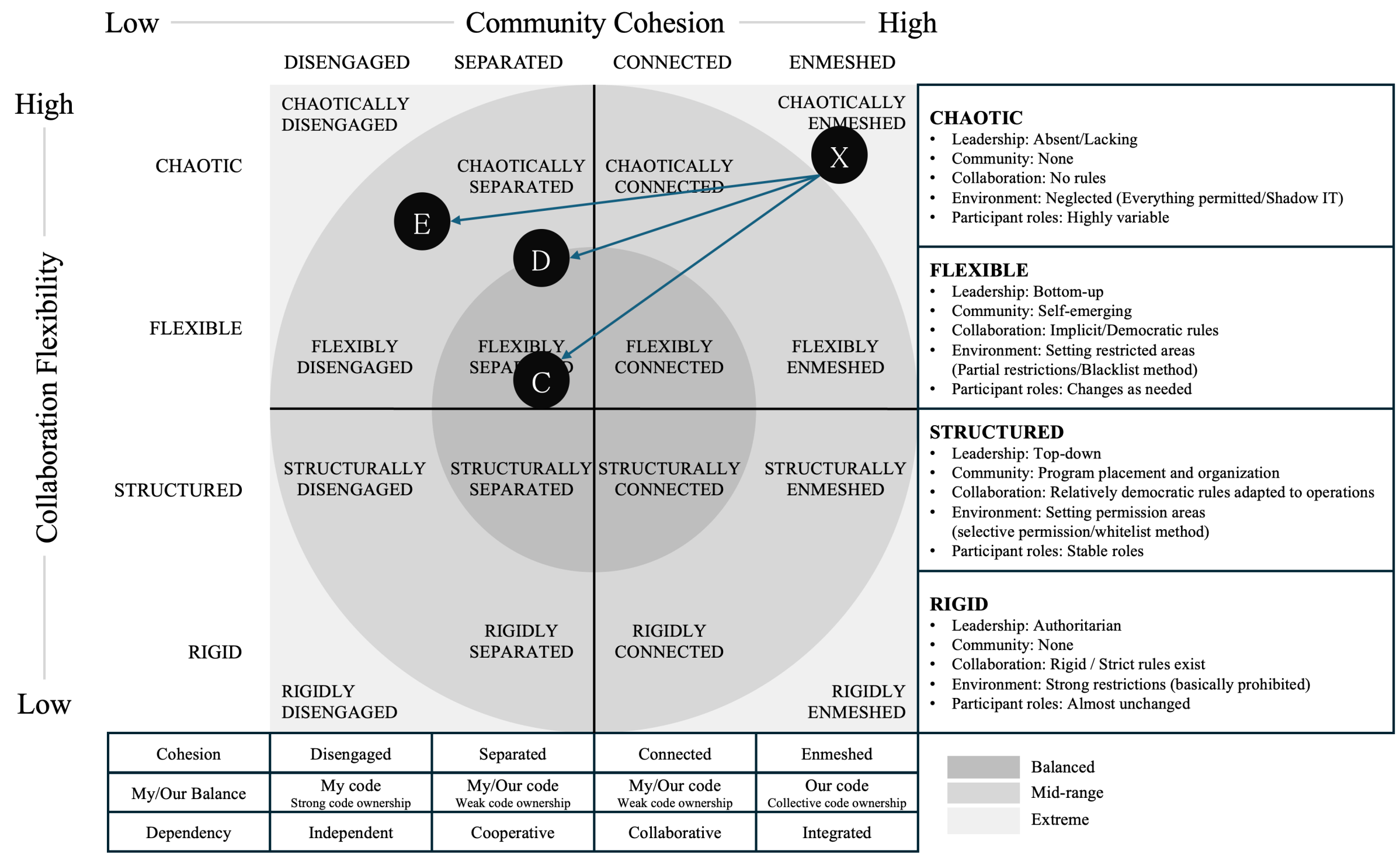

*Figure 10.5. Pathways by Which Startups Become DISENGAGED*

However, as a company grows in size, flexibility and cohesion tend to decline, creating a significant risk that it will ultimately shift toward low cohesion. Interviews with Company E cite examples in which an enterprise began with virtually no restrictions and enjoyed high degrees of collaboration, only to trend toward a more "separated" state once the organization expanded. Many startups that begin at point X (high flexibility, high cohesion) undergo multiple organizational restructures, ultimately transitioning to lower cohesion.

Preserving transparency and open communication through the “InnerSource” concept can serve as an effective means of preventing such decline. For instance, Company C and Company D are actively reconstructing communities and programs in ways that preserve the startup-like freedom present in early days while layering on rules required by a larger organization. Even enterprises currently at point X (small-scale, highly flexible, highly cohesive) may benefit from formalizing InnerSource early, thus maintaining transparency and avoiding a natural drift toward lower cohesion in the future.

## 10.2 Complementing Maturity Models with the Circumplex Model

While traditional maturity models have provided important guidelines for understanding InnerSource evolution, the circumplex model proposed in this study complements these models and enables richer interpretation. A particularly noteworthy aspect is the perspective that multiple "maturity levels" can coexist simultaneously within an organization.

In large enterprises especially, it is not uncommon for institutional design at the corporate level and team-level practices to exhibit different maturity levels. The development of InnerSource occurs topologically, with communities forming clusters that demonstrate varying degrees of maturity. In other words, even when an enterprise claims to be "practicing InnerSource" as a whole, the degree of implementation may vary significantly among internal communities. This suggests the difficulty in definitively stating that "a company practices InnerSource" based solely on examples from certain divisions of large technology companies.

The circumplex model enables clearer understanding of the diverse "maturity levels" present within organizational communities. Each community occupies a distinct position along the axes of flexibility and cohesion, following its own developmental trajectory. Therefore, rather than making a unidimensional assessment that "this company practices InnerSource," the model enables more nuanced analysis of "which communities practice InnerSource and in what manner."

Moreover, a crucial point is that the fundamental purpose of InnerSource programs is to guide the directional vectors of various clusters existing on this circumplex model toward more desirable states. Specifically, it aims to direct the diverse states of flexibility and cohesion existing on the two-dimensional plane toward the center as much as possible. This does not necessarily mean converging all communities to a single "ideal state." Rather, it emphasizes pursuing an appropriate balance of flexibility and cohesion while considering each community's characteristics and constraints.

This approach involves expanding collaboration opportunities while ensuring safety and improving areas requiring correction. Simultaneously, it means providing appropriate environments for areas with high security requirements while cultivating a collaboration-oriented mindset and enabling open source-like collaboration styles even within small clusters. I

To achieve these objectives, the maturity of InnerSource projects and programs must be enhanced as much as possible, and institutional flexibility must be secured to increase the "receptivity" to collaboration. For example, while Company A's consortium model provided organizational members with a "free environment," the next step requires building mechanisms to strengthen human connections.

To reach the mature stage, careful consideration must be given to the topology created by InnerSource communities and the role of champions. The goals to be set, messages to be conveyed, and collaborative environments to be built differ depending on each cluster's position on the circumplex model. This research provides tools for defining different maturity levels that various teams and their clusters should aim for, demonstrating new possibilities for maturity models.

The circumplex model functions as a framework that extends maturity models and enables more reality-based analysis. By utilizing both models complementarily, organizations can develop a deeper understanding of InnerSource development processes and formulate effective implementation strategies.

## 10.3 Toward a Redefinition of InnerSource: Adaptability to Diverse Organizations

From the outset, this study has explicitly stated that its purpose is not to establish a unified definition of InnerSource. However, through the proposed circumplex model, it has become evident that InnerSource's significance may resonate more broadly than generally recognized.

Considering how the concept of InnerSource is applied across diverse organizational contexts, it is clear that even when using the same term "InnerSource," the specific vision and state described by companies and champions can vary significantly. While some companies already operate close to an open source state, others partially implement InnerSource within their unique organizational cultures and rules. Attempting to evaluate all these variations using a single standard risk narrowing the understanding.

This suggests that InnerSource should not be viewed solely as an extension of open source, but rather as a flexible approach that can be adapted to address various challenges and growth phases within existing corporate environments. According to the proposed circumplex model, cohesion and flexibility vary significantly among companies, and fully open cultures are not the only functional form of InnerSource. Therefore, organizations must carefully consider the scope and degree of collaboration they wish to pursue.

For instance, in highly flexible and cohesive companies, particularly small startups, all members appear to have collective ownership of the code. While this might seem like they are "naturally practicing an open source-like development model," the essence of InnerSource lies in promoting code reuse and organic collaboration within organizations. Without appropriate governance, high flexibility could paradoxically lead to disorganized reinvention of the wheel. In such environments, InnerSource functions as a methodology for bringing order to unstructured internal development.

Even in highly cohesive environments, departmental barriers inevitably emerge with business expansion. Therefore, introducing the concept of "InnerSource" early on to establish a transparent, contributory environment holds significant value. This proactively prevents bureaucratic organization and barriers that might impede growth speed as the organization expands.

In companies with high flexibility but low cohesion, such as Company E, silos have emerged between development teams as startups undergo rapid growth. While source code management tools are flexibly configured, actual collaboration has declined. In such cases, it would be inappropriate to judge that "InnerSource is being practiced" merely based on tool settings. To achieve genuine InnerSource, as demonstrated by Company C and Company D, it is crucial to create mechanisms that remove barriers through structured programs.

In medium-sized companies with low flexibility but high cohesion, such as Company F, development teams are relatively small, but collaboration remains insufficient due to inadequate evaluation systems. In these environments, despite close personal relationships, organic cooperation is hindered, making it essential to implement appropriate programs and initiatives to mobilize talent.

Finally, companies with low cohesion and low flexibility arguably need InnerSource environments the most. However, when organizations have substantial size and established historical cultures, implementing InnerSource presents significant challenges. These companies face dual challenges: addressing institutional aspects such as rules, accounting, and transfer pricing while simultaneously transforming people's mindsets and behavioral patterns. As demonstrated by Company A and Company B, an effective strategy is to maintain areas of low cohesion while securing high-cohesion areas through community or consortium formation, enabling cross-organizational collaboration.

The circumplex model thus demonstrates that excessive flexibility or cohesion does not necessarily indicate an ideal InnerSource environment. What truly matters is generating organic collaboration while maturing InnerSource culture and achieving business efficiency. This requires appropriate champion selection for each implementation stage.

InnerSource implementation progresses gradually from ideation through pilot, early adoption, growth, and mature stages. Champions must fulfill different roles at each stage, flexibly engaging in multifaceted activities including internal advocacy, vision sharing, tool preparation, and evaluation system reform. This represents an endless journey of cultural transformation, where achieving both high development productivity and developer experience becomes possible by embedding a culture of "natural contribution" within the organization.

# Conclusion

## 11 Summary

As emphasized throughout this study, InnerSource is not merely a matter of providing a suitable environment; rather, its essence lies in cultivating diverse modes of collaboration within an organization. Accordingly, it is not appropriate to regard a state in which "the environment for InnerSource is fully prepared" as the ultimate goal. Given that InnerSource is inherently a model of collaboration, it is crucial not only to issue top-down directives to facilitate large-scale cooperation from the outset—through comprehensive infrastructural and policy arrangements—but also to begin steadily with localized, small-scale efforts, such as minor contributions to ongoing research and development or sharing documentation with adjacent teams.

The present research offers an integrated analysis of the dual nature of InnerSource implementation, which involves both organizational programs and organically emerging collaborations. The findings illuminate a multilayered maturation process that transcends conventional models. Consequently, the practice of InnerSource is shown to extend beyond top-down frameworks, such as environmental preparation and governance procedures, highlighting the pivotal role of grassroots or informal initiatives led by those on the ground.

The results indicate that a combined approach—encompassing both a comprehensive perspective on the organization as a whole and an individual lens focused on team-level practices—is effective for successfully implementing InnerSource. By reconciling these two perspectives, top-down measures such as phased institutionalization and governance improvements can be aligned with more flexible, bottom-up collaborative practices and contributions at the local level, thereby increasing the likelihood of fostering sustainable InnerSource development.

In the early stages of adoption, it is particularly beneficial to accumulate small-scale success stories before establishing a complete system, thereby making the value of InnerSource more visible. Even if early collaborations remain limited—such as partial attempts to promote cooperation across neighboring teams—these grassroots initiatives can serve as catalysts for disseminating the potential of InnerSource throughout the organization. Through these efforts, a community network that supports broader adoption can take shape.

Because InnerSource is fundamentally a collaboration model, a narrow or overly rigid definition of the "necessary settings" for its implementation carries the risk of hampering flexibility at the onset. In practice, many organizations begin by expanding networks through small-scale or localized projects, such as research and development work or documentation-related tasks. During this initial phase, even in the absence of formal plans or guidelines, iterative, on-the-ground exploration—driven by team members' own sense of purpose—often proves indispensable. Hence, it is vital to tolerate what may seem to be incomplete or informal stages of collaboration, incorporating lessons learned from these cumulative experiences into the organization's strategies as appropriate.

The circumplex model of InnerSource proposed in this study complements traditional maturity models by offering a more flexible perspective on the various stages or layers an organization may encounter. It recognizes that different projects or departments within the same organization can exhibit varying levels of maturity and underscores that aligning both the organic expansion of community networks and the timing of governance enhancements is essential. By doing so, organizations stand to maximize the benefits of InnerSource, such as increased productivity and cultural transformation.

Taken together, these considerations highlight the importance for InnerSource promoters to prioritize localized success stories in the initial stages while progressively building consensus and institutional structures at higher levels of the organization. Even if specific guidelines or access protocols for source code management are not yet fully in place, demonstrating early collaborative outcomes—supported by tangible data or concrete examples—can encourage broader organizational acceptance. Ultimately, this approach clarifies the activities required at each stage of InnerSource adoption, enabling companies to accelerate value creation in a manner attuned to their unique organizational culture and industrial context.

## 11.1 Contributions of This Study

In developing a staged understanding of the InnerSource adoption process, this study introduces a temporal viewpoint based on distinct implementation stages, moving beyond previous approaches that primarily analyzed parallel attributes. This new analytical framework enables a systematic grasp of how organizational perceptions evolve at each stage and how the challenges faced also shift over time.

Moreover, by focusing on specific phases of implementation at a high level of resolution, this study clarifies how practitioner groups at each stage recognize problems and take action. These insights are highly valuable for formulating stage-appropriate, effective strategies for introducing InnerSource.

Regarding existing maturity models, this research proposes a multilayered and multifaceted evaluation approach that goes beyond conventional growth indicators. Notably, it establishes a novel theoretical framework that comprehensively assesses the complexity of network structures within InnerSource programs and the maturity of the surrounding communities.

In a global context, this study is the first to systematically collect and analyze empirical data from Japanese corporations, thus adding a regional perspective to InnerSource research. By identifying characteristics of InnerSource adoption in regions with linguistic or cultural barriers, this work contributes new insights into the global applicability of InnerSource practices.

Furthermore, this study offers concrete solutions to the legal and accounting challenges often encountered when introducing InnerSource, providing an essential guide for organizations seeking to implement InnerSource from a practical standpoint.

Overall, these findings advance the theoretical understanding of InnerSource and offer valuable guidance for building real-world strategies to facilitate its adoption.

## 11.2 Future Outlook

For future research, expanding the range of organizations studied and collecting more multifaceted data will be necessary to derive more generalizable insights. While the present research focused on a relatively limited sample, examining InnerSource in companies of various sizes and across multiple industries will help validate the proposed theoretical framework and yield robust conclusions.

Accounting challenges related to InnerSource adoption continue to hinder inter-organizational collaboration. Establishing an accounting infrastructure that gives large enterprises sufficient confidence to engage in such efforts will be a key to advancing higher levels of collaboration in the future.

Although InnerSource Topologies is a highly useful concept for elucidating organizational collaboration structures and the intensity of cooperation, methods for quantitatively evaluating it remain underdeveloped. Particularly, future research should investigate how to apply these methods to more complex and large-scale networks, extending beyond the early-stage practices discussed in this study. Such exploration would offer practical guidance tailored to increasingly diverse organizational forms.

In theoretical models of InnerSource—particularly the circumplex model—there is a need for a mechanism to holistically evaluate numerous metrics, such as source code management tools, community maturity, and the frequency of cross-departmental pull requests. While some metrics are being partially developed, there is still ample room to refine the model through further empirical validation and methodological enhancements, which would aid in visualizing growth processes and systematizing adoption strategies.

Finally, within Japanese organizations, challenges such as delayed in-house development and limited sharing of source code have become more apparent. Given Japan's demographic shifts toward an aging society and the growing importance of AI, promoting productivity gains and organizational learning through InnerSource remains a pressing issue. Responding to these broader social demands will likely involve systematically developing Japan-specific approaches and guidelines for InnerSource implementation and fostering an environment in which both frontline teams and senior management can collaborate from a shared perspective. This endeavor is expected to be of great significance for future advancements in InnerSource.

# Acknowledgment

This research was made possible through the support and cooperation of many individuals. Special gratitude is extended to Professors Masuo Araki, Kunimaru Takahashi, and Ken Takeda of Aoyama Gakuin University for their thorough guidance throughout this study. Appreciation is also due to Ms. Clare Dillon of The InnerSource Commons Foundation for providing valuable information and advice. The members of the InnerSource Commons Japan community contributed significantly through their survey participation. Additionally, several companies provided practical insights through interviews that were instrumental to this work. This research could not have been completed without the support of these individuals and organizations. Sincere gratitude is expressed to all who contributed.

## State of InnerSource Japan Survey 2024

A comprehensive, in-depth survey was developed for this study, drawing on the global "State of InnerSource Survey" while adapting it to the current situation in Japan. This survey contains 26 questions focusing on the following areas:

- Respondent profiles
- Organizational characteristics
- Barriers to InnerSource adoption
- Actual inter-organizational collaboration
- Current status of InnerSource implementation

### Disclaimer

The following is an English translation of the survey items. The original survey was created in Japanese, and because it incorporates elements from a global reference survey, certain nuances might differ from original questionnaires. The question content has been adapted to align with Global one, but there may still be nuances that did not fully carry over into the English version. For more detailed insights, please refer to the original Japanese text.

# State of InnerSource Japan Survey 2024

This survey is being conducted by the InnerSource Commons Foundation (a 501(c)(3) non-profit) with operational support from Aoyama Gakuin University.
The goal of this survey is to comprehensively examine the current state of InnerSource practices, challenges, and success stories in organizational software development.
Your valuable insights will greatly contribute to the growth and adoption of InnerSource as well as to academic research.

All responses you provide will be treated as confidential.
Survey results will be reported in aggregated form only and will not identify any individual or organization.
The data from this survey will only be used for academic research and non-profit activities aimed at promoting InnerSource.
If you have any questions regarding this survey, please contact yuki[at]innersourcecommons.org (attn: Hattori).
For more details on our privacy policy, please visit:
https://innersourcecommons.org/about/privacy/

For the prize drawing, please provide your contact details below.
Please note that we may use this information for follow-up surveys.
By completing this questionnaire, you acknowledge that you have read the above information and voluntarily agree to participate.

## Contact Information

Email Address: ____________________

Organization Name: ____________________

Your Name: ____________________

# Your Profile

## Q1: Please state your total work experience in years:

◻ 0-2 / ◻ 3-5 / ◻ 6-10 / ◻ 11-15 / ◻ 16-20 / ◻ 21-25 / ◻ 26 or more / ◻ N/A or Don't Know

## Q2: Please indicate how you would best describe your gender:

◻ Male / ◻ Female / ◻ Non-binary / ◻ Don't want to say / ◻ None of the above; please specify: ______

## Q3: Please select the role that most closely matches yours:

Select <u>one</u> option that best applies.

◻ Developer (back-end / front-end / full stack / QA) / ◻ Specific technical specialist (e.g., DevOps specialist) / ◻ Software Architect /

◻ InnerSource Evangelist / Ambassador / ◻ InnerSource Program Officer / ◻ InnerSource Program Leader/Manager /

◻ Open Source Evangelist / Ambassador / ◻ Open Source Program Officer / ◻ OSPO Lead / Manager /

◻ Advisor / Consultant / ◻ Process improvement specialist / ◻ Agile coach / Scrum Master /

◻ Program Manager / ◻ Product Manager / ◻ Project Manager /

◻ Engineering Manager / ◻ Senior executive / VP / ◻ Business role /

◻ Other (please specify): ______ / ◻ N/A or Don't Know

## Q4: How is your role positioned in your organization?

Select <u>one</u> option that best applies.

◻ Research & Development / ◻ Software Development / ◻ System Design & Integration / ◻ Project Management & Consulting /

◻ Infrastructure & Operations / ◻ Quality Assurance & Support / ◻ User Experience / ◻ Data Science & AI / ◻ Security /

◻ Education & Training / ◻ Other (please specify): ______ / ◻ N/A or Don't Know

# About Your Organization

In this survey, "organization" refers to the entire company or group you belong to, not just a specific project or department.

## Q5: Please state your organizational sector:

Select <u>one</u> option that best applies.

◻ Technology / ◻ Financial services / ◻ Retail/Consumer/e-commerce / ◻ Industrials & Manufacturing / ◻ Education / ◻ Healthcare & Pharmaceuticals / ◻ Telecommunications / ◻ Government /

◻ Media/Entertainment / ◻ Insurance / ◻ Energy / ◻ Non-profit / ◻ N/A or Don't Know /

◻ Other, namely: ______

## Q6: How many people work in your organization?

Please count both regular employees and contract employees.

◻ <9 / ◻ 10-99 / ◻ 100-499 / ◻ 500-1,999 / ◻ 2,000-4,999 / ◻ 5,000-9,999 / ◻ 10,000-19,999 / ◻ 20,000-49,999 / ◻ 50,000+ /

◻ N/A or Don't Know

## Q7: How many software developers work in your organization?

*Here, "software developer" refers to anyone who writes code and commits to a source code management system, including testers, QA engineers, and DevOps engineers, etc.

◻ <9 / ◻ 10-99 / ◻ 100-499 / ◻ 500-1,999 / ◻ 2,000-4,999 / ◻ 5,000-9,999 / ◻ 10,000-19,999 / ◻ 20,000-49,999 / ◻ 50,000+ /

◻ N/A or Don't Know

# About InnerSource

### Q8: In your own words, how would you define or describe InnerSource?

### Q9: How would you describe the scale of InnerSource projects/programs in your organization?

Select one option that best applies.

- ☐ Ideas stage (there is no known InnerSource practice in my organization / we are speaking to people to generate interest)
- ☐ Pilot stage (we are piloting some InnerSource projects to demonstrate its value)
- ☐ Early adoption (we have a small number of InnerSource projects / it is considered a niche practice)
- ☐ Growth stage (we have a medium number of InnerSource projects / it is considered an accepted practice)
- ☐ Mature stage (we have a large number of InnerSource projects / it is considered a widespread practice)
- ☐ Other (please specify): ____________
- ☐ N/A or Don't Know

### Q10: When I think of InnerSource, I think of:

Select all that apply.

- ☐ Re-using software/source code
- ☐ Adopting open source practices within the organization
- ☐ Contributions to my project from those not in the project team
- ☐ Contributing to projects within my organization that I have a personal interest in, but am not formally involved in
- ☐ Contributing to projects within my organization that I or my team rely on, but am not formally involved in
- ☐ Connecting to people in my organization who are outside my team
- ☐ Learning from people in my organization who are outside my team
- ☐ Other (please specify): ____________

### Q11: Please indicate which of the following experiences apply to you regarding InnerSource:

Select all that apply.

- ☐ Contributed to an InnerSource project outside my team
- ☐ Worked on a project that has accepted guest contributions from people not in the team
- ☐ Been involved in rolling out or scaling InnerSource in my organization
- ☐ Advised or coached InnerSource practitioners
- ☐ Other (please specify): ____________
- ☐ N/A or Don't Know

### Q12: Which of the following are the most significant blockers or obstacles for InnerSource success, in your experience?

| Obstacle / Blocker | Critically High | High | Moderate | Low | Very Low | Unknown |
|---|---|---|---|---|---|---|
| Lack of overall awareness | ☐ | ☐ | ☐ | ☐ | ☐ | ☐ |
| Lack of executive management buy-in | ☐ | ☐ | ☐ | ☐ | ☐ | ☐ |
| Lack of middle management buy-in | ☐ | ☐ | ☐ | ☐ | ☐ | ☐ |
| Lack of interest from developers | ☐ | ☐ | ☐ | ☐ | ☐ | ☐ |
| Time constraints for reviewing code | ☐ | ☐ | ☐ | ☐ | ☐ | ☐ |
| Not getting timely feedback on your contribution | ☐ | ☐ | ☐ | ☐ | ☐ | ☐ |
| Unclear contribution guidelines | ☐ | ☐ | ☐ | ☐ | ☐ | ☐ |
| Not having enough time to contribute | ☐ | ☐ | ☐ | ☐ | ☐ | ☐ |
| Lack of project discoverability / findability | ☐ | ☐ | ☐ | ☐ | ☐ | ☐ |
| Chosen projects lack appeal | ☐ | ☐ | ☐ | ☐ | ☐ | ☐ |
| Lack of familiarity with InnerSource principles | ☐ | ☐ | ☐ | ☐ | ☐ | ☐ |
| Legal concerns | ☐ | ☐ | ☐ | ☐ | ☐ | ☐ |
| Issues related to transfer pricing / tax | ☐ | ☐ | ☐ | ☐ | ☐ | ☐ |
| Organizational culture or silo thinking | ☐ | ☐ | ☐ | ☐ | ☐ | ☐ |

### Q13: Please share more details about any obstacles or blockers you are facing:

## InnerSource in Your Organization

### Q14: How do members of your organization view or regard InnerSource?

| Question | Very Positive | Somewhat Positive | Neutral / No Interest | Somewhat Negative | Very Negative | Unknown |
|---|---|---|---|---|---|---|
| Executive management of the organization explicitly supporting the InnerSource initiative | ☐ | ☐ | ☐ | ☐ | ☐ | ☐ |
| The organization treating InnerSource as an important strategy | ☐ | ☐ | ☐ | ☐ | ☐ | ☐ |
| The organization affording time for contributors to work on InnerSource projects | ☐ | ☐ | ☐ | ☐ | ☐ | ☐ |
| People in the organization being able to choose projects based on their expertise or motivation | ☐ | ☐ | ☐ | ☐ | ☐ | ☐ |
| The organization offering rewards for contributing to other teams' projects | ☐ | ☐ | ☐ | ☐ | ☐ | ☐ |
| The organization defining criteria for career advancement based on InnerSource values | ☐ | ☐ | ☐ | ☐ | ☐ | ☐ |

### Q15: Which factors motivated your organization to participate in InnerSource?

Select all that apply.

- ☐ Employee satisfaction
- ☐ Talent retention
- ☐ Innovation
- ☐ Remove silos & bottlenecks
- ☐ Knowledge sharing
- ☐ Improve code quality
- ☐ Improve quality of processes
- ☐ Improve document quality
- ☐ Improve scope of testing
- ☐ Increase developer speed
- ☐ Creating reusable software
- ☐ Step on a path to open source readiness
- ☐ Eager to take part in the InnerSource trend
- ☐ InnerSource is named on my organization's roadmap
- ☐ Mitigate cultural differences through a shared engineering language
- ☐ Other (please specify): ____________
- ☐ N/A or Don't Know

### Q16: Is the source code of the project you work on visible to everyone in the organization?

Select one option that best applies.

☐ Yes, fully visible / ☐ Visible upon request / ☐ Not visible / ☐ Don't know / ☐ Other (please specify): ____________

## Q17: Thinking about the InnerSource project(s) you have been involved in within the last year, which of the following apply?

Select all that apply.

- ☐ The project can be found on an internal developer portal
- ☐ The project code is reused by other teams in the company who may need its functionality
- ☐ There are code contributions from outside the project team
- ☐ The project code is visible, but is not accepting inbound contributions
- ☐ The code is sufficiently documented so that guest contributors outside the project team can contribute
- ☐ The code is sufficiently modular for others to understand and make changes easily and safely
- ☐ All of the code is stored in a version control repository (such as GitHub or GitLab) that makes branches, pull requests, and integration easy
- ☐ The project has dedicated "trusted committers" or maintainers who handle guest contributions from outside the project team
- ☐ Other (please specify): ______________________
- ☐ N/A or Don't Know

## Q18: Which statements apply to the teams practicing InnerSource that you are involved with?

Select all that apply.

- ☐ The team is comfortable exposing code that might be less-than-perfect to people outside the team
- ☐ The team is open to receiving code contributions from outside contributors that might be less-than-perfect
- ☐ Team members are willing to do code reviews of submissions from guest contributors (developers outside the team)
- ☐ The team is willing to have difficult conversations with outside contributors about accepting or rejecting contributions
- ☐ The team actively refactors and modularizes code to encourage external contributions
- ☐ Team members are willing to participate in internal forums and patiently answer others' questions
- ☐ Team members are open to mentoring new guest contributors or learning how to mentor
- ☐ Some or all team members have participated in open source projects
- ☐ Team members create and maintain documentation for guest contributors, such as a "CONTRIBUTING.md" file
- ☐ Team members respond to bugs reported by people outside the project team
- ☐ There is a recorded (archival) mechanism for discussions so that all guest contributor questions and internal project team decisions are searchable (e.g., Slack, GitHub Discussions)
- ☐ Other (please specify): ______________________
- ☐ N/A or Don't Know

## Q19: Do members of your team have experience accepting guest contributions from outside the team?

Select one option that best applies.

☐ All members have experience / ☐ Some members have experience / ☐ No members have experience /

☐ Other (please specify): ______________________ / ☐ N/A or Don't Know

## Q20: Which statements best describe the InnerSource projects in your organization?

Select all that apply.

- ☐ Projects in my organization are InnerSource by default
- ☐ Platform projects
- ☐ Libraries and internal tools (commonly used artifacts)
- ☐ DevOps projects
- ☐ Documents as code
- ☐ Projects on a path to becoming open source in the future
- ☐ Inter-divisional projects
- ☐ Strategic projects
- ☐ AI projects
- ☐ Inter-organization / spanning legal boundaries
- ☐ Cross-regional projects / spanning geographical borders
- ☐ Grass-roots / community projects
- ☐ Club goods (where two or more organizations want to collaborate together without making the code fully public)
- ☐ Other (please specify): ______________________
- ☐ N/A or Don't Know

## Q21: In your view, how important is the introduction or expansion of InnerSource in your organization?

Select one option that best applies.

☐ It is a top priority / ☐ It is a short-term priority / ☐ It is a mid-term priority / ☐ It is a long-term priority / ☐ It is not a priority

## Q22: What is your outlook on how widely InnerSource will be adopted in your organization going forward?

Select one option that best applies.

☐ Fully adopted organization-wide / ☐ Adopted on a limited scale / ☐ No change expected from current state / ☐ Adoption is shrinking / ☐ Almost no longer used

### Q23: Thinking about your organization's approach to open source, please select statements that are true:

Select all that apply.

- ☐ My organization consumes or uses open source software
- ☐ My organization creates or publishes open source code
- ☐ My organization contributes to open source software projects
- ☐ My organization leads open source software communities
- ☐ N/A or Don't Know

## InnerSource Practice

### Q24: How long ago was InnerSource first adopted in your organization?

Select one option that best applies.

☐ <1 year / ☐ Between 1-2 years / ☐ Between 2-3 years / ☐ Between 3-4 years / ☐ Between 4-5 years / ☐ Over 5 years / ☐ N/A or Don't Know

### Q25: How was InnerSource introduced to your organization?

Select one option that best applies.

- ☐ Top-down: someone from management initiated the adoption
- ☐ Bottom-up: one/a few/one team of developers started the initiative
- ☐ A mix of both
- ☐ Other (please specify): ______________________
- ☐ N/A or Don't Know

### Q26: InnerSource has helped me and/or my team in the following ways:

Select all that apply.

- ☐ Reduce time to market because we could get features into a module my team relies on more quickly
- ☐ Share knowledge
- ☐ Get to know more people within the organization
- ☐ Retain talented developers
- ☐ Increase my enjoyment and/or excitement in my work
- ☐ Fix bugs more quickly
- ☐ Identify new use-cases of the software I (or my team) work on
- ☐ Improve the quality of the software I (or my team) work on
- ☐ Identify new features that I (or my team) hadn't thought of before
- ☐ Other (please specify): ______________________
- ☐ N/A or Don't Know

Thank you for completing this survey.

## Interview Records

To investigate the realities of InnerSource adoption, interviews were conducted with multiple companies. This appendix presents the interview records. As primary data sources, these records provide valuable insights into the status and challenges of InnerSource implementation. While maintaining participant anonymity, detailed information has been included to the extent possible.

The interviews were conducted with six companies:

- Company A (industry undisclosed)
- Global Manufacturing Company B
- Internet Industry Company C
- Internet Industry Company D
- Internet Industry Company E
- Information and Communications Industry Company F

Spanning manufacturing, internet services, and information and communications, this selection of companies ensures a comprehensive and balanced representation across different industries. During the interviews, questions covered the background of InnerSource adoption, current implementation practices, challenges encountered, and any benefits realized. The following sections present the interview contents for each company individually.

### Disclaimer

The followings are an English translation of the interview records. The actual interviews were conducted in Japanese, and these notes are based on the original Japanese transcripts. Consequently, certain contextual nuances may have been lost in translation. If you wish to explore the material in greater detail, please refer to the original Japanese text.

## Company A (Industry Undisclosed) – Translation

| Department | Company Size | Department | Position |
|---|---|---|---|
| Not disclosed | Over 50,000 employees | Technology-related Division | Not disclosed |

1. **Assigned Duties and Involvement in Open Source**

At Company A, my primary role involves leveraging software technologies to develop both products and services. This includes providing shared software components, offering software engineering expertise, and restructuring the software development process to deliver technical support and consultation across various internal divisions.

2. **History and Current Status of InnerSource Adoption**

I have been working to promote open source within the company. Over time, I observed a recurring phenomenon: when the same functionality was used in multiple products, the technology developed by Division A could also be utilized by Division B. Only after several years did I recognize this as an "InnerSource" framework—essentially bringing open source concepts inside the organization to dismantle internal silos. By collaboratively identifying and resolving common issues through shared open source usage, I believed we could further enhance the company. In cases where, for instance, "the right-hand division was doing the exact same thing," I sensed that InnerSource could facilitate improvements.

3. **Strategies for Promoting InnerSource**

My plan to disseminate InnerSource centered on three areas: culture, systems, and rules. First, culture. I sought to cultivate an open collaboration culture through educational programs that incorporate hackathons. Second, rules. From a top-down perspective, we established explicit rules stating, "it is permissible to engage in InnerSource." One such rule involves an InnerSource license that clarifies how far one can engage in collaborative development under InnerSource. For instance, research and development work is permitted under this license, but commercial products require a mechanism for compensatory payment. And finally, systems. We created a shared development platform accessible to everyone, where individuals can develop source code and engage in discussions.

4. **Current Outcomes and Challenges (Contributions and Benefits for Researchers)**

Although we have tried to promote InnerSource, only about ten instances can be concretely measured so far, and that number does not seem large. While download counts have increased, contributions to the code itself have not necessarily grown. Researchers, however, have expressed that making their research widely visible and obtaining feedback more quickly is advantageous. In some instances, these faster feedback loops have shortened the time to final product integration.

5. **Barriers to InnerSource Adoption: Lack of Transparency**

One major reason why InnerSource can be difficult to implement in Japan is the near absence of open documentation, which results in a lack of transparency. For example, project proposals often remain with specific divisions, and there is little effort to connect individuals across the organization. Without a foundation that enables early information sharing and transparent collaboration, the adoption of InnerSource faces significant hurdles.

6. **Silo Mentality, Cultural Obstacles, and the "Secret-Knowledge" Mindset**

A key cultural issue is the lack of enthusiasm for having "various people review your work and provide feedback." Many tend to hoard trade secrets without the mindset of improving them through openness. If an organization already embraces open source, everyone would normally work proactively to improve the shared code. However, that cultural perspective is lacking.

7. **Time Constraints, Divergent Perspectives Among Engineers and Managers, and "Stealth" Engineering**

Some engineers pursue InnerSource-related activities informally at the grassroots level. However, without organizational recognition, it is challenging to evaluate such efforts, which can lead to unsatisfactory outcomes for those engineers. Meanwhile, middle managers often voice a desire for an environment in which engineers can contribute code without concern for negative repercussions.

8. **Legal and Tax Issues**

Legal and tax considerations, particularly surrounding transfer pricing and taxation, often come under scrutiny. Even among consolidated subsidiaries, providing value free of charge across different legal entities could be interpreted as a form of profit transfer. Hence, collaborations under InnerSource must address valuation and balance, requiring proper coordination.

9. **Proposed Approaches: Balancing Through Commit Volume and Regulatory Methods**

We considered calculating each company's contribution ratio—using metrics such as commit counts or lines of code—and then billing for the difference. This approach would require a monitoring system, but accurately quantifying the value remains difficult. Even domestically, crossing corporate boundaries presents complex issues, especially when multiple parties are involved rather than just one-on-one transactions. We also contemplated a model in which all InnerSource resources are consolidated into the parent company, which then charges appropriate fees to the entire group. This might simplify matters, and some companies apparently adopt this consolidation strategy in practice. However, Company A has yet to implement such a model; in fact, we have not had the chance to explore it thoroughly at this point.

**10. InnerSource and the Consortium-Based Model**

Initially, InnerSource at Company A was limited to research and development. To bring it closer to practical application, however, we introduced a consortium-based approach. This model requires a membership fee, which grants access to certain resources. Some resources may also be made available by paying an additional fee to the owner. If members discover a bug, they are free to contribute a fix. We view this consortium approach as another variation of InnerSource.

**11. Consortium Membership Fee Model**

Similar to a subscription system, the membership fee is scaled according to the size of the participating organizations. Not only do various companies contribute resources they wish to share, but the consortium can also use membership fees to create new resources for common use.

**12. Rationale Behind the Consortium Approach**

A fully open form of InnerSource is suitable for visualizing research and development. However, in order to broaden the scope of use beyond R&D and anticipate commercialization, we decided to employ the consortium approach. One can view this as structuring multiple layers of InnerSource-like mechanisms, which we also regard as a valid form of InnerSource.

**13. Support Options and User Responsibility**

Within the Consortium Under the basic membership fee, no formal support is provided. Members are, however, free to use the initial resources at their own risk. Ownership of each resource is clearly defined, so should the user require support or modifications, they can pay the owner accordingly or engage in joint development. This structure is similar to open source: if one can fully handle the software independently, it remains free; if support is necessary, one compensates the organization offering it.

14. **Reciprocity and R&D Funding**

In scenarios where no formal guarantee is offered and the resource can be used freely, the user's side contributes through issue registration, feedback, or testing, which serves as a form of compensation to the provider. Thus, the relationship is balanced through reciprocal contributions. (Note: The original text ends mid-sentence here, but the context implies an emphasis on mutual benefit.)

15. Divergent Interpretations of InnerSource and the Introduction of Its Core Principles

(In response to the interviewer's observation that views on InnerSource vary, and some managers see it mainly as a technical concern.) Engineers who wish to contribute have a different perspective. Personally, I consider InnerSource to be a "development methodology," embodying the open source principles applied within the organization. Simply making the source code available internally or holding an InnerSource license is not, by itself, InnerSource. Rather, InnerSource is essentially about adopting the principles of open source development inside the company.

16. **Relationship Between InnerSource Licenses and Open Source Licenses**

When promoting InnerSource, it is sometimes necessary to establish formal rules and controls through an "InnerSource License." Such a license often draws on models like the Apache License, the Mozilla Public License, or the GPL. In Company A's case, we created our own InnerSource License to clarify the conditions for internal use within the corporate group. If the ultimate objective is to contribute back to open source, one could simply adopt an existing open source license. However, where collaboration is strictly confined to internal use, an InnerSource License becomes necessary. For a single corporation, an InnerSource License is not always mandatory.

17. **Reference Example of the InnerSource License and Its Rollout**

The InnerSource License used at Company A begins by stating its intent to foster collaborative software development within the corporate group. It defines terms such as "Company A Group," "company," "the repository that provides access to InnerSource," and "software," clarifying usage for internal or research purposes, as well as commercial applications.

Under the license grant, licensees within the group may obtain software free of charge and engage in its use, duplication, modification, combination, publication, or extended utilization. At the same time, they must preserve copyright notices and the stipulated conditions of use. Provisions are also included concerning assertion of copyrights for modifications and the rules governing source distribution. These attributes bear similarity to GPL-like clauses, while allowing freedom of use within the corporate group but restricting external release.

For internal purposes, this license treats usage as an in-house application. External usage or other forms of distribution require a separate contract. Additionally, the license clearly states that the software is provided “as is,” with no warranty or liability.

**18. Past Failures and Failure Patterns**

In a past endeavor similar to InnerSource, we faced a situation where aggregated assets were simply abandoned. Once there was no one to maintain them, they fell into disuse and became obsolete. This risk applies equally to InnerSource initiatives: if no one can assume the maintainer role, the asset ultimately becomes “shelved” and loses its value.

**19. Establishing Maintenance Guidelines**

We implemented a checklist for publishing code under InnerSource, requiring designating a maintainer, notifying relevant parties when a maintainer steps down, and explicitly declaring an End of Life (EOL) if no successor is available. These measures aim to ensure that the codebase remains functional and that maintenance responsibilities are clearly defined.

**20. InnerSource License and Royalty-Free Use Within Non-Commercial Boundaries**

Asked whether a consortium arrangement is always a prerequisite, we clarified that free-of-charge access is granted only under a non-commercial use scope. In other words, the software can be freely obtained, used, copied, or published for research and development, including proof-of-concept (PoC) phases and late-phase development before finalizing commercial viability. Because deciding when to transition a software project to commercial use can be postponed, a substantial range of R&D activities remain covered under the non-commercial usage framework.

**21. Feedback as Compensation and the “Give and Given” Principle**

To ensure that transactions appear fair and aboveboard even in R&D, we instituted a mechanism whereby some form of compensation is always provided. Conventionally, feedback or testing would be chargeable services, but we treat them as the user’s payment-in-kind. By making feedback exchanges a channel for value transfer, we aim to foster reciprocity among users. For simplicity in promotion, we referred to this arrangement as “Give and Given,” meaning that participants maintain equitable relationships through mutual contributions.

**22. Existence of Successful Cases and the Reality of Knowledge Sharing**

Asked whether there have been similar successes in InnerSource or knowledge-sharing, the answer within Company A is that no standout success stories have emerged yet. For instance, although we set up an internal wiki for employees, it has not grown into a widely utilized platform. While we do store data using Office-based platforms, there are almost no large-scale initiatives that successfully promote cross-organizational knowledge sharing. Certain divisions do develop specialized knowledge bases relevant to their respective domains, but this remains domain-specific and does not extend collaboration across different business lines. Conversely, there are small-scale InnerSource-like efforts, such as building a cross-departmental technology catalog using version control systems and soliciting feedback on broken links. However, these remain limited in scope.

**23. Variation in Interest by Organizational Role**

Within Company A, researchers appear to exhibit the greatest interest in InnerSource. They tend to be eager to explore new approaches. In contrast, managers often say, “It would be great if it works,” yet they are less likely to lead hands-on implementation. Managers also struggle to determine on what grounds they should endorse and support InnerSource activities. If no clear policy states “You are allowed to engage in InnerSource,” some engineers will take matters into their own hands, working quietly on intriguing projects behind the scenes. Currently, InnerSource is primarily considered for research and development, but if it is eventually to be scaled up for product sharing, more formal systems and guidelines will become necessary.

**24. Boundary Between Product Release and R&D Funding**

One approach we considered is to conduct main branch development under research and development, then create a separate release branch— “Release A,” for example—when transitioning to a commercial product. At that point, it is treated as an official product, and users are required to pay license fees. Hence, during the R&D phase, contributors can operate with the freedoms afforded by InnerSource, but once the software is commercialized, they enter a formal compensation framework. This clear differentiation helps maintain consistency as the project scales up.

**25. Future Expectations and the Drive for Clarity**

By establishing a logical framework and clearly organizing these rules, we aim to prepare for future scaling and official adoption. At the R&D stage, we treat user feedback as a form of compensation, allowing free internal usage. Once a product is released, we shift to a licensing fee model that clarifies accounting and valuation. We believe that constructing such a mechanism will lay the foundation for broader InnerSource utilization and shared product development in the future.

## Company B (Manufacturing) – Translation

| Department | Company Size | Department | Position |
|---|---|---|---|
| Manufacturing | Over 50,000 employees | Common Development Division (Holding Company) | Senior Technical Expert |

1. **Organizational Affiliation and Job Role**

   I am affiliated with the common development division of Company B's holding company, which functions as a cross-departmental development unit. This division offers development support to several subsidiaries in a concurrent manner. I also serve on the OSPO (Open Source Program Office) administrative team, which is composed of a community of roughly 100 to 200 people, although the administrative office itself consists of about five members. Additionally, some subsidiaries have similar open source–related initiatives, each with a small administrative office and representatives from each department, forming their own community structures.

2. **Current Status of Open Source Utilization and Contributions**

   Open source software (OSS) is utilized essentially across all of our business domains. In fact, our development processes would not function without open source software—one might even say we are "heavily reliant on it." Furthermore, certain business units also actively contribute back to various open source projects.

3. **Approaches Corresponding to InnerSource and Their Definition**

   Strictly speaking, we do not practice what the so-called "InnerSource Commons" would define as InnerSource in its narrow sense. However, we do engage in collaborative projects across multiple organizational units, which we regard as demonstrating characteristics of InnerSource in a broader sense. In other words, beyond a single department, cross-organizational projects and consortium-like collaborations can be interpreted as embodying InnerSource principles.

4. **Examples and Characteristics of Collaborative Projects**

   Many of these collaborative projects occur before a specific business area or product is clearly established. Often, the holding company's central organization joins forces with multiple business domains within International Company B—for instance, the X, Y, and Z divisions—resulting in cross-business projects that can support new business developments. These collaborations can involve structures in which certain departments collectively develop foundational systems; in some cases, the projects themselves are positioned as InnerSource–type endeavors.

5. **Project-Crossing Development Structures, Hierarchies, and Technology Sharing**

In such collaborative projects, each organization contributes specialized expertise, while the project leadership integrates these contributions. This approach is not limited to a particular layer such as middleware, platforms, or cloud infrastructures; rather, each specialized domain provides complementary components. Typically, it is not the case that everyone shares and modifies a single source code repository in a unified manner. Rather, each domain contributes its own portions, and mutual requests or bug fixes are communicated mainly through issue tracking systems.

6. **Source Code Handling and Feedback Methods**

In principle, we do not collaboratively modify the same source code across different groups. We do raise issues or provide small-scale feedback such as configuration changes or bug fixes to a few files. However, at the level of copyright ownership, we generally do not engage in joint ownership or joint modification of the code. Instead, we mainly offer feedback on requirements and bug reports, and this often remains the extent of our cooperation.

7. **Repository Management, Access Control, and Collaboration Systems**

We use enterprise-level source code management systems like GitHub Enterprise, where repositories are shared but with controlled access privileges. The holding company maintains this enterprise environment; subsidiaries can view relevant repositories as needed. In some cases, we set "outside collaborators" at the repository level, enabling other organizations to view and contribute. Decisions about which repositories to use often hinge on who bears the associated costs ("Will it be funded centrally?" and so on).

8. **Accounting Procedures, Outsourcing Contracts, and Cost Calculation Methods**

When we conduct collaborative projects, we typically establish outsourcing contracts because of potential issues with transfer pricing and profit allocation. In projects run by the holding company, for instance, subsidiaries can access enterprise-level repositories hosted by the holding company and contribute on a labor-hour basis. The project owner compensates them through outsourcing fees, and the resulting deliverables belong to the project. Within the same company, costs can be managed through internal cost-center accounting; but among group companies, formal outsourcing contracts are commonly used. These contracts are process-based, calculating costs as the number of labor hours multiplied by personnel expenses. For example, providing X00 or Y00 million yen worth of man-hours contributes to the project's development, with the resulting deliverables belonging to the project. Each participating organization's contribution is thus transparently compensated in terms of labor hours.

**9. Valuation of Feedback in R&D Phases versus Production Phases**

During the R&D phase, the valuation of deliverables is more flexible. Under Japanese tax regulations, if we are dealing with a development or lightweight or research version, mere "access" to the work can be compensated through feedback rather than strict monetary payments. Because R&D outputs have not yet generated clear monetary value, we may stipulate in the contract that "providing valuable feedback to the development process" satisfies the compensation requirement. In practice, access to the software under development is available to all internal stakeholders. If another department expresses interest in viewing it, as long as the project remains in an R&D context, it is possible to rely on such feedback-based value exchanges. However, once a product enters the production phase, we need to clearly set monetary compensation. In the production phase, payment in cash becomes a requirement, often through paid license agreements or support fees. Although feedback such as requirements definition or QA results is deemed sufficient compensation in the R&D phase, once the product is commercialized, monetary transactions are mandatory.

**10. License Agreements and the Existence of Contract Templates**

We use different contract templates depending on whether the software is for research and development or for commercial product distribution. Our administrative division provides these templates. For research and development licensing, using the software at an intermediate stage can be compensated primarily through feedback. In contrast, for formal product-level distribution, we utilize another contract template stipulating definitive monetary compensation (e.g., license fees or support costs).

Within the company, it is explicitly stated that if the software is for internal use or for a proof of concept (PoC) in an extended R&D setting, then feedback-based compensation is acceptable. But if it is to be embedded into a commercial product, payment is required. Such stipulations are clearly specified in the contract templates prepared by our administrative department. For example, if the software is used for research or pre-GM (general availability) development, we apply the R&D license template; whereas if it is at the RC (release candidate) stage, we shift to the product-oriented template. The availability of these standardized templates familiarizes all departments with the rules and facilitates compliance.

**11. Complex InnerSource Development and Special Contract Agreements**

For more intricate InnerSource–type development scenarios, our standard contract templates may not suffice. In such cases, we sometimes establish special agreements detailing ownership shares, intellectual property rights, and how to handle overlapping areas or mutual contributions, ensuring that both parties' benefit. These special agreements allow us to handle complex situations that cannot be adequately addressed under the regular contractual framework.

**12. Setting Support Fees and Options During R&D Phases**

Once a project transitions to the production phase, the default rule is to charge at least a basic support fee. Even during the R&D phase, we have the option to set such a fee when necessary. However, because the R&D phase allows non-monetary compensation, feedback and QA responses can also fulfill the role of compensation. By contrast, once the product is in commercial use, our policy is that compensation must take monetary form; the modes of compensation thus differ between the R&D and commercial phases.

**13. Using Architecture Diagrams in Contract Formulation**

In more complex agreements involving overlapping ownership, functionalities, or interests, we do not simply handle everything monetarily. Instead, we may clarify the ownership structure while granting usage rights. Frequently, an architecture diagram is appended to the front matter of such contracts. These diagrams typically specify which department develops which components, which department is responsible for other parts, how interfaces are defined, who retains ownership, and whether reuse is permitted. Such architectural stack diagrams facilitate clarifying issues of ownership, usage rights, interface requests, and reuse possibilities, making them essential in sophisticated InnerSource contractual arrangements.

**14. Rationale Behind Valuing Feedback as Compensation**

The idea that "merely viewing the software can be compensated by feedback" reflects the inherent challenges of valuation and pricing during early R&D. In this stage, business units provide feedback such as "It worked," "It did not work," or "We encountered an issue." Ordinarily, the party receiving the software would spend significant labor on testing and requirements definition. By having the business unit assume that workload, the resulting feedback is treated as a form of compensation. This arrangement helps avoid imposing monetary fees on business units that have not yet generated revenue from the product. Indeed, it is often a challenge to assign a monetary value at the trial stage, so the feedback-based approach emerged as a solution.

**15. Shifts in Usage and Compensation Between R&D and Production Phases**

When a business unit later decides to use a software product for either commercial or research purposes, the required compensation tier changes accordingly. In principle, for any additional development requests from a specific business unit, we collect fees under an outsourcing arrangement. At the same time, we maintain the rights for ongoing R&D or shared utilization in the contract. The rationale is that if modifications are made solely for a specific business unit, that code branch could diverge indefinitely, undermining the shared codebase. Therefore, the contract ensures that changes remain reusable across the organization.

**16. Maintenance Responsibilities and Support Fees**

Maintenance of the core functionality is generally handled by the provider, whereas any incremental portions are addressed through outsourcing, as part of support fees. Common components and custom-developed sections add to the maintenance burden, which we then reconcile in the following fiscal year's individual contracts. Even though it might resemble transactions between separate companies, we formalize these arrangements through contracts to uphold the organizational policy of maintaining a shared codebase. We have encountered issues in the past, and our current framework evolved from those lessons.

**17. Codebase Provision Formats at the Product Stage**

When releasing software in a product form, we often do not simply grant access to the entire repository. Instead, we may provide only a particular branch or even supply a tar archive. In the case of embedded systems or specific configurations, we may only provide executables along with a source code package for license compliance. Under Japanese tax regulations, internal-use software and software for sale are taxed differently, so we must adapt our approach when transitioning from R&D to product release.

**18. Internal-Use Software and the Issue of Fixed Asset Treatment**

If we straightforwardly designate something as "internal-use software" once specifications are finalized, the software typically becomes classified as a fixed asset. When we subsequently distribute or share such a fixed asset externally, we must charge fees at that point, nullifying the rationale for treating feedback as compensation. This necessitates careful maneuvering. Personally, most of my work ends up embedded in products, so I do not worry as much about fixed assets. However, for service-oriented software, distributing fixed assets in exchange for feedback-based compensation likely poses complications. In that case, we would need a license or subscription-based model, complicating how compensation is collected.

**19. International Transactions, Valuation, and Transfer Pricing**

The pricing model—whether licensing, subscription, or one-time purchase—varies by case, but once we reach the product stage, it is often more intuitive to charge support fees. The more significant challenge arises in international transactions, such as dealings with the United States or Europe, where we must consider how to calculate transfer pricing and determine the software's taxable value. Although I am not entirely sure how it is done, I know that domestic transactions are simpler, whereas international deals require more complex procedures and rules.

**20. Concerns When Transitioning Internal-Use Software to the Production Phase**

In my surroundings, most software eventually becomes a commercial product, so we rarely face the internal-use conundrum. However, for teams developing services, I wonder how they handle such transitions. With internal use, feedback-based compensation works fine; but once a production phase is reached, the software is "given away" as a fixed asset, negating the feedback-based logic. Consequently, a license or subscription model becomes necessary. Licensing contract terms vary, and some companies have accounting structures that allow subscription-based flexibility, while others resort to purchase models or separate support licenses. There are multiple scenarios, each requiring distinct approaches.

**21. Support Fee Models After Product Release and International Tax Regulations**

Ultimately, it is common for the project to treat feedback as valid compensation during the R&D phase and then shift to a model where the product requires support fee payments upon commercialization, as this approach tends to minimize bureaucratic hurdles. International transactions add another layer of complexity: we must formally value the software and establish compensation, particularly for dealings with the U.S. or Europe, necessitating careful adherence to transfer pricing rules. I do not have a thorough understanding of how we calculate transfer prices, but I know it is significantly more complicated than dealing solely with domestic transactions.

**22. Mechanism of Value Consolidation via Outsourcing and Transfer Pricing Avoidance**

In many cases, all development is formally treated as outsourced, and the software's value is effectively consolidated in Japan. Because the value remains in Japan, there is little "international transfer" of valuable IP. This structure largely circumvents the need for complicated transfer pricing. In other words, if no value is considered to move across borders, we can avoid having to calculate transfer prices in the first place.

**23. Overseas Bases, Value Retention, and Accounting Rules in the Competitive X Domain**

Our typical arrangement involves Japan commissioning the development, thereby consolidating IP value in Japan, while overseas divisions primarily supply labor. For instance, in the competitive X domain, the company invests its own funds and retains the resulting value, thus simplifying contractual and accounting processes. Even so, from accounting and tax perspectives, sharing software assets is permissible only within specified criteria. If a software module exceeds those criteria, we must pay compensation. Within the competitive X domain, high-value components could trigger transfer-pricing complications, so we have rules stating, "only code within this range may be shared." Anything beyond this range automatically incurs additional fees. For example, if a U.S. team develops high-value features for the X domain, and this is more than a mere labor-hour transaction, it may be seen as a transfer of valuable functionality, necessitating payment.

**24. Constraints on Sharing High-Value (Marketing) Code and External Explanations**

Particularly problematic are features with high commercial or marketing value. Such functionality can yield a competitive edge and heightened market reputation, so merely covering the labor cost under an outsourcing arrangement can appear, to external observers, as though we are transferring a valuable asset for free. Hence, these high-value functions inevitably require compensation. In fact, it is highly reasonable to charge for the transfer of a high-value feature. Marketing-focused functions, by their very nature, must be compensated in accordance with their substantial added value.

**25. One-Way Sharing of Open Source–Based Integrated Systems in Early R&D**

Products that integrate or build on open source software are relatively easy to share unidirectionally, particularly during R&D phases. Feedback loops in such one-way sharing structures are easier to manage. These are different from high-value features that might invite scrutiny from tax authorities. Overall, open source–based projects can be treated primarily as technical exchanges and thus do not tend to trigger the same level of concern about undervalued asset transfers.

**26. Locating and Evaluating Value**

The key issue is identifying which party holds the underlying source of value. Transfer pricing laws hinge on determining the origin of the intellectual property—where is the actual creation of value taking place? Even if our expenditures are on a time-and-effort basis, if the resulting functionality has marketing utility, we must pay for that added value. This principle must be clarified not only internally but also externally. Whether one counts lines of code or enumerates direct costs, if the developed function harbors additional marketing value, its ownership and fair compensation must be clearly established.

**27. Complexity of Imports/Exports and the Need for Export Control**

Regulations surrounding exports and imports complicate matters further. Japan must comply with U.S. export laws and other international regulations, and the default demand is “Please manage your exports responsibly.” If we took all regulatory mandates literally, we might never share source code at all. Hence, we define precisely what each repository contains and conduct export controls and export assessments within that scope, simplifying the process. For instance, by mutual agreement, any technology that could be construed as weapons-related is explicitly excluded from a particular repository. Once we confirm the repository’s content does not fall under restricted categories, we then operate under the rule “Do not add any functions that violate export control.” This approach obviates the need for painstaking reviews each time, although different divisions face different auditing intensities depending on the functions they manage.

**28. Defining Export Scope for Repositories and Restricting Feedback**

At the outset, a repository is labeled with a clear description of its intended functionalities. We verify that no export control restrictions apply, and we also agree not to include, for example, any weapon-related information. By clarifying the permissible export scope in this manner, we can ensure no feedback diverges into restricted domains, thereby enabling a simpler operational model.

**29. Procedures and Audits When Registering New Users or Creating New Repositories**

We typically assess potential export issues when creating a repository or adding new users. If we consider expanding a particular piece of software to offshore locations such as India or China, we submit an export application stating, "We plan to deploy this software to India and China." When only three or four colleagues within Japan are involved, we rarely go into such detail. Even if the initial repository setup is quite relaxed, the need for export compliance arises once we add users from outside Japan.

**30. Expansion to International Development Centers (e.g., India, China) and Regulatory Compliance**

When we plan to make internally developed software available to our facilities in India or China, it can become a significant undertaking. Company B has a policy against involvement in weapons development. Yet certain technologies might, depending on perspective, be convertible for weapons use, prompting more stringent reviews by certain divisions.

**31. Hierarchical Structures of Collaboration and How They Emerge**

Collaboration sometimes arises from the bottom up: if there is an immediate need, people are informally pulled in to address it, sometimes with limited regard for formal contracts or rigid rules. By contrast, there are also projects initiated when two departments formally agree to exchange technology, accompanied by appropriate contracts. At a higher organizational level, there may be communities focused on technology strategy, designed to identify cross-functional issues and recruit volunteers to solve them. This is closer to a top-down approach, albeit still community-based rather than purely management-driven. Above that, there are fully top-down directives— "Business unit A will collaborate with business unit B on a new project." Thus, collaboration in the company can be understood as existing in at least four layers, ranging from grassroots initiatives to top-level management mandates. In some cases, these technology strategy–focused communities form "V-teams" (virtual teams) that operate without formal departmental structures, while in other cases, they appear spontaneously without upper management's direct knowledge. In summary, multiple layers and modes of collaboration coexist.

**32. Overview and Role of Technology Strategy–Oriented Communities**

Communities dedicated to technology strategy emerge when cross-functional collaboration is needed in certain technical domains. For example, there might be a "software community" that tackles company-wide challenges, bringing together volunteers to find solutions. Within this software community, there might be efforts to establish open source compliance rules, adapt new development environments and cloud technologies, or build internal communities. Similar cross-functional groups exist for various other technical fields—perhaps around ten such communities, each hosting multiple subgroups. They often have a small amount of funding to cover incidental expenses, so members are not covering all costs from their own pockets.

**33. Incentives for Participation in Cross-Functional Communities**

Because these technology strategy–oriented communities work on cross-departmental issues, solving those issues benefits each participant's own department. For instance, if an open source compliance rule is formulated within such a community and adopted across the company, each department gains a streamlined procedure for open source usage. For individual employees, a key incentive is being able to work on areas of genuine interest. The community often produces a report detailing their achievements— "Here are the outcomes from this community"—which is provided to the participating departments. Departments, in turn, can factor these outcomes into the employees' performance evaluations. Thus, employees do not suffer a penalty for their cross-functional activities; on the contrary, their enthusiasm and contributions can enhance their performance evaluations. This system encourages individuals to pursue projects aligned with their professional interests.

**34. Cross-Departmental Activities and Engineer Evaluation Measures**

While an employee's direct supervisor has primary responsibility for performance evaluation, feedback from other departments is also considered. Therefore, involvement in cross-departmental projects can influence performance assessments. Participation in technology strategy–oriented communities or initiatives led by senior technical experts may yield favorable evaluations. Many other cross-functional projects exist, and employees can choose to participate in those that align with their skills or interests. Because these efforts address pressing company-wide challenges, they tend to be well-received by management and can positively affect an engineer's performance rating. Although there is no direct monetary bonus given to each employee, the communities release reports on their collective outcomes, which can be used to bolster the participants' evaluations. Ultimately, individuals benefit by pursuing projects they find meaningful, while departments gain from improved internal governance and knowledge sharing.

**35. Supplementary Explanation on Incentives and Evaluation**

The technology strategy–oriented communities also engage in educating junior software engineers, which indirectly serves as an incentive by improving skill levels across the organization. For employees, the main motivator is the ability to work on projects that interest them. Additionally, the communities provide outcome reports to departments, which help ensure that participating employees are not penalized in performance reviews and, in fact, may receive positive recognition. Hence, when we speak of "incentives," we mean a system in which employees benefit from personal development opportunities and positive performance evaluations, while the departments benefit from well-organized rules and consolidated knowledge.

**36. Approval for Cross-Departmental Participation and Reflection in Supervisor Evaluations**

Without a supervisor's endorsement, an employee who neglects their primary responsibilities to focus solely on cross-departmental activities will not be positively evaluated. Generally, employees need their supervisor's permission to join technology strategy–oriented communities or subgroups, or to participate in senior technical experts' projects. In some truly grassroots cases, the supervisor may not initially be aware, but most often managers do know. The communities themselves report individual contributions back to the managers— "This member accomplished these tasks within the project"— thereby incorporating cross-departmental achievements into formal evaluations.

**37. Presentation Opportunities for Bottom-Up Activities and Resulting Incentives**

A platform exists for employees to publicly showcase their grassroots activities, which in itself serves as an incentive. Sometimes awards are granted—these might include cash, or a budget allowing the project to be recognized as official departmental work for the following year. Winners may also be given the opportunity to exhibit at a large internal technology event held in December. Moreover, bottom-up initiatives can organize their own presentation sessions, disseminating announcements via the company's internal communications system— "We'll be holding this event"—and garner interest from potential collaborators.

**38. Bottom-Up Activities, Internal Events, Awards, and Funding Mechanisms**

Through community-building and presentations, bottom-up InnerSource–type activities can begin discreetly, attracting colleagues' interest— "This looks intriguing; let's join." Early on, these efforts are not formal "work tasks," so there is no barrier to sharing. However, once an initiative gains attention at a showcase event and receives funding, it transitions into a formal project and must align with higher-level frameworks. Indeed, many business units host annual presentations, providing a stage where employees can reveal informal projects they have been developing. Outstanding projects may receive awards, possibly cash or official project status, as well as promotion. Additionally, we have an in-house new-demand creation program, "Program A," and if an initiative is accepted there, it can be scaled and institutionalized. Hence, the company provides multiple pathways for bottom-up innovations to secure funding and support.

**39. Handling Small-Scale CI/CD Libraries and Integration into Technology Strategy–Oriented Communities**

Consider a small but valuable improvement—for example, a CI/CD library that may not warrant a large award but still confers benefits. The working group on development environments within the broader "software community" could adopt and incorporate it into the corporate infrastructure. Indeed, our internal ecosystem features multiple layers, including top-down projects and top-down communities that define activities internally, as well as genuine grassroots innovations. In all cases, there are various channels by which good ideas can be elevated and utilized across the organization.

**40. Integration into the Company-Wide Platform and the Role of Internal Team B**

Such initiatives may ultimately be integrated into a business unit's official scope, a technology strategy–oriented community, or into the corporate infrastructure. The question then arises: who manages maintenance? If the improvement involves internal platforms or base systems, coordination with an internal "Team B," handling platform engineering or common infrastructure, is crucial. For instance, Team B may already manage GitLab, and there was a discussion that "it would be inefficient to host your own repository for maintenance purposes." We have also considered how to handle a corporate GitHub contract, eventually leading to its integration into the broader internal infrastructure.

**41. Organizational C, Department A, OSPO Activities, and Multi-Layered Internal Structures**

Organizational C oversees IT infrastructure and security. Currently, I am in Department A, conducting IT-related development and handling OSPO-like functions. The company operates with multiple organizational layers that can support both personal and bottom-up activities. Rather than merely constraining them, the organization's culture—especially Company B's tradition of "pioneering interesting ideas behind the scenes"—facilitates multi-layered collaboration.

**42. Relationship to Evaluations, HR Policies, and Motivation of Younger Engineers**

How do these activities influence HR evaluations? The Human Resources department explicitly states that cross-departmental activities should be recognized positively, so there is no inherent conflict. I am not entirely sure how younger engineers perceive these opportunities, but because there are multiple large company-wide events and abundant information channels, those who are interested will likely find them appealing. Engineers with a strong drive also engage with external communities and conferences, then bring that energy back to form new communities within the company, ultimately fostering fresh technical collaborations.

**43. Defining InnerSource, Ensuring Transparency, and Meeting the Expectations of Overseas Branches**

Regarding InnerSource, merely proclaiming "we're open" internally does not necessarily guarantee real openness—transparency in decision-making and clarity in concepts such as "fairness" or "equality" are needed. Non-Japanese subsidiaries sometimes say they want to implement InnerSource because "Google does it," but fundamentally, they may be looking for genuine transparency within the corporate group. Since Company B's headquarters is in Japan, language and time-zone barriers can create silos, and overseas subsidiaries might feel the need for improved transparency. In particular, the B2C Technology Division Z is predominantly based overseas. Therefore, we not only hold technical events in Japan but also replicate them at other global sites, making it easier to detect hidden activities and incorporate them into structured processes with budgets and operational frameworks. Recently, there has been a new technology strategy–oriented community in the B2C domain, with a business-side owner communicating primarily in English, thereby achieving a more global perspective.

**44. Scope of InnerSource Within a Large Organization and Its Assessment**

(Responding to the interviewer's remark that Company B is highly complex, with multiple headquarters and different shared technologies in various domains, making a single unified approach unnecessary): Indeed, even within the competitive X domain alone, there is a large workforce, and that domain has sufficient scope to pursue sharing and innovation. In a narrower context—say, only for a specific business domain at Company B—the strategy might be to consolidate everyone onto a single GitHub Enterprise instance, ensuring an appropriate scale for implementing InnerSource. Certainly, there are subtle criteria, but overall, we adapt InnerSource to each domain or location, leading to cross-site technology exchanges and collaborations among communities focusing on content or technology strategy. This yields a mutually reinforcing environment across various regions and specializations.

## Company C (Internet) – Translation

| Department | Company Size | Department | Position |
|---|---|---|---|
| Internet | Several thousand | Development Productivity Division | Leader |

1. **Respondent's Career Background**

   Prior to joining Company C, the respondent worked in what is generally referred to as the system integrator (SIer) industry, taking part in multiple projects at various client sites. These projects included hybrid application development and the creation of internal banking applications, among others.

2. **InnerSource Culture at Company C and the Turning Point of Company-Wide GitHub Enterprise Deployment**

   Company C already had a culture of developing software on GitHub and sharing it across different departments, as well as a tradition of releasing projects as open source software (OSS), even before the respondent joined. The turning point came around 2020, when the company adopted GitHub Enterprise across the entire organization. This enabled the use of internal repositories, thereby accelerating the widespread sharing of corporate infrastructure, as well as collaborative development with adjacent teams. Building on this momentum, a new initiative known as "Program A" was launched in 2021. This initiative provides a framework for applying and reviewing technology assets—both internal and external—and offering medium- to long-term support and growth measures. Submissions are ranked based on their influence, and support is allocated accordingly.

3. **The Birth of Program A and Its Decision-Making Process (Relationship with "Meeting A")**

   Program A originated from a company-wide structure known as "Meeting A." Meeting A serves as a venue where a board director (acting as a leader) selects specific members who then propose new initiatives directly to the company president for approval. A "technical edition" of Meeting A led to the proposal of the original idea that would become Program A; once the idea was accepted in that forum, it was institutionalized. In the technical edition of Meeting A, the leader unilaterally selects members, who then refine their ideas and pitch them to the president. Because the president grants immediate approval in that setting, ideas brought up from lower levels of the organization are ultimately endorsed from the top down, leading to the creation of robust, influential programs. In other words, proposals flow from the bottom up and are formally recognized at the highest level, resulting in top-down implementation.

**4. Diverse Engineer Motivations and Approaches in InnerSource and OSS Publication**

Engineers at Company C have varied circumstances and perspectives. Some wish to rapidly release items required for daily work as OSS—so they can use these solutions within their own projects while also offering them to external users—whereas others limit sharing to internal use only. In essence, the balance between internal sharing and external release, along with each engineer's motivations, differs according to team policies and project objectives.

In certain cases, organizational technology selection includes building tools as OSS from the start. For example, in a foundational development project known as Product A, the team believed that if the product did not gain traction outside the company, it would not be widely adopted internally either. Hence, they adopted the strategy of releasing it publicly early on, attracting users and developers from the broader community and subsequently incorporating that feedback back into the organization.

Additionally, in certain domains—such as game development—multiple teams grapple with common challenges. In those scenarios, they may engage in InnerSource activities to collaboratively build a shared foundation that each team can utilize. Thus, the form and underlying philosophy of InnerSource and open source activities are diverse, and initiatives are often driven by unique visions within each team.

**5. Organizational Diversity and InnerSource Challenges (Large Number of Affiliated Companies and Reinventing the Wheel)**

Company C operates numerous subsidiaries, resulting in well over 100 different products. The overall picture is extremely complex. Under these varied organizational structures, reinventing the wheel tends to occur. Before the deployment of GitHub Enterprise, each subsidiary maintained its own GitHub environment, and there were no shared internal repositories. This lack of visibility sometimes meant that different teams were unaware of each other's work, even in adjacent divisions. Additionally, inter-team rivalry occasionally led to separate development of similar foundational assets. While the introduction of GitHub Enterprise has not completely eliminated these issues, it has contributed significantly to reducing duplication and improving collaboration.

## 6. Positioning and Scope of Program A, Updated Evaluation Criteria, Application Incentives, and Similarities to CNCF

Program A is an initiative spanning the entire corporate group, offering a common framework of collaboration that was previously inconsistent across teams. Over time, Program A has undergone iterative improvements. The application and review process has evolved to incorporate a more convincing set of evaluation criteria, such as the number of users, satisfaction levels, and GitHub star counts. These criteria are periodically revised to ensure fairness and transparency.

Motivations for applying to Program A are diverse. Some seek incentives (e.g., receiving resources or support), while others utilize Program A as an opportunity to enhance internal recognition of their projects. Program A applies a five-tier grading system, assessing multiple factors, including usage scope and recognition, and then assigns a corresponding rank.

A board oversees Program A, and the respondent is one of its members. Engineers from various organizational units participate, sharing responsibility for program management. Although the Organization for Development Productivity Improvement is an important stakeholder, Program A is jointly operated by multiple entities.

Program A encompasses a wide array of technology assets, from small- to medium-scale libraries to more recent developments in AI models. These are all tools that improve development or operational efficiency, owned by employees of Company C. They must be documented, have an established usage record, and meet specific basic criteria. Even if the OSS project has external contributors, so long as a Company C employee serves as a maintainer, the project is eligible for application.

It is uncommon for Program A itself to be the original driver of a project's development. More often, projects arise to solve a particular need, gain traction, and then seek Program A support—whether for additional resources, funding, or career advancement of the engineers involved.

Project ownership structures vary considerably. Some projects are driven by individual developers, others by a single division as a strategic initiative, and still others by multiple teams employing InnerSource collaboration (e.g., linting tools co-developed across divisions).

Conceptually, Program A is akin to the Cloud Native Computing Foundation (CNCF). Projects entering CNCF typically do not start out with the explicit goal of joining CNCF; rather, they emerge to solve a practical problem and later leverage CNCF's scale and support as the project grows. In much the same way, Program A serves as a platform to help promising projects expand and mature.

**7. The Open-Source Culture and Soil for Public Release**

When the respondent moved from the SIer industry to Company C, they noted a unique Web-centric atmosphere characterized by openness. Engineers commonly used open source tools, and it was not unusual to encounter committers of major OSS products among one's peers. Additionally, it was common for employees to engage in side-projects or launch new endeavors. This environment contrasts sharply with the respondent's prior experience at an SIer, where releasing creations publicly and leveraging external developments were far less prevalent.

**8. Freedom in Product Development and Technology Selection, and the Diversity of Business Areas**

A major factor that distinguishes Company C is its portfolio of in-house products. Spanning games, advertising, media, and more, each business area requires distinct technologies. Hence, technical decisions are highly decentralized; teams have considerable autonomy to adopt whatever technologies they deem optimal for their particular domain. There is also a company culture of launching subsidiaries, granting them ownership, and nurturing them to grow. While this could be described as a "siloed" approach, it can also be seen as empowering each unit to make independent decisions, both technically and in business. Within this culture, there is a natural tendency to "use whatever is already available", which promotes the evolution of a collaborative, open approach.

**9. "Siloed" Organizational Structure, Executive Jurisdiction, and Domain Segmentation**

Company C's organizational chart is segmented by domain, with executives assigned specific oversight. For instance, separate executives supervise the media business and the advertising business, each of which comprises various functions. Though engineers may not be particularly aware of these structural divisions in their everyday work, in practice, entire floors or buildings often align with specific business domains. Collaboration tends to be more active within a given executive's domain but is less evident across domains.

**10. Granularity of GitHub Organizations and Internal Repositories**

Typically, one GitHub Organization is established per product. However, numerous exceptions exist, such as a cross-domain organization dedicated to a single executive's jurisdiction, or a specialized organization set up at the outset for Product B. Because GitHub Enterprise enables the creation of internal repositories accessible to all enterprise members, visibility across the entire codebase is maintained. Even if product teams are organized under separate GitHub Organizations, access restrictions can be minimized, allowing for horizontal, enterprise-wide code discovery.

**11. Organization-Level Policies, Billing, and Security Management**

Within Company C, billing is sometimes managed at the GitHub Organization level, and security policies may be tailored for specific needs. For instance, teams in the gaming domain might require stronger security due to the competitive landscape. While certain baseline security measures are enforced enterprise-wide, each organization can manage additional rules autonomously. This setup preserves organizational self-governance while retaining the ability to share resources enterprise-wide via internal repositories.

**12. Asset Sharing Across Legal Entities and Treatment of Intangible Assets**

Practices such as providing complimentary resources among subsidiaries (separate legal entities) are not always clearly governed. Although billing for foundational development may be relatively well-defined, there seems to be no standardized policy at the corporate level regarding the valuation of intangible assets. According to the respondent, they have not encountered much discussion regarding how intangible assets should be tallied in financial statements. Each team or subsidiary presumably makes individual decisions about reflecting these items in their own profit-and-loss (PL) statements.

**13. Variations in Engineering Culture and the State of InnerSource Adoption**

Engineering culture at Company C is not homogeneous across the enterprise. Certain domains exhibit a more open mindset—owing to business priorities or historically established practices—while others face constraints that limit cross-team collaboration or the extent to which internal assets can be made "open." Even within the same building, the atmosphere can vary from a "classroom-like" environment with employees in suits to a far more "open source-inspired" setting with a casual dress code and a strong culture of openness.

**14. Lack of Dedicated Licenses or Scope Definitions for InnerSource Activities**

When engaging in InnerSource activities, the company does not maintain any formal, specialized license detailing the scope or permitted actions. However, a corporate guideline does exist for releasing projects as open source. This guideline outlines items that must be checked prior to publication to ensure compliance with security and branding standards.

**15. Program A and Trends in Target Projects (Relationship with OSS)**

Many of the projects submitted to Program A are indeed open source. Some are led by individual developers, while others are maintained by entire teams. One notable example is Product B, which has high visibility inside and outside the company. However, looking purely at usage metrics, there may be other technology assets used even more extensively than Product B. Depending on the specific evaluation criteria, different projects might be deemed "the most successful." Nevertheless, Product B is well recognized both internally and externally.

Additionally, there are cases in which employees have developed highly renowned personal OSS projects, which later gain renewed recognition through Program A. In many companies, assets developed during working hours are regarded as corporate property, but at Company C, there seems to be relatively little resistance to incorporating an engineer's privately initiated creation into the corporate technology stack, investing time and resources in it when beneficial.

**16. Flexible Evaluation Metrics and a Results-Oriented Culture**

Company C generally follows a relaxed evaluation philosophy that grants employees freedom so long as certain targets are met. While managers might be cautious about incorporating an OSS created by an individual engineer, the organizational culture is more receptive than traditional firms to using such assets—if they yield positive outcomes.

**17. Freedom, Accountability, and a Trust-Based Culture**

Although not explicitly documented, Company C is guided by a principle of "freedom and discretion." The notion is that autonomy is afforded to employees who, in turn, must shoulder responsibility. Developers have considerable latitude in choosing their development environments and tools, and engagement with external open source communities—be it through contributions or learning—is both accepted and encouraged. While security-related disclosures are restricted, employees are otherwise largely trusted to make sound judgments in their external technical communications.

**18. Loose Publication Policies and Minimal Review Processes**

Technological outreach activities, such as blog posts or conference presentations, are seldom subject to stringent internal reviews. Certain processes exist to safeguard the company's logos and brands, but as long as engineers understand these constraints, approval for publicizing technological content is relatively easy to obtain. The primary limitations concern sensitive data related to security, but most knowledge sharing and promotional materials can be disseminated without heavy oversight.

**19. The Silo Problem and the Rationale Behind Program A**

Company C frequently establishes subsidiaries, which can lead to silo formation. Program A was conceived to help break down these silos and foster cross-organizational collaboration. In parallel, the company as a whole is striving toward more active collaboration across domains.

**20. Recognition of Silo Problems After Becoming a Leader**

As an ordinary team member, the respondent felt the environment was open—simply raising one's hand could yield opportunities to tackle new challenges. However, upon assuming a leadership role aimed at making an impact across the entire corporation, the respondent began to perceive the extent of the silo problem—especially among different departments and subsidiaries. This issue becomes clearer at higher levels of the organizational hierarchy.

**21. Internal Promotion Efforts During the Introduction of GitHub Enterprise**

When GitHub Enterprise was being rolled out, the company prepared internal documentation and engaged in "sales-like" promotional activities encouraging employees to "join the Enterprise." Given that employees enjoy substantial autonomy, implementing a single, unified tool across the organization required clarifying the advantages to gain consensus. Thus, a significant internal advocacy campaign was undertaken.

**22. Challenges in Knowledge Sharing and Cross-Functional Organizations**

Certain cross-functional (horizontal) organizations at Company C aim to share knowledge and institutionalize best practices across the enterprise. In reality, however, achieving widespread adoption is often challenging. Even if a given practice is advantageous, it may not gain traction unless key individuals within various teams actively promote it. A single cross-functional group alone cannot fully ensure enterprise-wide implementation; thus, collaboration across multiple organizational boundaries is essential.

**23. Obstacles to Collaboration in Cross-Functional Organizations**

The respondent's department is itself a cross-functional group, dedicated to fostering company-wide knowledge sharing and best practices. Nonetheless, adoption of any practice heavily depends on involvement from each local team. Lacking a mechanism for outright enforcement, the cross-functional group struggles to unilaterally drive adoption across the entire enterprise, highlighting the limitations of this approach.

**24. Successes and Failures: The Challenge of Mandates and Grassroots Expansion**

Over time, there have been both successful and failed attempts to disseminate tools and practices. Mandates from the top generally prove ineffective, as decision-making power is dispersed across multiple domains and business units. Instead, successes typically stem from grassroots momentum supported by favorable internal buzz. When leaders such as the president or executives publicly endorse an initiative, creating an environment where people feel, “everyone is on board,” the practice often spreads more readily.

**25. GitHub Enterprise Deployment and Organic Adoption**

GitHub Enterprise is a classic example of organic adoption. It was initially implemented at the department or subsidiary level, and as its user base gradually grew, it transitioned to an enterprise-wide agreement. Around that time, security and account management concerns were intensifying, and GitHub Enterprise’s SSO integration aligned perfectly with these needs. This scenario— “we opened the lid to discover almost everyone was already using it”—is atypical but represents an ideal mode of technology adoption.

**26. Difficulties in Tool Selection and Common Failures (Challenges of Shifting the Entire Organization)**

Tool selection at Company C has not been free of failures. For instance, although G Suite is widely deployed company-wide, some employees argue for alternative tools. However, a total organizational switch is logistically difficult, compounded by the need to use Microsoft Teams in some collaborations with international affiliates. Similar complications have arisen around public cloud contracts, leading to contentious debates. Numerous attempts to standardize tools have been made, though many have been naturally phased out over time.

**27. Perspectives of Managers and Engineers Regarding Program A**

From a managerial perspective, Program A’s purpose of “evaluating existing technological assets” makes it relatively acceptable. Historically, the notion of building an enterprise-level shared foundation has been perceived by some managers as a “losing battle” without robust corporate support.

Program A thus serves as a workable solution, providing an equitable system for assessing and supporting assets once they are created. For engineers, it functions as a mechanism to have their projects recognized fairly.

**28. Evaluation Cycle of Program A and Forms of Support**

Program A operates on a one-year renewal cycle. Advancing to a higher tier after a year requires re-assessment and re-submission (e.g., going from Tier 1 to Tier 3), which serves as a beneficial incentive. While no direct monetary reward is provided, other forms of support—including technological PR assistance, covering cloud usage fees, social events, and conference participation costs—are available. For larger-scale projects, overseas conference participation can be considered. Thus, Program A enhances internal visibility and technical credibility for participating projects.

**29. Information Transparency and a Culture of Communication**

A significant barrier to expanding InnerSource is the difficulty of discovering relevant internal information—that is, knowing which teams face similar challenges or run comparable projects. Nonetheless, external outreach sometimes leads to internal collaboration. For instance, a post on Slack or an internal "Times" channel might spark cross-team partnerships. However, there is no single platform consolidating all project information for the entire organization, making such serendipitous connections sporadic and largely reliant on chance.

## Company D (Internet) – Translation

| Department | Company Size | Department | Position |
|---|---|---|---|
| Internet | Several thousand | Development Headquarters<br>(Core Group Business G) | Lead Engineer |

### 1. Self-Introduction

I am currently affiliated with the Development Headquarters, which functions as a cross-organizational technical unit. This unit consolidates various technical departments and requisite technologies under one umbrella. Within the group, we operate the core business (hereafter referred to as “Core Group Business G”), where my position is that of a Lead Engineer. A new division has recently been established to promote both the business and technological aspects in tandem. Lead Engineers here are considered to occupy higher-level roles primarily involving architecture design and code reviews. In addition, I am often selected to work on technically challenging features that pose high barriers to adoption.

### 2. Number of Employees and Company Scale

Under the Chief Technology Officer (CTO), the engineering organization is relatively flat, and we engineers maintain broad connections through the CTO. At a more hands-on level, we also cooperate extensively with other departments, acting as intermediaries when needed. According to our GitHub licensing data, the organization has about 1,500 seats. However, if we include non-regular employees, the total workforce is approximately 5,000 to 6,000 individuals. Roughly half of these employees are believed to be involved in development tasks.

### 3. Historical Context Leading to InnerSource Adoption

The starting point for InnerSource adoption can be traced to our longstanding culture of creating services through full participation by all relevant members. From infrastructure engineers to game developers, everyone has been developing within a single mono-repo. Although the company has grown and experienced various changes, the environment in which anyone can view and modify the codebase has persisted since our founding. The first initiative began when our CEO established the company as an independent entity. Subsequently, our CTO introduced version control systems and encouraged collaborative development. Notably, we have been using GitHub since its very early days.

4. **Evolution of Organizational and Repository Structures**

Most of Company D's primary services reside within a single GitHub organization. Historically, repositories and organizations were divided by budget lines and team boundaries, but following organizational restructuring, we are currently moving back toward centralization. While there are still vertical silos to some extent, we fundamentally maintain an environment where everyone has viewing permissions, thus ensuring that code and documentation remain widely accessible internally.

5. **Corporate Restructuring, Holding Company Model, and Code Sharing Among Subsidiaries**

We have multiple corporate entities, each funded by different budgets, and we invite members into respective GitHub organizations accordingly. In total, there are about 10 GitHub organizations, each corresponding to one of the separate legal entities. Members of the Development Headquarters, as a cross-functional unit, belong to nearly all of these organizations. From an accounting perspective, concern arises regarding profit distribution among these legal entities. Hence, contracts and NDAs are managed by the holding company to prevent unauthorized disclosure of sensitive information. Recognizing that anyone could potentially access the code, we adopt a relatively open stance while still demanding adherence to confidentiality obligations. Software assets are principally owned by the subsidiaries, while the holding company provides security and information systems management.

6. **Ownership and Management of Software Assets**

In principle, each subsidiary retains ownership of its own software assets. While the holding company is responsible for security, IT systems, and contract management, the business units primarily hold software ownership. At present, there appear to be no plans to centralize all software assets under the holding company.

7. **Provision and Utilization of Common Libraries and Accounting Adjustments**

Common assets are generally made internally available on the main GitHub organization, allowing other business units to be invited and to utilize those assets. When development outputs could pose legal issues if provided free of charge, the Development Headquarters (which was spun off as a separate entity) may assume control over asset management and, if necessary, charge a usage fee. While large transactions require approval from senior management, the usual flow involves the Development Headquarters aggregating libraries, offering them to other business units, and reporting the resulting revenue to the holding company.

8. **Management of Common Libraries and Infrastructure; Allocation of Assets**

The holding company is involved in financial audits and related procedures. Asset consolidation is an ongoing process: common components are concentrated in the Development Headquarters of the main corporate entity, while game-related assets are structured within the Game Business H. Thus, common libraries and infrastructure are provided by Core Group Business G, whereas the game components are maintained by Game Business H. This delineation continues to evolve.

9. **Distribution of Technical Know-How and Internal Rights**

Although no centralized knowledge portal currently exists, relevant information can generally be obtained by contacting the Development Headquarters. The executive team also relies on the Development Headquarters for technical inquiries, which are then directed to the appropriate department. Rights to libraries and infrastructure typically reside with the Development Headquarters, while game-related know-how remains with Game Business H. This division of responsibilities has emerged organically and functions smoothly.

10. **Architecture, Service Separation, and Infrastructure Provision**

From an architectural standpoint, we broadly categorize our systems into server-side and front-end components. We employ Kubernetes and provision single-tenant environments for each service. The shared departments provide infrastructure templates and fundamental configurations, while each business unit retains freedom to modify their application layers. This clear boundary delineates responsibilities: the platform is managed by Core Group Business G, whereas each service department manages its own custom settings.

11. **Collaboration Between the Platform Provider and Business Divisions**

In the past, the platform team successfully centralized a library previously maintained by Game Business H, thereby reducing management costs. By providing infrastructure and common libraries, we facilitate interdepartmental collaboration and optimize costs. In practice, the Development Headquarters takes on a cross-organizational role and frequently collaborates with various departments. However, each business unit often focuses primarily on its own domain. Consequently, the Development Headquarters frequently steps in to promote horizontal collaboration.

**12. Methods for Fostering Cross-Organizational Collaboration and Information-Sharing Tools**

Collaboration often arises when the Development Headquarters identifies common needs in the consultations and project requests it receives. Acting as GitHub administrators at the organizational center, the Development Headquarters can easily forge connections between business units. Internally, there is a dedicated Slack channel for GitHub support, which is open to all employees and thus enables transparent interdepartmental communication. Slack, Jira, and Confluence are widely used; documentation from the shared departments is publicly accessible throughout the company. This high level of transparency often triggers organic collaboration.

**13. In-House Study Sessions, Cultural Differences, and Efficiency Through Active Engagement**

The company promotes collaboration through internal study groups and large-scale exchange events held about twice a year. The corporate side coordinates these initiatives to strengthen cross-departmental ties, and we also host external tech conferences. While business units tend to focus narrowly on their own domains, the Development Headquarters invites them to adopt a broader perspective and facilitates interdepartmental collaboration. Culturally, business units operate with narrower but deeper expertise, whereas the Development Headquarters assumes a wider but more shallow coverage, achieving efficiency by "calling out" potential synergies and opportunities.

**14. Obstacles to Internal Information Sharing and Underlying Causes**

The primary obstacle is information restricted by intellectual property rights. In cases where third-party rights are involved, or when materials contain pre-release details, it becomes difficult to share information broadly within the company. Such legal and contractual issues often prevent open sharing until an official release. Once the product is released and internal sharing becomes permissible, most employees consult the corresponding repository. Our practice is to remain alert to such release timings so that we do not miss the opportunity for broader internal collaboration.

**15. Stream-Aligned Teams, Platform Teams, and the Potential for InnerSource**

Collaboration across stream-aligned teams can be challenging. However, InnerSource tends to work well in the context of platform teams and complicated subsystem teams, owing to platform-related concerns. Although the department in which I currently work is newly formed, we aim to expand team topologies and platform engineering from next year onward. Compared with direct business-unit collaborations, linkages between common components (or the platform) and stream-aligned teams are highly active and yield extensive feedback. At times, this high level of engagement leads to heated discussions, but our longstanding culture has been to freely offer suggestions for improvements. Given the interdependencies, feedback tends to arise naturally. The shared department consistently reviews the work of other departments, and the ongoing exchange of improvement proposals remains vibrant.

**16. Pull Requests, Concrete Examples of InnerSource, and the Underlying Culture**

Pull requests have become a standard development practice, enabled through permission management and the use of fork features. For instance, the shared department maintains lists used for authentication, which are updated via pull requests. The process is well-established: once tests pass, the person in charge merges changes into the main branch. While there was initially some resistance to issuing pull requests, it is now customary for anyone to open one. Persistent efforts to highlight the efficiency benefits have gradually encouraged even junior employees to submit pull requests. As the workforce shrank, the notion of shared operational responsibilities ("we are all in this together") became more prevalent, and the culture of accepting pull requests naturally took hold.

**17. Fluctuations in Collaboration Culture Due to Business Performance and the Impact of GitHub**

Some companies have experienced reductions in collaboration following business downturns, but at Company D, workforce contractions actually amplified a sense of collective effort ("let's all cooperate"), making pull requests and resource sharing even more commonplace. Everyone has embraced the philosophy that it is acceptable for someone else to implement necessary changes, provided it benefits the collective. GitHub has lowered collaboration costs, further encouraging cooperative development. Anyone can submit a pull request, and even procedural documents can be shared via pull requests. Although certain companies face difficulties with performance evaluations under such a scheme, we have been able to embed these practices quite naturally. The reduced friction for collaboration afforded by GitHub has encouraged more expansive thinking across teams.

**18. Engineer Evaluation System, Performance Goals, and Degree of Autonomy**

Our evaluation system uses a free-form approach in which each individual self-reports their progress against performance targets. A team with the relevant technical expertise then reviews pull requests and projects, sometimes sharing associated search queries as evidence of contributions. We deliberately avoid 360-degree evaluations to prevent undue manipulation. We emphasize freedom because developers in the gaming domain traditionally dislike micromanagement. Under this culture, individuals tend to handle their own tasks and produce measurable results. Failures are naturally weeded out, and there is virtually no room for those who merely await instructions. Ultimately, a relative assessment is conducted to determine bonus allocation. Nevertheless, this arrangement does not foster severe competition, thanks to systemic design that discourages employees from competing over tasks.

**19. Award Programs and Their Influence on Evaluations**

We have monthly, quarterly, and annual award systems. Project Managers (PMs) or other managers nominate candidates, and the CTO or department heads make the final decisions. Awards vary from small monetary tokens or meal stipends to more substantial incentives at the annual level. A strong record of receiving awards serves as a metric for being recognized as a Lead Engineer. We also hold an annual blogging award, featuring unique personal gifts from the CTO. We briefly attempted to incorporate external conference presentations as a formal evaluation metric but discontinued it once we realized that not everyone could present, and quality standards dropped. Today, presentations remain an optional bonus factor. Consequently, engineers showcase their achievements in these ways to garner recognition in the organization.

**20. Company Culture, Engineer Evaluations, and the Use of "End Credits"**

Given that our workforce includes both new graduates and mid-career hires, the cultural baseline often reflects standards from the gaming industry, which differ somewhat from typical software engineering cultures. In gaming, having created a successful or popular game is often highly regarded, akin to holding a prominent "badge." Moreover, products in the gaming industry typically face short life cycles. Once a game succeeds, attention swiftly shifts to the next title. There is a sense of pride and reward in seeing one's name in the game's end credits, and these credits can serve as valuable credentials when changing companies or taking on a new producer role. The shared departments occasionally receive recognition for their assistance, with a simple "thank you" and inclusion in the end credits—an informal but culturally important acknowledgment that differs from direct financial reward.

**21. Changes in the Award System and Nomination Process**

In our corporate culture, achieving concrete results from in-house development initiatives typically garners higher attention than simply having created something in isolation. This mindset also extends to whether pull requests are accepted or declined. Individual motivation greatly depends on the feeling of having contributed and produced visible outcomes. Recognition and mutual praise circulate widely in our culture. Although individuals derive a sense of satisfaction from their tangible contributions, the company's annual evaluations do not necessarily encourage a "me first" attitude. In fact, those who avoid "self-promotion" often accumulate significant achievements quietly, and the evaluation system is designed to ensure that genuinely deserving individuals eventually gain recognition. The nomination process for awards is open to PMs and similar roles. In principle, authentically valuable contributions are nominated, making the award system less susceptible to strategic manipulation.

**22. Emphasis on Integrity and Hospitality**

At Company D, it has been striking to see people genuinely nominating each other for recognition, offering a wide spectrum of award categories without contrivance. This practice is highly effective in sustaining a positive environment. Malevolent intentions are exceedingly rare. The CTO's guiding principles emphasize honesty and kindness, and the workforce largely shares this mindset, effectively minimizing harmful behaviors.

**23. Recruitment Policies and a Long-Term View of Personnel Selection**

In our hiring practices, we prioritize candidates who mesh well with our corporate culture. While we once recruited in large numbers during a boom period, we have since stabilized and now require employees who can persevere over the long run. Consequently, we value individuals who communicate effectively and can make collective decisions with team members. Perseverance is crucial given the cross-functional, platform-oriented nature of much of our work. Certain initiatives, such as migrating on-premises infrastructure to new technologies, must be gradually amortized over five years and cannot be shifted overnight. Tangible results in our domain may take three years to materialize, making resilience and interpersonal skills particularly important. This principle generally holds across our various corporate entities and business units.

**24. Objectives and Examples of Open Source Releases**

We maintain a minimal version of an "Open Source Program Office" that connects licensing reviews to our legal department. Staff in that office include a director from the corporate side who organizes tech conferences and manages human resources tasks. Their role is to ensure that any open sourced code does not harm customers or the company. If the team collectively deems a project beneficial, and the legal team or external legal counsel finds no violations, we typically open source it. While few open source projects remain from our more active "bubble era," there was a period when the CTO himself frequently published code. Projects that we do open source tend to be those where intercompany collaboration could be beneficial (e.g., game engine libraries), or those that are not directly tied to profits but have broader industry value. For resources not quite suitable for open sourcing but still worth sharing internally, we usually keep them as internal repositories to facilitate pull requests and lighten the editing burden. For instance, common application forms are managed in an internal repository and updated collaboratively via pull requests.

**25. Tools and Common Platform Development as InnerSource Initiatives**

Our InnerSource practices encompass the development of common platforms, shared libraries, new projects, and small hackathon-style endeavors. Overall, the organization does not overly fear changes or breakages. A common motto is "If something is not entirely satisfactory, accept it and move forward." This phrase encourages rapid deployment after sufficient testing, with the assumption that if a deployment fails, the team can quickly redeploy after making corrections. When something does break, an apology is typically sufficient, underscoring our belief that robust testing mitigates the risks associated with iterative deployment. This ethos also underpins our willingness to modify someone else's internal assets; if something fails, we simply offer a sincere apology and fix it.

**26. Maintaining Tools, Operational Burdens, and Organizational Support**

Reinventing the wheel is uncommon. Documentation is kept casual by simply writing a clear README.md, often including a note for prospective contributors on how to improve the project. The motivation for building in-house tools generally stems from dissatisfaction with external vendor solutions. For example, our CTO began writing an internal company search tool after being disappointed by the usability of a vendor product. Similarly, we developed our own frontend for a timekeeping system when existing tools proved cumbersome. Once these efforts prove beneficial, they are typically shared company-wide, with announcements posted on the SharePoint portal managed by the information systems team. The reception to such tools is typically positive ("This is exactly what we needed!"), occasionally leading to formal commendations.

**27. Agreement with Managers and Balancing Project Priorities**

Although maintenance can become an issue if a project's budget or department dissolves, the CTO's organization often takes over if a large user base still depends on the tool. The CTO maintains a dedicated budget to cover server costs and so forth. In one example, when HR needed a new hiring system, the Development Headquarters opted to fund and support it. Regarding priorities, our corporate strategy encourages us to address critical business needs first; however, if an initiative is genuinely beneficial to the company, we may reallocate resources or negotiate schedules. Sometimes individual engineers raise problems and serve as ad hoc project managers. For instance, if a single engineer proposes a major infrastructure change or feature removal, the community deliberates collectively before moving forward. We do have predetermined targets, but we also devote a considerable fraction of our capacity to self-initiated work. It is relatively easy to strike a balance: if managers urgently require something, we address it first as professionals and then proceed with our own projects.

**28. Linking Different Levels of the Organization and Supporting Internal Communication**

Engineers who struggle with communication often approach me for help in advocating their ideas to upper management. Written proposals sometimes fail to gain attention, so I step in to explain the merits in person to senior stakeholders, thereby smoothing the path. A representative example is introducing a novel deployment approach developed by junior staff to the leadership team. In this sense, I act as a bridge between executives and engineers, facilitating mutual understanding.

**29. Personal Open Source Activities and Their Treatment Within the Company**

Individual developers are free to pursue open source activities on their own time, and these can sometimes feed back into company projects. I myself maintain a personal open source project—albeit anonymously, due to policies at my previous employer prohibiting public disclosure of real names or affiliations. In contrast, some younger employees openly develop open source projects or author Requests for Comments (RFCs) for external organizations under their real names. In those cases, the company accommodates their activities by, for example, allowing them to attend global conferences or work late at night to align with time zones, providing travel support when necessary. Nonetheless, these activities usually fall under personal initiatives, and while benefits may occasionally flow back into the company, there is not yet any flagship open source project that the entire organization supports comprehensively.

**30. Repository Management, External Collaboration, and Mechanisms for Internal Openness**

InnerSource-like practices have existed within our organization long before the term "InnerSource" was coined. Newly hired employees occasionally experience initial surprises but quickly adapt as everyone around them naturally embraces open collaboration. A supportive environment, in which openness is the norm, fosters rapid assimilation of new hires and external contractors. Generally, contractors are granted the same visibility as employees, unless security concerns prohibit certain commits. We do not employ outside collaborators on GitHub because of the administrative complexities; instead, we bring them directly into our GitHub organizations, granting them a high degree of freedom. Every few years, we revisit our GitHub Cloud governance arrangements and discuss how to continue this open approach. Since no one wants fragmentation, we align on solutions acceptable to all. Precisely because everything is open, we are conscientious about properly safeguarding confidential information.

**31. International Collaboration and Handling IP and Pre-Release Code**

We do maintain private, internal repositories, but only for information that can be safely viewed by all entities within the group. We also have overseas members who participate in UI development or translation projects. Previously, each regional office stored assets independently, but we have centralized the process so that code is distributed from the head office. For compliance with export controls or intellectual property (IP) issues, we might fork and supply a specific revision of the repository instead of granting direct access to the main repository, ensuring only legally cleared versions are shared. Whenever there are no personal data or unresolved legal constraints, we can proceed without hindrance. If code pertains to pre-release projects, we protect confidentiality by providing blank placeholders in infrastructure sections or withholding certain details.

**32. Infrastructure, Documentation Management, and Internal Tools Adoption**

The concept of "InnerSource" is steadily gaining traction within the company, partly due to my ongoing efforts to promote it, and newly hired graduates or employees at subsidiary companies are starting to pay attention. During orientation, new graduates take courses in technical writing and documentation management. Mid-career hires typically receive abbreviated tool manuals from their teams and adapt these materials for their specific contexts.

## Company E (Internet) – Translation

| Department | Company Size | Department | Position |
|---|---|---|---|
| Internet | Several thousand | Accounting Department | Person in charge |

1. **Introduction and Work Responsibilities**

At the time in question, I was employed as an accountant at Company E in the internet industry. I had previously accumulated accounting experience at several startups, and upon joining Company E, I was again tasked with accounting responsibilities. Company E's core operations involve numerous transactions, which generate a large volume of financial data. Consequently, the accounting department needed a reporting system capable of extracting and processing relevant information from the company's service databases via SQL to facilitate accurate bookkeeping. My role involved reviewing system specifications and SQL queries to ensure that accounting processes proceeded smoothly behind the scenes.

2. **Audit Preparedness**

Company E was required to present accurate financial data in response to requests from external auditors. It was essential to extract transaction data from various services, process them under the correct accounting standards, and address any audit inquiries swiftly. Our team reinforced internal structures by establishing a reporting flow that could promptly deliver all necessary data for audit purposes.

3. **Background of Microservices Adoption**

As the organization expanded, the breadth of components under the development organization grew, thereby increasing the number of modules that each engineer needed to handle. When onboarding new engineers, they often faced a steep learning curve because the features they aimed to improve were tightly intertwined with other functions and datasets. This complexity made it difficult to understand how a small change might affect the broader system.

To mitigate these challenges, Company E began transitioning to a microservices architecture. Not only were the core services of Company E split into discrete units, but the newly launched Product A was also divided by service domain. Each team could then independently select technologies and build out the necessary infrastructure. This arrangement was expected to enable developers to improve particular services without worrying excessively about dependencies throughout the existing codebase.

**4. Rebuilding the Accounting System via Microservices**

With the shift to microservices, the accounting reporting system underwent a fundamental overhaul. Under the previous monolithic architecture, data was drawn directly from all-encompassing services. Going forward, however, each service would publish the accounting-relevant events (e.g., revenue, payments) through a Pub/Sub mechanism. The accounting system would then receive, aggregate, and process these events. Through this reconfiguration, the accounting reporting function became an independent microservice. While other development teams could innovate freely on their respective services, the new setup ensured the accounting team still received the information necessary for financial reporting and auditing.

**5. Existence and Expansion of Multiple Entities**

Company E's corporate group included multiple overseas subsidiaries beyond Japan. Additionally, new domestic subsidiaries were occasionally formed for launching new businesses, and some companies were integrated through mergers and acquisitions, resulting in a period with approximately five to ten entities under the corporate umbrella. With aspirations to expand to the United States and Europe, the group faced varying regulations and accounting standards across different jurisdictions, each requiring separate reporting and audit procedures.

**6. Accounting Process and the Necessity of Localization**

Although the basic accounting function in both Japan and the United States follows a similar three-step workflow, there are subtle differences that necessitate careful localization:

- **Initial Recognition of Business Events**: When a "business event" occurs (e.g., a sale, expense, or any transaction), it must be translated into the company's financial status. This process involves mapping each event to the appropriate account code and determining how it will appear in the financial statements.
- **Assigning Accounting Classifications**: Next, the accounting team determines whether an event represents revenue, expense, or some other account classification.
- **Final Aggregation and Reporting**: The data are then totaled and used for tax filings, shareholder reports, management decision-making, and payment processing.

While these steps remain consistent across different regions, the final aggregation and reporting stage often requires localization. For instance, U.S.-specific accounting rules may require distinct reports or different audit documentation. This might include demands such as, “Because we process accounting in this way in the U.S., please provide this specific form of report,” or “We need these particular data points for audit purposes.” Accommodating such locally targeted requests frequently compelled the Japanese and U.S. teams to maintain separate codebases and procedures.

7. **Diverging Codebases and Independent Approaches in Japan and the U.S.**

Initially, the same codebase served both Japan and the U.S. However, as the business expanded and microservices adoption proceeded—combined with diverging accounting and audit requirements across countries—the Japanese and U.S. codebases ultimately split. The U.S. side encountered a surge in localization needs, while the Japanese side continued to develop its own independent microservices. Ideally, a common platform that could be effortlessly inherited and integrated would have been preferable, but organizational changes and rapid business growth took place before such a platform could be fully established.

8. **Reusing the Reporting System in the Context of a Japan-U.S. Codebase Split**

Previously, there was a monolithic service that covered both Japanese and U.S. markets under a unified codebase. Japan subsequently began dividing its services into microservices for new business segments, separating the client, server, and infrastructure components. Meanwhile, the U.S. team continued to rely on the original shared codebase branched off from Japan.

In the early days, accounting-related reporting was generated by running SQL queries on the monolithic system, consolidating and processing data to fulfill accounting or audit requirements. After the microservices split in Japan, the older system could no longer be used there; however, the U.S. side continued leveraging this approach. Over time, the U.S. reporting system evolved independently but still fundamentally relied on the legacy Japanese source. Hence, once Japan implemented microservices, the U.S. team continued to expand its inherited monolithic-based reporting solution, resulting in two different trajectories for the two regions.

9. **Complexity of Capitalizing Software Development Costs under Accounting and Tax Regulations**

(Generally speaking) It is permissible, under accounting standards, to capitalize software development expenses. Nevertheless, many IT startups choose to expense these costs immediately rather than capitalize them. While capitalization would spread the cost of development over five years through amortization, resulting in the same total expense eventually recognized, the company must assess whether the software has been impaired or rendered obsolete each fiscal period, which is burdensome. Consequently, many companies opt to expense such costs from the outset to avoid continuous impairment testing.

On the other hand, tax regulations may mandate capitalization under certain conditions, even if a company expenses those costs for accounting purposes. At Company E, for example, when development efforts involved both Japanese and U.S. engineering resources, the tax team would collect data on how many work hours were spent by each country's engineers. Based on that information, they would determine which portion had to be capitalized as U.S. assets and which portion had to be recognized as Japanese assets to fulfill tax filing obligations.

**10. Cultural Differences with Manufacturing and the Treatment of Software Development Costs**

In the manufacturing industry, the future revenue generated by a product is typically easier to estimate, making it more common for development expenditures to be capitalized. If the software's business model resembles traditional manufacturing, capitalization can also be relatively straightforward. However, for IT startups, it is often challenging to rationally estimate how much future revenue a piece of code will generate. In manufacturing, there is a clearer link between production costs and future revenue, facilitating direct matching. In contrast, the high uncertainty in IT ventures discourages capitalization and leads many IT startups to expense software development costs.

**11. Reasons for Not Capitalizing Software and Its Tax Implications**

When a company capitalizes software expenditures for accounting purposes—say 1,000,000 JPY is spent on software development—it cannot then easily choose to capitalize, for instance, only 200,000 JPY. Moreover, if software is capitalized for accounting purposes, it typically must be capitalized for tax purposes as well, which may delay the timing at which such costs can be treated as expenses, potentially raising corporate tax burdens. Given that IT startups often carry substantial upfront investments and face uncertain future revenues, they would rather avoid the ongoing burden of impairment tests and the additional audit scrutiny associated with software capitalization. Consequently, many in the IT sector opt to expense such costs, a practice that appears to be quite common industry wide.

**12. Market Perceptions of Asset Expansion on the Balance Sheet and Examples from U.S. IT Companies**

(Generally speaking) If a company were to capitalize developer salaries, its assets on the Balance Sheet (BS) would balloon. External observers might then perceive these assets as underutilized if they cannot directly observe monetization, which is particularly common in IT startups holding substantial intangible assets on their balance sheets. For Company E, which operates a large-scale software platform where the code itself is a critical competitive advantage, capitalizing such costs would significantly raise the asset side of the balance sheet. However, the company often chose not to do so.

A similar trend is observed among many U.S. IT firms, which commonly do not capitalize software development costs in their financial statements, even though they presumably incur substantial personnel expenses. Under certain accounting standards—specifically tax-effect accounting—companies may opt not to capitalize on their financial statements, yet end up with mandatory capitalization for tax purposes. Many major IT companies do not capitalize internal software development costs, and this tendency is even more pronounced among startups.

**13. Partial Capitalization at Company E and Transfer Pricing**

Company E did, on rare occasions, capitalize specific software features in the past, but it largely avoided capitalization. Particularly in transfer pricing, when engineering hours in Japan and the U.S. are separately tallied, the tax team decides how much of each project should be treated as U.S. cost, U.S. asset, or Japanese asset. Even if a cost is expensed on the accounting side, it may still be capitalized for tax purposes. Therefore, it was indispensable to track engineer hours and then reassign classifications. The heightened scrutiny from international tax authorities regarding transfer pricing also contributed to this practice.

**14. Transfer Pricing Regulations and Standards for Software Asset Capitalization**

In compliance with transfer pricing regulations, Company E aggregated engineer hours in Japan and the U.S., classifying certain functionalities as expenses if they were not expected to generate direct revenue, while core features that showcased the company's unique strengths were sometimes considered capital assets. For instance, login features (which do not directly generate revenue) were typically expensed, whereas software logic enabling user transactions might be capitalized. Bug fixes or minor feature enhancements often lacked sufficient added value to warrant capitalization and were thus treated as expenses.

**15. Methods of Tracking Engineer Hours and Organizational Structure**

Accurately recording each engineer's working hours was a difficult task. While managers would often estimate work hours based on their impressions or use tickets and one-on-one discussions, these did not always yield objective documentation. Company E employed a matrix-like organizational structure, where some engineers might be involved in three or four projects concurrently, further complicating hour tracking. In many cases, managers resorted to non-quantitative approaches—such as observing Slack interactions or code review activity—to gauge the effort invested by each individual.

**16. Repository Structure and Project-Based Management**

Before and after the adoption of microservices, Company E's repository structure evolved significantly. Prior to microservices, the organization used a monolithic architecture from which accounting data were directly extracted through SQL queries for reporting. After transitioning to microservices, repositories were split per service. At times, the organization also maintained a shared repository to manage infrastructure configuration (e.g., Terraform), jointly administered by service teams and a platform team. In many cases, each project had its own repository for core functionality, while shared infrastructure configurations resided in a separate common repository—though the degree of separation or commonality varied case by case.

**17. Friction Arising from Microservices and Team Collaboration**

During the microservices transition, the casual practice of editing adjacent teams' code within a single monolithic repository became more cumbersome. Multiple repositories with clearly defined boundaries made it more challenging to submit code changes to other teams. New services were deliberately created with microservices in mind, whereas existing legacy services required careful discussions on how to slice them into microservices. During architectural design, multiple viewpoints—and "justifications"—vied for adoption, and high turnover among key proponents sometimes led to repeated "resets" in which a plan started over from scratch.

**18. Collaboration across Organizations and Companies and Easing Cultural Barriers**

Although Product A was technically a separate company, many of its staff were former Company E employees, which fostered a degree of familiarity and minimized organizational barriers. While cultural and technical divergences existed between the Japanese and U.S. offices, staff working on Product A had close internal relationships, often sharing code and infrastructure. This facilitated smoother collaboration compared to other parts of the company that had less overlap in personnel or technology stacks.

**19. Differences in Values and Career Aspirations among Japan- and U.S.-Hired Personnel**

Within Company E, differences in perspectives were sometimes observed between employees hired in Japan and those hired in the U.S. Personally, I believed that one's career is formed by venturing into problem areas that seem intellectually stimulating or challenging. With the company's expansion, the diversity of perspectives and working styles naturally increased. Some employees showed limited interest in networking or understanding others' work, preferring to focus strictly on their own areas. Such individuals were less inclined toward collaboration and sometimes demonstrated lower commitment to the company's overarching mission.

**20. Performance Evaluations and Behavioral Indicators Based on Corporate Values**

Company E employed a performance evaluation system aligned with the company's mission and values. Individuals periodically submitted self-evaluations, and managers conducted 360-degree feedback. A cross-departmental "calibration" process was then held among managers to ensure consistent application of evaluation criteria. Within the corporate divisions, managers from different departments similarly convened to harmonize performance standards, mitigating discrepancies in how particular behaviors or achievements were rated.

**21. Awards Program and MVP Selection Aligned with Corporate Values**

The company also had a recognition program in which employees were nominated for quarterly MVP awards aligned with corporate values. Managers recommended potential candidates, and executive leadership made the final decision. This practice incentivized cross-department collaboration and team spirit; employees who isolated themselves or demonstrated low mission alignment found it difficult to receive high evaluations or recognition.

**22. Managing GitHub Organization and Repository Structure at Company E**

Product A's team and Company E used the same GitHub organization rather than maintaining separate ones. As a result, numerous repositories coexisted under a single organizational umbrella. In principle, anyone could submit pull requests to any repository. However, in practice, cross-team pull requests were relatively infrequent, though it was fairly common to leave comments on someone else's repository. Instead of relying on GitHub Issues to request new features or improvements, employees often resorted to casual Slack messages such as "Could you implement this?" One reason for this informal process was the presence of non-engineering staff (e.g., customer service) who did not hold GitHub accounts, thus favoring more immediate communication channels. This approach reflected an InnerSource-like culture, where bottom-up, ad hoc collaboration complemented more formal processes.

**23. Differences among Entities, Access Controls, and Profit-Sharing Issues**

Although Company E and Product A existed as separate legal entities, there were minimal access restrictions on GitHub based on corporate affiliation. Employees and contractors collaborating on actual projects generally received access to relevant repositories. No significant issues concerning profit distribution or entity-based constraints emerged in my experience. Moreover, I was not aware of stringent export controls or sanctions-related checks for particular nationalities. Engineers from various countries worked at Company E, but I never encountered restrictive policies tied to nationality. Typically, joining the GitHub organization only required sharing one's GitHub account and receiving an invitation.

**24. InnerSource Culture and Bottom-Up Improvements**

Company E embraced an "InnerSource" philosophy, even though the term itself was not widely used in official documentation. Inspired by open source principles, the company sought to instill a culture of bottom-up, organic optimization. While formal processes involving Issues and Pull Requests existed in GitHub, employees often found it more efficient to raise requests casually via Slack. Nevertheless, GitHub provided a shared space where engineers could directly inspect and modify each other's code, fostering a collaborative environment. Although having formal guidelines and standardized processes might improve efficiency further, a bottom-up approach had already proven effective in encouraging spontaneous knowledge-sharing and innovation.

**25. Code Access and Information Disclosure Through a Shared Organization**

Because Company E and Product A used a single GitHub organization, most repositories were open to all engineers upon request. Non-engineering employees could also obtain access if they had a legitimate reason. To verify accounting queries, for example, I could view not only the repository for the accounting system itself but also any relevant code from Company E's main services. This open environment allowed motivated individuals to explore other teams' and domains' code, facilitating cross-functional collaboration and learning. From an organizational standpoint, everyone in the shared GitHub organization typically held read access to all repositories; however, practical concerns such as license costs and role-based requirements sometimes led to limiting access for certain non-engineering roles.

**26. Information Disclosure, Documentation, and Standardization of Processes**

Beyond broad code access, the effectiveness of collaboration depended on clear design documents and review structures. By openly sharing design documentation and establishing a review process, it became much easier to locate the information needed to understand a service's internal architecture or its APIs, thus laying the groundwork for spontaneous collaborative efforts. Such InnerSource practices are most beneficial when combined with comprehensive information-sharing, which supports widespread adoption of open source practices within the organization.

**27. Curiosity, Mission Orientation, and Cultural Background**

An open information environment attracts employees who feel strongly about the company mission and exhibit high levels of curiosity. Coupled with performance evaluations and awards that emphasize collaborative behavior, this open culture can foster a virtuous cycle in which employees naturally self-organize, share knowledge, and elevate overall business performance. However, any disturbance to this synergy—for example, if incentives become misaligned—might introduce vulnerabilities. In the initial growth phase at Company E, the open-code and open-information culture contributed significantly to the company's success.

**28. Open Access and Collaboration Mindset**

By managing all repositories under a single GitHub organization, Company E and Product A ensured that, when needed, even non-engineers were permitted to view relevant repositories, code, or documentation. Accounting specialists could review queries or data-processing logic by checking both the dedicated accounting repository and other core service repositories. This high level of transparency provided a platform for enthusiastic and curious employees to explore different service areas, thereby creating a culture of voluntary collaboration rather than one merely driven by cost reduction or duplication avoidance.

**29. Team Topologies and Platform Engineering**

Referring to the concept of "team topologies," Company E recognized that roles such as platform engineering and stream-aligned teams require close collaboration and broad code sharing. The presence of platform teams with multiple dependencies from various service teams reinforced the need for open, organization-wide repositories and a culture geared toward collaboration. This structural arrangement naturally promoted InnerSource-like activities, where teams could share core components and best practices without rigid boundaries.

**30. OSS Program Office (OSPO) and Open Source Activities**

Within Company E, there was an organizational group or initiative resembling an Open Source Program Office (OSPO). Renowned engineers occasionally released their tools openly and shared success stories of open source adoption in internal forums. Although I was not directly involved, by around 2020, these activities were increasingly being formalized and promoted. Embracing open source practices internally facilitated wider knowledge sharing and iterative improvements, thereby strengthening the InnerSource environment.

**31. Information Accessibility and the Proliferation of Tools**

Though accessing code on GitHub was straightforward, documentation and knowledge-sharing tools varied across projects and entities. Company E used a Markdown-based wiki, whereas Product A and the U.S. team often relied on Confluence. Information could also be scattered across Slack channels, Google Docs, or alternative platforms, causing duplication and fragmentation. While none of this was intentionally hidden, locating specific pieces of information could be time-consuming. Initiatives to unify tools—such as introducing a standard wiki—were occasionally undertaken, but new software or external requirement.

## 32. Efforts to Address Information-Sharing Issues and Remaining Challenges

Although there were few strict limitations on accessing information, challenges arose in determining where certain data or documentation was stored and which tool was used to organize it. Slack dialogues, project-specific document tools, outdated wikis, and a continuously shifting information landscape complicated consistent archival. Occasionally, corporate-wide initiatives sought to consolidate information on a single platform, but newly formed teams or subsidiaries often introduced different tools, eroding these standardization efforts. This proliferation of tools, coupled with an open but decentralized environment, expanded the horizons for proactive employees while making passive information retrieval more difficult.

## Company F (Information and Communication) – Translation

| Department | Company Size | Department | Position |
|---|---|---|---|
| Information and Communication | About 500 employees | Engineering department | Engineering Manager / Engineer |

1. **Introduction and Organizational Context**

Person A serves as an Engineering Manager at, which operates in the information and communications sector and has approximately 500 employees. Person B is an engineer affiliated with Organization An under Team A within the same company. Within the corporate structure of, there are three engineering departments, each with roughly 50 members, totaling about 150 engineers. The company emphasizes an organizational culture where engineers are encouraged to collaborate openly across teams.

2. **Emergence of InnerSource Initiatives**

The idea of pursuing InnerSource within first surfaced around October 2022. At that time, publicly available InnerSource resources were referenced to introduce the concept within the company. In 2022, after encountering the term “InnerSource,” the company appointed a key promoter, initiated a series of lightning talks (LTs) to disseminate the idea, and gradually familiarized employees with the concept. By 2023, following a trial period, the company began implementing InnerSource more concretely. As Engineering Manager, Person A designated a single promoter for an initial experiment and assigned Person B to carry it out.

3. **Preliminary Experiment**

In 2023, the company conducted a “simple” proof-of-concept experiment for InnerSource. Person B utilized a personal repository, which contained a simple tool aimed at solving a common issue in online meetings: presenters often failed to notice comments or messages in the chat function of web conferencing tools. This tool featured a single form on a web page that converted entered text into synthesized speech and played it in a machine-generated voice. As a result, the presenter could hear incoming comments in real time, increasing awareness and improving the flow of online presentations.

**4. Company-wide Rollout and Promotion Structure**

Around May 2023, began rolling out InnerSource company-wide, building on the success of the trial experiment. The company comprises three engineering departments (each with around 50 engineers, for a total of approximately 150 engineers). A dedicated InnerSource Promotion Group was formed, initially consisting of Person A and a few core members—three or four individuals in total.

These core members spanned two of the three departments. By coordinating with other Engineering Managers, they integrated InnerSource promotion into business plans, thereby involving relevant stakeholders. Because the Engineering Managers across the departments were closely connected, it was relatively easy to reach a shared understanding of “let’s try InnerSource.” Embedding InnerSource in both departmental plans and company-wide objectives significantly boosted frontline commitment and cultivated a broader sense of “let’s give it a go.”

**5. Contribution Evaluation, Reporting, and Guideline Development**

In or around June 2023, clarified how to count and evaluate contributions. The company has a web application for recording working hours; engineers log what they worked on and for how long. An agreement was reached among Engineering Managers that if someone contributed to another team’s product, the hours would be recorded under that product’s entry. This approach allowed contributions made outside one’s primary department to be recognized in the company’s managerial accounting system and thus be more fairly assessed.

Furthermore, internal guidelines were established to explain the benefits of InnerSource, how to participate, key terminology, and instructions on accessing the company’s InnerSource portal and communication channels. The guidelines also specified the steps for engineers to register their own repositories for InnerSource and how to report contributed hours, ensuring that engineers could engage without confusion. Through these guidelines and the establishment of clear reporting and evaluation mechanisms, the InnerSource initiative continued to gain momentum within the company, even as online meetings proliferated under pandemic conditions.

**6. Behavioral and Attitudinal Changes Among Engineers**

Shortly after InnerSource was introduced, only a limited set of tools and products were open for test contributions. However, other teams gradually started making their own products open to accept contributions. Because Engineering Managers were interconnected horizontally, integrated InnerSource into their plans, and had proper guidelines in place, a shift in atmosphere occurred—from "we could do this if we want" to "we really want to do this."

**7. Clarification of Evaluation Criteria and Impact on Motivation**

Before the evaluation criteria were clarified, engineers who made changes in other teams' codebases wondered if their efforts would be recognized. Once the InnerSource guidelines and the contribution logging rules were formalized, engineers were assured that efforts in other teams' codebases would be evaluated fairly. This clarity replaced mere permission to act ("it's okay to do this") with a sense of confidence that "if I do this, I will be recognized," which further spurred collaboration. As a result, visualizing the number of contributions and establishing a unified evaluation method boosted engineers' motivation.

**8. Collaboration Examples and In-house Efficiency**

Following the example of the voice notification tool, other collaborative cases emerged, such as Team A creating an infrastructure-related script and Team B refining it. Before InnerSource, most tools were locked within specific teams, but once opened, they evolved into more general-purpose solutions. This cross-team collaboration reduced redundant work and improved efficiency by allowing engineers to reuse and enhance existing solutions whenever another team had already tackled a similar problem. This success is attributed to fostering an InnerSource culture and establishing clear evaluation structures.

**9. Future Outlook and Expected Growth**

envisions that more products will become InnerSource projects, fostering a natural acceptance of external contributions from other teams. Another anticipated benefit is that new engineers will learn quickly from open in-house code, leading to smoother onboarding. Overall, the company expects these developments to strengthen its internal engineering culture.

**10. Cost Accounting and Time Management**

uses a web application to record working hours, specifying the product name or function on which each engineer spent time. Under InnerSource, managers reached a consensus that contributions to other teams' products should be fairly recognized through these records. Product or feature units are distinguished by an "order number," and time is generally logged in 30-minute increments.

Although logging work in such detail can be cumbersome for engineers, many do so at the end of the day or even the next day, approximating the time spent. For instance, the "cloud migration" initiative might be assigned a single order number, while all associated tasks are recorded under it. This system, which is run via SaaS, also handles time entries for InnerSource-related work.

11. **Performance Appraisal System and Engineer Action Plans**

employs a semiannual performance review cycle, revising plans every three months. During these review sessions, each engineer aligns with their manager on a plan that may include specific InnerSource contributions. At the time of evaluation, the extent and impact of these contributions are assessed. The InnerSource guidelines serve as a practical manual for engineers and also clarify to Engineering Managers what kinds of activities are encouraged. Since the guidelines have company-wide endorsement, they help form a consensus that “it’s not only allowed but encouraged.”

12. **Communication with Upper Management**

The InnerSource initiative has been communicated up to the department head level in the Systems Division, and the company president (Person F), who has a background in systems, is also generally aware of these efforts. When Person A proposed further steps, no significant objections arose, and the initiative progressed smoothly without major disagreements or debates.

13. **Purpose of InnerSource Promotion and Managerial Perceptions**

places great importance on creating an environment where engineers can genuinely enjoy their work. Encouraging them to contribute freely to other teams’ projects aligns with this cultural emphasis. At the managerial level, the focus is less on competitive advantages or resource efficiency and more on ensuring that engineers have fun and engage in active collaboration. The primary immediate objective is to enhance engineer satisfaction and harness a positive organizational energy through InnerSource.

14. **Engineer-centric Benefits and Sources of Motivation**

Engineers perceive multiple benefits from InnerSource, such as saving time by avoiding “reinventing the wheel” and fulfilling their technical curiosity through cross-team discussions and code sharing. For those who love technology, the opportunity to collaborate across teams is itself a motivating factor. Although faster lead times and higher motivation are welcome outcomes, simply having more opportunities for collaboration is highly valuable in its own right.

15. **Overcoming Organizational Silos and Enhancing Transparency**

Prior to InnerSource, technical discussions tended to remain confined within individual groups. The introduction of InnerSource has created a more outward-looking atmosphere. Many engineers are pleased to hear the term "InnerSource" circulating in areas previously unrelated to their work, and they have noted an overall increase in transparency and openness. Tools and repositories that were once limited to specific teams are now visible across the entire company, effectively lowering intra-organizational barriers.

**16. Repository Structure, Access Management, and Increasing Use Cases**

From the outset, had a single organization-level repository structure that was technically accessible to all engineers, but early on, visibility alone did not lead to actual collaboration. Once the company clarified how to contribute properly—by providing a CONTRIBUTING.md file and having an InnerSource portal automatically crawl and index eligible repositories—active participation began to rise. Initially, only one experimental project existed, but the company now has around 19 InnerSource projects. As these examples have multiplied, engineers have found it easier and more appealing to participate.

**17. Challenges of Applying Production-level Code**

Out of the 19 current InnerSource projects, only about five contain production-level code. Many repositories are geared toward useful tools or convenience scripts. The company aspires to increase production-facing code under InnerSource, but progress has been slow due to the diversity of technical stacks and the limited time engineers can devote to new initiatives. This situation is not unique to; many organizations struggle to dedicate sufficient resources to untested or emerging approaches.

**18. Diversity of Technical Stacks and the Difficulty of Standardization**

Each team independently chooses its own technology stack—such as Python or TypeScript—resulting in a highly heterogeneous environment. While this bottom-up freedom has historically served engineers' preferences, it now complicates unification efforts and adopting standard toolchains. The company has considered introducing more standardized projects and products, but challenges remain, such as configuration complexities. The path toward a fully integrated stack is under discussion but not yet decided.

**19. Time Constraints and Priority Issues**

Time constraints are a major barrier. Many legacy codebases lack even basic test commands, making collaboration and modernization difficult. Host teams are already occupied with production operations, while guest contributors from other teams must also prioritize their core tasks. Hence, engineers find it challenging to allocate time for new endeavors like InnerSource.

**20. External Project Evaluation and Related Surprises**

Currently, engineers are prohibited from working on company repositories outside of regular working hours. If this restriction were relaxed, more engineers might be inclined to contribute as a personal pursuit. However, this scenario is not yet realized.

**21. Relationship with Open Source**

At present, primarily consumes open source software and rarely contributes back. Although some engineers participated in open source projects during their student years, the company culture as a whole remains consumption oriented. A few engineers contribute on their own time, but these efforts are not formally recognized at the corporate level.

**22. Internal Feedback on Individual Learning Achievements**

When engineers use their personal time to acquire new skills and then translate those skills into internal outputs—such as new tools or improved processes—the company does recognize these contributions. However, if these skills are exercised only in external, private projects, they are not clearly acknowledged as contributions to the company. Thus, ensuring that personal learning gets “brought back” into the organization is key to formal recognition.

**23. Overview and Evolution of the Performance Evaluation System**

The current performance appraisal system resembles a Management by Objectives (MBO) model. About a year ago, the company improved the system, expanding the number of evaluation axes from three to five. Each semiannual cycle, employees set and revise goals, which are subsequently assessed based on their outcomes. These five axes include measures such as “innovation” and “being number one,” emphasizing uniqueness and ingenuity. This structure allows for clearer mapping between goals and overall direction, enabling managers to evaluate outcomes more effectively.

**24. Setting Individual and Team Goals, and the Review Process**

Every six months, engineers set specific goals in consultation with their Engineering Managers, and these goals may be adjusted every three months. During the review, managers assess achievements based on demonstrated impacts. For instance, an engineer might set goals related to development, cloud migration, or preventing recurring incidents. The process employs the five evaluation axes, enabling engineers to highlight the distinctiveness or innovative aspects of their achievements.

**25. Linking Compensation to a Point-based System**

The company's point-based assessment system ties directly into compensation. This common framework applies to all job categories, not just engineers. Accumulating points elevates an employee's rank, broadening the scope of their responsibilities and influence. Consequently, InnerSource contributions can become one subset of recognized activities that help an employee progress to higher leadership roles.

**26. Rank and Role Changes**

Person B is aiming for a higher rank (an "L-class" leadership role), which entails responsibilities extending beyond a single team and includes optimizing communication and organizational coordination across the entire engineering division. From this perspective, InnerSource—emphasizing cross-departmental collaboration—can serve as a valuable vehicle for demonstrating the leadership qualities expected at the L-class level.

**27. Award System and Incentives**

presents two key awards on a semiannual basis: the Creativity & Ingenuity Award and the President's Award. The former recognizes individual initiatives, while the latter typically honors projects with significant business impact. The Creativity & Ingenuity Award has three tiers, each accompanied by a monetary prize. There have been instances in which InnerSource promotion or technical improvement activities earned the top-tier prize. In contrast, the President's Award is more competitive, often requiring an endorsement from higher-level management, and is thus more challenging to obtain.

**28. Experiences with Innovation Awards and Reflections**

Person B has received the Creativity & Ingenuity Award on three occasions—more than any other individual in the company. Receiving a top-tier prize entails participating in an interview before the entire workforce and comes with a sense of privilege. Knowing their efforts "made the cut" instills pride and confidence in the awardee, particularly since other submissions do not always succeed. This recognition occurs at the company-wide meeting (the so-called "Information and Communications A Meeting"), held semiannually, where the award recipient is interviewed in front of all employees, significantly boosting overall motivation.

**29. Achieving Goals and Evaluating InnerSource Activities**

All InnerSource-related targets in the previous review cycle were successfully met. The group goals were closely aligned with individual goals, so achievements in InnerSource also contributed to collective success. This alignment between InnerSource outcomes and the company's performance appraisal system has effectively bridged engineers' technical interests with recognized metrics of success, thereby incentivizing participation.

**30. Team-based Incentives and Award Programs**

While individual incentives (e.g., points, Creativity & Ingenuity Awards) are well-established, team-level incentives are less formalized. In certain instances, the President's Award is granted to large-scale projects, sometimes involving 30–40 people across multiple departments. This enables large collaborative endeavors to be recognized, although such recognition is not specifically framed as a "team-only" incentive.

**31. Expansion into Platform Engineering**

Looking ahead, Person A aims to involve the Infrastructure Group—an independent entity—and work with the platform engineering department to Inner-Source additional production-ready systems. A new promoter has joined the effort, assisting in the intersection of InnerSource and platform engineering. With terminology and basic cultural awareness of InnerSource now well established, the focus shifts to generating tangible outcomes in day-to-day operations. By integrating InnerSource more deeply into routine workflows, the goal is to achieve a state where participation becomes "inevitable," thus facilitating natural adoption.

**32. Communication Among Engineering Managers and Consensus Building**

Because Engineering Managers shared similar challenges, they quickly reached a consensus that InnerSource might be a promising approach to addressing inefficiencies. In the initial stages, the rationale was more about excitement and curiosity— "This seems interesting; let's try it"—which helped secure their agreement. Minimal explanations were sufficient, and concrete examples provided by Person B proved especially compelling. Managers who were closely involved embraced the initiative almost immediately, whereas managers in more distant groups, such as infrastructure, needed only a brief presentation. Repeated explanations were rarely required; practical demonstrations and explicit statements of intent delivered greater impact.

**33. Reducing Lead Times and Enabling Self-service**

InnerSource has also reduced lead times in certain processes. For example, the team responsible for maintaining login system configurations (e.g., domain restrictions) had previously handled all modifications themselves. Through InnerSource, users can now create their own pull requests to edit and validate these configuration files, alleviating the maintainers' workload. Tasks that once took days or even a week have become more efficient. Although this change concerns configurations more than source code per se, it has significantly improved operations and motivated overloaded teams to embrace InnerSource.

**34. Limited Deep Contributions and Future Challenges**

As of the time of writing, most contributions revolve around configurations or other relatively lightweight tasks; profound contributions such as substantial feature additions or extensive refactoring of production-level code are still infrequent. The company's efforts to build a common library are ongoing but have not yet reached a mature state. Encouraging deeper, more robust forms of contribution remains an important next step.

**35. Profile of Contributors and Managerial Engagement**

Most current contributors are younger engineers who value the efficiency gains and reassurance of having official backing. Meanwhile, senior engineers and managers appear to hold a generally positive view, believing InnerSource can bring efficiency improvements. However, once InnerSource extends to more business-critical domains, concerns about sharing subordinates' time and resources may surface, potentially leading to greater friction. As of now, these issues have not yet emerged.

**36. Accounting and Contractual Concerns**

To date, no major accounting or contract-related issues have arisen, aside from basic cost management discussions. Since is a single-company entity with around 500 employees, legal complications such as cross-corporate benefit transfers or international transfer pricing do not generally apply.

**37. InnerSource Events and Contribution Initiatives**

In a bid to further promote contributions, Person B plans to assemble about ten potential contributors for a dedicated initiative focusing on small tasks like adding Docker support or fixing documentation. The aim is to lower the barrier to first-time collaboration, helping participants learn review processes in other departments and reducing uncertainty. Events that support such “baby steps” can accelerate the adoption of InnerSource, empowering engineers to venture into new domains and contribute across departments, thus further solidifying the InnerSource culture.

2024年度　青山学院大学　経営学研究科　修士論文

# インナーソース円環モデル:
# 日本企業の経験に基づく
# 組織横断的な開発者協働パターンの体系化

# InnerSource Circumplex Model:
# Mapping Cross-organizational Developer Collaboration Patterns
# with Insights from Japanese Corporate Experience

経営学研究科　経営学専攻
博士前期課程　服部 佑樹

# 要旨

本研究は，ソフトウェア開発の内製化や組織横断的なコード共有の重要性が増す現代の経営環境において，企業内部にオープンソース的開発手法を適用する「インナーソース（InnerSource）」の導入とその進化過程を明らかにすることを目的とする．まず日本企業とグローバル企業の比較分析を通じて，ソフトウェア共有の実態や認識，導入障壁の差異を示した．特に日本企業に見られる内製化の遅れやサイロ化の問題を抽出し，それらがインナーソースの初期導入を阻む要因として機能することを実証的に検証した．

次に，インナーソース導入プロセスを捉える際に，従来のプログラムやプロジェクトの進化に着目した単純な段階的化モデルだけでは捉えきれない「多層的・トポロジカルな進化」が存在する点を指摘し，本研究の主要な貢献として三つの新たな理論的枠組みを提示した．一つ目は，組織間連携構造を複数の形で概念化する「インナーソーストポロジー」であり，アクセス権や契約形態の違いなどを整理することで，企業内部の協働度合いを定義する．二つ目は，「重層的インセンティブモデル」である．個人やプロジェクト単位での金銭・非金銭的報酬を組み合わせ，多様なモチベーションを喚起する統合的な仕組みを設計する重要性を示した．三つ目は，「インナーソース円環モデル」である．このモデルによって，組織が直面する課題や目的に応じて多様なインナーソース形態を定義し，インナーソースの進化を単なる成熟度の直線的推移ではなく「円環状」にマッピングすることで，推進者が各実装段階で必要なフォーカスを適切に変化させられることを示した．これらの枠組みを通して，本研究では従来曖昧に捉えられてきたインナーソースの概念を，共有の範囲とコミュニティの成長という観点からより精緻化する．

これらの成果は，インナーソース導入の成功にはトップダウン型のプログラム整備と現場発の自発的協働が並行して進むことが不可欠であることを改めて示唆し，企業における持続的なイノベーション創出とソフトウェア共有文化の醸成に寄与する．特に円環モデルを含む複数の新たなフレームワークは，多様な文化的背景や組織規模を持つ企業に対して，インナーソースを柔軟に再定義・導入するための指針を提供し，今後の企業内ソフトウェア共有とイノベーション創出の一助となることが期待される．

# 目次

# 序論

## 第1章はじめに

現代のグローバルなビジネス環境において，ソフトウェア開発におけるコード共有や再利用は，開発プロセスの効率化，品質向上，イノベーション創出に資する重要な戦略的要素として注目を集めている．とりわけオープンソースの活用や，組織内における「インナーソース（InnerSource）」と称される内部的なオープンソース化手法は，国際的な企業を中心に広く実践されている．インナーソースとは，オープンソース開発手法を企業内部に適用することで，透明性・再利用性・共同性の高いソフトウェア開発プロセスを構築し，組織学習とイノベーション創出を支援するアプローチである．Gartner 社の分析によれば，インナーソースの導入により，組織はサイロ化を防ぎ，より強固で緊密なソフトウェア開発ライフサイクルを実現できるとされている [1]．実際に，2023 年の Gartner による「Hype Cycle for Software Engineering」においても重要なトレンドとして取り上げられており [2]，産業界における注目度の高さを示している．

ソフトウェア開発における知識共有とコードの再利用は，生産性向上とイノベーション創出の重要な推進力として国際的に認識されている．しかしながら，組織特性や文化的背景の違いにより，その推進度合いや実装方法は地域や企業グループによって顕著な差異が観察される．

本研究では，組織内でオープンソース的な開発手法を導入する「インナーソース」の実装プロセスに焦点を当て，異なる組織特性や文化的背景がソフトウェア共有行動に与える影響を実証的に検証する．特に，日本企業群をケーススタディとして取り上げ，グローバル企業との比較分析を通じて，組織文化，意思決定プロセス，雇用慣行などの制度的要因がコード共有行動に及ぼす影響メカニズムの解明を試みる．

質的調査手法を主軸とした実証研究を通じて，組織特性や文化的背景の違いがもたらすコード共有行動への影響を体系的に分析し，その促進・阻害要因を特定する．この分析結果に基づき，多様な組織特性や文化的背景を持つ企業が，それぞれの固有性を活かしながら効果的なコード共有文化を醸成するための理論的フレームワークの構築を目指す．

本研究の学術的意義は，組織行動論，技術経営論，ソフトウェア工学の学際的領域において，企業特有の組織的・文化的文脈を実証的に分析することで，既存の理論的枠組みを拡張・精緻化する点にある．さらに，実務的意義としては，本研究で提案するフレームワークを通じて，様々な組織特性を持つ企業がインナーソースを効果的に導入・実践するための具体的知見を提供することが期待される．

### 1.1　研究の背景 － ソフトウェアにかかる現代の経営環境

近年，デジタルトランスフォーメーション（DX）の進展により，製造業を含む多様な産業分野でソフトウェアが競争力の源泉として位置づけられている．この背景には，顧客ニーズの高速な変化への対応や，新規ビジネスモデルの創出，業務プロセスの最適化が，ソフトウェアを通じて可能になるとの認識が広がっていることが挙げられる．しかし，日本企業においては，ソフトウェア開発の内製化が依然として十分に進んでいないと指摘されている．独立行政法人情報処理推進機構（IPA）の調査によれば，日本の IT 人材

の 73.6%が IT 企業に所属しており，一方で米国では IT 企業への所属は 35.1%に留まっている [3]. この数値は，日本のユーザー企業における自社内でのエンジニアリングチーム形成が限定的であることを示唆している.

こうした状況において，大企業を中心に内製化の動きが進んでおり，同 IPA の調査によれば従業員 1,001 人以上の企業では 40.4%が「内製化を進めている」と回答している [4]. これらの企業では，アジャイル開発手法や継続的デリバリーなど，海外で広く採用されている迅速かつ柔軟な開発アプローチの導入の試みがなされることもあるが，製造業などの伝統的企業群では多様な制約要因によってこうした新たな開発手法の導入が困難であり，従来型のウォーターフォール開発手法が根強く残る状況にある.

さらに，企業内で育成したソフトウェア技術やコンポーネントの組織横断的な共有は，重複開発の回避や開発期間の短縮を通じ，製品競争力強化につながると期待される. しかし，現実には「サイロ化」と呼ばれる組織内分断が深刻な課題となることがある. 部門間連携が不十分で，情報の非対称性が存在する場合，同一機能を複数部署で独立に再開発する「車輪の再発明」が発生し，コスト増大や開発効率低下を招く. こうした弊害は大規模企業において特に顕著であり，社内の透明性や共創を阻害するリスクとなる.

このような組織的課題に対して，近年のソフトウェア産業では，組織文化の改革と開発者の働き方に関する様々なアプローチが提案されている. その代表的な例として，「開発者体験（Developer Experience; DevEx）」の概念が注目を集めている. 開発者体験は，ソフトウェアエンジニアが組織内で経験する環境・ツール・プロセス・文化の質的側面を総合的に捉える概念である. SPACE フレームワークは，開発者の生産性・満足度・効果を測定する統合的な枠組みを示しており，優れた開発者体験が，生産性および革新性の向上，さらには有能なエンジニアの獲得・定着に重要な要素であることを指摘している [5].

こうした背景を踏まえると，組織文化や構造そのものを見直し，柔軟性や透明性，さらには内製化能力に富んだプロダクト開発体制を構築することが，開発者体験の最適化と人材獲得・エンゲージメント向上の双方に寄与すると考えられる. このとき，オープンソースの原則や手法を組織内部に取り込むインナーソースの実践は，コードや知識の共有基盤を整え，チームや部門の垣根を越えてコラボレーションを促進するうえで，企業にとっても極めて有効である可能性がある.

インナーソースの導入は，製造業や金融業といった伝統的なセクターを含む，世界の様々な産業において着実に進展している. これは，組織の効率性向上や革新的な開発手法の確立を目指す企業のニーズと，インナーソースが提供する柔軟な協働モデルとの親和性の高さを示している. このような動きは，グローバルなソフトウェア開発エコシステムの形成を促進している. 日本企業においても，インナーソースの導入による恩恵を享受できる可能性があり，さらにその実践事例が他の地域における取り組みにも有益な示唆を提供することが期待される.

このような状況において，インナーソースは単なる技術的なフレームワークを超えて，組織変革の触媒として機能する可能性を持っている. 特に，日本企業が直面しているような内製化の遅れや組織のサイロ化といった構造的課題に対して，インナーソースは実践的な解決策を提供する. 開発プロセスの透明性向上，知識共有の促進，そして組織横断的な協働を通じて，持続可能なイノベーション創出基盤の構築に貢献することが期待される.

## 1.2 インナーソースの概念的枠組み

本研究において中核的な分析対象となるインナーソースとは，企業内部のソフトウェア開発プロセスにオープンソース開発モデル（協働的，透明性の高い分散型開発手法）の原則・手法を取り込む取り組みである．本来オープンソースは，地理的・組織的境界を超えた自発的コミュニティによる革新的なソフトウェア開発を特徴としており，その成果物は世界中で再利用可能な知的資産として蓄積されている．一方，インナーソースは，その理念を企業内部に適用し，社内の部門・組織単位間に存在する「サイロ」を解消し，コードや知識の共有および共同開発を促進することで，開発効率，品質，そして内製化能力の強化を図る戦略的アプローチである．

この概念は 2000 年に Tim O'Reilly によって"inner sourcing"という用語で初めて言及され，企業内でオープンソース開発手法を活用する取り組みとして紹介された [6]．その後，The InnerSource Commons Foundation をはじめとする団体や先進的な企業事例を通じて発展してきた．現時点では，インナーソースに対する厳密な学術定義は定まっていないものの，The InnerSource Commons Foundation は，以下の四つの原則を中核的な指針として示している [7]．

- **Openness（開放性）**：社内プロジェクトが社内コミュニティ全体にとって「発見可能」であり，十分なドキュメンテーションが整備され，特定部門に依存しないアクセスが容易であること．これにより，開発者は自発的に貢献し，ナレッジを相互補完することが可能となる．
- **Transparency（透明性）**：プロジェクトチームは，プロジェクトのビジョン，未解決の機能要件，進捗状況，意思決定プロセスを明確に開示し，他部門からの貢献者が能動的かつ有意義に参画できるようにする．
- **Prioritized Mentorship（指導と支援の重視）**：ホストチームは，ゲスト貢献者がプロジェクトのコードベースや設計原則を十分に理解し，円滑に変更を加えられるよう，適切な教育・メンタリングを提供する．これにより，インナーソースプロジェクトへの参加は，個々の開発者の成長やスキル強化にもつながる．
- **Voluntary Code Contribution（自発的なコード貢献）**：インナーソースへの参加は強制されず，開発者やチームは自律的に貢献する．これにより，自発的な協働が生まれ，長期的なコミュニティ形成が可能となる．

これらの原則が示唆するように，インナーソースは単なるコード共有にとどまらず，社内ナレッジフローの活性化，組織文化改革，開発者体験の改善を通して，企業内部のイノベーション基盤を強化する包括的な取り組みと位置づけられる．

インナーソースを導入する意義は様々であるが，主な焦点は企業が直面する競合環境や技術革新の加速に対応し，内製化能力の強化と開発効率の向上を同時に実現する点にある．代表的な利点として，(1)冗長なコードの重複開発を削減し，ソフトウェア資産の再利用性を高めることによる開発コスト削減，(2)部門間コラボレーション促進による知的資本の有効活用，(3)高スキル人材を惹きつけ・維持しやすい透明性と柔軟性に富んだ職場環境の創出，などが挙げられる．

これらの成果は，企業の持続的競争力強化に直結し，投資対効果の観点からも有望である．たとえば，共有基盤を通じた開発サイクルの短縮は，新製品・サービスの市場投入までのリードタイムを削減し，結果的に収益機会を増大させる．さらに，開発者コミュニティを社内に形成することで，知見や技術的卓越性が社内に蓄積され，外部委託依存からの脱却や人材確保においても有利に働くと考えられる．これらは，単なるコスト削減にとどまらない，戦略的な資本投下と位置づけることができる．

## 1.3 研究の目的

本研究の主たる目的は，インナーソース導入プロセスにおける実践的フレームワークの構築である．このフレームワークは，組織固有の文化的特性を考慮しつつ，効果的なソフトウェア資産の共有・再利用を実現するための段階的アプローチを提示する．

インナーソースの概念的定義の明確化も，本研究における重要な目的である．現状では，インナーソースの実践形態は，完全なオープン化から限定的な協働領域の設定まで多岐にわたり，その定義や範囲に関する統一的な見解が確立されていない．本研究では，組織の成熟度や文化的背景に応じた段階的な導入モデルを提示することで，この概念的曖昧さの解消を試みる．

最終的に，本研究はインナーソース導入に関する普遍的な知見の創出を目指す．日本企業の事例分析を出発点としながらも，得られた知見をグローバルなインナーソース・コミュニティに還元し，組織のデジタルトランスフォーメーションを促進する実践的示唆を提供することを究極の目的とする．

## 1.4 研究の意義

本研究は，インナーソース導入プロセスにおける実践的フレームワークの構築を通じて，以下の三つの主要な学術的・実務的意義を有している．

本研究の第一の意義は，新興市場や新規産業領域におけるインナーソース導入のための体系的なフレームワークを提供することにある．初期導入期の日本企業群と成熟期のグローバル企業群との比較分析を通じて，組織の成長段階に応じたインナーソース導入の状態遷移を明確化し，各段階における具体的な実装方法を提示する．これにより，インナーソースの導入を検討する組織に対して，その成熟度や組織特性に適合した段階的な導入アプローチを提供することが可能となる．

第二の意義は，インナーソース導入における組織の発展段階と状態遷移の実態を解明することにある．既存研究では，インナーソースの理想的な最終状態や段階的な成熟度は示されているものの，各段階における組織の具体的な状態や，それらの間の遷移プロセスについては十分な解明がなされていない．本研究は，詳細な事例分析を通じて，これらの状態とプロセスを明確化し，インナーソース導入の実態に即した理論的基盤を構築する．

第三の意義は，既存のインナーソースの成熟度を定義するインナーソースマチュリティモデルを補完し，その適用範囲を拡張することにある．Inner Source Capability Maturity Model [8]や InnerSource Patterns の Maturity Model [9]に代表される既存モデルは，段階的アプローチの基本的枠組みを提供している．しかし，実証分析を通じて，より詳細な観察と理論化が必要な領域が明らかになった．具体的には，インナーソースと通常の部門間協働の境界における組織的振る舞いや，戦略的意思決定以前のチャンピオ

ン[1]による非公式な活動などである．これらの「中間的状態」や「移行プロセス」を理論的に位置づけることで，インナーソースの実践的理解を深化させる．加えて，組織が追求するインナーソースの形態が単一ではなく，複数の異なる状態を並行して発展させる可能性を示すことで，より柔軟な戦略的選択肢を提示する．

[1] 「チャンピオン」とは活動の「旗振り役」を指し，主に現場で主体的に該当プロジェクト・プログラムを先導するリーダーを指す

# 本論

## 第2章 日本企業とグローバル企業におけるソフトウェア共有実態の比較

本章では，日本企業とグローバル企業におけるソフトウェア共有の実態を比較分析し，その差異を明らかにすることを目的とする．この比較分析を通じて，日本企業が直面する課題と，グローバル企業が実践している先進的な取り組みを浮き彫りにすることが可能となる．

本研究の中核概念である「インナーソース」に関する世界的な研究動向に着目すると，企業内でオープンソースの手法を適用するソフトウェア開発手法として，グローバルな文脈において多角的な研究が進展している．具体的には，移転価格税制に関する研究 [10]，インナーソースの成熟度を評価するマチュリティーモデル [8]に関する研究，実践の実態に関する研究，およびインナーソースがもたらす価値に関する研究 [11]など，多岐にわたる領域で学術的探究が行われている．

一方，日本におけるインナーソースの研究実態は極めて限定的であり，実践例も乏しい状況にある．しかしながら，「インナーソース」という用語を用いずとも，ソースコードの共有や再利用という観点では，日本においても一定の研究蓄積が存在することは言及に値する．特筆すべきは，日本の製造業におけるソフトウェアパーツの再利用に関する研究である．例えば，ソフトウェアプロダクトラインの適用 [12]と再利用部品整備が製造コスト削減や工程の効率化を通じて，ソフトウェア開発の生産性向上に寄与することが報告されている [13, 14]．これらの研究は主に製造コストの削減や工程の効率化を目的としており，ソフトウェア開発における生産性向上に寄与してきた．

しかしながら，日本企業におけるこれらの取り組みは，主としてコスト削減と製造工程の短縮化に焦点を当てており，共創や競争優位性の獲得，開発者体験の向上，さらには人材育成といった，より包括的な視点を欠いている傾向がある．このことは，グローバル企業が推進するインナーソースの概念と，日本企業が従来から実践してきたソフトウェア再利用の取り組みとの間に，本質的な差異が存在することを示唆している．

とりわけ，日本企業におけるソフトウェア共有・再利用の取り組みは，往々にして部門内または特定のプロジェクト内に限定されており，組織横断的な知識共有や協働の促進には至っていない．この状況は，日本企業の組織構造や文化的特性，さらには社内ルールや会計慣習などが，グローバルスタンダードとは異なる形で発展してきたことに起因すると考えられる．

一方，インナーソースの実践は単なるコード共有や再利用の枠を超え，組織文化の変革や知識の共有，イノベーションの促進，さらには開発者体験向上の手段 [15]として位置づけられている．インナーソースを採用した企業では，インナーソースを通じて部門間の壁を取り払い，社内の知的資産を最大限に活用することで，競争力の向上を図っている．

本章では，まず世界的なソフトウェア共有の動向と日本企業のソフトウェア共有の現状について，それぞれの背景や特徴を整理する．これにより，以降の章で展開される日本企業とグローバル企業の具体的な比較分析の基盤を形成する．特に，両者の組織構造や文化的特性，社内ルールや会計慣習の違いに着目し，その歴史的背景や発展過程を踏まえながら，インナーソースの実践における差異を生み出す要因を体系的に把握することを目指す．

## 2.1 世界的なソフトウェア共有の潮流

本節では，インナーソースの国際的な実践状況について考察する．以下の小節では，まず国際的な企業による実践事例とその示唆，続いて学術研究における動向および理論的・分析的な知見，さらに非営利団体 The InnerSource Commons Foundation による国際調査やその結果に基づく地域差や普及状況，最後に 2024 年時点での世界的動向の特徴的な側面について，順を追って詳述する．

## 2.2 業界における先行実践事例

インナーソースはソフトウェア開発をよりオープンで協働的なプロセスへ転換する一手法として，多様な業種・規模の企業で注目されている．特に SAP [16], Microsoft [17], IBM [18]といったテクノロジー分野の先進企業では，早期からインナーソースの概念を社内に取り込み，内部でのソースコードの再利用や，部門・拠点を越えた開発者間コラボレーションを促進している．これら企業では，明確なガイドラインやポータルサイトを用いて社内プロジェクトを可視化し，誰でもアクセス可能なコードリポジトリを通じて改良案やパッチを受け付けることで，組織内オープンソース的な環境が形成されている．

これらの企業はインナーソースを通じ，既存コード資産の再利用性向上だけでなく，組織内ナレッジマネジメントの高度化や，個々の開発者が社内外で培った知見を迅速に組織全体へ展開する仕組みを醸成している．結果として，内部コミュニティによるコードレビューやピアサポートが行われ，品質向上や開発リスクの低減が可能となる．また，リソースが不足する部門でも他部門の成果物を再利用することで開発コスト削減や納期短縮が実現でき，これが企業全体の競争力強化につながっている．

さらに，The Innersource Commons Foundation によるサミットワークショップにおいては，欧米を中心に多様な企業事例が共有されている [19, 20]や．そこでは，金融や製造，通信，ヘルスケアなど，テクノロジー企業以外の異業種であってもインナーソースが適用されているケースが報告されている．たとえば，Robert Bosch では，製品開発部門とサービス部門が共通のコードベースを用いることで，現場のフィードバックを迅速に反映した改善サイクルを回し，結果的に社内コラボレーションの効率化や製品品質の向上につなげている [21]．また，金融セクターでは，厳しいセキュリティ要件やコンプライアンス規制に対応しながら，内部でのコード共有とレビュー体制を整えることで，リスク軽減と開発効率改善の両立を図っている [22]．

一方で，インナーソースの定着には課題も認められる．文化的・組織的変革を伴うため，初期段階では管理職層や一部の開発者から抵抗を受ける可能性がある．また，社内貢献を可視化・評価する仕組み，ライセンスや知的財産に関するポリシー整備，品質保証プロセスの調整など，多面的な取り組みが求められる [23]．こうした課題に対して，The InnerSource Commons Foundation や，関連するコミュニティは，ガイドラインである InnerSource Learning Path [24]やケーススタディをパターンとしてまとめた InnerSource Patterns [25]を提供することで，企業のインナーソース実践を支援している．

総じて，産業界でのインナーソース先行事例は，従来の部門別・縦割りの開発アプローチから，よりオープンかつ協働的な開発文化へと移行する一助となっている．今後，世界各地の企業がこれら先行事例から学び，自社の文脈に適合した形でインナーソースを導入・拡大することが，企業革新の新たな指針として期待される．

## 2.3　学術的視点からの研究動向

学術的な文脈において，インナーソースは近年，ソフトウェア工学，知識マネジメント，情報システム研究など，複数の学際領域で関心を集めている．従来，オープンソースソフトウェアに関する研究は，その分散開発モデル，コミュニティガバナンス，品質保証プロセス，イノベーション創出メカニズムなどについて多くの知見が蓄積されてきた．一方，インナーソースはオープンソースの原則を組織内に導入することで，内部組織におけるリソース共有やイノベーションの創出過程にどのような影響を与えるかという新たな研究課題を提示している．

InnerSource Capability Maturity Model(IS-CMM) [8]は，インナーソースの成熟度を評価し，組織がどの段階で，どのような条件下でインナーソースを有効に導入・拡大できるかを示唆している．このモデルは，初期段階では限定的なコード共有から始まり，やがて社内コミュニティの形成，ガバナンスモデルの確立，品質管理プロセスの標準化，トップマネジメントによる支援などが進む中で，インナーソースが段階的に定着・発展するプロセスを明確にしている．

また，インナーソースには組織における生産性向上，コスト削減，知識共有の促進などの多面的な利点が存在し [26]，その価値を定量的手法によって分析する手法も提示されている [27]．これらの成果は，インナーソース導入が単なる形式的な手法導入に留まらずビジネス成果に大きく寄与しうることを実証的に示すものである．

プラットフォームベースの製品開発におけるインナーソースの導入効果に関する分析 [28]では，製品ユニットとプラットフォーム組織の構造的な分離が原因となる遅延，欠陥率の増加，冗長なソフトウェアコンポーネントの発生といった問題に着目している．また，インナーソースの導入により，知識共有の促進やコラボレーションの強化を通じて，企業内のソフトウェア開発プロセスが改善される可能性を示している．

さらに，大規模開発組織において，インナーソースと DevOps の実践が生産性向上や品質改善，アーキテクチャ上の課題解決に寄与する可能性が検討されている．Ericsson を対象にしたケーススタディでは，再利用可能な資産の開発が初期コストを伴うものの，長期的には品質や生産性，顧客体験の向上といったメリットをもたらすことが示されている [29]．

複雑化したソフトウェアシステムは，相互依存関係の増大や情報サイロ化を通じて技術的負債や保守コスト上昇を招く．この点に関して，インナーソースは社内リポジトリの可視化とオープンなコードレビュー慣行を通じ，開発者コミュニティによる透明かつ迅速な課題抽出・解決を可能にすることが指摘されている．

一方，インナーソース導入時には，個別チームやプロダクトレベルでの摩擦（例：既存開発プロセスとの衝突，リソース確保の困難さ，教育不足）に加え，組織全体としての戦略的導入・普及を支えるマクロ的要素も無視できない．インナーソースのプロジェクト管理における課題と成功要因について詳細な議論がなされており [30]，近年では「InnerSource Program Office（ISPO）」のような専任部門やガバナンス組織の設置が注目されている．このプログラムオフィスは，インナーソースの基本原則・ガイドラインを策定し，社内のトレーニングや教育プログラムを整備するとともに，各チーム間の利害調整や共通ツールの選定，インセンティブ設計など，包括的なサポートを提供する役割を担う [31]．

ISPO をはじめとする組織的イニシアティブは，経営陣からのトップダウンの支援や，コミュニティビルディング，評価基準（KPI）設計，ライセンス・知的財産への適切なルール設定など，インナーソース文化を全社的に醸成するための枠組みを整える．これにより，従来は個別プロダクトチームや技術者コミュニティに

閉じていた成功・失敗事例を組織横断で共有し，標準化されたプラクティスによってバリアを徐々に低減することが可能となる．結果として，インナーソースの効果が一部チームに限定されず，組織全体のソフトウェア資産管理・運用効率およびイノベーション能力向上へと波及する道筋が拓かれる．

会計および管理領域におけるプロセス変化に焦点を当てた分析では，企業内における新たなコミュニティ構築プロセスの必要性が指摘されている．また，インナーソースの発展を測定するための計算ツールと技術を検討した結果，既存のツールがインナーソースプロセスを十分に管理するには不適切であることが示されている [11]．

学術研究の将来的な課題としては，(1) インナーソースによる組織成果の定量評価手法の確立，(2) 不確実性が高いビジネス環境下での導入戦略やガバナンスモデルの最適化，(3) 多国籍・多文化な大規模組織における制度的バリアや適応要件の明確化，(4) 人材育成やキャリア形成への長期的影響評価などが挙げられる．すなわち，インナーソースを組織的に確立・拡大するためには，ミクロ（チーム／プロダクト）レベルの手法改善と同時に，マクロ（組織／ガバナンス）レベルでの制度・文化整備，指揮・調整機能の確立，組織学習メカニズムの強化が不可欠である．

総じて，学術的視点においてインナーソースは多面的な分析対象であり，テクニカルな側面からマクロ的な組織運営上の課題まで，幅広い領域で研究が深化している．今後は，こうしたマクロ的視点の研究が進むことで，インナーソースが組織全体の生産性や競争力強化にどのように資するのか，より包括的な知見が蓄積されていくことが期待される．

## 2.4　日本企業におけるソフトウェア共有の現況と課題

日本企業におけるソフトウェア共有の現状は，グローバルな動向と比較して，その取り組みが限定的であり，また異なる様相を呈している．前述した通り，日本企業では，「インナーソース」という用語を明示的に用いた取り組みは稀少であるが，ソフトウェアの再利用や共有に関する取り組みは，特に製造業を中心に一定の蓄積が存在する．製造業領域，特に自動車，家電，産業機器などの分野では，組み込みソフトウェアが不可欠なプロダクトが多く，企業内部には膨大なコード資産が蓄積されている．日本企業は従来，ハードウェア主導の製造業を成長エンジンとして発展し，これらの資産は厳格な品質保証プロセスや標準化された手続きの下で運用され，特定の部門内やサプライチェーン上のパートナー企業間で再利用されるケースが少なくない．この点で，日本企業は既に独自の「内なる」ソフトウェア共有基盤を構築しているともいえるが，この共有形態は往々にして部門横断的な透明性やオープンなコントリビューションモデルに乏しく，インナーソースが志向する「オープンソースコミュニティ原則の社内適用」とは様相を異にしている．特に品質管理やセキュリティ要件が厳格に設定される場合，コードの自由な閲覧や編集は制約を受けやすく，結果としてインナーソース的なコラボレーション文化が芽生える余地が限られる．

この背景には，いくつかの要因が存在する．第一に，日本企業の多くは，組織内での情報共有や意思決定プロセスが，欧米企業に比べて垂直的かつ階層的な傾向にある [32]．稟議制度や年功序列的な人事制度は，上下関係や部門境界を強固にし，横断的な知識共有や自発的なコントリビューションを促すインナーソース文化の醸成を阻害しうる．また，リスク回避的な性質が強い企業群では，新規の開発手法や組織変革への取り組みに慎重であることから，オープンコラボレーションに基づくインナーソース導入への心理的・制度的バリアが存在することが考えられる．

第二に，日本におけるソフトウェア開発は，長らく下請け構造を通じた垂直分業体制に依存してきた[33]．多くの場合，元請企業は要件定義や上流工程を担い，下請・孫請企業が実装・テストといった具体的作業を担うピラミッド構造が形成されている．このような外部委託前提の開発文化では，コードや知識は企業間境界で分断され，オープンな社内エコシステムの形成が難しい．一方，欧米企業が内部に多様な専門家コミュニティを抱え，組織内 OSS 的な文化を育んでいるのに対し，日本企業はサプライヤーネットワークに依存し，社内に統合的な「コミュニティ」を構築しにくい構造的課題が見受けられる．

第三に，ソフトウェア開発手法の革新に対する受容度の問題がある．アジャイル開発や DevOps などの開発プラクティスが世界的に普及する中，日本企業は依然としてウォーターフォール型開発や従来型プロセスへの依存が強いケースがある [34]．このような環境では，オープンなコントリビューションや改善サイクルを前提とするインナーソース手法が馴染みにくい可能性がある．

会計的側面においても，日本企業のソフトウェア共有には構造的な課題が存在する．特に製造業では，ソフトウェアは最終製品に組み込まれる無形資産として扱われ，その価値評価や原価管理は従来型のハードウェア中心の会計基準に依拠している．このような会計慣行は，部門横断的なソフトウェア共有や継続的な改善活動の価値を適切に反映できない可能性がある．日本の会計基準はソフトウェアの再利用特性を適切に反映しておらず，再利用特有の作業や費用の二重計上の回避についての考慮を欠いている．また，ソフトウェアのカスタマイズの程度によるケース分けが現実の開発状況を反映していない可能性があり，販売用ソフトウェアの再利用ケース設定が欠落していることが重大な問題として示されている[35]．

現代のソフトウェア開発では，納品後も継続的な改善や機能追加が行われる特性を持つため，従来の固定的な資産計上モデルとの不整合が生じている．特にサブスクリプションモデルやクラウドサービスなど，新しいビジネスモデルに対応した会計基準の確立は途上段階にある [36]．このような会計上の制約は，組織内でのオープンなソフトウェア共有や協働を促進する上での障壁となりうる．

しかし，日本企業においても，デジタルトランスフォーメーションの加速やグローバル競争環境の変化を受け，ソフトウェア開発のスピード，品質，柔軟性向上が喫緊の課題となりつつある．自動車産業では，コネクテッドカーや自動運転技術，スマートファクトリーなど新分野の登場が，ソフトウェアを中核とする開発力向上を要求している．また，金融や流通，小売，ヘルスケアなど非製造業でも IT 化が加速するなか，企業内部で共有可能なソフトウェア資産や，効率的なコントリビューションプロセスへの期待が高まりつつある．

さらに，日本でよく引用される知識創造理論における SECI モデル（Socialization, Externalization, Combination, Internalization）[37]は，暗黙知と形式知の往還を通じた組織的学習を重視する点で，インナーソースが指向する社内コミュニティ醸成と潜在的な親和性がある．実務的には，組織内部でのコラボレーション基盤整備，分散型バージョン管理ツール（例：GitHub Enterprise, GitLab）の導入，ドキュメンテーション文化の確立など，インナーソース的実践を後押しするための技術的・組織的措置が増加している．GitHub のような分散型バージョン管理ツールは日本でも主流化しつつあり．GitHub の報告によると 2024 年 11 月時点で日本の GitHub 利用者数は 350 万人を突破し，前年比で 23%の成長を遂げている[38]．これらの取り組みは，従来のトップダウン型ガバナンスモデルを補完し，エンジニアが自律的かつ横断的にコードベースへ貢献する「内なる OSS コミュニティ」形成への契機となりうる．

また，近年の国内政策動向やオープンイノベーション支援策，さらには国際的な標準化団体への日本企業の参加拡大なども，インナーソース的な手法の普及を後押しする下地となる可能性がある．例えば，Linux Foundation の OpenChain プロジェクトへの参加や，国内 OSS コミュニティとの連携強化などを通じて，日本企業はグローバルなオープンソースエコシステムとの接点を増やしつつある．トヨタの事例 [39] や日立製作所の事例 [40]など，主導的事例が存在し，ISO/IEC5230:2020(OpenChain 2.1)認証の取得に関しても複数の事例が観測された．こうした取り組みを通じて，日本企業が世界標準のソフトウェア開発手法や知識共有文化を取り込み，インナーソースを自社の戦略に組み込むことは十分に可能であろう．

総じて，日本企業におけるソフトウェア共有は，グローバルなインナーソース文化の成熟度と比較すれば発展途上であり，独自の組織文化・産業構造がその普及を阻んでいるといえる．しかしながら，DX 推進や国際競争環境，オープンソースコミュニティからの影響を受けつつ，インナーソース的手法を取り込み得る素地も存在する．

改めて，日本企業におけるインナーソース導入は，単なる技術的変革にとどまらず，組織文化・ガバナンス・人材育成手法の再構築を伴う包括的な変革プロセスとなる可能性が高く，その成否は日本企業がグローバルなソフトウェア開発エコシステムに円滑に参画し，持続的な競争力を維持する上での重要な鍵となると考えられる．

## 2.5 インナーソース導入国際比較の総括

本章では，インナーソースの世界的な動向と日本企業における現状について，産業界の実践事例，学術的研究，および日本固有の文脈から考察した．グローバルな視点では，テクノロジー企業を中心に，インナーソースが組織内のソフトウェア開発文化を変革し，イノベーション創出や効率性向上に寄与する手法として定着しつつある．特に，The InnerSource Commons Foundation を中心とした国際的なコミュニティの形成は，ベストプラクティスの共有や導入支援の基盤として機能している．

一方，学術研究の観点からは，インナーソースが組織学習，知識マネジメント，コミュニティ形成など，多面的な分析対象として扱われていることがわかった．特に，マチュリティモデルの確立や定量的効果測定，組織変革メカニズムの解明など，理論的・実証的な知見が蓄積されつつある．

日本企業に目を向けると，製造業を中心とした独自のソフトウェア共有文化が存在する一方で，グローバルなインナーソース潮流との間に顕著な乖離が認められる．この背景には，階層的な組織構造，垂直統合型の開発体制，リスク回避的な企業文化など，日本固有の要因が存在すると考えられる．しかしながら，デジタルトランスフォーメーションの進展や国際競争環境の変化は，日本企業にもインナーソース導入の機会を提供している．

これらの知見は，次章で詳述する定量的調査の前提となるものであり，インナーソースの国際比較分析において重要な示唆を与えている．

# 第3章グローバルと日本におけるインナーソースの潮流

インナーソースの国際的な動向と日本企業の現状を比較分析することは，グローバルなソフトウェア開発手法の普及における地域特性や文化的要因を理解する上で重要な示唆を与える．本章から第 6 章にかけ，世界規模で実施されている State of InnerSource Survey と，日本国内で独自に実施した調査結果を対比しながら，インナーソースの実践状況における共通点と相違点を明らかにする．なお，本研究では，2024 年度レポートの主たる著者である Clare Dillon 氏から許可を得て，サーベイの匿名データを受け取り，分析に使用している．このデータは，研究目的に限定して適切に利用されており，第三者への共有や不適切な利用は一切行われていないことをここに明記する．これに基づき，日本とグローバルのインナーソース実践の詳細な比較分析を行い，各地域の特有の課題と成功要因を明確化する．

本章の目的は，まず世界的なインナーソースの動向について，インナーソースに関する調査手法や収集データの特性を明確にし，後続の分析における解釈の基盤を提供することにある．特に，グローバルと日本の調査における方法論的な差異や，サンプル特性の違いについて留意点を示すことで，より精緻な比較分析を可能とする．

インナーソースの国際的な動向を定期的かつ体系的に把握するため，非営利団 The InnerSource Commons Foundation は，グローバル規模の調査である State of InnerSource Survey を主導している．この調査は 2016 年の初回実施以降，2020 年，2021 年，2023 年，2024 年と継続的に実施されており，各年度で異なる研究機関や企業スポンサーが主導的役割を担い，世界各地のインナーソース実践者や関係者を対象にアンケートを実施している．調査項目には，インナーソース導入状況，成功要因，直面する課題，実践におけるツールやプロセス，組織的サポート体制などが含まれ，これらを包括的に分析することで，国際的なベストプラクティスやトレンドを明確化している．

本研究において注目すべき点は，日本企業のこの国際調査への参加率が極めて低いことである．同財団への照会結果によれば，2021 年の調査で日本企業の回答者は 1 名のみ，さらに 2016 年，2023 年，2024 年には回答者が皆無という結果が得られている．一方，2024 年の調査回答者のうち 55%が欧州，40%が南北アメリカ地域からの参加であり，欧米企業を中心にインナーソースが広範に普及している現状が浮き彫りになった [41].

この地理的な偏在性は，西欧圏以外の地域におけるインナーソースへの認知度や関心の低さを示唆している．特に日本においては，グローバルなソフトウェア開発動向から乖離するリスクが存在する中，日本固有の組織文化や慣行がインナーソース導入の障壁となっている可能性が高い．このような状況は，日本企業におけるインナーソース適用可能性の探求と，その阻害要因の特定・分析の重要性を一層高めている．

## 3.1　2024 年時点におけるグローバルインナーソースの潮流と主要課題

2024 年時点において，世界的なインナーソース実践は，その多様性および成熟度の向上が顕著となっている．レポートによれば，インナーソースは依然として「ソフトウェア／ソースコードの再利用」を中心概念として認識されているが，組織内部でオープンソース的な文化や手法を適用する取り組みとして，その範囲と深度は拡大し続けている．

まず概念面では，インナーソースが単なるツール導入やコード共有に留まらず，開発組織の文化転換，学習組織の形成，そしてエンジニア同士のネットワーキング強化に資する包括的なアプローチとして捉えられている点が注目される．2024 年の調査結果では，知識共有における測定可能な進展がその前年の 68%から 33%へと大幅減少していることが示され，インナーソース実践が広がる一方で，その成果測定や評価は未だ確立途上にあることがわかる．また，インナーソースが適用されるプロジェクトタイプとしては，ライブラリや内部ツール，DevOps プロジェクト，プラットフォームプロジェクトが主流であり，組織横断的な汎用コンポーネントの生成やツールチェーン整備が進行している．一方で，ドキュメンテーションプロジェクトに関する取り組みが徐々に増加している点は，インナーソース初期導入段階にある企業がドキュメンテーションを足掛かりとして文化的転換を図る傾向を示している [41].

こうした動向は，インナーソースが今後さらに進化していく可能性を示す一方で，いくつかの課題も明らかにしている．たとえば，知識共有測定指標の確立や，中間管理層の積極的な巻き込み，品質とスピードの両立，グローバルなエンジニアリングチーム間での文化的調整などが依然として改善余地のある領域とされている [41]．また，技術進歩が速い現代のソフトウェア開発環境においては，インナーソース活動を支える開発ツール，アーキテクチャ指針，ドキュメント標準化，コミュニティ形成施策の整備が不可欠である．

総じて，2024 年における世界的インナーソース動向は，実践範囲・深度の拡張と成熟化傾向を示す一方で，測定可能な成果指標や文化的・組織的課題の克服が求められる「過渡期」にあると位置づけられる．今後の発展においては，学術研究と産業界の実務知見を統合し，持続的な組織学習とイノベーション創出に資するインナーソース環境の確立が期待される．

## 3.2 2024 年時点における日本企業のインナーソース適用状況

日本企業におけるインナーソースの現状を包括的に把握するため，文化的背景や組織的特性を考慮した調査を実施した．この調査は，単なる技術的側面だけでなく，組織文化や推進方法といった多面的な観点からインナーソースの実態を捉えることを目的としている．

調査手法として，業界の第一線で活躍する専門家の知見を収集し，それに基づいた包括的なサーベイを作成した．さらに，初期採用者および関心の高い層を対象としたカンファレンスを開催し，現場の声を直接収集する機会を設けた．

### 3.2.1 国内インナーソース動向調査の設計と実施

本研究では，前述した State of InnerSource Survey という世界規模の企業を対象とした調査を参考にしつつ，日本の現状に即した包括的かつ精緻なサーベイの策定を行った．サーベイ作成にあたり特に留意した点は，米国および欧州諸国においてはインナーソースを実践する企業が多数存在し実践面に焦点が当てられている一方，日本においては浸透の途上段階にあるという現状認識である．

この認識に基づき，本サーベイでは日本の現状と 2024 年までの世界調査との比較が可能となるよう設計し，具体的な実践の側面を維持しつつも，インナーソース推進における障壁に焦点を当てることで，導入の関心が高い層に向けた情報収集を可能にした． 具体的には，以下の項目に関する 26 の設問を含むサーベイを作成した．

- アンケート回答者のプロフィール
- 所属組織の特性
- インナーソース導入に係る障壁
- 組織間協働の実態
- インナーソース実践の現状

### 3.2.2 「InnerSource Gathering Tokyo 2024」の開催と回答の収集

2024 年 8 月 8 日に，日本初のインナーソースカンファレンスである「InnerSource Gathering Tokyo 2024」を開催した．本カンファレンスは，初期採用者および当該分野に対して高い興味関心を有する層を対象とし，知見共有および議論の場，及びサーベイ回収の場としての機能を果たすことを企図して企画・実施された．

本カンファレンスは，The InnerSource Commons Foundation が主催し，当該財団の理事を務める筆者が運営面で中心的な役割を担った．また，The Linux Foundation からのスポンサーシップ，KDDI 株式会社およびニフティ株式会社からの運営支援を得たことにより，企画運営上の基盤が強化された．参加者は小規模企業から大企業まで多岐にわたり，合計 74 名が参集した．その内訳には，日本国内で既にインナーソースを実践している企業，同概念に強い関心を有する先進的なユーザ企業，製造業や大手 IT 系メガベンチャーなどが含まれ，参加者層は極めて多様であった [42]．このような多様な参加者層を有する集会は，インナーソース導入・推進状況が異なる組織間の比較や異分野・異業種間の知識移転を可能とし，議論の質・幅を広げる上で有意義であると考えられる．

本カンファレンスは，一方向的な情報提供に終始しないよう，参加者間の双方向的な知見交換を促進する設計を採用した．具体的には，各種企業による事例紹介セッションに加え，グループディスカッションセッションを組み込み，インナーソース導入における成功要因や課題点，適用戦略などについて，参加者同士が直接議論できる機会を提供した．これらの試みにより，インナーソースに関する理解の深化，実践的知見の共有，および将来的な取り組み指針の探索が促進されたものと考えられる．

### 3.2.3 サーベイ回収とデータ集計の手法

本研究においては，上記カンファレンスの参加者を対象にアンケート調査（サーベイ）を実施した．74名の参加者のうち，52 名から回答を得ることができたこの回収率は約 70%に達するものである．サーベイ回答者は，すでに日本国内でインナーソースを実施している，もしくはその分野に高い関心があるグループであることを考えると，この値は本調査の信頼性を担保する十分なサンプルサイズであると考えられる． 回答者の属性分析から，日本におけるインナーソースへの関心が強い層に対していくつかの特筆すべき傾向が明らかになった．第 4 章以降では，これらの知見を国際比較の文脈で詳細に検討する．

## 3.3 インナーソース国際比較分析におけるフレームワーク

本研究では，2024 年度の State of InnerSource Survey 2024 から得られた 59 件の有効回答を主たる分析対象として採用した．有効回答の判定においては，質問項目の 50%以上への回答完了と，自由記述欄における具体的かつ詳細な記述内容を基準として設定し，データの質的担保と分析の信頼性を確保した．

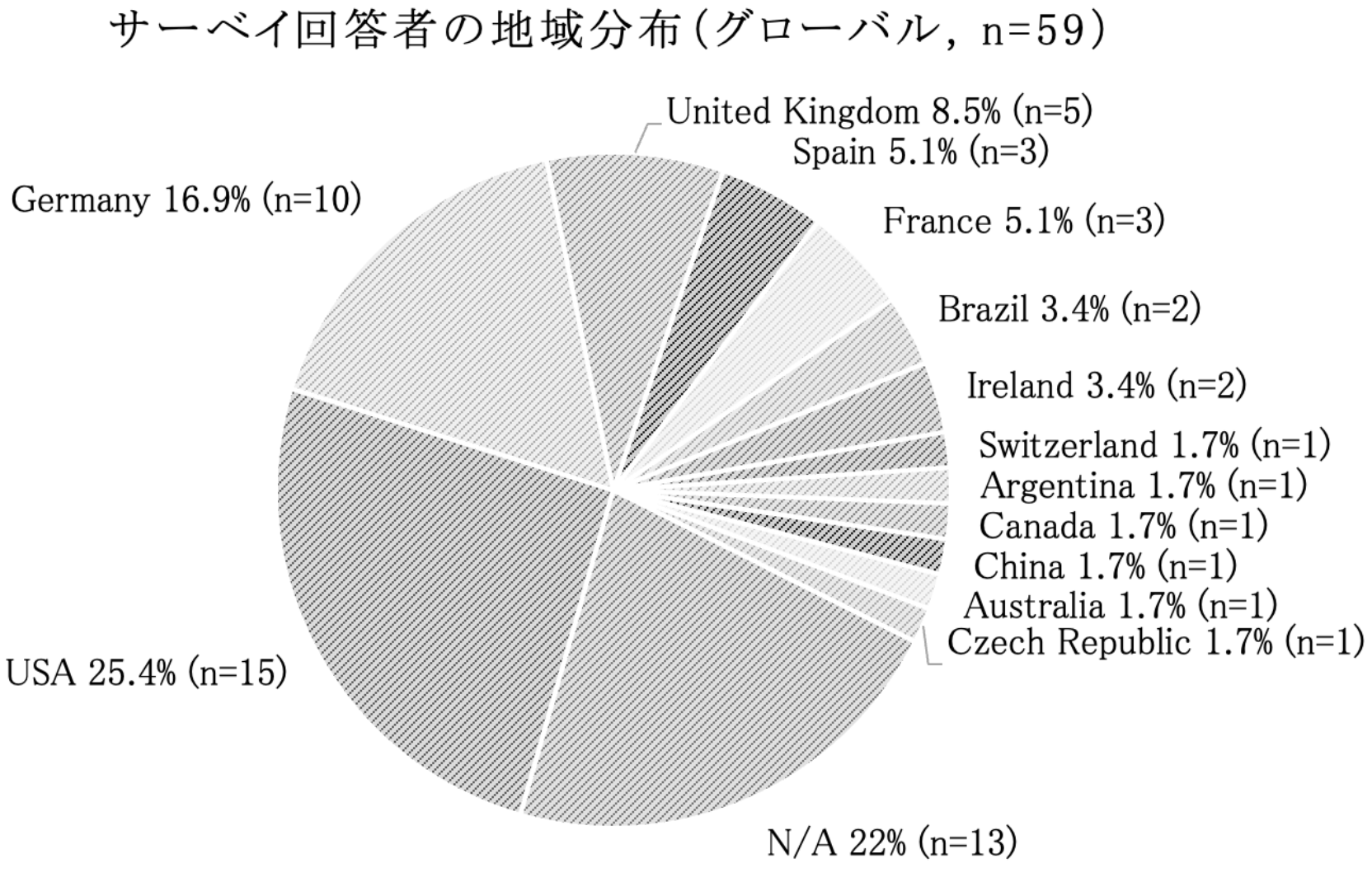

*図 3.1 State of InnerSource Survey 2024 回答者の地域分布*

グローバルサーベイの回答者分布は，北米 25%，欧州 30%を中心として，南米 5%，アジア・オセアニア 4%と広範な地理的多様性を示しており，インナーソースの実践が地域を超えた普遍的な現象として確立されつつあることを示唆している．これに対し，日本の調査データは，インナーソース導入に先進的に取り組む組織からの記名式回答を中心としており，回答の質的水準において特徴的な性質を有している．

調査手法の観点では，日本のサンプルが国内カンファレンス参加者を対象とした直接的なアプローチを採用したのに対し，グローバルサンプルはオンラインを通じた国際的な募集方式を採用している．この方法論的差異は，両サンプルともにインナーソースへの高い関心を持つ層を捕捉しつつも，グローバルサンプルにおいてより成熟した実践コミュニティの包含を可能にしている．

両調査は，設問構造に部分的な相違があるものの，インナーソースの基本概念，実践手法，組織的課題などの核心的要素において共通の分析枠組みを提供している．この統合的な分析アプローチにより，日本企業特有の文化的背景や組織的特性を，グローバルな文脈における位置づけとともに明確化することが可能となっている．

## 3.4 日本グローバル調査比較の総括

本章では，インナーソースに関する国際的な動向調査と日本国内の独自調査について，その調査手法，データ特性，および比較における留意点を詳細に検討した．特に，The InnerSource Commons Foundation による世界規模の調査と，日本初のインナーソースカンファレンスを通じて収集された国内データの特性を明らかにし，両者の比較分析における方法論的基盤を確立した．

グローバルサーベイと日本の調査データは，サンプリング手法や回答者層に一定の差異が存在するものの，インナーソースの基本概念や実践手法，組織的課題などの核心的要素において共通の分析枠組みを提供している．特に 2024 年という同時期に実施された両調査の時間的一致性は，地域間比較における信頼性を高める重要な要素となっている．

第 4 章では，属性に関する比較を通じて，インナーソースが日本とグローバル，すなわち新興地域と発展的グループそれぞれにおいて，どのような特性を持ち，その進展度合いがどのような状況にあるかを比較分析する．続く第 5 章および第 6 章では，両サンプル群（N=111）を統合したデータセットを用いて，推進度の段階における群間差異に留意しながら，より包括的な知見の導出を試みる．これらの分析を通じて，日本企業におけるインナーソース導入の特徴と課題を，グローバルな文脈の中で明確化することを目指す．

# 第4章インナーソース導入における日本とグローバルの比較検証

インナーソース実践の現状分析において，日本とグローバルの比較研究は，新興市場におけるソフトウェア開発手法の導入プロセスを理解する上で重要な示唆を提供する．本章以降では，日本のサンプル(n=52)とグローバルサンプル(n=59)の定量的分析を通じて，実践経験，導入段階，経験年数，役割分布，職務の位置づけ，業種，そして組織・開発者規模の観点から，両者の特徴と差異を体系的に検証する．

組織的成熟度の観点から見ると，日本企業における導入初期段階の特徴と，グローバル企業における成熟段階の特徴には顕著な差異が存在する．この差異は，インナーソース導入における地域特性や文化的要因の影響を理解する上で，重要な分析視点を提供している．

本研究の知見は，インナーソースの導入を検討している新興市場や発展途上地域に対して，実践的な示唆を提供する可能性を持つ．特に，アジア太平洋地域やラテンアメリカなど，組織構造や開発文化が欧米とは異なる地域において，インナーソースをどのように適応させ，発展させていくべきかという課題に対する実践的な指針となり得る．

この比較分析アプローチは，将来的に様々な地域でインナーソースを導入する際の参照モデルとして活用できる可能性を持つ．組織規模，業種特性，開発者の役割分布などの要因が，インナーソース導入の成功にどのように影響するかについての理解を深めることで，地域固有の課題に対応したインナーソース導入戦略の構築が可能となる．

本研究の分析フレームワークは，グローバルなソフトウェア開発コミュニティにおける知識移転モデルとしても機能し得る．特に，新興市場における組織変革の過程で直面する課題や，それらを克服するための具体的なアプローチについて，実証的な知見を提供することが期待される．

## 4.1　インナーソース実践の成熟度に関する国際比較

インナーソースの実践において，日本とグローバル市場の間には顕著な成熟度の差異が観察される．本節では，個人の実践経験，組織的な導入段階，そして実務者の経験年数の三つの観点から，この差異を詳細に分析する．

グローバル市場，特に欧米を中心とする地域では，インナーソースの実践が既に成熟期に入っており，組織全体での展開や戦略的な活用が進んでいる．これらの地域では，熟練した実務者による体系的な知識移転と，確立されたガバナンスフレームワークに基づく実践が一般的である．

一方，日本市場におけるインナーソースの導入は初期段階にあり，多くの組織が探索的なアプローチを採用している．しかし，この発展段階の違いは，むしろ日本市場特有の組織文化や開発プラクティスに適応したインナーソース導入モデルを構築する機会として捉えることができる．

以下では，定量的調査に基づき，両市場における実践の現状を詳細に比較分析する．この分析を通じて，日本市場におけるインナーソース導入の課題と可能性を明らかにするとともに，今後の発展に向けた示唆を導出する．

### 4.1.1 個人レベルで捉えた日本・グローバル間のインナーソース成熟度差

日本サンプル(n=52)における調査結果では，インナーソース関連プロジェクトへの直接的関与経験について「該当なし／わからない」が 53.8%と過半数を占めた一方，46.2%は何らかの形でインナーソースに携わった経験があると回答している．具体的な実践例としては，チーム外プロジェクトへの貢献が 23.1%ゲスト貢献の受け入れが 26.9%，組織内での展開・スケーリングが 19.2%，そして実践者へのアドバイスやコーチングが 11.5%に上っている．

これに対し，グローバルサンプルでは，89.8%がいずれかのインナーソース実践経験を有しており，特に組織全体の展開やスケーリング，他者へのアドバイス・コーチングといった戦略的関与を担う割合が 66.1%に達する点が顕著である．これらの数値は，日本の 19.2%や 11.5%と比較すると 3 倍以上に相当し，グローバル環境におけるインナーソース推進リーダーシップの存在を示唆している．

これらの差異は，日本においてインナーソースが導入初期もしくは検討段階にあることを示すと同時に，グローバルではすでに一定の成熟度に達している組織が多い点を示唆する．日本では，チーム外からの貢献の取り込みや組織横断的な知識共有といった初歩的な取り組みは広がりつつあるものの，大規模な展開や指導的役割(コーチング等)を担う人材の層は依然として限定的である．

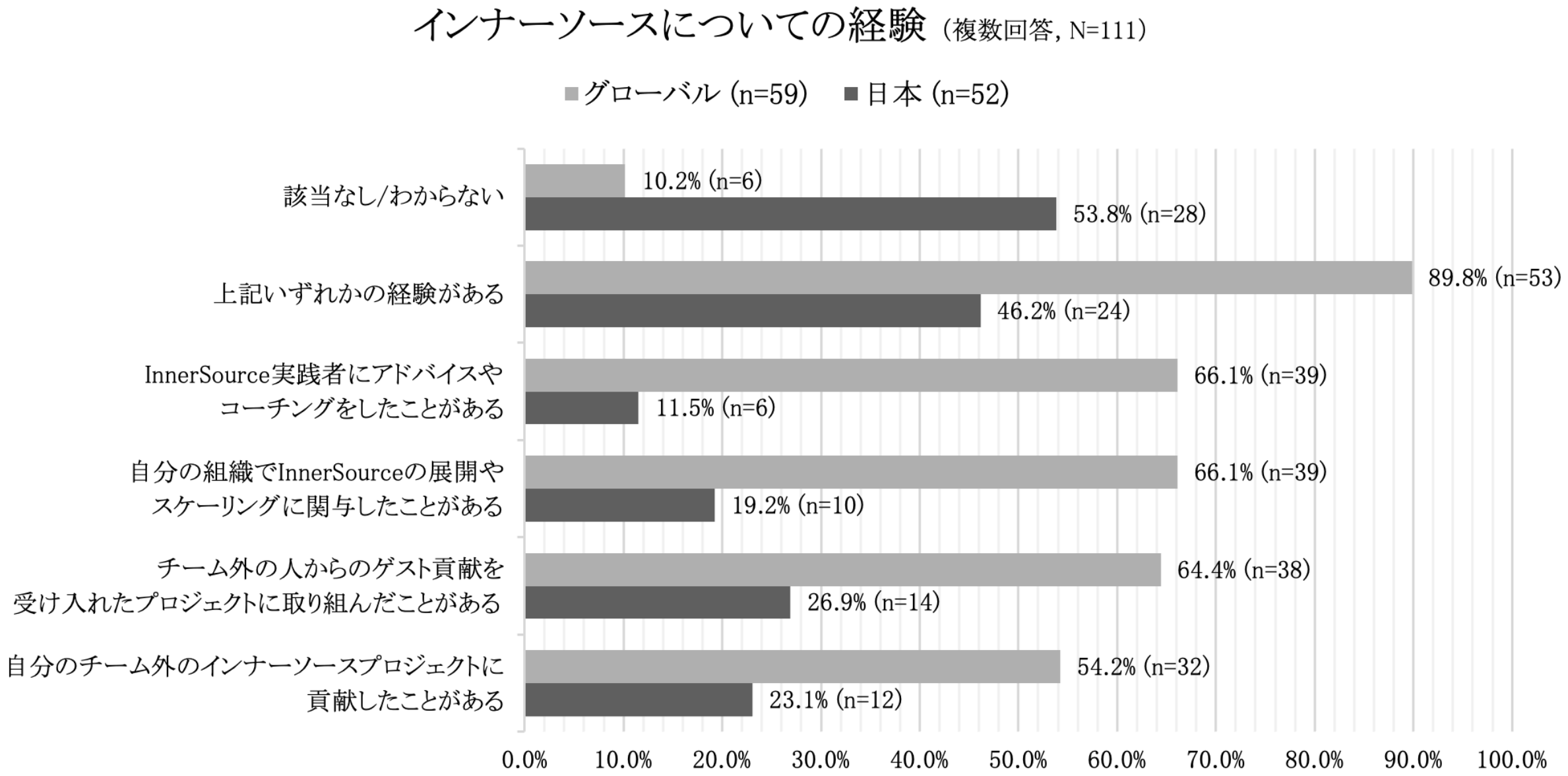

*図 4.1 インナーソースについての経験の比較*

前述した通り，両サーベイともインナーソースへの関心が高い層を対象としているが，グローバルサンプルにおいては，より成熟した実践コミュニティを内包する傾向が観察された．ただし，グローバルサンプル内にも導入初期段階の組織や経験の浅い実践者が含まれており，日本サンプルにおいても将来的なリーダーシップ人材の潜在的な存在が示唆される．

これらの分析結果は，日本におけるインナーソースの導入が初期段階にあることを示すと同時に，欧米を中心とするグローバル環境との間に顕著な成熟度の差が存在することを明確に示している．特筆すべきは，グローバル環境においては既に実務レベルでの活用と知見の蓄積が進んでいる一方で，日本で

は関心層を対象とした調査においても実践事例が限定的である点である．この成熟度の格差は，日本におけるインナーソースの普及と定着における課題を浮き彫りにしている．

### 4.1.2 経験年数分布の比較：日本の均質性とグローバルの熟練層集中

日本国内におけるインナーソース導入段階は「アイデア段階」が44.2%と最も多く，次いで「パイロット段階」（19.2%）が続いている．この結果は，回答企業の過半数が依然として最初期段階にとどまり，インナーソースの実践が萌芽的または探索的な性質を帯びていることを示唆する．

一部の先進的企業では，初期採用段階（13.5%），成長段階（5.8%），成熟段階（3.8%）と，より進んだ段階への移行を実現している．これらの組織では，インナーソースの体系的な展開と組織への統合が着実に進展していると解釈できる．しかしながら，成熟度の高い段階に到達している組織は依然として少数派であり，日本の組織全体としてはインナーソースの本格的な定着には至っていない現状が明らかとなった．

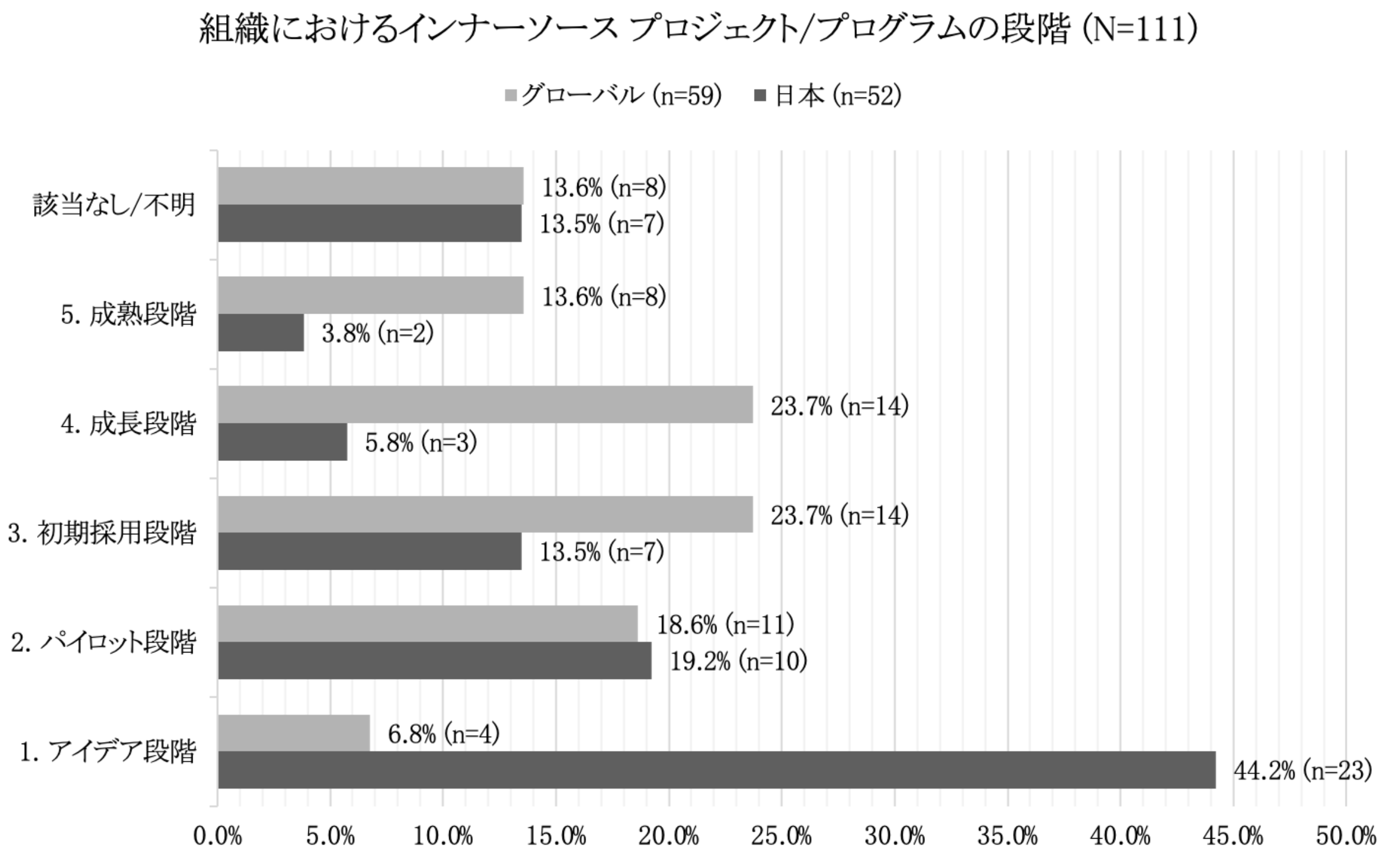

*図 4.2 組織におけるインナーソースプロジェクト/プログラムの段階*

グローバルサンプルとの比較分析からは，顕著な成熟度の差異が観察される．インナーソースへの関心を有する組織という共通の母集団において，グローバル企業では初期採用段階から成長段階に位置する組織の割合が日本企業と比較して明確に高い．一方で，グローバル企業においてもパイロット段階に位置する組織が成熟段階の組織を上回っている．この結果は，インナーソース導入が世界的に現在進行形で展開されており，その成熟過程において多様な発展段階の組織が共存していることを示唆している．

### 4.1.3 経験年数分布が示す日本の均等性とグローバルの熟練層集中

グローバルサンプルは，11～15 年(14.3%)，16～20 年(40.5%)，26 年以上(28.6%)と，いずれも一定以上の経験を持つ層に集中しており，この傾向からはインナーソース導入に際して中堅から熟練層が主導的役割を担っている可能性が示唆される．

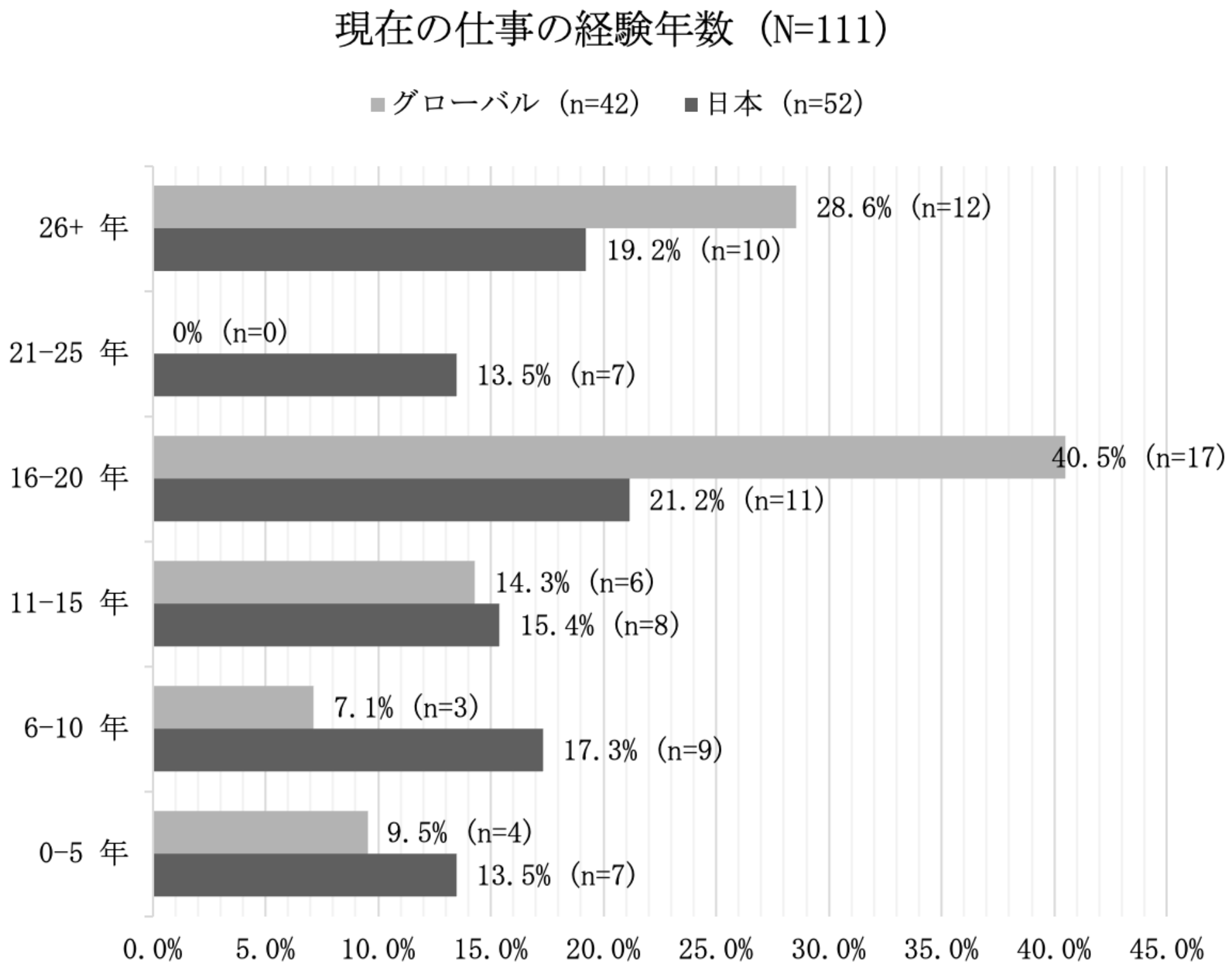

*図 4.3 現在の仕事の経験年数*

一方で日本サンプルでは，回答者が全ての層に比較的均等に分布していることが特徴的である．この幅広い層が関心を持っているという現象は，インナーソースが社内で「特定の経験年数や役職者だけのテーマ」にとどまっていないことを示唆し，若手を含む多様な層が積極的に関与し得る基盤が整っていることを意味し得る．若手が早期からインナーソース活動に興味を抱きやすい社内風土や教育の仕組みが存在する場合，イノベーションの発芽が分散的に広がり，特定のリーダーに依存しない導入モデルが成立する可能性もある．

また，経験年数の分布が均等な一方で，日本のインナーソース導入実態が初期段階に留まっていることを示す結果とあわせて考えると，ベテラン層の存在が必ずしも導入の加速要因として十分機能していない可能性がある．すなわち，グローバルでは熟練層が積極的に組織内オープン文化の定着を支援しているのに対し，日本では熟練層が存在しても，それがコミュニティ全体の成長をリードする構造にはまだつながっていないと考えられる．

結果として，グローバルサンプルにおいては実績ある熟練層が積極的にガバナンスやコミュニティ形成をリードしやすい一方，日本ではより幅広い層の参加が期待されるものの，実際の大規模展開や運用設

計のノウハウが手探りの段階にとどまる組織も多いと推察される．今後の普及や実践活性化に向けては，経験豊富な層が積み上げた知見と，若手も含めた多様な層の柔軟性・発想力をいかに連動させるかが重要な課題となるだろう．

## 4.2 インナーソース実践を担う人材構成と産業特性の総合分析

インナーソースの実践において，その担い手となる人材の役割や所属する産業の特性を理解することは，この手法の普及状況や今後の展開を考える上で重要な示唆を与える．本節では，日本およびグローバルにおけるインナーソース実践者の職務役割，業務領域，および所属産業の分布を分析し，その特徴的な差異を明らかにする．

日本のインナーソース実践では，開発者を中心としながらも管理職やアジャイル関連職種など，幅広い職種からの参画が確認された．一方，グローバルな文脈では，インナーソースやオープンソースに特化した専門職種の存在が顕著である．この対比は，両地域におけるインナーソースの成熟度や組織文化の違いを反映している．

また，産業分布においては，両地域ともにテクノロジー企業が主導的な立場にある一方で，日本では製造業，グローバルでは金融・ヘルスケアなど，非テクノロジー分野への展開にそれぞれ特徴的な傾向が見られる．これらの差異は，各地域における産業構造やデジタルトランスフォーメーションの進展度合いと密接に関連している．

次に，これらの観点について詳細な分析を行い，インナーソース実践の地域性と普及状況について考察する．

### 4.2.1 役割の分布が示すインナーソース実践の地域特性

回答者の役割の分布に関して，日本のサンプル（n=52）においては，最も多くを占めたのが「開発者（29%）」であり，続いて「管理職（25%）」「アジャイル関連（15%）」が多い割合を示した．

この結果は，インナーソースが組織内の「現場」すなわちソフトウェア開発工程に直接関与する層によって強く支持・関心を集めていることを示唆する．開発者を中心としながらも，管理職や幹部が 25%以上を占めている点は，インナーソースが単なる技術的手法にとどまらず，戦略的な導入判断や組織マネジメント上の考慮事項として認識されていることを示す．また，アジャイル開発関連職種が全体の 15%と比較的多く含まれていることは，インナーソースがアジャイル開発手法と親和性を有する可能性を示唆する．これは，組織内部で迅速なフィードバックサイクルや継続的な改善を重視するアジャイル文化が，インナーソースによる透明性・コラボレーション志向の価値観と整合し，相補的な効果を生む可能性を示している．

一方で，グローバルサンプル（n=59）では，インナーソース関連（24%）と開発者（19%）が主要な職種として挙がったほか，アーキテクト（14%），オープンソース関連（10%）が続いた．これに対し，管理職（7%）やPM（5%）は比較的少なく，アジャイル関連職種は 1 名も報告されなかった．ここからは，グローバルな文脈では，既にインナーソースやオープンソースに直接関与する職種が抽出されていることが示唆される．つまり，インナーソース・オープンソース活用の文化がある程度成熟している組織や，グローバルな環境での職務定義が明確な人材が多く参加している可能性がある．

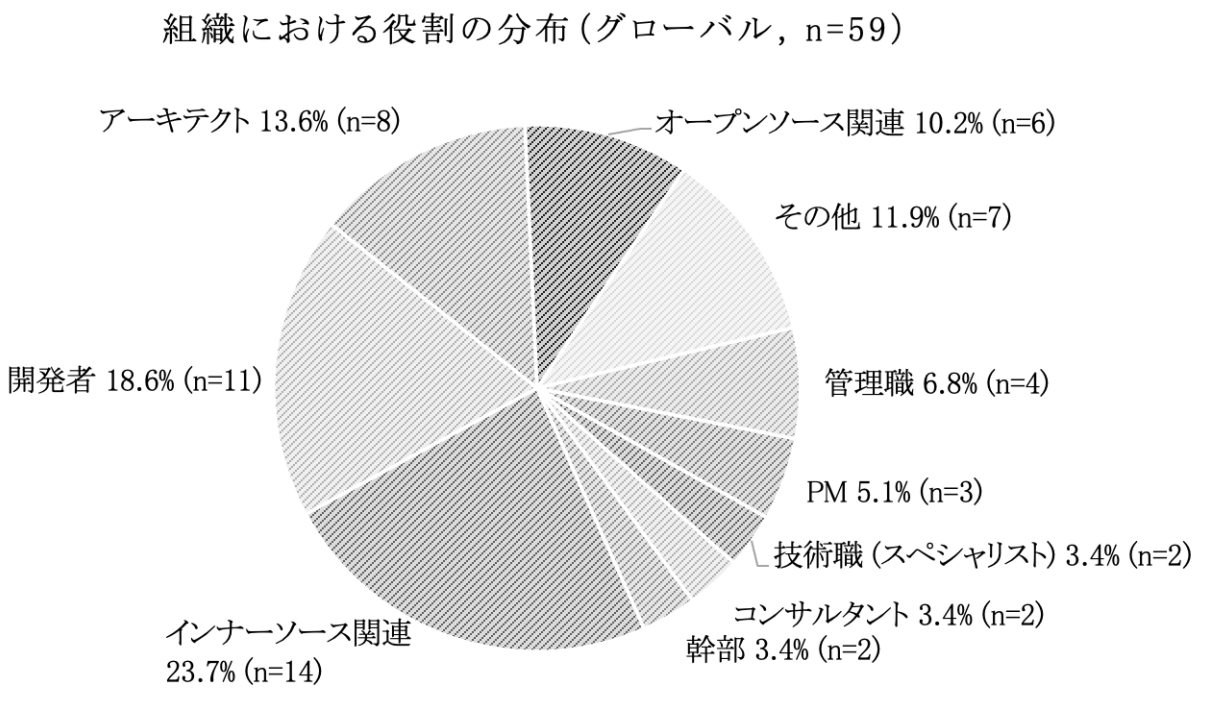

図 4.4 組織における役割の分布（グローバル）

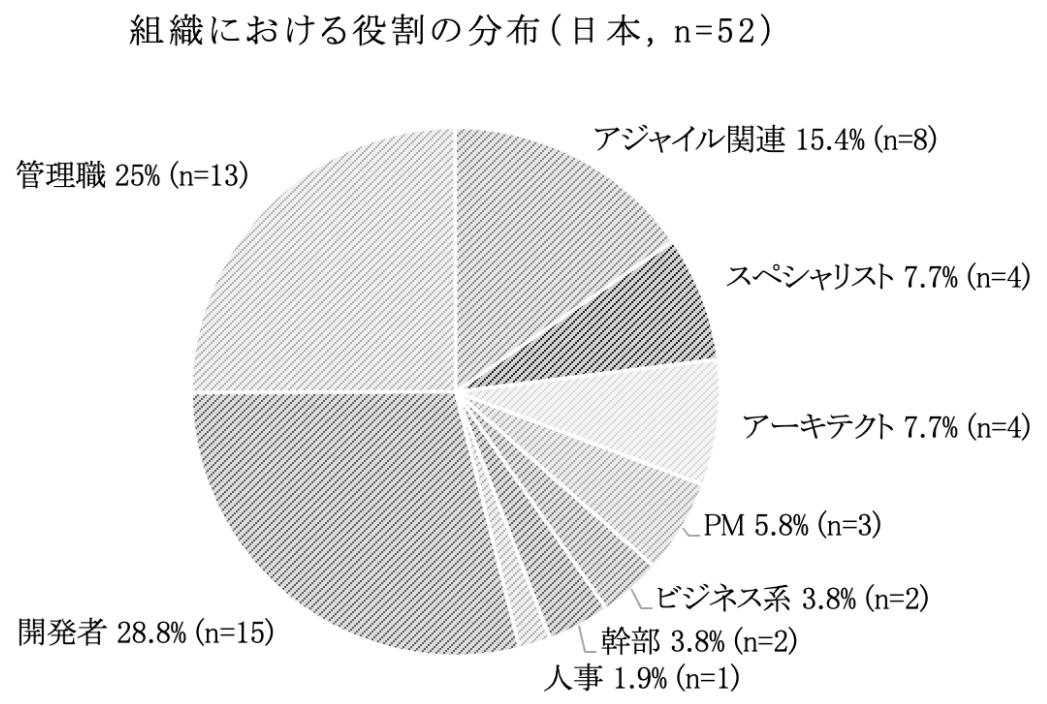

図 4.5 組織における役割の分布（日本）

日本のサンプルでは，アジャイル関連職種を含む幅広い職種からインナーソースへの関心が示されている点が特筆される．これは，日本企業が従来の開発プロセスからアジャイル手法やオープンなナレッジ共有の実践へと移行しつつある状況を反映していると考えられる．日本企業においては，「管理職」が25%を占める一方で，「インナーソース関連」と明示的に分類される職種が見られない点が特徴的である．日本では，地理的・文化的背景や長期雇用慣行により，企業内での役割分担や組織構造，職種定義が独特の展開を見せている．例として，ジョブローテーションや総合職のようなキャリアパスが一般的であり，特定分野を専門とする職種を細分化する文化が海外に比べて希薄である可能性がある．また，インナーソースの認知度もまだ広がりの途上にある．その結果，インナーソースへの関心は，明確な「インナーソース関連職種」として区別される以前の段階にあり，開発者やアジャイル関連専門家，管理職層など，比較的広範な職務領域に広がっていることが示唆される．

一方，グローバルサンプルでは，既に「インナーソース関連」や「オープンソース関連」という職種分類が存在しており，専門家集団が牽引する形で成熟した知識共有やガバナンスモデルが形成されていると推測される．これは，OSS 文化が歴史的に根付く海外企業やコミュニティにおいて，インナーソースを単なる開発スタイルではなく，戦略的差別化要因や組織横断的な協働の基盤として捉え，そのための専門人材を配置していることを意味している．また，インナーソースの普及・定着がある程度進んだ環境では，特定の専門スキルや知識を持つ人材が役割を担い，組織横断的なイノベーションや技術移転を容易にする ISPO や OSPO などの基盤が形成されると考えられる．

以上の分析から，日本では明示的なインナーソース専門職は少ないものの，現場と管理層双方を巻き込んだ幅広い関与が見られ，アジャイル文化との親和性も相まって新たな開発手法としての定着が期待される．一方，グローバルな文脈では既に専門職種や明確なガバナンスモデルが確立されており，組織横断的なオープンコラボレーションを支える仕組みとしてインナーソースが高度に制度化されつつある．このような役割分布の差異は，インナーソースの導入段階や歴史的背景のみならず，企業文化や職務定義の違いが導入・普及プロセスに大きく影響していることを示唆している．

### 4.2.2 日本企業における職務別インナーソース関心傾向

回答者の職務について，グローバルサンプルにおいては，前節で示したように特定の役職や専門領域がインナーソース導入初期に主導的役割を果たす可能性が示唆された．しかし，その詳細な職務上の位置づけ，すなわち業務の具体的領域や貢献分野については，グローバルレベルのデータが本調査時点では十分得られていない．そこで，本研究では，より詳細な洞察を得るための一環として，日本において特別に「業務の位置づけ」を問う追加的な調査項目を設けた．これにより，インナーソースが実際にどのような実務領域で関心を集め，どのような業務分野に対して有用性が認識されているかを把握し，ひいてはグローバルな展開時に考慮すべき職務・職能上の示唆を得ることが可能となる．

ソフトウェア開発業務に従事する参加者が最多であることは，インナーソースが主として社内開発プロセスの改善や標準化，コード再利用による生産性向上に強く関与していることを示す．研究開発部門の参加が 2 割近くを占めることは，インナーソースがイノベーション環境の醸成や新技術の社内実装を支援し得ることを示唆する．研究開発業務に従事する専門家は，組織内部で発生した新規アイデアや試行的プロジェクトをインナーソース型のリポジトリやコミュニティを通じて発信・共有し，他部門からのフィードバックや貢献を得ることで，技術的知見の統合や迅速な実用化を図る可能性がある．

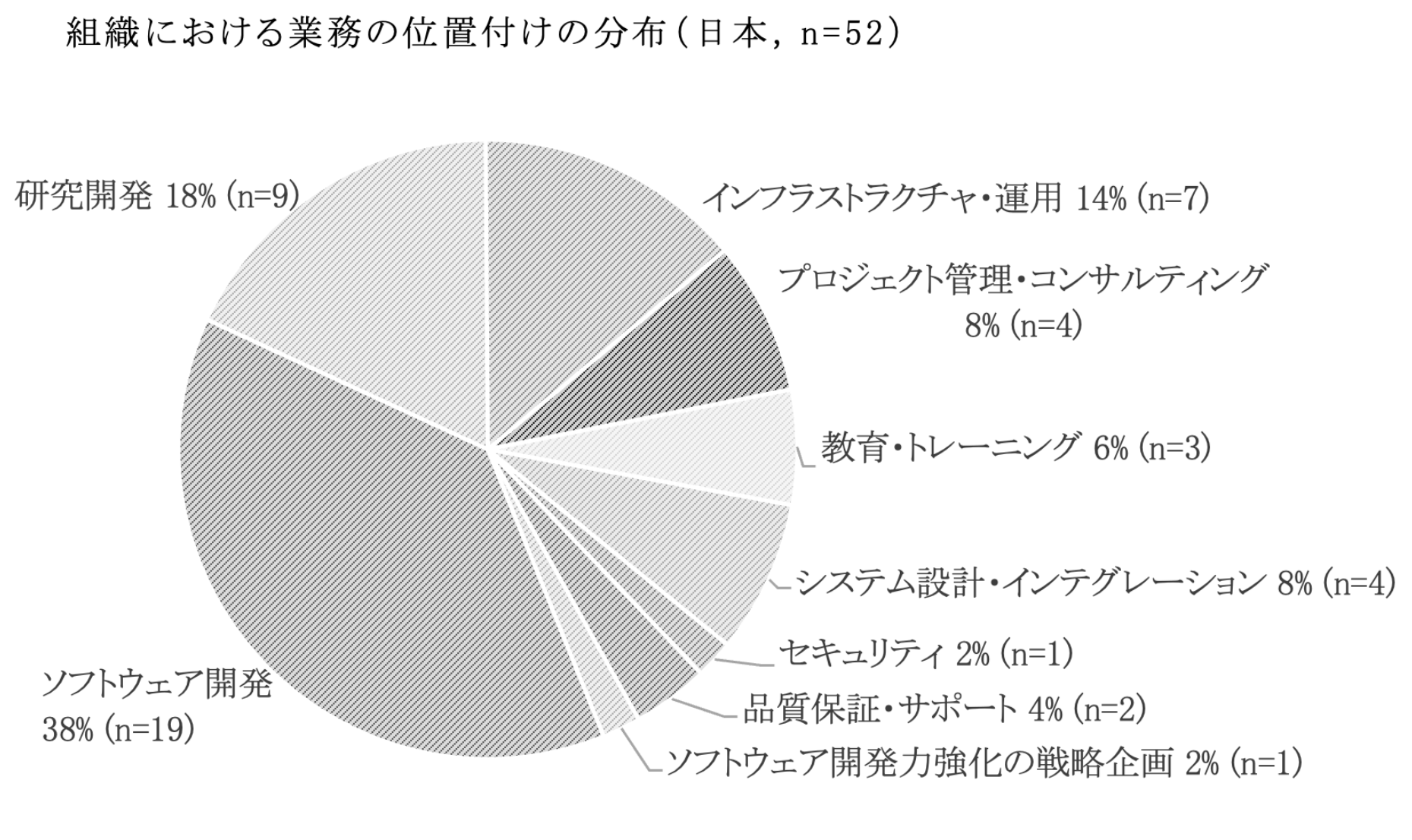

*図 4.6 組織における業務の位置付けの分布*

今回の分析は日本のサンプルに限定されたものであり，グローバルデータとの直接的比較は行っていない．しかし，この詳細な職務領域分析によって，インナーソースが開発担当者以外にも幅広い職務カテゴリに興味・関心を喚起していることが確認された．このように，日本では職務上の位置づけが多様な領域にわたっており，インナーソースが組織内部の幅広い機能群と接続し，互いに影響を与え合う複合的エコシステムとして機能しうる可能性が見出される．この知見は，将来的なグローバルサンプル分析において，地域文化的要因，職務定義の差異，国際労働市場の動向などを考慮した包括的な検証を行う上で，重要な参照点となり得る．

前小節で言及した欧米の ISPO や OSPO の専門職制度の確立は，インナーソースの組織的位置づけを考える上で示唆に富む．これら先行事例との比較分析は，組織内での確立・発展プロセスや，地域特性が職務分布に及ぼす影響について，より体系的な理解をもたらすだろう．

### 4.2.3 業種別にみるインナーソース導入の広がりと特徴

日本における分布では，テクノロジー分野が 65%と極めて高い割合を示しており，この傾向はインナーソースが従来から最も注目を集めてきた IT セクターで引き続き優勢であることを裏付ける．テクノロジー企業はソフトウェア開発および関連するイノベーション創出が中核的事業要素であり，インナーソースを活用することで組織内部での知識共有や再利用性向上を図る戦略的価値が高い．その結果，テクノロジー領域でのインナーソース定着は，自然な帰結と解釈できる．

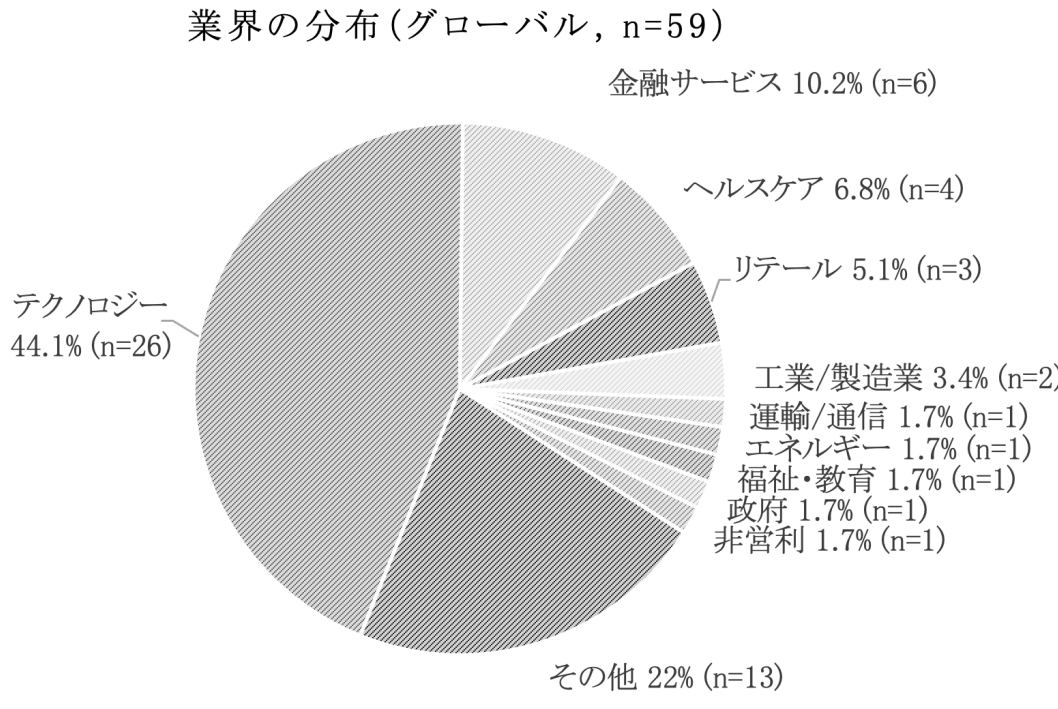

図 4.7 業界の分布（グローバル）

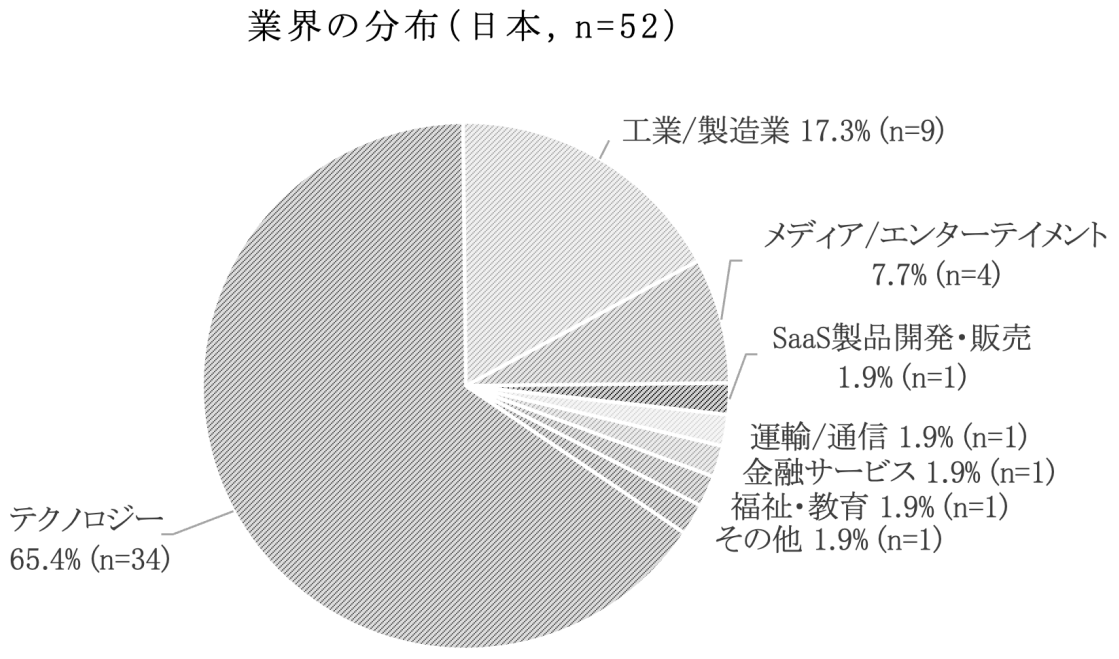

図 4.8 業界の分布（日本）

一方，日本における工業・製造業（17%）の相対的な割合の高さは，従来のハードウェア中心モデルからソフトウェア主導の価値創出への転換，すなわちデジタルトランスフォーメーションの進展を端的に示している．これにより，製造業界でも設計・生産プロセスやサプライチェーン管理などにソフトウェアを組み込み，インナーソースを活用することで社内の開発リソースを効率的に活用する動きが顕在化していると考えられる．

他方，グローバルな視点ではテクノロジー分野が引き続き主要領域（44%）を占めるものの，日本との比較において工業・製造業セクターの相対的比率は低い．その代わり，金融サービス（10%）やヘルスケア（7%）など，非テクノロジー分野が一定割合を示している点が注目に値する．これらの分野は近年，規制や顧客ニーズの多様化，そしてデータ駆動型サービスへの移行により，ソフトウェア開発能力の内製化が戦略的課題となっている．その結果，インナーソースを活用することで組織内部でのノウハウ共有や開発基盤の標準化を推進し，既存のビジネスモデルに付加価値をもたらそうとする動向が確認される．

総じて，テクノロジー企業を主軸としつつも，それ以外の業種にも徐々に波及している事実は，インナーソースが業種横断的に活用され得る汎用的なアプローチであることを示す．日本ではデジタルトランスフォーメーションやアジャイル開発の潮流と相まって工業・製造業の割合が顕著に表れており [43, 44]，グローバルでは規制産業 [45]を含む幅広い分野でインナーソースが定着しつつある．このような産業間の温度差は，各組織が直面する競争環境や技術移行のタイミングによって左右されるが，いずれにして

もインナーソースはソフトウェアを軸とした価値創造を加速する有力な手段となりつつあると結論づけられる.

## 4.3 企業規模とインナーソースの実践形態

本節では，企業規模がインナーソース実践に与える影響について，従業員数と開発者数の両面から分析を行う．特に，日本企業とグローバル企業では，組織構造や規模に起因する特徴的な差異が観察された．これらの差異は，インナーソース活用の目的や期待される効果にも密接に関連している．以降では，組織規模の観点から見られた特徴的な傾向を詳細に検討し，その意味するところを考察する．

### 4.3.1 組織の従業員数による導入文脈の相違

組織の従業員について日本サンプルでは，100～499 名規模の中堅企業が約 29%を占めるとともに，500～4,999 名規模を含む中小～中堅組織が高い割合を示している．さらに日本企業では，グループ会社を細分化する傾向があり，システム子会社や特定ドメインの子会社など，組織単位が比較的小規模であるケースが多い．そのため，必ずしも「サイロ解消」を主要な目的としてインナーソースを導入しているわけではないと考えられる．

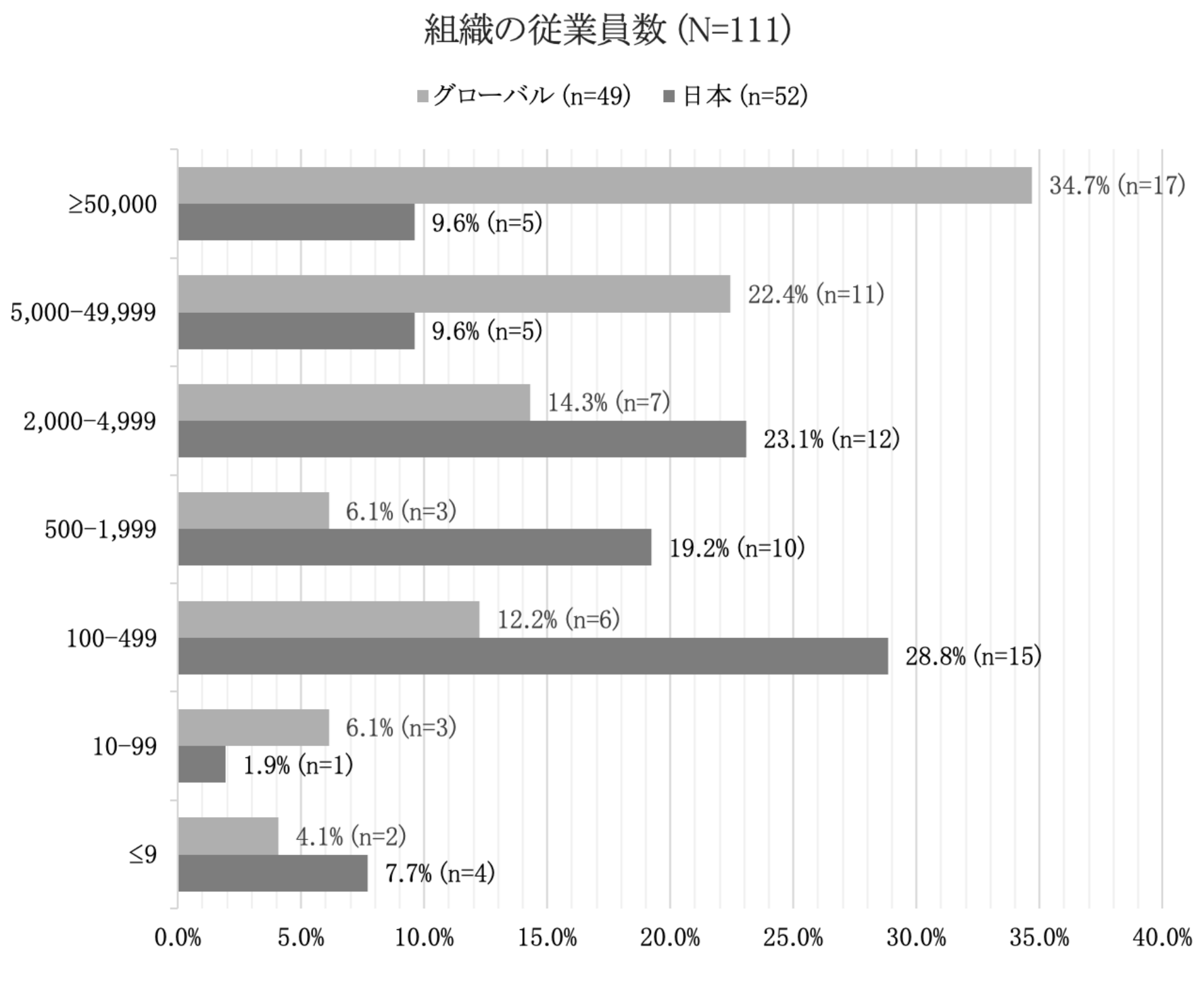

*図 4.9 組織の従業員数*

一方，グローバルサンプル（n=59）には 5,000 名以上，特に 5 万人を超える超大規模組織が数多く含まれており，部門間の断絶（サイロ化）が深刻化しやすい．このような組織では，サイロ解消を主要な目的としたインナーソース活用の必要性が高いと推察される．

グローバルサンプルの企業群は，いわゆる超大規模企業への偏りが顕著であるため，インナーソース導入の背景や動機は「地理的および組織的に拡大した開発現場のサイロ解消」が中心的な議題になりやすいと考えられる．その一方で，日本の場合はインナーソースが必ずしもサイロ解消を主目的として導入されているわけではなく，リソース不足やノウハウ共有など，より機能的な面を重視する可能性がある．この点については第 5 章において，導入の背景や障壁を分析する際に詳しく議論する．

さらに日本では，システム子会社や特定ドメイン領域の子会社など，多様なグループ会社が比較的独立した組織単位を形成しやすい．そのため，名目上の従業員規模だけでなく，各社が扱うプロジェクト領域の明確な棲み分けや意思決定の独立性も考慮しなければ，インナーソース実践の狙いや効果を正確に評価することは難しいと考えられる．日本企業を連結ベースで見れば，大規模なコングロマリットの一角を担うケースもあるものの，世界的な巨大企業に比べると一つひとつの組織単位が小規模な傾向は否めない．そのため，インナーソースを活用してもグローバル企業とは異なる課題や動機が表面化する可能性がある点については，引き続き検討したい．

## 4.3.2 組織の開発者数による導入文脈の相違

開発者数規模別の分布を比較すると，日本とグローバルの回答間で明確な差異が浮かび上がる．日本の回答では，開発者数 100-499 名の規模が最も多く(40.4%)，次いで 500-1,999 名(27.7%)，10-99 名(17%)が続いている．一方，グローバルでは 5,000〜49,999 名(28.9%)や 2,000〜4,999 名(13.3%)，さらには 5 万名を超える(8.9%)極めて大規模な開発者集団を擁する組織が存在する点が特徴的である．

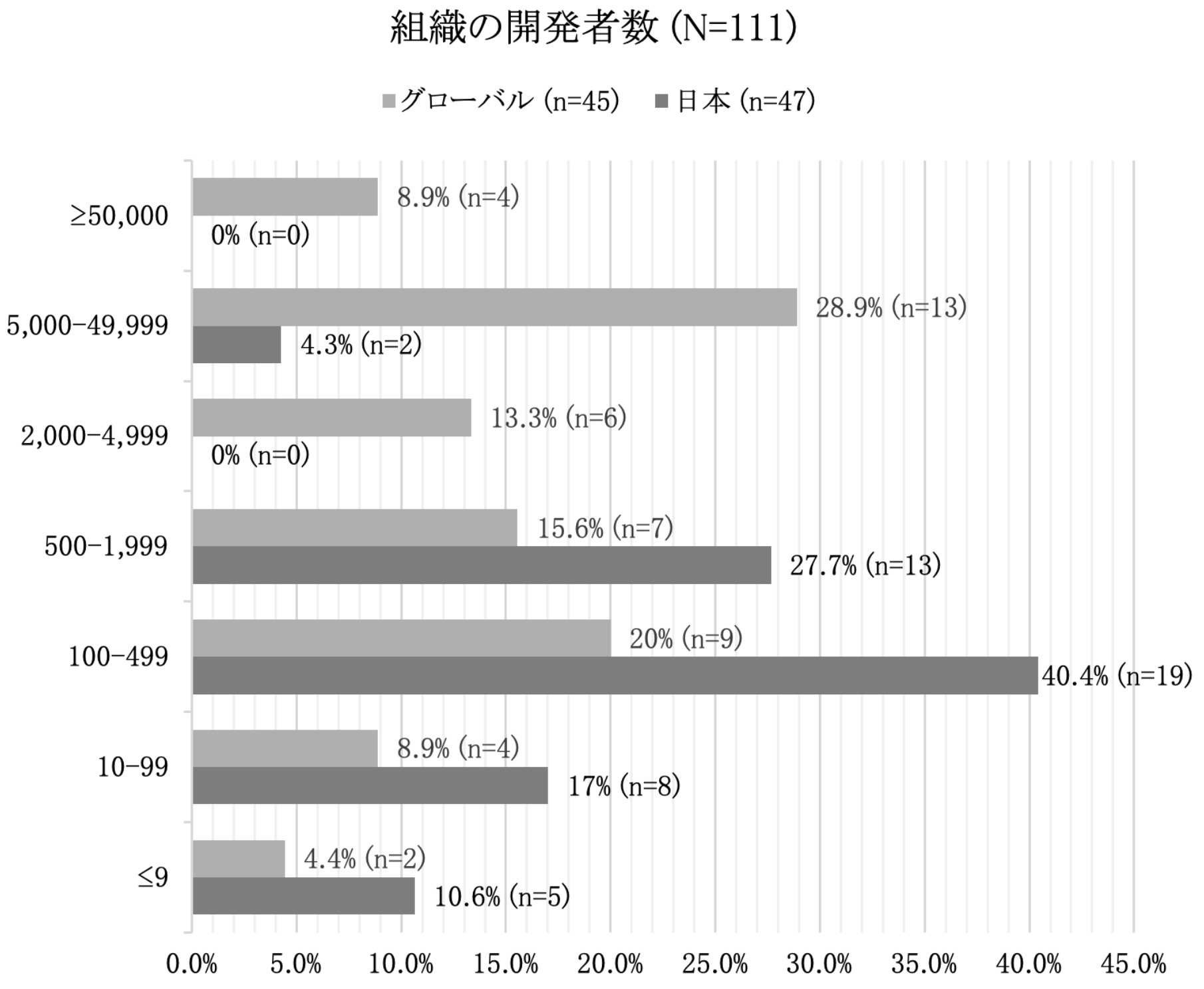

*図 4.10 組織の開発者数*

この差異は，前述した組織規模や文化的背景との整合性を示唆する．日本のサンプルでは，グループ子会社などの細分化された組織形態を含めて，中小～中堅規模の企業が相対的に多いため，開発者数も数百名～2,000 名程度にとどまるケースが一般的である．その結果，インナーソースは比較的コンパクトな開発コミュニティ内での情報共有や知識流通を強化する手段として機能しやすいと考えられる．仮にサイロ化が生じたとしても，その規模や複雑性が大規模組織ほど深刻化しにくいため，インナーソース導入による課題解消がより迅速かつ直接的に成果を生み出せる可能性がある．

一方，グローバルサンプルでは 5,000 名を超える巨大なエンジニアリング組織が多数含まれており，多地域・多部門にわたる複雑な開発体制を背景として，インナーソースがサイロ解消と効率的なナレッジ再利用のための「構造的問題解決手段」として期待されている．すなわち，膨大な規模の開発組織をいかに「一つのコミュニティ」として機能させるかが主要な課題となり，その戦略的ソリューションとしてインナーソースが定着しやすい可能性がある．

総合的にみると，開発者数規模の比較は，日本とグローバルにおけるインナーソース導入の文脈の違いを際立たせる．組織における役割の分析と併せて考察すると，日本では中小～中堅規模の開発組織がアジャイル文化や透明性重視の姿勢を背景に，インナーソースを新たな文化的価値観の定着や強化の手段として活用する傾向が見受けられる．一方，グローバルの巨大かつ複雑な開発組織では，オープンソースやインナーソース関連の専門職を設置する企業もあり，すでにオープンソース文化を内包したうえで，サイロ解消や大規模ナレッジ統合を目的とする戦略的「問題解決ツール」としてインナーソースが位置づけられやすい．このような差異は，国や地域による文化的成熟度やオープンソース文化の受容歴の違いだけでなく，グローバルサンプルにおける「インナーソースの早期導入関心層」と日本のような「後期受け入れグループ」という導入段階の相違にも起因していると考えられる．その結果として，組織のスケールや歴史的成長段階によって，インナーソース活用の意義や導入効果が大きく変化し得ることを示す好例といえよう．

## 4.4　属性別の導入差異の総括

日本とグローバルにおける比較分析を通じて，インナーソース導入には地域的・組織的な属性の差異が存在することが明らかになった．本研究の知見によれば，欧米圏の企業では長く浸透してきたオープンソース文化を背景として，インナーソースは既存のコミュニティ思考やオープンな開発スタイルをさらに内製化・強化する手段として機能していると推測される．一方，日本ではオープンソース文化が十分に成熟していない組織が多く，またグループ会社が細分化される傾向も相まって，一つの大企業内に複数の中規模組織が存在する場合が珍しくない．そのため，インナーソースは「パッケージ化された手法」として取り入れられ，アジャイル文化や開発の透明性を重視する新たな風土を生み出す試みとして捉えられることが多いと考えられる．

組織規模が大きいか小さいかにかかわらず，部門間の断絶や情報共有の不足は普遍的に生じうる問題であるため，中規模以下の企業でもインナーソースに対する関心は高い点が注目される．グローバル事例では，超大規模組織が多拠点化やサイロ化によるコミュニケーションコストの増大を克服するためにインナーソースを活用している．一方，日本の中小～中堅企業では比較的小規模のサイロであっても，

それを解消しつつアジャイル文化を育成し，開発現場と管理職が協働してプロセスを改革する意図が強い．こうした動機の差異が，地域別や規模別の調査結果からうかがえる．

インナーソースに関心を寄せる属性や職種においても，地域と組織風土の違いが色濃く表れている．グローバルの調査対象では，「インナーソース関連」や「オープンソース関連」といった専門役職が明確に存在しており，熟練エンジニアやコーチによる大規模かつ横断的な導入が進んでいる．一方，日本においては管理層やアジャイル関連職種が興味を示すケースはあるものの，「インナーソース専門家」という位置づけはまだ一般的ではなく，開発者層も若手からベテランまで広く関与しながらも実践者の数は限られている．こうした状況は，地域や企業におけるインナーソースの成熟度によって「インナーソース」の持つ意味合いが微妙に異なる可能性を示唆する．

しかしながら，「OSS プラクティスの内部化」と「協働のためのパッケージ手法」という二種類のインナーソースが明確に区分されると断定するのは時期尚早である．むしろ，導入目的，企業の成熟度，歴史的経緯などの要因が複合的に影響し合っている状況として捉えるべきである．実際の導入背景や課題，直面する障壁の種類を詳細に検討することで，インナーソースの機能する文脈がより鮮明になると考えられる．

第 5 章以降では，本章までに示した地域差や企業規模の差が導入阻害要因に与える影響を分析し，それらが生み出す成功モデルを具体的に考察する．これにより，インナーソースの機能や位置づけについて，より総合的な議論を展開する．

# 第5章インナーソースの認知・導入動機に関する多角的分析

インナーソースの組織的導入は，企業のソフトウェア開発プロセスと組織文化に大きな変革をもたらす可能性を秘めている．2020 年代に入り，特にグローバル企業を中心にインナーソース導入の取り組みが加速する中，その認識や採用動機は地域や組織の特性，そしてインナーソース文化・プロセスの受容度合いによって多様な様相を見せている．本章では，日本企業とグローバル企業におけるインナーソースの認識と採用動機について，包括的な分析を行う．

まず，インナーソースの定義に関する共通認識と解釈の多様性を探り，次に組織におけるインナーソースの認識パターンを日本とグローバルの比較を通じて明らかにする．さらに，インナーソースへの参加動機を詳細に分析し，最後に導入フェーズごとの採用動機の変遷を追跡する．これらの分析を通じて，インナーソース導入における組織的・文化的な課題と，その効果的な推進に向けた示唆を導出することを目指す．

## 5.1 インナーソースの基本概念と定義上の多様性

インナーソースの定義に関する自由記述回答の分析から，その認識と期待される効果について興味深い傾向が明らかになった．回答者の多くは「オープンソースの社内適用」という基本的な枠組みを理解しているものの，その解釈や期待される効果には特徴的な偏りが観察された．

*表 5.1 インナーソースの意味の定義における頻出単語数*

| 正規化済単語 | 出現回数 |
|---|---|
| open-source | 41 |
| organization | 29 |
| collaboration | 25 |
| practices | 21 |
| development | 19 |
| code | 13 |
| company | 12 |
| contribution | 12 |
| InnerSource | 12 |
| software | 12 |
| culture | 10 |
| internal | 10 |

具体的には，「コラボレーション」「プラクティス」「貢献」「文化」といったキーワードが頻出する一方で，「イノベーション」や「創造性」に関する言及は相対的に少数であった．この傾向は，インナーソースが主として既存リソースの活用や社内連携の強化といった実務的な文脈で理解されており，オープンソースコミュニティに特徴的な創造的価値創出の側面が，必ずしも定義の中核として認識されていないことを示唆している．

このような認識の偏りは，企業内でのインナーソース導入において，オープンソースコミュニティに見られる自発的な問題解決や新規価値創出の側面が十分に重視されていない可能性を示唆している．

インナーソースの概念的理解は組織や個人によって多様に解釈されており，その本質的要素を体系的に把握することが喫緊の課題となっている．本研究では，テキストマイニングによる共起ネットワーク分析を用いて，インナーソースに関する様々な定義や目的の記述から，その本質的要素の可視化・構造化を試みた．分析の結果，「オープンソース手法の適用」「企業内部への適用」「ナレッジシェア」「車輪の再発明防止」「車輪の再発明防止」という五つの主要な概念クラスターが特定された．

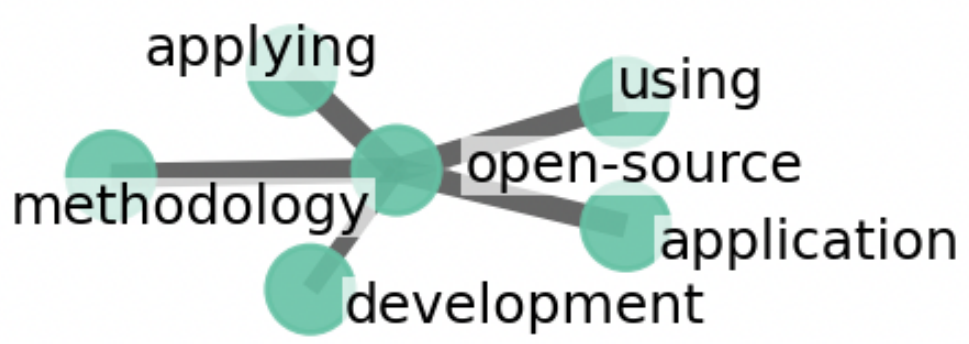

図 5.1 共起分析の結果“open-source”

**オープンソース手法の適用**
e.g. InnerSource is enterprise in-house open-source methodology on product development and collaboration.

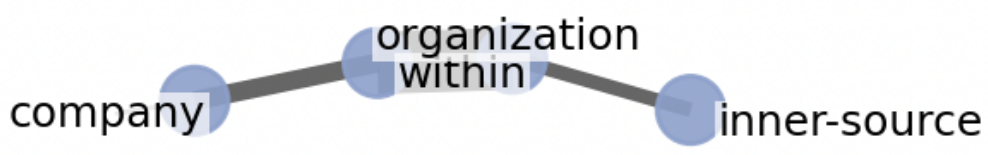

図 5.2 共起分析の結果“within organization”

**企業の内部への適用**
e.g. It's like open-source, just everything stay within the organization.

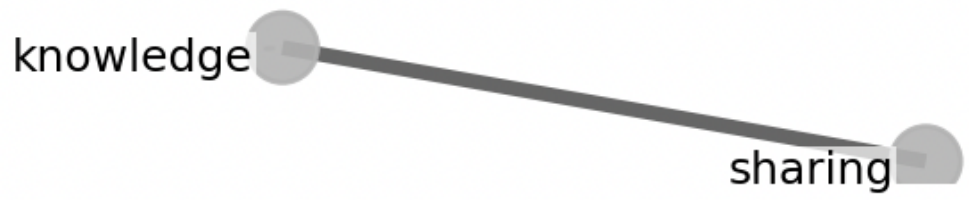

図 5.3 共起分析の結果“knowledge-sharing”

**ナレッジシェア**
e.g. Promoting culture of collaboration, fostering voluntary contribution within the organization, and enabling knowledge sharing.

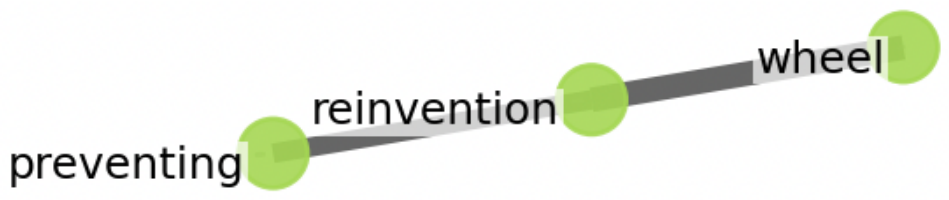

図 5.4 共起分析の結果 “preventing wheel reinvention”

**車輪の再発明防止**
e.g. Collaboration across the organization and preventing the reinvention of the wheel.

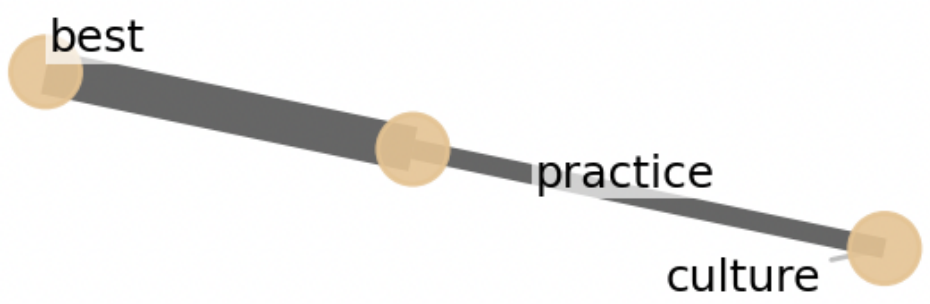

図 5.5 共起分析の結果“culture best practice”

**文化変革のプラクティス**
e.g. Application of open-source best practices, culture and methodology for Enterprise-internal projects and collaboration.

つまり，インナーソースは技術的手法の転用を超えて，組織文化の変革やナレッジマネジメントの革新を含む，より包括的な組織変革のメカニズムとして機能することが期待されているようである．ただし，本研究のサンプルサイズによる制約から，抽出された共起関係の網羅性には一定の限界が存在することに留意が必要である．

## 5.2　インナーソースに対する認識構造と地域別差異

インナーソースに対する認識調査の分析結果から，日本とグローバルの間で顕著な差異が観察された．この差異は，インナーソースの概念理解と期待される価値において，地域特有の特徴を反映している．

グローバルサーベイの回答者は，インナーソースを実務的かつ具体的な価値創出の手段として認識している傾向が強い．特に「ソフトウェア／ソースコードの再利用」に対する関心は 98.3%と極めて高く，「正式には関与していないが依存するプロジェクトへの貢献」といった実践的な側面を重視している．これは，オープンソースの実践から自然な発展として，インナーソースを組織内のソフトウェア開発効率化のための具体的なメカニズムとして捉える認識が定着していることを示唆している．

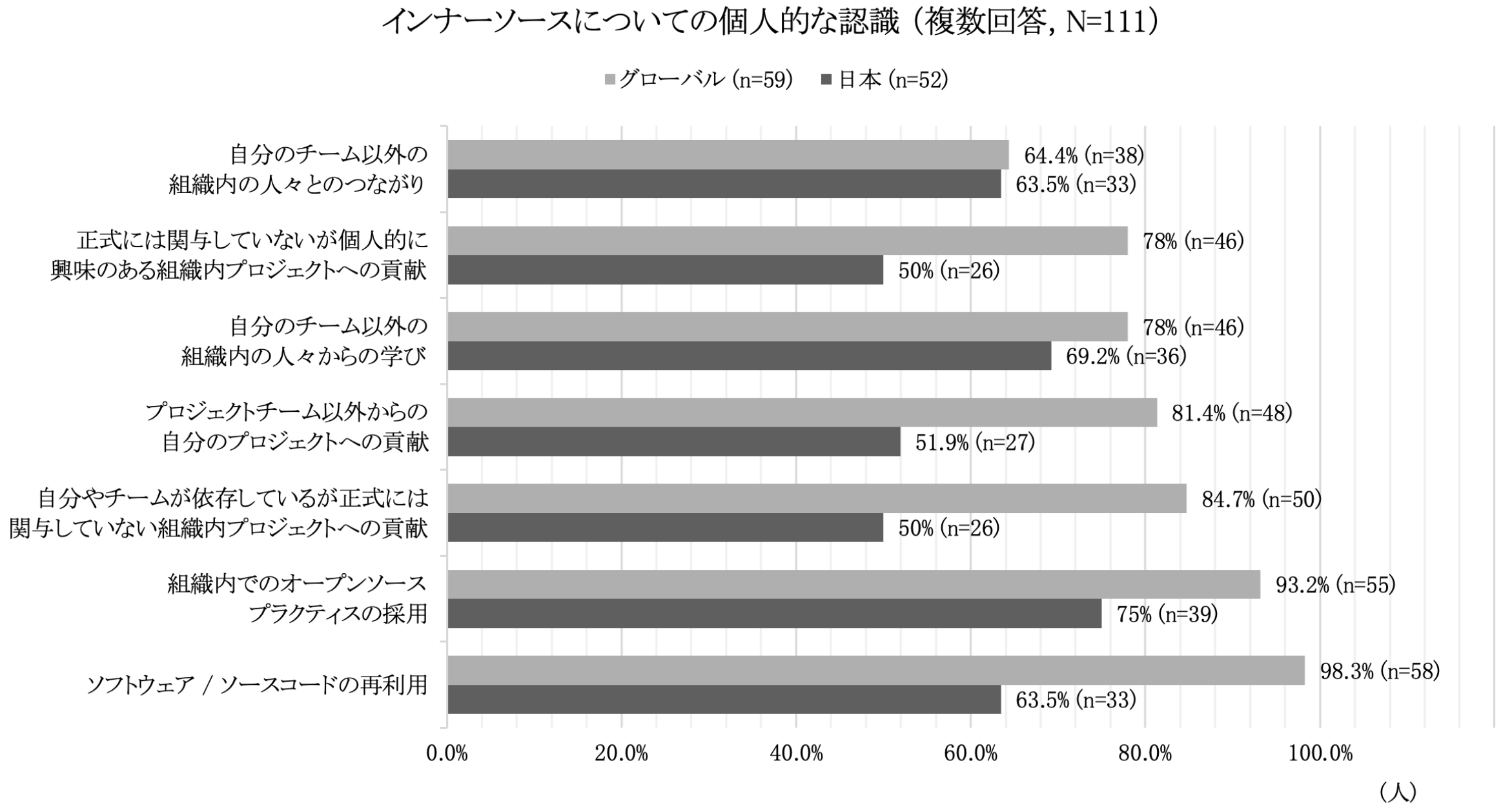

*図 5.6 インナーソースについての個人的な認識*

一方，日本の回答者は「学び」や「つながり」といった抽象的・関係性重視の要素を相対的に高く評価する特徴が見られた．特筆すべきは，「組織内の人々とのつながり」に対する関心がグローバルとほぼ同等の水準を示した点である．しかし，「プロジェクトへの貢献」に対する評価は相対的に低く，これは組織間の壁を取り除きたいという願望は存在するものの，実際の貢献や協働に関する具体的なイメージや実践への移行が十分に確立されていない可能性を示唆している．

第 4 章の終わりで言及したように，この現象は「オープンソースプラクティスの内製化フェーズ」としてのインナーソースと，「パッケージ化された手法」としてのインナーソースという二つの位相の存在を裏付けている．日本における認識パターンは，後者の特徴を強く示しており，これは 2022 年以降のインナーソース関連情報の翻訳・普及過程とも時期的に整合している．

特に注目すべきは，「オープンソースプラクティスの社内適用」という概念に対する認識の差異である．この要素はインナーソースの本質的定義と密接に関連するにもかかわらず，日本の回答者間でその重要性の認識にばらつきが見られた．これは，インナーソースを組織変革やイノベーションの手段として抽象的に捉える傾向が強いことを示唆している．

このような認識の差異は，インナーソースの導入プロセスが段階的な発展を必要とすることを示している．日本のコンテキストでは，まず組織の壁を取り払い，オープンな協働文化を醸成することが初期段階として位置づけられている可能性が高い．これは，具体的な成果を追求する以前の，文化的基盤の構築フェーズとしても解釈できるだろう．

## 5.3 インナーソースへの参加動機の包括的分析

インナーソースへの参加動機に関する包括的な定量調査を実施した結果，地域を超えた普遍的な傾向と，地域特有の特徴的な差異が明確に観察された．最も顕著な共通要素として，「知識共有」への高い期待が日本およびグローバル双方において突出して確認された．これに続いて「組織的サイロとボトルネックの排除」も両地域で高い評価を得ており，この結果は組織横断的な情報流通の促進というインナーソースの基本的価値提案が，地理的・文化的な境界を超えて広く認知されていることを示唆している．

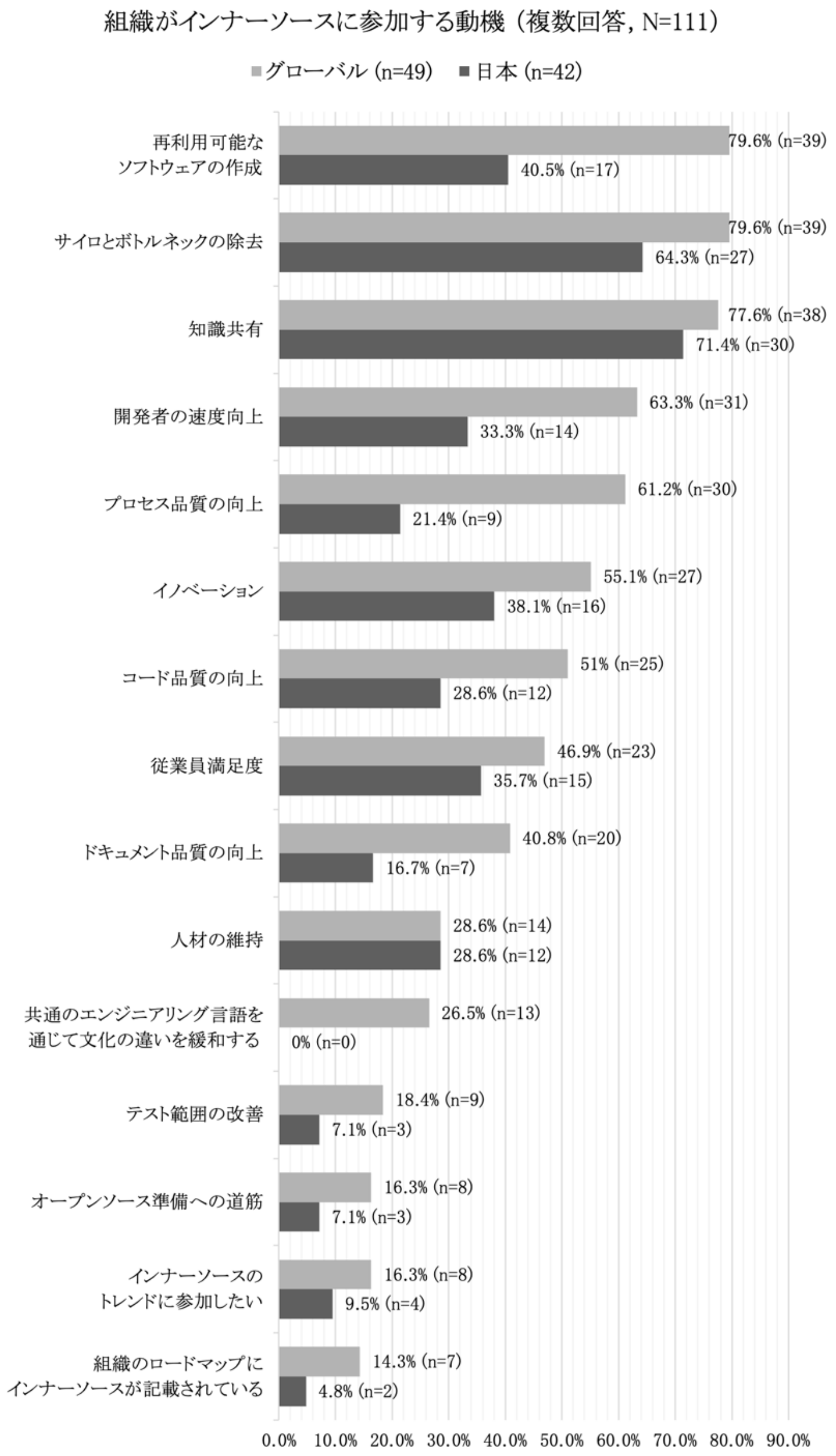

一方で，特筆すべき差異として，「再利用可能なソフトウェアの作成」に対する評価の乖離が観察された．成熟グループに属するグローバル企業の回答者は，インナーソースを通じたソフトウェア再利用の促進，開発速度の向上，コード品質の改善など，定量的に測定可能な技術的成果に直結する要素を強い動機として選択する傾向が顕著であった．これは，インナーソースが技術的資産の最適化や生産性向上を実現する具体的な手段として認識されていることを示している．対照的に，初期段階グループである日本企業の回答者は，こうした定量的な技術的メリットよりも，知識共有やサイロ除去といった組織文化の変革に関連する効用を重視する傾向が明確であった．

この結果は，「インナーソースへの個人的認識（図 5.6）」で示された分析とも整合性を持つ．すなわち，初期段階グループにおいては，インナーソースが直接的な生産性指標の改善手段というよりも，組織内のコミュニケーションとコラボレーションを深化させ，文化的変革をもたらす触媒として認識されている．この認識の差異は，組織の成熟度や開発文化の違いに起因すると考えられ，インナーソース導入における多様なアプローチの必要性を示唆している．

さらに注目すべき点として，オープンソース実践への準備をインナーソース導入の主たる参加動機とした回答が，両集団において極めて限定的であった点が挙げられる．この結果は，インナーソースが必ず

しもオープンソースコミュニティへの参画に向けた中間段階として位置付けられているわけではないことを示している．むしろ，インナーソースは組織内部での効率化や文化的改善のための独立した戦略的施策として認識されていると解釈できる．

総合的な分析として，インナーソースへの参加動機は，知識共有と組織内コラボレーション強化という共通基盤を持ちながら，初期段階グループでは文化的・関係改善重視，成熟グループでは技術的・業務効率的な指標重視という異なる重点化が観察された．この相違は，組織の成熟度や開発文化，さらには経営戦略上の優先順位の違いを反映していると考えられる．これらの知見は，インナーソース導入において，組織の成熟度や文化的背景に応じた柔軟なアプローチの必要性を示唆している．

## 5.4　導入段階別にみるインナーソース採用動機の変遷

インナーソースの組織的導入において，各導入段階（アイデア段階，パイロット段階，初期採用段階，成長・成熟段階）に着目すると，組織の成熟度とともに採用動機が段階的に変容する構造が見出される．ここでは，日本とグローバル両方の回答データを相互補完的に参照し，2024 年度における包括的なインナーソース採用動機の動態を捉えることを試みた．日本側のデータは成長・成熟段階の回答数が限定的であり，グローバル側はアイデア段階が少ないという制約を伴うが，両者を統合することで，各導入段階を連続的に俯瞰できる利点が生まれる．結果として，初期段階では主に知識共有やサイロ解消といった基礎的メリットの期待が中心に位置づけられ，後期段階に進むほどコード品質やドキュメント品質など，より具体的かつ実務的な利点に対する意識が高まるという全体像が描き出された．

こうした変遷過程は，導入段階の進行に従ってインナーソースが担う役割が段階的に拡大し，企業文化や開発プロセス全体に浸透していく様子を反映していると考えられる．次の小節では，グローバルおよび日本のデータをそれぞれ詳説し，最後に統合的見地から考察を行う．以降のレーダーチャートでは，いずれかのデータセットで 40%を超えた主要項目の推移を示す．

### 5.4.1 グローバル企業における導入段階と採用要因の特徴

グローバルの回答データ（n=45）を俯瞰すると，インナーソース導入の動機はパイロット段階から初期採用段階，そして成長・成熟段階にかけて顕著に増大する．今回，アイデア段階のサンプル数が 3 件に限られていたため統計的に一般化できるほどの代表性は担保されていないが，パイロットから成長・成熟への移行期において，サイロやボトルネックの除去，ソフトウェア再利用，知識共有などの動機が高い割合で維持されつつ，プロセス品質，コード品質，ドキュメント品質などの技術的要素への期待が増大する傾向が確認された．これは，インナーソースを導入する初期段階では，業務効率の向上や既存リソースの有効活用といった即時的メリットが優先される一方，組織内での実践経験が蓄積するにつれ，長期的に開発品質を高める仕組みづくりを模索するようになるというプロセスを示唆している．

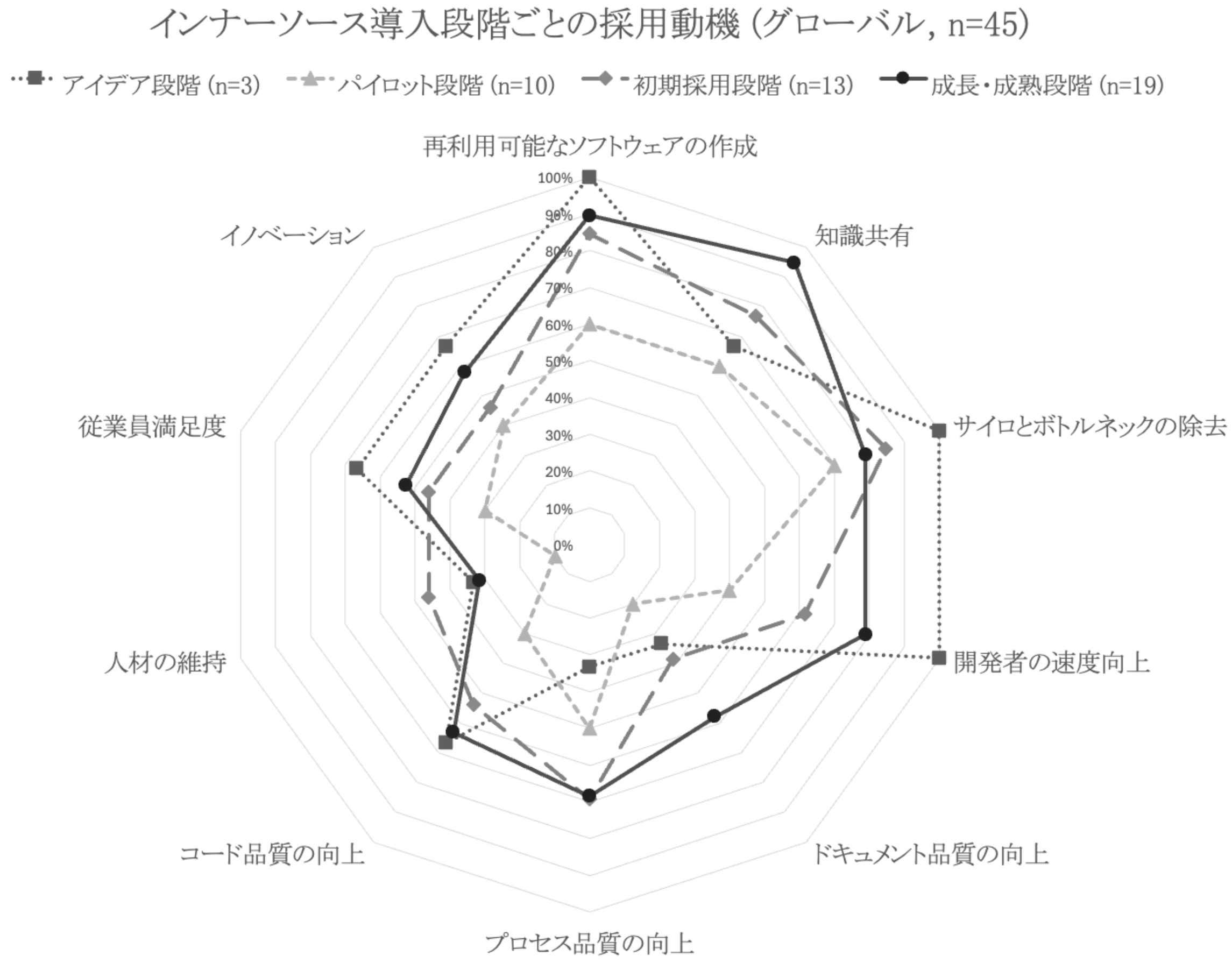

*図 5.8 インナーソース導入段階ごとの採用動機(グローバル)*

*表 5.2 インナーソース導入段階ごとの採用動機(グローバル)*

| アイデア段階 (n=3) | パイロット段階 (n=10) | 初期採用段階 (n=13) | 成長・成熟段階 (n=19) |
|---|---|---|---|
| サイロとボトルネックの除去 (100%) | サイロとボトルネックの除去 (70%) | サイロとボトルネックの除去 (84.6%) | 知識共有 (94.7%) |
| 再利用可能なソフトウェアの作成 (100%) | 再利用可能なソフトウェアの作成 (60%) | 再利用可能なソフトウェアの作成 (84.6%) | 再利用可能なソフトウェアの作成 (89.5%) |
| 開発者の速度向上 (100%) | 知識共有 (60%) | 知識共有 (76.9%) | サイロとボトルネックの除去 (78.9%) |
| イノベーション (66.7%) | プロセス品質の向上 (50%) | プロセス品質の向上 (69.2%) | 開発者の速度向上 (78.9%) |
| インナーソースのトレンドに参加したい (66.7%) | イノベーション (40%) | 開発者の速度向上 (61.5%) | プロセス品質の向上 (68.4%) |
| コード品質の向上 (66.7%) | 開発者の速度向上 (40%) | コード品質の向上 (53.8%) | コード品質の向上 (63.2%) |
| 従業員満足度 (66.7%) | | イノベーション (46.2%) | イノベーション (57.9%) |
| 知識共有 (66.7%) | | 人材の維持 (46.2%) | ドキュメント品質の向上 (57.9%) |
| | | 従業員満足度 (46.2%) | 従業員満足度 (52.6%) |

表 5.2 は，グローバルにおいて各導入段階で 40%以上の回答を得た採用動機を抜粋したものである．パイロット段階から成長・成熟段階に進むほど項目全体の数値が上昇するとともに，関心を持つ要素の幅も拡大することが見て取れる．

今回得られたグローバルサンプルを導入段階別に比較した際，レーダーチャート上でパイロットから初期採用，さらに成長・成熟へと移行するにつれて評価項目の面積が拡大する現象が観察された．具体的には，アイデア段階ではインナーソースの有用性が漠然と期待されるのみだったとしても，パイロット段階で特定の領域における実践的な検証が進むと，特定のビジネス課題（サイロの解消やソフトウェア再利用など）をターゲットに取り組みが狭まる傾向にある．その後の初期採用や成長・成熟段階では，導入に成功したプロジェクトが他の部門へ波及して全社的なプロセス改善やコードレビュー文化の構築が促され，結果として品質向上やプロセス最適化への意識が大きく高まると推測される．

一方で，アイデア段階の一部回答者がインナーソースに対して広範な期待を述べている点は注目に値する．組織全体でアジリティ向上やイノベーション喚起を図るべく，早期からあらゆる可能性を見出したいというトップダウンの意志が働いているケースでは，パイロット段階で急激に導入意義が絞られていく現象が起こりやすい．つまり，実証可能な目標にフォーカスするために動機が部分的に限定されるが，その後，プロジェクトが拡充するとともに再度認識が広がり，多角的なメリットを重視する局面が訪れると考えられる．このような導入初期の揺らぎをより精緻に理解するためには，日本側のサンプルとの比較によって導入段階間の相違点を解釈することが有益である．

### 5.4.2 日本企業の導入実態と特徴的な動機

日本の回答データ（n=42）は，アイデア段階の回答が最も多く，成長・成熟段階の数が非常に少ないという逆の分布を示している．つまり，導入初期段階でインナーソースに対する漠然とした期待や関心が集まる一方，実際に成長・成熟まで達した組織事例が極めて限られている．さらに日本のサンプルでは，パイロットや初期採用段階におけるインナーソース動機の回答率がグローバルよりもやや低く，知識共有やサイロ解消など特定の項目が高めに現れる一方で，コード品質やドキュメント品質，開発速度といった技術的・実務的領域は必ずしも大きなモチベーション要因とはなっていない．

この背景には，国内企業の多くが組織文化や既存の品質管理プロセスを重視し，インナーソース導入を必要最低限の範囲で試験運用する傾向があることが考えられる．日本特有の職務分掌やチーム体制，あるいは長期的なジョブローテーション文化のもとでは，インナーソースの利点を的確に測定しづらい場面があるため，まずは知識共有や部門間連携の一部改善を目的とした小規模導入から始める企業が多いと推測される．その結果，パイロット段階や初期採用段階で知識共有やサイロ解消などのメリットは一定の肯定感を得るものの，より広範な技術的メリットにまで認識が及ぶ前に足踏みしてしまうケースが散見される．

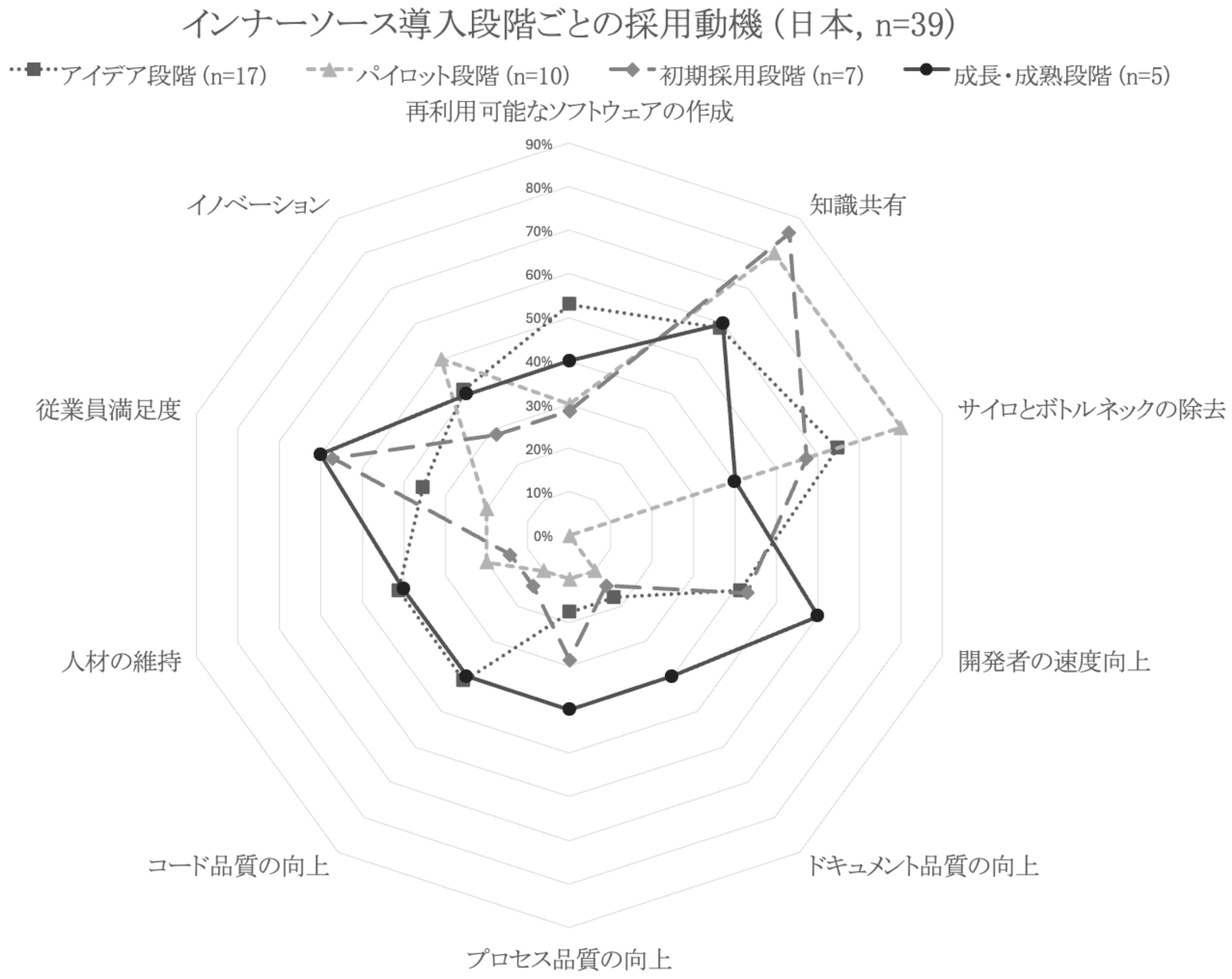

*図 5.9 インナーソース導入段階ごとの採用動機（日本）*

表 5.3 は日本の段階別データをまとめたもので，グローバルの傾向と併せて参照すると，アイデア段階が多い日本ではインナーソースの価値が必ずしも広く理解されていないものの，導入初期では一定のモチベーション要因が絞り込まれている様子が読み取れる．さらに成長・成熟段階では社員の意欲向上に着目が集まる一方，再利用などの技術的メリットはあまり顕在化していない可能性がある．

*表 5.3 インナーソース導入段階ごとの採用動機（日本）*

| アイデア段階（n=17） | パイロット段階（n=10） | 初期採用段階（n=7） | 成長・成熟段階（n=5） |
|---|---|---|---|
| サイロとボトルネックの除去（64.7%） | サイロとボトルネックの除去（80%） | 知識共有（85.7%） | 従業員満足度（60%） |
| 知識共有（58.8%） | 知識共有（80%） | サイロとボトルネックの除去（57.1%） | 知識共有（60%） |
| 再利用可能なソフトウェアの作成（52.9%） | イノベーション（50%） | 従業員満足度（57.1%） | 開発者の速度向上（60%） |
| イノベーション（41.2%） | | 共通のエンジニアリング言語を通じて文化の違いを緩和する（42.9%） | イノベーション（40%） |
| コード品質の向上（41.2%） | | 開発者の速度向上（42.9%） | コード品質の向上（40%） |
| 人材の維持（41.2%） | | | サイロとボトルネックの除去（40%） |
| 開発者の速度向上（41.2%） | | | テスト範囲の改善（40%） |
| | | | ドキュメント品質の向上（40%） |
| | | | プロセス品質の向上（40%） |

成長・成熟段階に達した企業の回答を分析すると，従業員満足度の向上や企業文化の変革といった社会的・文化的側面での効果が顕著に表れている．これは，インナーソースの導入が進んだ組織において，技術的側面よりも組織変革や人材マネジメントの文脈で評価される傾向があることを示唆している．ただし，成長・成熟段階に至った企業のサンプル数が限定的であることから，この傾向を一般化することには慎重を期す必要がある．また，パイロットを経て初期導入した段階でも技術的手法としての利点が十分認知されにくいことを踏まえると，導入段階が進んだ段階で改めて技術的メリットを享受できる仕組みを整備する必要があると考えられる．これらの要因については，より詳細なケーススタディが求められる．

### 5.4.3 日本・グローバル統合データからみる全体的な動向分析

日本とグローバル双方のデータを統合することで，アイデア段階から成長・成熟段階までの包括的な採用動機の推移を把握できる．アイデア段階の豊富な日本のサンプルと，成長・成熟段階でのグローバルデータを組み合わせた結果，インナーソースの導入期には知識共有やサイロの除去といった基本的な組織メリットに期待が集まり，導入が進むにつれてプロセス品質やコード品質，ドキュメント品質などの高度な技術的要素に対する評価が高まるプロセスが明確になった．これは，導入段階を経るごとにインナーソースが新たな部門やプロジェクトへ広がり，多様なユースケースで効果が検証される過程を示している．

レーダーチャートの面積が導入段階進行に従って拡大するという観察結果は，導入当初に局所的なニーズを満たすための手段として注目されていたインナーソースが，経験の蓄積とともに品質管理や開発スピードの向上にまで貢献するようになる変遷を示唆する．日本では技術面よりも従業員満足度やイノベーション意欲など組織文化的な指標が強調される一方，グローバルのデータでは開発速度やコード品質といったエンジニアリング指標への意識が後期段階において顕著に高まりやすい．この違いは，企業文化や組織マネジメントの方針に起因する可能性が高く，大規模開発経験の豊富なグローバル企業ほど技術的要因を定量的に評価しやすい環境にあると考えられる．

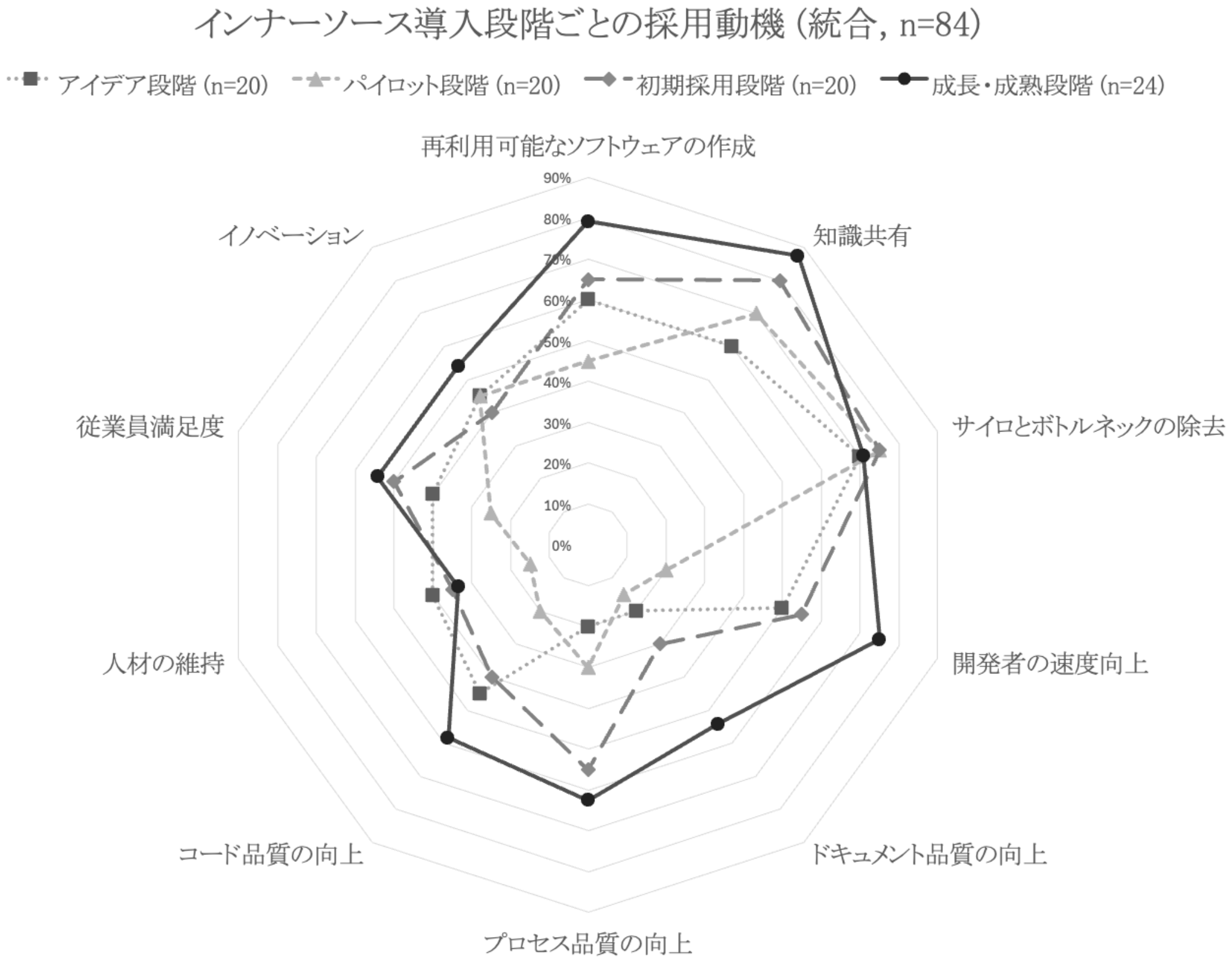

*図 5.10 インナーソース導入段階ごとの採用動機（全体）*

表 5.4 は，両データを統合した際の各導入段階別回答である．最終的に成長・成熟段階に到達した組織ほど広範なベネフィットを感じており，浸透度が高まるにつれインナーソースの実質的効果が開発速度や品質向上にまで及ぶという動態が示唆される．

*表 5.4 インナーソース導入段階ごとの採用動機（全体）*

| アイデア段階（n=20） | パイロット段階（n=20） | 初期採用段階（n=20） | 成長・成熟段階（n=24） |
|---|---|---|---|
| サイロとボトルネックの除去（70%） | サイロとボトルネックの除去（75%） | 知識共有（80%） | 知識共有（88%） |
| 再利用可能なソフトウェアの作成（60%） | 知識共有（70%） | サイロとボトルネックの除去（75%） | 再利用可能なソフトウェアの作成（79%） |
| 知識共有（60%） | 再利用可能なソフトウェアの作成（45%） | 再利用可能なソフトウェアの作成（65%） | 開発者の速度向上（75%） |
| 開発者の速度向上（50%） | イノベーション（45%） | 開発者の速度向上（55%） | サイロとボトルネックの除去（71%） |
| コード品質の向上（45%） | | プロセス品質の向上（55%） | プロセス品質の向上（63%） |
| イノベーション（45%） | | 従業員満足度（50%） | コード品質の向上（58%） |
| 人材の維持（40%） | | コード品質の向上（40%） | ドキュメント品質の向上（54%） |
| 従業員満足度（40%） | | イノベーション（40%） | 従業員満足度（54%） |
| | | | イノベーション（54%） |

統合分析の最終的な示唆として，インナーソースは導入段階の進行にともない技術的メリットや組織的メリットが段階的に拡大すると捉えられるが，その具体的な内訳や重点領域は各国・各組織文化に応じて

大きく異なる傾向がある．このため，アイデア段階からパイロット段階への移行時には導入目的を絞り込み，プロジェクト成功事例を積み上げげる戦略が重要である一方，初期採用から成長・成熟段階に至る段階では，開発品質や組織改革など多角的な視点で価値を捉えていくことが，最終的な導入効果を最大化する鍵になると考えられる．日本では従業員満足度や人事戦略との関連が相対的に重視される傾向があり，グローバルでは技術的効率や品質管理が鮮明に捉えられるため，それぞれの狙いや組織特性を踏まえた導入計画を策定する必要があるといえる．

### 5.4.4 採用動機と組織属性との関連性検証

グローバル回答者群(n=49)と日本回答者群(n=42)のそれぞれについて，インナーソースの採用動機と組織規模や開発チーム規模などの属性との関連を明らかにするため，クロス集計表を作成し，各項目間の独立性をカイ二乗検定によって評価した．クロス集計は，インナーソース導入を促す要因(例としてサイロとボトルネックの除去やイノベーションなど)を行別，組織規模や開発チームの大きさなどの属性を列別に配置し，回答者の分布に偏りがあるかどうかを視覚的に比較する目的で行った．カイ二乗検定では，いずれの動機項目と属性間の関連が統計的に有意と言えるかを確かめ，p 値の大小をもとに採用動機が組織特性と強く結びついているかどうかを判定した．

グローバルの回答者群については，「サイロとボトルネックの除去」が組織規模と有意に関連していることが明らかになった．具体的には，従業員数 5000 人以上の大規模組織ほど，この動機を強く支持する傾向が統計的に示唆され($\chi^2$=8.39, p<.01)，部門間の情報格差を解消して全社的な知の循環を促す必要性が高いと認識されている可能性が高い．

このグローバル分析では，イノベーション創出に関して，組織のインナーソース経験の有無(パイロット段階以降)や開発チーム規模(500 名以上)との間に若干の関連(p 値が.10 レベル)も観察された．この結果は，大規模開発組織を擁する企業やインナーソース経験を蓄積してきた企業ほど，インナーソースをイノベーション促進の手段として活用する見込みが高まる可能性を示している．

一方，日本の回答者群における「サイロとボトルネックの除去」と組織規模の関連は弱い傾向にとどまった($\chi^2$=3.02, p≈.08)．これは国内の大企業においても，部門間連携の必要性は意識されているものの，グローバル企業ほど明確な定量的アプローチでインナーソースを組織課題の解消に活用していないか，あるいは導入段階によるバラつきが大きい可能性を示唆する．

日本において特に注目されるのは，「イノベーション」が開発チーム規模(500 名以上)と有意な関連を示した点であり($\chi^2$=7.17, p<.05)，大規模開発チームを持つ企業ほどインナーソースによる新規技術や新事業の創出を期待しやすい傾向が明確に表れた．こうした結果は，国内の一部企業はインナーソースを組織改革や人材活性化だけでなく，より直接的な事業革新のエンジンとして位置づけている可能性を裏付ける．

*表 5.5 インナーソース採用動機にかかる属性感の独立性*

**インナーソースの採用動機の該当有無と属性間の独立性をカイ二乗検定（$\chi^2$）で評価 (N=111)**

| | 人材 | | 組織 | | | 効率化 | | | 品質 | | | 戦略 | | | |
|---|---|---|---|---|---|---|---|---|---|---|---|---|---|---|---|
| | 人材の維持 | 従業員満足度 | サイロとボトルネックの除去 | 知識共有 | 共通のエンジニアリング言語を通じて文化の違いを緩和する | 再利用可能なソフトウェアの作成 | 開発者の速度向上 | テスト範囲の改善 | コード品質の向上 | プロセス品質の向上 | ドキュメント品質の向上 | オープンソース準備への道筋 | イノベーション | インナーソースのトレンドに参加したい | 組織のロードマップにインナーソースが記載されている |
| **グローバル (n=49)** | | | | | | | | | | | | | | | |
| 組織の規模 *1 | 0.54 | 1.05 | 8.39 | 0.23 | 0.52 | 0.32 | 0.01 | 0.04 | 1.18 | 1.04 | 0.64 | 4.76 | 0.03 | 0.56 | 0.65 |
| (sig P,* ≦ .05, † ≦ .10) | 0.46 | 0.30 | **0.00*** | 0.63 | 0.47 | 0.57 | 0.91 | 0.83 | 0.28 | 0.31 | 0.42 | **0.03*** | 0.86 | 0.46 | 0.42 |
| 開発組織の規模 *2 | 0.04 | 0.00 | 0.13 | 0.13 | 0.01 | 1.41 | 1.22 | 0.13 | 0.04 | 0.04 | 1.16 | 2.80 | 2.78 | 0.20 | 0.36 |
| (sig P,* ≦ .05, † ≦ .10) | 0.83 | 1.00 | 0.72 | 0.72 | 0.91 | 0.23 | 0.27 | 0.72 | 0.85 | 0.84 | 0.28 | 0.09† | 0.10† | 0.65 | 0.55 |
| 開発者か否か *3 | 0.22 | 1.72 | 0.02 | 0.81 | 4.77 | 0.02 | 3.11 | 0.11 | 0.09 | 1.27 | 0.99 | 0.22 | 0.60 | 0.28 | 0.09 |
| (sig P,* ≦ .05, † ≦ .10) | 0.64 | 0.19 | 0.88 | 0.37 | **0.03*** | 0.88 | 0.08† | 0.74 | 0.76 | 0.26 | 0.32 | 0.64 | 0.44 | 0.60 | 0.76 |
| **日本 (n=42)** | | | | | | | | | | | | | | | |
| 組織の規模 *1 | 1.71 | 0.91 | 3.02 | 0.23 | 0.30 | 0.07 | 0.64 | 0.27 | 0.23 | 0.72 | 0.25 | 0.27 | 1.48 | 1.21 | 1.02 |
| (sig P,* ≦ .05, † ≦ .10) | 0.19 | 0.34 | 0.08† | 0.63 | 0.58 | 0.78 | 0.43 | 0.60 | 0.63 | 0.39 | 0.61 | 0.60 | 0.22 | 0.27 | 0.31 |
| 開発組織の規模 *2 | 1.30 | 0.13 | 1.10 | 0.06 | 1.08 | 2.06 | 1.61 | 0.02 | 1.30 | 0.29 | 0.13 | 0.02 | 7.17 | 0.33 | 0.16 |
| (sig P,* ≦ .05, † ≦ .10) | 0.25 | 0.72 | 0.29 | 0.81 | 0.30 | 0.15 | 0.20 | 0.90 | 0.25 | 0.59 | 0.71 | 0.90 | **0.01*** | 0.56 | 0.69 |
| 開発者か否か *3 | 0.79 | 8.28 | 0.00 | 0.01 | 0.01 | 1.08 | 0.06 | 1.15 | 2.77 | 0.09 | 0.62 | 1.15 | 0.74 | 0.00 | 0.75 |
| (sig P,* ≦ .05, † ≦ .10) | 0.37 | **0.00*** | 0.96 | 0.91 | 0.92 | 0.30 | 0.80 | 0.28 | 0.10† | 0.76 | 0.43 | 0.28 | 0.39 | 0.95 | 0.39 |

*1 組織規模が5000人以上である
*2 開発チームの規模が500名以上である
*3 職務が「開発者」であるか

グローバルとの比較では，大規模組織がサイロ解消の必要性を強く感じている様子が海外企業のデータから浮かび上がる一方，日本では組織規模の影響よりも開発チームの規模や技術者コミュニティのマネジメントに起因するイノベーション機能への期待が顕在化している．これは，国内企業が従業員満足度や企業文化の改善を重視しがちな一方で，大規模な技術者集団のなかに新規開発領域を切り拓く潜在力を強く見出している構図ともいえる．

これらの定量分析の知見は，インナーソースが導入される過程で企業がどのような動機づけをもとに意思決定を行うかを理解するうえで重要な補足情報となる．特に大規模組織や大規模開発チームでは，当初からサイロ解消やイノベーションの創出といった具体的なニーズがはっきりしており，それらがインナーソース採用の初期段階において強い推進力を持つことが示唆された．

### 5.4.5 インナーソース導入要因の総括

インナーソースの導入プロセスにおける採用動機は，アイデア段階からパイロット段階，初期採用段階，そして成長・成熟段階へと進むにつれて，組織課題の解決から始まり，ソフトウェア開発上の技術的効果，さらにはイノベーションや従業員満足度などの多角的ベネフィットへと焦点が移り変わることが明らかになった．初期段階ではサイロ解消や知識共有といった緊急度の高い組織課題に強い関心が寄せられる一方，パイロット段階を経ることで効果が実感されると，品質や速度といった開発効率への期待が拡大し，成長・成熟段階では組織文化や人材マネジメントまで含む包括的な価値向上を見据える動機へと発展する．

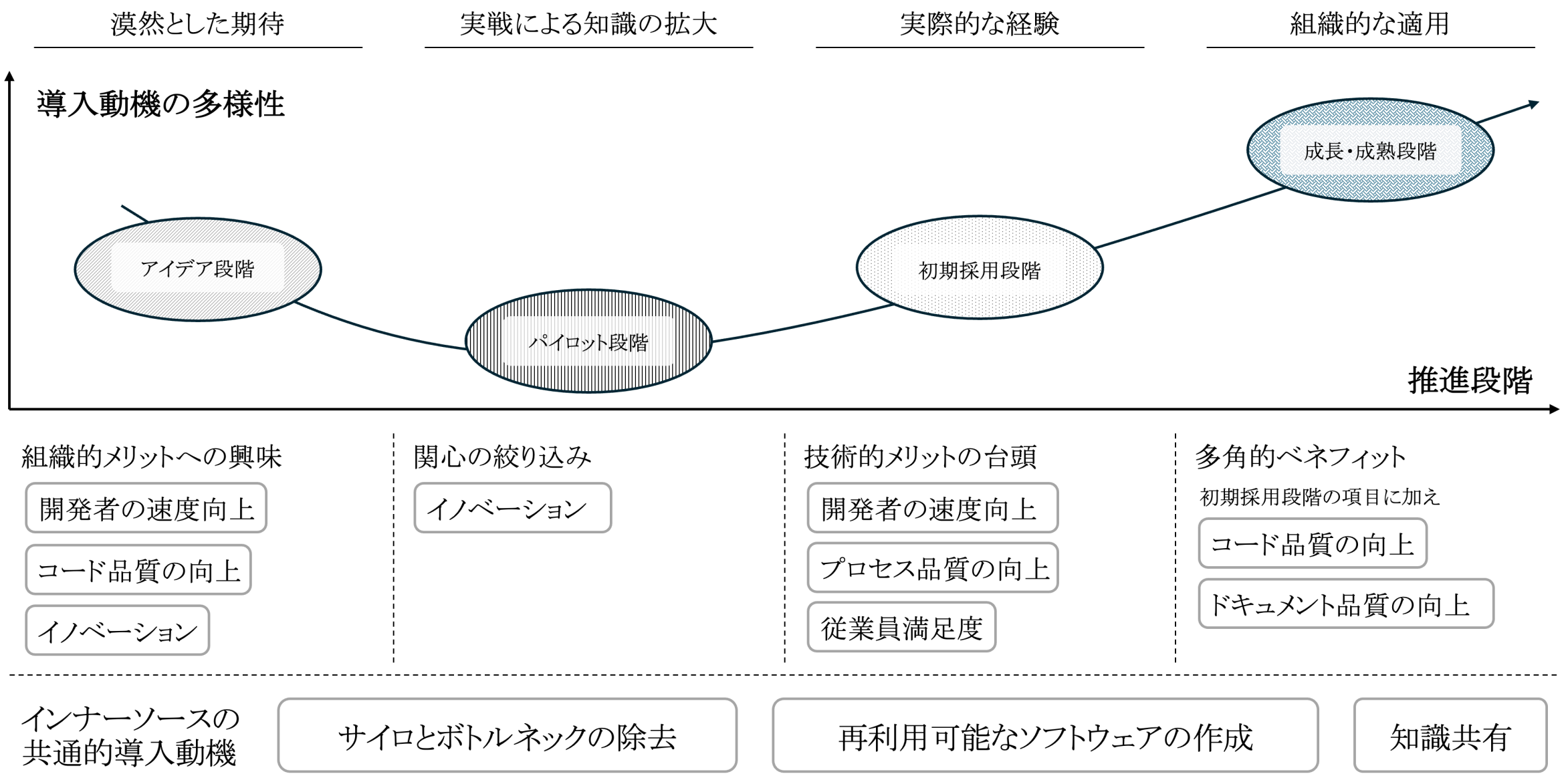

*図 5.11 インナーソース採用動機の変遷と経験による多様な動機づけの出現*

こうした段階的な変化は，企業の属性によっても異なる側面を見せる．大規模組織ではサイロやボトルネックの除去が特に重要視される傾向があり，大規模開発チームを擁する企業ほどイノベーションを得る手段としてインナーソース導入を検討しやすい．日本では管理職層の比率が高い回答群に従業員満足度への関心が多く表れた一方で，純粋な開発者層だけでは必ずしもそれが重視されないという構造が示唆された．

さらに，クロス集計とカイ二乗検定を通じて，オープンソースへの準備段階を大企業や大規模開発組織が重視するのはグローバルで顕著な動向であり，日本では企業規模と直接的に結びつく傾向が見られない点が浮き彫りとなった．これは，オープンソースの利活用や外部連携を積極的に進める海外企業と，必ずしもそれを優先課題としない日本企業との温度差を反映していると考えられるが，いずれの国や地域においても，導入段階の異なる企業間で動機や期待値が大きく変化することに変わりはない．

最終的に，インナーソース採用の背景には企業規模や開発チーム規模，導入をリードする人物の職責や視点といった多様な要因が複合的に影響し合うため，導入目的や期待効果を明確にした上で導入段

階に応じた戦略設計を行う必要がある．属性差と導入段階の進行度を照らし合わせながら，組織内でコンセンサスを形成し，インナーソースの多面的なメリットを最大化するための手法を検討することが，インナーソースの有効な導入推進の鍵となる．

# 第6章インナーソース導入における組織的障壁

インナーソース導入プロセスにおける組織的障壁の定量的分析は，効果的な導入戦略の策定において重要な示唆を提供する．本章では，日本企業とグローバル企業における導入障壁の比較分析を通じて，その特徴的パターンと組織的含意を明らかにする．

測定手法の妥当性を担保するため，データの標準化プロセスを実施した．具体的には，日本企業から収集した五段階リッカート尺度データを二値データへと変換し，グローバル企業から直接収集した二値形式データとの比較可能性を確保した．この変換過程では，「かなり大きい」「決定的に大きい」という上位 2 段階の回答を，重大な導入障壁として認識されているものと操作的に定義した．

統計的特性の検証結果は，両データセット間の比較可能性を支持している．日本企業データとグローバル企業データの平均値（0.351 対 0.397）および中央値（0.372 対 0.398）は近似した分布を示している．ただし，標準偏差の差異（日本 0.116，グローバル 0.180）は，日本企業データにおける変動幅の抑制を示唆している．これは，リッカート尺度から二値データへの変換過程で生じた情報の集約効果によるものと解釈できる．

このような測定特性の違いは認められるものの，両データセットの数値レンジに顕著な乖離は観察されなかった．したがって，本研究では，これらのデータセットを用いて，インナーソース導入プロセスの各段階における組織的障壁を，統計的に頑健な方法で比較分析することが可能であると判断した．この分析枠組みは，文化的背景の異なる組織間における導入障壁の共通点と相違点を明らかにする上で，有効な視座を提供するものである．

## 6.1　インナーソース導入における主要障壁の概観

調査結果からは，組織の規模や地域性を問わず，インナーソース導入における障壁が主として文化的・組織的な要因に起因することが示唆された．特筆すべき点は，特に導入初期について技術的な課題よりも，組織文化やマインドセットの変革に関連する課題が上位を占めている点である．この傾向は，従来型の階層的組織構造から，より開放的で協調的な組織形態への移行における本質的な困難さを示唆している．

このような文化的・組織的要因が特に強く影響を及ぼす理由としては，インナーソースの導入が単なるツールや手法の変更ではなく，既存の組織体制や業務プロセスを再構築することに近い取り組みである点が挙げられる．たとえば，トップダウンによる従来型の指示命令系統を維持しながら，エンジニア個人の自発的な貢献や協働を促すという二重構造を整合的に運用するのは容易ではない．一方で開発者同士がコードをオープンに共有し合うためには，評価やインセンティブの基準も見直しが求められるため，導入担当部門だけでなく，人事部門や法務部門などを巻き込んだ調整が不可欠となる．

### 6.1.1　グローバル企業における導入障壁の特徴

グローバル企業において，「組織文化やサイロ思考」「認知の不足」「中間管理職の理解不足」はインナーソース導入時の主要な障壁として認識されている．これは，開発文化の内製化を目指す組織において，従来の部門間の壁や階層性が機能不全を引き起こしていることを示唆している．

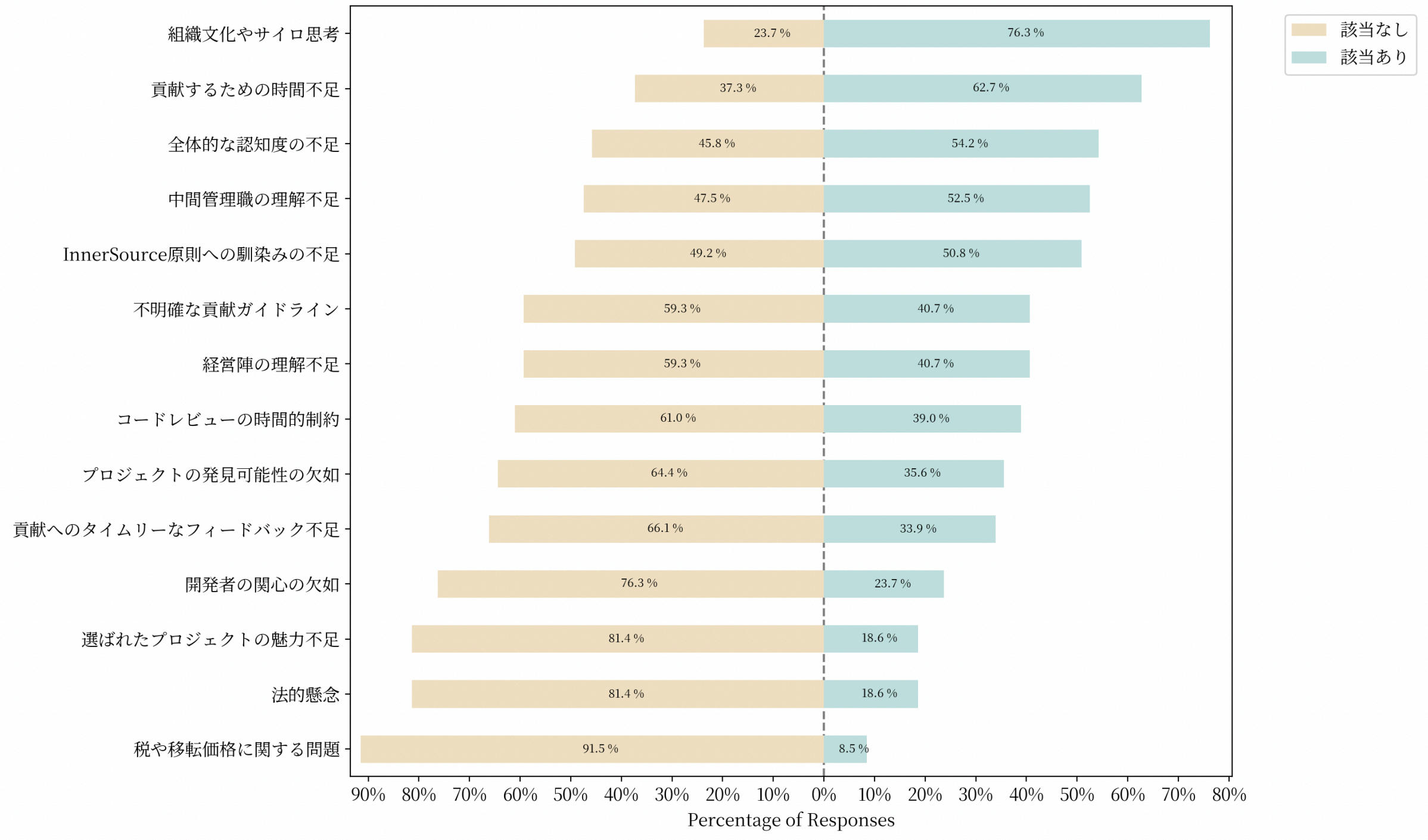

*図 6.1 インナーソースの成功に対する重大な障壁(グローバル)*

グローバル企業の特徴的な点として,「開発者の関心の欠如」が主要な障壁としてあまり認識されていない傾向が確認された. この背景には, エンジニアの高い流動性と, キャリア形成における新技術や方法論への積極的な関与を重視する文化的要因が存在すると考えられる. 日本で見られがちな終身雇用の意識や固定化されたキャリアパスが存在しにくい環境では, インナーソースがむしろ個人のスキルアップや実績蓄積の手段として捉えられやすく, 開発者が自発的に貢献するモチベーションが保たれやすいという現象が推察される.

税務上の問題や法的リスクについては, 全体的に小さな障壁と見なされている一方で, 特定の業界や国際的規制の厳しい産業では深刻な制約となりうる. 知的財産権や移転価格税制に関する複雑性は, 一部の国や地域の制度と相まってインナーソース活動に不確実性をもたらす可能性がある.

### 6.1.2 日本企業における導入障壁の特徴

リッカート尺度に基づく分析によれば, 日本企業におけるインナーソース導入で特に大きな障壁として浮上したのは, 全体的な認知度の不足, 組織分化やサイロ思考, 開発者の関心の欠如, そして中間管理職の理解不足である. 「かなり大きい」「決定的に大きい」と回答した割合は, いずれも過半数を超えている.

注目すべき点として, 日本企業では「開発者の関心の欠如」が大きな障壁として第三位に挙げられ, そのうち「決定的に大きい」と認識した回答は 26%に達した. この傾向はグローバル調査との顕著な差異を示している. サイロ化した組織構造の中で開発を行っているエンジニアが, 新たなコラボレーション手法や技術導入に興味を持てなくなっている現状が示唆される. この背景には, 長期にわたるプロジェクト型開

発やシステムインテグレーション企業の業態などが影響していると推測される．自社プロダクトではなく顧客企業や親会社のプロジェクトを中心に受託する場合，インナーソースに基づく貢献活動を行うだけの余裕がなく，慣習的な開発手法に縛られたままになってしまう可能性が高い．こうした状況では，開発者個人の意欲が高くても，組織的な枠組みやプロセスがそれをサポートできないため，関心やモチベーションが低下する悪循環に陥りやすい．

また，中間管理職が障壁として認識された背景には，日本企業における意思決定や業務推進が中間管理職の合意形成に大きく依存している構造があると考えられるが，一方で，経営陣の理解不足も同様に重要な課題として指摘されており，組織全体における認識の欠如が根本的な問題として浮き彫りとなっている．

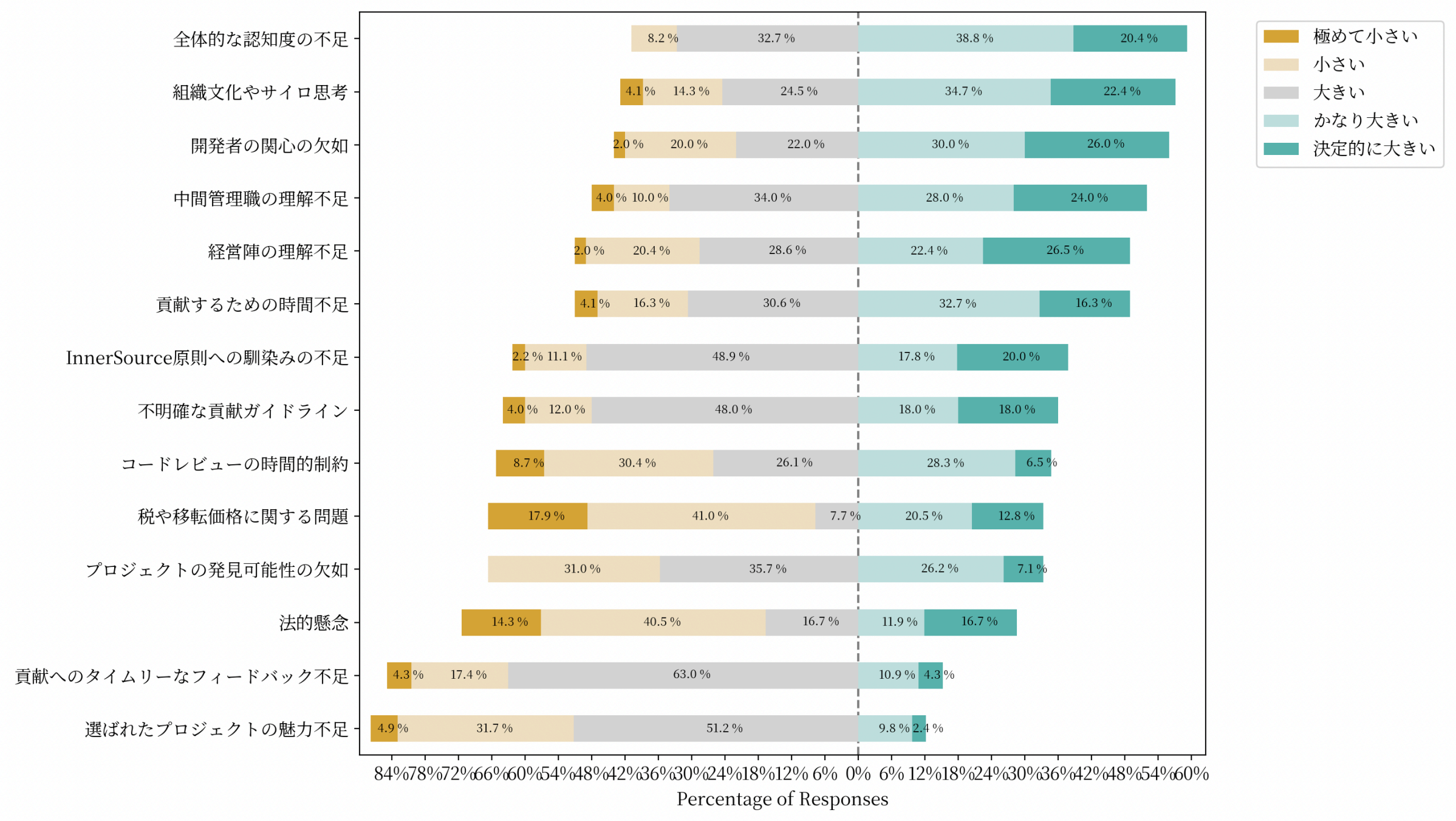

*図 6.2 インナーソースの成功に対する重大な障壁（日本）*

また「貢献するための時間不足」も比較的大きな障壁であると指摘されている一方で，貢献後に生じる「コードレビューの時間的制約」や「貢献へのタイムリーなフィードバック不足」は比較的低い割合にとどまり，日本企業におけるインナーソースの実践がまだ初期段階にあることを裏付けている．初期段階ではまず貢献そのものが実現しづらく，その先にあるレビューやフィードバックに関する問題が顕在化していない可能性が高い．

税務や法的問題は比較的小さな障壁として認識されているものの，その重要性は看過できない．インナーソース活動の拡大に伴い，海外拠点や異なる部門間の協働が増加した場合，知的財産権の取り扱いや移転価格税制に関する法的リスクが顕在化する可能性がある．特筆すべきは，これらの課題に対する認識が二極化している点であり，一部の回答者にとっては重大な懸念事項となっている一方で，他の回答者は軽微な問題として捉えている．

## 6.2 導入段階別に見る障壁の変容

インナーソースの導入プロセスにおいて，組織が直面する障壁は静的なものではなく，導入段階の進展に伴って大きく変容する．本研究では，グローバルサンプル（n=51）と日本サンプル（n=43）を比較し，インナーソース導入における段階別の障壁を分析した．

特に，日本のサンプルはアイデア段階に偏り，グローバルのサンプルは成長・成熟段階に多いことを考慮すると，両者を同列に比較するのは単純ではない．しかしながら，段階ごとの共通点と相違点を比較することで，導入プロセス全体を通した障壁の変遷を解明できる．また，日本とグローバルのサンプル構造は異なるが，それぞれが補完的に示す知見から，段階的な克服が必要な課題と最終的に残り続ける課題とが区別できる．

本研究の結果からは，初期段階では組織文化や認知度といった概念的障壁が顕著である一方，導入が進むにつれて実務レベルの課題が前面化していく傾向が示唆された．さらに，日本サンプルとグローバルサンプルの障壁内容を比較すると，同様の文化的要因が認められつつも，各段階の特徴には差異があることが確認された．以降のレーダーチャートでは，いずれかのデータセットで 40%を超えた主要項目の推移を示す．

### 6.2.1 グローバル企業における段階別障壁

グローバルサンプルのアイデア段階は回答数が少ないものの，インナーソースのメリットが十分に可視化されていないことから，概念への理解不足や組織ルールの未整備を大きな阻害要因として認識する回答が多い．具体的なプロジェクトが未始動の段階では，どのような運用指針を定めるかや，どのような成果が期待できるかが不透明であるため，多種多様な潜在的困難が想定されている可能性が高い．

パイロット段階になると，経営陣の理解不足やインナーソース原則の浸透度の低さといった障壁が表面化する．アイデア段階では抽象的だった問題が，実際に試験運用に踏み切ることで具体化しやすくなるため，組織上層部との合意形成や啓発活動の不足が鮮明に浮かび上がる．

初期採用段階に進んだ際には，多様な課題が同時並行的に発生しやすくなる．ガイドラインの不備や開発者の関心の低さが問題視され，中間管理職との協調体制の構築も不十分なまま拡大を図ることで，導入へのモチベーションが散逸するリスクが高まることが示唆される．

成長段階や成熟段階においては，貢献に対するタイムリーなフィードバックが不足したり，コードレビューの時間が制約されたりするなど，実務的・運用的な負荷が顕在化する．一定のプロジェクト数が同時稼働している組織では，インナーソースの利点であるコラボレーションが進む一方で，リソース配分の逼迫やタスク調整の困難さが具体的な障壁として浮上する．

一方で，サイロ思考や中間管理職の理解不足，経営陣の理解不足などの課題は成長・成熟の進行とともに比率が低下しており，インナーソースによる価値が組織内で可視化されるにつれ，理解促進が進んでいることが示唆される．つまり，インナーソースの効果が具体的に現れるにつれ，トップやミドル層の抵抗が減少し，それらが大きな障壁として認識されにくくなると推察される．

総合すると，グローバルデータからは，アイデア段階やパイロット段階では経営層・中間管理職・開発者それぞれの理解が十分に得られず，導入が具体化すると具体的な作業負荷やガイドライン不足が問題

化し，成長・成熟段階ではフィードバックのやり取りや時間的制約といった運用上の障壁が大きくなるというプロセスが推測される．

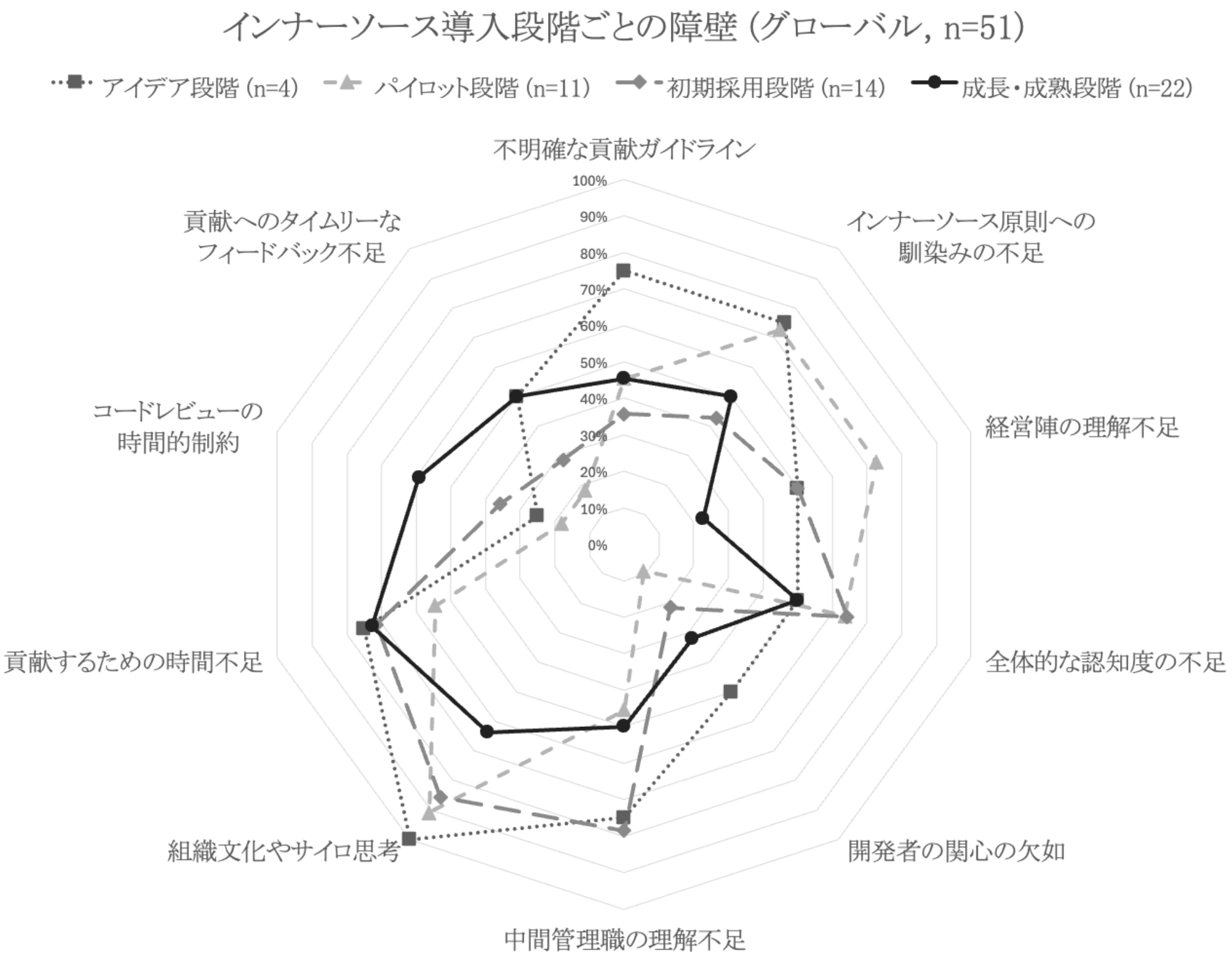

*図 6.3 インナーソース導入段階ごとの障害や障壁（グローバル）*

*表 6.1 インナーソース導入段階ごとの障害や障壁（グローバル）*

| アイデア段階（n=4） | パイロット段階（n=11） | 初期採用段階（n=14） | 成長・成熟段階（n=22） |
|---|---|---|---|
| 組織文化やサイロ思考（100%） | 組織文化やサイロ思考（91%） | 組織文化やサイロ思考（86%） | 貢献するための時間不足（73%） |
| 不明確な貢献ガイドライン（75%） | インナーソース原則への馴染みの不足（73%） | 中間管理職の理解不足（79%） | 組織文化やサイロ思考（64%） |
| インナーソース原則への馴染みの不足（75%） | 経営陣の理解不足（73%） | 貢献するための時間不足（71%） | コードレビューの時間的制約（59%） |
| 中間管理職の理解不足（75%） | 全体的な認知度の不足（64%） | 全体的な認知度の不足（64%） | インナーソース原則への馴染みの不足（50%） |
| 貢献するための時間不足（75%） | 貢献するための時間不足（55%） | 経営陣の理解不足（50%） | 全体的な認知度の不足（50%） |
| 経営陣の理解不足（50%） | 不明確な貢献ガイドライン（45%） | インナーソース原則への馴染みの不足（43%） | 中間管理職の理解不足（50%） |
| 全体的な認知度の不足（50%） | 中間管理職の理解不足（45%） | | 貢献へのタイムリーなフィードバック不足（50%） |
| 開発者の関心の欠如（50%） | プロジェクトの発見可能性の欠如（45%） | | 不明確な貢献ガイドライン（45%） |
| 貢献へのタイムリーなフィードバック不足（50%） | | | |
| プロジェクトの発見可能性の欠如（50%） | | | |
| 選ばれたプロジェクトの魅力不足（50%） | | | |

総合すると，グローバルサンプルからは，アイデア段階やパイロット段階では経営層・中間管理職・開発者それぞれの理解が十分に得られず，導入が具体化すると具体的な作業負荷やガイドライン不足が問題化し，成長・成熟段階ではフィードバックのやり取りや時間的制約といった運用上の障壁が大きくなるというプロセスが推測される．つまり，導入各段階で異なる性質の障壁が顕在化し，それらが段階的に解消されるか，新たな課題に置き換わるプロセスをたどる点が特徴的である．ある程度導入が進んで経営層や管理職が成果を目にすると文化的抵抗は急速に低減し，実務レベルの課題が前面化してくる．これは，インナーソースが「やればわかる」性質を帯びており，早期から小規模パイロットを試すことで組織的な合意形成が加速することを示唆している．もっとも，運用面の問題がスムーズに解決されるかどうかは，開発者や管理職のコミットメントの度合いに左右されるため，一定の継続的リソースを保障する経営判断が欠かせない．

### 6.2.2 日本企業における段階別障壁

日本の回答を導入段階別に分析すると，アイデア段階では「全体的な認知度の不足」や「開発者の関心の欠如」「組織文化やサイロ思考」「貢献するための時間不足」といった抽象度の高い障壁が際立つ．ここでは，インナーソースが組織外から新規輸入された概念であるがゆえに，従来型のプロセスや評価指標に馴染まないことが，無意識の抵抗や関心欠如を生み出している可能性が高い．これは，インナーソース自体への理解が浅い段階では，コードレビューの運用やガイドラインの詳細以前に「そもそもインナーソースとは何か」「なぜ参加する意義があるのか」を共有できていないことがボトルネックになっていると推測できる．さらに，日本特有の労働慣行や上下関係の強い組織文化が，個々の開発者が自発的にプロジェクトを越えてコラボレーションする動機を損なっている恐れもある．

パイロット段階に進むと，全体の障壁総量はアイデア段階より減少することが確認されたが，それはあくまで「動き出してみたものの，範囲が限定的なために大きな問題が顕在化しにくい」ことを意味していると推測できる．この段階では「中間管理職の理解不足」や「貢献するための時間不足」が依然として目立ち，特定のチームや有志開発者のイニシアティブだけでは解消困難な領域が残っている．日本企業特有の階層的組織構造において，中間管理職が合意しなければ成功事例が全社に展開しないリスクがあるため，パイロット段階でトップダウンの後押しをいかに確保するかが勝負の分かれ目となる．

初期採用段階では，パイロット段階で見えなかった問題として「不明確な貢献ガイドライン」や「開発者の関心の欠如」が再度大きく浮上し，部分的導入時には熱意をもって取り組んだ開発者でさえ，より広範な組織との連携を図るにつれて規定やルールの曖昧さに戸惑う状況が推測される．これは，日本社会が重視する形式的合意や手続きの整合性が十分に確立されないままに拡大段階へと移行してしまうことが原因と考えられ，初期採用段階でのガイドライン整備や説明責任が不足するなら，熱量が一時的に高くても継続性が損なわれるおそれがある．こうした組織文化的な抵抗は，トップダウンの方針転換だけでは不十分で，実際の開発現場と密接に連携しながら解きほぐす必要がある．

成長・成熟段階に至った日本企業のサンプルは限られるが，最終的に成長・成熟段階に到達すると，コードレビューやフィードバック時間の不足など，開発実務の最適化に関連した課題が顕在化する．しかし，日本のデータに特徴的なのは，この最終段階においても「不明確なガイドライン」や「中間管理職の

理解不足」が残存している点であり，導入が進む過程でも完全には解消されない文化的・制度的課題が根強く存在していることを示唆する．これは，組織レベルでの成功実績が蓄積されても，ガイドラインの正式化や管理者の評価制度の更新といった根幹の仕組みづくりが追いついていないことを示唆する．結局のところ，インナーソースの導入規模が拡大しても，管理職や経営陣との円滑な協業体制が欠落していると，部分的な成功が組織全体に展開しにくい構造が維持されてしまう．インナーソースを正式に評価制度へ組み込むかどうかが将来的な分岐点となる可能性がある．

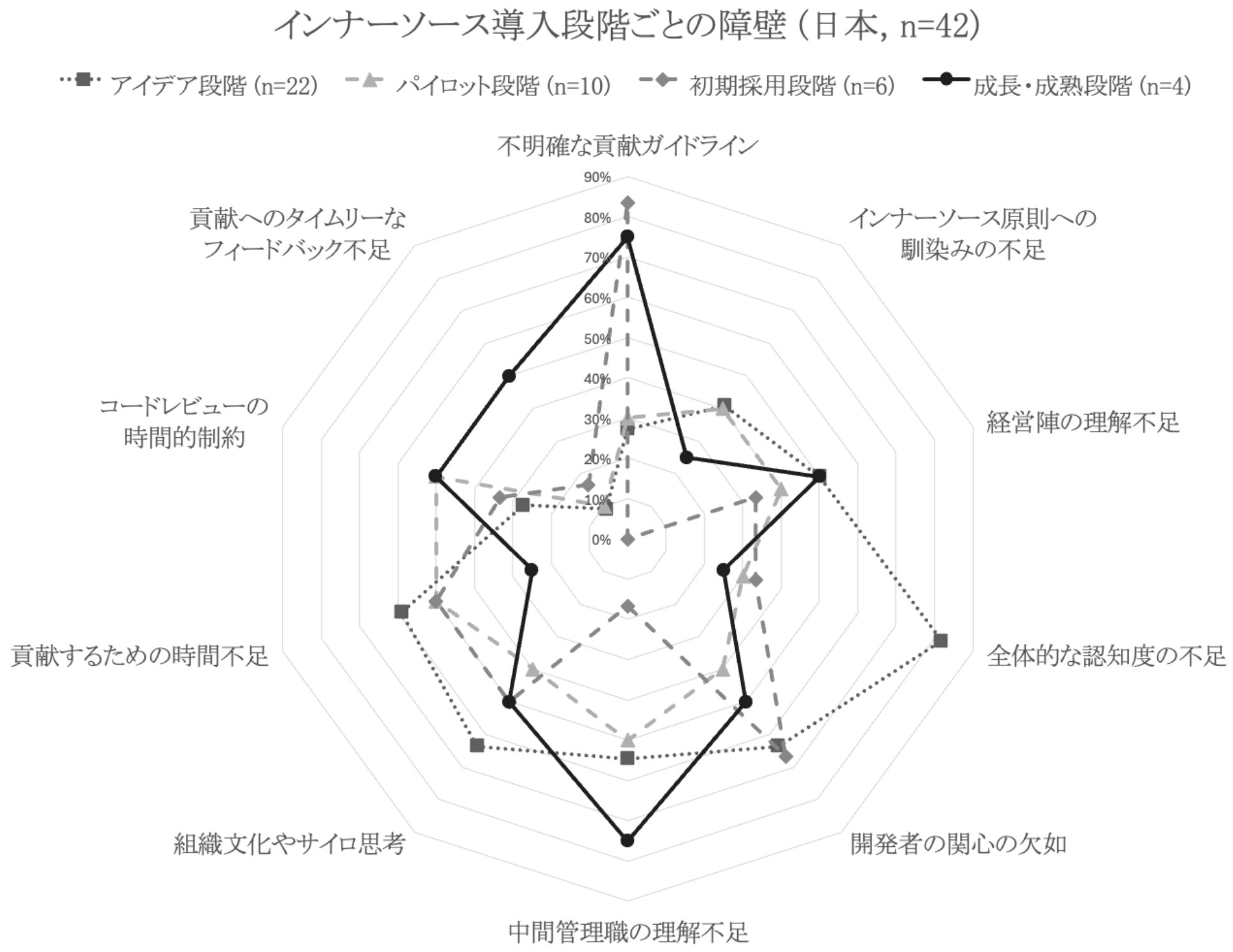

*図 6.4 インナーソース導入段階ごとの障害や障壁（日本）*

*表 6.2 インナーソース導入段階ごとの障害や障壁（日本）*

| アイデア段階（n=22） | パイロット段階（n=10） | 初期採用段階（n=6） | 成長・成熟段階（n=4） |
|---|---|---|---|
| 全体的な認知度の不足（82%） | 中間管理職の理解不足（50%） | 不明確な貢献ガイドライン（83%） | 不明確な貢献ガイドライン（75%） |
| 開発者の関心の欠如（64%） | 貢献するための時間不足（50%） | 開発者の関心の欠如（67%） | 中間管理職の理解不足（75%） |
| 組織文化やサイロ思考（64%） | コードレビューの時間的制約（50%） | 組織文化やサイロ思考（50%） | 経営陣の理解不足（50%） |
| 貢献するための時間不足（59%） | 法的懸念（50%） | 貢献するための時間不足（50%） | 開発者の関心の欠如（50%） |
| 中間管理職の理解不足（55%） | インナーソース原則への馴染みの不足（40%） | | 組織文化やサイロ思考（50%） |
| 経営陣の理解不足（50%） | 経営陣の理解不足（40%） | | コードレビューの時間的制約（50%） |
| インナーソース原則への馴染みの不足（41%） | 開発者の関心の欠如（40%） | | 貢献へのタイムリーなフィードバック不足（50%） |
| | 組織文化やサイロ思考（40%） | | プロジェクトの発見可能性の欠如（50%） |
| | 税や移転価格に関する問題（40%） | | |

### 6.2.3 各導入段階共通の傾向と差異：導入プロセスの俯瞰

日本とグローバルを統合的に捉えた全体のデータ(n=93)の分析から，インナーソース導入の各段階で特徴的な障壁が存在することが明らかになった．アイデア段階では認知不足や関心欠如が主要な課題となり，パイロットから初期採用段階ではガイドラインの不備や管理職の説得が最も顕著な問題として浮上する．さらに，成長・成熟段階に至ると，実務レベルでのレビュー・フィードバックやリソース不足という運用面での課題が顕在化する．

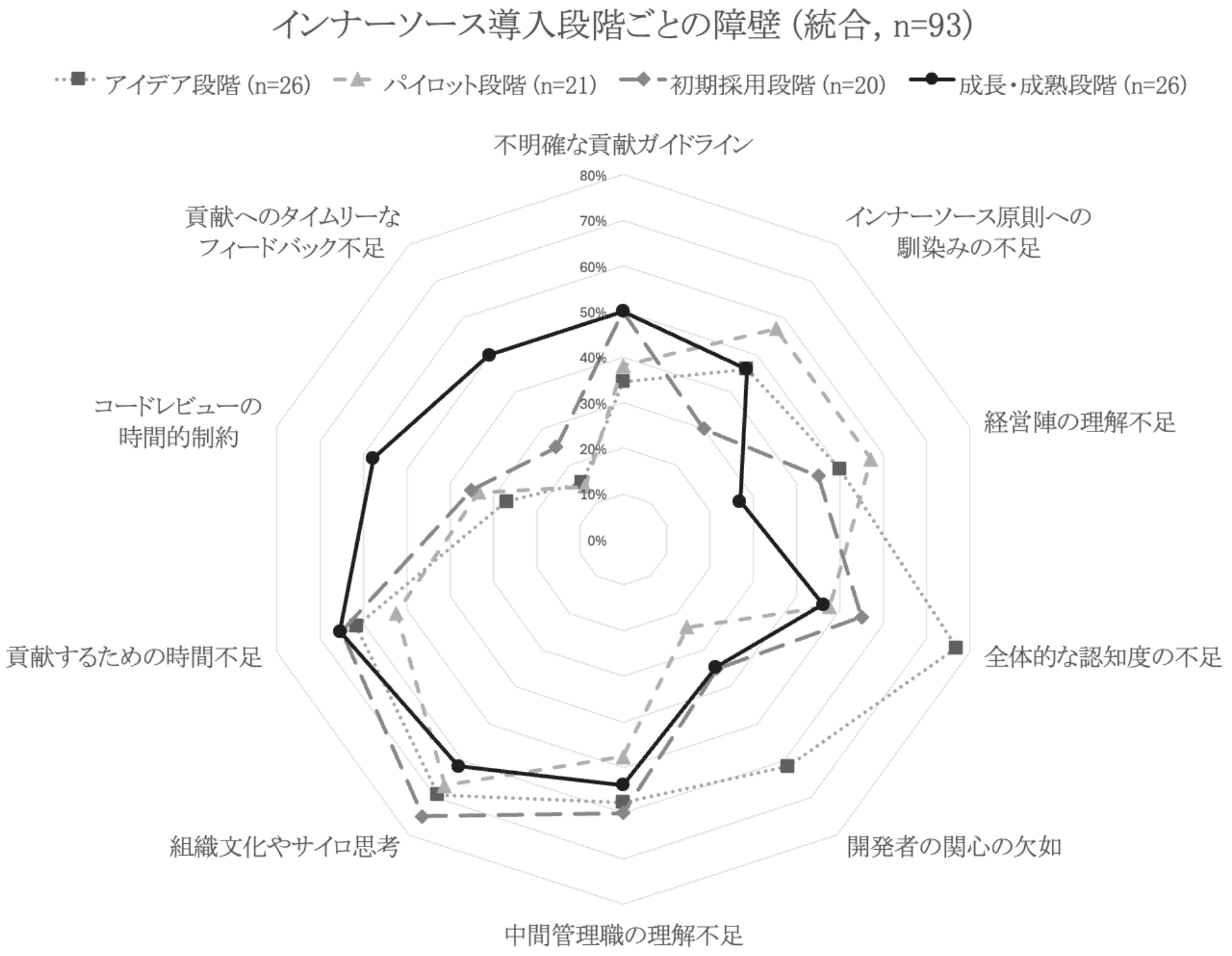

*図 6.5 インナーソース導入段階ごとの障害や障壁（全体）*

*表 6.3 インナーソース導入段階ごとの障害や障壁（全体）*

| アイデア段階 (n=20) | パイロット段階 (n=20) | 初期採用段階 (n=20) | 成長・成熟段階 (n=24) |
|---|---|---|---|
| 全体的な認知度の不足 (77%) | 組織文化やサイロ思考 (67%) | 組織文化やサイロ思考 (75%) | 貢献するための時間不足 (65%) |
| 組織文化やサイロ思考 (69%) | インナーソース原則への馴染みの不足 (57%) | 貢献するための時間不足 (65%) | 組織文化やサイロ思考 (62%) |
| 開発者の関心の欠如 (62%) | 経営陣の理解不足 (57%) | 中間管理職の理解不足 (60%) | コードレビューの時間的制約 (58%) |
| 貢献するための時間不足 (62%) | 貢献するための時間不足 (52%) | 全体的な認知度の不足 (55%) | 中間管理職の理解不足 (54%) |
| 中間管理職の理解不足 (58%) | 全体的な認知度の不足 (48%) | 不明確な貢献ガイドライン (50%) | 不明確な貢献ガイドライン (50%) |
| 経営陣の理解不足 (50%) | 中間管理職の理解不足 (48%) | 経営陣の理解不足 (45%) | 貢献へのタイムリーなフィードバック不足 (50%) |
| インナーソース原則への馴染みの不足 (46%) | | | インナーソース原則への馴染みの不足 (46%) |
| | | | 全体的な認知度の不足 (46%) |

つまり，全体としては導入段階が進むにつれて障壁の焦点が変化し，導入初期の概念的・文化的な抵抗から，成長期以降の実務的・運用的な問題へとシフトしていくという大きな流れが見てとれる．企業や組織がインナーソースを本格導入する際には，この二段階の壁をすり抜けるための包括的な施策と迅速な組織学習が求められる．

このように，アイデア段階からパイロット，初期採用，そして成長・成熟に至るまでの連続的なプロセスを丁寧に設計し，それぞれの段階で想定される課題を先取りして対応しなければ，インナーソース導入のメリットが限定的にとどまる危険性がある．特に，日本では「不明確なガイドライン」や「管理職の理解不足」が後期段階でも残存しやすい構造が示唆されており，これは組織が自然に学習していくモデルだけでは問題が解決されにくいことを意味する．文化的かつ制度的要因が強く影響すると見られる場合，たとえ最終段階に到達しても，中間管理職の理解不足や組織文化的なハードルが依然として残りやすい構造が推測される．これは，海外の企業が初期段階で大きく舵を切って改革を促進するのに対し，日本の企業はより段階的かつ緩やかな意思決定を踏むため，障壁の克服が部分的に遅れがちになることを示している．

本研究においては，日本とグローバルのサンプル構成が必ずしも同程度の成熟度を持つ企業群を対比する形では集まっていないため，直接比較の際には留意が必要である．しかし，全体としては，段階的な障壁の変容過程が共通して確認できるため，各段階で予想される課題を先回りして対応するマネジメント上の重要性が示唆される．

これらの課題に対する解決策については第 7 章以降で詳述するフレームワークを通して示される．管理職をふくむステークホルダーに対する正しいインナーソースの理解を促進するためのガイドラインの整備が不可欠であり，それが実践的な取り組みのスタートポイントとなる．インナーソースや組織改革を抽象的に理解するだけでは，結果として障壁を取り除けないまま進むことになりかねない．そのため，インナーソースの定義を正確に捉え直し，各関係者がスコープの特定やゴール設定を行うことが必要である．

最終的には，トップダウンのアプローチによる制度整備と，ボトムアップによる現場主体のイノベーションを両立させ，文化的抵抗から運用上の負荷まで多段階にわたる障壁を統合的に乗り越えることが，インナーソース導入を成功に導く鍵となる．特に日本の場合，後期段階でのガイドライン整備や管理職の理解促進が追いつかないリスクが大きいため，早期のパイロット段階から評価制度や合意形成のプロセスを視野に入れた計画策定が不可欠である．

## 6.3　導入段階と関連要因の統計的検証

インナーソース導入段階と障壁認識の関連を把握するため，導入経験の有無や組織特性と各種障壁項目との間の独立性をカイ二乗検定によって検証した．クロス集計表を作成し，導入段階(パイロット以降であるか否か)と回答者の組織規模，開発チーム規模，職務属性などを列別に配置して可視化した．

分析の結果，「税や移転価格に関する問題」が，日本およびグローバル企業の双方において，組織規模との間に統計的に有意な関連を示した．具体的には，日本企業では組織規模($\chi^2$=5.85，p<.05)および開発チーム規模($\chi^2$=6.47，p<.01)との有意な関連が確認され，グローバル企業においても組織規模で同様の傾向($\chi^2$=4.18，p<.05)が観察された．

*表 6.4 インナーソース成功の障壁にかかる属性感の独立性*

**インナーソース採用に対する重大な障壁の該当有無を二値化し属性間の独立性をカイ二乗検定($\chi^2$)で評価　(N=111)**

| | 関心および理解 | | | | | リソース配分 | | | 環境 | | | ガイドライン | | |
|---|---|---|---|---|---|---|---|---|---|---|---|---|---|---|
| | インナーソース原則への馴染みの不足 | 全体的な認知度の不足 | 経営陣の理解不足 | 中間管理職の理解不足 | 開発者の関心の欠如 | 貢献するための時間不足 | コードレビューの時間的制約 | 貢献へのタイムリーなフィードバック不足 | 組織文化やサイロ思考 | プロジェクトの発見可能性の欠如 | 選ばれたプロジェクトの魅力不足 | 不明確な貢献ガイドライン | 法的懸念 | 税や移転価格に関する問題 |
| **グローバル (n=59)** | | | | | | | | | | | | | | |
| 組織のインナーソース経験の有無 *1 | 0.51 | 0.96 | 0.34 | 2.23 | 0.01 | 2.85 | 1.24 | 0.53 | 0.01 | 0.74 | 0.40 | 0.34 | 0.40 | 1.30 |
| (sig P,* ≤ .05, † ≤ .10) | 0.48 | 0.33 | 0.56 | 0.14 | 0.91 | 0.09† | 0.27 | 0.47 | 0.91 | 0.39 | 0.53 | 0.56 | 0.53 | 0.25 |
| 組織の規模 *2 | 10.89 | 5.74 | 1.36 | 7.04 | 0.59 | 2.87 | 0.85 | 0.19 | 6.70 | 1.31 | 0.73 | 0.34 | 1.92 | 4.18 |
| (sig P,* ≤ .05, † ≤ .10) | **0.00*** | **0.02*** | 0.24 | **0.01*** | 0.44 | 0.09† | 0.36 | 0.66 | **0.01*** | 0.25 | 0.39 | 0.56 | 0.17 | **0.04*** |
| 開発チームの規模 *3 | 1.11 | 1.61 | 0.05 | 1.14 | 0.00 | 1.21 | 0.05 | 4.73 | 4.11 | 0.45 | 0.63 | 0.00 | 2.50 | 2.20 |
| (sig P,* ≤ .05, † ≤ .10) | 0.29 | 0.20 | 0.83 | 0.29 | 1.00 | 0.27 | 0.83 | **0.03*** | **0.04*** | 0.50 | 0.43 | 1.00 | 0.11 | 0.14 |
| 開発者か否か *4 | 0.07 | 0.00 | 1.08 | 0.27 | 0.23 | 2.11 | 3.45 | 2.57 | 0.23 | 0.41 | 0.66 | 1.08 | 3.10 | 1.25 |
| (sig P,* ≤ .05, † ≤ .10) | 0.79 | 0.98 | 0.30 | 0.60 | 0.63 | 0.15 | 0.06† | 0.11 | 0.63 | 0.52 | 0.42 | 0.30 | 0.08† | 0.26 |
| **日本 (n=52)** | | | | | | | | | | | | | | |
| 組織のインナーソース経験の有無 *1 | 1.84 | 11.91 | 1.09 | 1.21 | 1.77 | 1.04 | 1.11 | 0.63 | 3.06 | 0.79 | 0.04 | 3.34 | 1.16 | 0.46 |
| (sig P,* ≤ .05, † ≤ .10) | 0.18 | **0.00*** | 0.30 | 0.27 | 0.18 | 0.31 | 0.29 | 0.43 | 0.08† | 0.37 | 0.85 | 0.07† | 0.28 | 0.50 |
| 組織の規模 *2 | 0.09 | 0.99 | 0.19 | 0.06 | 0.59 | 0.19 | 0.46 | 2.01 | 0.01 | 0.64 | 0.00 | 0.34 | 4.09 | 5.85 |
| (sig P,* ≤ .05, † ≤ .10) | 0.76 | 0.32 | 0.66 | 0.81 | 0.44 | 0.66 | 0.50 | 0.16 | 0.92 | 0.43 | 0.98 | 0.56 | **0.04*** | **0.02*** |
| 開発チームの規模 *3 | 1.34 | 0.42 | 1.61 | 0.10 | 0.19 | 0.04 | 0.00 | 1.11 | 0.00 | 0.25 | 0.23 | 0.02 | 5.35 | 6.47 |
| (sig P,* ≤ .05, † ≤ .10) | 0.25 | 0.52 | 0.20 | 0.75 | 0.67 | 0.85 | 0.97 | 0.29 | 1.00 | 0.61 | 0.63 | 0.89 | **0.02*** | **0.01*** |
| 開発者か否か *4 | 0.00 | 0.01 | 1.05 | 0.55 | 2.61 | 0.01 | 0.58 | 0.06 | 0.00 | 1.39 | 2.60 | 0.15 | 0.53 | 0.92 |
| (sig P,* ≤ .05, † ≤ .10) | 0.95 | 0.94 | 0.31 | 0.46 | 0.11 | 0.93 | 0.45 | 0.80 | 1.00 | 0.24 | 0.11 | 0.70 | 0.47 | 0.34 |

*1 インナーソース推進ステージが「パイロット段階」以降であるか
*2 組織規模が5000人以上である
*3 開発チームの規模が500名以上である
*4 職務が「開発者」であるか

大規模組織，特に上場企業や従業員数 50,000 名以上の企業において，税制や移転価格に関する問題は深刻な障壁として認識されている．この傾向は，企業規模の拡大に伴う移転価格税制や知的財産権保護などの制度的対応の必要性を反映している．さらに，グローバルな事業展開や複雑な子会社構造を有する大規模組織では，法務・税務リスクの管理がより重要な課題として位置づけられている．

一方で，法的懸念や税務上の問題と導入経験の有無との間には統計的独立性が確認された点は，重要な示唆を含んでいる．この結果は，インナーソース導入の経験を重ねた組織においても，法務・税務に関する不安要素が必ずしも解消されない可能性を示している．多くの場合，税務上の問題は当局の指摘まで判明しないため，社内財務部門との緊密な連携があっても不安が残存する．また，全体的なガバナンスの維持と抜け漏れの排除を両立させることの困難さも存在する．ただし，過度に厳格な運用は柔軟性を損なう可能性があり，適切なバランスを模索する必要がある．

*表 6.5 企業規模ごとの税や移転価格に関する障壁認識有無の該当*

税や移転価格に関する問題が
インナーソース成功の障壁である
（日本及びグローバル，n=88）

| 企業規模（人） | 該当なし | 該当あり |
|---|---|---|
| ～ 10 | 5 | 0 |
| 10 ～ 100 | 3 | 1 |
| 100 ～ 500 | 14 | 0 |
| 500 ～ 2000 | 8 | 3 |
| 2000 ～ 5000 | 14 | 3 |
| 5000 ～ 50000 | 12 | 3 |
| 50000 ～ | 14 | 8 |
| 合計 | 70 | 18 |

（以上 ～ 未満）

補足的な分析としては，税や移転価格に関する問題の重要性認識には，職種による顕著な差異が観察された．日本のサンプルにおいて，当該問題を重要課題と位置付けた回答者の約 46%が研究開発職に従事している一方，中程度以下の障壁（5 段階評価で 3 以下）と認識した回答者のうち，研究開発職の割合は 8%程度にとどまった．また，この問題を重要課題と認識した日本の回答者は，テクノロジーおよび工業/製造業に集中しており，技術が競争優位性の源泉となる業界において特にこの種の課題が顕在化する可能性を示唆している．

移転価格税制に関する諸問題とその解決策については，先行研究において詳細に論じられており，インナーソースの推進において克服すべき重要な課題として位置付けられている．本研究では，移転価格税制や社内会計ルールが局所的な課題であることを認識した上で，第 8 章においてより包括的に論じる．これらの課題は，貢献ガイドラインや社内のガバナンスモデル，プロセス構築という広範なコンテキストの中で取り扱われるべきであり，インナーソースの全体論を論じる上では，この特定の問題に過度に焦点を当てた場合，全体像の把握を妨げる可能性がある．

職種による障壁認識については，グローバルのサンプルにおけるコードレビューに要する時間的制約に関してのみ有意傾向（$\chi^2=3.45, p<.10$）が観察された．特に開発者は，マネジメント職やビジネス職と比較して，この課題をより切実な問題として認識している．これは開発者が実際のソフトウェア開発プロセスに直接携わり，品質確保のための作業負荷を日常的に経験していることに起因すると推察される．

これらの分析結果から，インナーソースの障壁は重層的な構造を持つことが明らかになった．認知度や経営層・中間管理職の理解不足などの一般的な障壁に加え，税制や法的側面に関連する特定の企業群において顕在化する障壁の存在が確認された．このような多層的な障壁の構造を理解し，各組織の特性に応じた適切な対処を行うことが，インナーソース推進の成功には不可欠であると結論付けられる．

## 6.4 障壁の進展と組織への影響に関する総括

インナーソースの導入段階において，組織が直面する障壁は段階的に変化することが観察される．図 6.6 で記載した通り，初期段階では技術的な課題が主であるのに対し，導入が進むにつれて中間管理職の理解獲得，ガイドラインの整備，リソース確保，実務的な便益の具体化といった組織的な課題へと移行する．

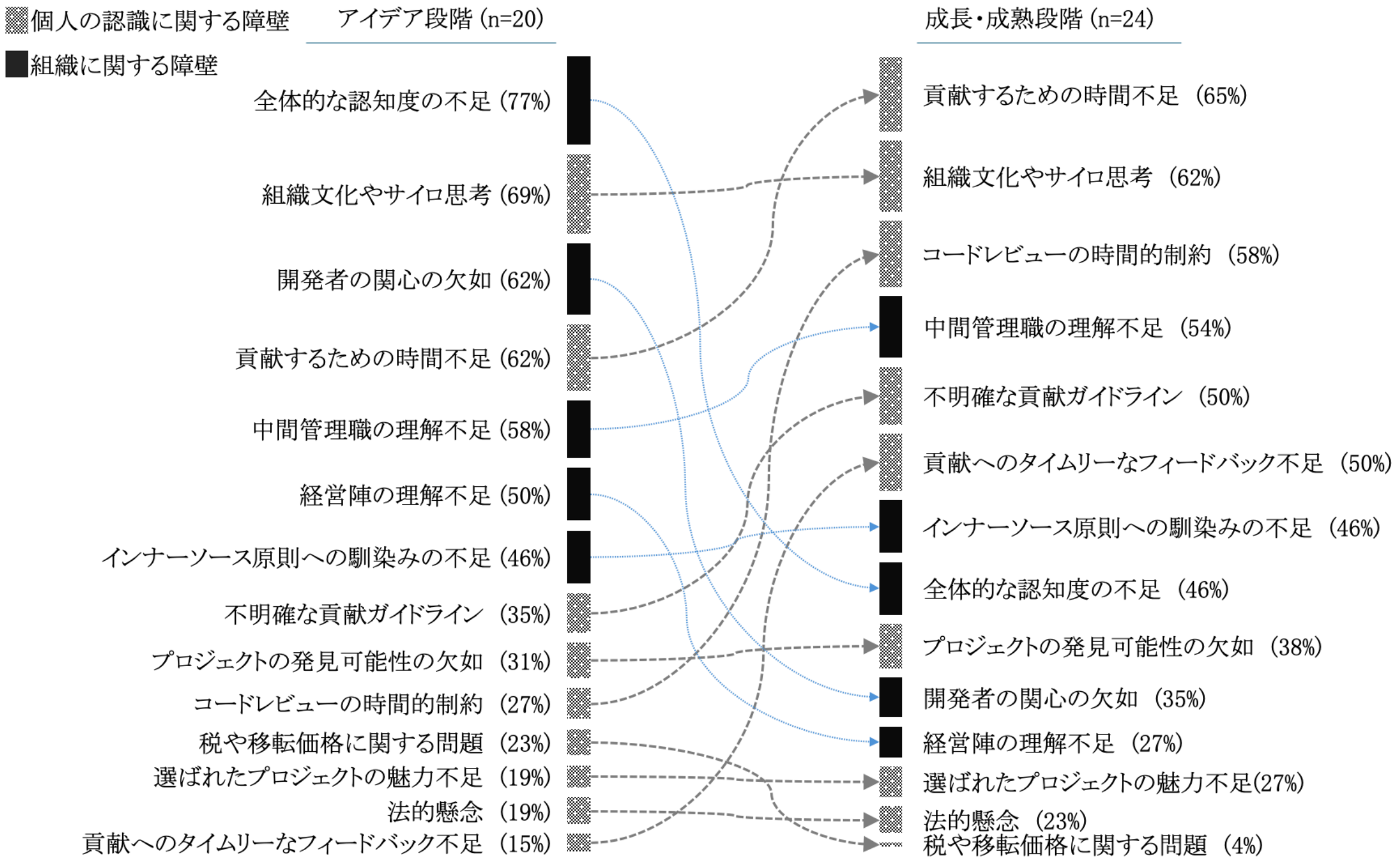

*図 6.6 インナーソース導入段階における障壁タイプの遷移*

既存のインナーソースのマチュリティモデルは導入における発展段階を示す指標としては有用であるものの，導入プロセスにおける具体的な障壁の特定や，その克服方法および事例を扱うことをその目的としていない．

これまでのインナーソース研究は，個別のプロジェクトや特定の技術的枠組み（例えば ISPO などの取り組みやガバナンスモデル，ストラテジー）に焦点を当てた分析が主流である．しかしながら，具体的な実装プロセスについての包括的な分析や，発展途上段階にあるインナーソース活動の取り扱い，また組織内に散在する複数の個別活動の統合方法についての検討が不足している．特に，インナーソースの導入体制が整っていない企業や，プログラムの構築段階，経営層からの承認獲得プロセス，さらには個人レベルでインナーソースの価値実現を目指す初期段階における具体的なガイダンスは限定的である．

インナーソース活動の推進者（チャンピオン）は，必ずしも OSPO 担当者や開発生産性エンジニア，GitHub の管理者といった公式な役割を持つ者とは限らない．むしろ，組織内の様々な立場から，インナーソースの価値に共感した個人が自発的に活動を開始するケースが多く観察される．本研究は，既存の

フレームワークを補完し, こうした多様な推進者の視点を包含する新たな分析枠組みの構築を目指している. 第 7 章以降では, これらのチャンピオンが直面する具体的な障壁とその克服方法について詳細な分析を展開する.

# 第7章 既存マチュリティモデルの総合的再検討

本研究における調査の分析によれば，日本企業ではインナーソースのアイデア段階から初期導入期にかけて，技術的ハードルよりも文化的障壁や組織的障壁が顕在化しやすいことが指摘されている．また，成長・成熟期には貢献に関するルール整備の遅れが拡大フェーズを停滞させる要因となっていると考えられる．

実際，日本の多様な企業が内製化の加速を目標にインナーソースを導入し始めているものの，組織的に成熟した形で定着している事例はまだ限られている．その背景として，中間管理職や経営層との合意形成の停滞や，組織横断的な評価基準の不備，既存の外注型開発慣行からの脱却が進まない構造的問題などが複合的に作用していると推察される．

一方，グローバルにおける先進事例では，ISPO といった専任チームやコミュニティマネジメント体制を整備して，自発的な貢献文化とガバナンスの両立を図る企業が少なくない．そこでは評価モデルやライセンス管理，報酬インセンティブといった体制を統合し，自然発生的なコラボレーションを経営戦略レベルで支える基盤を構築している．しかし，日本企業においてはトップダウン型の意思決定とボトムアップの自発的貢献意欲を結びつける仕組みが決定的に不足しており，この点を見誤ると企業全体への浸透が進まず，局所的成功にとどまる危険性が高い．

さらに，本研究の結果からは，日本企業におけるインナーソース導入の動機として「組織文化の変革」を重視する傾向が強い一方，ソフトウェア再利用や開発効率改善といった技術的メリットを明確に打ち出している事例は相対的に少ないことが浮き彫りになった．こうした状況は，エンジニアリング上の定量的な成果指標の不足を招き，社内の説得や投資判断の難度を高める一因になっていると考えられる．

加えて，具体的なインセンティブ設計や社内会計ルール，大規模組織における可視化手法の確立など，多くの分野で課題が十分に解明されていない実態も認められる．インナーソースにおけるチャンピオンがどの程度導入効果を左右するのかを実証的に検証する研究もなお不十分であり，これらの要素を包括的に解き明かすことがインナーソース導入における成功パターンの確立につながると考えられる．

以上の分析を踏まえたうえで，本章では，これまでに浮き彫りになった課題や要素を整理し，次章以降（第 8 章・第 9 章・第 10 章）で詳しく論じる「インナーソース・トポロジー」「重層的インセンティブモデル」「インナーソース円環モデル」を検討するための前提を確立することに焦点を当てる．これによって，次章以降で展開するインナーソースのスコープや評価手法，推進モデルに関する議論をより深め，最終的には企業特有の条件に即した持続的なインナーソース推進策を提示する道筋を明らかにする．

## 7.1 既存マチュリティモデルへの補強と適用拡張

本研究は，アイデア期から成長・成熟期に至るまでの各段階における優先的な課題を可視化するとともに，具体的な解消プロセスを提示することに重点を置いており，インナーソース導入プロジェクトが直面する段階的なタスクと論点を整理し，組織的変革を計画的に進めるための実践的枠組みを提供する．

既存の InnerSource Maturity Model としては， Inner Source Capability Maturity Model（IS-CMM） [8] や，The InnerSource Commons Foundation によるアセスメントフレームワーク [9]が知られている．しかし，これらのモデルは段階的な成熟度を測定する指標として機能する一方で，「導入をどこから始め，どのよ

うなスコープで着手すべきか」というプライオリティや，組織全体を対象とするトランスフォーメーションの視点が十分に明示されていない．

既存モデルにおいては，導入段階でのプライオリティが曖昧なために，組織的側面やプロダクト的側面，プロセス的側面をどのように区別して動機付けや阻害要因を捉えればよいかという課題が残されている．また，スコープの欠如によって，プロジェクト単位の特性と組織単位の戦略的観点が混同されやすい点や，ガバナンスに関する具体的スキームが不足している点も指摘される．

さらに，ガバナンスに関する議論の具体性が乏しく，会計管理やライセンス観点，評価制度の整合性を検討する枠組みが明確に示されていない点も指摘できる．企業ごとの内部慣行や会計制度，セキュリティガイドラインなど，一般に外部に公開する必要がない領域に関するベストプラクティス事例は限られている．特にオープンソースの導入経験が少ない組織においては，導入最初期段階で顕在化する同様の障壁が大きく，初期段階の障壁を体系的に除去する術が未確立である．これらは単に「ルールをつくる」という一方向的な観点では捉えきれない．現場の実態や企業の文化的背景に即した対話と折衝が繰り返され，そこでどのような合意形成が行われたかを解き明かす必要がある．

また，これまでの調査から，企業がインナーソースを採用する動機は多様であり，日本企業とグローバル企業の間で，また同じ企業内でも「パッケージとしてのインナーソース」「OSS プラクティスの内部化としてのインナーソース」など，その位置づけに違いが見られることが明らかになっている．このことは，マチュリティモデルにおいて，各企業が理想とする「成熟」状態を明確に定義できる枠組みが必要であることを示唆している．単に「オープンソースのカルチャー実践」という抽象的な目標を掲げるだけでは，その解釈は多岐にわたり，具体的な方向性を見失う可能性がある．

このような状況下で，企業内でどのようなインナーソースが実践されているのかを理解することは極めて重要である．とりわけ，先導者であるチャンピオンがどのように企業内部で位置付けられ，実際にどの程度の影響力を行使しているかは，アンケート調査だけでは見えにくい．自発的なボランティア的情熱に依存しているケースなのか，それとも組織として正式に責任範囲を定めているケースなのかによって，導入段階を乗り越えるうえで必要なガバナンスやインセンティブ設計が大きく異なりうる．

実務に即した視点を深く得るためには，初期導入期に主導的役割を果たした開発者や管理職，経営層，パイロットプロジェクト参加者，企業全体への普及を指揮した担当者などへの横断的なインタビューが有効である．これにより，「どの段階でどのような判断がなされ，どのような対話や調整が奏功したのか」「インナーソースが制度として確立する転機はどのフェーズで訪れたのか」など，組織変革の核心部分を掘り下げることが可能になる．さらに重要なのは，「彼らが『インナーソース』という言葉を使って何を目指しているのか」という本質的な目的の解明である．

実際の現場では，導入を進めるにつれて，プロジェクト初期は有志エンジニアの献身に頼って立ち上がるケースが多いものの，制度面の不備や組織内調整の難しさなどに阻まれて失速する事例が頻繁に認められる．

こうした状況から見ても，それぞれの導入段階で，推進者たちが実際にはどこに向かっているのかを共有し，立ち位置を確かめるモデルが改めて必要な理由は明白である．特にアイデア期やパイロット期の施策は，その後の拡大・成熟に大きな影響を及ぼすため，初期の小規模プロジェクトで成功事例を創出し，経営層や中間管理職を巻き込むプロセスを明確に設計する必要がある

インナーソースの導入プロセスを初動の過程に着目して論じる意義は，実務的にも学術的にも大きい．従来のマチュリティモデルが示す「成熟状態」は多岐にわたる解釈が可能であり，それに至るまでの「導入開始から軌道に乗せるまでの具体的手立て」を補完的に考察することで，インナーソース全体像をより総合的に理解でき，さらには既存のツールセットをさらに活用できる可能性が高まる．

## 7.2 インナーソース・チャンピオンの発生経路と組織内ダイナミクス

今回のアンケート結果によれば，日本企業においてはボトムアップによるインナーソース導入事例が際立っていた．これは一見すると「現場主導のイノベーションが活発」という好意的な解釈が成り立つが，実際には合議制や意思決定プロセスが階層的で合意形成に時間を要することから，小規模ユニットで着手せざるを得ないという側面もある可能性がある．大企業ほど管理部門や上層部との調整が煩雑になり，総意を得るハードルが高い一方で，現場レベルではエンジニアたちが日々の課題を解決するために新手法を導入しやすい土壌があり，これがボトムアップのインナーソースとして可視化されている．

これに対し，米国企業においては全てのインナーソース導入段階においてトップダウンによる取り組みが一定数観測された．この傾向は，導入の初期段階から経営層の合意を得ていることを示唆している．さらに，インナーソース導入における障壁分析からは，段階が進むにつれて「トップやミドルの合意形成」から「時間的制約」へと主要な課題が移行していることが確認された．このことから，グローバルサンプルにおいては，組織的な取り組みとしてのインナーソースが比較的円滑に進展していることが示唆される．

一方，日本企業においては導入の後期段階に至っても中間管理職の抵抗が主要な障壁として残存していることが確認された．この状況を打開するためには，早期からの組織的な関係者の巻き込みが重要である．しかしながら，トップダウンのみでは現場の納得感を醸成することが困難であり，文化的障壁やリソース不足が顕在化した際に導入が停滞するリスクが高い．他方，ボトムアップ型のアプローチのみでは，制度改革や全社的なルール整備を要する段階において権限不足という課題に直面する可能性が高い．これらの課題を克服するためには，トップダウンとボトムアップを組み合わせた「ミックス型」の導入プロセスが有効であり，これにより組織文化の柔軟性を高め，インナーソースの効果を迅速に実現できると考えられる．

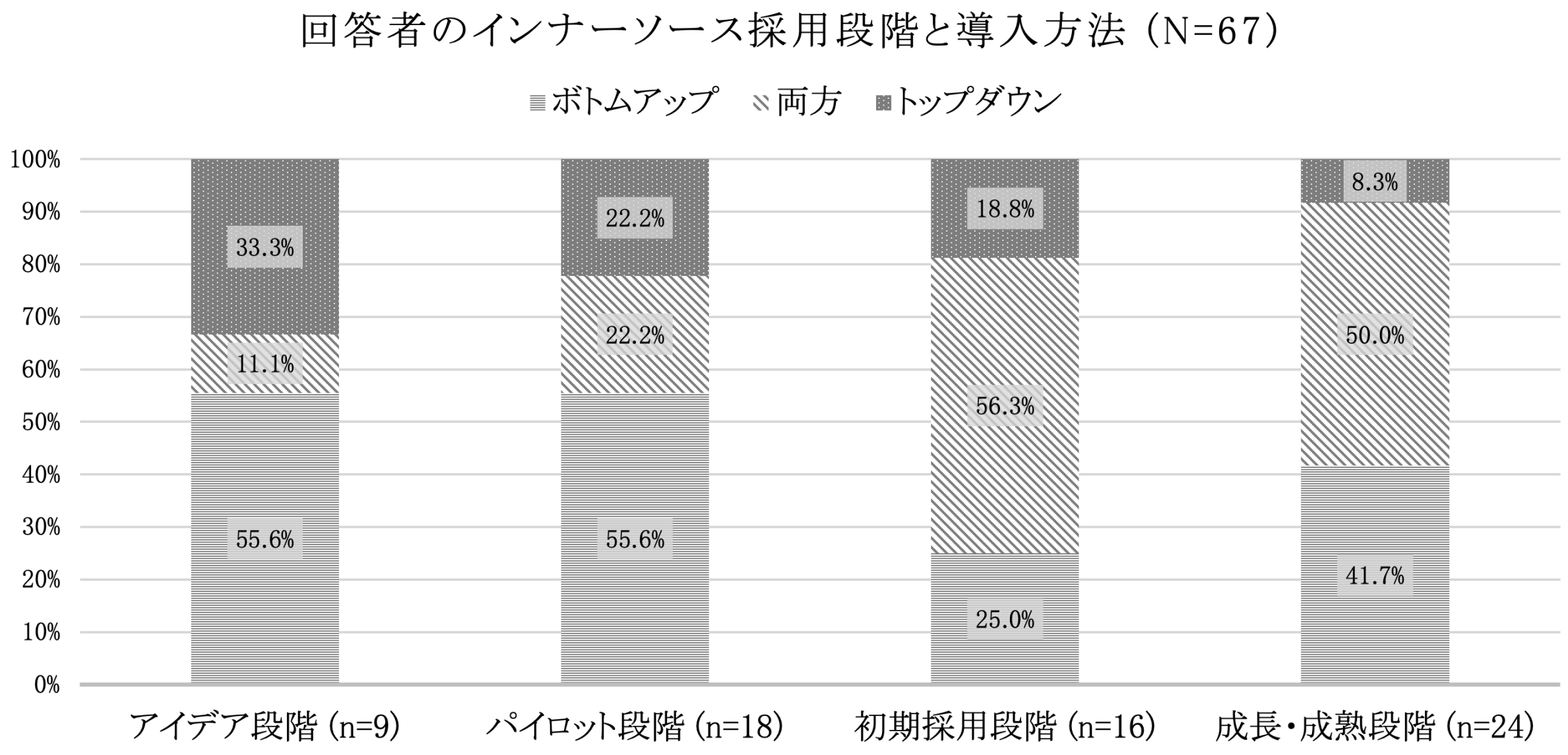

*図 7.1 インナーソースの採用段階と導入方法（全体）*

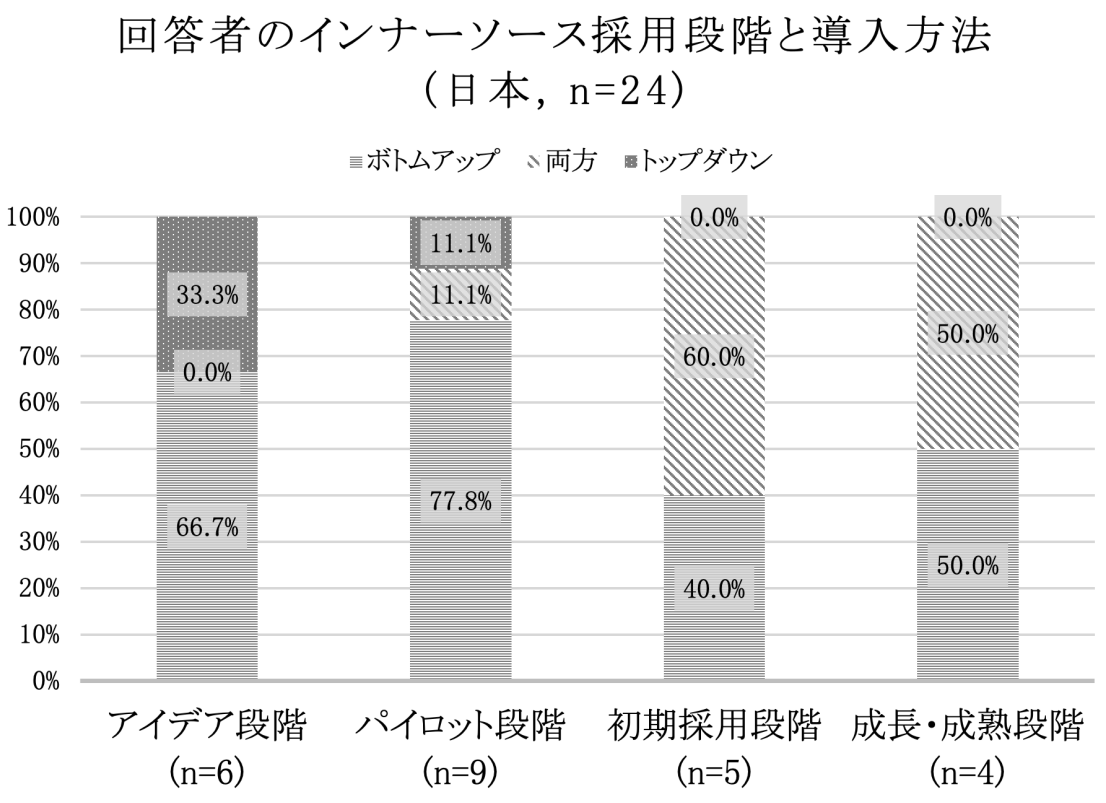

*図 7.2 インナーソースの採用段階と導入方法（日本）*

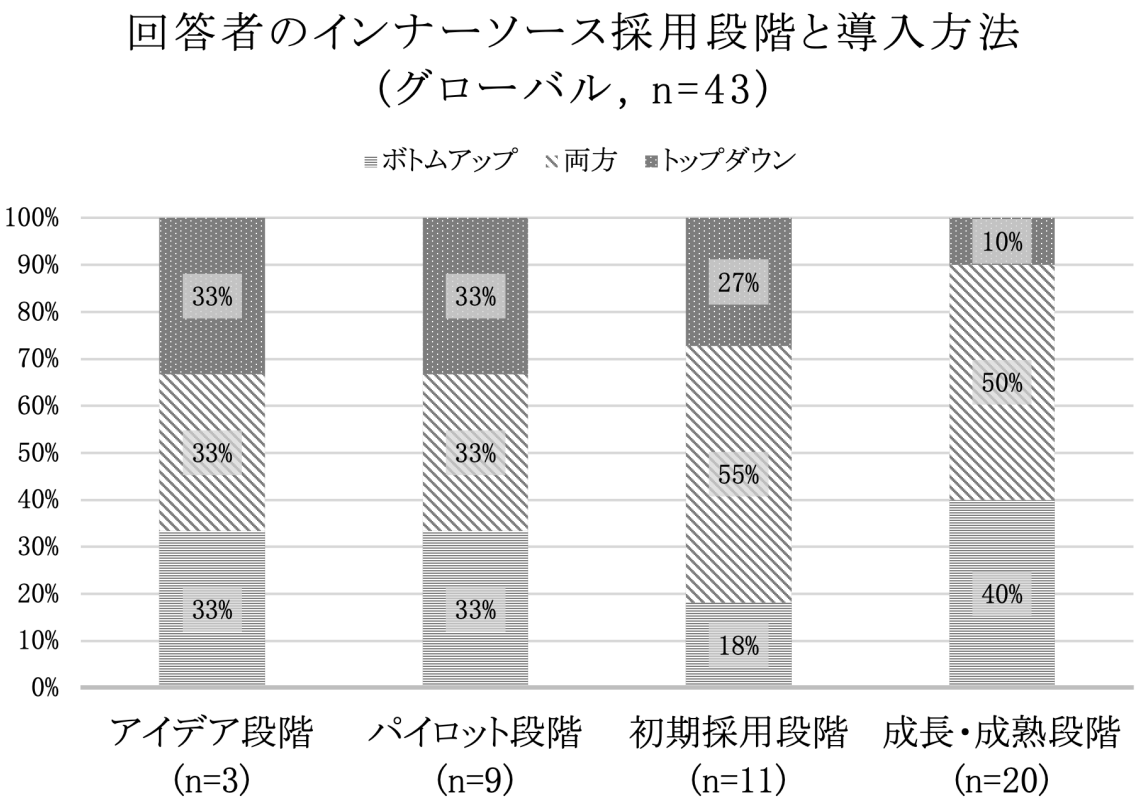

*図 7.3 インナーソースの採用段階と導入方法（グローバル）*

ただし，チャンピオンがボランティアベースで活動するだけでは，制度改革や全社的な合意形成が求められる局面で限界に直面しやすい．たとえば，OSS に由来するライセンスや知的財産の管理方法と既存ルールの整合を検討しなければならないのに，十分な権限やリソースが与えられないままでは，初期段階から導入が停滞してしまう懸念がある．日本の合議制組織では，非公式なチャンピオンが法務・会計・経営陣を巻き込んで調整を進めるのは容易ではなく，その過程で当初の熱意が失われるリスクも高い．

このような導入時のリスクを回避するためには，インナーソースを単なる「OSS の社内適用の延長」ではなく，企業独自の人事や品質管理，会計慣行にも適合する「組織改革のパッケージ」として設計する発想が重要になる．チャンピオンは，このパッケージ設計をリードする立場として，技術部門だけでなく経営企画や人事，コンプライアンス部門など多様なステークホルダーと横断的に連携し，導入段階ごとに必要な制度・ルールの改訂や合意形成を担わなければならない．

## 7.3 実務的知見獲得のためのインタビュー調査

本研究では，複数企業への詳細インタビュー調査を実施した．具体的には，インナーソース導入に成功して大規模展開へ移行した企業，パイロット段階で挫折を経験した企業，初期導入期は完了したものの会計処理や評価モデルの整備に苦戦している企業など，さまざまな事例をカバーするよう対象を選定し，担当者から管理職，場合によってはリーダーシップ層に至るまで多角的な視点を収集している．

調査の際は，第 6 章や本章前半で提起した障壁と仮説を起点に，「どのように障壁を取り除いたのか，あるいはどこで失敗に至ったのか」「チャンピオンは実際にどのような役割を果たしたのか」「評価モデルはいつ・どのように確立し，インセンティブをどう設計したのか」「会計・コンプライアンス上の課題を具体的にどう解決したのか」といった問いを中心にヒアリングを行った．こうして得られた回答は，これまでのマチュリティモデルや海外の事例研究では必ずしも十分に扱われてこなかった，具体的な導入戦略や障壁排除の手法を体系化するうえで貴重な資料となる．本研究で実施したインタビューの詳細は，論文末尾の付録（Appendix）にまとめており，以降の章や議論において適宜参照されたい．本付録では，企業名は匿名化しつつも，なるべく詳しい情報を開示することで客観的かつ深い分析に役立つよう配慮している．製造業からインターネット業界，情報通信業界まで，幅広い業種の企業を対象としたことで，業界横断的に見たインナーソースの実態を把握するうえで高い網羅性とバランスを確保できたと考えている．読者が多角的な視点からインナーソース導入の実態を検証するうえで有用な資料となることを期待している．

インタビュー調査は，以下の 6 社を対象に実施した．

*表 7.1 実施したインタビュー対象者の概要*

| 企業名 | 業界 | 企業規模 | 所属 | 役職 | 導入段階 |
|---|---|---|---|---|---|
| **企業 A 社** | 無回答 | 50,000 人以上規模 | 技術関連部門 | 無回答 | 初期採用期 |
| **国際的製造業 B 社** | 製造 | 50,000 人以上規模 | 開発共通部門 (持株) | 上級技術専門家 | 初期採用期 |
| **インターネット業界 C 社** | インターネット | 数千人規模 | 開発生産性向上組織 | リーダー | 成熟期 |
| **インターネット業界 D 社** | インターネット | 数千人規模 | 開発本部 (グループ内中核企業) | リードエンジニア | 成熟期 |
| **インターネット業界 E 社** | インターネット | 数千人規模 | 経理部門 | 担当 | 成熟期 |
| **情報通信業界 F 社** | 情報通信 | 500 人規模 | 開発部門 | マネージャー / エンジニア | 初期採用期 |

本調査対象企業は，製造業からインターネット業界，情報通信業界まで，幅広い業種をカバーしており，業界分布の観点から見ても十分な網羅性とバランスを確保している．各企業におけるインナーソースの取り組みについて，導入の経緯，実施状況，直面している課題，得られた効果などを詳細に聴取した．

次章以降では，これらのインタビューをもとに具体的事例や得られた知見を整理し，日本企業におけるインナーソース導入モデルの精緻化を試みる．そこで浮かび上がる成功要因・失敗要因や，チャンピオンの活動様式，会計・評価モデルの設計手法などを明らかにし，先行研究で十分に触れられてこなかった導入時の課題解決策を提示していく．

## 7.4　導入段階別に見る主要論点と実務上の考察

本節では，インナーソース導入プロセスにおける各導入段階の主な検討事項を俯瞰し，次章以降で詳細を掘り下げる．

アイデア期では，社内にコラボレーションの価値を認識させると同時に，会計ルールや知財管理を一挙に整備するのではなく，影響範囲を限定して段階的に取り組む姿勢が重要である．まずは狭いスコープで試験的にプロジェクトを進めることで，複数の小規模活動が同時並行で行われてもリスクを最小化できる．こうした初動の取り組みが，のちの大規模展開に向けた基盤となる．

次のパイロット期では，実際のリポジトリ運用によって得られたデータを可視化し，管理職や横断的部門を説得するための根拠を構築することが要となる．パイロットが成功し，一定の証拠とメリットを示せれば，会計処理や評価制度にインナーソースを統合する具体的な下地が整い始める．情報通信業界 F 社の事例によれば，パイロットの成功事例が広まると，試しに導入してもよいという段階から積極的な実践へと認識が変化しやすく，横の連携やガイドラインの明記によってエンジニアリングマネージャーの理解と合意を得ることが容易になる．

パイロット期から初期導入期への移行に際しては，抵抗や障壁が再び顕在化する点に注意が必要である．この段階で明確な評価基準が提示されないまま拡大を図ると，推進の勢いが削がれてしまうリスクが高まる．パイロットで得られたデータを活用しつつ，評価モデルや権限設計を明確に示すことが初期導入期の円滑化に寄与する．

初期導入期が軌道に乗る段階では，ガバナンスやセキュリティ，コンプライアンスへの対応を組織的に整備する必要が生じる．大企業の場合，評価の仕組みや会計ルール，社内規定などが複雑に絡み合うため，早期に管理部門を巻き込みつつ基盤を固めないと，コミュニティが拡大期に入る前に失速する可能性がある．情報通信業界 F 社では，インナーソースガイドライン整備前は評価制度が不透明で「他チームへ貢献しても報われるか分からない」という声が多かったが，評価軸や計上ルールの設計後はエンジニアが安心してコラボレーションに参加できるようになったという．

初期導入期には，エンジニアや中間管理職の協力体制をどう構築するかが大きな課題となる．企業 A 社の事例にあるように，当事者の熱意だけでは組織的バックアップが追いつかず，エンジニアが「隠れてやる」状況に陥る場合もある．したがって，部門長や管理部門との連携を強化し，エンジニアが自信を持って貢献できる制度や評価体制を早急に確立することが欠かせない．

初期導入期を乗り越え成長・成熟段階に差しかかるころには，組織的な運用体制や権限管理が整備され，プロジェクト数が増加するとともにコラボレーションの裾野が広がりやすくなる．情報通信業界 F 社では，全社的にリポジトリ参照権限が既に存在していたにもかかわらず，当初は誰もアクションを起こさなかった．しかし，インナーソースの適切なコラボレーション手順や CONTRIBUTING.md のようなガイドが普及すると，最初の実験的取り組みから派生して約 19 の新規プロジェクトが創出された．こうした成功体験が重なるほど，他部署との連携は加速度的に強まる．

これらの導入段階を通観すると，モチベーションを維持する担当者を明確にし，公正な評価モデルを段階的に具体化することが，最終的な成熟期への道筋を拓くために極めて重要である．企業 A 社や情報通信業界 F 社の事例からもうかがえるように，評価制度が不透明なままではエンジニアや管理職の合意

形成が難航し，導入初期の拡大段階で停滞するおそれが高い．したがって，導入段階の進行に伴い制度設計と部門間の理解形成を連動させる工夫が求められる．

## 7.5 本章の総括

本章では，インナーソース導入における複数のアプローチについて，企業規模や組織文化に応じた詳細な検討を行った．特に，トップダウン型とボトムアップ型の双方のアプローチが持つ特徴や課題を明らかにし，それぞれの状況下での効果的な導入戦略について分析を進めた．

これらの分析をより具体的に進めるため，製造業からインターネット業界，情報通信業界まで，幅広い業種の企業を対象とした詳細なインタビュー調査を実施した．インタビューでは，インナーソース導入に成功した企業や課題に直面している企業など，様々な段階にある事例を取り上げ，各社の具体的な取り組みや直面した課題，その解決プロセスについて詳細な情報を収集した．

これらのインタビュー結果をもとに，インナーソースの導入段階における特徴を整理し，各段階で必要となる施策や考慮すべき要点について体系的な分析を行った．特に，アイデア期からパイロット期，初期導入期，そして成長・成熟期に至るまでの各段階で必要となる組織的な取り組みや，それぞれの段階で直面する課題への対応策について，実践的な知見を得ることができた．

アイデア期には，インナーソースの必要性が十分に理解されておらず，「本当にやっていいのか」という曖昧さが障壁となりやすい．初期には熱意あるチャンピオンが社内啓蒙を進め，小規模でも早期に成功事例をつくり出し，軽量なガイドラインを提示することで「まずは試してみる」という環境を整える必要がある．パイロット期では，組織を横断するリポジトリの試験運用を通じて，コミット数やレビュー状況といった定量指標に加え質的評価も活用し，管理職との合意形成を図る体制を構築することが望ましい．初期導入期では，会計・法務・人事といった管理部門との連携を強化し，推進役であるチャンピオンを正式に位置付けるなど，拡大に耐えうる組織的基盤の整備が不可欠となる．会計ルールや評価モデルの素案を策定し，ガバナンスガイドラインを明文化することで，企業全体が「インナーソースをやってよい」から「やるべきである」という認識に至るプロセスを後押しする．成長・成熟期に移行したあとも，コードレビューや運用管理の工数が増大し，個々の貢献をどう評価するかという新たな課題が生まれやすい．専任コミュニティマネージャーや ISPO 機能を整備し，定量指標と定性指標を組み合わせた総合的な評価モデルを導入するとともに，標準化ガイドラインやツールの導入によって管理負荷を軽減する仕組みを整えることが必要である．

このような段階的アプローチを踏むことで，企業が成熟期に至ったときには，複数の部署が相互に連携しながら自主的なコラボレーションを展開し，新たなイノベーションや技術開発を加速させる文化が形成される可能性が高まる．特に，初期導入期の阻害要因を詳細に把握し，その解消プロセスを体系化する視点は，本論文の中核を成す．連続的な課題認識と組織横断的な合意形成を同時に進めることで，導入初期から成熟期までの道筋を明確に描くことができる．

第 8 章ではコラボレーションのスコープ設定と貢献ガイダンスを扱う「インナーソーストポロジー」を示し，第 9 章ではインナーソース推進におけるインセンティブ設計を 6 類型に整理する．第 10 章ではインナーソース円環モデルの提示を通じ，インナーソースの成長プロセスを多面的に評価し，より柔軟な推進策を検討するためのフレームワーク構築を目指す．最終的に第 11 章でそれらの知見を総括することで，段階

的導入の障壁克服と持続的なコラボレーション基盤の構築に関する包括的な視座を提供し，組織規模や文化に応じた多様なインナーソース導入戦略を提案する．

# 第8章 ネットワークとしての表現：インナーソース・トポロジー

本章では，インナーソースの多義的な定義と，その適用範囲の多様性に着目する．従来の研究や実践知では，インナーソースの本質を「組織内でのオープンソース的開発手法」として単一に捉えがちだが，実際には企業ごとに異なる認識や運用上の要請があり，広義のインナーソースから狭義のインナーソースまで，複数の概念が並存する可能性がある．こうした多層的な定義の存在は，企業規模，事業構造，リスク管理体制などに応じて，インナーソースの取り入れ方が可変的であることを示唆している．

さらに，本章の分析では，インナーソースの成熟度が単純な段階論として直線的に高まるという見方だけでなく，コミュニティが形成される過程で予期しないネットワーク構造が生じるという，いわばトポロジカルな進化に着目する．インナーソースプログラムが組織変革の過程で一定の段階を踏んで拡張していく一方，個々の開発者コミュニティやチーム間の連携が複雑に絡み合いながら，新たなコラボレーション形態を形成するからである．

最終的に本章では，インナーソースがもたらすネットワーク構造を明らかにすることを目的とする．インナーソースが企業内でどのように実装され，どのようにノード（人や組織単位）が結びつき，どのような契約や会計上の整理がエッジ（関係性）として構成されるのかを解明することで，持続的で円滑なコラボレーションを実現するための指針を提示したい．

インナーソースを導入する際には，既存組織の大規模な変革を一気に推し進めることが難しく，会計制度やトップダウンの指示体系との不整合など「やらない理由」が多く存在する．したがって，実務上は企業の事情に合わせてどの範囲を対象とし，どの程度の透明性を許容するかの「線引き」を慎重に検討せざるを得ない．仮に企業が「すべてをオープンにするか，まったくオープンにしないか」という二択で考えると，実際にはインナーソース導入を断念せざるをえなくなるケースが多いだろう．

また，企業間や部門間の人員構成・規模や企業構造の違いから，インナーソースの範囲は多様に変化し得る．たとえば，5 万人規模の企業において，その 10%にあたる 5000 人の部門がインナーソースを実践する場合と，2,000 人企業の単一企業が，全員でインナーソースに取り組む場合のいずれもがインナーソースと見なされ得るように，定義上は「すべてがオープンか否か」だけで二分されるものではない．そのため，導入段階での柔軟な透明化レベルの設定こそが持続的な発展の鍵となる．規模や事業構造によっては，たとえば子会社の内部だけで区切りを設けたり，研究開発部門のみでパイロットを実施するなど，部分的にインナーソースを試すケースも多い．

インナーソースは一般に「組織内のオープンソース」を指し，透明なコラボレーションを特徴とするが，その適用範囲が企業全体に及ぶ必要はない．完全にオープンソースに近い形態を狭義とする一方で，一部の領域に限定して透明性を担保する「広義のインナーソース」も十分に成立し得ると考えられ，企業側が自社の制度やリスク許容度を踏まえて柔軟に選択できる．

以上を踏まえると，インナーソースのコラボレーションは「どの程度の透明性と可視化された貢献関係を保持しているか」で評価するのが妥当である．実際には国際的製造業 B 社が適用する，限られたメンバーのみがアクセスできる「個別契約型」の種類や，企業 A 社が適用している会員のみがアクセスできる「コンソーシアム型」などが存在しうることは本章の後半でも言及するが，特定のグループ内でのみ契約や予算を確保する手法が取られる場合もある．それでもコラボレーションスタイルにインナーソース的要素が含まれていれば，広義のインナーソースとして考えられる．

例えば，個別契約型の取り組みは，一見すると形式的な共同プロジェクトにすぎないようにもみえる．しかし，専用のインナーソースポータルを整備したり，SDK やインタフェース情報を公開したりすることで，インナーソース的なメリットを享受できる場合がある．つまり，会計やセキュリティ上の要件からソース全体を公開できなくても，透明性の一部を確保しつつ貢献者を募る状況が生まれれば，それはインナーソースの一形態として成立し得る．同一企業内とはいえコントリビューションの可視化やコードのアクセス範囲は多様な形態をとる．

また，現場の状況を見ると，インナーソースを推進する主目的が「オープンソースの社内適用」から「透明性を促進する新たなコラボレーション様式の導入」へと変化している例が散見される．このような企業では，社内の全コードを完全公開するのではなく，必要な部分のみを共有することで，効率的かつ柔軟な協働環境の構築を目指している．そのため，インナーソースの要素を部分的に取り入れた形態を採用する傾向がある．

インナーソースの定義や，これらの取り組みをインナーソースとして受容するかという議論は本論文の主題ではないため，詳細には立ち入らない．しかし，インナーソースの成熟過程において，このような議論や中間的なコラボレーションが存在することは事実である．また，企業や社員がインナーソースの能力を獲得する過程でこれらの段階を経る可能性を考慮すると，こうした多様なコラボレーション形態も包含して検討する必要がある．理想的には全面的な可視性と編集権限の付与が望ましいが，企業の会計制度や組織構造，セキュリティ要件を考慮すると，様々なコラボレーション形態が存在し得る．会計やセキュリティの都合で個別に切り出される場合でも，貢献を募集する状態や協働者が増加する状況については，本稿ではインナーソースの一形態として扱う．

このようなインナーソース導入過程で生じるアクセス権限や権利のレイヤーを整理するために，インナーソース・トポロジーという枠組みを用いて議論を展開する．企業内のチームや個人によるリポジトリへのアクセスと双方向の貢献を可能にする方法は，ボトムアップな開発における需要と経営戦略が交錯する場となるため，多様なトポロジーが同時に存在する可能性がある．

## 8.1 インナーソース・トポロジーを活用した組織内コラボレーションの定義

インナーソースの展開は単純なツリー構造に還元できないため，チーム間のコラボレーションや投資対効果（ROI），経営戦略の介在によって多様な結合関係が生じると考えられる．完全なボトムアップとも言い切れない構図は，クリストファー・アレグザンダーの議論 [46]に通じるセミラティス的な連結を示唆し，複数の軸で組織が相互に結びつく可能性を示す．

ボトムアップの活動が需要を喚起しながら，そこに計画性が加わることで，インナーソースの構造は自律性と統制が並存する独自のネットワークとなる．こうした観点では，ノードを個人やチーム，エッジをリポジトリへのアクセス権など，コラボレーションの方向とみなすことで，片方向や双方向の貢献が複数のパターンとして可視化できる．

従来，インナーソースの取り組みやオープンソースの依存関係をネットワークとして表現する試みは様々な形で行われてきた．本研究では，インナーソースにおけるチーム間の連携の方向性を捉えるだけでなく，その連携，すなわち「エッジ」の接続性をより精緻に把握することを主要な議論として展開する．これまでの研究では，依存関係やアクセスの有無といった単純な接続性のみが注目されてきたが，改めて接続の定義や種類を整理し，実際の連携形態を体系化することには大きな意義がある．

表 8.1 は，インタビュー調査を通じて発見されたトポロジーを類型化したものである（インタビューの詳細は Appendix を参照）．ここでは，ポイント・トゥ・ポイント型，スター型，部分メッシュ型，フルメッシュ型といった形態が，インナーソースの実務においてどのように現れうるかを概説している．注目すべき点は，これらすべてのトポロジーが「インナーソース」もしくは「インナーソース的」な実践として認識されていたことである．原理主義的な解釈では，インナーソースは完全なオープン性を前提とするものとされがちだが，実際には部分的なオープン性や，2 チーム間の限定的な連携であっても，インナーソース的なコミュニケーションが発生し得ることが明らかとなった．

実務における導入事例では，これらの形態が単独で存在することは稀であり，同一企業内で複数のパターンが併存するハイブリッド型が一般的である．多くの企業は最終的に複雑なチーム間ネットワークを構成してハイブリッド形態に至るが，個々の関係者によって，どのトポロジー領域（クラスター）をインナーソースと認識するかは異なる場合がある．

*表 8.1 インナーソーストポロジーの概要*

| 名前 | 解説 |
|---|---|
| **ポイント・トゥ・ポイント型** | 個別契約だけで成立し，全面的オープンとまでは言えないが，コミュニケーションや検索可能性を部分的に高めることでインナーソース的効果を生む．国際的製造業 B 社事例では個別契約のテンプレートが存在する． |
| **スター型** | 特定リポジトリに複数の個人が貢献する形態．小規模でも大規模でも成立し，最終的に部分メッシュ型やフルメッシュ型へ進化する可能性が高い． |
| **部分メッシュ型** | 複数の個人やチームが相互アクセスを保持しているが，すべてをフルオープンにはしない形態．既存の制約下で比較的多くのコラボレーションを実現できる． |
| **フルメッシュ型** | 全個人・全チームが，それぞれ相互にアクセス権限を持ち，コラボレーションを制限なく行う形態 |
| **ハイブリッド型** | 共通のインナーソース組織やチームでコラボレーションを行う一方，その他の領域では個別スタイルの貢献が並存する状態．実務ではこの形が多く観察される． |

### 8.1.1 ポイント・トゥ・ポイント型

ポイント・トゥ・ポイント型は，企業内で個別契約が結ばれたチームや個人間でコラボレーションが成立する状況を指す．インナーソース的な要素が加わらない限り，単に限定的な共同作業にとどまりやすいが，専用ポータルを整備して README ファイルや，ドキュメント，もしくは SDK やインターフェース情報を一定範囲で共有すれば，インナーソースへの入り口となる可能性がある．メリットとしては，既存の会計制度や管理フローと抵触せずに着手しやすく，秘密情報を限定された範囲で共有できる安全策となる点が挙げられる．反面，当事者間にしか可視化されないため，企業全体への波及効果が低く，インナーソ

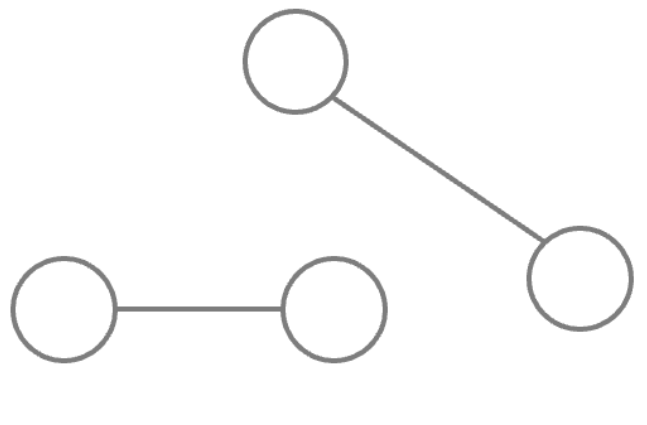

*図 8.1*
*ポイント・トゥ・ポイント型*

ースとしての広がりに乏しい場合が多い．事例として国際的製造業 B 社の個別契約テンプレートでは，契約対象プロジェクトだけを相互共有し，参加者のコミュニティを小規模に維持する仕組みが見られる．こうした形式は心理的安全性が高く始めやすいものの，より大きなコラボレーションへ波及するには，追加のガバナンスやポータルの整備が不可欠となる．

### 8.1.2 スター型

スター型は，特定のリポジトリやプラットフォームを中心に，複数のチームや個人がコンシューマーとして集まり，それぞれがコントリビューションを行う形態である．小規模なケースでは，特定の基盤チームが提供するライブラリやツールに対し，数チームがフィードバックや改修リクエストを行う関係として成立する．中規模のケースでは，基盤チームと製品チームが分離し，基盤チーム（ホスト）側が開発基盤を整備し，製品チーム（ゲスト）側がその上にサービスを展開しながら改良を提案する形をとる．一方で大規模になると，巨大なモノリスに多数のエンジニアが貢献する状況が生まれ，最終的には後述する部分メッシュ型やフルメッシュ型へと進化しやすい．

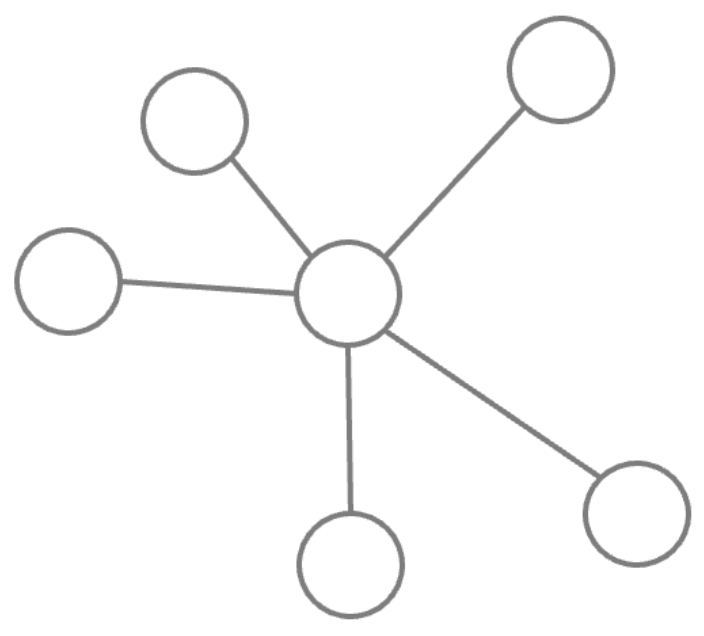

図 8.2 スター型

スター型のメリットは，中心となるリポジトリが整備されていればゲスト側が貢献を開始しやすい点にあるが，周辺のゲスト間連携は当初あまり活性化しないため，協働がホスト依存になりがちである．インターネット業界 E 社における事例では，マイクロサービス移行前に一つのプラットフォームリポジトリを中心として多くのチームが関わっていたが，やがてそれぞれが改修の自走能力を高めるにつれて，局所的なスター型から広範なメッシュ型へと変化していった．

### 8.1.3 メッシュ型

メッシュ型は，チームや個人が相互にアクセス権を持ち，複数のリポジトリを縦横に連携させる形態の総称である．概要としては，ボトムアップの活動が進むにつれて自然発生する場合もあれば，組織的戦略として特定のチーム間をつなぐ計画が進められる場合もある．メリットとしては，複数のチームが相互にコントリビューションしあうことで，重複作業を削減し，多様な観点を取り込んだイノベーションが生まれやすい．デメリットとしては，すべてのチームがフラットにアクセスできるわけではないため，閲覧権限や書き込み権限の設計が複雑化しやすく，調整コストが増大するリスクがある．

事例としては，情報通信業界 F 社がリポジトリの閲覧権限を全社的に設定していたものの，実際のコラボレーションはごく一部に留まっていたケースが該当し，そこから小さな成功事例を積み上げて段階的にメッシュ型へと発展させた．留意事項としては，Mesh 型は単線的に変化するのではなく，部分的に仮想メッシュ型になったり，状況によってフルメッシュ型に近づいたりと，段階を踏みながら変容する動態的プロセスを経る点が特徴となる．

### 8.1.3.1 部分メッシュ型 / 仮想メッシュ型

部分メッシュ型は，組織やチームの一部が互いに個別，またはチーム・組織向けにアクセス権を付与し合いながらも，全体としてはフルオープンになっていない状態を指す．概要としては，企業が既存の秘密情報や会計上の制約を尊重しつつ，特定領域だけインナーソース的に共有するケースが該当する．メリットとしては，外部へ露出したくないコードを守りながら，同じ組織内の複数チームで協働を促進できる点が挙げられる．デメリットとしては，アクセス範囲をどう線引きするかが複雑になりやすく，部署間の境界が増えることで権限管理やポリシー調整に手間がかかる．事例としては，先述したように情報通信業界Ｆ社のように漸進的にコラボレーションを拡大した例が考えられる．

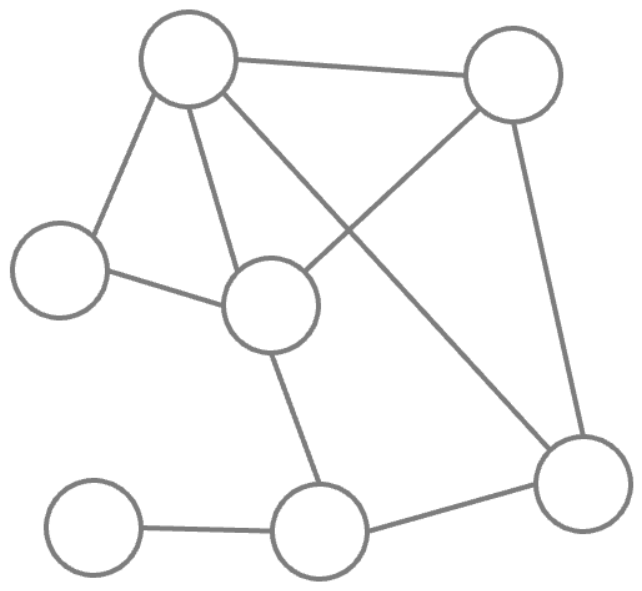

図 8.3 部分メッシュ型

留意事項としては，部分メッシュ型は企業の現実的なニーズに即した導入が可能な一方で，フルメッシュ型的な効果を得るにはさらなる調整や合意形成が必須となる．一方で，フルメッシュ型を採用していた小規模企業が成長に伴い部分的に見えない領域を設定するために部分メッシュ型を採用する可能性もある．

また，企業のソースコード管理ツールの設定によっては，仮想的な「全社員共有チーム」といった権限を付与する可能性もある．こうしたものは仮想メッシュ型と呼ぶこともできる．既存のソースコード管理ツールや組織構造を大幅に変更することなく，特定のインナーソースチームやプロジェクト空間を仮想的に設定し，その内部ではフルメッシュに近いコラボレーションを許容する手法である．メリットとしては，企業の主要システムを大きく改変せずに導入可能であり，限られた領域で高度なコラボレーションを試行し，成功事例を横展開する余地が生まれやすい．デメリットは，仮想空間の外側では依然として従来の制約が残るため，全社的な相乗効果に至るには追加の調整や拡張が求められる点である．例としては，企業がソースコード管理プラットフォーム上でインナーソース参加者を集めた「インナーソースチーム」を定義し，その定義されたグループにアクセス権を付与するはほぼフルメッシュに近いコラボレーションを実践するケースが挙げられる．

ただし，これらの設定はソースコード管理ツールの設定に依存するため，部分的なコラボレーションを促進するものとして，本稿では部分メッシュ型と同義のものとして扱う

### 8.1.3.2 フルメッシュ型

フルメッシュ型は，すべてのチームや個人が互いのリポジトリへ直接アクセスし，双方向のコントリビューションが許容される構造であり，最もオープンソースに近い形態といえる．企業内の情報を完全公開することに等しいため，会計ルールやセキュリティの観点で難易度が高いが，一度確立されれば異なる部門やプロジェクトを横断した迅速なコラボレーションが期待できる．原理主義的にはフルメッシュ型こそがインナーソースの理想形と言えるが，必ずしもこれを目指す必要はない．実務ではインナーソース用の特別 フルメッシュ型組織をソースコ

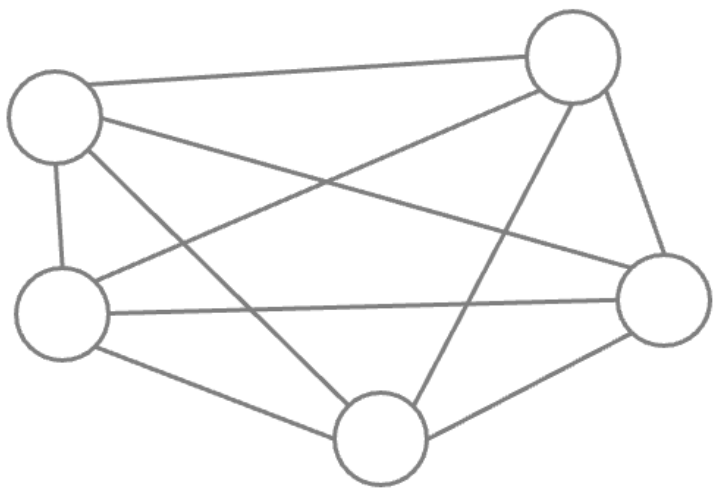

図 8.4 フルメッシュ型

ード管理ツールに別途作ることで部分的な フルメッシュ型を実現することができる. これは運営上はハイブリッド型構成とみなせる.

### 8.1.4 ハイブリッド型

ハイブリッド型は, 上記の複数形態が企業内で並行して運用され, 状況や目的に応じて柔軟に切り替えが行われる状態を指す. 概要としては, 企業全体を完全なフルメッシュにするのではなく, セキュリティ要件や事業ドメインに応じてポイント・トゥ・ポイント型や部分メッシュ型など異なる手法が併存する. メリットとしては, 各領域のニーズに合わせた最適化が可能となり, 段階的にインナーソースを拡張していく際の衝突を最小化できる. デメリットとしては, 複数の権限設定や契約形態が混在するために管理が複雑化しやすく, 組織全体を統合するビジョンが不明確になりがちである. 事例としては, 企業 A 社が一部にフルメッシュ型に近い環境を持ちながら, 他部門ではポイント・トゥ・ポイント型的な契約を継続するなど, 複数のコラボレーションモデルを使い分けるケースが挙げられる. 最終的に Hybrid を包括する形でインナーソース推進を行う企業が多いため, 全社横断の方針とローカルな最適化をいかに調整するかが重要な課題となる.

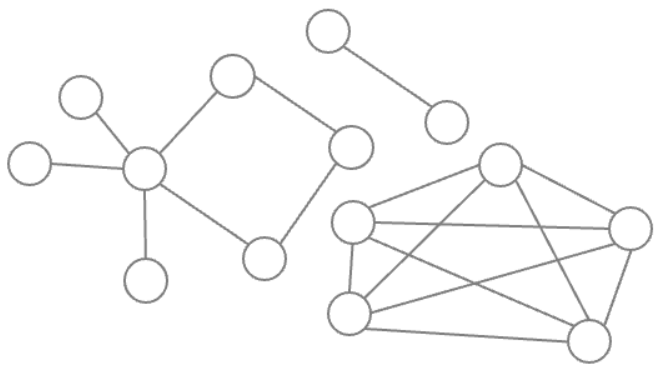

図 8.5 ハイブリッド型

## 8.2 インナーソースのコミュニティ形成と段階的拡張

インナーソースを実際に推進する際には, 担当者や推進役のチャンピオンがどのようなコラボレーション形態を志向し, どの程度の影響力を行使できるかを明確にすることが重要になる. 企業 A 社のように, キーマンが包括的な権限を持ち, ステークホルダー全体の利害を見渡せる場合は, コンソーシアム型の部分メッシュ型を構築して範囲を限定しつつインナーソースを実践する手法が取りやすい. 一方, 情報通信業界 F 社のようにエンジニアリングマネージャーから始動するケースでは, まずはスター型の小規模コラボレーションを試行し, そこから横断的な連携を広げて部分メッシュ型やハイブリッド型へと成長させていく過程が見られる. もともとソースコードの閲覧権限を大部分の社員に付与している場合, 実質的には部分メッシュ型の環境があるにもかかわらず, 実際のコントリビューションは限定的だったという事もありうる. こうした状況で「インナーソースをやる」と明言し, 小さな成功事例を積み重ね, 他部署のマネージャーとも連携しながらメッシュ型へと展開するプロセスは, インナーソース導入時の典型的なパターンといえる.

Krebs らによれば, コミュニティは「つながり」の上に成り立ち, ネットワーク構造が発展するにつれて新たな機会が生まれるとされる. 具体的には, 最初に「ばらばらに散在するクラスター(Scattered Cluster)」をとり, 次に「ハブ・アンド・スポーク(スター型)」をとり, 「複数のハブ(Multi-hub)」へ成長し, 最終的には「中心と周辺(Core Periphery)」の構造を形成するプロセスが, コミュニティネットワークの進化として示唆されている [47]. インナーソースにおいても同様に, 時間の経過とともに自然にネットワークが拡大していく. インナーソースというのは, 「コラボレーションをする箱をつくる」ことではなく, 本質的にはチームのコラボレーションを活性化させることであるため, いかにインナーソースプログラムが優れていてもコミュニティやチームがこうした複雑なネットワークを構築しない限りは絵に描いた餅になる. したがって, 段階的かつ意図的に人々をつなげ, ネットワークを編む姿勢が不可欠となる.

その意味では，国際的製造業 B 社のようにポイント・トゥ・ポイントのコラボレーションを積極的に奨励し，限定的な連携から企業内の縦割り構造に風穴を開けるアプローチも有効である．この場合，企業自体がまだ成熟段階に達していなくとも，小さなコラボレーションが社内ネットワークを刺激し，インナーソースとしてのさらなる発展に寄与する．特に日本企業の場合，組織的変革を一足飛びに成し遂げるのは難しく，まずは安全圏内での連携を重ねる中で，徐々にマネジメント層やセキュリティ担当者の合意形成を図る現実的プロセスが多くみられる．

さらに，インナーソースを個人のボランティア的活動に任せるのではなく，ビジネス活動と結び付けながら制度的に位置づけることが大きなカギとなる．どの段階でどの程度のオープン化やアクセス権限の拡張を行うかは，トップダウンの承認とボトムアップの自発性との兼ね合いが必要であり，段階ごとにガバナンス体制や評価制度を調整しながら進めることが推奨される．インナーソースの定義を厳密に定めることは本論文の射程外であるが，フルメッシュやバーチャルメッシュなど高度な連携形態を構築するには，段階的な試行と評価を繰り返すアジャイルな手法が効果的である．企業によっては，複数のトポロジーが同時並行で稼働し，ポイント・トゥ・ポイント型からスター型，さらに部分メッシュ型に至るまで多様な連携方式が混在するケースも見受けられる．最終的にどの形態を採用し，どの程度までオープン化を推進するかは，企業ごとの文化や会計制度，人事評価の仕組みなどと深く結びついているため，導入推進者がこれらの要素を的確に見極め，段階的なスケールアップを図ることが成功の鍵となる．

## 8.3　アクセス権と利用権限の区別：多層的制御の実態

企業のインナーソース環境において，あるメッシュ構造を単一の名称で定義することが難しいのは，ノード間をつなぐエッジの種類によって協働の様相が変化するためである．チーム間に強い相互依存がある密結合型のインナーソースであっても，ライセンスや利用権限の制約が厳しい場合は部分的な連携にとどまり，実質的には ポイント・トゥ・ポイント型に近い運用しかできないケースも観察される．

今回の調査では，企業 A 社と国際的製造業 B 社の事例を通じ，単なる Read/Write の権限だけでなく，アクセス権のみ，研究開発利用，商用利用という三つの利用形態が実務上の重要な区分を形成していることが判明した．これは，日本の会計制度で研究開発費として計上可能なコスト区分のソフトウェアと，資産化が要請される商用利用との区別に加え，単にマーケットプレイスとして閲覧のみを行うアクセスという利用パターンが存在するためである．

これらの区分を整理すると，表 8.2 のような構成となり，同じ Read/Write 権限であっても利用目的が異なることで実質的な制約が変化する．ツール上の設定では Write が付与されていても，会計上のルールや契約条件により実際には利用不可能な場合が生じ得ることは，インナーソース運用の複雑性を象徴している．詳細に関しては，次節のインナーソース契約にて扱う．

*表 8.2 アクセス権と利用権の整理*

| | Read | Write |
|---|---|---|
| **アクセス権のみ** | 社内リポジトリを閲覧するだけで改変や商用利用は伴わず，最小限のリスクで情報を共有する状態．試験的にコードを参照し合う過程でナレッジを広げるが，コントリビューションの活発化にはつながりにくい．基本的に利用は認められず，利用したい場合は研究開発<br>利用か商用利用を選ぶ必要がある． | 基本的な閲覧や軽微な編集が認められるものの，利用には踏み込めない．利用したい場合は研究開発<br>利用か商用利用を選ぶ必要がある． |
| **研究開発利用** | 研究開発目的にのみ閲覧・利用権限が与えられ，ソースコードを参照しながらアイデアを検討したり，技術的フィージビリティを評価したりできる．商用化を視野に入れる場合は別途契約や予算承認などが必要になるケースがある． | 研究開発目的での閲覧と改変が可能であり，試作実装を通じた技術検証や，将来的な商用化に向けた実験を進めやすい．<br>R&D 段階におけるコード提供の対価として何らかのフィードバックが求められることから，Write 権限は多くの場合で必要になる．<br>ただし商用段階に移行する際には，費用配分の見直しが必須となり，社内手続きが増加する． |
| **商用利用** | 商用利用を前提とした閲覧が許可され，実際のビジネスプロセスへのソース組み込みが可能な状態．ただし改変を伴わないため，バグ修正や機能追加は別の権限を要する場合が多い．<br>個別契約になる可能性が高く，またこの状態においてはホストチームに対してサポート契約もまかれる場合がある． | 商用目的での閲覧と改変が可能な完全権限であり，企業として製品化する段階でコードを組み込むことが想定される．この区分では最も厳格なリスク管理と費用配分が行われる．<br>個別契約になる可能性が高く，またこの状態においてはホストチームに対してサポート契約もまかれる場合がある． |

上記のように，ソースコード管理ツールの表面的な権限設定だけでは把握しきれない利用権限の階層が存在するため，実際にメッシュ構造を評価する際には，こうした多層的な制御を考慮する必要がある．

インナーソース環境を準備する際，多くの場合，表 8.2 のシステム権限の列（縦軸）に注目が集まりがちである．これは，GitHub などのツールが基本的に Read または Write という可視性オプションをシステムとして提供しているためである．しかし，これらの権限設定だけでは根本的な問題は解決されない．なぜなら，会計上や税制上の要件，あるいは可視化の範囲に関する明確な判断基準を提供していないためであり，結果として制限的なアクセス設定がデフォルトとして選択されることになりかねない．

有意義なインナーソースのガイダンスを提供するためには，表 8.2 の利用権限の行（横軸）に注目する必要がある．特に日本のような環境では，会計実務上，研究開発費と商用利用との間に明確な区別が存在する．表 8.2 ではコストと商用利用の二分法を特に強調しているが，利用権限の行は輸出管理制限，移転価格税制，利益供与の防止など，様々な条件に対応できるよう拡張することができる．

結果として，システムレベルで Read/Write 権限によってチーム間が完全なメッシュ接続を持っているように見える環境であっても，実際の利用は他の要因によって大きく制限される可能性があり，表面下では部分的なスター型やポイント・トゥ・ポイント型の構成となっていることがある．組織の要件は，ソースコード管理ツールが提供する単純な Read/Write 権限を超えて広がっていることを認識することが重要である．

進化の過程をみても，当初は ポイント・トゥ・ポイント型の個別契約によってわずかなコラボレーションを実験的に進め，一定の成果を得た後で複数チームが部分的にアクセス権を共有する スター や 部分メッシュ型 へ展開する事例が見受けられる．一方で，大規模なフルメッシュ型環境を整備したあとに，特定の秘匿領域だけを局所的に ポイント・トゥ・ポイント型で扱うなどの動きも存在する．こうした複合的な連携形

態は，組織の構造が変化するにつれ，チーム単位の新たなコラボレーションや 仮想メッシュ型の重層が生成されることでますます多様化する．

最終的に，ノード間の契約形態や技術的権限の設定を詳細に分析することで，インナーソース導入時に直面する障壁と，そこから得られる潜在的メリットがより明確になる．本研究では，こうした分析がインナーソースの本質を解明する手がかりになると考えられるが，重要なのはツール上の権限設定を超えて，参加者同士がどう有機的に協働を生み出すかという点である．

上記の利用権限区分に関しては，ライセンスや個別契約の形態が認められる．次節ではこの契約について扱う．

## 8.4 インナーソース契約：ライセンス・会計・税務の複合的検討

前節のとおり，インナーソースの導入現場では，ノードとノード，つまりチームとチームが複雑につながり合い，個別に別々の利用形態を利用し得る．その際に形成される契約形態として，本稿では以下の3種に類型した．

- 暗黙的インナーソースライセンス
- 明示的インナーソースライセンス
- 個別契約

インナーソースライセンスは，同一組織内の異なる法人間でソフトウェアソースコードを共有する際に生じる法的責任や会計処理上の課題に対応するための枠組みである [48]．このライセンスは，組織内でのソースコード共有に関する再利用可能な法的フレームワークを提供し，新たな協働の選択肢を創出するとともに，関係する法人間の権利と義務を明確化する役割を果たす．

インナーソースライセンス自体は新たな概念ではないが，日本企業における具体的事例や会計・ライセンスとの整合を検証する機会はまだ十分に整っていない．本研究では，それらの事例を俯瞰し，どのように契約形態がインナーソース環境を支えているのかを論じる．

### 8.4.1 暗黙的インナーソースライセンス：事実上の合意とコラボレーション慣習

インナーソースのライセンスを明示的に策定せず，事実上の合意や慣習に基づいてコラボレーションを進める例は少なくない．インターネット業界 C 社や企業 D などのデジタルネイティブ企業やオンラインサービス事業者では，会計やライセンスの取り決めを厳密に詰める前段階として，エンジニア間の黙示的合意のもとでソースコードの共有や改変が実質的に認められている．これらの企業では「経理がうまく処理を行う」「社内の合意がある」といった前提のもと，プロジェクト横断的なコラボレーションが自然発生的に成立している．スタートアップやインターネット系企業で特に顕著なこのアプローチは，開発がコストとして計上されることが多く，商用利用範囲が不明瞭な段階では厳密なライセンス文面を用意せずとも支障が生じにくい点が背景にある．

こうした暗黙的ライセンスのあり方は，利用可能な範囲や貢献できる機能を明示していないため，多くのエンジニアにとっては「何をしてよいのか分からない」という問題を抱える場合もある．その一方で，実際に

使っても差し支えないリポジトリを明確に示すライセンス文書を置くことで，エンジニアが積極的にコラボレーションの機会を見つけやすくなる可能性がある．したがって，暗黙的ライセンスが理想的な解ではない場合もあり，企業がインナーソースを制度化しようとする際には，最低限の利用範囲や改変ルールを示す明示的なライセンス文面を提供するメリットが生じ得る．

こうした暗黙的ライセンスに基づくコラボレーションと会計上の工数評価を結び付けるために，情報通信業界 F 社のように貢献時間を正当に計上する仕組みに関する合意を正式に行う事例も存在する．情報通信業界 F 社では Web アプリを用いた工数管理が行われており，オーダー番号でプロダクトや機能を区別し，三十分単位で作業時間を記録する運用が定着している．エンジニアが一日の終わりや翌日にまとめて記録するケースもあるが，それでも他部門や他チームへのコントリビューションが可視化され，正当な評価につながりやすい環境が醸成されている．SaaS 型の工数管理ツールを活用しているため，インナーソース関連の作業時間も同一システムで管理できる点が特徴である．

インターネット業界 E 社のような，大規模なグローバル企業では移転価格税制やソフトウェア資産の会計処理が重要視されるが，そこで必ずしも厳密なライセンス文書が先行して定義されているわけではない．たとえば，収益と直接的に結び付かない機能は費用として処理し，売買ロジックなどインターネット業界 E 社の中核的価値を生む機能は資産計上するなど，会計上の切り分けを行っている．このような複雑な会計要件を満たしつつ，マネージャーによる非定量的な工数把握が多用される一方で，暗黙のうちに社内のボトムアップ文化が成立している事実は，インナーソースにおけるライセンスやルールの整備が必ずしも必須条件ではないことを示唆している．

最終的には，特定のライセンス文書を用意せずともエンジニア間の合意と企業文化がコラボレーションを活性化させる例もあれば，明示的なライセンスを設定して利用範囲を示すほうが貢献意欲を高める場合もある．いずれにしても，企業内での黙示的合意がボトムアップの開発様式を支えている実態を踏まえると，インナーソースが成熟する過程で，暗黙的ライセンスの運用と明示的ライセンスの導入とが併存し，会計や評価制度の整備と連動する多様なパターンが生じることが推察される．

### 8.4.2 明示的インナーソースライセンス：コンソーシアムモデルの事例

複数法人格を抱える大企業においては，単純な無償共有は利益供与や移転価格の問題に抵触しやすく，海外拠点を持つ場合には輸出管理や海外税制など追加の制約が生じる．このような状況では，インナーソースの利用範囲を一定のルールに基づいて制限する設計が求められる．そのための手段として，インナーソースライセンスの策定が挙げられるが，往々にして実務では GPL 系ライセンスの延長線上のようなごく簡易的な文書を作成しているにとどまり，会計面や連携スキームについて深く言及されることは少ない．

実際には，既存の事例として公表されているインナーソースライセンスをそのまま日本企業に導入しても，会計や税制，企業文化と整合性が取れない場合がある．また，なぜそのライセンスの条文になったのかなどの背景情報や，そのライセンスを取り入れることによって解決される事柄，例えば会計に関する条件などはそもそもライセンス条文には含まれていない．そのためライセンス単体ではなく，バックグラウンドとなるスキームや期待されるコラボレーション様式まで一体的に検討する必要がある．

企業 A 社の事例では，グループ内で複数法人が緩やかに連携するモデルを採用し，「コンソーシアムモデル」と呼んでいる．親会社の特定部門がプラットフォームを運営し，参加企業はサブスクリプション的

な年会費を支払って一定範囲のソースコードやツールにアクセスできる仕組みである. 企業 A 社は過去に本社が全てを買い上げて明確な使用料を分割するモデルを検討したが, 採用難易度が高く結局採用しなかった. 該当モデルは負担と恩恵を柔軟に分配する点が特徴である.

#### 8.4.2.1 三階層コンソーシアムモデル:閲覧・貢献・商用利用の段階設定

コンソーシアムモデルでは「閲覧」「貢献」「商用利用」の三階層を設け, 研究開発段階での閲覧や自由なコントリビューションを許容しながらも, 商用利用に移行する際はライセンスフィーを徴収する流れを確立している.

*表 8.3 コンソーシアムモデルの三つの階層*

| | **閲覧 Tier** | **貢献 Tier** | **商用利用 Tier** |
|---|---|---|---|
| **閲覧** | 可 (研究開発段階) | 可 (研究開発段階) | 可 |
| **ライセンス** | インナーソースライセンス<br>(非商用範囲) | インナーソースライセンス<br>(非商用範囲) | 開発部門との個別相談 |
| **利用形態** | トライアル利用 | 自由利用<br>(研究開発段階での利用に限定) | 商用利用 |
| **利用と改変の自由** | N/A<br>(原則利用のみ) | 改変・共有可能<br>(研究開発段階での利用に限定) | 改変可<br>(契約に依存) |
| **サポート** | N/A | N/A | サポート契約可<br>(契約に依存) |
| **メンテナンス** | N/A | N/A | メンテナンス提供有り<br>(契約に依存) |
| **金銭のやりとり** | N/A | 運営への会費支払い | 対価支払必須<br>(明確な費用発生) |
| **利用者提供価値** | フィードバック・テストなどの非金銭的対価 | フィードバック・テスト・プロジェクト貢献などの非金銭的対価 | N/A |

閲覧層は研究開発目的での資産閲覧・利用に限定されており, 特許出願前や新規技術検討段階の PoC に最適化されている. この段階では各参加企業が自由にソースコードを閲覧, テスト, フィードバックをすることができる. ここでの無償利用は研究開発目的に限定され, 税務上・会計上も「研究開発費用」として内部処理がされる. グループ内での価値交換が「テスターとしてのフィードバック」など非金銭的な形で行われる. 「Give and Given」という考え方により, 金銭面における無償でのやりとりを研究開発活動として合理化できる. 企業 A 社は社内に対する全オープンでのソース公開が困難であった. ここでは, 商用利用手前の段階に限定することでリスクを抑えつつ, 技術検証を円滑化する点に意義がある. また, 研究段階の時間を比較的長く取ることで, 内部的な利用範囲を最大化し, 結果として開発コストを研究開発費として取り扱うことができるメリットがある.

続いて貢献層では, 引き続き非商用範囲の利用を前提としながら, 参加企業間でコードコントリビューションが許される. こちらは軽微な会費で参加可能なため, コンソーシアムモデル的な相互利用関係が強まる. ここでの会計上の扱いは引き続き研究開発費として計上可能であるが, 追加的に各社が機能改良やバグ修正を行い, これを共有することで, フィードバックだけでなく実質的な資産の高度化が進む. ある種の「アセットの相互増殖」が起こり, 企業間コラボレーションが深化する. 貢献層を設けることで, 単なる閲覧に留まらず, 各社が主体的に改良を加える誘因が生じる. これにより, コンソーシアム全体としてのアセット品質・有用性が向上し, 最終的な商用化や社内横展開に向けた素地が整う. ここで依然として研究開発費処理が可能なことは, 財務的柔軟性を保ちながら技術コアの成熟化を図る上で有効である.

最後に，商用利用層では，商用利用に踏み込む．ここから先は明確なライセンスフィーが発生し，製品化されたソフトウェアやコンポーネントを使用する場合は，個別契約や有償ライセンスに移行する．この段階で，インナーソース的な「オープン」から一転，明確な収益確保メカニズムが働き，内部資産を外販または別事業体に提供する際の移転価格や税務処理が適正化される．最終的に商用利用を目指す場合，研究開発フェーズでのコラボレーションから一転し，法的・会計的な整理が必要となる．商用利用層では有償契約によって移転価格問題をクリアにし，対価徴収を正当化する．こうして対価関係が明確化され，純然たる市場取引として整理される．

#### 8.4.2.2 コンソーシアムモデルの利点

コンソーシアムモデルは，複数法人格の並立に伴う法務・税務対応の柔軟性，組織内部におけるコラボレーション文化の醸成，そしてスケール段階での収益モデル確立という三つの点で優位性を示す．法人格が分かれている企業間でも研究開発段階での無償コラボレーションを正当化でき，商用化フェーズにおいてはライセンスフィーを通じた対価徴収が可能になる．閲覧やコントリビューションの段階で自由度が高いため，ハッカソンや教育プログラムとも親和性が高く，インナーソース文化を醸成しやすい．スケールアップの際には，研究開発費で柔軟に活用していたアセットを商用ライセンスに移行するロードマップが明確なため，単なる無償共有では終わらない持続的なビジネスモデルを描きやすい．結果として，インナーソース導入企業が内部統制を維持しつつ価値創造を促進するうえで，有力な選択肢となり得る．

以上のように，コンソーシアムモデルを含むインナーソースライセンスの運用は，分社体制やグローバル拠点を抱える大企業において，インナーソースを段階的かつ戦略的に推進する強固な仕組みを提供する．

### 8.4.3 個別契約：業務委託型協働モデルの事例

調査対象となった国際的製造業 B 社は，ホールディングカンパニー本社のもとに複数の事業会社を擁する大企業グループであり，本社には共通開発部門が設置されている．本社は社内の OSPO 事務局と協力して各子会社との共同開発を主導し，グループ全体としてオープンソース的な手法を社内展開している．ここでは，必ずしもすべての開発行為が完全にオープンなリポジトリで行われるわけではなく，広範なインナーソースコラボレーションだけでなく，より厳密な意味合いでは「業務委託型協働」に該当するコラボレーションの両軸が同居しうる．これは明確に業務委託契約を結び，工数ベースで対価を支払うモデルである．

これはインナーソースの既存の事例では一部の部門の従業員を受託業者に見立てて，部内でお金を出し合う事例が報告されているが [49]，この事例はより包括的にグループ内企業と企業の間の価値移転を可能にする契約として昇華されている．この場合 R&D 段階ではフィードバックを対価とみなし，製品化段階になるとライセンスフィーやサポートフィーを徴収する二段構えが基本的な流れとなっている．

#### 8.4.3.1 業務委託型協働モデル：研究開発と商用化を繋ぐ二段階設計

業務委託型協働モデルは，以下のような要素で構成される．まず，グループ内組織間で業務委託契約を交わし，「工数×人件費」で対価を算出する仕組みを整える．たとえば，本社が保有するエンタープラ

イズレポジトリに子会社のエンジニアがアクセスし，開発工数を提供する代わりに，本社から子会社に委託費が支払われる形である．その成果物の所有権は本社またはプロジェクトオーナーに帰属し，R&D フェーズと製品化フェーズでそれぞれ異なる対価計算を行う．

研究開発フェーズでは，開発成果がまだ確定していないことから金銭的対価のやりとりが困難であるため，要件定義やバグ報告，QA（品質保証）対応など，いわゆる「フィードバック」を対価と位置づける．実際など，フィードバックはテスト工数相当の価値がある．そのため，無償譲渡にはらならず，等価交換として処理することになる．

一方で，製品化フェーズでは商用利用が可能になるため，ライセンスフィーやサポートフィーなどを設定し，契約フォーマットも製品提供用のものに切り替える．こうした二段階設計によって，研究開発段階の不確実性を吸収しつつ，後の商用化フェーズでは明確な収益化スキームに移行できる点が特長である．さらに，標準化された契約フォーマットを活用し，権利帰属や再利用範囲，インターフェース仕様などをアーキテクチャ図で明示することで，契約交渉や将来的な混乱を最小化する努力がなされている．

*表 8.4 個別契約における契約フォーマットのフェーズ別対価形式*

| フェーズ | 対価形式 | 契約フォーマット例 | 会計・税務上の扱い |
|---|---|---|---|
| **R&D 段階** | フィードバック(非金銭) | 研究開発用ライセンス契約書 | 工数・要件定義労力相当として説明可能 |
| **製品化・商用段階** | 金銭（ライセンス費，サポート費） | 製品提供用ライセンス契約書，業務委託契約書 | 移転価格算出対象，国際取引時は追加検討要 |

## 8.4.3.2 移転価格と高付加価値コード共有：グローバル展開への実務的視点

グローバルな事業展開に伴って，海外拠点間での無償ソースコード共有が移転価格税制上の問題と直結する例は多い．特に高い市場価値やマーケティング効果をもつコア機能は，工数だけで説明がつかないためライセンス費や買い切り費用，サブスクリプション費用を設定している．ここでいう高い市場価値の一例として，CM（コマーシャル・マーケティング）価値を有する高付加価値機能が挙げられる．さらに，一般的な契約フォーマットでは対処しきれない権利帰属・機能配分の複雑なケースでは，特別契約を締結し，移転価格対応の詳細を規定する場合もある．

一部の企業では，国内（たとえば日本）にコア開発機能を集約し，海外拠点はあくまで労働提供の役割にとどめるなど，組織設計面での工夫によって移転価格問題を軽減する戦略が採用されている．このように，インナーソース的なコード共有が引き起こす国際税務上のリスクに対しては，契約の仕組みを厳格化するだけでなく，事業体の配置や拠点機能の設計から見直す取り組みが有効である．

## 8.4.3.3 標準化とテンプレート運用：契約・会計・税務フローの統合管理

研究開発から製品（無形資産の「ソフトウェア」）への会計上の段階移行に従い，契約書を研究開発用から製品提供用へ切り替える．また，共通管理部門が用意した標準契約フォーマットを用いることで，担当部署は指示に沿うだけで会計上・税務上の整合性を確保できる．これによりルール逸脱や社内交渉負担を低減し，自然なコンプライアンス遵守が実現する．以下に契約書フォーマット例を示す

*表 8.5 契約テンプレートの主な種類*

| 種類 | 対象フェーズ | 対価形態 | 主な条項 |
|---|---|---|---|
| **研究開発用ライセンス** | R&D フェーズ | フィードバック対価 | 要件定義・QA 報告の義務，共同 R&D 利用範囲 |
| **製品提供用ライセンス** | 商用フェーズ | 金銭対価（ライセンス費等） | 製品化後サポート費用，所有権・再利用条項 |
| **特別契約** | 複雑・高付加価値案件 | ケースバイケース | 権利帰属，移転価格対応，機密保持，再利用可否 |

なお，グローバル展開を行う際，ソースコードを海外拠点に共有する行為が輸出管理対象となる場合がある．一般論として兵器転用の可能性や制裁対象国への技術流出リスクを踏まえ，レポジトリ開設時やユーザー追加時に輸出管理手続きを実施することが求められる．インナーソースを推進する場合も，各リポジトリの機能範囲や利用目的を明記し，問題のある機能を含まないポリシー運用でクリアリングを行うなど，繰り返し確認するプロセスが必須となる．

### 8.4.3.4 業務委託型協働モデルの利点

本モデルは，以下の点で有用性が高い．まず，R&D 段階ではフィードバックを対価とすることで，不確実性の高い研究開発に柔軟に対応できる．その一方，製品化段階では金銭対価を要求し，実際のライセンス契約やサポート契約に切り替える二段構造により，研究フェーズから商用フェーズへのスムーズな移行と収益化が可能となる．次に，契約フォーマットやアーキテクチャ図の整備，会計・税務における説明ロジックの標準化によって，移転価格や輸出管理など複雑な国際要件への対応コストを軽減できる．最後に，ボトムアップのイノベーションとトップダウンの統制が両立しやすいため，グループ内の潜在的コラボレーション機会を制度的に取り込める枠組みとなり得る．

この業務委託型協働モデルは，単なる「社内版オープンソース」の導入にとどまらず，法務・会計・税務の各側面を踏まえて設計された包括的なフレームワークである．今後の展開としては，業種や企業規模が異なるケースや，高付加価値ソフトウェアの移転基準をさらに厳密化する必要がある分野での活用が考えられ，比較研究や実装事例の集積が期待される．結果として，インナーソース推進企業において，内部統制を損なうことなく柔軟なコラボレーションを実現するための有力な手段として位置付けられるだろう．

このモデルの会計上のコードの扱いは企業 A 社と類似する．異なる点は個別の複雑な契約を含めたテンプレートを前提としたコラボレーションとするか，コンソーシアム型の会費を前提としたコラボレーションになるかである．これらのモデルは，単なるオープンソース的振る舞いの社内展開ではなく，会計・税務・法務上の要請を踏まえた体系的なフレームワークである．一方で社内公開性という意味ではこうした取り組みに被せる形で，社内のインナーソースポータルや，ドキュメントの公開など，個別契約を超えた「インナーソース的コラボレーションをどう誘引するか」という観点が欠かせない．インナーソース自体がどこまでの範囲を含み得るかについて，この論文では触れないが，こうした個別契約をどのように全体的なコラボレーションへと派生させるかなど，高度化の手法を探求することで，インナーソース導入の一層の発展が期待される．

## 8.5 総括:インナーソースガイドラインの策定による柔軟な統制

インナーソースを企業内に導入する際には, 単に会計や税務のルールを整備すればよいというわけではなく, それ以上に, どのようなコラボレーションを実現したいのかを全社的に共有しつつ, 必要に応じたルールを設けることが重要となる. 前半で示したインナーソース・トポロジーの多様性を理解し, 各組織が置かれた状況に合わせて透明性や編集権限の範囲を調整できる仕組みが不可欠である.

日本企業を対象とした先行調査からは, 会計や税務, さらには法務上の懸念に対し, 非常に敏感に反応する企業と, まったく気にしない企業とがはっきり二極化している実態が確認されている. 企業 A 社や国際的製造業 B 社のような厳格な管理体制は必ずしもすべての企業に求められるわけではないが, インターネット企業 E 社やインターネット企業 F 社へのインタビュー調査が示すように, 多くの企業では会計上の取り扱いや部門間調整に関する暗黙的なプロトコルが既に確立されている. このため, 価値やリソースの移転に関する基本的なガイドラインは, インナーソース導入時に必ず整備すべき要素と言える. 心理的安全性やどこまでコラボレーションが許されるのかという境界を曖昧にしたまま進めると, 結局は中間管理職や監査部門などが導入を阻む形になりやすい. その際, 単なるソースコードへのアクセス権限だけでなく, 実際の利用許諾の範囲まで含めた包括的な管理の枠組みを検討する必要がある.

インナーソースの核心は, 社内のコード共有や相互のコントリビューションを通じて新しい価値を創出することにあり, その過程で部門間の垣根を取り払っていく点に大きな意義がある. ただし, 日本企業特有の稟議制度や品質保証の文化のもとでは, ルールが見えないまま外部部門のエンジニアがコードを修正する状況に対して, 管理側が強い抵抗感を示す可能性は否めない. そこに会計や輸出管理などの法令リスクが加わると, さらに懸念が高まることになる.

一方で, これらの懸念事項を導入初期から全面的に取り扱うのは, 当事者や利害関係者を必要以上に増やす要因になり, かえって導入プロセスが難航するという矛盾が生じる. インナーソースの魅力を生かすためには, まずは限られた範囲で心理的安全性を確保し, 小さな成功事例から社内に浸透させる段階的アプローチが有効である. 初期フェーズでは最低限のガイドラインを示しつつ, 一定の実績が積み上がってから会計やセキュリティに関する精緻な制度設計を進めるほうが, 現場のモチベーションを失わずに拡大を図りやすい.

また, 企業 A 社の言及にもある通り, 会計ルール, コンプライアンス, 知的財産権の取り扱いなどを「初期導入期におけるルール整備のモデルとして扱う」ことで, むしろインナーソース文化を段階的に浸透させる手がかりとなる. この点は一見逆説的だが, 細かい制度面をあえて明示することで, 「インナーソースの取り組みによって生じる版権や社内会計などの整理はこうするのだ」と最初から定義すれば, 各部門の管理職も「そういう仕組みがあれば混乱は起きにくい」と納得しやすいからである. たとえば, 社内で使うライセンス体系を設計し, 利用・改変・配布の範囲を明文化することで, 初期段階の懸念を事前に解消する. 日本企業に特有の「暗黙的合意形成」の慣習を踏まえると, 特定キーパーソンとの密な擦り合わせを行いながら, 早期に各部門の意向を吸い上げ, チャンピオンや伝道者が全社的に情報共有するプロセスが障壁回避に有効である. 繊細なコミュニケーション設計が求められる一方, いったん合意が形成されれば安易に覆らないため, 大規模なスケールアウトが期待できるというメリットもある.

上記をふまえると, インナーソースガイドラインは「企業全体で共通化すべき部分」と「各部門・プロジェクトが柔軟に決められる部分」を明確化することが大切である. すべてを統一しすぎるとイノベーションを

阻害しかねないが，まったく指針がないと不安に駆られた管理者が導入を拒絶する恐れがある．したがって，企業のリスク許容度や既存の人事評価制度，さらには技術スタックとの兼ね合いを見ながら，どの段階で何をガイドラインとして提示するかを丁寧に設計する必要がある．

総じて，インナーソースガイドラインの策定は，企業が最終的にどのようなネットワーク構造とコラボレーションの広がりを実現したいのかを明確化し，そのためにどの程度の透明性や編集権限を与えるかを段階的に定義する行為にほかならない．単に会計や税務上のルールを整備するだけではなく，心理的安全性を高めて社内の自発的コントリビューションを促し，最終的にボトムアップとトップダウンが折り合う形でインナーソースが成熟する状態をつくり出すことが，ガイドライン策定の真の意義である．

# 第9章 多層的インセンティブ設計と企業内導入の要諦

本章では，インナーソース活動を企業内で定着させるために活用される多様なインセンティブの仕組みを6種類に類型化して論じる．これらの類型は各企業の組織文化やビジネスモデル，従業員の意欲設計などと複合的に結びついて形成されており，選択や組み合わせ方によってインナーソースの活性度や継続性が大きく左右されると考えられる．具体的には，個人に焦点を当てるものとプロジェクトに焦点を当てるものがあり，さらにそれぞれ金銭的還元と功績顕彰（活動認知）という二つの評価軸が存在する．以下の表が示すように，これらが組み合わさって六つのモデルが構成される．インタビューを通じ，モデルの採用実態が明らかになった．

インナーソースプロジェクトにおけるインセンティブプログラムの有効性については，Huaweiの事例研究を通じて既に実証的な知見が得られている [50]．本章では，これに加えて日本企業における事例分析を通じ，インセンティブプログラムの有効性を体系的に整理する．さらに，プログラムの制度化以前のインセンティブ形成過程に着目し，各発展段階において採用されうるインセンティブモデルを分析する．特に，初期フェーズと成熟フェーズにおいて，異なるインセンティブモデルが要請されることを明らかにする．

インセンティブモデルの有効性は広く認識されているものの，組織へのインナーソースの定着度が低い段階では，体系的な表彰プログラムなどの制度的施策を即座に導入することは困難である．そこで本章では，導入初期段階における企業の実践可能なインセンティブモデルを類型化し，その段階的な導入経路を提示する．これにより，インナーソース活動の発展段階に応じた適切なインセンティブ設計の指針を示すことを目指す．

*表 9.1 インセンティブモデルの6類型*

| | 個人ベース | プロジェクトベース |
|---|---|---|
| Monetary<br>（金銭還元） | **個人評価反映型**<br><br>給与やボーナスへの直接反映．情報通信業界 F 社のようにマネージャーとの面談で定期評価と連動している事例がある． | **部門間金銭授受**<br><br>インナーソース活動で開発したツールやライブラリの利用実績をもとに，所属部門の予算に還元する，もしくはやりとりを実施する．国際的製造業 B 社のように明示的に契約を結ぶ例がある． |
| Acknowledgement<br>（功績顕彰/活動認知） | **表彰制度型**<br><br>エンジニア個々の功績を称え，個人表彰やゲームのエンドロールへの名前掲載によって顕彰する方式．企業 D では「名前を載せる」文化がエンジニアのモチベーションを高める． | **活動正式化型**<br><br>エンジニア個々の功績を称え，個人表彰やゲームのエンドロールへの名前掲載によって顕彰する方式．企業 D では「名前を載せる」文化がエンジニアのモチベーションを高める．（実質的に稼働が裏で調整されて予算がつく状況も生まれうる） |
| Hybrid（両軸） | **賞与を伴う表彰制度型**<br><br>表彰制度や社内アワードで周知度を高め，同時に金銭的報酬も付与する方式．情報通信業界 F 社は創意工夫賞や社長賞を用意している． | **ファンデーション型**<br><br>CNCF（Cloud Native Computing Foundation）のような仕組みを社内に取り入れ，個人やチームの技術貢献を名誉とともに拡大予算や業務リソース拡充で支援するモデル．インターネット業界 C 社の事例が該当する． |

以上の 6 種類はそれぞれ長所と短所をもち，インナーソースを導入する企業では状況に応じて部分的に組み合わせて運用するケースが多い．単一の評価軸では捉えきれないコラボレーションや文化的貢献を正当に認めるには，複数の報酬要素を組み込んだ評価設計が不可欠である．

どのタイプを選択するにしても，その評価基準や運用プロセスを組織全体で共有し，「この会社でインナーソースに貢献すれば，このような見返りが得られる」というイメージをエンジニアが具体的に持てる状態を作ることが重要である．もし評価制度があいまいなままでは，貢献意欲の高いエンジニアに負荷が集中したり，制度が形骸化したりする恐れがあるため，インナーソース推進時には評価モデルの設計と運用を慎重に行う必要がある．

## 9.1.1 対個人金銭還元型 / 個人評価反映型

対個人金銭還元型，あるいは既存の個人評価制度への反映型は，日本企業の人事制度に比較的導入しやすいと考えられ，現にインナーソースを導入した情報通信業界 F 社の事例では大きな役割を果たしている．エンジニア個人が他部門や他チームに積極的に貢献した場合，既存の業績評価制度の枠組みに沿って報酬やボーナスへ反映される仕組みが整えられており，エンジニアの意欲を支える要因になっている．

従来の評価制度を活用しつつも，インナーソース特有の貢献度をどのように加点するかは，管理職の方針や裁量に大きく左右される．情報通信業界 F 社では，エンジニアリングマネージャーが半期ごとに目標をすり合わせる際に「他部門リポジトリへの貢献をどれだけ行うか」を事前に共有し，期末に成果を数値化して給与へ反映する仕組みを導入している．

この運用を機能させるには，インナーソースへの貢献実績を客観的に把握できる可視化ツールや工数管理システムが不可欠である．情報通信業界 F 社では，社内の Web アプリを通じてエンジニアが作業内容を入力し，エンジニアリングマネージャーがそれを参照して評価資料を作成するフローを定着させた．これにより，エンジニアは「コントリビューションが数値として評価される」という実感を得やすくなった．

もっとも，既存の個人評価反映のアプローチだけでは，難易度の高い課題の解決や部門間をつなぐ調整力など，質的な貢献を見落とす危険がある．そこで情報通信業界 F 社では，管理職がエンジニア本人や関係者への聞き取りを重視し，主観的な評価要素も加味して最終判断を行うプロセスを採用している．

インナーソース導入期には，管理職が「本来の業務に支障が出ないか」を懸念する場合が多いが，情報通信業界 F 社ではチーム目標とインナーソース活動の両立を明示することでこれを緩和している．エンジニアが本業プロジェクトの目標を達成しながら余剰工数でインナーソース活動に参加できる二段階目標を設定し，どちらを優先すべきかを上長と事前に相談できる仕組みが整備されている．

賃金還元型を採用する利点としては，組織全体の人事制度を大きく変革しなくてもインナーソースを試行できる点が挙げられる．情報通信業界 F 社のように目標管理制度が既に存在する場合は，その評価項目に「インナーソース関連の貢献度」を追加するだけで導入コストが低減し，経営陣の抵抗感も小さくて済む．

一方，管理職の方針に依存しやすい評価モデルは，人によって評価度合いがばらつくリスクをはらむ．ある管理職はインナーソースを積極的に評価するが，別の管理職はそうでない場合，エンジニア間で評価に不公平感が生じる可能性がある．情報通信業界 F 社ではエンジニアリングマネージャー同士の連携

を強化して評価基準をなるべくそろえようとしているが，現場レベルのばらつきはまだ完全には解消されていない．

評価の公正さを担保するには，定量的評価（プルリクエスト数や課題対応数など）と定性的評価（コミュニケーション貢献度や他部門からの感謝の声など）のバランスを取ることが重要である．情報通信業界 F 社は，こうした多面的アプローチを通じて，単なる作業量だけでなく革新的なアイデアや指導力も評価対象とする余地を確保している．

結論として，賃金還元型と既存評価型の組み合わせは，企業文化を大きく変えずにインナーソース活動を評価できる有効な手段である．ただし，管理職の姿勢への依存度合いや定性的評価の組み込み方など，運用上の配慮が不可欠であるため，単なる KPI（重要業績評価指標）評価に終わらせず，エンジニアの学習・創造・自律性を後押しする仕組みへ成長させることが望ましい．

### 9.1.2 対個人功績顕彰型 / 表彰制度型

対個人功績顕彰型，あるいは表彰制度型とは，インナーソース貢献者に対して金銭的見返りではなく，称賛やステータスシンボルの付与を重視する評価のしくみである．ゲーム事業を中核とする企業 D の事例では，エンドロールへの名前の掲載が象徴的な功績顕彰策として機能し，開発者の自己肯定感と誇りを高める効果を生んでいる．

例として，企業 D のようなインターネット業界では「面白いものを作りたい」という動機が強く，例えばゲーム事業の場合は売れたゲームの実績や看板を誇りにする文化が根付いている．このような背景のもと，インナーソースでも「自分がここに関わった」という痕跡を公式に残す行為は，エンジニアのモチベーションを高めるうえで大きな意味をもつ．ゲーム開発では，エンドロール掲載という功績顕彰が次の作品のチャンスやエンジニアのキャリアにも直結すると考えられている．

企業 D では，共通ライブラリやインフラツールなどの整備に貢献した開発者の名前をゲームや社内ドキュメントに刻むことで，「あの人がいたからこの作品が完成した」という評価を可視化している．こうした名誉型の評価は，金銭的報酬とは異なる次元で，技術者のプライドや相互敬意を高める効果を発揮するため，長期的に見れば組織全体の技術力や学習意欲を底上げする可能性がある．

功績顕彰型のメリットは，企業側が余分な予算を投じなくてもエンジニアの意欲を引き出せる点にあるといえる．ただし，プロジェクトが拡大し，数十人から数百人単位の貢献者が増えてきた場合，功績顕彰の粒度や範囲をどうコントロールするのかという新たな課題が生じる．一定の基準を設けないまま無制限に名前を載せると，やがては掲載の意味が希薄化する可能性がある．

他方，名誉が過剰に持ち上げられるあまり，意図的に目立つコミットを狙う開発者が増えるリスクにも留意すべきだろう．形だけのコミットや，急いで目立つ成果を出そうとするあまり，レビュー体制がなおざりになるようでは，本来の品質向上や学習効果が損なわれてしまう．

名誉による還元を評価制度全体にどう位置づけるかは企業ごとに異なるが，企業 D のように，ゲーム基準の価値観が社内に浸透している場合，エンジニアも名誉を強く求める傾向がある．その結果として，賃金還元型や予算還元型よりも，エンドロール掲載や社内アワードといった名誉重視の制度が強く機能する．組織文化がいかに評価モデルとフィットするかが，インナーソース導入の成否を左右するといっても過言ではない．

総じて，功績顕彰型はコストをかけずにインナーソースを活性化する有力な手段になり得る一方，運用面での課題や他の評価軸とのバランス調整を慎重に考慮しなければならない．名誉以外の要素を軽視すると，エンジニアのキャリアパスや長期的な成長への配慮が十分になされず，結果的に人材流出やモチベーションの停滞を招く懸念があるため，あくまで企業全体の文化と整合的に設計することが重要である．

### 9.1.3 対個人ハイブリッド型 / 賞与を伴う表彰制度型

賃金と功績顕彰を組み合わせるハイブリッド型は，給与やボーナスなどの金銭的報酬と，名誉的な表彰や称号付与という二つの報酬軸を同時に機能させる評価モデルとして位置づけられる．表彰制度が既に定着している情報通信業界 F 社では，推薦制度（ノミネーション制度）をインナーソース活動の評価に取り込み，このハイブリッド型を円滑に運用している．

情報通信業界 F 社では毎月や四半期ごとに表彰を実施し，エンジニアが「いかに価値のある貢献を果たしたか」を社内に周知できる．特にインナーソースに関わる成果が認められた場合，少額の金銭ボーナスや昼食代などが授与されるため，名誉的メリットと金銭的還元が同時に得られる仕組みとなっている．

功績顕彰をベースにしながら，高評価を得たエンジニアには給与査定や昇進にも好影響が及ぶよう設計されている．たとえば受賞回数に応じてリードエンジニアに認定される可能性が高まり，そこから昇給や役職テーブルの引き上げにつながるなど，段階的にステップアップできる制度が用意されている．

この制度は推薦プロセスに特徴があり，情報通信業界 F 社ではプロジェクトマネージャーやチームリーダーが推薦権を持つ．これによって受賞が実績や行動に基づくものであると評価されやすく，エンジニア同士も公正感を抱きやすい．一方で，推薦プロセスが形骸化すれば真の貢献者が埋もれるリスクがあるため，制度の運用には注意が求められる．

賃金と名誉の両輪を走らせる利点は，エンジニアの多様な価値観に対応できる点にある．金銭より自己実現や承認を重視するエンジニアもいるため，組織として多様な人材を巻き込みやすくなる．この特徴は特に若手エンジニアや学習意欲の高いエンジニアにも好評である．

ただし，評価が二軸で走る分，運用コストが増える恐れもある．たとえば，いつ功績顕彰を行い，その結果がいつ給与やボーナスに反映されるのかがあいまいだと，現場で混乱が起こり，制度そのものへの不信感を招きかねない．情報通信業界 F 社は四半期サイクルと連動させるなど，評価フローを明確化して混乱を抑えている．

推薦制度や表彰制度は，エンジニア同士が互いを称賛する文化を生む点でも効果がある．情報通信業界 F 社では表彰事例を社内掲示や全社会議で発表し，他のエンジニアから賞賛やコメントが届く仕組みを整えている．これにより，インナーソース活動に対するポジティブな空気が醸成され，エンジニア間の心理的安全性が高まる．

しかしながら，名誉を求めるあまり過度に自己アピールに走る開発者や，行き過ぎた管理を行うリーダーが出現するリスクがあることも認識すべきである．情報通信業界 F 社では，チーム目標や相対評価も適度に組み合わせ，名誉の過剰競争を抑制している．

こうしたハイブリッド評価モデルは，多様な人材の動機づけを可能にしながら，組織としても着実に成果を回収したいという企業の思惑を反映している．インナーソース活動が増えると，技術革新だけでなく，エ

ンジニアの自発的な行動や他者への貢献を推奨する企業文化が形成されるため，企業内イノベーションの土壌を耕すうえで有効な手段といえる．

### 9.1.4 対プロジェクト金銭還元型 / 部門間の金銭授受

本モデルは，インナーソース活動から生まれた成果や共通基盤の利用状況に応じて，関係部門やチームに予算を還元する方式，あるいは利用に対してライセンスを支払う方式を指す．社内で開発された共通ライブラリやツールを他部門が利用し，その結果としてコスト削減などの具体的なメリットが得られた場合，貢献元チームへメリットの一部を再配分したり，利用に応じて部門間で金銭のやり取りが発生する点が特徴である．このような仕組みを導入した企業としては，企業Ａ社や国際的製造業Ｂ社が挙げられ，コンソーシアム型の導入や個別契約の締結，サポート契約による金銭徴収などが具体例となる．

本モデルの核心は，インナーソース貢献による経済的効果を定量化し，その対価を部門間で可視化する点にある．共通ライブラリやプラットフォームを整備する部門は，経済的見返りが明示されることで継続的なメンテナンスや拡張に注力しやすくなる．従来は「裏方」のように扱われていた共通基盤管理部門も，金銭的リターンの明示化によって社内での地位や発言権が向上する可能性がある．

なお，こうした取り組みは企業組織においては協働の基本的姿勢とみなされる場合も多いが，8 章で言及した企業 A の「ギブ・アンド・ギブン（Give and Given）」の考え方のように，単なる金銭のやり取りとして処理するのではなく，インナーソースプログラム全体の文脈の中でどのように位置づけるかが重要である．

対プロジェクト金銭還元型は，個々人のコミット数のみならず，プロジェクトや部門単位での貢献度を評価対象とするため，部門間のコラボレーションを活性化しやすい．たとえば，「自部門でもこの機能を使いたいので開発コストを一部支援する」といった動きが促進される可能性が高く，インナーソースを「参加者によって支えられる仕組み」として組織文化に定着させるきっかけとなりうる．

金銭をどのように還元するかについては，技術的指標と財務的指標を組み合わせて設計する必要がある．コミット数だけを根拠とすると，表面的な貢献に偏る危険性がある一方，完成したプロダクトの利益貢献度にのみ着目すると，初期段階の地道なインフラ整備が正当に評価されにくいという問題点が生じる．

さらに，還元のタイミングや手法もエンジニアの意欲に大きく影響を与える．たとえば，年度末に一括で還元する方式の場合，プロジェクト進捗の評価サイクルと整合しない可能性があり，導入時には慎重な検討が求められる．

金銭還元型には，法人間の取引や会計処理の問題も伴う．大規模企業においては，子会社間での無償の成果共有が移転価格税制と抵触するリスクがあるため，工数の算定や契約形態などを厳密に取り決めなければならない．

したがって，予算反映型を徹底するには，経理や法務といったコーポレート部門との協働，および全社横断的な指標設計が不可欠であり，導入には相応のコストがかかる．一方で，導入に成功した場合には，エンジニア個人のみならず部門単位でもコミュニティ形成が進み，インナーソースを長期的に維持する強力な原動力となり得る．結果として，社内リソースの効率的活用や新規事業の迅速化にも貢献しうる点が大きな意義である．

総括すると，本モデルは「コラボレーションが経済的リターンに直結する」という明確なメッセージを組織にもたらすが，その円滑な運用には高度な調整力が必要となる．また，本モデルにおけるインセンティブ

はプロジェクトや部門に対するものであり，個人が認められるかは該当組織の中での評価に依存する可能性があり，個人に対する追加的考慮が必要である．最終的には，組織の成熟度や経営層の強いコミットメントが，導入成功の成否を左右すると考えられる．

### 9.1.5 対プロジェクト功績顕彰型 / 活動正式化型

このモデルは，部門をまたぐコラボレーションが成果を上げた際，グループ全体として功績を顕彰するだけでなく，組織内の正式プロセスとして認定して継続支援や評価を行う方式を指す．国際的製造業 B 社の事例では，ボトムアップで生まれた活動や中間的な共同体による自主的連携が上位組織に認められ，全社共通の基盤へ統合される仕組みが確立されている．

インタビューによると，B 社にはコラボレーションを促す複数の階層構造が存在している．最も下位のレベルではエンジニア同士が必要に応じて自由に連携するボトムアップ的な取り組みがあり，上位のレイヤーには部同士が合意して進める正式プロジェクトや技術戦略にかかわる共同体がある．ここで「活動正式化」が鍵となり，契約書の整備や資金配分などが実施される．

この「技術戦略共同体」はソフトウェア開発やオープンソース利用といった全社的課題に取り組む仕組みとして機能しており，各部門のエンジニアが自主的に参加して課題解決を探る．優れたアウトプットが得られた場合は，グループ全体で功績を顕彰し，さらに上層部からの予算やリソースを投入して「活動正式化」へ進める．

共同体に参加するエンジニアは，「自分がやりたいことを正面から評価され，本業の評価も下がらない」というインセンティブを得られる．たとえばオープンソース活用のルール整備や，開発環境の標準化を共同体が進める際には，貢献度合いをレポート化できるため，部門側も「このエンジニアは共同体でこれほどの成果を上げた」と評価しやすくなる．

横断的活動が人事評価にどう反映されるのかを明確にすることで，エンジニアがボトムアップの活動に安心して参加できる．インタビューでは，「横断活動側から『この人はこれだけの貢献をしてくれた』と情報が上がり，評価が向上する」という仕組みが社員のやる気を高めていると指摘されている．

さらに国際的製造業 B 社では，ボトムアップ活動を発表するイベントを年に一度開催し，選ばれた事例には現金報酬や追加予算，正式プロジェクト化といった特典が与えられる．こうして「草の根活動」が正式化され，上位の技術戦略共同体や部門から支援を受けられるようになるため，インナーソースが社内に広く根付いていく．たとえば，小規模に始まった CI/CD ライブラリの開発が共同体や部門のワーキンググループを経由して「これを全社インフラにしよう」という提案に発展し，最終的にプラットフォームエンジニアリング部門が正式管理する事例がある．こうした流れこそ，ボトムアップ活動を「正式プロジェクト」へ昇格させる活動正式化型の本質と言える．

企業 B における人事施策との関連では，「横断的活動を評価対象に含める」と明記されているため，若手エンジニアにも大きなチャンスが広がっている．有志で外部イベントや社内コミュニティに参加し，OSPO（オープンソースプログラムオフィス）に相当する部門と連携するなど，新たな技術コラボが連鎖的に起こりやすい．こうしてインナーソースの多層ネットワークが国際拠点まで波及している．

総じて，対グループ功績顕彰型 / 活動正式化型モデルは，ボトムアップで生まれる活動を積極的に肯定しつつ，優れた成果を上位組織で正式承認・支援することで組織文化に深く根付かせようとする狙いがある．技術戦略共同体を軸に横断活動を束ね，人事評価にも結びつけることで，埋もれていた取り組

みや地道な改善が顕在化し，多層的なコラボレーションの継続を促す原動力となる．結果として，エンジニアは挑戦を続けやすくなり，企業としても新たな技術基盤や革新の機会を得られる．

### 9.1.6 対プロジェクト功績顕彰型 / ファンデーション型

対グループ功績顕彰型，いわゆる「予算名誉反映型」は，インナーソースの貢献を金銭面（予算）と名誉面（表彰・称賛）の両軸で評価し，特にグループやプロジェクト単位の功績を大きく取り上げるモデルである．インターネット業界 C 社が実践している，いわゆるファンデーション型は，CNCF（Cloud Native Computing Foundation）のような外部基金の仕組みを社内に取り入れ，複数事業やチームを横断して活動するエンジニア・プロジェクトを財政的かつ名誉的にサポートする点が大きな特徴といえる．

インターネット業界 C 社では，自分たちが取り組みたい技術テーマや OSS 的な開発を社内で実践し，一定の成果を上げた場合に，そのチームを公式に称賛すると同時に翌年度の開発予算やイベント参加費用を支援する枠組みを整えている．このファンデーション型と呼ばれる制度を導入した背景には，インターネット業界 C 社が複数の事業領域を抱えており，エンジニアが本業の合間を縫いながら新たな技術アイデアを持ち出し開発する風土があったものの，それらが本来の MBO や KPI に取り込まれにくく，評価面で埋もれていたという問題がある．

こうした課題感から誕生したファンデーション型では，社内の任意チームが審査を申請すると，プロジェクトの利用者数，満足度，技術的完成度など多面的な指標をもとに審査が行われる．インタビューによれば，この審査基準は制度導入以来，バージョンを重ねるごとに見直しが進められ，より納得感の高い仕組みとしてアップデートされてきたという．これらの評価指標は，GitHub のスター数やドキュメント整備度，事業部間のコラボ実績など，多様な要素を総合的にチェックすることで，単純なコミット数に依存しない公平性を確保している．

*表 9.2 ファンデーション型のインセンティブモデルごとの概要*

| フェーズ | 1. 準備・申請 | 2. 審査 | 3. 評価決定・ランク付与 | 4. 定期更新申請（再審査） |
|---|---|---|---|---|
| **概要** | 技術資産を制度 A に申請する準備を行い，必要な条件を満たした上で申請する | 提出された申請内容を基に技術資産を評価する | ランクを決定し，インセンティブや支援内容を割り当てる | 前回の評価を基に，技術資産の状況を再確認し，ランク更新または維持を決定する |
| **主要活動** | 導入実績，ドキュメント整備などの必要条件を確認し，プロジェクトの申請<br><br>（小規模・中規模のライブラリや，最近では AI モデルなどが例として該当） | 審査基準（利用者数，満足度，GitHub スター数など）に基づき評価<br>必要に応じて申請者へのヒアリングを実施<br><br>（審査にたいしては回数を重ねるごとに見直し・改善を図っている） | 評価結果を基に 5 段階のランクを決定し，支援内容を確定<br><br>（技術広報サポート，費用負担が含まれる） | 更新申請を受け付け<br>アセスメントを実施<br>ランクの更新または維持を決定 |
| **関係者** | 申請者 | ボードメンバー（各管轄からエンジニアが参加） | ボードメンバー<br>技術広報担当チーム | 申請者<br>ボードメンバー |
| **関係者の意図** | 社内外の認知度向上<br>サポートや費用負担などのインセンティブを得るため，あるいはキャリア形成の一環 | 技術資産の公正な評価 | 社内外の認知度向上，<br>エンジニアリング文化の発信 | 支援内容の継続または拡大 |

事実として，インターネット業界 C 社のエンジニアはこの制度（企業内では「制度 A」と呼ばれている）を活用することで，モチベーション向上の糸口を得ている．インタビューによれば，制度 A への申請理由はさまざまで，インセンティブを得たい人から社内認知度を高めたい人までニーズは多様だという．また，制度 A は複数のグレード（5 段階）を設けており，一定条件件を満たせばステップアップの申請が可能である．これにより，プロジェクトの継続的な成長や，エンジニアのキャリア形成にも寄与する仕組みになっている．

名誉面では，審査を通過したチームや個人が社内イベントや社内ブログで大々的に取り上げられる．これによって，同社の縦割り構造（ゲームや広告，メディアなど多様なドメインを含む）を横断するかたちでプロダクトや基盤技術が周知され，開発者個人の評価やブランド力の向上に直結する．自社プロダクト開発の自由度が高い文化や，「自分が作った OSS を社内外に広めたい」というエンジニア気質とも相性が良く，インナーソースを活性化する大きな原動力となっている．

一方で，予算と名誉を両立させるファンデーション型には，審査の透明性や公平性をどのように確保するかという課題が常につきまとう．インターネット業界 C 社では，一定規模以上のプロジェクトでは他部署から技術リーダーや有識者を含む審査委員会を組成し，より客観的かつ総合的な判断を下す仕組みづくりを進めている．

このように予算を獲得したプロジェクトは，実験的アプローチや試作品リリースがしやすくなる点が大きなメリットだ．これは管理職サイドにとっても，「全社的な支援があれば基盤を育てられる」という安心感につながり，インナーソースの展開を後押ししている．

制度 A の評価サイクルは 1 年単位で更新される形をとり，ランクアップの申請時期なども定期的に設定されている．金銭的報酬こそ限定的だが，技術広報サポートやカンファレンス参加費用，あるいはコンテナ基盤などのインフラ利用費など，多面的な支援が施される点が特徴的だ．海外カンファレンスへの渡航費を一部負担してもらえるなど，表彰とは別の形でエンジニアが恩恵を受けられる設計が組み込まれている点は，CNCF の助成プログラムを想起させる．

総括すると，インターネット業界 C 社の「ファンデーション型（対グループ功績顕彰型）」は，CNCF さながらのコミュニティ育成モデルを社内に展開し，複数事業間のサイロ化を緩和するとともに，エンジニアの主体的なチャレンジを後押しする重要な仕組みとして機能している．名誉と予算を両立させ，かつインナーソースの成果をフェアに評価するためには，審査基準のアップデートや審査委員会の多角的視点が欠かせない．だが，その難しさを乗り越えた先では，企業全体の横断的なコラボレーションと技術イノベーションが一層加速し，インナーソース文化の深い定着が期待できる．

## 9.2　重層的な支援体制による活動支援メカニズム

インナーソース活動を支えるインセンティブモデルは，支援の仕組みを多層の網のように張り巡らせることでより効果的に機能する．国際的製造業 B 社の事例では，トップダウンでもボトムアップでもない「中間共同体」的な組織が横断的連携を担い，さまざまなレイヤーで個別のプロジェクトを支援する網のように張り巡らせた受け入れ体制が整備されている．また情報通信業界 F 社ではマネージャーとのすり合わせに加え表彰制度があるほか，企業 D でも個別の感謝やエンドロールなど多様な還元の手段が用意されている．

この網のように張り巡らせた仕組みでは，エンジニアたちが所属部門の本業に加えて，興味をもつプロジェクトに自由に参加できる風土が尊重される．ただし，単なる自由放任ではなく，正式な役割やコミュニティの枠組みが用意されており，参加者には活動を報告する場や進捗を共有する場が与えられる．

例えば国際的製造業 B 社では，トップダウンの指令で無理に横断活動を押し付けるのではなく，中間層のチャンピオンや有志コミュニティが自身のプログラムやプロジェクトを立ち上げ，興味のあるエンジニアを募るスタイルをとっている．これによって，「やりたい人が集まり，成果が出れば自然に評価される」仕組みが回りやすくなり，会社としても大掛かりな改革を強要する必要がなくなる．

さらに，インナーソースの本質的な社内発展を考えた場合，インターネット企業 C 社のようなファンデーション型のモデルの採用が有効な選択肢となる．導入初期段階においては，既存の人事評価モデルを維持しながら，個人評価を基点とした仕組みを採用することも現実的な選択である．ただし，その場合でも，個人の裁量や熱意に依存するのではなく，組織的にサポートできる体系的なメカニズムの構築が不可欠となる．注目すべきは，これらのファンデーション型で表彰されるような仕組みも，突き詰めれば主たる業務に貢献するプロダクトに関連しており，それが最終的には個別の人事評価にも反映されうる点である．このように，様々な接続点を持ち，包括的に扱えるインセンティブ設計が重要となる．

## 9.3 総括と実践への指針

これまで，賃金還元・功績顕彰・予算配分・ハイブリッドなど多様な視点から，インナーソースを支えるインセンティブモデルの六つの類型を検討してきたが，それぞれの導入パターンは企業文化や事業形態によって変化し，一つの絶対的解が存在するわけではないという点が明らかになった．

個人単位のインセンティブを強調するアプローチ（個人評価反映型，表彰制度型，賞与を伴う表彰制度型）は，エンジニア一人ひとりの創造性やモチベーションを引き出すうえで有効だが，組織横断的な連携が手薄になりやすい側面がある．一方，グループ単位を重視するアプローチ（予算反映型，活動正式化型，ファンデーション型）は，大規模プロジェクトや部門間コラボレーションの促進に向いているが，導入には各方面との調整が必要となることから，両者のバランスが重要である．

賃金のみならず名誉や予算を組み合わせるハイブリッド型が注目されている背景には，エンジニアのモチベーションが給与面だけで語れないほど多様化している現状がある．名誉中心の設計では実利面での不足が懸念され，金銭中心ではエンジニアの創造性や学習意欲を十分に引き出せない場合があるため，複数の報酬要素を併用できる仕組みこそが柔軟性をもたらすと考えられる．

さらに，大企業ほど法人間取引や移転価格税制などの制約が複雑化するため，予算配分を伴うモデルの導入には，法務や経理部門との調整が不可欠になる．トップダウンの強力な支援を得つつ，こうしたハードルを乗り越えることが最終的な成功を左右すると言えよう．

インナーソース推進においては，時間やコミュニケーションなどの「自由化」をどの程度認めるか，また優れた成果を公式に承認し支援する「正式化」をどのように制度化するかも極めて重要である．賃金や名誉といった直接的な報酬だけではなく，エンジニアが開発時間を確保しやすい体制や，社内で横断的につながりやすい仕組みを整えることが，現場の意欲を持続させる鍵になる．

日本企業の場合，インナーソース自体がまだ発展途中であるため，初期段階では比較的導入しやすい個人評価反映型や表彰制度型といったアプローチから始め，成果が出始めてから予算反映型やファン

デーション型へ段階的に拡張していく戦略が適切と考えられる．こうした漸進的な導入によって，現場の混乱を最小限に抑えつつインナーソース文化を育むことができる．

インナーソースが成長期に入った企業では，インセンティブモデルが単なる数値評価の仕組みを越えて，コミュニティや学習，イノベーションを支える基盤へと発展する．エンジニアが他部門のコードに気軽に貢献し，それが正しく評価されるサイクルが回り始めると，社内全体の開発速度が上がり，新たな製品やサービスを生み出す源泉になりうる．

総括すると，企業ごとに異なる人事制度や事業環境を踏まえつつ，六つの類型を必要に応じて組み合わせ，現場からのフィードバックをもとに継続的にアップデートしていくプロセスが求められる．いずれのモデルも一長一短があるため，トップダウンとボトムアップを組み合わせながら，企業の成長段階にふさわしいインナーソース活動の評価枠組みを段階的に整備することが望ましい．

最終的には，インナーソースの本質が単なるコード共有やコスト削減にとどまらず，学習とイノベーションの創発にあることを見失わないことが肝要である．インセンティブモデルの設計はエンジニアの創造性と組織的成果を両立させる中核的役割を担うため，企業は長期的な事業戦略やエンジニア文化に照らして，その進化を継続的に支援していく必要がある．

# 第10章 インナーソース円環モデル:複雑性を捉える二次元アプローチ

本研究では，インナーソースの実践において多様なチャンピオンがそれぞれ異なる動機をもって活動していることが確認された．これは前章で示した定性的調査と定量的調査からも示唆されるように，インナーソースの解釈や目的意識が固定的ではなく多様性を帯びていることを意味する．さらに，各チャンピオンが「インナーソース」という用語から想起する到達点が異なるため，組織内で目指す方向性も一様ではなく，インナーソースの発展が単純な段階的成長では説明できない複雑なプロセスを含むことが明らかになった．

従来の研究では，インナーソース導入や成熟度を単一の段階的マチュリティモデルで捉える傾向が強かったが，本章ではより複雑な二次元のマトリクスモデルとして把握する必要性があることを指摘する．この新たな視点に基づき，マチュリティモデルとインナーソース円環モデルを併用することで，インナーソースの進展を従来よりも高い解像度で分析でき，異なる組織文化や動機をもつ複数のチャンピオンの活動様式を精密に把握できる点に大きな意義がある．

本研究ではこの議論をさらに深めるため，オルソン円環モデル（Circumplex model）[51]をインナーソースの文脈に適用した．企業を一つの家族にたとえると，多様なメンバーやチーム同士の関係性を円環モデルで捉えることにより，組織内部のコラボレーション様態を柔軟性と凝集性の両軸から正確に把握できることが判明した．

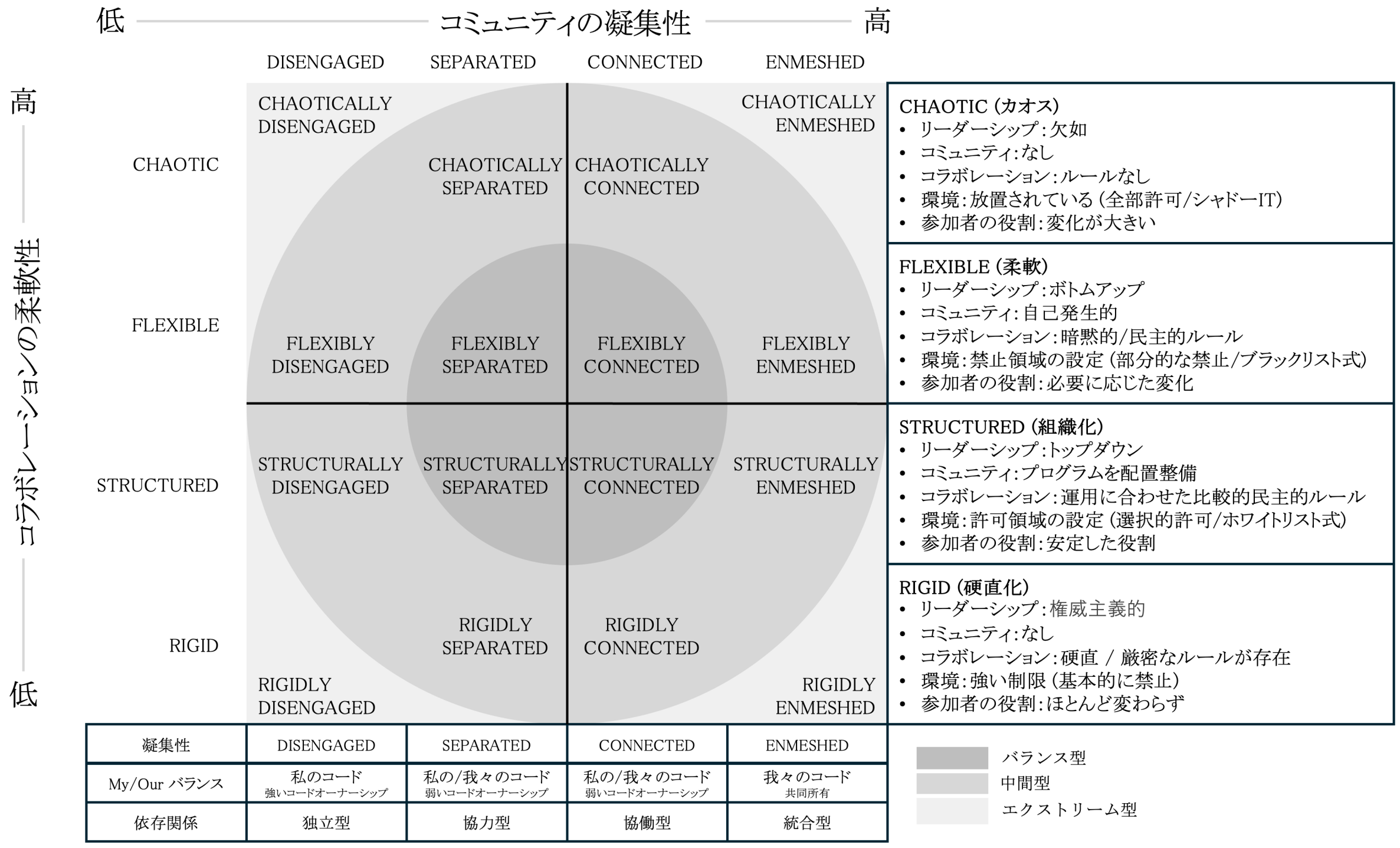

*図 10.1 インナーソース円環モデル*

図 10.1 はインナーソースの円環モデルを示している．インナーソースチャンピオンの意図は，この円環モデル上で多様な軌跡を描くが，必ずしも組織内のあらゆるコードやリソースを全面的に共有するわけで

はない．実務的制約やビジネス上の要請を踏まえながら，インナーソースの利点を部分的に活用する動機が観測されることがその理由である．

たとえば，企業が RIGIDLY DISENGAGED（硬直した低凝集性・低柔軟性）状態にある場合，リーダーシップやチーム間連携の欠如から車輪の再発明が繰り返されるなど，先進的なオープンソース型アプローチを導入しても人材やコミュニティが未成熟で協力者が得られにくいリスクが高い．

こうした場面では，まず STRUCTURALLY DISENGAGED（少し組織化された環境）を整備し，特定のプロセスやルールを導入することで段階的にチーム間の協働を促すアプローチが現実的といえる．ある程度の制限下でも組織内の文化変化を促しながら協働を実現していくことで，最終的に FLEXIBLY CONNECTED（高い柔軟性と高い凝集性）へ近づけるプロセスを描ける点が，本モデルの重要な意義である．

さらに，この円環モデルが示す特徴として，マチュリティレベルを直線的に高めるだけが唯一の道筋ではないことが挙げられる．すなわち，硬直した低凝集性の組織が段階的な専用環境を導入する一方で，特定のプロジェクトや範囲だけ先進的な柔軟環境を創出し，柔軟性と凝集性の違う「インナーソース」を同時並行的に運用することも可能となる．

インナーソースについて十分な知識を得る前や，チャンピオンになったばかりの段階では，オープンソースと同様に「すべてを一度にオープン化する」イメージを抱きがちだが，実際には短期間で完全にオープン化することは難しい．企業内の文化やプロセスを着実に変革していく必要があり，フルオープンを望まないメンバーが一定数存在する点も，インナーソース導入の難しさとして認識すべきである．

こうした状況を踏まえると，インナーソースのエッセンスのみを抽出して段階的な運用をすることが企業にとって現実的な戦略となりうる．本研究が提案する円環モデルは，そうした段階的・多様的な活用の様相を可視化し，チャンピオンが自らの立ち位置を正確に把握できるようにする効果をもつ．以降の節では，事例を通して各移行経路の具体的な内容について議論する．

### 10.1.1　低凝集性・低柔軟性からの移行：大企業におけるプロセス改革型導入

既存プロセスの変革を主たる動機としたインナーソース導入は大企業において特に多く見受けられるパターンであり，会計処理や法務，複数部門にわたる意思決定構造などの障壁を打開するためにインナーソースが活用される．表面的にはトップダウンに映る事例でも，実際にはオープンソースに精通するエンジニアや中間管理職が下地を作り，財務や法務との折衝を繰り返すことでインナーソース化を推進しているケースが確認されている．

大企業では，財務や法務，経営企画の承認を得て既存ルールを改変し，社内で新たなコラボレーションを認める制度を整える必要がある．部門間連携にともなう移転価格税制の適用や複雑な手続きを処理しなければならないため，現場エンジニアだけの熱意では突破できない壁が存在し，これを乗り越えることがプロセス改革型インナーソース導入の核心的特徴となっている．

このようなアプローチでは，制度設計やルール整備が中心課題となるため，エンジニアリング自体への関心だけでなく，合議制の組織文化を深く理解し，各部門の利害やリソース配分に配慮できる人材が推進役として求められる．国際的製造業 B 社でも，財務部門や法務部門との折衝を粘り強く行う担当者が存在し，それが部門横断の協働を実現する要因となっていた．インナーソースによる技術的な活用を足掛かりに，組織の意思決定プロセスを抜本的に変革する活動が本質であるため，プロセス変革型導入に

は大きな調整力が求められる．企業 A 社と国際的製造業 B 社を円環モデルの視点で捉えると，両社はインナーソースを活用する複数の環境を保持している可能性が高い．

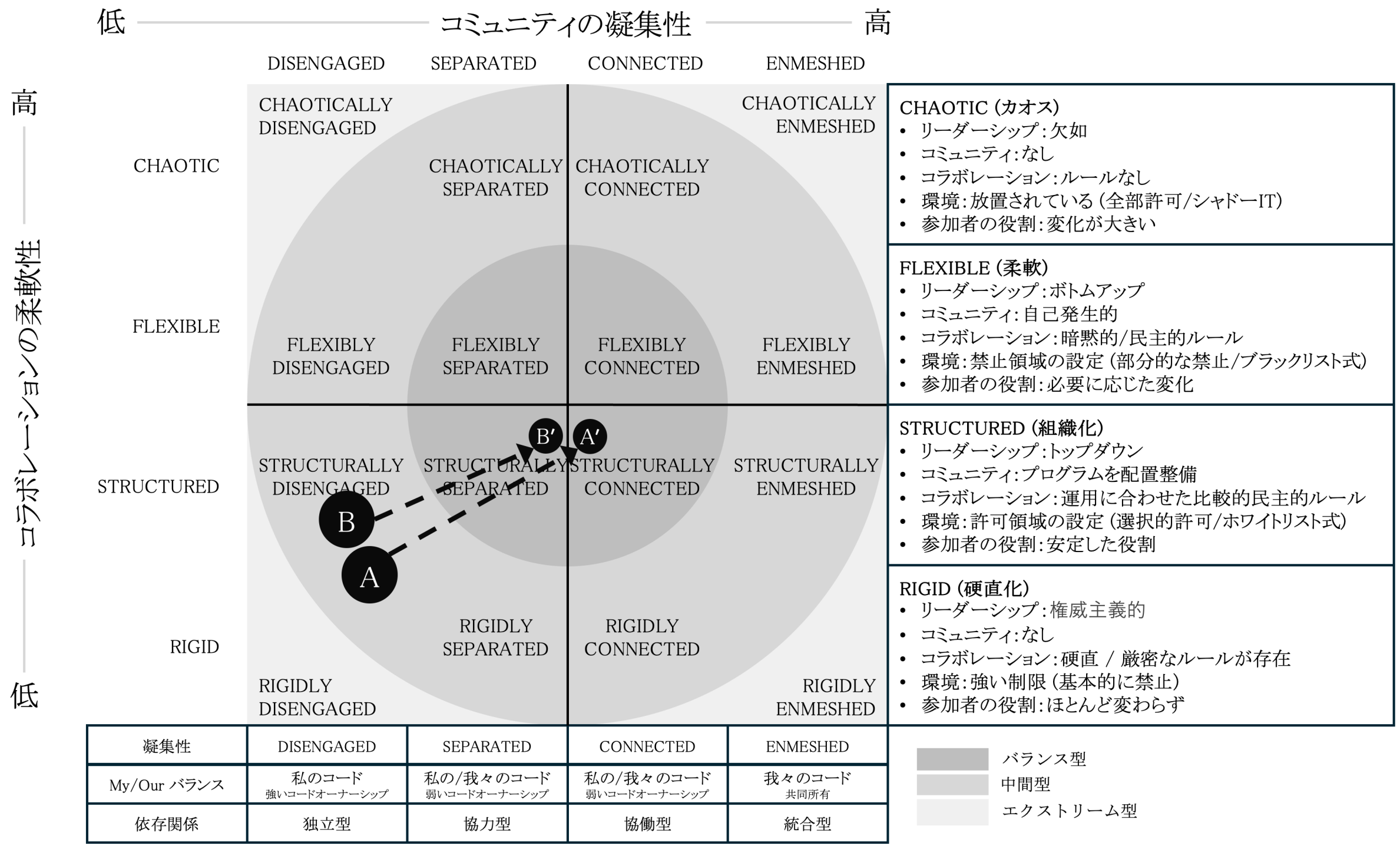

*図 10.2 企業 A 社と国際的製造業 B 社のインナーソース円環モデルにおけるマッピング*

第 9 章で述べたように，企業 A 社は「コンソーシアムモデル」を採用し，会費システムを利用して選択的にアクセス許可を与える仕組みを整え，コラボレーションが生じやすい環境を構築している．国際的製造業 B 社では既存の技術コミュニティが少しずつ拡大しており，契約テンプレートの整備などによってシームレスな協働が生まれ始めているものの，ソースコード管理ツールの設定がプロジェクト単位にとどまるなど，コンソーシアムとしての機能が十分には浸透していない部分もある．

一方で，企業 A 社がコンソーシアムモデルを活用して新たな柔軟環境（A'）を志向しているのに対し，国際的製造業 B 社は個別契約を重視してある程度分離した状態（B'）を志向している可能性もあり，会計上は似たインナーソース形態を実施していても，実現手法によって文脈がわずかに変わりうる．

このように分析を深めると，インナーソース導入に伴う制度やルール変更が企業文化や部門間の利害関係に及ぼす影響を広い視野で把握できる．パイロット期から初期導入期に移行する段階では，試験的な協働の成果を全社制度設計へ発展させる必要があるため，組織的障壁の度合いに見合った認知変革とシステム整備が求められる．企業 A 社のように，パイロットで成果を示したのちに正式スキームを構築し，管理職や財務部門を巻き込むかたちで制度改革を進める方法論は，プロセス改革型インナーソースの典型的進め方だといえる．

### 10.1.2 低柔軟性・高凝集性からの移行：組織文化変革への段階的アプローチ

低柔軟性と高凝集性をもつ企業は，比較的規模が小さい中小企業や組織に見られ，既にチーム内でのコラボレーションが成立している一方，早期から厳格なルールが存在することでエンジニアの自由度が低い場合がある．

たとえばシステムインテグレーションや SES 系企業のように，プロジェクトや工数が厳密に管理され，メンバーの役割が固定されている場合には，インナーソース的な取り組みが自然発生しにくいと推測される．このケースではトップの権限が強く，制約が多いため，エンジニアの開発者体験が低下しやすいことが推察される．組織や開発チームはコードを所有しているものの，決められたルールに従う文化が根強い場合，横の連携や相互貢献が定着せず，コミュニケーションが乏しいことも少なくない．

こうした企業は，円環モデル上で STRUCTURALLY CONNECTED を目指して段階的にコラボレーションを広げる方法が望ましいと考えられる．企業規模が拡大すると，自然に RIGIDLY DISENGAGED（硬直化）の状態へ進むリスクがあるため，インナーソースという言葉を用いてコラボレーションを促進し，完全に凝集性が失われる前に組織内の連携や透明性を強化する施策が重要となる．

情報通信業界 F 社は，500 人規模の比較的小規模組織でありながら，拡大にともなって自然なコラボレーションを阻む要素が増えた点で RIGIDLY CONNECTED の状態に位置すると推察される．情報通信業界 F 社は柔軟性を高めようとする好例であり，F’の FLEXIBLY CONNECTED（高柔軟性・高凝集性）への移行を目指していることがうかがえる．

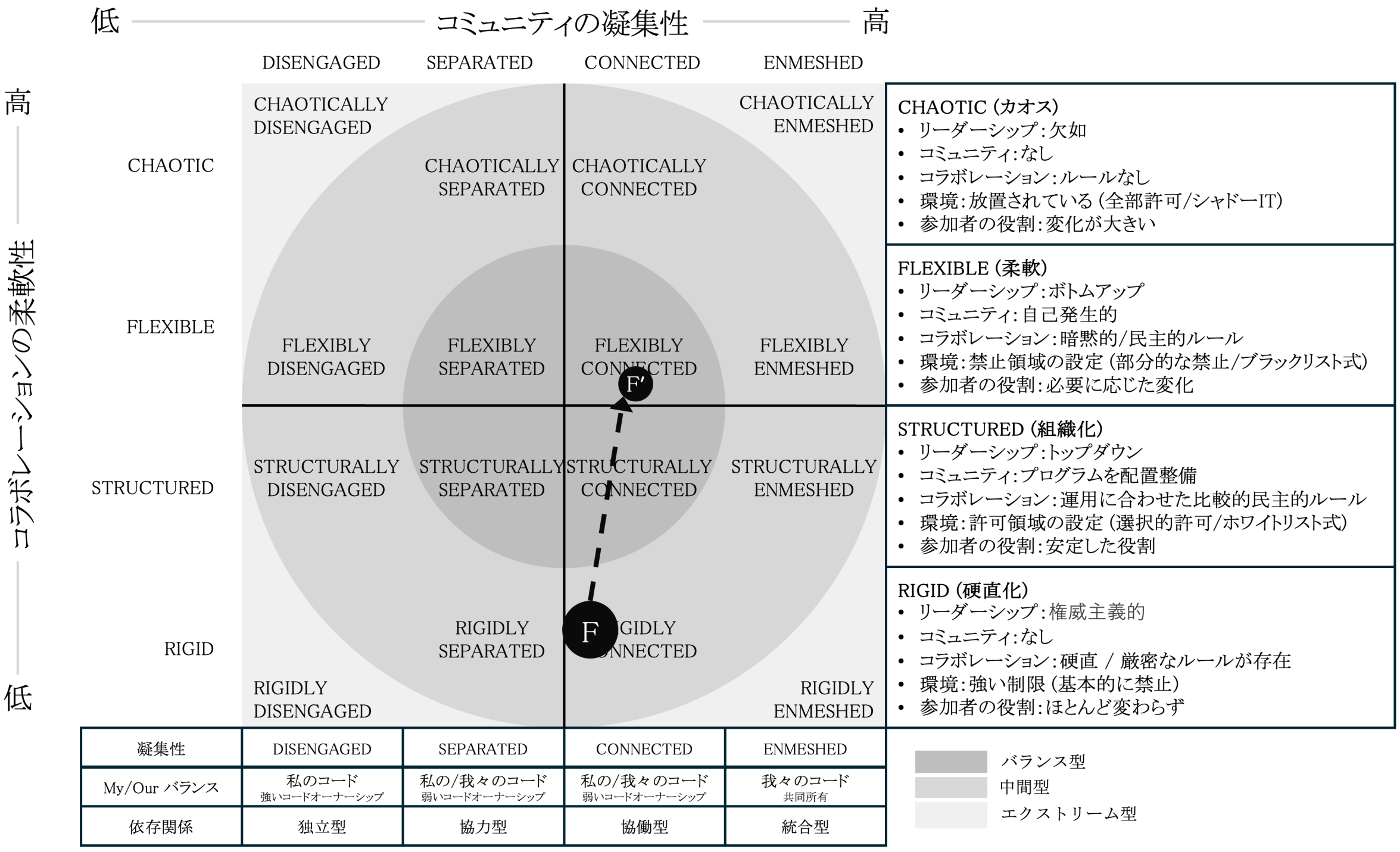

*図 10.3 情報通信業界 F 社のインナーソース円環モデルにおけるマッピング*

もともとリポジトリ閲覧やコントリビューション権限が比較的オープンで，複雑な法人間会計調整が不要な規模であったため，大規模なプロセス改革をせずともエンジニアの横連携を促進できる下地が整って

いた．情報通信業界 F 社の場合，エンジニアが創造的に活動できる職場環境や従業員満足度の向上を主眼に置き，その手段としてインナーソースを位置づけた点に特徴がある．

こうした企業では，組織文化の醸成が導入の主なモチベーションであり，会計や法務といった手続き上の問題が大きな障壁になる可能性は低い．そのため初期導入段階では，エンジニア同士のコラボレーションが有益であることを示すパイロットプロジェクトを立ち上げ，そこからの成功体験を全組織に波及させていく動きが中心になる．情報通信業界 F 社はプラットフォームやツールの利便性を小規模事例で立証し，管理職に成果を訴求することで全社展開へ徐々に拡張していった．

組織文化変革の際には，人材の動機づけとインセンティブ設計が重要視される．情報通信業界 F 社は創意工夫賞や社長賞といった既存表彰制度の枠内でインナーソースを評価し，社内インタビューや賞金制度を用いて個人の達成感と組織の相乗効果を高めている．ただし，まだインナーソース専用の評価設計には至っておらず，評価は管理職の裁量に大きく依存する．インセンティブも短期的なものが中心であるため，活動をスケールさせて既存評価プロセスを変革するには，今後さらに制度的整備が必要になると考えられる．

情報通信業界 F 社の導入経路を振り返ると，当初はマネージャー同士の話し合いから試行が始まり，その後エンジニアの働きかけを得て具体的な事例を積み上げ，これらが評価制度と結びつく形で全社規模へ広がっていくというプロセスをたどっており，今後はボトムアップの動きがより自然発生的に進行することや，さらに大きなプロダクトへインナーソースを展開するステップが想定される．

日本企業においては，トップダウンから明示的に評価設計を改革しないかぎり，インナーソースへの貢献がどの程度給与や昇進に反映されるかは依然として曖昧なケースも少なくない．エンジニア同士が楽しんで取り組める環境が整っていても，事業部間の合議や人事制度の硬直性が障壁となり，十分な認知やリソース配分が得られないこともある．情報通信業界 F 社では，このリスクを回避するために，まずはパイロットで成果を上げ，全社的に認知される土壌を作ったうえで，チャンピオンを育成して組織の応援を取り付けるという段階的アプローチを採っている．日本企業の一部では経営層がインナーソース を導入する方針を打ち出しても，現場では「評価制度が追いつかない」「ライセンスポリシーが曖昧」「ソースコードへのアクセス権限が不明確」といった不満が噴出し，導入期の盛り上がりがしぼむ可能性がある．チャンピオンが先回りしてパイロットプロジェクトをモデルケースに選定し，小さな成功事例を積み上げながら全社展開を図るアプローチをとれば，トップダウンによる支援とボトムアップの動きが連動し，導入初期の失速を回避しやすくなる．

組織文化変革は，従業員の主体性をどこまで引き出せるか，また経営層がそれをどのように評価するかが成功の分かれ道となる．エンジニアのボトムアップな情熱をいかに可視化し，経営層や管理職の支援と結びつけるかが課題であり，チャンピオンには「人を動かす」リーダーシップ能力が問われる．コミュニケーションを重視する文化が定着すれば，インナーソースが単なる開発手法を超えた組織的イノベーションの原動力となる可能性を秘めている．

これはインナーソースの採用動機の調査が裏付けている内容と符合する．初期導入気に至るにつれて，再度従業員満足度の向上などが動機づけになる．つまり，文化的な変革や働きやすさのようなことに焦点が当たることになる．企業 D の証言からも，成長から成熟段階へ至るためには，チーム同士のフィードバックやコミュニケーションが既に活発な文化であることが望ましいと分かる．そこにインナーソースが媒介として導入されれば，従業員満足度とイノベーションが同時に高まる可能性がある．こうした組織文化

変革型アプローチは，開発者の情熱をいかに活かし，社内コミュニケーションをどう促進するかが最大のテーマであり，チャンピオンには技術的知見だけでなく，人材マネジメントや対話の設計力が求められる．

### 10.1.3 高柔軟性・低凝集性からの移行：生産性拡張を狙うインナーソース活用

インナーソースが一定程度成熟してくると，特定のプロダクトやサービスの開発効率向上を主目的としてインナーソースを導入する動機が生まれる．インターネット業界 C 社では，社内プロジェクトが大規模化・複雑化して開発者リソースの不足や同じ機能の重複実装が頻発するなか，その解消策としてインナーソースを利用している．こうしたケースでは，会計や組織文化の問題がある程度クリアされており，技術的コラボレーションを活発化すればすぐに成果が出るという利点がある．

また，インターネット業界 E 社のように「インナーソースを行う」という明示的意図がなくとも，かつて同じチームで働いていたエンジニアが部署異動後も相互にバグ修正や機能追加を行い合うことで，結果的にインナーソース的な開発様式が自然に確立される場合がある．これは開発生産性を優先する思惑がエンジニア間で自然発生し，正式なライセンスや制度を経ずとも「助け合い」が促進されたパターンといえる．

企業 D では，プラットフォームチームやユーザーチーム間の連携を強化し，サービス全体の品質と開発スピードを底上げしている．生産性拡張型アプローチでは，既にある程度法務や会計リスクへの対策が整い，プラットフォーム整備の重要性が認識されている企業が多いため，インナーソース化が加速しやすい側面がある．ただし，特定のプロダクトで成果が上がっても，それが全社規模のインナーソース文化に波及しない可能性があるため，インターネット業界 C 社や企業 D のように成功事例を全社へ意図的に共有・展開する取り組みが重要になる．

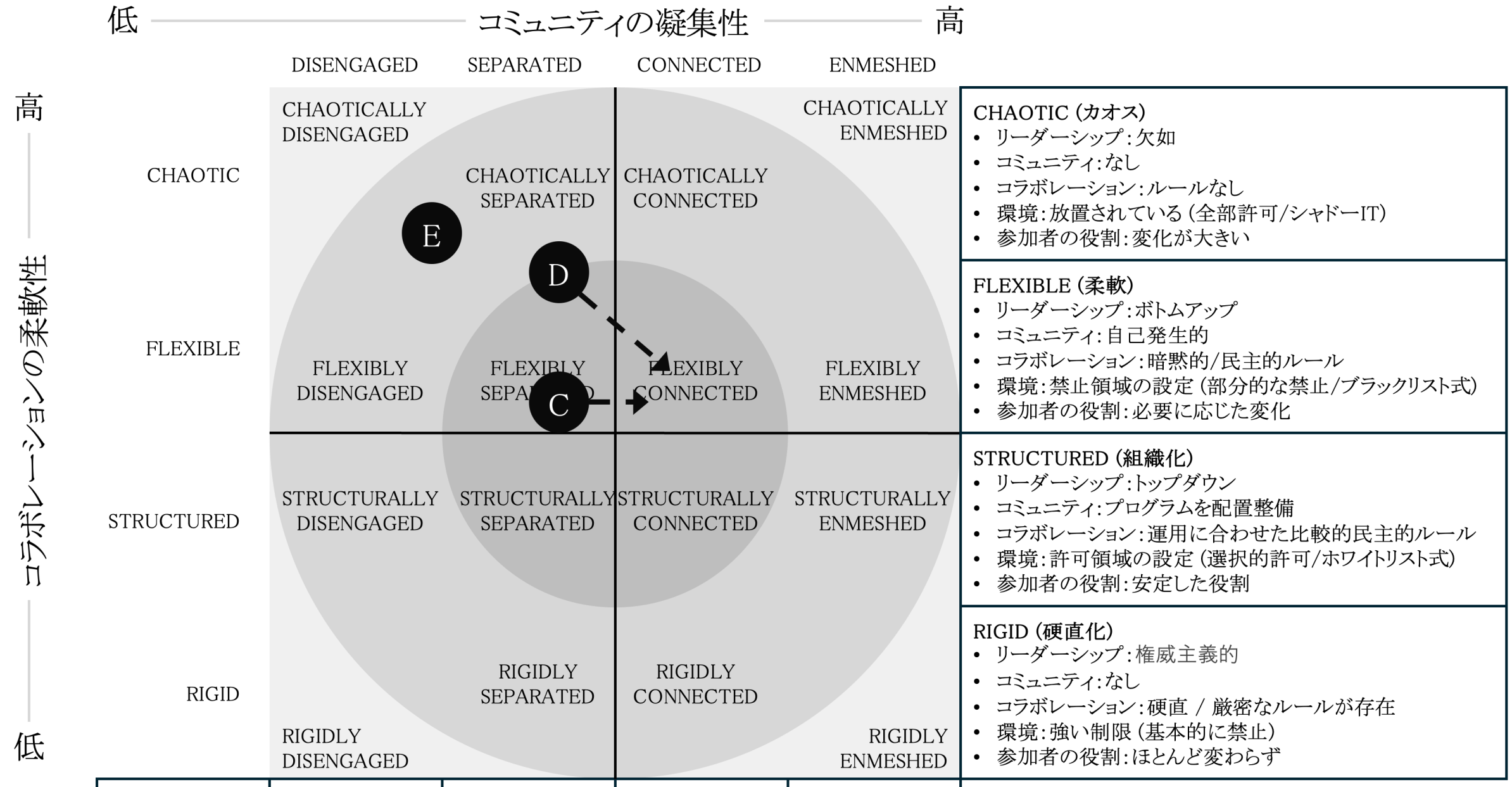

| 凝集性 | DISENGAGED | SEPARATED | CONNECTED | ENMESHED |
|---|---|---|---|---|
| My/Our バランス | 私のコード<br>強いコードオーナーシップ | 私の/我々のコード<br>弱いコードオーナーシップ | 私の/我々のコード<br>弱いコードオーナーシップ | 我々のコード<br>共同所有 |
| 依存関係 | 独立型 | 協力型 | 協働型 | 統合型 |

バランス型
中間型
エクストリーム型

*図 10.4 インターネット業界 C 社・D 社・E 社のインナーソース円環モデルにおけるマッピング*

インターネット業界 C 社のインタビューによると，統合開発プロジェクト向けの評価プログラムやネットワークが整備されており，他部門との連携を促す仕組みがある．一方，企業 D には制度化されたプログラムは存在しないものの，リーダーの存在と自然発生的な連携によって成果が出ている．インターネット業界 E 社では一定のルールはあるが，全社的にインナーソースが統合されているわけではなく，部分的なコラボレーションにとどまっているため，今後さらなる加速にはリーダーシップやルール化が不可欠と考えられる．

このように，高柔軟性であってもコラボレーションがエンジニアに委ねられており，凝集性をまだ高める余地があるケースにおいては，インターネット業界 C 社のようにインナーソースをプログラムとして位置付けて運用することに意義が認められる．このように既存制度の障壁が低くエンジニア同士の横つながりが生まれやすい環境では，初期段階から目立った成果が得られる．しかし，プロダクト単位の成果に終始すると全社的な浸透が限定される懸念もあるため，トップダウンによる合意形成や評価制度の更新などの後押しが不可欠である．企業が成長・成熟段階へ進む段階では，インナーソースの有用性を組織全体で認知し，チャンピオンを包括的に支援する仕組みが重要となる．

### 10.1.4 高柔軟性・高凝集性からの移行：スタートアップ的自由度と制御の両立

高柔軟性・高凝集性はスタートアップ企業に典型的で，既にソースコード管理ツールが整備され，単一の製品や小規模コミュニティで柔軟なコラボレーションが成立している状況が多い．この状態は，オープンソースに極めて近く，厳格なルールがない代わりに，高い透明性や迅速な意思決定が維持されているといえる．こうした企業では「インナーソース」という言葉自体がほとんど使われない場合があり，オープンソースと似た文化が小さな組織単位で自然に確立していることが理由の一つであると言えよう．

一方，企業が大きくなるにつれ，柔軟性と凝集性は低下し，最終的に低い凝集性へ移行してしまうリスクが高い．インターネット業界 E 社のインタビューでも，創業当初はほとんどルールがなく自由なコラボレーションが行われていた企業が，組織拡大に伴いセパレーテッドな状態に向かう例が挙げられた．多くのスタートアップは，創業初期に X のような高い柔軟性・高い凝集性の状態を持ちながら，度重なる組織変更の末に低い凝集性へ移行する．

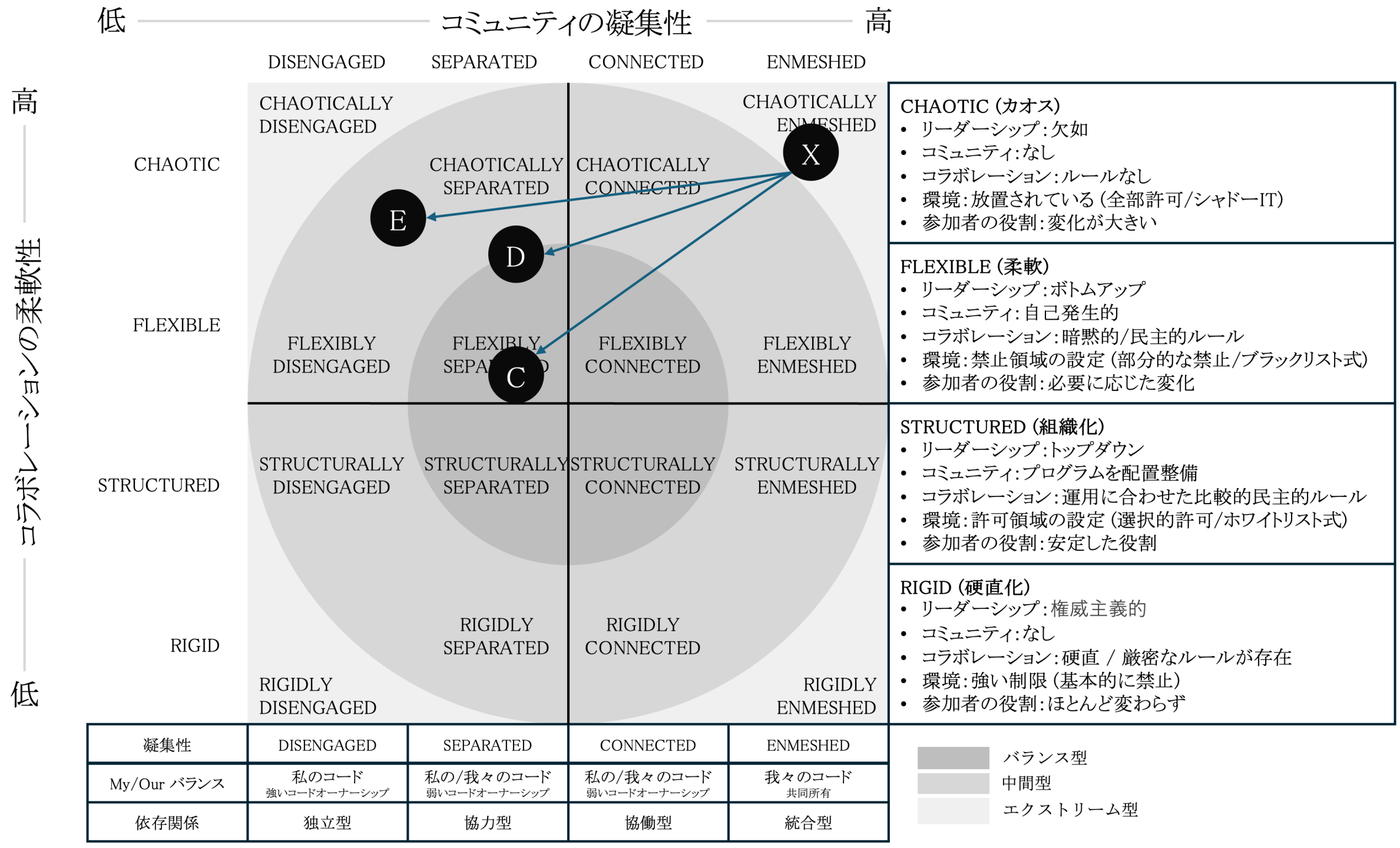

*図 10.4 スタートアップが DISENGAGED へと移行する経路*

この変化を食い止めるには,「インナーソース」というコンセプトを活用して透明性を保つコミュニケーションスタイルを組織全体に定着させることが有効である. 実際にインターネット業界 C 社や企業 D が行っているのは, コミュニティやプログラムを組織的に再構築し, 成長した組織でも創業当初に近い自由なコラボレーションと, 大企業ならではのルールを両立させる試みだといえる. 現在 X の状態にある小規模企業でも, 将来的に凝集性が低い状態へ自然と移行するのを防止するために, 早期からインナーソースをプログラム化することが重要になる.

## 10.2 円環モデルによるマチュリティモデルの補完

インナーソースの発展過程を理解する上で, 従来のマチュリティモデルは重要な指針を提供してきた. しかし, 本研究で提示した円環モデルは, そのマチュリティモデルを補完し, より豊かな解釈を可能にする. 特に注目すべきは, 組織内で複数の「成熟度」が同時に存在しうるという視点である.

特に大規模企業においては, 企業レベルの制度設計とチームレベルの実践が異なる成熟度を示すことが珍しくない. インナーソースの発展は, トポロジカルにコミュニティが形成され, そのクラスターごとに異なる成熟度を持つことが観察される. つまり, 企業全体として「インナーソースを実践している」と言明されても, その内部のコミュニティによっては実践度合いが大きく異なる可能性がある. このことは, 大手テクノロジー企業の一部の事例をもって「その企業がインナーソースを実践している」と断定することの難しさを示唆している.

円環モデルは, 組織内に存在する多様なコミュニティの「成熟度」を, より明確に把握することを可能にする. 各コミュニティは, 柔軟性と凝集性という二つの軸上で異なる位置づけを持ち, それぞれが独自の発展経路を辿る. したがって,「この企業はインナーソースを実践している」という一元的な評価ではなく,

「どのコミュニティが，どのような形でインナーソースを実践しているか」という，より精緻な分析が可能となる．

さらに重要な点として，インナーソースプログラムの本質的な目的は，この円環モデル上に存在する様々な状態のクラスターが持ちうるベクトルの方向を，より望ましい方向へ導くことにある．つまり，円環モデルの二平面上に存在する多様な柔軟性および凝集性の状態を，可能な限り中心に向かわせることにある．これは必ずしも全てのコミュニティを同一の「理想的な状態」へ収束させることを意味しない．むしろ，各コミュニティの特性や制約を考慮しながら，適切な柔軟性と凝集性のバランスを追求することが重要となる．

つまり，コラボレーションの安全性を確保しつつ是正すべき点は改善し，協働の機会を拡大していく取り組みといえる．同時に，セキュリティ要件の高い領域に対しては適切な環境を提供しながら，コラボレーション志向のマインドセットを育成し，小規模なクラスター内でもオープンソースに類似したコラボレーションスタイルを実現することを意味する．

このような目標を達成するために，インナーソースのプロジェクトやプログラムのマチュリティは可能な限り向上させる必要があり，制度面での柔軟性を確保することでコラボレーションの「受容度」を高めることが求められる．例えば企業 A のように，コンソーシアムモデルを定義することで組織メンバーは「自由な環境」を獲得したが，次のステップとして人的つながりを強化するための仕組みの構築が重要となる．

成熟段階に到達するためには，インナーソースのコミュニティが生み出すトポロジーに注目し，チャンピオンの役割を慎重に検討する必要がある．各クラスターが円環モデル上のどの位置にあるかによって，設定すべきゴール，伝えるべきメッセージ，構築すべき協働環境は異なってくる．本研究は，企業の各チームとそれが織りなすクラスターが目指すべき異なる成熟度を定義するためのツールを提示し，マチュリティモデルの新たな可能性を示したと言える．

円環モデルはマチュリティモデルを拡張し，より現実に即した分析を可能にする枠組みとして機能する．両モデルを相補的に活用することで，インナーソースの発展過程をより深く理解し，効果的な導入戦略を立案することが可能となるのである．

## 10.3 インナーソースの再定義に向けて：多様な組織への適応可能性

本研究は当初から，インナーソースの統一的な定義を確立することを目的としていないことを明確にしてきた．しかし，提案した円環モデルを通じて，インナーソースの重要性が一般的に認識されている以上に広範な共感を得られる可能性があることが示された．

インナーソースの概念が多様な組織状況において適用されている点を踏まえると，同じ「インナーソース」という用語を使用していても，企業やチャンピオンが示す具体像や状態が大きく異なることが明らかになった．すでにオープンソースの状態に近い企業もあれば，独自の組織文化やルールの中でインナーソースを部分的に運用している企業も存在し，それらをすべて同一の基準で捉えようとすると，かえって理解を狭める危険が生じる

インナーソースをオープンソースの延長線上でのみ捉えるのではなく，既存の企業環境が抱える多様な課題や成長段階に合わせて柔軟に導入する余地があるという点が示唆される．提案した円環モデルによれば，企業ごとに凝集性や柔軟性が大きく変動し，必ずしもフルオープンな文化だけがインナーソー

スとして機能するわけではないため，どの範囲でコラボレーションをどの程度推し進めるのかを検討する必要がある．

例えば，高柔軟性・高凝集性の企業，特に小規模なスタートアップでは，全員がコードに対する集合的オーナーシップを持っているように見える．一見すると「生まれながらにしてオープンソース的な開発モデル」を実践しているように映るかもしれないが，インナーソースの本質は組織内でのコード再利用の促進と有機的なコラボレーションの実現にある．適切なガバナンスがなければ，高い柔軟性が却って無秩序な車輪の再発明を招く可能性がある．このような環境では，インナーソースは無秩序な社内開発に秩序を与えるための手法論として機能する．

また，高凝集性の環境であっても，事業拡大に伴って必然的に部門間の壁が生まれてくる．そのため，拡大前の早期段階から貢献し合う透明な環境を意識づけるために「インナーソース」という概念を導入することには大きな意義がある．これにより，組織の成長に伴って自然と持ち込まれがちな官僚的な組織化や，成長の速度を妨げるような障壁を予防的に排除することが可能となる．

高柔軟性・低凝集性の企業，例えばインターネット業界 E 社のような事例では，スタートアップが急成長する中で開発チーム間にサイロが生まれている．ソースコード管理ツールは柔軟に設定されているものの，実際のコラボレーションは低下している状態だ．このような場合，単にツールの設定を見て「インナーソースを実践している」と判断するのは適切ではない．真のインナーソースを実現するためには，インターネット業界 C 社や企業 D のように，プログラムを作って壁を取り除くための仕組みづくりが重要となる．

低柔軟性・高凝集性の企業，例えば情報通信業界 F 社のような中規模企業では，開発チームの規模は比較的小さいものの，既存の評価システムの不備などによってコラボレーションが十分に生まれていない．このような環境では，人的距離は近いにも関わらず有機的な協働が阻害されているため，適切なプログラムを用意して人材を動かしていく施策が重要となる．

最後に，低凝集性・低柔軟性の企業は，ある意味で最もインナーソース的な環境を必要としている．しかし，十分な組織規模があり，歴史的な文化が根付いている場合，インナーソースの導入には大きな困難が伴う．これらの企業では，ルールや会計の理解，移転価格などの制度的側面への対応と同時に，人々の意識や行動様式を変革するという二重の課題に直面する．企業 A 社や国際的製造業 B 社の事例が示すように，低凝集性の領域を抱えながらも，別途コミュニティやコンソーシアムを形成して高凝集性のエリアを確保することで，組織横断的なコラボレーションを可能にする戦略が有効である．

このように円環モデルは，過度に高い柔軟性や凝集性が必ずしも理想的なインナーソース環境を意味しないことを示している．真に重要なのは，有機的なコラボレーションを生み出しながら，インナーソース文化を成熟させ，ビジネス効率化を実現することである．そのためには，導入の各段階に適したチャンピオンの設定が極めて重要となる．

インナーソース導入はアイデア期からパイロット期，初期導入期，さらには成長・成熟期へと段階的に進むが，それぞれの段階でチャンピオンが担う役割は異なり，社内啓蒙やビジョン共有，ツール整備，評価制度改革など，多面的な活動を柔軟に展開することが求められる．これは終わりのない文化的な変革の旅であり，「当たり前に貢献する」という文化を組織に根付かせることで，高い開発生産性と開発者体験の両立が実現されるのである．

# 結論

## 第11章 総括

本論文で重視したとおり，インナーソースは単に環境を整備するだけではなく，組織内の多様な協働モデルを育むことに本質があるため，「インナーソースのための環境が整っている」状態をゴールとみなすのは適切ではない．インナーソースが本質的に協働モデルであることを考慮すれば，当初から全面的に環境と設定を整備してコラボレーションを促すような指示をするだけではなく，研究開発の小規模な貢献や隣接チームへのドキュメント協力など，実際に近接した局所的な協働から着実に始める必要がある．

本研究は，インナーソースの導入における組織的プログラムと自然発生的コラボレーションの二面性を統合的に分析し，既存のモデルでは把握できない多層的な成熟過程の存在を明らかにした．これにより，インナーソースの実践形態は環境整備やガバナンスの手順といったトップダウン型の枠組みに限定されず，現場発の部分的な協働や非公式な取り組みが重要な役割を果たすことが示唆された．

インナーソースの成功には，組織全体の能力を測る包括的視点と，チーム単位の実践度を評価する個別的視点を併用するアプローチが有効であることが示された．これらの視点の両立により，段階的な制度化やガバナンス整備といった上層部主導の施策と，現場レベルでの柔軟な協働や貢献の促進が連携し，より持続可能なインナーソースの発展へとつながる可能性が高まる．

導入初期段階においては，完全な仕組みを整備する前に小規模な成功事例を積み重ね，インナーソースの価値を可視化していくことが有用である．本研究で言及した発展途上のコラボレーションや，隣接するチーム間の限定的な協働であっても，インナーソースの可能性を組織内に伝播する触媒として機能し得る．このような草の根的な取り組みを通じて，多様な局面での導入を支えるコミュニティネットワークが形成される．

インナーソースが本質的に協働モデルである以上，環境整備や設定要件を狭義に定義しすぎると，実践の初期段階で柔軟性を失うリスクが高まる．実際には，研究開発プロジェクトや文書作成への貢献など，少人数や局所的な活動からネットワークを拡張していくことが多く，当初は計画やガイドラインが不在の状態でも，現場の主体的な意義付けを軸とした試行錯誤が重要となる．したがって，一見不完全な段階を含めた多様な協働形態を許容し，そうした積み重ねの成果を適宜組織の施策へ反映させていく必要がある．

本研究が提案したインナーソースの円環モデルは，既存のマチュリティモデルが示唆する進化論を補完し，組織が直面する段階や層をより柔軟に捉える視点を提示する．このモデルでは，同一組織内でもプロジェクトや部署ごとに異なる成熟度が併存することを前提とし，コミュニティネットワークの自然な拡大とガバナンス強化の時機を適切に捉えることで，インナーソースがもたらす生産性向上や組織文化変革の効果を最大化できる．

以上の考察は，インナーソース推進者が初期段階の局所的成功体験を重視しながら，段階的に上層部の合意形成や制度整備を進める戦略の重要性を示唆する．具体的な指針やソースコード管理権限が未整備でも，先行的に協働を実践し，その成果をデータや事例として示すことで，より広範な組織的受容

を促し得る. 結果として, インナーソース導入の各段階で求められる活動が明確化され, 企業ごとの特有の組織文化や産業構造に適合した形で, インナーソースの価値創出を促進できることが示された.

## 11.1 研究の貢献

インナーソース導入プロセスの段階的理解において, 従来の並列的な属性分析を超えて, 導入段階という時系列的な視点を導入した. この新しい分析フレームワークにより, 各段階における組織の認識変化や直面する障壁の変遷を体系的に把握することが可能となった.

また, 特定の導入段階に焦点を当てた高解像度の分析により, 各段階における実践者グループの具体的な課題認識や行動様式を明確化した. この知見は, 段階に応じた効果的な導入戦略の構築に重要な示唆を提供する.

既存のマチュリティモデルに関して, 本研究は段階的な成長指標を超えた多層的・多面的な評価アプローチを提示した. 特に, インナーソースプログラムにおけるネットワーク構造の複雑性や, コミュニティの成熟度を包括的に評価する新たな理論的枠組みを構築した.

グローバルな文脈において, 本研究は日本企業の実態データを初めて体系的に収集・分析し, インナーソース研究に地域的な多様性の視点を付加した. 特に, 言語的・文化的な障壁を持つ地域におけるインナーソース導入の特徴を明らかにし, グローバルな適用可能性に関する新たな知見を提供した.

さらに, インナーソース導入における法的・会計的課題に関して, 具体的な解決アプローチを提示した. これは, 実務的な観点から組織がインナーソースを導入する際の重要な指針となる.

これらの知見は, インナーソース研究の理論的深化に貢献するとともに, 実践的な導入戦略の構築に有用な示唆を提供する.

## 11.2 今後の課題と展望

今後の研究拡張に向けては, 調査対象の拡大と多角的なデータ収集を通じて, より一般化可能な知見を得る必要がある. この研究では限定的なサンプルを主とした分析を実施したが, 規模の異なる企業や多様な業種にわたってインナーソースの実践状況を把握することで, 理論的枠組みの妥当性を検証し, 説得力のある結論を提示できると考えられる.

インナーソースの推進に関連する会計上の課題は, 企業間コラボレーションを阻害する要因として依然として残存している. 特に, 大企業が安心して積極的に参入できるような会計基盤を整備することが, 将来的なコラボレーションの高度化を実現する上で重要な鍵となる.

インナーソース・トポロジーは, 組織内の協業構造やコラボレーションの強度を解明する有用な概念であるにもかかわらず, その定量的評価手法には未確立の部分が多い. とりわけ, 本研究で示したような初期段階の実践例にとどまらず, 複雑化・大規模化するネットワークへの適用方法を探求することで, 多様化する組織形態に応じた具体的なガイダンスを得られる可能性がある.

円環モデルをはじめとするインナーソースに関する理論モデルでは, ソースコード管理ツールやコミュニティ成熟度, 部門横断的プルリクエストの頻度など, 多数のメトリクスを複合的に評価する仕組みが求められる. 現時点では, 一部のメトリクスが局所的に整備されつつあるものの, さらなる検証とモデルの洗練を通じて, 成長過程の可視化や導入戦略の体系化を進める余地が残されている.

最後に，日本の組織では，内製化の遅れやソースコード共有の局所性が課題として浮き彫りになっているが，少子高齢化や AI 活用の重要性を踏まえると，インナーソース導入による生産性向上と組織学習の促進は喫緊の課題となる．こうした社会的要請の高まりに対して，日本市場に特化したインナーソースの導入手法やガイドラインを体系化し，現場と経営層が共通の視点でコラボレーションを推進できる環境を整備することが，今後の展望として大いに期待される．

# 謝辞

本研究を進めるにあたり，多くの方々からご支援とご協力を賜りました．青山学院大学の荒木万寿夫教授，高橋邦丸教授，竹田賢教授には，研究全般にわたり丁寧なご指導を賜りました．心より感謝申し上げます．The InnerSource Commons Foundation の Clare Dillon 氏には，貴重な情報とアドバイスをいただきました．深く感謝いたします．また，InnerSource Commons の日本コミュニティの皆様には，サーベイにご協力いただき，誠にありがとうございました．インタビューにご協力くださった企業の皆様からは，実践的な知見を共有していただきました．心より御礼申し上げます．

これらの方々のご支援なくして本研究は成し得ませんでした．ここに記して深謝の意を表します．

# 付録

## 参考文献一覧

## InnerSource 動向調査 2024

本研究では，State of InnerSource Survey という世界規模の企業を対象とした調査を参考にしつつ，日本の現状に即した包括的かつ精緻なサーベイの策定を行った．

以下の項目に関する 26 の設問を含むサーベイを作成した．

- アンケート回答者のプロフィール
- 所属組織の特性
- インナーソース導入に係る障壁
- 組織間協働の実態
- インナーソース実践の現状企業 A 社（業界無回答）

# InnerSource動向調査2024

本調査の実施主体はInnerSource Commons Foundation(501(c)(3))であり，青山学院大学が実査支援を行っています．
この調査の目的は，組織内でのソフトウェア開発におけるInnerSourceの実践状況，課題，成功事例を包括的に調査することです．
皆様の貴重な洞察は，InnerSourceの成長と普及，ならびに学術的議論の充実に大きく寄与します．

なお，ご提供いただくすべての回答は機密情報として扱われます．
調査結果は統計的に処理され，個人や組織が特定されない形式でのみ報告されます．
本調査から得られたデータは，**学術研究およびInnerSourceの普及を目的とした非営利活動にのみ使用されます．**
本調査研究に関するご質問は，yuki[at]innersourcecommons.org（担当：服部）までお寄せください．
プライバシーポリシーの詳細については，以下のリンクをご参照ください：
https://innersourcecommons.org/about/privacy/

プレゼント抽選のため，各項目にご記入をお願いします．
なお，頂いた情報をフォローアップ調査に利用させていただくことがあります．あらかじめご了承ください．
アンケートのご回答に際しては，上記の情報を読み，自発的に参加することに同意頂いたものとさせていただきます．

## 連絡先情報

メールアドレス: __________

組織名: __________

お名前: __________

# あなたのプロフィールについて

## Q1: 現在の仕事の経験年数を記入してください

☐ 0-2 / ☐ 3-5 / ☐ 6-10 / ☐ 11-15 / ☐ 16-20 / ☐ 21-25 / ☐ 26 以上 / ☐ 該当なし/わからない

## Q2: あなたの性別を最もよく表す選択肢を選んでください

☐ 男性 / ☐ 女性 / ☐ ノンバイナリー / ☐ 回答を控える / ☐ その他(具体的に記入してください): ____

## Q3: あなたの役割を教えて下さい

最も近いものを**ひとつ**選択してください

☐ 開発者 (バックエンド/フロントエンド/フルスタック/QA) / ☐ 特定技術のスペシャリスト(e.g. DevOpsスペシャリスト) / ☐ ソフトウェアアーキテクト /

☐ InnerSource エバンジェリスト/アンバサダー / ☐ InnerSource プログラムオフィサー / ☐ InnerSource プログラムリード/マネージャー /

☐ オープンソース エバンジェリスト/アンバサダー / ☐ オープンソース プログラムオフィサー / ☐ OSPOリード/マネージャー /

☐ アドバイザー/コンサルタント / ☐ プロセス改善スペシャリスト / ☐ アジャイルコーチ/スクラムマスター /

☐ プログラムマネージャー / ☐ プロダクトマネージャー / ☐ プロジェクトマネージャー /

☐ エンジニアリング部門のマネージャー / ☐ 上級幹部/VP / ☐ ビジネス関連のロール /

☐ その他(具体的に記入してください): ____ / ☐ 該当なし/わからない

## Q4: あなたの組織において，あなたの職務はどのように位置づけられていますか

最も当てはまるものを**ひとつ**選択してください

☐ 研究開発 / ☐ ソフトウェア開発 / ☐ システム設計・インテグレーション / ☐ プロジェクト管理・コンサルティング /

☐ インフラストラクチャ・運用 / ☐ 品質保証・サポート / ☐ ユーザーエクスペリエンス / ☐ データサイエンス・AI / ☐ セキュリティ /

☐ 教育・トレーニング / ☐ その他(具体的に): ____ / ☐ 該当なし/わからない

# ご所属の組織について

このサーベイにおける「組織」とは，あなたが所属する会社，団体，または組織全体を指します
特定のプロジェクトや部門についての質問ではないことにご注意ください

## Q5: あなたの組織の業種を記入してください

当てはまるものを**ひとつ**選択してください

☐ テクノロジー / ☐ 金融サービス / ☐ 卸売/小売/eコマース / ☐ 工業/製造業 / ☐ 教育 / ☐ ヘルスケア & 製薬 / ☐ 運輸/通信 / ☐ 政府 /

☐ メディア/エンターテイメント / ☐ 保険 / ☐ エネルギー / ☐ 非営利 / ☐ 該当なし/わからない /

☐ その他，具体的に: ____

## Q6: あなたの組織には何人の従業員がいますか

該当法人の正社員だけでなく，契約社員なども含めた全体の数を記載してください

☐ <9 / ☐ 10-99 / ☐ 100-499 / ☐ 500-1,999 / ☐ 2,000-4,999 / ☐ 5,000-9,999 / ☐ 10,000-19,999 / ☐ 20,000-49,999 / ☐ 50,000+ /
☐ 該当なし/わからない

## Q7: あなたの組織には何人のソフトウェア開発者がいますか?

*ソフトウェア開発者とは，コードを書き，ソースコード管理システムにコミットする人々を指します
プログラマーだけでなく，テスターやQAエンジニア，DevOpsエンジニアなど，幅広い文脈における「ソフトウェアエンジニア」であることにご留意ください

☐ <9 / ☐ 10-99 / ☐ 100-499 / ☐ 500-1,999 / ☐ 2,000-4,999 / ☐ 5,000-9,999 / ☐ 10,000-19,999 / ☐ 20,000-49,999 / ☐ 50,000+ /
☐ 該当なし/わからない

# InnerSourceについて

## Q8: あなたが考えるInnerSourceの意義とは何ですか

## Q9: 組織におけるインナーソース プロジェクト/プログラムの段階をお答えください

当てはまるものをひとつ選択してください

- ☐ アイデア段階 (組織内で知られているインナーソースの実践はない / 関心を集めるために人々と話をしている)
- ☐ パイロット段階 (価値を実証するためにいくつかのインナーソースプロジェクトをパイロット実施している)
- ☐ 初期採用段階 (組織内に少数のインナーソースプロジェクトがある / ニッチな実践と考えられている)
- ☐ 成長段階 (組織内に中程度の数のインナーソースプロジェクトがある / 受け入れられた実践と考えられている)
- ☐ 成熟段階 (組織内に多数のインナーソースプロジェクトがある / 広く普及した実践と考えられている)
- ☐ その他(具体的に記入してください): ______
- ☐ 該当なし/わからない

## Q10: InnerSourceに関してあなたが個人的に思っている認識として当てはまるものを選んでください

当てはまるものを全て選択してください

- ☐ ソフトウェア/ソースコードの再利用
- ☐ 組織内でのオープンソースプラクティスの採用
- ☐ プロジェクトチーム以外からの自分のプロジェクトへの貢献
- ☐ 正式には関与していないが個人的に興味のある組織内プロジェクトへの貢献
- ☐ 自分やチームが依存しているが正式には関与していない組織内プロジェクトへの貢献
- ☐ 自分のチーム以外の組織内の人々とのつながり
- ☐ 自分のチーム以外の組織内の人々からの学び
- ☐ その他(具体的に): ______

## Q11: InnerSourceについてのご経験について当てはまることを教えて下さい

当てはまるものを全て選択してください

- ☐ 自分のチーム外のInnerSourceプロジェクトに貢献したことがある
- ☐ チーム外の人からのゲスト貢献を受け入れたプロジェクトに取り組んだことがある
- ☐ 自分の組織でInnerSourceの展開やスケーリングに関与したことがある
- ☐ InnerSource実践者にアドバイスやコーチングをしたことがある
- ☐ その他(具体的に): ______
- ☐ 該当なし/わからない

## Q12: InnerSourceの成功に対する最も重大な障害や障壁について，あなたが考える影響度を教えて下さい

| 影響のある項目 | 決定的に大きい | かなり大きい | 大きい | 小さい | 極めて小さい | 不明 |
|---|---|---|---|---|---|---|
| 全体的な認知度の不足 | ☐ | ☐ | ☐ | ☐ | ☐ | ☐ |
| 経営陣の理解不足 | ☐ | ☐ | ☐ | ☐ | ☐ | ☐ |
| 中間管理職の理解不足 | ☐ | ☐ | ☐ | ☐ | ☐ | ☐ |
| 開発者の関心の欠如 | ☐ | ☐ | ☐ | ☐ | ☐ | ☐ |
| コードレビューの時間的制約 | ☐ | ☐ | ☐ | ☐ | ☐ | ☐ |
| 貢献へのタイムリーなフィードバック不足 | ☐ | ☐ | ☐ | ☐ | ☐ | ☐ |
| 不明確な貢献ガイドライン | ☐ | ☐ | ☐ | ☐ | ☐ | ☐ |
| 貢献するための時間不足 | ☐ | ☐ | ☐ | ☐ | ☐ | ☐ |
| プロジェクトの発見可能性の欠如 | ☐ | ☐ | ☐ | ☐ | ☐ | ☐ |
| 選ばれたプロジェクトの魅力不足 | ☐ | ☐ | ☐ | ☐ | ☐ | ☐ |
| InnerSource原則への馴染みの不足 | ☐ | ☐ | ☐ | ☐ | ☐ | ☐ |
| 法的懸念 | ☐ | ☐ | ☐ | ☐ | ☐ | ☐ |
| 税や移転価格に関する問題 | ☐ | ☐ | ☐ | ☐ | ☐ | ☐ |
| 組織文化やサイロ思考 | ☐ | ☐ | ☐ | ☐ | ☐ | ☐ |

### Q13: 具体的に直面している障害や障壁について，より詳しい情報をお聞かせください

## 組織におけるインナーソースの状況

### Q14: あなたの組織のメンバーはインナーソースをどのように捉えていますか

| 質問項目 | とても肯定的 | どちらかというと肯定的 | 関心がない | どちらかというと否定的 | とても否定的 | 不明 |
|---|---|---|---|---|---|---|
| 組織がインナーソースイニシアチブを明示的に支持すること | ☐ | ☐ | ☐ | ☐ | ☐ | ☐ |
| 組織がインナーソースを重要な戦略と位置づけること | ☐ | ☐ | ☐ | ☐ | ☐ | ☐ |
| 組織がインナーソースプロジェクトに貢献する時間を与えること | ☐ | ☐ | ☐ | ☐ | ☐ | ☐ |
| 組織の人々が自分の専門知識や動機に基づいて取り組むプロジェクトを選択すること | ☐ | ☐ | ☐ | ☐ | ☐ | ☐ |
| 組織が他チームのプロジェクトへの貢献に報酬を提供すること | ☐ | ☐ | ☐ | ☐ | ☐ | ☐ |
| 組織がインナーソースの価値観に基づいてキャリア昇進基準を定義すること | ☐ | ☐ | ☐ | ☐ | ☐ | ☐ |

### Q15: あなたの組織がインナーソースに参加する動機は何ですか?

当てはまるものを<u>全て</u>選択してください

☐ 従業員満足度
☐ 人材の維持
☐ イノベーション
☐ サイロとボトルネックの除去
☐ 知識共有
☐ コード品質の向上
☐ プロセス品質の向上
☐ ドキュメント品質の向上
☐ テスト範囲の改善
☐ 開発者の速度向上
☐ 再利用可能なソフトウェアの作成
☐ オープンソース準備への道筋
☐ インナーソースのトレンドに参加したい
☐ 組織のロードマップにインナーソースが記載されている
☐ 共通のエンジニアリング言語を通じて文化の違いを緩和する
☐ その他(具体的に記入してください): ________
☐ 該当なし/わからない

### Q16: あなたが取り組んでいるプロジェクトのソースコードは，組織内で参照可能ですか

当てはまるものを<u>ひとつ</u>選択してください

☐ 参照可能 / ☐ 要求に応じて参照可能 / ☐ 参照不可能 / ☐ わからない / ☐ その他(具体的に): ________

## Q17: 過去1年間に関わったInnerSourceプロジェクトについて教えて下さい

当てはまるものを<u>全て</u>選択してください

- ☐ プロジェクトは内部開発者ポータルで見つけることができる
- ☐ プロジェクトのコードは，その機能を必要とする可能性のある会社内の他のチームによって再利用されている
- ☐ プロジェクトにはプロジェクトチーム外からのコード貢献がある
- ☐ プロジェクトコードは可視化されているが，プロジェクトは外部からの貢献を受け付けていない
- ☐ コードは，ゲスト貢献者(プロジェクトチーム外)が貢献できるように十分に文書化されている
- ☐ コードは，他の人が簡単かつ安全に理解し変更できるように十分にモジュール化されている
- ☐ すべてのコードは，ブランチ，プルリクエスト，統合を容易にするバージョン管理リポジトリ(GitHubやGitLabなど)に保存されている
- ☐ プロジェクトには，プロジェクトチーム外の人からのゲスト貢献を処理する専任の「Trusted Committer」やメンテナーがいる
- ☐ その他(具体的に): ____________
- ☐ 該当なし/わからない

## Q18: あなたが関わっているチームの組織としての傾向を教えて下さい

当てはまるものを<u>全て</u>選択してください

- ☐ チームは完璧ではない可能性があるコードを外部に公開することに抵抗がない
- ☐ 外部からの不完全かもしれないコードの提供を受けることに前向きである
- ☐ 外部貢献社のコードを喜んでレビューする
- ☐ 外部協力者からの貢献の受け入れや拒否に関する難しい対話も厭わない
- ☐ 外部からの貢献を促すためにコードを積極的にリファクタリングおよびモジュール化している
- ☐ 社内フォーラムで他の人の質問に辛抱強く答える意思がある
- ☐ 新しいゲスト貢献者をメンタリングしたり，メンタリングを学んだりすることに前向きである
- ☐ オープンソース活動の経験がある
- ☐ "CONTRIBUTING.md"ファイルなど，ゲスト貢献者向けのドキュメントを作成し維持している
- ☐ プロジェクトチーム外の人々から報告されたバグに対応している
- ☐ ゲスト貢献者の質問とチームの決定が，新たに参加するゲスト貢献者によって検索可能となるように，議論を記録する仕組みがある(e.g. Slack / GitHub Discussion)
- ☐ その他(具体的に): ____________
- ☐ 該当なし/わからない

## Q19: あなたのチームメンバーは，チーム外からのゲスト貢献を受け入れた経験がありますか

当てはまるものを<u>ひとつ</u>選択してください

☐ 全員経験あり / ☐ 一部のメンバーは経験あり / ☐ 誰も経験がない /

☐ その他(具体的に): ____________ / ☐ 該当なし/わからない

## Q20: あなたの組織にはどのようなインナーソースプロジェクトがありますか

当てはまるものを<u>全て</u>選択してください

- ☐ 私の組織のプロジェクトはデフォルトでインナーソースプロジェクトである
- ☐ プラットフォームプロジェクト
- ☐ ライブラリと内部ツール (一般的に使用されるアーティファクト)
- ☐ DevOpsプロジェクト
- ☐ ドキュメントをコードとして扱うプロジェクト
- ☐ 将来的にオープンソース化を目指すプロジェクト
- ☐ 部門間プロジェクト
- ☐ 戦略的プロジェクト
- ☐ AIプロジェクト
- ☐ 組織間 / 法的境界を越えるプロジェクト
- ☐ 地域を越えるプロジェクト / 地理的境界を越えるプロジェクト
- ☐ 草の根 / コミュニティプロジェクト
- ☐ クラブ財（特定組織間での共有プロジェクト）
- ☐ その他(具体的に記入してください): ____________
- ☐ 該当なし/わからない

## Q21: インナーソースの導入や普及に関する重要度はどの程度だと思いますか

当てはまるものを<u>ひとつ</u>選択してください

☐ 優先課題である / ☐ 短期的な課題である / ☐ 中期的な課題である / ☐ 長期的な課題である / ☐ 課題にはならない

## Q22: 組織のインナーソースに関する将来的な普及度について，どのような見込みをお持ちですか

当てはまるものを<u>ひとつ</u>選択してください

☐ 完全に浸透している / ☐ 限定的に浸透している / ☐ 現状と変わらず / ☐ 採用範囲が狭まっている / ☐ ほぼ利用されなくなっている

### Q23: あなたの組織の「オープンソース」へのアプローチについて当てはまる文を選択してください

当てはまるものを全て選択してください

- ☐ 私の組織はオープンソースソフトウェアを消費または使用している
- ☐ 私の組織はオープンソースコードを作成または公開している
- ☐ 私の組織はオープンソースソフトウェアプロジェクトに貢献している
- ☐ 私の組織はオープンソースソフトウェアコミュニティをリードしている
- ☐ 該当なし/わからない

## インナーソース実践の状況

### Q24: あなたの組織でInnerSourceが最初に採用されてからどのくらい経ちましたか

当てはまるものをひとつ選択してください

☐ 1年未満 / ☐ 1年以上 2年未満 / ☐ 2年以上 3年未満 / ☐ 3年以上 4年未満 / ☐ 4年以上 5年未満 / ☐ 5年以上 / ☐ 該当なし/わからない

### Q25: InnerSourceはあなたの会社にどのように導入されましたか

当てはまるものをひとつ選択してください

- ☐ トップダウン: 経営陣の誰かが採用を開始した
- ☐ ボトムアップ: 1人/数人/1チームの開発者がイニシアチブを開始した
- ☐ 両方の混合
- ☐ その他(具体的に): ____________________
- ☐ 該当なし/わからない

### Q26: InnerSourceは私や私のチームに以下の点で役立ちました:

当てはまるものを全て選択してください

- ☐ 私のチームが依存しているモジュールに機能をより迅速に取り入れることができたため，市場投入までの時間を短縮できた
- ☐ 知識を共有できた
- ☐ 組織内のより多くの人々を知ることができた
- ☐ 優秀な開発者を維持できた
- ☐ 仕事の楽しさや興奮を高めることができた
- ☐ バグをより迅速に修正できた
- ☐ 私(または私のチーム)が取り組んでいるソフトウェアの新しい使用事例を特定できた
- ☐ 私(または私のチーム)が取り組んでいるソフトウェアの品質を向上させることができた
- ☐ 以前は思いつかなかった新機能を特定できた
- ☐ その他(具体的に): ____________________
- ☐ 該当なし/わからない

ありがとうございました．アンケートは以上です．

# インタビューの記録

本研究におけるインナーソースの実態解明のため，複数の企業に対してインタビュー調査を実施した．本付録では，そのインタビューの記録を収録する．これらの記録は，インナーソースの導入状況や課題を把握する上で重要な一次資料として位置付けられる．インタビュー対象企業の匿名性を保持しつつ，可能な限り詳細な情報を記載している．

インタビュー調査は，以下の 6 社を対象に実施した．

- 企業 A 社（業界無回答）
- 国際的製造業 B 社
- インターネット業界 C 社
- インターネット業界 D 社
- インターネット業界 E 社
- 情報通信業界 F 社

本調査対象企業は，製造業からインターネット業界，情報通信業界まで，幅広い業種をカバーしており，業界分布の観点から見ても十分な網羅性とバランスを確保している．各企業におけるインナーソースの取り組みについて，導入の経緯，実施状況，直面している課題，得られた効果などを詳細に聴取した．以下，各社のインタビュー内容を個別に記載する．

## 企業 A 社（業界無回答）

| 業界 | 企業規模 | 所属 | 役職 |
|---|---|---|---|
| 無回答 | 50,000 人以上規模 | 技術関連部門 | 無回答 |

1. **担当業務，オープンソースへの関与**

   ソフトウェア技術を企業 A 社のプロダクト開発やサービス開発に活かすことを行っており，主に共通のソフトウェアを提供すること，ソフトウェアエンジニアリング技術の提供，ソフトウェア開発プロセスを整理して各部門に技術支援やコンサルベースで提供している．

2. **インナーソース導入検討の経緯と現状**

   私はオープンソースを社内に広める活動をしてきたが，同じ機能を複数プロダクトに使うと，A 部門で得た技術が B 部門でも使えるようになる現象が起こっていた．数年たってから，これはインナーソースという枠組みに当てはまると気づいた．インナーソースとは，オープンソースの考え方を社内に導入し，社内の壁を取り払うものだと理解した．同じオープンソース利用がきっかけで共通課題を見つけ，それを共有・解決できれば会社をより良くできると思った．「右の部門も同じことをやっていた」というケースで，インナーソースを使えば改善が可能だと感じた．

3. **インナーソースを広めるための戦略**

   文化，システム，ルールの三つを整備しようと考えた．文化面ではハッカソンを用いた教育プログラムでオープンコラボレーション文化を醸成を目指している．ルール面は，トップダウンでの対応として「インナーソースをやっていい」というルールづくりを行った．その一つがインナーソースライセンスだ．どこまでの範囲であれば，インナーソースの枠組みで開発のコラボレーションが可能かを定めた．研究開発の範囲は OK だが，商用プロダクトとなれば対価が発生する形を整えた．最後のシステムについては，誰からもアクセス可能な共通開発プラットフォームを立ち上げて，そこでソースを開発したり議論できる状態を作った．

4. **現状の成果と課題（コントリビューション，リサーチャーへの利点）**

   インナーソースが広まったかというと，まだ測定できるのは 10 件程度．多いとは感じていない．ダウンロード数は増えているが，コントリビューションが増えたわけではない．しかし，リサーチャーからは「自分のリサーチが早く多くの人に見られフィードバックが得やすい」という声はある．結果的に搭載までの期間短縮につながった例もある．

5. **インナーソース導入の障壁（透明性の欠如）**

   日本でインナーソース導入が難しい理由として，オープンドキュメントなどがほとんど行われず，情報が透明にならない点がある．企画書などは特定部門にあるが，人をつなごうとする努力をしていないので全体が見えない．早期に情報をオープンにして透明性のあるコラボレーションをする素地がないことが大きな障壁だと感じる．

6. **障壁としてのサイロ思考，文化的問題，秘伝志向**

   「いろんな人に見てもらいフィードバックを得ることを善としない文化」が問題だと考える．秘伝的なノウハウを抱え込み，オープンで良くする発想が乏しい．オープンソースをやっていれば皆で見て良くしていくことを前向きに考えるはずだが，その文化がない．

7. **時間不足，エンジニアと管理者の視点差異，隠れてやるエンジニアの存在**

   エンジニアの場合，草の根活動としてやっている人もいるが，草の根活動のままだと成果評価が難しく，結果としてあまり幸せにならないと思う．中間管理職からは，エンジニアが安心してコードコントリビューションできる環境が欲しいという声を聞く．

8. **法的・税務的問題**

   これは一般的な話だが，特に税や移転価格が問題視される．連結法人間でも法人が違うと無償でのやりとりが利益供与になる可能性がある．インナーソースでのコラボレーションにも価値算定やバランス確保が必要になり調整が必要だ．

9. **検討した案：コミット量によるバランス確保案と，ルール上でのバランス維持の試み**

   コミット数やコミットライン数で，どのくらいの割合を各社がコミットしたかを算出し，その差額を請求するという方法も検討した．モニタリング手段が必要な一方で，価値の算出ができない．国内でも法人を超えればやはり難しい問題だ．1 対 1 ならまだしも，1 対多，多対多になるとさらに複雑になる．インナーソースのリソースを全て本体に集約し，全体が適正価格でそれを使うという案も考えたことがある．これなら単純化できる可能性がある．集約方式の実現可能性がどうなのかは不明だが，実際にやっている会社があるらしい．企業 A 社では，この方式を採用できていない．現時点では，検討できていないというのが正しい．

10. **インナーソース的アプローチとコンソーシアム型の採用**

    インナーソースは研究開発目的に限定していたが，これを実用に近づけるためにコンソーシアム型の手法を取っている．会費制で，会費を払っていれば，ある範囲のリソースを使える．モノによっては，オーナーに対して対価を支払うことで利用可能なものもある．このコンソーシアムに入っていれば，バグを見つけた時に修正をコントリビュートすることもできる．これも一つのインナーソース的なやり方だと思っている．

11. **コンソーシアム型会費制モデルの実態**

    サブスクリプションのような形で規模に応じて決められる．いろいろなリソースがいろいろな会社から共用したいものとして出されるだけでなく，会費を原資としてコンソーシアム内で共通利用するものも作ることもできるようにしている，

12. **コンソーシアム型導入の経緯**

    インナーソースの全オープンの形は研究開発の見える化として利用できるが，もう少し利用の範囲を広めて商用化を見据えて利用する場合にも対応できるようにしたのがコンソーシアム型．インナーソース的な仕組みを階層立てて作っているとも言える．これもインナーソースの一形態だと考えている．

13. **コンソーシアムにおけるサポート有無と自己責任利用**

    会費の範囲ではサポートは基本的になく，最初のリソースはサポートなしで自由に使える．ただ，持ち主が明確なので，サポートや改修が欲しい場合，費用負担する形でサポートを依頼したり，一緒に開発することはできる．オープンソースと似ていて，自己責任で使い切れるなら無料で使えばいいし，サポートが必要ならサポートしている会社にお金を払えばいい，とのような感じ．

14. **相互提供によるイーブンな関係と研究開発費範疇**

無保障だが自由に使う場合においては，提供する側は一方的にコントリビューションするだけでなく，利用側からのイシュー登録やフィードバック，テスターとしての利用などが対価となる．だから

15. **インナーソースに対する認識のばらつきと原則の導入**

（質問者の「インナーソースが何であるかについては，認識がばらついている．管理職層はより技術的な内容に重きを置いた解釈を示すことがある」という話の中で）一方でエンジニアとしてコントリビューションしたい人たちは別の観点を持つ場合がある．自分自身はインナーソースを一種の「手法」だと考えている．オープンソースの開発原則を社内に持ち込むことがインナーソースの本質だ．単に社内でソースコードを公開していることや，インナーソースライセンスを有していることがインナーソースなのではなく，オープンソースの原則を社内で適用するというやり方そのものがインナーソースだと理解している．

16. **インナーソースライセンスとオープンソースライセンスの関係**

インナーソースの取り組みを進める際，制御やルールづけのためにインナーソースライセンスが用いられることがある．Apache ライセンスや Mozilla Public ライセンス，GPL などを参考にして作成している．企業 A 社グループ会社として内部目的での利用条件を明示するために，インナーソースライセンスという形を取った．最終的にオープンソースへのコントリビューションバックを目指すなら，オープンソースライセンスをそのまま利用すればよいという話もある．ただ，社内でのコラボレーション用途に限定される場合は，インナーソースライセンスが必要になる．1社でインナーソースをやるのであれば，インナーソースライセンスは必ずしも不可欠ではない．

17. **インナーソースライセンスの参考事例と社内展開**

インナーソースライセンスには，まず企業 A 社グループ内でのソフトウェア開発におけるコラボレーション促進を目指す趣旨が記されている．続いて，「企業 A グループ」や「会社」，「インナーソースにアクセスするリポジトリ」，「ソフトウェア」などの用語が定義されており，内部目的やリサーチ目的での使用，営利目的での使用など，利用目的別の定義が示されている． ライセンスの許諾範囲として，ライセンシー企業がソフトウェアを無償で取得し，利用，複製，改変，結合，公開，利用の拡大を行う権利が認められている．ただし，著作権表示や利用許諾条件の維持が義務づけられ，改変による著作権の行使や，ソースで配布する場合のルールなどが規定されている．これらは GPL ライセンスに近い性質を持ちつつ，グループ企業内であれば自由に使えるが，外部への公開には制限がかかる．内部目的での利用は社内実用として扱われ，社内利用や外部利用に関しては別途契約を結ぶ必要がある． 保証については，このソフトウェアが無保証で提供されることが明示されており，損害が発生しても責任を負わないという免責条項が存在する．許諾条件はこのような形で締めくくられている．

18. **過去の失敗例と失敗パターン**

過去に類似の取り組みが失敗した例としては，収集した資産が放置され，誰もメンテナンスをしなくなって有用性が失われる，という状況があった．メンテナンスを引き継ぐ人がいないと，その資産は使われなくなり，最終的に死蔵されてしまう．こうした失敗は，インナーソース的な取り組みにおいても危惧されている．

19. **メンテナンスガイドラインの整備**

インナーソースとしてコードを公開する際にはチェックリストを用意している．メンテナーを明記すること，メンテナーがいなくなる場合の通知，交代不可能な場合はEOL（End of Life）を明示するなど，放置されないためのルールを設けている．これにより，コード資産が利用され続け，メンテナンスが途切れないようにする仕組みを整えている．

20. **インナーソースライセンスと商用利用不可範囲での無償利用**

許諾範囲について，コンソーシアムが前提かどうか問われたが，こちらは商用利用不可の範囲で限定することで，無償でソフトウェアを取得し，利用，複製，公開できる状況を作っている．研究開発の一環として PoC（概念実証）段階や，実用利用を判断する前の開発後期まで広く適用できるようにした．ソフトウェア開発では実用利用をどのタイミングで決めるか後ろ倒しできるため，かなりの範囲を研究開発扱いで使えるようにしている．

21. **対価としてのフィードバックと「Give and Given」の建前**

客観的に見て正当な取引であるように，研究開発においても対価が必ず発生するようにしている．この場合は，本来であれば有償であるべきフィードバックやテスターとしての貢献を対価として扱う．お互いにフィードバックをもとにした価値交換を実現するべくこのような仕組みにした．普及時には容易性のために「Give and Given」と呼ばれていた．利用者どうしがなんらかの貢献を行うことで，対価交換の関係性を確保する仕組みである．

22. **成功例の有無とナレッジシェアの実態**

インナーソースや知見共有に関して，似たような成功例が存在するのかという問いに対して，社内ではまだ顕著な成功例はない状態だ．たとえば，社内向けのウィキを用意して皆が使えるようにしたことはあるが，実際にそれが広く活用されて大成功しているわけではない．ナレッジシェア全般において，Office 系の基盤を利用してデータを蓄積するケースもあるが，大規模なイニシアチブによって横断的な知見共有がうまくいった例はほとんどないと感じている．部門単位で特定ドメインに特化したナレッジベースを作ることはあるが，それはあくまでドメイン知識であり，事業を超えた規模での共有やコラボレーションはうまく機能していない．共通のホームページを用意していても，それは特定の担当者がメンテナンスしているに過ぎず，コラボレーション的な活動にはなっていない．一方で，部門共通の技術カタログをコード管理的なやり方で蓄積し，リンク切れに対するフィードバックを受けるなど，一部でインナーソース的な試みがなされている例もある．

23. **インナーソースへの興味の関心層ごとの違い**

企業 A 内部でインナーソースに関心を持つ層は，リサーチャーが多い印象だ．リサーチャーは新しい手法に興味を持ち，試してみようとする姿勢がある．一方，管理職は「うまくいけばいいな」という期待感はあるが，自ら実践する立場にはあまりない．管理職はサポートをする際に何を根拠にするか戸惑うことが多く，「インナーソースをやっていいんだ」と明文化されたよりどころがない状況だと，草の根活動として見えないところでおもしろいことをやってしまう人が出る．現時点で研究開発としてのインナーソース活用は考えられているが，将来的にスケールさせてプロダクトを共有する段階になると，さらに明確な制度やルールが求められるはずだ．

24. **プロダクトリリースと研究開発費の境界**

開発中のメインブランチは研究開発で進め，ある段階で事業にする場合にはリリースブランチを切って「リリース A」といった形でプロダクト扱いとすることで，リサーチとプロダクト利用の境界を明確にすることを考えた．そのリリース時点でプロダクトと見なされるため，利用する者はちゃんとライセンスや利用料を払う必要がある．つまり，リサーチ段階ではインナーソース的な自由を享受しつつ，プロダクト化されたら正式な対価関係に移行する仕組みだ．こうしたルールによって，リサーチとプロダクト化のプロセスを明確に分け，スケール段階での整合性を担保している．

25. **将来への期待と明確化への意欲**

このような運用で理屈づけを行って，わかりやすく整理することで，今後のスケールや公式化に備えている．研究開発段階ではフィードバックを対価とする建前や，社内で自由に使える条件を設け，プロダクト段階ではライセンス費用を徴収し，会計処理を明確化していく．こうした仕組み作りが，将来より広範なインナーソース利用やプロダクト共有に向けての土台になると考えている．

## 国際的製造業 B 社

| 業界 | 企業規模 | 所属 | 役職 |
|---|---|---|---|
| 製造 | 50,000 人以上規模 | 開発共通部門 (持株) | 上級技術専門家 |

1. **所属組織と職種・担当業務**

   所属は 製造業 B 持株会社の開発共通部門だ. ここはいわゆる共通開発組織で, 複数の子会社へ開発的な支援を兼務する形で行っている状態だ. OSPO 事務局も担当しており, そこには 100 名から 200 名ほどのコミュニティが存在する. 事務局は5人程度の実組織である. また, 一部の子会社にもオープンソース関係で似たような取り組みを行う部門があり, そこでも少数の事務局と各部署からの代表のコミュニティを形成している.

2. **オープンソース利用状況とコントリビューション**

   オープンソースは基本的にどの事業領域でも活用している. オープンソースなしでは開発が成り立たない状況であり, 「バリバリ使っている」という状態だ. さらに, 一部のビジネスユニットはオープンソースへのコントリビューションも行っている.

3. **インナーソースに相当する取り組みと定義**

   いわゆる「インナーソースコモンズ」が狭義で定義するインナーソースは行っていない. ただし, グループ間での連携プロジェクトや, 複数組織が共同で取り組む形があるので, 広義のインナーソース的状況と呼べるものは存在する. 特定の部門だけでなく, 横断的なプロジェクトやコンソーシアム的な協働をインナーソース的なものとしてとらえることが可能だ.

4. **共同プロジェクトの事例と性質**

   共同プロジェクトには, まだ既存事業として明確に定まっていない段階のものも多い. 本社共通組織や, 国際的製造業 B 社の X 事業, Y 事業, Z 事業といった複数の事業領域をまたぐ形で, 本社との共同プロジェクトが行われることがある. これらは, ビジネスを横断し, 新規事業的な意味合いを持つこともある. 特定部門間での持ち合い的な仕組み開発や, 共同プロジェクトそのものをインナーソース的な取り組みとして位置付けることもある.

5. **プロジェクト横断的な開発形態・階層と技術共有**

   共同プロジェクトでは, それぞれの専門領域を持つ組織が得意分野を提供し合い, それを統括するプロジェクト側がインテグレーションしている. ミドルウェアやプラットフォーム, クラウドといった特定の階層に限定されず, 各専門領域が相互補完し合う構造だ. 全員で一つのソースコードを共有・改変するような形ではなく, 各領域が自前の部分を持ち寄り, 要求やバグ修正などをイシューを通してやりとりするレベルが多い.

6. **ソースコードの取り扱いとフィードバック様式**

   ソースコード自体を共同で改変することは基本的に行わない. イシューを上げたり, 数ファイル程度のコンフィグ修正やバグフィックスのフィードバックを行うことはある. ただ, ソースコード著作権レベルでの共同所有・共同改変は行わず, 要件定義やバグ報告といった段階でのフィードバックにとどまることが多い.

7. **リポジトリ管理・アクセス制御・コラボレーション体制**
   GitHub Enterprise のようなソースコード管理システムでアクセス権を制御しつつ，リポジトリを共有している．本社がエンタープライズ環境を保持し，そこにあるリポジトリを必要に応じて子会社に見せる．場合によってはリポジトリ単位でアウトサイドコラボレーターを設定し，他組織が閲覧・コントリビューションできるようにする．このとき，どの組織がコストを負担するか（「セントラルでお金を出す」など）によって，どのリポジトリを利用するかが変わる．
8. **会計処理・業務委託契約とコスト計算手法**
   共同プロジェクトを進める際は，移転価格税制や利益配分の問題があるため，業務委託契約を結ぶ形が多い．本社プロジェクトであれば，本社が保持するエンタープライズ上のリポジトリに子会社がアクセスし，工数ベースでコントリビュートする．その際，プロジェクトオーナーが業務委託費を払い，成果物はプロジェクト側に帰属する．社内であれば単にコストセンター間のやりとりとして管理会計上処理するが，グループ会社間では業務委託契約を結ぶことが一般的だ．こうした業務委託はプロセス型の委託であり，工数×人件費という形でコスト算定する．たとえば，X00 万円や Y00 万円相当の人件費を何時間か提供することで，そのプロジェクトに貢献することになる．成果物の所有権はプロジェクト側にあり，各組織は工数という形でコントリビューションを行うことで費用対価関係を明確にしている．
9. **研究開発フェーズと製品フェーズでの対価設定**
   研究開発の段階では対価の設定が緩く，日本の税制上，開発版・軽量版・研究版のフェーズにある場合，見せるだけの対価としてはフィードバックで成立する立て付けになっている．研究開発フェーズではまだ金銭価値を生み出していないため，対価の算定を現金以外の，開発に対して価値のあるフィードバックを必ず行うという契約としている．実際，この段階では開発物は全員が見られるようになっているが，それをさらに別の部門が見たいとなった場合も研究開発的な位置付けであれば，フィードバック的な価値交換で成立する． 一方，製品として扱う段階になると，明確に対価設定が必要になる．製品フェーズでは，現金で対価を支払うことが求められ，有償のライセンス契約やサポートフィーの設定を行う．研究開発フェーズであれば要件や QA 結果のフィードバックなどを対価として認めるが，製品化後は現金での支払いが要求される構造になっている．
10. **ライセンス契約と契約書フォーマットの存在**
   研究開発用に使う場合と，プロダクトとして販売・提供する場合とで契約書が分かれている．管理部門が用意した契約書フォーマットが存在し，研究開発用ライセンスであれば，開発行為の途中段階での利用に対して，フィードバック程度の対価で良いと定義している．一方で，正式に製品として提供する場合は，別の契約書フォーマットを用いて，明確な対価（現金やサポートフィー）を求める．社内利用や POC（Proof of Concept）的な，研究開発の延長線上で使う場合にはフィードバックでよいが，プロダクトに組み込むなら金銭的な対価を求める，そのような取り決めは，各部署が物を渡す際に使用する管理部門作成の契約書フォーマットに明示されている． たとえば，「これは研究開発用に使う」「プロダクトの GM（General Availability）前の開発行為に使うライセンスである」といった場合は，研究開発用の契約書フォーマットを用いる．一方で，「これは RC（Release Candidate）になります」といったフェーズにおいては，製品化を前提とした別の契約書フォーマットを用いることになる． このよ

うなテンプレートが存在することで，社内はそのフォーマットに従い，慣れやすく，自然にルールを守れるようになっている．

11. **複雑なインナーソース的開発と特別契約**
より複雑なインナーソース的な開発を行う場合，一般的な契約書フォーマットではカバーしきれない場面が出てくる．そのような場合は特別契約を結ぶことがある．どの部分がどの部門の持ち分なのか，どこが権利を有するのか，相互に干渉する部分の扱いをどうするかなどを詳細に定め，相互メリットを明確化する契約が存在する．この特別契約によって，通常のフォーマットでは扱い切れない複雑な状況に対応している．

12. **サポートフィーの設定と研究開発フェーズでのオプション**
製品のフェーズに入った場合，最低限サポートフィーを取ることが基本的なルールになっている．研究開発フェーズでも，必要に応じてサポートフィー的な対価を設定するオプションは存在する．ただし研究開発中は，現金以外の対価でも許容されるため，フィードバックや QA 対応などが対価として見なされることもある．商用利用時には現金で対価を支払うルールになっているため，研究開発と商用利用で求められる対価の形式に差がある．

13. **アーキテクチャ図を用いた契約書策定**
相互の持ち物や思惑，パーツが重なるような部分がある場合，単純に金銭で処理するのではなく，帰属はこうだが利用権は与える，といった工夫を行う． そうした契約においては，契約書の前段に添付される資料としてアーキテクチャ図が用いられることが多い．アーキテクチャ図には，「この部分はこの部署が開発している」「この部分は別の部署が担当」「インターフェイスはこのリクエストにもとづき，帰属はこうで，再利用は可能」といった情報がまとめられている． こうしたアーキテクチャスタック図は，所有権や利用権，リクエスト定義や再利用の可否を明示するのに役立っており，複雑なインナーソース契約で頻繁に利用される．

14. **フィードバックを対価とする考え方の背景**
「見せるだけならフィードバックを対価とする」という社内理解の背景には，価値算定の難しさと金額設定の困難さがある．研究開発中は事業部側が「動いた」「動かなかった」「問題があった」といったフィードバックを提供する．このフィードバックは，本来であれば受け取る側でテストや要件定義を行う際に必要な工数だ．その前提を踏まえ，事業部側がそれを負担することで，対価として成立する．研究開発中は事業部側が「XXの条件でも動いた」「YYの条件では動かなかった」「ZZの問題があった」といったフィードバックを提供することを必須としている．試用段階でまだ事業部にお金が発生していない段階で金銭的対価を設定するのは難しいためこのような形を取ることに至った．

15. **研究開発フェーズと製品フェーズ間での利用・対価設定**
商品や研究開発用として後から使いたい時，研究と製品で対価レベルが変わる．追加開発を行う場合は基本的にお金をもらう．事業部単体のリクエストによる開発は業務委託として行うが，そのうえで研究開発や共通利用の権利は引き続き保有し，契約で担保している．これは，オリジナルのコードに改造を加えた場合，その改造部分が特定の事業部専用になってしまうとブランチが増え続け，共通利用が崩壊するからだ．契約上，共通に再利用できる形を保っている．

16. **メンテナンス責任とサポートフィー**

メンテナンスは基本的に提供側がコア部分を担うが，差分部分は受託として対応する．これはサポートフィーの一環だ．共通部分開発と，受託で開発した部分のメンテナンス負荷は，次年度の個別契約で対応する．グループ内でも完全業務委託で，まるで別会社どうしのやりとりをするような形になっているが，共通化したい意図やポリシーがあるので，その点は契約で担保している．過去に問題が起きたこともあるが，その経験を踏まえてこうした仕組みになった．

17. **製品化時のコードベース提供形態**
実際には製品用に出す時にリポジトリそのものを見せるわけではなく，別の手段を使うことが多い．特定のブランチだけを切り出して見せることもあれば，単に tar アーカイブを渡したりする場合もある．組み込み系や特定のシステム構成によっては，実行用バイナリとライセンスコンプライアンス用のソースコードパッケージだけを渡すこともある．日本の税制上，内部用ソフトウェアと販売用ソフトウェアの税制が異なるため，切り替え時期に合わせた対応が求められる．

18. **内部用ソフトと固定資産化の問題**
内部用ソフトをストレートに扱うと，仕様決定後は全て固定資産化されてしまう．固定資産を他社に渡す，シェアするとなると，その時点でお金を取らなければいけない．その結果，先ほど説明したフィードバック対価のロジックが通らなくなってしまう．このため，うまくやり繰りする必要がある．自分の周囲では製品に載るものが多いのであまり心配しないが，サービス系だと固定資産配布でフィードバック対価ロジックが成立しにくくなるはずだ．そうなるとライセンスやサブスクリプションでお金を取らなければならず，ケースによっては対価の取り方が難しくなる．

19. **国際取引と評価額，移転価格への対応**
ライセンスやサブスクリプション，買い切りなど，金銭の取り方はケースバイケースだが，製品化以降の段階でサポートに対価を払うモデルが心理上スムーズになることが多い．問題は国際取引で，対アメリカやヨーロッパとなると，移転価格の算出やソフトウェア評価額を決めた上で対価設定をしなければならないなど，より面倒な条件が出てくると聞いている．移転価格の算出方法については，実際どうしているのか（回答者は）分からないが，日本国内で完結する場合と国際的なやりとりではルールや手続きが違い，複雑な対応が求められる．

20. **内部利用ソフトと製品フェーズへの移行における懸念**
周囲では内部利用というより製品に載るものが多いので，あまり心配はしていない．ただ，サービス系のものを開発している人たちはどう対応しているか疑問がある．内部利用であればフィードバックを対価として扱うロジックが通るが，製品に組み込む段階になると，固定資産を配ってしまう形となり，フィードバック対価の理屈が通らなくなる．そうなるとライセンスやサブスクリプションなど，お金を取る仕組みが必要になる．ライセンス契約はお金の取り方がケースバイケースで，サブスクリプションのように柔軟に取れる会計上の仕組みが社内にある場合もあれば，買い切りや別途サポートライセンス用意など，いくつものパターンが存在する．

21. **製品化後のサポート対価モデルと国際的な税務対応**
最終的には評価時点（研究開発段階）ではフィードバックを対価として扱い，製品化後にサポートに対して対価を払うモデルの方が審理障壁が低いことが多い．だが国をまたぐと面倒な条件が増え，ソフトウェアの評価額を決めて対価を設定しなければいけないなど，対アメリカやヨーロッパ向けでは国

内とは異なる対応が必要になる. 移転価格に関してはどうやって算出しているのか, 明確な知識はないが, 国際的なやりとりは一筋縄ではいかない.

22. **業務委託による価値集約と移転価格回避の仕組み**

多くの場合で, 開発は全て業務委託として構成され, ソフトウェアの価値は日本に集約される. 価値が日本に集まる以上, 国をまたいだ価値移転は発生しない. つまり, 価値を移転せず, コストを日本が集約する形を取ることで, 移転価格課題を回避している. 価値の移転がないため, 移転価格算出の必要性が生じにくい構造になっている.

23. **海外拠点と価値保有構造, 競争領域 X 事業での経理上のルール**

日本からの業務委託で日本に価値を生成する形を取り, 他国側は労働力提供にとどめる. たとえば, 競争領域 X 事業は自社内で完結する形でお金を使い, 価値を自前で保有するため, 契約や経理処理が単純化されている. ただし, 経理や税務的な観点から, 共有可能なソフトウェアにはクライテリアが存在する. その範囲を超えるものをやりとりする場合には, 対価を支払う必要がある. 競争領域 X 事業内でも, 特に高付加価値であると認定されるものについては移転価格税制に抵触する可能性があるため,「この範囲のソフトウェアしか共有できない」といったルールが定められている. 範囲を超える場合は金銭的対価が発生する仕組みになっている. たとえば, US のチームが競争領域 X 事業側に高付加価値な機能群を開発・提供している場合, その機能がただの工数移転ではなく, 付加価値を生み出す機能移転とみなされる可能性がある.

24. **高付加価値(マーケティング価値)を持つコードの共有制約と対外的なの説明**

CM(コマーシャル, マーケティング)に該当するような高付加価値な機能は特に問題となる. こうした機能は単なる技術的要素ではなく, 競争優位性を生み出し, 市場価値を向上させる要因であるため, 業務委託として開発にかかるコストだけを取り扱った場合, 外より「高付加価値なものを無償で移転している」という疑念が生じる可能性がある. したがって, こういった高付加価値機能には必ず対価が必要になる. 外観として「移転するなら対価を払う」というのは極めて妥当な要求だ. マーケティング的価値を前提にしている機能であれば, その付加価値に見合う支払いが生じるのは当然と考えられる.

25. **オープンソースベース統合物での一方向共有と R&D 段階での低負荷**

オープンソースをベースにインテグレーションした成果物などは, 比較的ハードルが低く, 一方向的な共有はやりやすい. R&D 段階での一方向的な共有やフィードバックの循環は比較的容易に行える. この点は高付加価値機能のように税務署が口を出す類のものとは異なり, あくまで技術的なやりとりとして成立しやすい状況がある.

26. **価値の所在と価値評価**

問題の本質は, 価値の源泉がどちら側にあるかという点にある. アイデアの原点がどちらにあったか, 価値創造がどこで生まれたかが移転価格税制の評価軸となる. たとえ工数ベースで支出をしていたとしても, そこから生まれるマーケティングに活用可能な価値(価値)があるならば, それに対して対価を支払わなければならない. こうした点は対外的にも整理する必要がある. コード行数ベースで語ろうが, コストを説明しようが, 付加価値やマーケティング的価値を伴う機能の移転は, 価値の所在を明確にし, それに対する適正な支払いを求められる, ということだ.

27. **輸出入関連の複雑性とエクスポートコントロールの必要性**

いわゆる輸出入関連はもっと大変である. アメリカの法律など, 日本も守らなければいけない規則があるので, 基本的には輸出管理をしなさいと言われてしまう状況がある. ただ, それを全て文字通り受け取った場合柔軟性がなくなり, どのようなソースコードを共有しようにも「全く共有できない」ことになる. そうした中で, リポジトリに対して「このリポジトリで扱うものはこういうものだ」と定義し, その範囲でエクスポートコントロールを行い, 輸出判定をする. 参加者のフィードバックは, その範囲を逸脱しないというルールで, 簡略化して回している.

たとえば, (一般例として) 兵器に関する情報は, そのリポジトリでは扱えないようにする合意がある. 機能定義を行い, エクスポートコントロールの分類で問題ないことを一度確認したら, その後は「エクスポートコントロール上引っかかるような機能は絶対に入れない」という方針を敷く. そうすることで, 毎回厳密な審査をしなくても緩く監査するルールになっている. 事業会社 A の扱いうる機能により, その監査の軽重は異なる.

28. **リポジトリへのエクスポート範囲定義とフィードバックの制限**

リポジトリには最初に「このリポジトリはこういう機能があります」と書く. その際, エクスポートコントロールのさまざまな分類を確認し, 問題ないと判断したら, そこにはエクスポートコントロール上の要注意機能は入れない. これにより, (たとえば極端な話 – 一般的な話として仮に) 兵器関連の情報はそのリポジトリで扱えないようにする合意を取る. こうすることで, エクスポート範囲を明確に定め, その範囲を逸脱したフィードバックは出さないという運用を可能にしている.

29. **新規ユーザー登録・リポジトリ作成時の手続きと監査プロセス**

どのタイミングで参加者をチェックするのかというと, リポジトリを作ってユーザーを登録する時などに判断する. インドや中国などオフショア先への展開を考える時は, 「こういうソフトをインドと中国に出します」と出すことでエクスポートの手続きを行う. 日本国内の仲間内 3, 4 人でやる場合にそこまで意識されることはない. リポジトリを起こすまでは緩くても, ユーザーを追加する時にエクスポート処理が必要になってくる.

30. **国際的開発拠点 (インド, 中国など) への展開と規制対応**

内部開発しているものをインドや中国などの拠点に出す時には大騒ぎになる. 国際的製造業 B 社は会社のポリシーとして武器や兵器関連はやらない方針がある. ただ, 見方によっては兵器に転用可能な技術などがあるかもしれないため, 事業会社によってはチェックが厳しいところもある.

31. **コラボレーションの階層構造と起こり方**

ボトムアップのコラボレーションもある. ボトムアップのものは「必要があったからその場で人を巻き込む」といった柔軟性のあるものが多く, 正式な契約書やルールの調整よりも柔軟性を優先して動いているケースも存在するようだ. 一方で部と部が技術をオファーし合って合意して始まるプロジェクトは, 部どうしが握り合って進めるため, 契約書などを含めてしっかり対応する. さらに上の層には技術戦略にかかる共同体があり, そこでは横連携が必要な技術エリアに関して課題を見つけ, 有志で集まって解決する. これはトップダウンに近い形だが, マネジメントラインではなくコミュニティベースで起きているコラボレーションになる. そのうえにはもっと上層から「あそことあそこでプロジェクトを起こす」といった完全なトップダウンが存在する. つまりコラボレーションには, ボトムアップからトップダウンまで 4 階層くらいの形がある. 技術戦略にかかる共同体のような場では, V チーム (バーチャルチーム) を

作って動くこともある．知らないところで勝手に立ち上がっている場合もある．要するに，コラボレーションには複数の階層と発生様式が混在している．

32. **技術戦略にかかる共同体の概要と役割**

技術戦略にかかる共同体とは，いくつかの技術エリアで横連携が必要な状況において立ち上がる組織的枠組みだ．ソフトウェアにかかる共同体が例であり，そこでは全社横断で課題を見つけて有志が集まり，解決する動きをする．ソフトウェアにかかる共同体の中には，オープンソースのコンプライアンスルールを作る動きや，開発環境やクラウド技術を利用した取り組み，コミュニティ作りなどがある．同様のグループは，さまざまな技術領域で横断的に存在している．これらはだいたい 10 種ほどあり，その中に複数のサブグループが存在している．多少の予算もあり，完全な持ち出しにはなっていない．

33. **共同体参加におけるインセンティブ**

技術戦略にかかる共同体が扱うのは横断的な課題で，それが解決されることが部門に対するインセンティブになる．たとえば，オープンソースコンプライアンスのルールが共同体内で作られ，それが社内ルール化される．部門側はそのルールに従えばよく，手続きが整うことでメリットを得る．社員に対するインセンティブは，興味があることややりたいことを実現できる点が大きい．共同体は，協力してくれた部門に対して「この共同体ではこういう成果が出ました」というレポートを返す．その結果，部門側はその社員の評価をプラスアルファで行うことができる．つまり社員は活動によって評価が下がらないし，やりたいことに取り組める，これがインセンティブになっている．

34. **横断的な活動とエンジニア評価施策**

エンジニアの評価は基本的に上長が行うが，業務先からの評価値も反映する必要がある．横断的な活動による評価もあり，技術戦略にかかる共同体などに参加することで，その取り組みが評価される可能性がある．これ以外にも上級技術専門家が起こしている活動など，横断的な動きは多様に存在する．社員は，こうした横断的活動に参加することで，評価が下がらないようになっているし，部門横断で必要とされる課題解決が進む．これがエンジニアの評価にも良い影響を与える．インセンティブという言い方をすると，社員個々に直接報酬を渡すわけではないが，共同体で達成した成果がレポートされ，それが評価にプラスされる構造になっている．結果として，社員は自分のやりたい課題に取り組むチャンスを得るし，部門は社内ルールやナレッジ整備といったメリットを受ける．

35. **インセンティブと評価に関する補足説明**

ソフトの若い人の教育に関しても技術戦略にかかる共同体を通じて行われ，それが結果的に教育効果となり，インセンティブととらえることができる．社員へのインセンティブは基本的に「興味があってやりたいことができる」という点だ．また，技術戦略にかかる共同体から部門に対して「この共同体ではこういう成果が出ました」と報告し，それによって評価がプラスされる．社内共同体活動をしていることで評価が下がらず，自分がやりたい課題に取り組めるのが社員側のインセンティブになる．インセンティブという言葉で表現すると，そういった形で社員に対してメリットが戻ってくる構造になっている．参加していること自体がそのまま評価やモチベーションにつながるわけだ．

36. **横断活動への参加許可と上長評価への反映**

上長から認められていないうえに本職をしっかりやっていないのに横断的活動ばかりしていても評価はされない．基本的には技術戦略にかかる共同体やそのサブグループ，上級技術専門家のプロジェクトに参加するには上長の許可が必要だ．本当に草の根の活動は勝手にやっているので上長が

知らない場合もあるが，基本的には上長は横断活動への参加を把握している．横断的活動側から「この人はこういう活動をしてくれた」「このプロジェクトでこういうことができた」という情報が伝わるようになっており，それが評価へと反映される仕組みになっている．

37. **ボトムアップ活動の発表機会とインセンティブ**

隠れて行われているボトムアップ活動を発表できる場が設定されており，その発表機会自体がインセンティブになっている．その場で賞が与えられることもあり，それは現金だったり，その活動を部のオフィシャルな仕事として 1 年間行うための予算だったりする．また，12 月に開催される技術イベントへの出展権利が与えられることもある．さらにボトムアップ活動は自分たちで発表の場を作り，会社の広報システムで「こんなイベントがある」と告知することが可能だ．

38. **ボトムアップ活動と社内イベント・賞・ファンディング**

コミュニティビルディングや発表を通じて，ボトムアップのインナーソース的な活動が社内でこっそり始まり，「おもしろそうなことをやっている，どうだろう」という形でメンバーが集まる．その段階では業務ではないので，シェアしても問題はない．ところが，その活動が発表会で認められ予算が付くと，正式なプロジェクトとして扱わなければならなくなり，そこから上位の仕組みに組み込まれることになる．実際，さまざまな事業体で年 1 回の発表機会を用意し，裏でこそこそやっていた活動を披露し，賞をもらい現金や公式プロジェクト昇格が得られることがある．また，プログラム A という新規需要創出プログラムもあり，そこに通れば組織化されて継続開発が可能になる．こうしたボトムアップ活動を吸い上げてファンドするシステムは社内に多数存在している．

39. **小規模な CI/CD ライブラリ等の扱いと技術戦略にかかる共同体への吸い上げ**

たとえば CI/CD のライブラリのように，価値はあるが大きなインパクトや賞を与えるほどでもない小さな改善は，ソフトウェアにかかる共同体の開発環境ワーキンググループが拾い上げる必要がある．その結果，そうしたシステムが社内インフラに組み込まれていくことがある．全体として多層的な構造があり，トップダウンのプロジェクトもあれば，トップダウンでコミュニティを立ち上げて中で勝手に活動を定義するケースもあるし，本当にボトムアップから吸い上げるケースも存在する．いずれにせよ多様な経路が用意されている．

40. **社内共通基盤への統合と組織 B チームの役割**

吸い上げ先は事業部かもしれないし，技術戦略にかかる共同体かもしれない．内製インフラに組み込まれる場合は，最終的に誰がメンテナンスをするかといった問題が出てくる．プラットフォームエンジニアリングや共通基盤チームに相当する社内組織 B チームとの調整が必要になる．たとえば社内組織 B チームが管理している GitLab の存在もあり，もともとは開発環境ワジの人々が「ここで勝手にリポジトリを立ち上げてメンテナンスするのはナンセンスだろう」という議論があった．そのうえで GitHub のコーポレート契約をどうするかといった点も提言し，最終的にそういった形で社内インフラに組み込まれていった．

41. **組織 C，部署 A，OSPO 的活動と社内多層構造**

組織 C は IT インフラやセキュリティなどを担っている．回答者は今部署 A という部署に所属しており，IT 系の開発を行い，OSPO（オープンソースプログラムオフィス）に近い機能もそこで開発している．社内には多層構造があり，各層で個人活動やボトムアップ活動を受け止める土壌がある．単に抑制

するだけでなく，「隠れておもしろいことをやる」という製造業 B 文化と合わさって多層的なコラボレーションが成立している．

42. **評価との関係と人事施策，若手エンジニアのモチベーション**

こういった活動は人事評価にどうつながるのか．人事が「横断的活動を評価せよ」と明記しているので問題ない．若手エンジニアがどう思っているかは回答者には定かではないが，巨大イベントや社内の活動情報が目に触れる機会は多いので，関心がある者は響くだろう．情熱があるエンジニアは外部イベントにも積極的に参加し，社内外のコミュニティに関わっている．そうした人たちが社内でコミュニティを起こすことで，新たな技術的コラボレーションが生まれている．

43. **インナーソースの定義・透明性・海外拠点の期待**

インナーソースについても，社内で「オープンだ」と言っても意思決定が透明でなければ単なるオープンとは言えないし，公平なのか平等なのかといった言葉の定義が必要だ．非日本系の下部組織は，「Google がやっているから」などの理由でインナーソースをやりたがるが，結局は巨大グループ企業内での透明性を求めているのだろう．製造業 B は本社が日本にあり言語やタイムゾーンのバリアがあるため，そうした下部組織がサイロ化を感じ，透明性を欲していることも考えられる．特に BtoC 技術領域 Z の拠点は海外が中心なので，日本で技術イベントを開催するだけでなく，他の世界の拠点でも類似のイベントを実施している．そうすることで，グローバルな視点で隠れた活動もキャッチしやすくなり，整理された歴史を踏まえて予算を付け，オペレーションする座組が可能になる．最近では BtoC 技術領域の扱う技術戦略にかかる共同体ができており，事業会社の人物がオーナーで英語ベースのコミュニケーションをとっている．

44. **巨大組織内でのインナーソース適用範囲と評価**

（質問者が「国際的製造業 B 社は事業体が複雑で，本社が各所にあり，ドメインによって共有技術が異なるため，全体で統一する必要がない」と指摘したのに対して，回答者は）その通りだと応じる．競争領域 X 事業だけでも大量の人員がおり，その中で共有やイノベーションは十分可能だ．製造業 B 特定事業領域だけに言及しても GitHub Enterprise を一つにまとめるなどの戦略があるので，その範囲でインナーソースを実現できる．もちろん微妙なクライテリアはあるものの，ドメインごと，拠点ごとの特性に合わせたインナーソース活用が行われており，その結果として海外拠点を巻き込んだ技術エクスチェンジやコンテンツ技術戦略にかかる共同体などが相互補完的に動いている．

# インターネット業界 C 社

| 業界 | 企業規模 | 所属 | 役職 |
|---|---|---|---|
| インターネット | 数千人規模 | 開発生産性向上組織 | リーダー |

1. **回答者の経歴**

   インターネット業界 C 社入社以前 いわゆる SIer と呼ばれる業界で，複数の企業に派遣・常駐する形で，ハイブリッドアプリ開発や銀行内で使用される社内アプリケーション開発など，さまざまなプロジェクトに携わっていた．

2. **インターネット業界 C 社におけるインナーソース文化と GitHub Enterprise 全社導入の転機**

   もともと，インターネット業界 C 社には GitHub 上で開発し，他部署でも利用してもらう文化や，OSS として公開する風土が入社前から存在していた．転機となったのは，2020 年ごろに GitHub Enterprise を全社導入したことである．これによりインターナルリポジトリが利用可能となり，社内基盤を社内全体で広く共有したり，隣接チームと共同開発を行う動きが加速した．その背景を受け，2021 年には「制度 A」と呼ばれる施策が開始された．これは，グループ内で利用されている技術資産を中長期的に支援・育成する制度であり，社内・社外を問わず一定の影響力を持つ技術資産を申請・審査し，ランク付けして支援する仕組みが整えられた．

3. **制度 A の誕生と意思決定プロセス（会議 A との関係）**

   制度 A は，社内に存在する「会議 A」と呼ばれる仕組みから生まれた．「会議 A」とは，取締役がリーダーとなり，選出したメンバーが新たな施策を社長へ提案し，その場で決裁を得る場である．その技術者版会議 A において，ある社員が制度 A の原型となるアイデアを提案し，採択されたことで制度化が実現した． 技術者版会議 A では，リーダーが独断でメンバーを選び，メンバーはアイデアをまとめて社長に提案する．社長が即時に決裁するため，現場（ダウン）から出たアイデアがトップで承認され，強い決定力と推進力を持つ制度が生まれる．つまり，下から上へと提案が行われ，最終的にトップダウンで制度化される仕組みである．

4. **インナーソース活動と OSS 公開におけるエンジニアの多様な動機・手法**

   エンジニアごとに状況や考え方は異なる．業務上必要なものを迅速に OSS として公開し，自身のプロジェクトで活用しつつ外部にも提供したいと考える者もいれば，社内利用にとどめるケースもある．要するに，社内共有と社外公開のバランスやエンジニアの動機は，各人や各チームの方針・目的によってさまざまである．また，組織的な技術選定を含めて OSS として構築しているものも存在する．たとえば，プロダクト A などの基盤開発では，外部で流行らなければ社内で利用されにくい状況があり，最初から社外に公開することでユーザーや開発者を集め，そのフィードバックを社内に還元する流れをとるケースもある．さらに，特定の領域（例：ゲーム領域）内で複数チームが共通課題を抱える場合，インナーソースによって共通基盤的なものを開発し，各チームに活用してもらう動きがある．このように，インナーソースやオープンソース活動の形態は多様であり，それぞれが独自の思想で進められている．

5. **社内組織構造の多様性とインナーソースの課題（社数の多さ・車輪の再発明）**

   社数は非常に多く，プロダクト数でいえば 100 以上に及ぶと考えられる．全貌が複雑である．こうした多様な組織構造の中では，車輪の再発明も起こりやすい．GitHub Enterprise 導入以前は，各社が

独自の GitHub 環境を使っており，インターナルリポジトリの概念が存在しなかったため，隣の事業部が何をしているのか見えなかった．また，隣接チームどうしが競合関係にあることもあり，それぞれが独自に基盤を作ってしまい，結果として同じようなものが複数生み出されていた．GitHub Enterprise 導入後も，こうした状況が完全に解消されたわけではないが，ある程度改善されつつある．

6. **制度 A の位置付け，適用範囲，評価基準の更新，申請モチベーション，CNCF との類似性**

   制度 A は，全社・グループ全体を対象とした取り組みである．協力体制はチームごとにバラバラであったが，制度 A は共通の枠組みとして機能している． 制度 A は少しずつアップデートされており，申請や審査内容の改善を通じて，より納得感のある基準づくりが進んでいる．たとえば，制度 A 申請時の審査基準として，利用者数，満足度，GitHub のスター数など，多面的な指標を取り入れ，バージョンを重ねるごとに見直し・改善を図っている． 制度 A への申請モチベーションは多様である．インセンティブを得るために申請する者もいれば，社内での認知度を高めるために利用するケースもある．また，制度 A は 5 段階のグレードを設けており，利用数や認知度など複数要素を総合的に判断してランク付けを行っている． 制度 A にはボードメンバーがおり，回答者自身もその一員である．各管轄からエンジニアが参加し，共同で制度を推進している．開発生産性向上組織 A が主体というわけではなく，複数のメンバーで運営を持ち合っている． 申請対象となる技術資産は幅広く，小規模・中規模のライブラリや，最近では AI モデルなど多岐にわたる．いずれも社内開発・運用を効率化する技術資産であり，インターネット業界 C 社社員がオーナーであること，導入実績やドキュメント整備があることなどが申請の最低条件となっている．外部がコントリビュートする OSS であっても，インターネット業界 C 社社員がメンテナーであれば申請可能である． 制度 A 自体が開発モチベーションの出発点になることはまれであり，もともと他の目的で開発されたプロジェクトが，より多く使われたり，サポートや費用負担などのインセンティブを得るため，あるいはキャリア形成の一環として申請されるケースが多い． 申請されるプロジェクトにはさまざまな形態がある．個人で作っているもの，事業部の一つのチームとして戦略的に作ったもの，複数チームがインナーソース的に共同開発したリンターなども申請例として挙げられる． 制度 A の立ち位置は CNCF（Cloud Native Computing Foundation）に近いイメージである．CNCF に参加するプロジェクトは，CNCF 入りを目標に立ち上げられたわけではなく，ある課題を解決するために作られ，スケール拡大やサポート強化が必要になった段階で CNCF の力を借りて大きくしていく．このように，制度 A も同様の役割を担い，プロジェクトをより大きく育てていくためのプラットフォームとして機能している．

7. **オープンソース文化と公開の土壌**

   SIer 時代からインターネット業界 C 社へ移った際に感じたこととして，Web 系の独特な雰囲気がある．オープンな文化があり，オープンソースを日常的に利用する環境で，さまざまな OSS プロダクトのコミッターが身近に存在する．また，副業や新たなプロジェクト立ち上げが当たり前のように行われている．何かを作って公開したり，他者が作ったものを活用することが当たり前であり，以前所属していた SIer の環境とは大きく異なる．

8. **自社プロダクト開発・技術選定の自由度と事業の多様性**

   インターネット業界 C 社は自社プロダクトを持っている点が大きい．ゲーム，広告，メディアなどさまざまなドメインで事業を展開しており，それぞれが必要とする技術は異なる．よって技術選定についても自由度が高く，その領域で最適な技術を自分たちで決められる．子会社を設立してオーナーシップ

を委ね，事業を成長させていく文化も根付いている．悪く言えば縦割りだが，良く言えば各所がオーナーシップを持ち，技術もビジネスも自主的に決めていく風土があり，その中で「使えるものは使っていく」文化が自然と醸成されている．

9. **「縦割り」と組織構造，役員管轄とドメイン分化**
   インターネット業界 C 社はドメインごとに分かれており，担当役員がいる形で組織を構成している．たとえばメディア事業，広告事業といった単位で分かれ，その下にさまざまなファンクションが割り振られる．エンジニアにとってはあまり意識しない部分だが，組織的には各フロアやビルごとに，そうした単位で分かれていることも多い．管轄内では比較的連携が活発だが，管轄を超えるとそうでもない場合がある．
10. **GitHub オーガニゼーションの粒度とインターナルリポジトリ**
   基本的には 1 プロダクト 1 オーガニゼーションという粒度が多いが，例外は非常に多い．また「1 管轄 1 オーガニゼーション」という横串オーガニゼーションも存在することがある．プロダクト B などは特殊で，最初から独立したオーガニゼーションを用意する場合もある．GitHub Enterprise 上のインターナルリポジトリを利用すれば，エンタープライズメンバー全員がアクセス可能であり，可視性を担保できる．インターナルで横断的なソースコード閲覧が可能なため，オーガニゼーションが分かれていても，アクセス不能という問題は小さい．
11. **オーガニゼーション単位の方針・請求・セキュリティ管理**
   インターネット業界 C 社ではオーガニゼーション単位で請求先が割り振られたり，その単位でセキュリティポリシーを独自に設定したりすることがある．たとえばゲーム領域では競合状況を考慮したセキュリティ強化などが求められる．エンタープライズレベルで最低限の強制事項はあるが，それ以上は各オーガニゼーションが独自に管理する．これにより自治権を確保しつつ，インターナルリポジトリで横断的な利用も可能な状態を保っている．
12. **法人間の資産共有・無形資産の扱い**
   会社間（法人間）のやりとりについて，たとえば無償でリソースを付与しているような状態がどう扱われるかは不明である．基盤開発に関する請求は比較的しっかりしているが，インターナルな資産をどう評価するかという点は明確でないようだ．少なくとも回答者が知る限り，全社レベルで無形資産をどう計上するかという議論はあまり聞いたことがない．各チームが自分たちの PL（損益計算）上でどのように扱うかを判断していると考えられる．
13. **エンジニアカルチャーのばらつきとインナーソース促進状況**
   エンジニアカルチャーは必ずしも全社的に均一ではない．オープンな文化が比較的浸透している領域もあれば，ビジネス上の理由などで自由度が低く，横のつながりが持ちにくい，あるいはオープンにしにくいチームも存在する．フロアやビルごとに色合いが異なり，スーツ姿が目立つ職員室のような雰囲気のフロアもあれば，よりカジュアルで「オープンソース的」な文化が色濃いフロアもある．
14. **インナーソースに関する明確なライセンスや範囲の定義**
   インナーソース活動を行う際に，その範囲や行為を明文化する特別なライセンスは用意していない．ただし，オープンソースとして社外公開する場合のガイドラインは社内に用意されており，公開前にチェックすべき項目が定められている．

15. **制度 A と対象プロジェクトの傾向（OSS との関連）**

制度 A に申請されるプロジェクトには，オープンソースのものが多く含まれる．個人がオーナーとなっているプロジェクトもあれば，チームでオーナーシップを持つものもある． たとえばプロダクト B は認知度の高い事例である．しかし導入数で見ると，プロダクト B よりも多く利用されている技術資産も存在する．評価基準によってどれが「最も成功しているか」は異なるが，プロダクト B は社内外で比較的知られた存在である．また，個人が開発しているオープンソースの中にも非常に有名なものがあり，それらが制度 A を通じて社内で再評価されることもある． 多くの企業では，会社時間で作ったものは会社の所有物とみなされることが多いが，インターネット業界 C 社では個人が作ったものを会社に取り入れ，それに対して時間やリソースを投資することにさほど抵抗がないように思う．

16. **緩やかな評価指標と成果重視の文化**

インターネット業界 C 社には「この成果さえ出せば他は自由」という緩やかな評価基準がある．マネージャーが「個人が作った OSS を取り入れるべきか」と迷う状況は想像できるが，インターネット業界 C 社の場合，それを受け入れる土壌が比較的整っている．

17. **自由と自己責任，信頼にもとづくカルチャー**

インターネット業界 C 社には名文化されていないが「自由と裁量」という考え方が根付いている．自由を与えられているからには責任を持つべき，という発想である．開発環境やツール選択の自由度は非常に高く，それに伴いオープンソースコミュニティへの発信や学びを得ることも奨励されている．技術的な発信にあまり細かい制約はなく，社員は基本的に信用されている．

18. **発信ポリシーとチェック体制の緩やかさ**

技術的な発信については，特別なチェックプロセスはほとんどない．ロゴやブランド関連に関しては厳密な確認が行われるが，技術的な情報発信はエンジニアが理解しているという前提で比較的緩やかである．セキュリティ情報は当然出せないが，技術的なナレッジや広報につながる情報は自由に発信できる傾向がある．

19. **サイロ化と制度 A 施策の背景**

インターネット業界 C 社には子会社を多数立ち上げる文化があり，それによる「サイロ化」が課題として存在する．制度 A はそのような課題感から生まれ，社内横断的なコラボレーションを促進するための仕組みとして機能している．また，組織全体でより積極的なコラボレーションを求める機運が強い．

20. **リーダー就任後に見えたサイロ問題**

一般的なメンバーとして働いているときは，手を挙げればいろいろなことに挑戦でき，風通しが良いと感じていた．しかし，リーダーとして全社的な影響を狙おうとすると，各事業部や子会社間に目に見えないサイロを感じるようになった．上位階層に行くほど，このようなサイロ状況を認識しやすい．

21. **GitHub Enterprise 導入時の営業的アプローチ**

GitHub Enterprise を導入する際には，社内で資料を用意し，「Enterprise に入ろう」という営業的な活動を行った．自由と裁量がある環境下で，統合的なツール導入を進めるには，エンジニアに対するメリットを明確に示し，社内合意を形成する努力が必要だった．

22. **ナレッジ共有や横軸組織の課題**

横軸組織は，全社的なナレッジ共有やプラクティス浸透を目的としているが，現実には難しい面が多い．あるプラクティスを全社に広めようとしても，各チームに協力者がいなければ定着は難しい．単一

の横軸組織だけで全社横断的な改革を推進するには限界があり，組織間コラボレーションの必要性を日々痛感している．

23. **横軸組織の課題とコラボレーションへの障壁**
私の所属部署は横軸組織であり，日々，全社的なナレッジ共有やプラクティスの普及に取り組んでいる．しかし，各チームに協力者がいないと，どれほど優れたプラクティスでも実践に至らないことが多い．横軸組織のみで全社へ浸透させるには限界があり，強制が難しい状況である．

24. **成功例と失敗例（強制の難しさと草の根的な広がり）**
過去には数多くの成功例・失敗例が存在する．強制的な手段は機能しにくく，裁量権が各部門・ドメインに分散しているため，「一律でこれをやれ」という方針は通りにくい．成功例の特徴としては，草の根的な広がりと雰囲気づくりが重要である．社長や役員が後押しし，「みんながやろうとしている」という空気を醸成することで普及が進む．

25. **GitHub Enterprise 導入と自然発生的な普及パターン**
GitHub Enterprise 導入は，自然発生的な普及の典型例である．最初は部署や子会社単位で個別に導入が進行し，最終的に多数が使うようになってから全社的なエンタープライズ契約へ移行した．当時はセキュリティやアカウント管理の課題感が高まっており，SSO 連携などがタイミングよくマッチしたことも成功要因の一つである．このような「蓋を開けたら，もうみんな使っていた」という状態は極めてレアケースだが，理想的な展開であった．

26. **ツール選定の困難さと失敗例（全社切り替えの難しさ）**
ツール選定に関しては失敗も多い．たとえば G Suite が全社導入されているが，別のツールがよいのではないかという議論も出る．しかし，全社一斉切り替えは難しく，海外メンバーとの連携で Microsoft Teams が必要なケースもあるなど，状況は多様だ．また，パブリッククラウド契約の取り扱いを巡って紛糾することもある．こうした試みは非常に多く，自然淘汰で消えていくものも少なくない．

27. **制度 A と管理職・エンジニアの視点**
制度 A は，すでに存在する技術資産を「評価する」仕組みとして，管理職側にも受け入れやすい面がある．管理職は「全社基盤は負け戦だ」と見る向きもあり，全社的な支援なしに基盤を育てる難しさが歴史的に見られる．その中で制度 A は，作られたものを公正に評価・支援する落とし所であり，エンジニアにとっては自身のプロジェクトを正当に評価してもらう機会となり得る．

28. **制度 A の評価サイクルと支援内容**
制度 A は 1 年更新である．1 年後にステップアップするには再度アセスメントと申請が必要であり，1→3 へ昇格するなど，ランク向上はうれしいインセンティブとなる．金銭的報酬はないが，技術広報サポートやクラウド利用量負担，懇親会費用，カンファレンス参加費用負担など，多面的な支援が提供される．規模に応じて海外カンファレンスも検討可能であり，社内での認知度向上や技術的評価を高める手段として機能している．

29. **情報透明性，発信文化**
インナーソースを拡大する上での障壁は，情報が見えづらいことである．類似プロジェクトや課題を抱えるチームがどこにあるか分からないと，協力が生まれにくい．一方で社外向け発信が社内のコラボレーションにつながることもある．Slack や Times への投稿がきっかけでチーム間コラボが生まれる場合もあるが，全社的な情報集約は難しい．

## インターネット業界 D 社

| 業界 | 企業規模 | 所属 | 役職 |
|---|---|---|---|
| インターネット | 数千人規模 | 開発本部 (グループ中核事業 G) | リードエンジニア |

1. **自己紹介**

   現状，開発本部という横断的な技術組織に所属している．ここは技術部門や必要な技術をまとめる横断組織だ．グループの中では中核事業を行う(以下，グループ中核事業 G)．その中でリードエンジニアというポジションになっている．新しい部署ができて，事業と技術をセットで推進する状況にある．リードエンジニアは上位クラスとして扱われ，設計やレビューが主業務だ．レビューや，導入が難しい技術が必要な際には抜擢されることが多い．

2. **社員数と会社規模**

   エンジニアは CTO 以下みんなフラットにやっている．技術系エンジニアは CTO を通していろんな人とつながっている．下の方では自分たちも他部門と広く仕事をしているので間に入ってつないだりしている．従業員数はライセンスで見ると GitHub のライセンスは 1500 だが，非正規の社員を入れると 5000〜6000 人ぐらい．半分ぐらいは開発に関わっていると思う．

3. **インナーソース導入の歴史的経緯について**

   インナーソース導入の起点は，全員参加でサービスを作ってきた文化にある．インフラエンジニアからゲーム開発者まで，全員が同一のモノレポで開発してきた．会社が大きくなる中で色々と変化はあったが，誰でも見て触れる状況が設立当初より続いている． 最初のイニシアチブは社長が独立して会社を起こした際に始まった．その後 CTO がバージョン管理を導入し，皆でやろうとなった．GitHub もかなり初期から使っている．

4. **オーガニゼーションとリポジトリ構成の変遷**

   インターネット業界 D 社の主たるサービスは基本的には一つの GitHub オーガニゼーションに集約している．予算やチーム区分でリポジトリやオーガニゼーションが分かれたが，組織の再編にともない，また集中化の方向に向かっている．縦割り・サイロ化はあるものの，基本的に全員が閲覧権限を持ち，社内で広く閲覧できる環境を維持している．

5. **会社再編，持株会社，子会社間のコード共有・会計処理**

   予算ごとに分けられた複数の法人があり，GitHub のオーガニゼーションに人を招待する形で構成している．GitHub のオーガニゼーション数は約 10 ほどで，それぞれの法人に対応する．開発本部メンバーは横断組織としてほぼ全オーガニゼーションに属している．会計上，法人間で利益供与の問題があるため，契約や NDA は持株会社側で結び，情報を外に出さないようにしている．コードは誰でも見られる可能性があるため緩く構えつつ，守秘義務を求める形をとっている．ソフトウェア資産は基本的に子会社所有で，持株会社はセキュリティや情報システム管理を担っている．

6. **ソフトウェア資産の所有権・管理体制**

   ソフトウェア資産は子会社ごとに所有する形が主流だ．持株会社はセキュリティや情報システム，契約管理を担うが，ソフトウェア資産は事業部側が主体で持っている．今の所，持株会社に集約する考えはないようだ．

7. **共通ライブラリの提供・利用と会計面の調整**
   共通アセットは主たる GitHub オーガナイゼーションなどでインターナル公開し，他事業部は招待することで利用できる．無償提供が問題になるような開発対象物は，分社化された開発本部側が資産管理を引き取り，必要に応じて利用料を発生させることもある．大きな金額の場合は上層部決裁が入るが，基本的には開発本部でライブラリを集約し，他事業部へ提供，その売上を持株会社へ報告する流れだ．
8. **共通ライブラリ・インフラ管理と資産配分の整理**
   会計監査などは持株会社側が関与する．資産は整理が進行中で，共通部品は主たる法人にある開発本部に集約され，ゲーム側はゲーム事業 H で整理されている．このように共通ライブラリやインフラはグループ中核事業 G が提供し，ゲームはゲーム事業 H といった切り分けが進んでいる．
9. **技術ノウハウの分配と社内権利関係**
   ナレッジは完全なポータル化はされていないが，開発本部に問い合わせれば情報が得られる．経営層も技術相談は開発本部経由で関連部署へ振り分ける．技術的なライブラリやインフラは開発本部が権利を持ち，ゲームのノウハウはゲーム事業 H に属すなど，自然な分業が成立している．
10. **アーキテクチャとサービス間の切り分け・インフラの提供**
    アーキテクチャ的にはサーバーサイド，フロントエンドなどで大まかに分かれる．Kubernetes などを用い，シングルテナントでサービスごとに環境を立てている．インフラのテンプレートや基本構成は共通部門が提供し，各事業部はアプリケーション部分を自由に改変する．その境界を明確にし，プラットフォームはグループ中核事業 G が，サービス部は独自の設定を持つ構造で整理が進んでいる．
11. **プラットフォーム提供会社と事業部の連携**
    以前，ゲーム事業 H が持っていたライブラリを集約し，管理コストを削減した実績がある．インフラや共通ライブラリの提供で事業部間連携を容易にし，費用最適化を図っている．開発本部は横断的な役割を持ち，他部門と組んで仕事をするのが日常だが，事業部は自部門に集中しがちだ．そのため開発本部が間に入って横の連携を支えることが多い．
12. **横断的コラボレーションの促進方法・情報共有ツール**
    コラボレーションは，開発本部に集まる相談や仕事依頼の中で共通項を見いだして発生する．GitHub 管理者として中央に位置することで，他事業部を繋ぎやすい．社内には GitHub サポート用 Slack チャンネルがあり，全員参加可能で他事業部間のやり取りもオープンだ．Slack，Jira，Confluence を使い，共通部門ドキュメントは社内でパブリックに閲覧可能な状態だ．こうした透明性高い情報共有が自然なコラボを誘発している．
13. **社内勉強会・文化面の違い・声かけによる効率化**
    コラボレーション促進の仕組みとして社内勉強会や，年 2 回ほどの大規模交流イベントがある．コーポレート主導で横の繋がりを強化し，外向けにはテックカンファレンスを開催している．事業部側は自分の領域に集中しがちだが，開発本部が声をかけ，横断的な視野で連携を補助する．文化的に，事業部は狭く深く，開発本部は広く浅くをカバーし，声かけで効率化を図っている．
14. **社内で情報を共有できない障壁と背景**
    最大の障壁は権利関係で共有できない情報だ．他社権利を借りている場合やリリース前の情報を含む場合は社内公開が難しい．こうした権利的な問題で公開ができず，オープンになるまで交流でき

ないことがある．それは仕方ないと思っている．リリース後に社内オープンになれば，皆が見に行く．そうしたタイミングを逃さないよう心がけている．

15. **ストリームアラインドチームとプラットフォームチーム，インナーソースの可能性**
ストリームアラインドチーム間でのコラボレーションは難しいが，プラットフォームチームやコンプリケイテッドサブシステムなど，プラットフォームの文脈でインナーソースが成り立つ可能性がある．現在の部署はできたばかりだが，来年からチームトポロジーやプラットフォームエンジニアリングを広める予定だ．事業部との関連よりも，共通パーツやプラットフォーム側とストリームアラインドチームとの連携は活発で，フィードバックも多い．活発すぎて言い合いになることもあるが，昔から「こうした方が良い」と意見し合う文化がある．依存関係があるのでフィードバックは自然発生しやすく，共通部門は他部門のレビューを常に行い，改善提案で活発なやり取りが続いている．

16. **プルリクエストのやり取り，インナーソースの具体例と文化**
プルリクエストは権限付与やフォーク機能の活用で一般的になった．共通部門で認証用リストなどがあり，プルリクエストで更新する流れが定着している．テストを通して責任者がマージし反映させる仕組みだ．かつてはプルリクエストを出すことに抵抗があった時代もあったが，今は誰もが出せる風土だ．効率面でメリットがあると説得してきた結果，若手も積極的にプルリクエストを出すようになった．人が減り，運用はお互い様という意識が強まり，プルリクエストが自然に受け入れられるようになった．

17. **業績や社内環境変化によるコラボレーション文化の変動，GitHub の効果**
他社では業績悪化でコラボレーションが減った例もあるが，ここでは逆に人が減った時こそ「皆で協力しよう」という流れが強まり，プルリクエスト受け入れや共有が当たり前になった．運用はお互い様で，変更は他人が行っても良いという意識が根付いている．GitHub のおかげでコラボレーションコストが下がり，自然に共同作業が促進された．プルリクエストは誰でも出すことができ，手順書もプルリクエストで共有可能になった．評価制度の問題でうまくいかない会社もあるが，こちらは自然に定着している．GitHub による低コストなコラボレーションで考え方が広がった．

18. **エンジニア評価制度，業績目標，評価の自由度**
評価制度は業績目標の達成率を自由記入でまとめる形式だ．技術がわかるチームがプルリクエストやプロジェクトを評価し，検索クエリを貼って見せることも可能だ．360 度評価などはあえて行っておらず，思惑通りに動かそうとすると人材が思惑に縛られるため，自由度を重視している．ゲーム開発者は指示を嫌う風土があり，基本的に自分で考えて数字を上げる文化だ．失敗は自然淘汰され，指示待ちの仕事は存在しない．最終的な評価は相対評価でボーナスが振り分けられる．激しい競争はなく，仕事の取り合いを避ける制度になっている．

19. **表彰制度と評価への影響**
表彰制度は毎月，四半期，年次であり，PM やプロジェクトマネージャーの推薦で CTO や部長が決める．金銭（少額）や，ランチ代，年次であれば比較的多額のインセンティブがある．表彰実績はリードエンジニアとして認められる指標にもなる．ブログ執筆でも年 1 回表彰があり，CTO の個人的な贈呈などユニークな取り組みもある．イベント登壇を評価軸にしたこともあるが，全員が出られるわけでなく質が下がったためやめた．登壇はプラスアルファ程度にとどめている．こうした背景でエンジニアは成果を示し，評価を得ている．

20. **社内文化・エンジニア評価とエンドロール文化**

例えば，新卒の人もいれば中途の人もいる中で，会社の文化的なところはどうやって統一がなされているかというと，ゲーム業界基準なところが結構あるかなと思っている．ゲーム業界基準というのは，普通のソフトウェアエンジニアと少しだけ違う点がある．良いゲームや売れたゲームを作ったという看板は結構聞く．OSS の話よりも，売れたゲームを作った実績を重視するイメージがある．かつ，賞味期限が短い．売れたら次のゲームに行かなきゃいけない．そのあたりの誇り感というか褒美感みたいなところは一応あって，エンドロールに名前を載せるという文化もある．エンディングのエンドロールに名前を入れてもらうのは他社に転職するときや新しいプロデューサーになるときにも履歴や証拠として価値が高く評価されているかもしれない．共通部門でも，手伝ったら「ありがとう」と言われてエンドロールに載せてくれたりもする．それはお金じゃない評価の仕組みだ．

21. **表彰制度とノミネーションプロセスの変化**

社内で何かを作りましたという話よりも，自分たちで作った結果が出たという話の方が意識が高い．文化の違いとして一番大きく反映されている部分だと思う．これがプルリクエストしたり，受け入れたり，拒否したりするところにも採用されていると思う．社員としてのモチベーションは，自分が貢献している，自分がやったという満足度が得られる部分が大きい．それをまた表彰されたり，お互いに褒め合ったりする文化がある．褒め合いの方が多い．そうした文化が回っているので良いところだと思う．このミクロな視点で自分がやったという満足度は高いが，年度評価みたいなものがある場合，なんでもかんでも「自分が自分が」，という風にはならないかというと，「自分が自分が」と言わない人ほど積み重ねているし，評価は公平に行われている．ちゃんと表彰される人は上に来るし，そのバランスは取れていると思う．表彰にはノミネーションがあり，PM みたいな人たちにも権利がある．本当にいいと思うものがノミネーションされる．作為的な表彰ではない．

22. **インテグリティとホスピタリティを重視する文化**

インターネット業界 D 社に来て思ったのは，みんなが本当に褒め合うためにノミネーションしていることで，バリエーションも豊かで作為的なものがない．これがうまくいっているところだ．悪意的なものはほとんどない．CTO が掲げていた標語に誠実さや優しさなどを大事にしている．そういった思想に共感している人が仕事しているので，悪意的なものは軽減されていると思う．

23. **採用方針・長期的な視点での人材選考**

採用では，やはり話が合う人を優先して取っている感じがする．バブルだった頃は人手不足であらゆるスキルの人を取っていたと思うが，今は落ち着いて長く粘り強く仕事する必要がある．だからコミュニケーションが取りやすく，メンバーとまとめて意思決定できそうな人材が重視されている感じだ．粘り強いというのは，プラットフォーム的な横断系の仕事が多く，急に変われない．オンプレからの移行には 5 年間の償却期間があるなど，急に新しいものに置き換わることはない．大きな評価が出るのが 3 年後かもしれない．粘り強い人の方が向いている．ゲームも一発勝負といえど 3 年くらいかかる．3 ヶ月や 1 年でできるわけじゃない．ギャンブル感があるが，それを粘り強く 3 年我慢できるか，忍耐力やコミュニケーション，人間関係が大事だ．今は長期的にやれる人が評価を得やすい．法人や事業体ごとでも共通していると思う．

24. **オープンソース公開の目的と事例**

オープンソースプログラムオフィスみたいなものは一応ある．ライセンス確認や法務とつなげてくれたりする人たちがいる．その人たちはコーポレート側の部長さんで，テックカンファレンスを主催したり，人事系のこともしている．ライセンス確認は，顧客や自分たちに迷惑がかからないかを見る程度．みんながいいと思うものは出す．リーガルの人か委託弁護士事務所がチェックしてくれる．コーポレート部長やリーガルを巻き込んで問題なければ出す．オープンソースに出したものは正直今あまり残っていないが，バブルの頃には CTO 自身が書いて出していた流れがある．オープンソースとして出すものは，他社や共同開発しているものや，ゲームエンジンのライブラリを他社にも使ってほしい，業界的に意味があるもの，儲かっているわけじゃないものは公開する傾向がある．オープンソースに出すまでもないが社内で共有したいものは，普通にインターナルで公開して，プルリクエストで編集の負荷を減らしたりしている．共通の申請系などをインターナル化してプルリクエストを送り合っている．

25. **インナーソースにおけるツール・共通基盤の開発例**

インナーソースで言うと，共通基盤や共通ライブラリ，新しいプロジェクトやハッカソンプロジェクト的な小さなものなどがある．あまり変化や壊れることを恐れていない．「気に入らないところがあっても受け入れて前に進もう」という文化があり，合言葉になっている言葉もある． ある程度テストしたらもう出すしかないときに，「気に入らないところがあっても受け入れて前に進もう」というノリでデプロイする文化だ．確認のしようがないところはやってみて壊れたらもう一度デプロイし直せばいいという考え方がある．壊れても「ごめんごめん」で終わる．それを可能にするためにテストを頑張っているので，一回の気持ちが楽になっている．他人の社内のものへの変更も壊れても大したリスクと考えていない．壊れたら謝ればいいという感じだ．

26. **ツールの維持・運用負担と組織サポート**

車輪の再発明はあまり起きない．カジュアルに README.md を書く．圧をかけず，使うなとか言わず，貢献者向けに「こうしてくれるとうれしい」と書く．車内ツールを作るモチベーションは，自分たちでやりたいことを邪魔されたくないからだ．社内検索のやつも，ベンダーに任せたら全然使えなかったので頭にきて CTO が書き始めたりしている．勤怠管理システムも，既存のツールが使いにくいのでフロントエンドを自分たちで書いて，ボタン一発でできるようにしたりしている．良くなったら社内で広めたり，情報システム部が用意する SharePoint のポータルで告知したりしている．反応は当たり前のように受け入れられている．「これだよ欲しかったのは」という感じで表彰までいく場合もある．

27. **マネージャーとの合意形成・プロジェクトとの優先度バランス**

メンテナンスは予算の問題があるが，部門がなくなったり予算がなくなったら CTO 以下の組織が引き取る．使っている人が多ければコストをかける意味があるからだ．CTO は予算枠を持っていて，そこからサーバー代などを出したりして支援する．人事の採用システムなどは人事が頼むと，開発本部が費用を持ったりしている．優先度の話では，本業とのバランスを考える必要がある．だが，本当に会社に効果があるなら他のリソースを呼んできたり，間に入って調整したりする．エンジニアは問題提起して，自分が PM 的に動くこともある．例えばインフラでツールの特定の機能を廃止して新規機能開発するとき，1 エンジニアが「こういう風にしたい」と言い，みんなで検討して実行する．目標があり，やるべきことがあるが，それと同じくらいの比率で自分たちのやりたい仕事もある．折り合いはつきやす

い．マネージャーがどうしてもこれを先にやってほしいと言えば，プロとしてそれを片付け，その上で自分たちの仕事もやる．

28. **上下関係を繋ぐ役割・組織内コミュニケーションの支援**

コミュニケーション苦手なエンジニアは自分のところに来て「助けてくれ」と言うので，自分が上層部に説明して回る．資料を書いても読んでもらえないこともあるから，自分が間に入って説明すると，他がやりやすくなる．若者たちが考えた新しいデプロイ方式を上層部に理解してもらうために，自分が間に入って説明するなど，上下をつなぐ役割をしている．

29. **個人でのオープンソース活動と会社内での扱い**

個人が開発しているものや，個人で自分の時間を使ってオープンソースにコミットしているようなものの扱いは，基本的に個人で好きにやれという話だ．だが，それが会社のタスクや仕事に還元される機会も時々ある．自分自身も実は会社としてやっていない OSS 活動がある．公開していないが，昔から続けている匿名のものだ．前職では本名や所属を出すことが禁止されていたため，当時から匿名で活動していた．一方，会社内でオープンソースや外部組織 A の RFC を書くような活動に関しては，若手などが自分の名前でやっていることもある．その場合，業務的に世界的なカンファレンスへ参加したり，時差で夜中に活動したりといったことに会社も対応し，出張やサポートを行う．ただ，それらは基本的に個人の話として切り出されることが多い．仕事きっかけで OSS 活動がフィードバックとして返ってきたらラッキーだが，毎回そううまくいくわけではない．全力で一つの OSS を盛り立てようという流れに社内全体で深く関わるようなプロダクトは今のところない．

30. **レポジトリ管理・外部コラボレーション・社内オープン化の仕組み**

インナーソース的な活動はインナーソースという言葉が出る前から社内に存在していた．新しく入社した人たちは当初ギャップがあるかもしれないが，周囲が当たり前にオープンに接していれば自然と慣れる．空気が大事で，当たり前を作ってしまえば，すぐ順応していく． 社内では当たり前のようにオープンな文化があるため，業務委託の人にも一度やり方を見せれば，次からは普通にやってくれる．リポジトリは契約内容にもよるが，基本的に社員と同等に見られる．セキュリティ的に禁止するコミット以外は全てオープンだ．外部コラボレーターを使わず，オーガニゼーション内に入ってもらう形をとる．アウトサイドコラボレーターは管理が難しいため完全に切り捨て，その代わり中に入った人は自由にできる仕組みだ．GitHub Cloud 運用については何年かに 1 回節目で話し合い，「みんなで続けるにはどうするか」を軸に設定を決めている．誰も分断したいとは思っていないので，全員が受け入れられる方向で調整する．秘匿情報の管理が重要であり，すべてがオープンだからこそ秘密はきちんと隠す文化がある．

31. **海外とのコラボレーション，IP やリリース前コードの取り扱い**

インターナルリポジトリも使っている．インターナルな場合，グループ企業全員が見られる前提で問題ないものを選択している．海外メンバーもいる． UI や翻訳で海外の人が入ることもある．輸出管理は，昔は各拠点で資産を分けていたが，今は本社側に集約して外に出す流れなので細かくはなっていない．制裁国への対応や，IP 関連の問題がある場合は，リビジョン単位でフォークしたリポジトリを渡すことが多い．本家リポジトリをそのままコラボせず，権利がクリアになった段階までのバージョンを切り出して提供する．個人情報を伴わず，権利や秘密保持をクリアできれば問題ない．リリース前のものや秘密情報は適切に隠し，インフラもブランクな状態で渡すことが多い．

32. **インフラ，ドキュメント管理，社内ツール利用状況**

インナーソースという言葉は社内に広まりつつあり，自分が言いまくったことで若手や子会社の新卒も気にしている．オンボーディング時には新卒研修でテクニカルライティングやドキュメント整理の講座を行う．中途入社は各チームで簡易なツール資料を渡し，それを自分たちで補足しつつ運用している．

## インターネット業界 E 社

| 業界 | 企業規模 | 所属 | 役職 |
|---|---|---|---|
| インターネット | 数千人規模 | 経理部門 | 担当 |

1. **自己紹介と当時の業務内容**

   当時はインターネット業界 E 社で経理として採用されていた．もともと複数のスタートアップで経理の業務経験があり，インターネット業界 E 社に入社した際も経理担当として携わっていた．インターネット業界 E 社では多くの取引が発生する．その結果，それらを会計処理するために，サービスのデータベースから必要な情報を SQL で取得・加工し，経理向けにレポーティングするシステムが求められていた．自分はそのシステム仕様を確認し，クエリのレビューを行うなど，経理業務が円滑に進むための裏方的な役割を果たしていた．

2. **監査対応**

   インターネット業界 E 社では監査人からの要求に応じて正確な会計データを提示することが求められていた．サービスから抽出されるトランザクションデータを正しい会計基準で処理し，監査の問い合わせに素早く対応する必要があった．自分たちは，必要なデータを速やかに提供できるレポーティングフローを整えて社内体制を強化していった．

3. **マイクロサービス化の背景**

   組織が拡大すると，開発組織全体で把握すべき範囲が広がり，一人のエンジニアがカバーしなければならないコンポーネント数が増えていった．新規エンジニアが開発に参加しようとすると，改善したい機能がほかの機能やデータと複雑に絡み合っていて，理解が困難になる問題があった．

   ここでマイクロサービス化が進められた．インターネット業界 E 社本体のサービスはもちろん，新たに始まるプロダクト A の機能群をサービス単位で切り出し，それぞれのチームが独立して技術選定やインフラ整備を行えるようにした．これにより，ある特定のサービスの改善を行う際に，広範囲にわたる既存コードとの依存関係を気にしすぎずに開発できるようになることが期待された．

4. **マイクロサービス化による会計システム再構築**

   マイクロサービス化によって，会計のレポーティングシステムも抜本的な変更が求められた．従来はサービス全体から直接データを引き出していたが，今後は各サービスが会計に必要なイベント（売上や支払いなど）を Pub/Sub 経由で送信し，会計側がそのイベントを受け取って集計・加工する流れに変えた．これにより，会計用のレポーティングサービスは独立したマイクロサービスとして機能し，他の開発チームがサービス単位で自由に動きながらも，会計が必要とする情報を確実に取得できる仕組みを整えた．

5. **複数エンティティの存在と展開**

   インターネット業界 E 社グループには，日本以外にも海外に複数の子会社が存在した．また，国内で新規事業立ち上げを目的とした子会社や M&A で取得した企業を含めると，合計で 5 社から 10 社前後のエンティティが存在した時期もあった．US やヨーロッパへの展開を狙う中で，規制や会計処理が国ごとに異なり，それぞれ別のレポーティングや監査対応が必要だった．

6. **会計処理プロセスとローカライゼーションの必要性**

   基本的には日本も US も会計として同じような業務を行っているが，微妙な違いがある．会計は大きく三つのステップで考えられる．

一つ目は「ビジネス上のイベント」が発生した際，それをどう財政状態として捉えるかという基本設計だ．ここにはイベントに勘定科目をマッピングして，ビジネスイベントが財務諸表上どこに反映されるかを決める．二つ目は，そのビジネスイベントに対して会計的な色づけを行う段階だ．ここでビジネスイベントが売上なのか，費用なのか，どの科目として計上すべきかを決める．三つ目は，最終的にそれらを集計し，申告書や株主報告，経営判断資料，支払い関連資料として使用する段階だ．これらの基本フローは同じでも，最後の集計・報告段階でローカライゼーションが必要になる．たとえば，US では特定の会計処理ルールが求められるため，現地要件に沿ったレポートや監査対応が求められる．米国で「こういう会計処理をしているから，こういうレポートをくれ」とか「監査対応上必要だからこのデータを提示してくれ」といった要求が発生する．こうしたローカライズされた要望に応えるため，結局は日本向けと US 向けでコードや処理手続きが分かれ，独自の仕組みが必要となっていった．

7. **コードベース分岐と日本・US の独立対応**

もともとは日本と US で同じコードベースを使っていたが，事業拡大とマイクロサービス化に伴う要件の分岐，国別の会計・監査要件の相違によって，最終的には日本側と US 側でコードベースが分かれていった．US は US 向けのローカライズ対応が増え，日本は日本で独立したマイクロサービスを整備し，米国側は独自の担当者が別の流れで対応する状況になった．本当は US 向けにも同じ基盤を活用し，容易に継承・統合できるような仕組みが望ましかったが，そうした基盤を構築する前に組織変更や事業拡大が進んでしまった．

8. **日本・US 間のコードベース分岐とレポーティングシステムの流用**

もともと日本で巨大なモノリシックなサービスが存在した段階があった．それは日本と US が同じコードベースを共有し，同じ仕組みを使っていた時期のものだった．日本側はやがて別事業の開始でマイクロサービス化を進め，クライアント，サーバー，インフラといった構成要素を分離していった．しかし，US 側は日本から枝分かれした当初のコードベースをそのまま使い続ける状況になった．元々，会計向けのレポーティングは，巨大なモノリスから SQL でデータを集計して加工し，経理・監査対応のためにレポートを出す仕組みだった．日本ではサービス分割後にその旧来の仕組みが使えなくなったが，そのまま US に持っていき調整して進化させる形がとられた．結果として，日本がマイクロサービス化で進んだ後，US 側は旧来の日本ソースを基にしたレポーティング手法を継続し，独自にそれを発展させる構図になっていた．

9. **ソフトウェア資産計上の会計・税務上の扱いと面倒さ**

（一般論として）会計上，ソフトウェア開発費を資産として計上することは可能だが，多くの IT ベンチャーは会計上は費用処理することが多い．資産計上すると 5 年間で均等焼却し，最終的には 5 年かけて費用化されるのでトータルの費用額は変わらないが，毎期ごとに減損の判定や価値の陳腐化がないかをチェックする必要が出てくる．これが大変で，資産計上して将来毎期大変な作業をするよりは，はじめから費用処理してしまう方が手軽だった．一方，税務上は会計と定義が異なり，税務上のルールに従うと，ソフトウェア開発費を資産として計上しなければならないケースがある．会計上は費用処理しているが，税務上は資産として扱うことになると，国ごとの費用配分や資産計上が発生する．インターネット業界 E 社の場合，日本と US のエンジニアがどれくらいの工数を使ったかをマネージャー経由で集計し，税務チームへ報告した．税務チームは報告をもとに，この部分は US 資産，この部分は日本資産，といったように付け替えを行い，税務申告上の損金計上や資産計上を調整していた．

10. **製造業との文化的差異とソフトウェア開発費の扱い**
    製造業では開発した製品が将来一定の収益を生むことが見込みやすいため，資産計上しやすい文化がある．ソフトウェアでも製造業的なモデルなら比較的容易に資産計上できる．しかし IT ベンチャーは，エンジニアが書いたコードがどれだけ将来収益に寄与するか合理的に見積もりにくい．製造業は作ったものを売る構図が明確で，工数計算を行えば将来収益との対応も立てやすいが，IT ベンチャーでは不確実性が高いため，資産計上が困難であり，費用処理する慣行が広まりやすい．
11. **会計上資産計上しない背景と税務上の影響**
    会計上資産計上すると，例えば 100 万円をソフトウェア開発に使った場合，そのうち 20 万円だけ資産化したいといった柔軟な対応は難しくなる．会計上資産計上すれば，税務上も資産計上せざるを得ず，結果的に損金として落とせるタイミングが遅れる．法人税を考えると，はじめから費用処理してしまったほうが損金計上が早く，税負担を抑えられる可能性がある．IT ベンチャーは先行投資が多く，将来収益を合理的に見積もることが難しいため，会計上資産計上することで生まれる減損判定や監査対応の手間を考えると，費用処理してしまうほうが合理的だった．周りの IT 企業も同様で，資産計上を避けている会社が多い状況があった．
12. **資産計上による BS 拡大への市場の見方と米国 IT 企業の実例**
    （一般論として）ソフトウェア開発者の人件費を資産計上すれば，BS（バランスシート）の資産が膨らみ，PL（損益計算書）に比して資産が大きく，十分に活用できていないように市場から見られる可能性がある．これは IT ベンチャー特有の「あるある」で，巨大な無形資産が BS に載っていても，収益化や実体的な裏付けが見えづらいため，外部からは資産を有効活用していないと思われがちだ．インターネット業界 E 社はソフトウェア市場を運営し，コード自体が競争優位性となる．エンジニアが多く，資産計上すれば BS に大きくその価値が表れるが，それを避ける傾向があった．一般的に US の企業も会計上は資産計上を行わないところが多く，人件費が発生しているはずなのに BS に資産が載っていない．税効果会計という会計上資産と税務上資産を比較する会計基準に照らせば，会計上は計上せず，税務上は資産計上している企業が少なくない．大手 IT 企業もソフトウェア開発費用を会計上資産化しないことが多く，IT ベンチャーではこれが一般的な傾向になっている．
13. **インターネット業界 E 社内での一部資産計上例と移転価格税制**
    インターネット業界 E 社内でも過去に特定の機能について資産計上したことがあったが，基本的には資産計上を避けてきた．特に移転価格税制への対応では，日本と US で発生した工数を正確に集計し，税務チームが「これは US の費用」「これは US の資産」「これは日本の資産」というように付け替え，最終的な税務申告での損金・資産計上を調整した．会計上は費用処理しても，税務上は資産計上しなければならないケースがあるため，このような工数集計と付け替え作業は欠かせなかった．一般的に，複数国で事業展開している企業に対する法人税の税務調査で移転価格は注視されがちという背景もあると思う．
14. **移転価格税制とソフトウェア資産計上の基準**
    移転価格税制への対応として，日本と US でどれだけ工数を使ったかを集計し，収益に直接つながらない機能はコストとして，一方で「インターネット業界 E 社らしさ」を生み出すコア機能を資産として計上するなどの切り分けを行っていた．具体的にはログインなどの収益に直結しにくい機能は費用として処理し，一方でユーザーが実際に売買を行うロジック部分の機能に関しては資産計上することも

ありえた. バグフィックスや軽微な追加開発などは, 重いものを積み上げるほどの価値がないと判断し, 費用扱いになっていたことがあった.

15. **工数トラッキング方法と組織構造**
    エンジニアがどれだけの工数を費やしたかを正確に記録するのは難しかった. マネージャーが肌感覚で把握したり, チケットや 1on1 などを通じてエフォートを積算することもあったが, 必ずしも客観的な記録が残るわけではなかった. マトリクス型の組織だったため, エンジニアが同時に三つ, 4 つのプロジェクトに関わることもあり, 複雑な状況になっていた. 少数精鋭で広いカバレッジを求められるケースもあり, マネージャーが Slack やレビューの動きを見て工数を把握するなど, 非定量的なやり方が多かった.
16. **リポジトリ構成とプロジェクト単位の管理**
    マイクロサービス化前と後でリポジトリの構成は変化し, マイクロサービス化前は巨大なモノリスから SQL で集計・レポートを行っていたが, マイクロサービス化後はサービスごとにリポジトリが分割された. インフラ設定をまとめた共通リポジトリ(たとえば Terraform の設定を集約する場所)もあり, 各サービスチームとプラットフォームチームが共同で管理する場合があった. プロジェクトごとにリポジトリがあり, サービスを動かすためのインフラは共通リポジトリを使う形で, 個別に分割される場合もあれば, 共通化される部分もあった.
17. **マイクロサービス化によるフリクションとチーム間コラボレーション**
    マイクロサービス化の過程で, モノレポによる隣接チームへのコード修正がカジュアルだった環境が変わった. 複数のリポジトリができ, 明確な境界が生まれることで, コード修正を他チームに送る際に一定の障壁が生じることがあった. 新しく作られるサービスは最初からマイクロサービス化する方針で始まり, 旧来から運用されていたサービスはどのように区切ってマイクロサービス化するかが議論の的になった. アーキテクチャ設計時にはいろいろな「正義」が主張され, 一方で推進者が辞めてしまうなど, ゼロから再スタートを繰り返すような状況もあった.
18. **組織間・会社間のコラボレーションと文化的な障壁の軽減**
    プロダクト A などは別会社であっても, そこにいる人々は元々インターネット業界 E 社で働いていたメンバーが多く, 知り合い同士が柔軟にコラボレーションしていたため, 会社をまたぐことで大きな壁を感じることは少なかった. 日本と US 間では文化的・技術的分岐があったが, プロダクト A に関しては内部的な人間関係が密接であり, 知っている人同士がコードを共有したり, 共通のインフラ基盤やリポジトリを使ったりすることで, 比較的スムーズに連携できていた.
19. **日米採用人材間の価値観の違いとキャリア志向**
    インターネット業界 E 社では, 日米で採用した人材の間に価値観の違いがあったように思われる. 私自身は, 問題解決が面白そうな領域を探し, そこへ飛び込むことで結果的にキャリアが形成されるという考え方を持っていた. 会社の成長とともにメンバーの考え方も多様化し, メンバー同士で考え方について多少のギャップが生じたように思われる. 中には社内ネットワークを広げたり, 他者の取り組みを理解しようとする行動が少ないメンバーもおり, そうした人々は自らの仕事領域に閉じこもり, 他者や他チームとのコラボレーションを積極的に行わない傾向があった. ミッションへの共感についても, 人によって差が生じた.

20. **人事評価とバリューに基づく行動指標**

人事評価は会社が掲げるミッションやバリューに沿った行動をどれだけ実行したかで判断される仕組みがあった．個人は定期的に自己評価を提出し，360 度評価や周囲からのフィードバックを踏まえた上で，マネージャー間でカリブレーション（調整）が行われていた．コーポレート部門でも他部署のマネージャーが集まり，評価基準を横に揃え，同様の行動に対して評価のばらつきが生じないようにしていた．

21. **表彰制度とバリュー軸での MVP 選定**

バリューに紐づけた表彰や，四半期ごとの MVP 選定も行われていた．マネージャーたちが候補者を推薦し，経営陣が最終決定を下す流れがあった．こうした仕組みは，他部署との連携やチームワークを評価する仕掛けになっていた．他者との関わりがなければ評価が得られにくく，結果として自分勝手な行動やミッションへの共感が低い人は評価面で不利になっていた．

22. **プロダクト A とインターネット業界 E 社間の GitHub オーガニゼーション運用とリポジトリ構成**

プロダクト A とインターネット業界 E 社間で GitHub のオーガニゼーションは分けずに一緒に利用していた．結果として，オーガニゼーション内には多くのリポジトリが並んでいる状態だった．プルリクエストは，理論上どのリポジトリへも送ることができる環境だったが，実際に他のチームへ頻繁にプルリクエストを投げるケースはあまり見なかった．コメントを付ける行為は比較的行われていたが，基本的には自分が関わるプロジェクト内での開発が中心だった．機能要求や改善提案に関しては，GitHub 上のイシューを使うよりも，Slack で「これが欲しい」と伝えるなど，よりカジュアルなやり方が多かった．これについては，E 社ではプロダクト開発者だけでなくカスタマーサービスなど GitHub アカウントを持たない・使った経験がない職の方々からも機能改善や要望の提案がされていたことが背景としてあるかもしれない．明確なワークフローやガイドラインが厳密に整備されていたわけではなく，カジュアルなコミュニケーションを通じて開発を進めていた．こうした状況は，インナーソース的な取り組みを行う際にも，必ずしも正式なプロセスやワークフローに頼るのではなく，ボトムアップで自然発生的に問題解決や改善が行われる文化を反映していた．

23. **エンティティの違いとアクセス制御，利益共有の問題**

インターネット業界 E 社，プロダクト A の別会社といったエンティティの違いは存在していたが，それによって GitHub 上のアクセス制御に大きな制約があったわけではなかった．社員や業務委託など，実際に仕事を一緒に行う人物であれば，基本的にリポジトリへのアクセスは可能だった．利益供与やエンティティ差異による複雑なルールは，自分の知る限りでは特に存在せず，あまり問題として浮上していなかった．輸出管理や制裁対象国に関する厳密なチェックも，自分が知る範囲では行われていなかった．海外出身のエンジニアが在籍することもあったが，特定の国籍や所在によるアクセス制限が厳しく設けられていた記憶はない．入社時に GitHub アカウントを提示し，招待を受けるというシンプルな手続きが基本であった．

24. **インナーソース的な文化とボトムアップの改善**

インターネット業界 E 社内部には，いわゆる「インナーソース」を意識した取り組みがあった．GitHub を用いることで，OSS の文化を内製開発にも取り入れ，ボトムアップで自然な形で最適化が進むことが期待されていた．イシューやプルリクエストでフォーマルに手続きを踏むよりは，Slack で「これ欲しい」と依頼し，必要なものから解決されていくような流れが見られた．「インナーソース」という名前自体

は（私の知る範囲では）使われていませんでしたが，コードや情報を可能な限り広く共有して協力し合おうという空気は全社的にあった．給与や人事・個人情報以外は原則的にオープンだった．もちろん，正式なガイドラインやプロセスが確立すれば，さらに効率的になる可能性はあっただろうが，自分が把握している範囲では，そのような整備はそれほど進んでいなかった．それでも，GitHub の存在が，エンジニア同士がソースコード上で直接やりとりし，コラボレーションしやすい環境を整えていたのは確かだった．

25. **オーガニゼーション共有によるコードアクセスと情報公開**
インターネット業界 E 社とプロダクト A の別会社は GitHub のオーガニゼーションを分けず，同一オーガニゼーション下に全てのリポジトリを集約していた．これにより，エンジニアでなくとも申請次第で全リポジトリを閲覧できる環境が整っていた．会計システムが発行するクエリの整合性を確認する際には，当該システムのリポジトリのみならず，関連するインターネット業界 E 社本体のソースコードに直接アクセスし，内部構造や関連機能を俯瞰的に理解できた．こうしたオープンな環境は，やる気がある人材が自発的に他チームや他ドメインのコードを読んだり，コラボレーションを行うことを容易にした．コードへのオープンアクセスは，知的好奇心を持つ人材にとって一種の福利厚生のようなものであり，優秀なエンジニアがどのようなコードを書いているかを直接学ぶ機会となった．組織としても，オーガニゼーションに所属している社員であれば基本的にリポジトリへのアクセス権が与えられ，すべてを見放題という状態だった．ただし，ライセンスコストや合理性を考慮し，必要性がない職種へのアクセスは制限するなど，一定のバランスを保つ運用も存在した．

26. **情報公開とドキュメンテーション，プロセスの型化**
コードアクセスが自由なだけでなく，設計ドキュメントやデザインレビューが整備されていれば，他のサービスやプロジェクトの全体像を把握しやすくなり，コラボレーションが自然に生まれる下地となる．デザインドキュメントの公開やレビュー体制の存在は，必要なときに必要な情報へ自由にアクセスし，他チームのコード構造や API 仕様を理解し，コラボレーションを容易にする．インナーソース的な取り組みはこうした情報共有とセットになることで，ボトムアップで自然な最適化が進み，オープンソース文化を社内に取り込む際のベネフィットを最大限に活用できる．

27. **好奇心・ミッション志向と文化的背景**
このようなオープンな情報環境は，ミッションやバリューへの共感が強く，好奇心旺盛な人材を引き付けやすかった．評価制度や表彰制度，コラボレーション促進の仕組みと合わさることで，社員同士が自発的に連携し，ビジネスの成果を底上げする好循環が生まれる．逆に，何らかの歯車が狂えば，文化や仕組み全体にほころびが生じる可能性があったが，少なくともこのオープンなコードアクセスと情報共有の文化は，インターネット業界 E 社の成長初期において効果的に機能していた．

28. **オープンアクセスとコラボレーション志向**
インターネット業界 E 社やプロダクト A では，GitHub のオーガニゼーションを分けず，全て同一のオーガニゼーション下でリポジトリを管理していた．このオープンな環境は，特定の理由があればエンジニアでなくともリポジトリ閲覧が許可され，コードや関連資料に自由にアクセスできる状態を生んでいた．会計システムのクエリを確認したい場合，当該システムのリポジトリのみならず，本体のソースコードや他の関連するコンポーネントにも直接アクセス可能だった．こうしたオープンな状態は，やる気と好奇心を持つ人材が自発的に他チームや他のサービス領域へと踏み込み，コラボレーションを発生

させる土台となっていた．単なるコスト削減や重複作業の防止といったトップダウンの目的というよりは，自然発生的な協働や知識共有を促す文化が背景にあった．複数のプロダクトを持たないシングルプロダクト企業であったこともあり，このようなオープンアクセスが当たり前とされていた．

29. **チームトポロジーとプラットフォームエンジニアリング**

 質問者が示す「チームトポロジー」に関しては，プラットフォームエンジニアリングやストリームアラインドチームなど，役割や責務に応じたチーム編成の考え方がある．プラットフォームチームや多くの依存関係を抱えるチームほど，オーガニゼーション内のコードや基盤を共有し，コラボレーションを行う必然性が高まる．インターネット業界 E 社内でも，その点に同意する考え方があった．コラボレーションを前提とするチーム編成や文化が，コード共有やインナーソース的な活動を自然に引き出していた．

30. **OSS プログラムオフィス（OSPO）とオープンソース活動**

 インターネット業界 E 社内部には，オープンソースプログラムオフィス（OSPO）に相当するような組織や活動を行う人々がいた．著名なエンジニアが自分の作ったツールを公開していたり，オープンソースを活用した事例が社内で共有されていた．自分が直接的な関わりを持っていたわけではないが，2020 年頃にはそうした活動が整理され，正式な形で推進されていた話を聞いたことがある．オープンソース的な文化やプラクティスの取り込みは，インナーソース文化にも通じ，社内で自然な知見共有と改善が行われる下地となっていた．

31. **情報アクセス性とツールの乱立**

 GitHub 上のコードアクセスは容易だったが，ドキュメントやナレッジ共有のためのツールはプロジェクトやエンティティごとに異なっており，必ずしも統一されていなかった．インターネット業界 E 社本体は Markdown ベースの Wiki ツールを使っていたが，プロダクト A は Confluence，US チームも Confluence を利用するなど，情報が分散しやすい環境だった．情報自体が意図的に遮断されている感覚はなかったが，どの情報がどこにあるか把握するのが難しくなりがちだった．Slack には無数のチャンネルが存在し，会議記録や仕様書は Google Docs やサイト，Wiki，Confluence など様々なツールに点在した．あるチームの定例ミーティング資料が Google Docs にあっても，他のチームは別のツールを使っている場合があり，情報探索には手間がかかった．新 CTO の着任時に Markdown ベースの Wiki を導入するなど，情報統合の試みもあったが，次第に Confluence など別のツールが増え，再び分散が起きるなど，前社共通の情報共有基盤の確立は難航した．

32. **情報共有問題への取り組みと課題**

 アクセスできない情報は少なかったが，「この情報はどこにあるのか」「どのツールにまとまっているのか」といった問題が発生しやすかった．Slack 上でのやりとりや，プロジェクトごとに異なるドキュメントツール，更新が止まった Wiki など，情報資産は常に変動し続けていた．一度は前社共通ツールを目指すイニシアチブが立ち上がっても，別のチームやエンティティが異なるツールを使い始めるなど，統一化が定着しづらい状況があった．こうしたツール乱立と情報分散の課題は，巨大な組織が多様な開発プロセスを内包していることの表れだった．最終的に，誰もがアクセスできるオープンな環境と，分散した情報基盤が混在した結果，能動的に情報を探索する人材にとっては行動の幅が広がり，一方で受動的な情報取得が難しくなる側面も生んでいた．

## 情報通信業 F 社

| 業界 | 企業規模 | 所属 | 役職 |
|---|---|---|---|
| 情報通信 | 500 人規模 | 開発部門 | エンジニアリングマネージャー / エンジニア |

1. **自己紹介と所属組織**

   人物 A は情報通信業 F 社 の役職はエンジニアリングマネージャーだ. 人物 B は組織 A のチーム A に所属しているエンジニアである.

2. **インナーソースへの着目時期と背景**

   情報通信業 F 社内でインナーソースの取り組みが話題になったのは 2022 年 10 月ころだ. 公開されているインナーソースの資料にもとづいて社内で説明を行うことでインナーソースという概念を広めようと考えた. 2022 年にインナーソースという言葉と出会い, 推進者を決め, まずは LT(ライトニングトーク)によって言葉を社内に浸透させることを行った. 2023 年にはお試し実験を経て, インナーソースを具体的に進めるようになった. 私(人物 A)としては, 1 人の推進者を決めてお試し実験をさせる形をとり, 担当者(人物 B)をアサインして取り組ませた.

3. **お試し実験の内容**

   2023 年に行ったお試し実験は,「簡単なものでいい」という方針で進めた. その実験として, 人物 B が個人で持っていたリポジトリを活用した. このリポジトリには, 当時オンライン会議中に発表者がコメントや Web 会議ツールでのチャットに気づきにくい問題を解消するための簡易なツールがあった. このツールは Web 画面上に一つフォームがあるだけのシンプルな構成で, 入力されたテキストを音声合成し, 機械音声で流す仕組みだった. これによって, 発表中でもコメントを音声で割り込ませることができ, 発表者が気づきやすくなる効果があった.

4. **社内展開と推進体制の整備**

   2023 年 5 月ころ, このお試し実験の結果を示して社内展開を始めた. 情報通信業 F 社内にはエンジニアの部署が三つあり, それぞれ約 50 人, 合計で 150 人ほどのエンジニアがいる. その中でインナーソース推進グループを立ち上げ, 私含め 3〜4 人のコアメンバーで進める体制を作った. このグループは 3 部署のうち 2 部署をまたぎ, 他のエンジニアリングマネージャーにも呼びかけることでインナーソース推進を事業計画に組み込み, 関係者を巻き込んでいった. 社内のエンジニアリングマネージャーどうしは横でつながっているため,「インナーソースをやっていい」という共通認識を作ることが容易だった. 計画や会社目標にインナーソース推進を組み込むことで, 現場のコミット感が増し,「やってみよう」という雰囲気が醸成された.

5. **コントリビューションの評価・計上方法とガイドライン整備**

   2023 年 6 月ごろ, コントリビューション数の計上や評価軸の明確化を行った. 情報通信業 F 社には工数を計上するための Web アプリがあり, エンジニアは何時間, 何を行ったかを入力する. そこで, 他チームのプロダクトにコントリビュートした場合, そのプロダクトとして計上する運用ルールをエンジニアリングマネージャー間で合意した. こうすることで, 他部門への貢献が管理会計上見える化され, 正当に評価できるようになった. また, 社内ガイドラインを整備した. ガイドラインには, インナーソースのメリット, 参加方法, 用語解説, 社内インナーソースポータルやコミュニケーションチャンネルの案内をまとめた. コントリビューターとして参加する場合や, 自分のリポジトリをインナーソース化する場合

の手順，工数計上方法も明記し，エンジニアが迷わず動けるようにした．このようなガイドライン整備や評価・計上ルールの確立により，コロナ禍でオンライン会議が増える中でもインナーソース取り組みが社内に徐々に浸透していった．

6. **インナーソース導入後のエンジニア行動・意識変化**

   インナーソースを導入した直後は，特定のツールやプロダクトでテスト的に取り組んでいた段階だったが，徐々に他のチームも自分たちのプロダクトをオープンにしてコントリビュートを受け入れ始めた．エンジニアリングマネージャーどうしが横でつながり，計画にも明記し，ガイドラインを整備したことで，「やってもいい」という雰囲気から「やろう」という積極的な意識変化が生まれたように感じている．

7. **評価軸の明確化とモチベーションへの影響**

   評価軸を明確化する前は，他チームのコードベースに手を入れても「それは自分の評価につながるのか」という不透明さがあった．しかし，インナーソースガイドラインと計上ルールが整い，他チームへの貢献が正当に評価されるようになると，エンジニアは別プロダクトへのコントリビュートを積極的に行いやすくなった．「やっていい」だけでなく「やれば評価される」という安心感がコラボレーションを後押しした．結果的に，コントリビュート数の可視化と評価方法の統一がエンジニアのモチベーションを底上げした．

8. **コラボレーション事例と社内効率化**

   音声読み上げツールのような例に続き，インフラ関連のスクリプトを A チームが作り，B チームが改良するといったケースが生まれた．これまでは各チーム内で閉じていたものがインナーソースでオープンになることで汎用的なツールへと発展していく．こうしたコラボレーションによって社内での車輪の再発明が減り，同様の問題をすでに他チームが解決している場合はそのコードを流用・改良することで，効率が向上した．これらはインナーソース文化を根付かせ，評価体系を整備した結果だと考えている．

9. **将来の展望と期待される発展**

   今後は，さらに多くのプロダクトがインナーソース化され，他チームからの貢献を自然に受け入れる文化が成熟することを期待している．また，新規入社エンジニアがオープンな社内コードから学びやすくなり，オンボーディングがスムーズになる効果も見込んでいる．こうした循環が社内エンジニアリング文化のさらなる強化につながると考えている．

10. **コスト計上と工数管理ルールの運用実態**

    社内には工数を計上する Web アプリがあり，何時間どのプロダクトに関わったかを記録する．インナーソース下では他チームプロダクトへのコントリビュートが正当に評価されるよう，この計上方法をエンジニアリングマネージャー間で合意した．オーダー番号でプロダクトや機能単位を識別し，30 分単位で作業時間を計上する運用に落ち着いている．現場としては細かく記録することに面倒を感じることもあるが，1 日単位より細かな 30 分単位で計上する運用が現状定着しつつある．たとえばエンジニアが，1 日の終わりに「このくらいの時間をかけた」とまとめて記録することも多く，場合によっては翌日にまとめて計上することもある．オーダー番号は機能や取り組み単位で割り当てられ，たとえばクラウド移行が一つのオーダー番号になっている．これらのツールは SaaS であり，インナーソースに関わる工数もこのシステムで管理している．

11. **評価制度とエンジニア行動計画**

情報通信業Ｆ社では半期に１回の評価サイクルがあり，３ヶ月ごとに計画修正を行う．その場で「こういうコントリビューションをする」という計画を上長（エンジニアリングマネージャー）と擦り合わせ，評価時にその実績を効果ベースで評価する仕組みだ．インナーソースガイドラインはエンジニアにとっての行動指針であり，エンジニアリングマネージャー側にも「こういうことが推奨される」という認識を与えるものである．会社の方向性として，このガイドラインが「やっていい」「奨励されている」という合意形成を可能にしている．

12. **上層部への共有状況と反応**

インナーソース施策はシステム部門内の部長クラスまでは伝わっており，社長である人物 F もシステム側出身であるため，この取り組み自体を知っていると考えている．上長に「こうしていきたい」と話を上げた際に大きな異論はなく，比較的スムーズに通っている状況だ．今まで特に強い反対意見や大きな議論は起きていない．

13. **インナーソース推進の目的と上長側の認識**

情報通信業 F 社はエンジニアが「楽しい」と感じる文化を重視しており，他部門に自由にコミットしていいという風土を大切にしている．上長層は競争優位性やリソース効率化よりも，まずエンジニアが楽しみながら活発なコラボレーションを行うことを重視している．インナーソースを通じてエンジニアの満足度を高め，組織的なエネルギーを引き出すことが現段階での主な狙いとなっている．

14. **エンジニア個人の魅力・モチベーション源**

エンジニアにとってインナーソースの魅力は車輪の再発明防止による時間削減効果や他チームとの議論・ソースコード閲覧など技術的好奇心が満たせる点だ．技術好きなエンジニアにとっては他チームとのコラボはモチベーションになる．モチベーション向上やリードタイム削減もうれしいが，コラボレーション機会を増やすこと自体が有益だ．

15. **サイロ状態と透明性向上**

以前はグループ内に閉じていたが，インナーソースにより外向きに話が展開される雰囲気が生まれた．「インナーソース」という言葉が自分の知らないところで使われていることがうれしく，透明性やオープンさが増している．関心のあるエンジニアにとってはカルチャー改革的な感覚もある．ツール類もグループごとから全社で見られる形に変わり，グループ間の壁が薄くなった．

16. **リポジトリ構造・権限管理と増える事例**

最初から全社一つのオーガナイゼーション内でリポジトリが参照可能だったが，当初は見えているだけで行動がなかった．インナーソースで正しいコラボレーションのしかたがわかり，CONTRIBUTING.md（貢献のしかたを記した簡易ガイド)を記載すればインナーソースポータルがクロールして掲載する仕組みができた．実験プロジェクト一つだけだったところから，今では約 19 個のインナーソースプロジェクトが存在する．事例が増えたことで分かりやすくなり，参加が促進された．

17. **プロダクションコード適用の難しさと障壁**

19 個中，本番で使えるコードは 5 個程度しかなく，多くは便利ツール系だ．プロダクション向けコードを増やしたいがなかなか広がらない．技術スタックがバラバラで，時間不足も要因となっている．他社でも同様だろうが，新しい取り組みに割く余裕が不足している．

18. **技術スタックの多様性と標準化の難題**
Python や TypeScript など技術選定がチーム任せでバラバラなため，統一が難しい．ボトムアップで現場が使いやすい技術を選び続けてきた結果，今は自由さが足かせとなっている．標準プロジェクトやプロダクト導入に取り組もうとしているが，コンフィグが難しいなど課題が残る．今後統合する方針は検討中だが，まだ確定はしていない．
19. **時間不足と業務優先度の課題**
時間不足は大きな障壁だ．レガシーコードが多くテストコマンドすら整っていないリポジトリもある．ホスト側も本番運用などで手いっぱいで，ゲスト側も業務優先で動きづらい．日々の業務を優先せざるを得ず，新規取り組みに時間を割きにくい実情がある．
20. **オーガナイゼーション外プロジェクト評価への驚き**
現状，業務時間外にリポジトリを触ることは禁じられている．もし許可されれば，業務外でのコーディングを楽しむエンジニアも増える可能性がある．
21. **オープンソースへの関わり方**
基本的には消費がメインで，コントリビュートはほとんど行っていない．学生時代に参加経験がある人もいるが，社内としては消費中心．エンジニアの中には積極的なメンバーもいるが，個人時間で行っているケースが多い．
22. **個人学習成果の社内還元と評価認識**
個人時間でスキルアップした成果を社内で形にすれば評価対象になるが，個人で公開しただけでは会社の貢献として見なされにくい．社内への還元が評価のポイントとなっている．
23. **評価制度の概要と変遷**
現在は MBO に近い形をとっている．評価制度は約 1 年前に改善して，評価軸が三つから五つになった．半期ごとに計画と評価を行い，目標を管理する仕組みになっている．五つの軸を持つ指標を用いている．それぞれ「イノベーション」や，「ナンバーワン」などの，独自性や革新性を示す指標になっている． これによってどの目標がどの方向性を持っているかが分かりやすくなり，評価時に効果ベースで判断できるようになっている．大きな目標（開発やクラウド移行など）を設定し，その中で個人が達成すべき内容を明確にすることで，行動指標と成果指標を組み合わせた評価を行っている．
24. **個人・チーム目標の設定と振り返りプロセス**
半期ごとに計画と評価を行い，3 ヶ月ごとに計画修正を行う．計画時にエンジニア個人が「こういうコントリビューションをする」とエンジニアリングマネージャーとすり合わせ，評価時に効果ベースで成果を振り返る．このプロセスでは，指標 A 軸を用いて目標がどの程度イノベーティブで独自性があるかを示し，達成度合いを判断する．エンジニアは開発やクラウド移行，障害再発防止などの目標を立て，期末に振り返る．
25. **給与への反映とポイント制評価**
評価はポイント制で行っている．1 ポイント単位で給与へ反映される仕組みがある．評価制度はエンジニアだけでなく全職種共通で，同じフォーマットで評価する．ポイントを積み上げ，ランクアップしていくと責務範囲や影響範囲が増え，より上位のロール（より上位のリーダークラス，つまり組織を牽引するエース的存在）への挑戦が可能になる．インナーソースの活用もこれら目標の一部となり得る．

26. **ランクと役割の変化**

人物Bは，より上位のクラスへのチャレンジを行おうとしている．リーダークラスはエンジニア部門全体を視野に入れ，コミュニケーションや組織連携を最適化する役割が期待される．インナーソースは部門間コラボレーションを促進する施策として，L クラスを目指すエンジニアにとって有用な手段となる可能性がある．

27. **表彰制度とインセンティブ**

半期に 1 回，創意工夫賞と社長賞が用意されている．創意工夫賞は個人の活動に対して，社長賞は大きなビジネスインパクトをもたらした案件などに対して与えられる．創意工夫賞は三つのランクがあり，それぞれ賞金が出る．インナーソース推進や技術的な改善活動で金賞が取得された実績もある．社長賞はさらに大きな成果を出した案件が対象となり，上長の推薦も必要なのでハードルが高い．

28. **創意工夫賞受賞経験と受賞時の感想**

人物Bは創意工夫賞を3回ほど受賞しており，社内で最も多く受賞している人物だ．金賞を取ると全社員の前でインタビューがあり，特別感がある．受賞すると「通ったんだ」という素直なうれしさがあるし，他にも応募して通らない人がいる中で選ばれたことは大きな自信になる．半期に一度開かれる全社会議（情報通信業 A 会議）で表彰され，全社員が見る中でインタビューを受けるため，社内的なモチベーション向上につながる．

29. **目標達成状況とインナーソース活動の評価**

インナーソース関連の目標は前回全て達成されている．グループ目標と連動させてクリアしており，これが評価時にプラス要因となる．インナーソース活動が実績として認められることで，評価制度とエンジニアの行動が結び付いている．結果的に技術的好奇心や社内連携が評価指標に組み込まれ，エンジニアの行動を促進する要素になっている．

30. **チーム単位のインセンティブと表彰制度**

個人へのインセンティブはポイント制や創意工夫賞などで明確だが，チーム単位のインセンティブとしては直接的な仕組みはない．ただ，社長賞で案件全体が表彰されることがあり，その場合は 30～40 人規模で表彰対象となることもある．これによって，チームや複数部署が関わった大規模な取り組みが評価される機会がある．

31. **インナーソースとプラットフォームエンジニアリングへの展開**

今後は別のグループであるインフラグループを巻き込み，プラットフォーム提供部署と連動して業務で使える仕組みをインナーソース化したいと考えている．現在 1 人新たな推進メンバーが加わっており，その協力を得てプラットフォームエンジニアリングとインナーソースを交差させる取り組みを模索している．用語や文化的認知は進んできたので，次は実務に直結した成果が求められる段階だと認識している．業務へ組み込むことで「やらざるを得ない」状況を作り，自然な広がりを狙っている．

32. **エンジニアリングマネージャーへの働きかけと合意形成**

エンジニアリングマネージャーどうしは同様の課題や悩みを共有していたため，「インナーソースは良さそうだ」という共通認識を得やすかった．当初はリソース効率化をきっかけに「おもしろそうだからやってみよう」と合意できた．周囲への説明は最小限で済み，人物 B による実例提示や「これをやる」という明確な宣言がエンジニアリングマネージャーたちの納得を得るのに有効だった．近い距離のエン

ジニアリングマネージャーはすぐに納得し，少し離れたインフラグループなどには一度説明する機会を設けた程度で済んだ．繰り返しの説明より実例提示と明確な推進方針が大きな効果を生んだ．

33. **リードタイム削減とセルフサービス化の事例**

インナーソースによって，依頼から対応までのリードタイムが短縮された例がある．ログインシステム関連の設定ファイル（ドメイン制限などの情報定義ファイルなど）をメンテナ側が全て登録していたが，インナーソース化により利用者側が自ら変更や動作確認の Pull Request を出せるようになった．これによってメンテナ側の作業負荷が軽減され，以前は数日から 1 週間かかる作業が効率化された．コードそのものというより，コンフィグ的要素をセルフサービス化して分散管理することで，運用負担が多いチームへのインナーソースの動機づけにもなった．

34. **本格的なコントリビュート不足と今後の課題**

インタビュー時点では，セルフサービス化や設定ファイルの変更といった軽度の貢献がメインであり，本質的な機能強化やプロダクション向けコード拡張といった深い貢献はまだ少ない．共通ライブラリの整備などは道半ばであり，今後より本格的な貢献を促していく必要がある．

35. **コントリビュート層の特徴とシニア・マネージャー層の関心**

コントリビューターは若手が多い印象があり，彼らはインナーソース化による効率化に前向きで安心感を得ている．一方で，シニアやマネージャー層は「インナーソースで効率的になる」と考えている可能性があるが，インタビュー時点では抵抗はあまり表面化していない．ビジネスに直結する領域でインナーソースが本格化すれば，部下のリソースを他チームが使うことに対する懸念や，さらに大きな摩擦が生じる可能性はあるが，まだその段階にはいたっていない．

36. **会計・契約面での懸念と規模の影響**

会計や契約面での問題は今のところ発生していない．コスト管理程度の話があるくらいだ．（質問の「外部からの納品物扱いなど法的な面では，グループ企業間で法人が異なると，ただでのやりとりが利益供与になる懸念がある．また，国をまたぐと移転価格税制が関係し，どの程度の価値をどこに付けるかといった問題が大きくなる．」という問いに対しては）情報通信業 F 社は 1 社 500 人規模なので，そういった問題が少ない状況にある．と答えた．

37. **インナーソース活用イベントとコントリビューション促進**

人物 B がコントリビューター候補を 10 人ほど集めて社内ソースにコントリビュートさせる企画を考えている．Docker 導入やドキュメント修正など，小さな一歩から始めて他部門のレビュー方法を知ることで，次へのハードルが下がる．こうした初動をサポートするイベントは有益であり，最初のハードルを下げて，エンジニアが新しい領域や他部門へのコントリビュートに踏み出しやすくする取り組みは，インナーソース文化定着に効果的だと考えている．